\documentclass{article}
\usepackage{emulateapj,apjfonts}

\usepackage{psfig}

\newcommand{\etal}{et~al.\ }

\def\ts{\thinspace}
\def\gapprox{$_>\atop{^\sim}$} 
\def\lapprox{$_<\atop{^\sim}$}

\newdimen\sa  \def\sd{\sa=.1em \ifmmode $\rlap{.}$''$\kern -\sa$
                               \else \rlap{.}$''$\kern -\sa\fi}

\reversemarginpar

\begin{document}

\lefthead{Structure and Formation of Elliptical and Spheroidal Galaxies}

\righthead{Kormendy et al.}

\centerline{\null}\vskip -87pt

\title{Structure and Formation of Elliptical and Spheroidal Galaxies\altaffilmark{1,2,3}}

\author{
John Kormendy\altaffilmark{4,5,6}, 
David B.~Fisher\altaffilmark{4,5,6},
Mark E.~Cornell\altaffilmark{4}, and
Ralf Bender\altaffilmark{4,5,6}
}

\altaffiltext{1}{Based on observations made with the NASA/ESA {\it Hubble Space 
                 Telescope}, obtained from the Data Archive at STScI,
                 which is operated by AURA, Inc., under NASA 
                 contract NAS 5-26555.  These observations are associated with program
                 numbers 5999, 6107, 6357, 6844, 7868, 8686, 9401, and 10558.} 

\altaffiltext{2}{Based on observations obtained at the Canada-France-Hawaii 
                 Telescope (CFHT), which is operated by the National Research 
                 Council of Canada, the Institut National des Sciences de l'Univers
                 of the Centre National de la Recherche Scientifique of France, 
                 and the University of Hawaii.} 

\altaffiltext{3}{Based in part on observations obtained with the Hobby-Eberly Telescope,
                 which is a joint project of the University of Texas at Austin, the 
                 Pennsylvania State University, Stanford University, 
                 Ludwig-Maximilians-Universit\"at M\"unchen, and 
                 Georg-August-Universit\"at G\"ottingen.} 

\altaffiltext{4}{Department of Astronomy, University of Texas, Austin,
                 Texas 78712; kormendy@astro.as.utexas.edu, dbfisher@astro.as.utexas.edu,
                 cornell@astro.as.utexas.edu}

\altaffiltext{5}{Universit\"ats-Sternwarte, Scheinerstrasse 1,
                 M\"unchen D-81679, Germany; bender@usm.uni-muenchen.de}

\altaffiltext{6}{Max-Planck-Institut f\"ur Extraterrestrische Physik,
                 Giessenbachstrasse, D-85748 Garching-bei-M\"unchen, Germany; 
                 bender@mpe.mpg.de}

\pretolerance=15000  \tolerance=15000

\begin{abstract} 

      New surface photometry of all known elliptical galaxies in the Virgo cluster is combined with 
published data to derive composite profiles of brightness, ellipticity, position angle, isophote shape, 
and color over large radius ranges.  These provide enough leverage to show that S\'ersic 
$\log{I} \propto r^{1/n}$ functions fit the brightness profiles $I(r)$ of nearly all ellipticals 
remarkably well over large dynamic ranges.  Therefore we can confidently identify departures from 
these profiles that are diagnostic of galaxy formation.  Two kinds of departures are seen at small radii.  
All 10 of our ellipticals with total absolute magnitudes $M_{VT} \leq -21.66$ have cuspy cores -- 
``missing light''~-- at small radii.  Cores are well known and naturally scoured by binary black holes 
formed in dissipationless (``dry'') mergers.  All 17 ellipticals with $-21.54 \leq M_{VT} \leq -15.53$ 
do not have cores.  We find a new distinct component in these galaxies: All coreless ellipticals in 
our sample have extra light at the center above the inward extrapolation of the outer S\'ersic profile.  
In large ellipticals, the excess light is spatially resolved and resembles the the central components 
predicted in numerical simulations of mergers of galaxies that contain gas.  In the simulations, the gas 
dissipates, falls toward the center, undergoes a starburst, and builds a compact stellar component that, 
as in our observations, is distinct from the S\'ersic-function main body of the elliptical.  But ellipticals 
with extra light also contain supermassive black holes.  We suggest that the starburst has swamped core 
scouring by binary black holes.  That is, we interpret extra light components as a signature of formation 
in dissipative (``wet'') mergers.

     Besides extra light, we find three new aspects to the (``E{\ts}--{\ts}E'') dichotomy into two types
of elliptical galaxies.  Core galaxies are known to be slowly rotating, to have relatively anisotropic 
velocity distributions, and to have boxy isophotes.  We show that they have S\'ersic indices $n > 4$
uncorrelated with $M_{VT}$.  They also are $\alpha$-element enhanced, implying short star formation 
timescales.  And their stellar populations have a variety of ages but mostly are very old.  Extra light 
ellipticals generally rotate rapidly, are more isotropic than core Es, and have disky isophotes.  We show 
that they have $n \simeq 3 \pm 1$ almost uncorrelated with $M_{VT}$ and younger and less 
$\alpha$-enhanced stellar populations.  These are new clues to galaxy formation.  We suggest that extra 
light ellipticals got their low S\'ersic indices by forming in relatively few binary mergers, whereas 
giant ellipticals have $n > 4$ because they formed in larger numbers of mergers of more galaxies at
once plus later heating during hierarchical~clustering. 

      We confirm that core Es contain X-ray-emitting gas whereas extra light Es generally do not.
This leads us to suggest why the E{\ts}--{\ts}E dichotomy arose.  If AGN energy feedback requires
a ``working surface'' of hot gas, then this is present in core galaxies but absent in extra light galaxies.
We suggest that AGN energy feedback is a strong function of galaxy mass: it is weak enough in 
small Es not to prevent merger starbursts, but strong enough in giant Es and their progenitors to
make dry mergers dry and to protect old stellar populations from late star formation.

      Finally, we verify that there is a strong dichotomy between elliptical and
spheroidal galaxies.  Their properties are consistent with our understanding of their 
different formation processes:~mergers for ellipticals and conversion of late-type 
galaxies into spheroidals by environmental effects and by energy feedback from supernovae.

      In an Appendix, we develop machinery to get realistic error estimates for S\'ersic parameters even
when they are strongly coupled.  And we discuss photometric dynamic ranges necessary to get robust results 
from S\'ersic fits.

\end{abstract}


\section{Introduction}

\pretolerance=15000  \tolerance=15000

\centerline{\null} \vskip -15pt

      This is the first of a series of papers in which we study elliptical galaxies 
by combining new surface photometry with published data to construct composite 
brightness profiles over large radius ranges.  This approach has two strengths.
Combining data from many sources allows us to reduce systematic errors arising 
(e.{\ts}g.)~from imperfect sky subtraction.  Having accurate profiles over large 
radius ranges provides leverage necessary for reliable conclusions about 
profile shapes and what they tell us about galaxy formation.

      What is at stake?  We have a formation paradigm.  We believe that galaxies grow 
as part of the hierarchical clustering that makes all structure in the Universe.  
Ellipticals form  in violent galaxy mergers that often include gas dissipation 
and star formation (Toomre 1977; White \& Rees 1978; Joseph \& Wright 1985; Schweizer 1989; Kauffmann
\etal 1993; Steinmetz \& Navarro 2002).  What questions remain unanswered?

      We focus on two well-known dichotomies.  We confirm that there is a physical difference
between elliptical and spheroidal galaxies.  This has been much criticised in recent literature.  
With photometry over large dynamic ranges, we find that elliptical and spheroidal galaxies have
very different parameter correlations.  This result is consistent with our 
understanding of their differenct formation processes.  Spheroidals are not low-luminosity 
ellipticals but rather are defunct late-type galaxies transformed by internal and environmental 
processes.  A second dichotomy is the main focus of this paper.
Why are there two kinds of elliptical galaxies?  We suggest an explanation~-- that the 
last major mergers that determined the present-day structure either did or did not involve cold
gas dissipation and starbursts.

\begin{figure*}[b] 

\vskip 3.9truein
 
\includegraphics{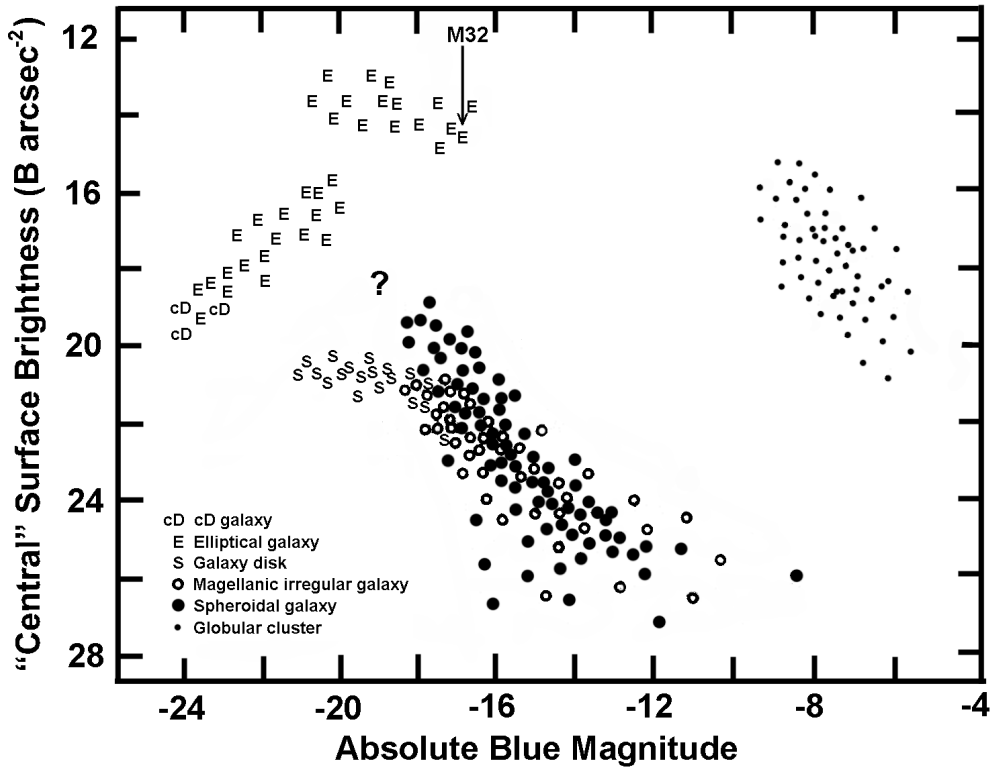}

\figcaption[]
{Schematic illustration of the dichotomies discussed in this paper.  The figure 
sketches the correlation between total absolute magnitude and central surface
brightness (for spheroidal and irregular galaxies, galaxy disks, and globular
clusters) or the highest surface brightness resolved by the {\it Hubble Space 
Telescope\/} (for elliptical and cD galaxies).  Surface brightnesses apply to the 
main bodies of the galaxies; that is, nuclear star clusters and active galactic
nuclei are omitted.  This figure is adapted from Binggeli (1994) but with the 
dichotomy between ``core'' and ``power law'' ellipticals -- i.{\ts}e., the discontinuity
in E points at $M_B \sim -20.5$ -- added from Faber et al.~(1997).  M{\ts}32 is one 
of the lowest-luminosity true ellipticals; the arrow points from the maximum 
surface brightness observed at a distance of 0.8 Mpc to the lower limit that would be 
observed if the galaxy were moved to the Virgo cluster.  M{\ts}32 resembles the faintest 
ellipticals in Virgo.  The distribution of Sph and S$+$Im galaxies is disjoint from 
that of ellipticals.  Sph and S$+$Im galaxies have similar global parameters at low 
luminosities, but the most luminous spheroidals ``peel off'' of the distribution of 
late-type galaxies toward higher surface brightness.  Spheroidals with $M_B$ \lapprox
\ts$-18$ are rare, so the degree to which the Sph sequence approaches the E sequence
is poorly known ({\it question mark\/}).  Note: Binggeli (1994) and some other authors
call spheroidal galaxies ``dwarf ellipticals''~(dEs).  We do not do this, because
correlations like those in this figure and in Figures 34{\ts}--{\ts}38 and 41, as well 
as the considerations discussed in \S\ts2.1 and \S\ts8, persuade us that they are not 
small ellipticals but rather are physically related to late-type galaxies.
}
\end{figure*}

\section{Two Dichotomies}

\subsection{Elliptical Versus Spheroidal Galaxies}

       In a pioneering paper, Wirth \& Gallagher (1984) suggested that compact
dwarf ellipticals like M{\ts}32 and not -- as previously thought -- diffuse 
``spheroidal'' dwarfs like NGC 205 are the extension to low luminosities 
of the family of giant ellipticals.  This was based on the identification 
of several free-flying M{\ts}32 analogs, implying that the compactness of the 
best known dwarf Es -- M{\ts}32, NGC 4486B, and NGC 5846A (Faber 1973) -- 
is not due only to tidal pruning by their giant galaxy neighbors.  Wirth and 
Gallagher hypothesized that ellipticals and spheroidals form disjoint families
overlapping for $-15$ \gapprox \ts$M_B$ \gapprox \ts$-18$ but differing in mean 
surface brightness at $M_B = -15$ ``by nearly two orders of magnitude''.  This
implied that the luminosity function of true ellipticals is bounded and that 
M{\ts}32 is one of the faintest examples.  The latter result was confirmed for 
the Virgo cluster by Sandage et al.~(1985a, b) and by Binggeli et al.~(1988).

      Kormendy (1985a,{\thinspace}b, 1987b) used the high spatial resolution of 
the Canada-France-Hawaii Telescope to obtain surface photometry of the cores of
bulges and elliptical galaxies.  He showed in larger
galaxy samples that ellipticals form a well defined sequence in parameter
space from cD galaxies to dwarfs like M{\ts}32.  Lower-luminosity
ellipticals are more compact; they have smaller core radii and higher central
surface brightnesses.  These are projections of the core fundamental plane 
correlations (Lauer 1985b).  Kormendy found a clearcut dichotomy between E and
Sph galaxies.  Fainter spheroidals have lower central surface brightnesses.
In fact, spheroidals have almost the same parameter correlations as 
spiral-galaxy disks and Magellanic irregular galaxies.  These results are 
most clearly seen in correlations between central properties, but they are
also evident in global properties (Kormendy 1987b; Binggeli \& Cameron 1991; 
Bender \etal 1992, 1993).  The brightest spheroidals ``peel off'' 
of the correlations for late-type galaxies and approach the E sequence, but they 
are rare, and the two sequences remain distinct (Kormendy \& Bender 1994).  
The E{\ts}--{\ts}Sph dichotomy is illustrated in Figure 1.

      Kormendy (1985b, 1987b) concluded that E and Sph galaxies are distinct types
of stellar systems with different formation processes.  Spheroidals are 
physically unrelated to ellipticals; Figure 1 hints that they are related to 
S+Im galaxies.  They may be late-type galaxies that lost their gas or 
processed it all into stars.  Relevant evolution processes include supernova-driven 
energy feedback (Saito 1979a, b; Dekel \& Silk 1986; Navarro \etal 1996; 
Klypin et al.~1999; Veilleux \etal 2005); ram-pressure gas stripping (Lin \& Faber 1983; 
Kormendy 1987b; van Zee \etal 2004a, b), stochastic starbursts (Gerola 
et al.~1980), and galaxy harrassment (Moore et al.~1996, 1998).  

\begin{figure*}[t] 

\vskip 2.2truein

\includegraphics{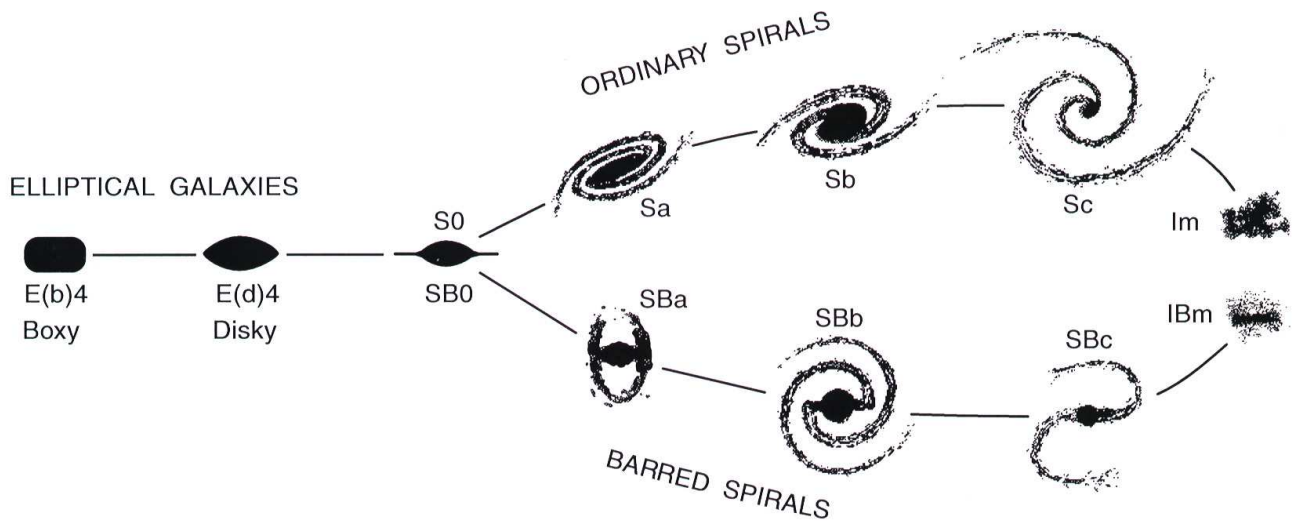}

\figcaption[]
{Revision of Hubble's (1936) morphological classification scheme proposed by Kormendy 
\& Bender (1996).  Here ellipticals are not classified by apparent flattening, which
in large part encodes our viewing geometry.  Rather, they are classified according to 
whether they show boxy or disky isophote distortions.  This is also the dichotomy 
between ellipticals that do and do not have cuspy cores (Fig.~1); it is the one
summarized in \S\ts2.2.  Boxy-core galaxies tend to rotate less and to be more 
dominated by velosity dispersion anisotropies than are disky-coreless galaxies.  
Therefore the revised classification orders galaxies along the Hubble sequence by 
physically fundamental properties, i.{\ts}e., by the increasing importance from left 
to right of ordered rotation as compared with random internal velocities.}
\end{figure*}

      Additional differences diagnostic of formation processes include luminosity 
functions (\S\ts8) and rotation properties.   Faint Es are rotationally supported, and
some Sph galaxies are, also 
(Pedraz \etal 2002;
van Zee \etal 2004b),
but many are non-rotating and anisotropic 
(Bender \& Nieto 1990;
Bender \etal 1991;
Held \etal 1992;
Geha \etal 2002, 2003, 2006;
Thomas \etal 2003, 2006).
Possible explanations include galaxy harrassment (Gonz\'alez-Garc\'\i a \etal 2005)
and rapid expansion after baryonic blowout 
(Dekel\ts\&{\ts}Silk 1986;{\ts}Hensler{\ts}et{\ts}al.{\ts}2004).  

      The dichotomy has been challenged by Jerjen \& Binggeli (1997); Graham \& Guzm\'an 
(2003); Graham \etal (2003); Trujillo \etal (2004), and Gavazzi \etal (2005).  They argue 
that Sph parameters are continuous with those of low-luminosity Es, while bright Es 
deviate from these correlations only because scouring by binary supermassive black holes (BHs)
excavates cores.  Another argument is that the correlation between brightness profile shape 
and galaxy luminosity 
is continuous from Es through Sphs.  Recently, Ferrarese et al.~(2006a)\footnote{Ferrarese 
et al. (2006a) argue against both dichotomies discussed in this paper.
We discuss our procedures and analysis in the main text and save a 
comparison of our differences with Ferrarese \etal (2006a) for Appendix B.  
Our paper and theirs are independent looks at the 
same science using similar analysis techniques.} 
argue forcefully against the E -- Sph dichotomy based on their {\it HST\/} 
photometry of Virgo cluster galaxies.~For these reasons, and because we need
to settle the controversy in order to define our sample of ellipticals, we return
to the issue in \S\ts8.  Because the fundamental plane of ellipticals is so thin 
(\S\ts3), we will find that E and Sph galaxies are cleanly distinguishable.

      At stake are the different formation mechanisms of small ellipticals and
big spheroidals.~We have good reasons to believe that ellipticals form via
galaxy mergers.  We also know that many spheroidal satellites of 
our Galaxy are defunct irregulars.  Their intermediate-age stellar populations 
(see Da Costa 1994 for a review) tell us that dIm galaxies have gradually 
converted themselves into dSph galaxies via episodic star formation. 
For example, the Carina dSph is made up of two stellar populations:  
15\ts--\ts20\ts\% of the stars are 12\ts--\ts15 Gy old, but $> 80$\ts\% of the 
stars are 6\ts--\ts8 Gy old.  Kormendy \& Bender (1994) emphasize that there must 
have been gas $\sim 7$ Gyr ago to make these stars.  Gas-rich, star-forming 
dwarfs are Magellanic irregulars.  We know less about the formation of spheroidals
in Virgo, although additional signs that Sph galaxies are related to late-type
galaxies are observations of spiral structure (Jerjen \etal 2000) and star 
formation (Lisker \etal 2006).  It is natural to expect that galaxy harrassment 
would convert larger late-type galaxies into Sphs in Virgo than in the Local Group.
Moreover, one effect is to concentrate gas toward the center before star formation 
happens (Moore et al.~1996, 1998).  This provides a natural explanation for why the
Sph sequence peels off the S$+$Im sequence at high galaxy luminosities (Figure 1).  

      If E and Sph galaxies formed a continuous family, it would be
surprising to conclude that different formation processes dominated
at high and low luminosities, with major mergers making ellipticals but not
spheroidals (\S\ts13, Tremaine 1981).

\subsection{The Dichotomy Into Two Kinds of Elliptical Galaxies}

      There are two kinds of elliptical~galaxies: (1) Normal- and low-luminosity 
ellipticals rotate rapidly; they are relatively isotropic, oblate-spheroidal, and 
flattened (E3); they are coreless, and they have disky-distorted isophotes.  Most bulges 
of disk galaxies are like low-luminosity ellipticals.  In contrast, (2) giant ellipticals 
are essentially non-rotating; they are anisotropic and triaxial; they are less 
flattened (E1.5); they have cuspy cores, and they have boxy-distorted isophotes. These 
results are established in
Davies \etal (1983);
Kormendy \& Illingworth (1982);
Bender (1987, 1988a);
Bender, D\"obereiner, \& M\"ollenhoff (1987);
Bender et al.~(1989); 
Nieto \& Bender (1989);
Nieto \etal (1991);
Kormendy \etal (1994, 1996a); 
Lauer \etal (1995);
Kormendy \& Bender (1996);
Tremblay \& Merritt (1996);
Gebhardt \etal (1996); 
Faber et~al.\ (1997); 
Rest \etal (2001),
Ravindranath \etal (2001);
Lauer \etal (2005, 2007b);
Emsellem \etal (2007), and
Cappellari \etal (2007).
The differences between the two kinds of ellipticals are fundamental.  They motivated 
Kormendy \& Bender (1996) to suggest that the Hubble sequence be revised (Figure 2)
so that rotation increases in importance and random motions decrease in importance 
along the Hubble sequence from boxy Es through Scs.  The ``E{\ts}--{\ts}E dichotomy'' is 
the main subject of this paper.

\section{Regularity in the Structure of Elliptical Galaxies}

      Why do we think that surface brightness profiles can tell us about the formation
of elliptical galaxies?

      Our picture of hierarchical clustering implies that different galaxies are the 
products of different merger histories in which different progenitor morphologies and 
encounter geometries produce a variety of results.  It is remarkable that the remnants
of such varied mergers show regularity that we can interpret.  In fact, ellipticals show 
surprising regularity in structure.  Interpreting these regularities -- and departures 
from them -- has been a profitable way to study galaxy formation.
      
      A well known example is the ``fundamental plane'' of elliptical galaxies.  Their 
half-light radii\ts$r_e$, effective surface brightnesses\ts$\mu(r_e)$,~and velocity
dispersions $\sigma$ interior to $r_e$ lie in a tilted plane in parameter space 
(Djorgovski \& Davis 1987; 
Faber \etal 1987;
Dressler \etal 1987; 
Djorgovski \etal 1988; 
Bender \etal 1992, 1993), 
$ r_e \propto \sigma^{1.4 \pm 0.15} ~ I_e^{-0.9 \pm 0.1}$,
whose scatter is similar to the parameter measurement errors (Saglia \etal 1993; 
J\o rgensen \etal 1996).  This is a consequence of the virial theorem and the fact that 
ellipticals are nearly homologous over a wide range in luminosities $L$.  Slow variations 
with $L$ in density profiles, velocity structure, and mass-to-light ratio 
$M/L \propto L^{0.2}$ combine to give the fundamental plane slopes that are slightly 
different from the virial theorem prediction, $r_e \propto \sigma^2 ~ I_e^{-1}$, 
for exactly homologous galaxies.  

    The part of the near-homology that concerns us here is the slow variation of profile
shape with $L$.  Kormendy (1980), Michard (1985), and Schombert (1986, 1987) found that the
de Vaucouleurs (1948) $r^{1/4}$ law fits ellipticals best at $M_B \simeq -20.2$
($H_0 = 70$ km s$^{-1}$ Mpc$^{-1}$; Komatsu et al.~2008).  More (less) luminous ellipticals
have brighter (fainter) outer profiles than the extrapolation of the best-fitting $r^{1/4}$
law.  Schombert (1986, 1987) provides a nonparametric illustration by deriving average
profiles for ellipticals binned by luminosity.  {\it Nothing guarantees 
that any simple parametrization of profile variations describes the results of 
mergers and dissipative starbursts.}~However, Nature proves to be extraordinarily~kind.  
{\it The theme of this paper is that S\'ersic (1968) $\log{I(r)} \propto r^{1/n}$ functions 
fit most ellipticals remarkably well.  The result is that local departures from the fits and 
correlations involving the fit parameters provide new insights into galaxy formation.}

    Caon \etal (1993) were the first to prove that $r^{1/n}$ 
functions fit ellipticals better than~do~$r^{1/4}$ laws.  This is not a surprise~--
$r^{1/n}$ laws have three parameters while $r^{1/4}$ laws have two.  Kormendy 
(1980, 1982) and Kormendy \& Djorgovski (1989) emphasized that elliptical galaxy profiles are
close enough to $I \propto r^{-2}$ power laws -- which have only one parameter -- so that accurate 
photometry over a large radius range is required to derive even two parameters.  
Three-parameter fits can involve so much parameter coupling that the results are useless.
This was true in the era of photographic photometry (see Fig.~12 in Kormendy 1982 for an example).
It is no longer true, because CCDs provide more accurate photometry and 
because the {\it Hubble Space Telescope\/} ({\it HST\/}) has greatly increased the dynamic 
range by providing PSF-corrected  photometry inward to radii $r \simeq 0\sd1$.  Improved 
data now support three-parameter fits, and Caon and collaborators argue convincingly that 
the S\'ersic index $n$ has physical meaning.  For example, $n$ correlates with the effective 
radius $r_e$ and total absolute magnitude $M_B$ of the elliptical or bulge.  
These correlations have been confirmed by 
D'Onofrio \etal (1994); 
Graham et al.~(1996); 
Graham \& Colless (1997); 
Graham (2001); 
Trujillo \etal (2001, 2002), 
Ferrarese \etal (2006a),
and others.

      This rapid progress slowed down as the easy results enabled by
CCDs were derived.  Now, however, an important iteration in quality is within reach.  
The shortcoming of most CCD photometry is limited field of view.  Many published
profiles do not reach large radii and may be affected by sky subtraction errors. 
However, images are now available from a variety of wide-field, mosaic detectors and
surveys such as the 2MASS survey (Jarrett \etal 2003; Skrutskie et al.~2006) and the Sloan
Digital Sky Survey (Stoughton et al.~2002; Abazajian et al.~2003, 2004, 2005).

      Our aim is to exploit the significant improvements in the dynamic
range of brightness profiles that can be gained by combining data from
a variety of telescopes.  Intercomparison of these data allows us to
reduce systematic errors.  Confirming previous work, we find that
S\'ersic functions fit many ellipticals over large radius ranges.  As a
result, we can derive more accurate values of $r_e$, $\mu_e$, and
S\'ersic index~$n$.  This allows us to improve the derivation of parameter correlations.
Most important, the robust detection of S\'ersic profiles over large radius ranges
allows us reliably to see departures from these profiles that are diagnostic of galaxy 
formation mechanisms.

      One purpose of this paper is to expand on a result summarized in \S\ts4.2.  We 
enlarge the sample on which it is based by measuring all known elliptical galaxies in the 
Virgo cluster as listed in Binggeli \etal (1985) and as confirmed by radial velocities.
The sample and the new photometry are discussed in \S\S\ts5~and~6.  Tables of composite 
profiles are included. Section 7 illustrates these composite profiles of all 
the galaxies. Sections \hbox{8{\ts}--{\ts}13} discuss our conclusions.

\section{Cuspy Cores and ``Extra Light'' at the Centers of Elliptical Galaxies}

\subsection{A Digression on Analytic Fitting Functions}

\def\b{\kern -1pt}

      ``Cuspy cores'' are defined to be the region interior to the ``break radius'' 
$r_b$ where $I(r)$ breaks from a steep outer power law,
$I \propto r^{-\beta}$, to a shallow inner cusp, $I \propto r^{-\gamma}$.  This region of
the profile can conveniently be parametrized as:
$$ I(r) = I_b~2^{{{\beta - \gamma} \over \alpha }}~
   {             \biggl({r \over r_b}\biggr)^{\b\b\b-\gamma}  }
   {  \biggl[1 + \biggl({r \over r_b}\biggr)^{\b\b\alpha}\ts\biggr]^{{{\gamma -
                                           \beta}\over\alpha}}  },  \eqno{(1)}
$$
where $I_b$ is the surface brightness at $r_b$ and $\alpha$ measures the sharpness of the
break (Kormendy \etal 1994; Lauer \etal 1995; Byun \etal 1996; cf.~Lauer \etal 1992b;
Ferrarese et al.~1994 for earlier, simpler versions). 

      Since Equation (1) is asymptotically a power law at~large~$r$, it does not fit
S\'ersic profiles, nor was it devised to~do~so.  Rather, it was devised to fit central
profiles in the vicinity of the break radius in order to derive core parameters.  This was
done in Byun \etal (1996) and in Lauer \etal (2005, 2007b) and used to study core parameter
correlations in Faber \etal (1997) and in Lauer \etal (2007a). 
Graham \etal (2003, 2004) and  Trujillo et al.~(2004) advocate replacing Equation (1) with an
analytic ``core-S\'ersic function'' that becomes S\'ersic at 
large $r$. This is a plausible idea, but making it uncovers a problem with any attempt to fit 
cores and outer profiles with a single analytic function.  Analytic functions are
stiff. Their core and outer parameters are coupled in a way that depends 
on the chosen fitting function.  This is why Trujillo \etal (2004) get slightly 
different parameter values than those derived using Equation (1).  Core parameters 
inevitably depend on the parametrization; Lauer \etal (2007b) provide further discussion.
The solution is to avoid fitting functions that are complicated enough to result
in large, coupled errors in the derived parameters.  

      Therefore, we do not use one fitting function 
to parametrize all of a profile whose form is nowhere analytic and whose underlying 
distribution function is controlled by different physics at different radii.  Rather, 
we fit the profile piecewise.  That is, we fit the outer profile using a 
S\'ersic function over the radius range where it fits well (\S\ts7.2; Appendix A).  Departures 
from these fits are measured non-parametrically.

\vfill\eject

\subsection{``Extra Light'' at the Centers of Elliptical Galaxies}

      One new result of this paper is confirmation in a larger sample of galaxies of
an effect seen by Kormendy (1999). It is illustrated in Figure 3. NGC 4621, NGC 3377, and
M{\ts}32 are normal ellipticals with absolute magnitudes $M_V = -21.54$, $-20.18$,
and $-16.69$, respectively.  Their main bodies are well fitted by S\'ersic functions.  
At small radii, the behavior of the profile is opposite to that in a core galaxy -- there
is extra light compared to the inward extrapolation of the outer S\'ersic fit.

\vskip 7.8truein

\includegraphics{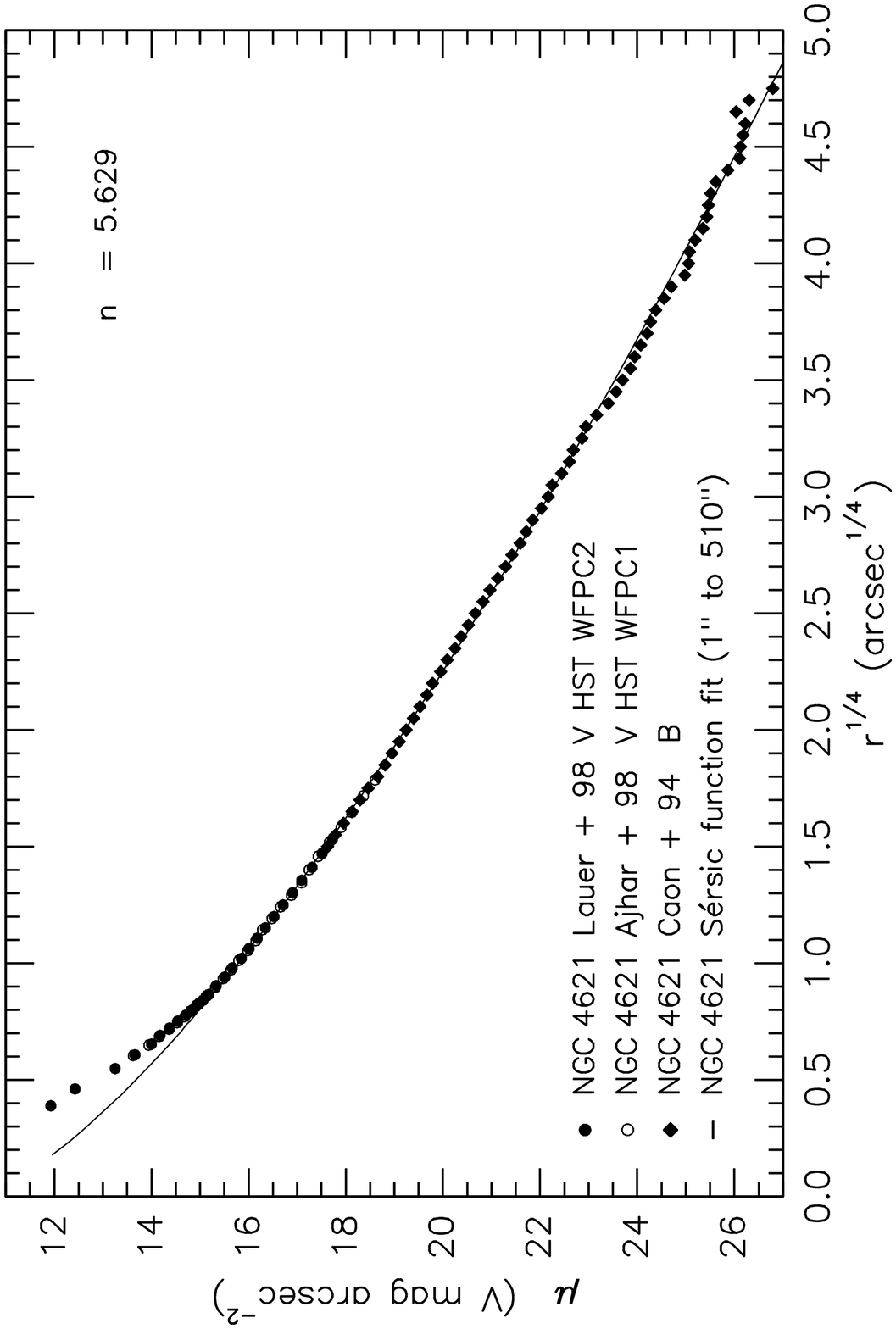}
\includegraphics{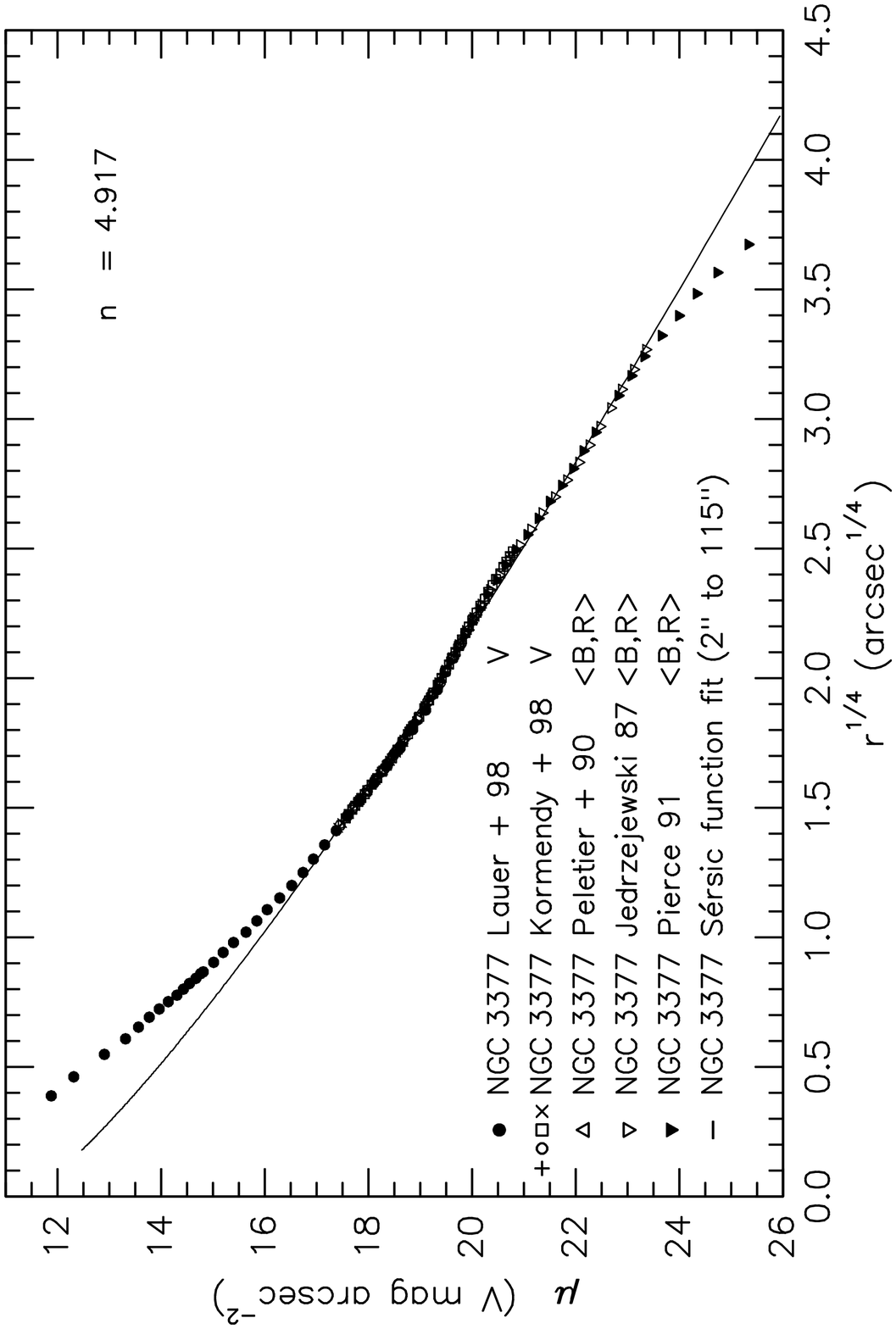}
\includegraphics{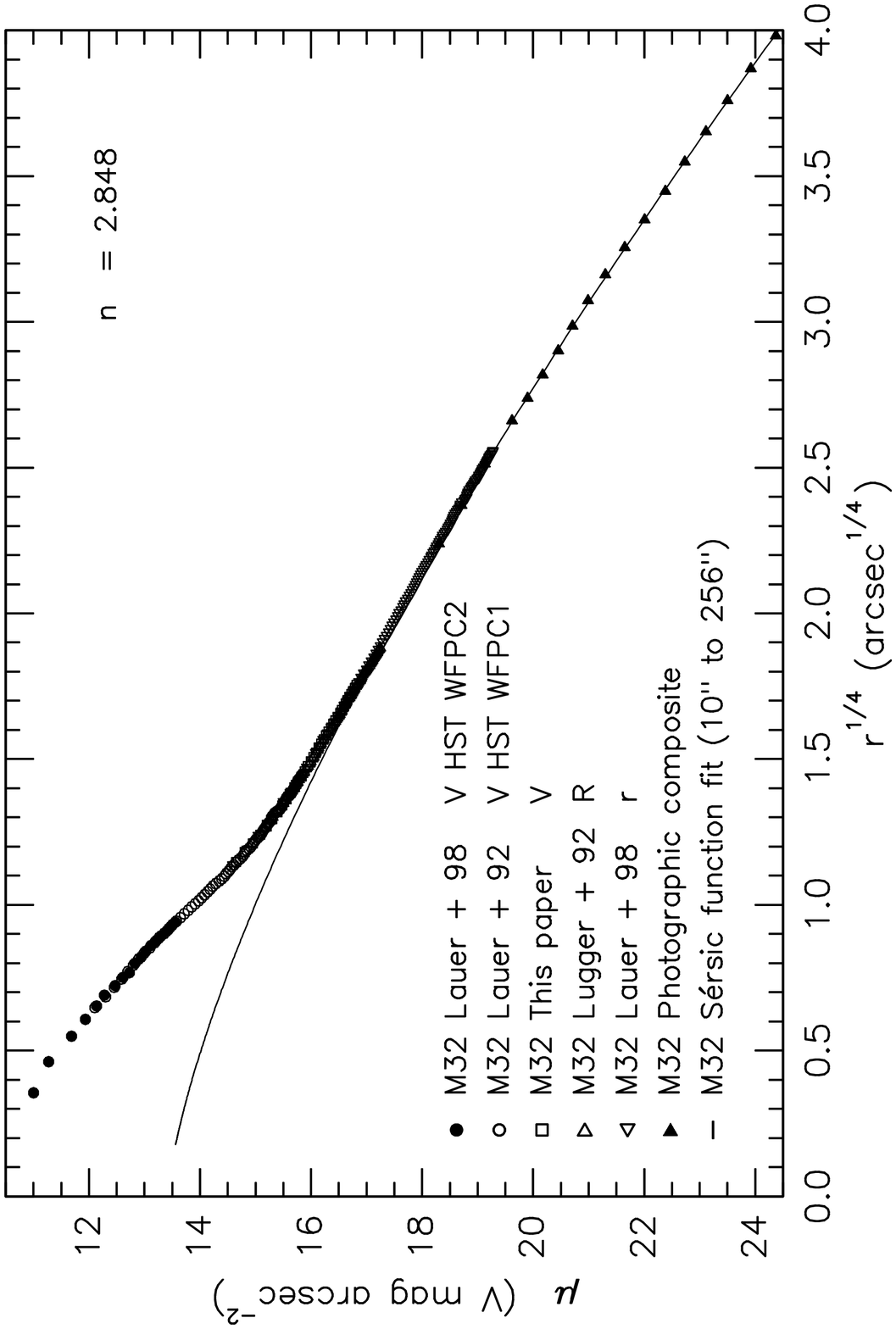}

\figcaption[]
{Composite major-axis brightness profiles of coreless elliptical 
galaxies fitted with S\'ersic functions ({\it solid curves\/}) with index $n$ ({\it
see the key}).  This figure is from Kormendy (1999).
\label{}}

\lineskip=-4pt \lineskiplimit=-4pt

      Kormendy (1999) pointed out that the extra light is similar to predictions by Mihos
\& Hernquist (1994) of high-density centers produced by dissipative mergers (Figure 4).
In their simulations, the excess light is a result of rapid inward transport of gas 
during the merger followed by a starburst.  The transition from starburst center to outer
profile occurs at $\sim 4$\thinspace\% of the effective radius $r_e$. The radii of the
observed breaks from the $r^{1/n}$ laws bracket 0.04 $r_e$ in Figure 3.  The observed
transitions are less sharp than the ones in the simulations, but the numerical
prescriptions used for star formation and energy feedback were approximate. Interestingly,
the observed departures from S\'ersic function fits are larger in smaller galaxies;
observations imply more dissipation at lower galaxy luminosities
(e.{\thinspace}g., Kormendy 1989).  It was too early to be sure of an interpretation, but
Kormendy (1999) noted that the observations are suggestive of dissipative starbursts. 
We will reach the same conclusion. 

\vfill

\includegraphics{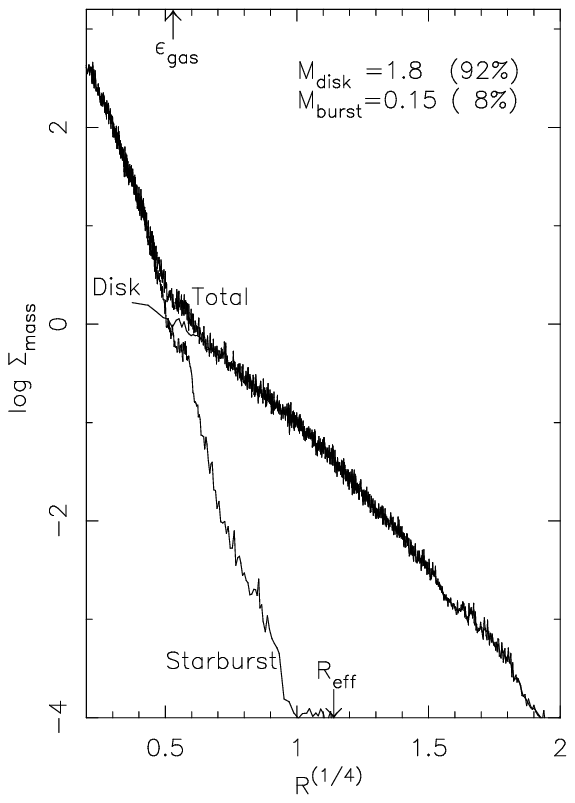}
\includegraphics{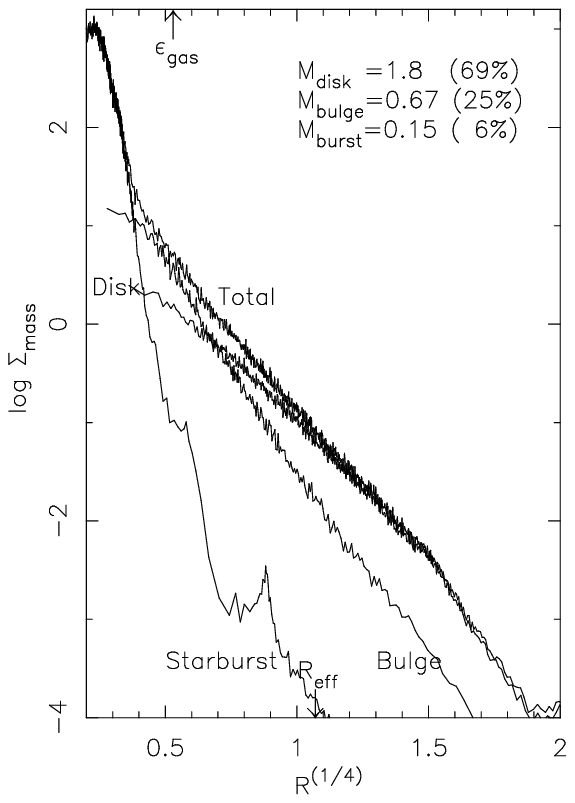}

\figcaption[]
{Luminous mass density profiles of merger remnants for progenitor
galaxies consisting of ({\it top\/}) a disk and a dark halo and ({\it bottom\/})
a disk, a bulge, and a dark halo.  These results are based on $N$-body 
simulations with gas.  During the merger, the gas falls to the center and
produces the ``Starburst'' density distribution.  Note that the outer profiles
are better described by S\'ersic functions than by $r^{1/4}$ laws.  This is 
Figure 1 from Mihos \& Hernquist (1994).
\lineskip=-4pt \lineskiplimit=-4pt
}

\eject

\centerline\null

\vfill

\includegraphics{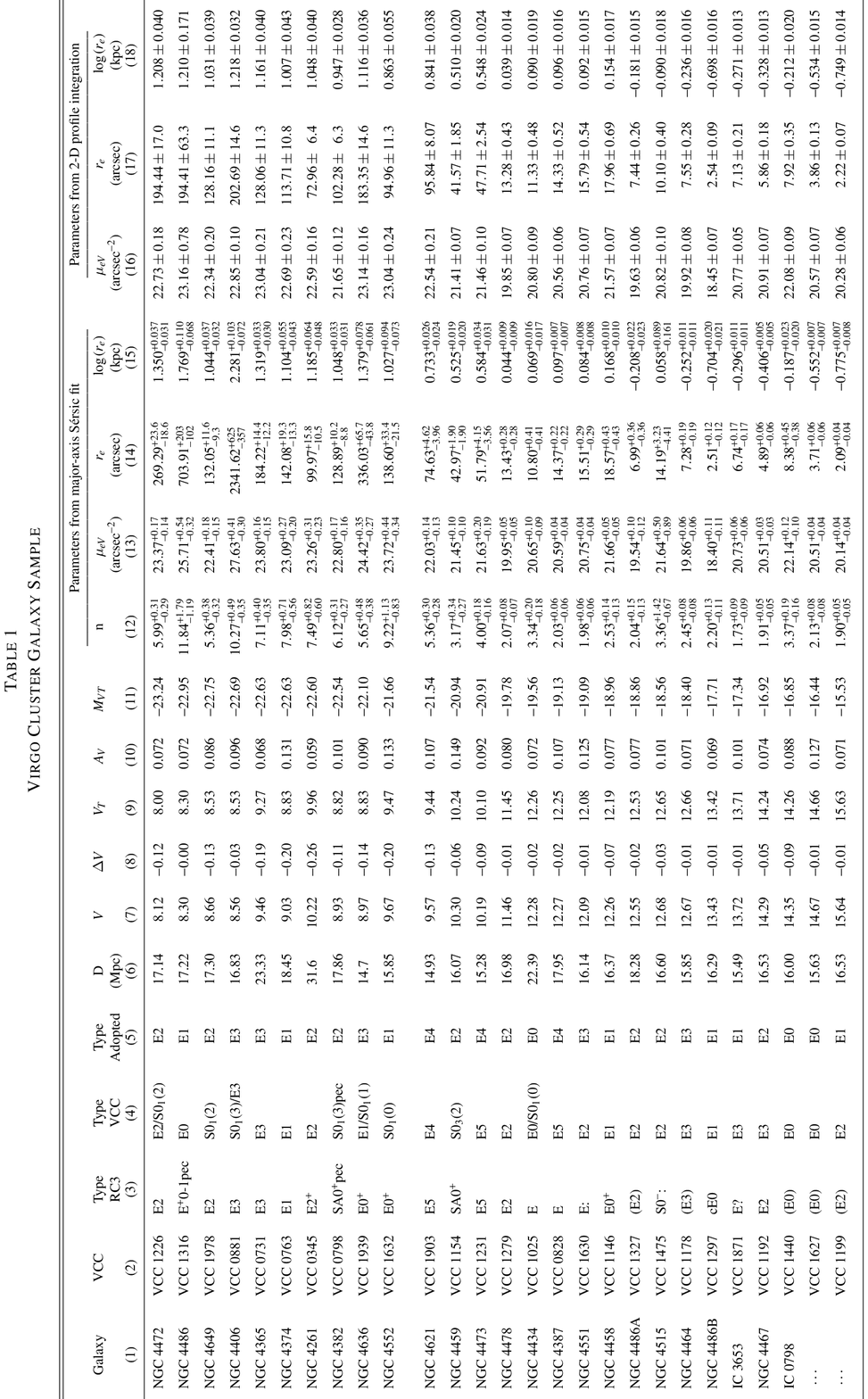}

\centerline\null

\eject

\centerline\null

\vfill

\includegraphics{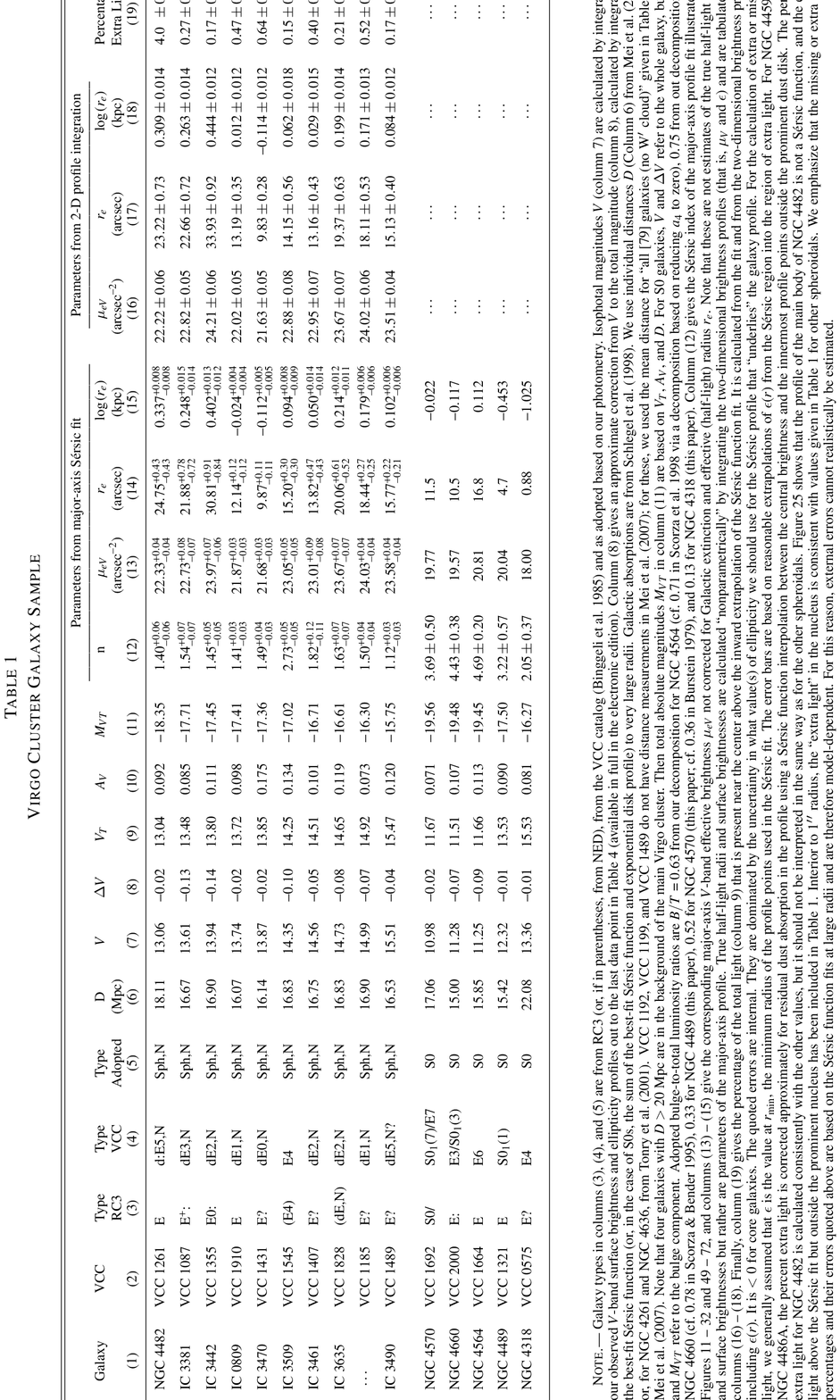}

\centerline\null\eject

\section{Galaxy Sample}

      Table 1 lists our sample ordered by total absolute magnitude $M_{VT}$ (column 11)
determined from our photometry.  The Virgo cluster has depth along the 
line of sight, so we use individual galaxy distances from Mei \etal (2007) or from
Tonry \etal (2001).  Galactic extinctions are from Schlegel et al.~(1998). 

      We wish to study all elliptical galaxies in the Virgo cluster.  
Distinguishing elliptical (E), S0, and spheroidal (Sph) galaxies is
nontrivial but important, because different types of galaxies are
likely to have different formation processes.  To construct a pure
sample of ellipticals, we erred on the side of caution and included
galaxies with uncertain classifications (e.{\ts}g., E/S0).  We then
used the photometry to resolve problem cases.  How we distinguish E and
Sph galaxies is discussed in \S\ts8.  How we distinguish E and S0
galaxies is discussed here.

\subsection{The Distinction Between Ellipticals and S0 Galaxies}

      If we want our classification to distill clean physics, we should 
not mix disks with ellipticals.  When both are present, as in an S0 galaxy, 
we need to make a photometric decomposition and analyze bulge and disk 
separately.  However, the distinction between Es, which by definition
are supposed not to contain disks, and S0s, which by definition do contain disks,
has been blurred in recent years by the recognition of ``disky ellipticals'' whose
isophotes are distorted from ellipses by \hbox{$\sim$\ts1\ts--\ts2\ts\%} as 
they would be if they contained embedded disks (Carter 1978; Lauer 1985c; 
Bender \& M\"ollenhoff 1987; Bender \etal 1987, 1988; Franx \etal 1989a; Bender 
\etal 1988,~1989; Peletier \etal 1990).~Photometric decompositions imply that 
the difference between an underlying, exactly ellipsoidal galaxy and the observed,
disky-distorted object is typically $\sim$ 10\ts\% and sometimes as much as 
40\ts\% (Scorza \& Bender 1995).  This does not prove that the disky distortions 
formed like the disks of spiral galaxies.  Disky distortions could instead 
be a natural consequence of gas-rich mergers, if stars rain out of the gas 
distribution while dissipation causes it to flatten.  In simulations, even 
dissipationless mergers can make disky ellipticals (Naab \etal 1999; Naab \& 
Burkert 2003).  On the other hand, the above ``disk fractions'' are well within
the range of disk contributions in S0s (Simien \& de Vaucouleurs 1986).  
Also, bulge-dominated S0s are easily recogized when seen edge-on (e.{\ts}g.,
NGC 3115) but not when seen face-on.  Then their disks perturb the 
bulge profile by only small amounts at intermediate radii (Hamabe 1982).  
Capaccioli \etal (1991) even suggest that NGC 3379, often called a prototypical 
elliptical, is a bulge-dominated S0.  In our sample, NGC 4636 may be such a galaxy
(Figure 55).

      Distinguishing E and S0 galaxies is therefore tricky.  We are saved
by our result (\S\ts9.1; Appendix A) that ellipticals are accurately described
by S\'ersic functions except near their centers;
only a few galaxies with extra halos compared to the outward extrapolation of 
inner S\'ersic fits require interpretation.  To recognize S0s, we use 
the ellipticity and isophote distortion profiles as discussed in \S\ts7.  
Disks should be more flattened than bulges, and they should -- except when
nearly face-on -- be disky by \gapprox \ts{a} few percent.  And S0 disks
live at large radii; {\it nuclear\/} disks do not disqualify a galaxy
from being an elliptical.

      Fortunately, distinguishing ellipticals from bulge-dominated S0s is not
critical to our results, because the Hubble sequence is continuous between 
them (Kormendy \& Djorgovski 1989; Kormendy \& Bender 1996).  The bulge-dominated
S0s that are most easily confused with ellipticals behave like ellipticals of similar 
luminosity.  They reinforce our conclusions. 

\subsection{Construction of Galaxy Sample}

      Our sample was constructed as follows.  We started with the 30 galaxies that
Binggeli \etal (1985) classify as E and list as Virgo cluster members.~We
added M{\ts}32 analogs from Binggeli's Table XIII after eliminating S0, Sph, and
background galaxies, provided that {\it HST\/} photometry is available.
We added S0s with $M_V \lesssim -21.5$ and checked which are ellipticals 
using our photometry.  The tendency to classify giant Es as S0s results mainly from 
the $M_V$ -- $n$ correlation.  Giant ellipticals have S\'ersic $n > 4$; i.{\ts}e., 
shallow brightness profiles at large radii.  Absent quantitative photometry, these 
halos look similar to S0 disks when galaxies are seen not nearly edge-on.  
Ellipticals can also get misclassified as S0s when they contain prominent nuclear 
dust disks (NGC 4459) or asymmetries diagnostic of unfinished mergers (NGC 4382).  
We obtained photometry of the combined sample plus the most elliptical-like Sph galaxies 
(called dE in Binggeli \etal 1985) as identified by previous authors in parameter correlations.
We then identified S0 and Sph galaxies based on our photometry.  However, we retain 
Sph and S0 galaxies in Figures 34\ts--\ts38 to illustrate how we distinguish the 
different types.  This procedure resulted in the sample of 27 elliptical galaxies in 
Table 1.  Three are now known to be background galaxies; we keep them but do 
not include them in Virgo statistics.

      Clearly we cannot be sure that we found all Virgo ellipticals.
Some omitted galaxies that Binggeli \etal (1985) list as possible
members will prove to be members.  Some spheroidals listed by Binggeli
may turn out to be
misclassified ellipticals.  We describe our sample as ``all known 
Virgo ellipticals'', recognizing that future work may find a few more.  
We defined our sample carefully and tried not to omit galaxies 
with special properties whose lack would bias our conclusions.

\section {Surface Photometry}

      Throughout this work, our aim is to improve the accuracy of galaxy 
photometry as much as possible.  For each galaxy, we combine photometry from a wide
range of sources to provide independent consistency checks and thereby to reduce 
systematic errors.  The sources include published data, our photometry of images 
available in public archives, and our photometry of images from our own observing 
programs.  All magnitude zeropoints come from {\it HST\/} images, but many have been 
checked against ground-based sources.  Both our relative brightness profiles and our 
zeropoints should be substantially more accurate than data available in the literature.
We cannot, of course, exclude the possibility that a small number of errors that are
larger than our error estimates have ``slipped through the cracks''.  But for most 
galaxies, the results have survived more consistency checks and comparisons of 
independent data sources than other photometry in the literature.

\subsection{Sources}

      Data sources are listed in Table 2 and cited in the keys to Figures 11 -- 32 
(\S\ts7).  Comments on individual sources follow.

\def\p{\phantom{.}}
\def\0{\phantom{0}}

\begin{deluxetable}{llccccc}
\tablenum{2}
\tablefontsize{}
\tablewidth{560pt}
\tablecaption{Data Sources}
\tablehead{
\colhead{No.}                     &
\colhead{Telescope~~~~~~~~~~~~~~} &
\colhead{Filter}                  &
\colhead{Scale}                   &
\colhead{\0Field of View}         &
\colhead{References}              &
\colhead{Number of}               \nl
\colhead{}                        &
\colhead{and Instrument~~~~~}     &
\colhead{}                        &
\colhead{(arcsec pixel$^{-1}$)}   &
\colhead{\0(arcmin)}              &
\colhead{}                        &                
\colhead{Galaxies}                }
\startdata
\01 & CFHT AOB Pueo   & $K$                                & 0.035  &\0 \00.15  $\times$\0\0 0.15 & \01,8     &\01 \nl
\02 & HST WFPC1 PC    & F555W, F785LP                      & 0.043  &\0   1.1 \0$\times$\0\0  1.1 & \0\p6,10,11 & 15 \nl
\03 & HST WFPC2 PC&\llap{F555}W, F675W, F702W, F\rlap{814W}& 0.046  &\0   0.6 \0$\times$\0\0  0.6 & \0\p1,12  & 20 \nl
\04 & HST ACS         & F475W, F850LP $\rightarrow$ $V$    & 0.049  &\0   3.5 \0$\times$\0\0  3.4 & \01       & 40 \nl
\05 & HST NICMOS      & F160W, F205W                       & 0.075  &\0   0.3 \0$\times$\0\0  0.3 & \01       &\02 \nl
\06 & CFHT HRCam      & $V$, $I$                           & 0.110  &\0   1.9 \0$\times$\0\0  1.2 & \01       & 18 \nl
\07 & CFH12K          & $R$                                & 0.21\0 &\p42\p\0 \0$\times$\0 28\p\0 & \01       & 23 \nl
\08 & CFHT Cass       & $V$                                & 0.22\0 &\0   7.0 \0$\times$\0\0  7.0 & \01       & 21 \nl
\09 & ESO/MPI 2.2 m   & $B$                                & 0.351  &\0   3.0 \0$\times$\0\0  1.9 & \03       &\08 \nl
 10 & KPNO 2.1 m      & $B$, $R$ $\rightarrow$ $V$         & 0.38\0 &\0   3.2 \0$\times$\0\0  2.0 &  15       &\04 \nl
 11 & SDSS            & $g$, $z$ $\rightarrow$ $V$         & 0.396  &          \nodata            & \01       & 31 \nl
 12 & Lick 1 m        & $R$                                & 0.43\0 &\0   3.6 \0$\times$\0\0  3.6 & \09       &\01 \nl
 13 & ESO 1.5 m Danish& $B$                                & 0.463  &\0   4.0 \0$\times$\0\0  2.5 & \04       &\05 \nl
 14 & KPNO 4 m        & $C$, $T_{1}$ $\rightarrow$ $V$     & 0.48\0 &  \p16.4 \0$\times$\0 16.4   & \07       &\01 \nl
 15 & Hawaii 2.2 m    & $B$, $R$ $\rightarrow$ $V$         & 0.595  &\0   5.1 \0$\times$\0\0  5.1 & \02       &\03 \nl
 16 & KPNO 0.9 m      & $B$, $R$ $\rightarrow$ $V$         & 0.86\0 &\0   7.3 \0$\times$\0\0  4.6 & \0\p5,15  & 11 \nl
 17 & McDonald 0.8 m PFC & $V$                             & 1.36\0 &\p46\p\0 \0$\times$\0 46\p\0 & \01       & 31 \nl
 18 & CWRU 0.6 m Burrell Schmidt & $M$ $\rightarrow$ $V$   & 1.45\0 &\p90\p\0 \0$\times$\0 45\p\0 &  14       &\02 \nl
 19 & Hawaii 0.6 m    & $B$, $R$ $\rightarrow$ $V$         & 1.6\0\0&  \p13.3 \0$\times$\0 13.3   & \02       &\09 \nl
 20 & NAO China 60 cm Schmidt & various $\rightarrow$ $R$  & 1.7\0\0&\p58\p\0 \0$\times$\0 58\p\0 &  13       &\01 \nl
\enddata
\tablecomments{References:
 1. -- This paper; 
 2. -- Bender \etal (2008);  
 3. -- Caon \etal (1990); 
 4. -- Caon \etal (1994); 
 5. -- Davis \etal (1985);
 6. -- Ferrarese \etal (1994); 
 7. -- Kim \etal (2000); 
 8. -- Kormendy \etal (2005);
 9. -- Lauer (1985a); 
10. -- Lauer \etal (1992a); 
11. -- Lauer \etal (1995);
12. -- Lauer \etal (2005);
13. -- Liu \etal (2005);
14. -- Mihos \etal (2005); 
15. -- Peletier et al.~(1990).}
\tablecomments{
The Caon \etal (1990, 1994) CCD data at small $r$ were augmented by photographic data at large radii taken
with the 1.8 m UK Schmidt telescope.  Most Caon \etal (1990) galaxies were observed with the ESO/MPI 2.2 m 
telescope, but 5 of 33 galaxies were observed with the ESO 1.5 m Danish telescope (entry 13).  The paper does 
not specify which galaxies were observed with which telescope, so all Caon \etal (1990) galaxies are credited 
to the ESO 2.2 m telescope.  Similarly, 6 of 19 Virgo galaxies discussed in Caon \etal (1994) were observed 
with the Steward Observatory 2.3 m telescope (scale = 0\farcs3 pixel$^{-1}$; field size 1.9 $\times$ 2.0 arcmin), 
but the paper does not specify which ones.  All Caon \etal (1994) galaxies are therefore credited to the 
ESO 1.5 m Danish telescope.  The uncertainty in telescope is unimportant here, because Caon data are used 
only at intermediate and large radii; the large-radius data are in any case dominated by the photographic 
results.  Further discussion is given in Appendix A3, which discusses the same photometry.
\vskip -20pt
}
\end{deluxetable}

\def\ts{\thinspace}
\def\d{$\phantom{.}$}
\def\0{\phantom{1}}
\def\z{\phantom{11}}

\begin{deluxetable}{ccccccccccc}
\tablenum{3}
\tablewidth{550pt}
\tablecaption{NGC 4486 = M{\thinspace}87 Composite Surface Photometry}
\tablehead{
\colhead{Galaxy }    &
\colhead{$r$    }    &
\colhead{$\mu_V$}    &
\colhead{$\epsilon$} &
\colhead{PA}         &
\colhead{}           &
\colhead{Galaxy }    &
\colhead{$r$    }    &
\colhead{$\mu_V$}    &
\colhead{$\epsilon$} &
\colhead{PA}         \nl
\colhead{}                  &
\colhead{(arcsec)}          &
\colhead{(mag arcsec$^{-2}$)}&
\colhead{}                  &   
\colhead{(deg E of N)}      &
\colhead{}                  &
\colhead{}                  &
\colhead{(arcsec)}          &
\colhead{(mag arcsec$^{-2})$}&
\colhead{}                  &
\colhead{(deg E of N)}      }
\startdata
NGC4486  &  \00.017 &  16.266  & \dots &$  \dots $& &  NGC4486 &  \z26.318 &  19.560 &  0.050 &$ -17.72 $ \nl
NGC4486  &  \00.044 &  16.358  & 0.160 &$ 163.60 $& &  NGC4486 &  \z29.040 &  19.716 &  0.051 &$ -18.93 $ \nl
NGC4486  &  \00.088 &  16.511  & 0.161 &$ 163.60 $& &  NGC4486 &  \z31.750 &  19.860 &  0.052 &$ -18.86 $ \nl
NGC4486  &  \00.176 &  16.589  & 0.161 &$ 187.60 $& &  NGC4486 &  \z35.015 &  20.016 &  0.059 &$ -19.82 $ \nl
NGC4486  &  \00.220 &  16.646  & 0.161 &$ 146.70 $& &  NGC4486 &  \z38.371 &  20.161 &  0.064 &$ -19.03 $ \nl
NGC4486  &  \00.264 &  16.700  & 0.162 &$ 136.60 $& &  NGC4486 &  \z42.073 &  20.318 &  0.072 &$ -21.15 $ \nl
NGC4486  &  \00.308 &  16.746  & 0.162 &$ 125.50 $& &  NGC4486 &  \z45.779 &  20.455 &  0.076 &$ -22.77 $ \nl
NGC4486  &  \00.352 &  16.788  & 0.162 &$ 119.00 $& &  NGC4486 &  \z50.855 &  20.631 &  0.078 &$ -21.66 $ \nl
NGC4486  &  \00.396 &  16.838  & 0.138 &$ 118.40 $& &  NGC4486 &  \z56.040 &  20.794 &  0.082 &$ -23.00 $ \nl
NGC4486  &  \00.440 &  16.889  & 0.131 &$ 118.40 $& &  NGC4486 &  \z61.094 &  20.936 &  0.086 &$ -22.84 $ \nl
NGC4486  &  \00.484 &  16.927  & 0.118 &$ 114.00 $& &  NGC4486 &  \z67.531 &  21.097 &  0.096 &$ -23.61 $ \nl
NGC4486  &  \00.548 &  16.953  & 0.109 &$ 110.95 $& &  NGC4486 &  \z72.277 &  21.217 &  0.100 &$ -23.49 $ \nl
NGC4486  &  \00.604 &  16.966  & 0.097 &$ 118.40 $& &  NGC4486 &  \z77.179 &  21.331 &  0.099 &$ -23.19 $ \nl
NGC4486  &  \00.660 &  16.996  & 0.094 &$ 118.40 $& &  NGC4486 &  \z84.918 &  21.499 &  0.109 &$ -24.58 $ \nl
NGC4486  &  \00.727 &  17.031  & 0.084 &$ 126.35 $& &  NGC4486 &  \z93.972 &  21.693 &  0.114 &$ -24.82 $ \nl
NGC4486  &  \00.795 &  17.062  & 0.090 &$ 147.40 $& &  NGC4486 & \0104.954 &  21.912 &  0.128 &$ -25.22 $ \nl
NGC4486  &  \00.867 &  17.091  & 0.087 &$ 147.60 $& &  NGC4486 & \0116.011 &  22.116 &  0.139 &$ -24.22 $ \nl
NGC4486  &  \00.950 &  17.114  & 0.079 &$ 147.60 $& &  NGC4486 & \0127.938 &  22.317 &  0.153 &$ -25.73 $ \nl
NGC4486  &  \01.038 &  17.134  & 0.075 &$ 111.15 $& &  NGC4486 & \0139.798 &  22.515 &  0.157 &$ -25.18 $ \nl
NGC4486  &  \01.147 &  17.165  & 0.071 &$\068.83 $& &  NGC4486 & \0154.170 &  22.714 &  0.171 &$ -24.06 $ \nl
NGC4486  &  \01.254 &  17.195  & 0.072 &$\062.30 $& &  NGC4486 & \0166.341 &  22.870 &  0.185 &$ -24.52 $ \nl
NGC4486  &  \01.365 &  17.210  & 0.049 &$\063.83 $& &  NGC4486 & \0180.926 &  23.019 &  0.206 &$ -24.98 $ \nl
NGC4486  &  \01.515 &  17.241  & 0.030 &$\067.50 $& &  NGC4486 & \0200.909 &  23.220 &  0.222 &$ -24.33 $ \nl
NGC4486  &  \01.669 &  17.270  & 0.023 &$\053.80 $& &  NGC4486 & \0222.587 &  23.420 &  0.237 &$ -23.90 $ \nl
NGC4486  &  \01.825 &  17.290  & 0.015 &$\057.60 $& &  NGC4486 & \0242.103 &  23.573 &  0.254 &$ -23.52 $ \nl
NGC4486  &  \01.998 &  17.318  & 0.007 &$\094.75 $& &  NGC4486 & \0265.053 &  23.742 &  0.275 &$ -24.12 $ \nl
NGC4486  &  \02.196 &  17.346  & 0.018 &$\094.62 $& &  NGC4486 & \0293.990 &  23.934 &  0.293 &$ -23.47 $ \nl
NGC4486  &  \02.419 &  17.371  & 0.015 &$\086.82 $& &  NGC4486 & \0321.366 &  24.096 &  0.303 &$ -23.70 $ \nl
NGC4486  &  \02.640 &  17.399  & 0.012 &$ 115.62 $& &  NGC4486 & \0346.737 &  24.257 &  0.313 &$ -23.83 $ \nl
NGC4486  &  \02.835 &  17.418  & 0.008 &$\091.40 $& &  NGC4486 & \0381.651 &  24.441 &  0.329 &$ -24.78 $ \nl
NGC4486  &  \03.218 &  17.470  & 0.005 &$\059.60 $& &  NGC4486 & \0419.276 &  24.658 &  0.337 &$ -25.59 $ \nl
NGC4486  &  \03.823 &  17.538  & 0.012 &$\025.40 $& &  NGC4486 & \0462.914 &  24.820 &  0.348 &$ -23.58 $ \nl
NGC4486  &  \04.546 &  17.613  & 0.010 &$\020.40 $& &  NGC4486 & \0502.343 &  25.011 &  0.370 &$ -23.56 $ \nl
NGC4486  &  \05.413 &  17.715  & 0.017 &$\012.50 $& &  NGC4486 & \0541.377 &  25.090 &  0.381 &$ -23.84 $ \nl
NGC4486  &  \06.092 &  17.790  & 0.021 &$\012.38 $& &  NGC4486 & \0593.608 &  25.288 &  0.388 &$ -24.66 $ \nl
NGC4486  &  \07.118 &  17.913  & 0.028 &$\010.27 $& &  NGC4486 & \0653.131 &  25.486 &  0.398 &$ -25.78 $ \nl
NGC4486  &  \07.780 &  17.991  & 0.023 &$ \z5.66 $& &  NGC4486 & \0719.449 &  25.697 &  0.427 &$ -27.03 $ \nl
NGC4486  &  \08.610 &  18.086  & 0.028 &$ \z5.82 $& &  NGC4486 & \0794.328 &  25.917 &  0.447 &$ -26.86 $ \nl
NGC4486  &  \09.441 &  18.183  & 0.020 &$\d-1.00 $& &  NGC4486 & \0878.348 &  26.100 &  0.454 &$ -26.75 $ \nl
NGC4486  &   10.304 &  18.277  & 0.026 &$ \z0.12 $& &  NGC4486 & \0946.237 &  26.328 &  0.447 &$ -26.57 $ \nl
NGC4486  &   11.552 &  18.409  & 0.030 &$\d-1.06 $& &  NGC4486 &  1046.325 &  26.620 &  0.457 &$ -26.88 $ \nl
NGC4486  &   12.322 &  18.489  & 0.026 &$\d-7.05 $& &  NGC4486 &  1145.513 &  26.848 &  0.464 &$ -27.50 $ \nl
NGC4486  &   13.715 &  18.622  & 0.030 &$\d-5.10 $& &  NGC4486 &  1230.269 &  26.995 &  0.454 &$  \dots $ \nl
NGC4486  &   15.109 &  18.749  & 0.030 &$\d-5.23 $& &  NGC4486 &  1336.595 &  27.180 &  0.443 &$ -29.80 $ \nl
NGC4486  &   16.615 &  18.879  & 0.032 &$\d-8.61 $& &  NGC4486 &  1479.109 &  27.305 &  0.439 &$  \dots $ \nl
NGC4486  &   18.249 &  19.009  & 0.036 &$\d-9.85 $& &  NGC4486 &  1621.810 &  27.535 &  0.436 &$  \dots $ \nl
NGC4486  &   19.971 &  19.143  & 0.036 &$ -12.95\d$&&  NGC4486 &  1778.279 &  27.715 &  0.433 &$  \dots $ \nl
NGC4486  &   21.945 &  19.284  & 0.040 &$ -15.19\d$&&  NGC4486 &  1995.262 &  27.755 &  0.429 &$  \dots $ \nl
NGC4486  &   23.961 &  19.415  & 0.043 &$ -17.19\d$&&  NGC4486 &  2443.700 &  28.045 &  0.422 &$ -34.10 $ \nl
\enddata
\tablecomments{Radius $r$ is measured along the major axis.  In the electronic table, the profile labeled
NGC4486A is the $V$-band profile of NGC 4486A.  Profile NGC4486AK is an alternative profile of NGC 4486A with 
$V$-band zeropoint and $V$-band data used at large radii but with the CFHT deconvolved $K$-band profile 
(brown points in Fig.~20) substituted at $r \leq 1\farcs4$ to minimize the effects of dust absorption.}
\end{deluxetable}

      {\it HST\/} WFPC2 data provide the highest spatial resolution (Lauer \etal 2005) with
scale = 0\farcs0456 pixel$^{-1}$ for the Planetary Camera (hereafter PC).  All WFPC1 and WFPC2 PC 
profiles from Lauer \etal (1995, 2005) are based on PSF-deconvolved images.  They allow us reliably 
to identify central departures from S\'ersic functions fitted to the main body of each galaxy.  
However, the PC field of view is small, so it is important to supplement {\it HST\/} data 
with wide-field photometry.

      The ACS Virgo cluster survey by C\^ot\'e \etal (2004) provides 
high-quality, archival images of almost all of our sample galaxies. Because it is
uniform in quality, it is our best source of color profiles.  Good resolution 
(scale 0\farcs05  pixel$^{-1}$) means that it provides an important supplement to 
the WFPC1 and WFPC2 photometry of the brighter galaxies and the best photometry of 
the centers of faint galaxies that were not previously observed by {\it HST\/}. 
The ACS images have high signal-to-noise (S/N) and a reasonably large field of view,
so they also yield the deepest profiles for some of the smallest galaxies in our sample.

      We have {\it HST\/} WFPC1, WFPC2, or ACS profiles for all of our galaxies.
Note, however, that we did not carry out PSF deconvolution of the ACS images.
Therefore the ACS profiles have slightly lower spatial resolution than the
WFPC profiles.  For many of the fainter galaxies, we have {\it HST\/} profiles
only from ACS.  The lower resolution affects how well we do or do not spatially
resolve any extra light or nuclei.  But it does not compromise our estimates
of the amount of extra light, and it has no effect on any conclusions in this paper.

      {\it HST\/} NICMOS images allow us to correct the optical {\it HST\/}
profiles of NGC 4261 and NGC 4374 for dust absorption.  Comparison of the NICMOS
F160W or F205W profiles and ACS $z$-band profiles shows that any residual absorption in the 
near-infrared is small.  NGC 4261 and NGC 4374 both have cuspy cores.  The NICMOS 
profiles are used only at small radii; they affect
our calculation of the total amount of light ``missing'' because of the presence 
of the core (\S\ts10.1), but they do not affect the S\'ersic fits or the determination
of global parameters.

     Adaptive optics observations obtained in $K$ band with the Canada-France-Hawaii 
Telescope (CFHT) and PUEO (Arsenault \etal 1994) were used to minimize 
absorption seen in NGC 4486A.  This is a small elliptical galaxy with an
edge-on stellar disk that is bisected by a strong dust lane (Kormendy \etal 2005).  
Again, use of a central infrared profile improves our estimate of the amount of extra 
light in the galaxy, but it does not affect the determination of global parameters.

     We include the CFHT photometry obtained in \hbox{1982{\ts}--{\ts}1994} by Kormendy
with the Cassegrain CCD camera and the High Resolution Camera (HRCam:~Racine \& McClure 
1989; McClure \etal 1989).~Kormendy \& McClure (1993) discuss image reduction.  HRCam 
includes tip-tilt image stabilization.  We also measured images obtained by Wainscoat 
and Kormendy in 2000 -- 2002 with the CFHT  12K CCD mosaic. 

      For as many galaxies as possible and especially for all of the largest 
galaxies, we obtained $V$-band images using the 
McDonald Observatory 0.8 m telescope.  These data generally provide the 
deepest profiles and thus are important for constraining the S\'ersic fits.  
We reach especially low surface brightnesses with the 0.8 m telescope 
because we can accumulate long exposures and because the wide unvignetted
field (46\arcmin $\times$ 46\arcmin) allows accurate sky subtraction.

      When papers published profiles or archives contained images in two 
bandpasses that bracket $V$, we used the bracketing profiles to calculate a 
$V$ profile using standard calibrations.

\subsection{Surface Photometry}

      Most profile calculations are based on isophote fits using the algorithm 
of Bender (1987), Bender \& M\"ollenhoff (1987), and Bender, D\"obereiner, \& 
M\"ollenhoff  (1987, 1988) as implemented in  the ESO image processing system 
{\tt MIDAS\/} (Banse \etal 1988) by Bender and by Roberto Saglia (2003, private 
communication).
The software fits ellipses to the galaxy isophotes; it calculates the ellipse
parameters and parameters describing departures of the isophotes from
ellipses.  The ellipse parameters are surface brightness, isophote center 
coordinates $X_{\rm cen}$ and $Y_{\rm cen}$, major and minor axis radii, and hence
ellipticity $\epsilon$ and position angle PA of the major axis. 
The radial deviations of the isophotes from the fitted ellipses are 
expanded in a Fourier series of the form, 
\vskip -5pt
$$ 
\Delta r_i = \sum_{k=3}^{N} 
             \left [ a_k \cos (k \theta_i) + b_k \sin (k \theta_i ) 
             \right ].\eqno{ (2) } 
$$
\vskip -5pt
The most important of these parameters is $a_4$, expressed in the
figures as a percent of the major-axis radius $a$.  If $a_4 > 0$,
the isophotes are disky-distorted; large $a_4$ at intermediate or large 
radii indicates an S0 disk.  If $a_4 < 0$, the isophotes are boxy.
The importance of boxy and disky distortions is discussed in Bender (1987);
Bender \etal (1987, 1988, 1989); Kormendy \& Djorgovski (1989); Kormendy \& 
Bender (1996), and below.

      Some profiles were measured using Lauer's (1985a) program 
{\tt profile\/} in the image processing system {\tt VISTA} (Stover 1988).  The
interpolation scheme in {\tt profile\/} is optimized for high
spatial resolution, so it is best suited to high-$S/N$ images of galaxy centers.
The isophote calculation is Fourier-based, so it is less well suited to measuring 
outer parts of galaxies, where low $S/N$ results in
noisy isophotes or where star removal or limited field of view results in 
incomplete isophotes.

      Some profiles were calculated with the isophote ellipse fitting program
{\tt GASP} (Cawson 1983; Davis et al.~1985).  {\tt GASP\/} does not provide isophote
distortion parameters, but it is the most robust of our isophote fitters at
low $S/N$, and it handles non-monotonic brightness profiles without problems.
Therefore it was sometimes the program of choice at large radii.

      Finally, in some cases (e.{\ts}g., NGC 4486A), it was impossible to
calculate reliable ellipse fits because of dust absorption or because of
overlapping galaxies or bright foreground stars.  In these cases, we calculated
cut profiles by averaging the surface brighness in one- to several-pixel-wide
cuts through the galaxy center.  Cut profiles are identified in the keys to
Figure 11 -- 32.

      Some profiles showed a few glitches produced, for example, by
imperfectly masked foreground stars.  By this, we mean that one value
of $\mu$ (rarely), $\epsilon$, or PA among a set of smoothly varying values was
much different from the adjacent values.  These values were replaced
by the average of the adjacent points when it was clear that they were
measurement errors.  

\subsection{Photometric Zeropoints}

      All zeropoints are based on {\it HST\/} images.  When available, WFPC1 or WFPC2, 
F555W zeropoints were used.  For most galaxies with these zeropoints, the keys to
Figures 11\ts--\ts32 list Lauer \etal (1995, 2005) as data sources.  Then $V$-band 
profiles were taken directly from these papers.  For a few galaxies, we measured 
and zeropointed WFPC2 images ourselves.

\lineskip=-4pt \lineskiplimit=-4pt

      We have a particularly good external check of the WFPC1 and WFPC2 zeropoints.
Many Virgo galaxies were observed during an excellent, seven-night observing run with 
the CFHT (1984, March 6/7 -- 12/13).  The entire run was photometric.  We 
observed large numbers of $V$- and $I$-band standard stars to tie our
photometry to Landolt (1983).  Most standards were in M{\ts}67 (Schildt1983).  
The CFHT and {\it HST\/} zeropoints agree very well.  In obvious notation, the 
mean difference in zeropoint for 3 WFPC1 values
is $V_{\it HST} - V_{\rm CFHT} = +0.004 \pm 0.002$ mag arcsec$^{-2}$ ($\sigma/\sqrt{3}$).
The mean difference in zeropoint for 11 WFPC2 values
is $V_{\it HST} - V_{\rm CFHT} = -0.009 \pm 0.004$ mag arcsec$^{-2}$ ($\sigma/\sqrt{11}$).

      All galaxies in our sample that do not have zeropoints from WFPC1 or WFPC2 were
observed in the Virgo cluster ACS survey.  However, the profiles are not tabulated in 
Ferrarese \etal (2006a).  We remeasured the $g$- and $z$-band images using the Bender 
code to ensure consistent $a_4$ values.  Zeropoints were taken from Sirianni \etal 
(2005):\footnote{The currently adopted ACS zeropoints 
({\tt http://www.stsci.edu/hst/acs/analysis/zeropoints}) 
are different from the above.  These changes have no effect on the present paper: the 
zeropoint that we adopted for each galaxy is the one that we calibrated to $V$.  However, 
readers who wish to use Equation (3) to calibrate current photometry using updated {\it HST\/}
zeropoints need to correct it for the changes in zeropoints from Sirianni \etal (2005) values.}
$g = -2.5\log{{\rm ADU}} + 26.168$ and
$z = -2.5\log{{\rm ADU}} + 24.326$, 
where ADU represents counts in the F475W or F850LP band, as appropriate..  
The $g$ and $g - z$ profiles were converted to $V$ as follows.  

\figurenum{5}

\vskip 3.65truein

\includegraphics{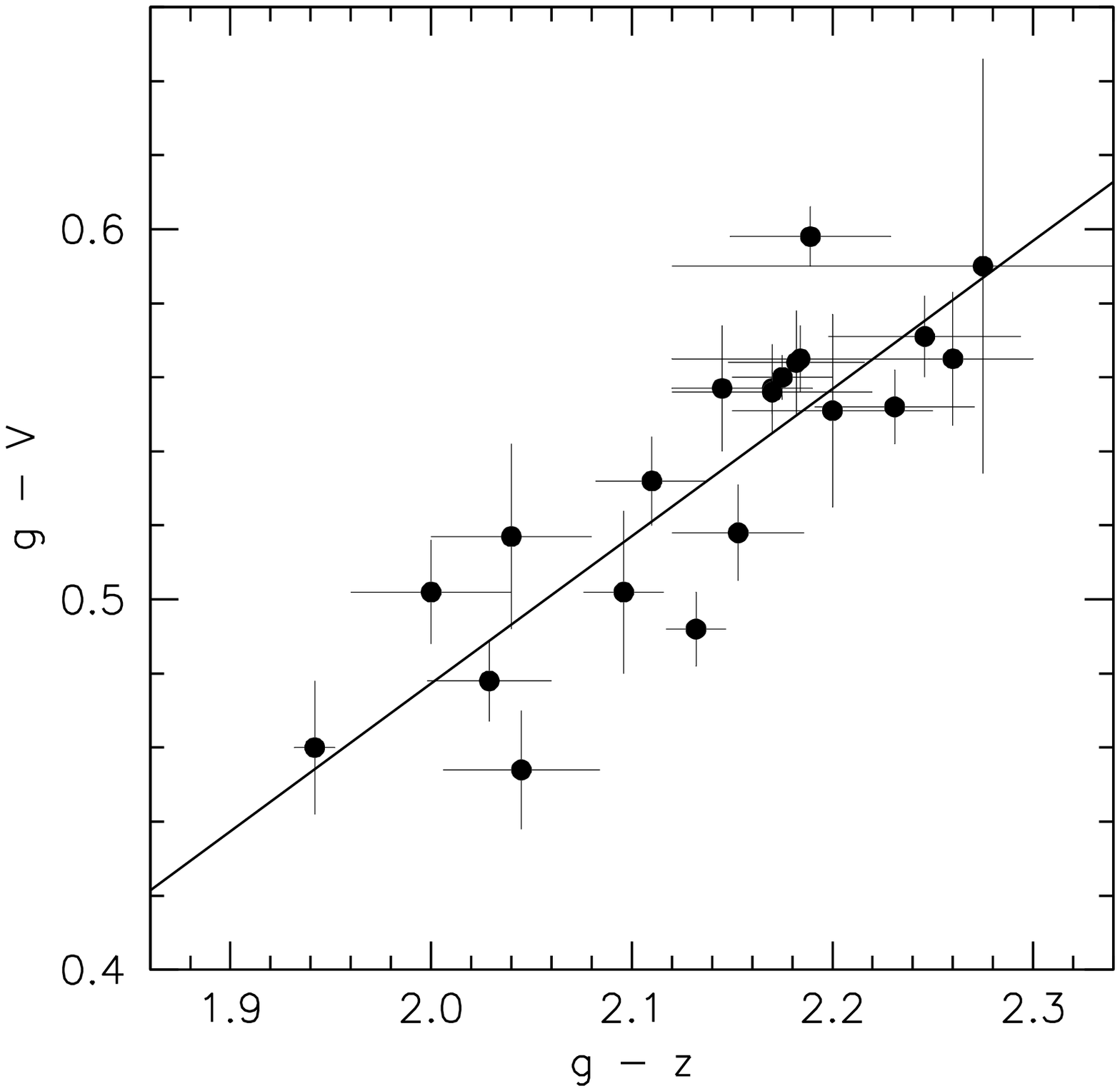}

\figcaption[]
{Calibration of {\it HST\/} ACS F475W $g$ and F850LP $z$ magnitudes to WFPC1 and 
WFPC2 $V$ band.  The ACS magnitude system used is VEGAmag with zeropoints from 
Sirianni \etal (2005).  Each point represents one galaxy for which we can compare 
the $g$ profile from ACS with a $V$ profile from Lauer \etal (1995, 2005).  The 
least-squares fit to the points ({\it straight line\/}) is our adopted 
transformation, Equation (3).
}
\vskip 10pt

      Most galaxies with WFPC1 or WFPC2 zeropoints were also observed with ACS.  
We calibrated $g$ and $g-z$ against $V$ by comparing our $g$ profiles to Lauer's 
$V$ profiles.  The results are shown in Figure 5.  Our adopted transformation is,
$$    V = g + 0.320 - 0.399\ (g - z).   \eqno{(3)} $$
Similar calibrations have been derived using standard stars (Smith \etal 2002;  Sirianni 
\etal 2005), but Equation~(3) is more relevant here, because it is based on the 
composite, old, metal-rich stellar populations that make up elliptical galaxies.  
The scatter in Figure 5 is 0.021 mag arcsec$^{-2}$ in $g - V$. Fig.~5 does not reach 
the bluest colors of our galaxies; some extrapolation is required.  We have an
external check of our $g - z$ colors: after converting their AB magnitudes to VEGAmag, 
we can compare Ferrarese \etal (2006a) color measurements $(g - z)_{\rm VEGA,F}$ to 
ours $(g - z)_{\rm VEGA,KFCB}$ over the radius range $1^{\prime\prime} \leq r \leq
16^{\prime\prime}$.  For 34 E $+$ Sph galaxies, the mean difference~is 
$$(g - z)_{\rm VEGA,F} - (g - z)_{\rm VEGA,KFCB} = +0.015 \pm 0.004~~(\sigma/\sqrt{34}. \eqno{(4)}$$
The dispersion, $\sigma = 0.024$ mag arcsec$^{-2}$, includes our errors in measuring
Ferrarese colors by hand in their published~plots.
Figure 5 suggests no reason to believe that ACS $g$ zeropoints converted to $V$ are less
accurate than WFPC1 and WFPC2 zeropoints.  Galaxies with ACS zeropoints are identified 
in Figures 11 -- 32: the keys list ``ACS V'' but not Lauer \etal (1995, 2005) 
as a data source.

      How accurate are our zeropoints?  The answer is notoriously difficult to determine.
Our comparison of WFPC and CFHT photometry was reassuring, but the agreement was
fortuitously good.  The ground-based standard star system was uncertain by several 
percent (e.{\ts}g., Joner \& Taylor 1990).  The same is true of {\it HST\/}.  Photometric 
standards and science targets are observed at different times, and the telescope plus
instruments show short-term instabilities and long-term sensitivity trends of a few 
percent or occasionally more 
(Baggett \& Gonzaga 1998;
Heyer \etal 2004;
Biretta 2005;
Bohlin 2007).  
Aperture effects are nontrivial (Holtzman \etal 1995).  Ground-based, WFPC, and ACS standard 
star measurements are made within apertures of different sizes, but the total amount of 
light at large radii in a PSF can be surprisingly large (King 1971; Kormendy 1973).  The 
outer PSF halo is often unmeasureably faint, but its light is taken away from the 
central profile, so it affects the zeropoint.  Given 
these considerations and our tests, we estimate that the random errors in our zeropoints are 
$\pm 0.03$ mag arcsec$^{-2}$ and the systematic errors are $\lesssim$ 0.05 mag arcsec$^{-2}$.  
These are better than the science requirements of this paper.

\subsection{Construction of Composite Profiles}

      Composite profiles were constructed from as many data sources as possible
(Table 2), including our own and published photometry.  Our emphasis was on accuracy.  
E.{\ts}g., almost all photographic profiles and many early CCD results proved not to 
be accurate enough to add weight to modern CCD data. 

      To construct composite profiles, we began with {\it HST\/} profiles,
including zeropoints.  We then added profiles one at a time, starting with the 
highest-accuracy ones measured with the highest spatial resolution.  Each profile 
was shifted in surface brightness to minimize the scatter with the previous composite 
over the largest possible radius range.  This must be done ``by hand'', because at this
stage, the deviations of individual profiles from the composite reveal systematic errors.  
Only a few of these can be anticipated.  E.{\ts}g., ground-based profiles ``peel~off''
 the {\it HST\/} profiles near the center when atmospheric seeing or telescope 
aberrations become important.  But it is not obvious {\it a priori\/} -- although it
becomes clear in carrying out the exercise -- that ellipticities are more sensitive 
to seeing than are surface brightnesses.  Position angles are most robust.  Another 
problem was that WFPC1 profiles are generally not accurate at large tabulated radii.  
In general, it quickly became clear that some profile sources (e.{\ts}g., Peletier 
\etal 1990) are more reliable than others (e.{\ts}g., our CFHT Cassegrain camera profiles,
which are excellent at small radii, but which have poor sky subtraction at large radii
when the field of view is too small for the galaxy).  Since we have many data sources 
at most radii in most galaxies, we were draconian in our pruning of 
individual profiles that did not agree with the means.  The final composite profiles 
are the means of the individual $\mu$-shifted profile points that were not pruned; 
i.{\ts}e., the data identified by asterisks in the keys to Figures 11 -- 32.  The
averages were carried out in $\log{r}$ bins of 0.04.  These profiles are illustrated 
in Figures 11 -- 32 and used in all analysis.  They are published in 
the electronic edition of ApJS.  Table 3 provides a sample.

      Some profile data are plotted in Figures 11 -- 32 but were not included in the
averaging.  They are not accurate enough to add significantly to our results, 
but they provide important consistency checks.  These are identified in Figures 11 -- 32: 
the keys do not have asterisks at the end of the source references.

      The accuracy of the final profiles is difficult to estimate.
However, we have many external checks.  The residual plots in Figures 
11 -- 32 illustrate with an expanded $\mu$ scale how well the individual profiles
agree with each other.  At small radii, our composite profiles should be accurate 
to a few percent or better.  At large radii, the number of independent data sources 
decreases.  It is even possible that, among (say) three sources, two agreed 
fortuitously but were less accurate than the third.  The agreement of different 
data sources provides a guide to the accuracy at large $r$, but it is not bomb-proof. 
When we discard a few points from the S\'ersic fits at large radii, this implies that
we do not trust the sky subtraction.  In general, we believe that our profiles are 
accurate to $\lesssim 0.1$ mag arcsec$^{-2}$ at large radii. 

\section{Photometry Results}

\subsection{Composite Brightness Profiles and Photometric Data}

      Figures 11 -- 15, 16 -- 24, 25 -- 29, and 30 -- 32 illustrate the photometry 
of the core ellipticals, the extra light ellipticals, the spheroidals, and the S0 
galaxies, respectively. 

      The bottom three panels show the $V$-band, major-axis brightness profile $\mu$, 
the isophote ellipticity $\epsilon$, and the major-axis position angle PA.  The next 
two panels are the isophote shape parameters $a_4$ and $a_3$.  Parameter $a_3$ 
shows that isophotes have reasonably pure boxy or disky distortions that are aligned
with the major axes; they generally have no triangular ($a_3$) or rotated ($b_n$) 
components.  Second from the top is the $g - z$ color profile from the {\it HST\/} 
ACS and SDSS surveys.  The top panel shows the deviations of the individual profiles 
in the bottom panel from the adopted S\'ersic function fit shown by the black curve.
The S\'ersic index $n$ is given in the key.  

\subsection{S\'ersic Function Fits to the Profiles}

      Appendix{\ts}A discusses our S\'ersic~fits.  Figures 49\ts--\ts72 show all of the
fits and the $\chi^2$ hyperellipses~of~the~three~fit~parameters.  They show that 
the parameter errors are often strongly coupled.  In this situation, parameter errors
can only be estimated from the maximum half-widths of the $\chi^2$ hyperellipses.  
Appendix~A also explores the dependence of the fit parameters on the radial range in 
which we make the fit.  We show that the parameters are robust provided that the fit 
range is large enough. This is why we aim to measure profiles that are reliable over 
large dynamic ranges.  No conclusions of this paper are vulnerable to small changes in 
fit ranges.  To aid users of S\'ersic functions, Appendix A presents guidelines on 
dynamic ranges needed to get reliable fits.  
Parameters of our fits including error estimates are listed in Figures 49\ts--\ts72 
and in Table 1.

      We fit S\'ersic functions over the largest radius ranges over which the fit 
residuals are (i) not systematic and (ii) roughly in agreement with our profile 
measurement errors.  The median RMS of the 27 E fits is 0.040 $V$ mag arcsec$^{-2}$, and 
the dispersion in RMS values is 0.019 $V$ mag arcsec$^{-2}$.
One of the main conclusions of this paper is that {\it S\'ersic functions fit the main 
parts of the profiles of both elliptical and spheroidal galaxies astonishingly well over 
large ranges in surface brightness.  For most galaxies, the S\'ersic fits accurately 
describe the major-axis profiles over radius ranges that include $\sim 93$\ts\% to 99\ts\% 
of the light of the galaxies\/} (see Figure 41).  

      At small $r$, all profiles deviate suddenly and systematically from the
best fits.  This is the signature of a core or extra light.  Including either 
one in the fit produces large systematic residuals that are inconsistent with our 
measurement errors.  Figure 64 (Appendix A) shows an example.  We emphasize in \S\ts4.1 
that we choose not to use fitting functions that combine (say) a central core with a 
S\'ersic envelope: the resulting parameters are too strongly coupled.  Our fits are 
robust descriptions of the main bodies of the galaxies.  In later sections we measure 
and interpret the amount of extra or missing light with respect to the inward 
extrapolation of the fits.  

\subsection{Galaxy Magnitudes}

\lineskip=0pt \lineskiplimit=0pt

      Galaxy apparent magnitudes $V$ (Table 1, Column 7) are calculated by 
integrating the two-dimensional mean brightness profiles including ellipticities
$\epsilon(r)$.   That is, $V$ is the  magnitude interior to the outermost 
nearly-elliptical isophote for which we have data.  These magnitudes, after 
conversion to $B$ using total ($B - V$) colors, are compared to Hyperleda 
total magnitudes in Figure 6.  Our iophotal magnitudes $B$ are slightly 
fainter than Byperleda total magnitudes $B_T$.  For ten core galaxies, the average 
difference is
$<$\null$B - B_T$\null$> = 0.045 \pm 0.035$ mag; for 15~coreless ellipticals (omitting
                                                 NGC 4486A) and two Sphs,
$<$\null$B - B_T$\null$> = 0.087 \pm 0.031$ mag, and for five S0 galaxies, 
$<$\null$B - B_T$\null$> = 0.180 \pm 0.060$ mag.
It is not surprising that our magnitudes are fainter, because they certainly
do not include all of the light of the galaxies.  Our limiting surface brightnesses
are 25.5 -- 28 $V$ mag arcsec$^{-2}$ for E and Sph galaxies and about 1 mag arcsec$^{-2}$
brighter for S0s.  The galaxy's surface brightnesses do not drop suddenly to zero outside 
these isophotes.  The corrections to total magnitudes are not very large, because the surface 
brightnesses that we fail to reach are faint.  But the corrections are not negligible, 
either, because the area of the outer isophotes is large.

\figurenum{6}

\centerline{\null} \vfill

\includegraphics{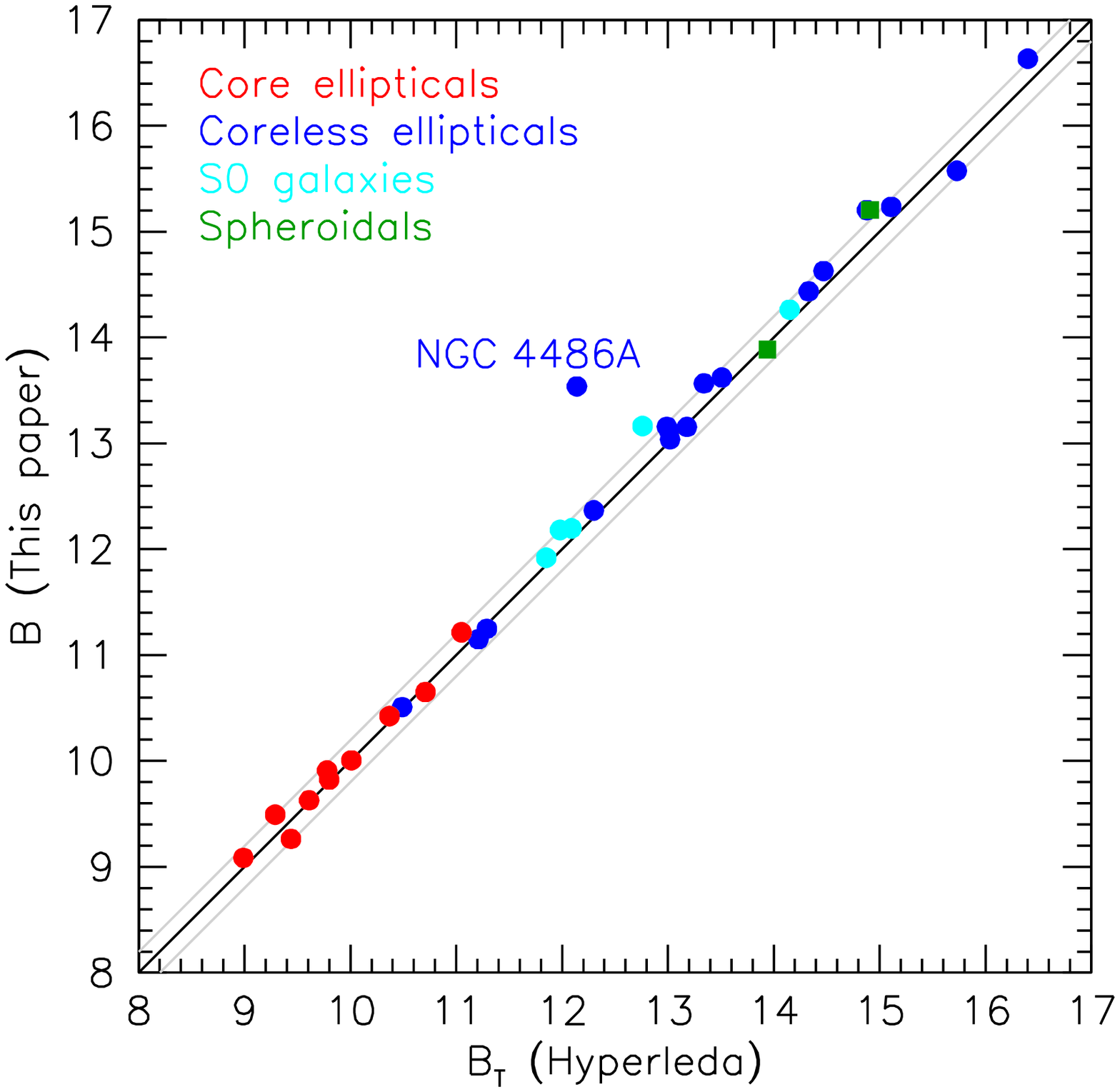}

\figcaption[]
{Comparison of our galaxy magnitudes with total $B_T$ magnitudes from Hyperleda
(Paturel \etal 2003:~their ``integrated photometry'' values).  Our $V$ magnitudes from Table 1 are 
converted to $B$ using total $B - V$ colors from RC3 when possible or colors within 
the effective radius from Hyperleda in a few cases.  Galaxy classifications are
from Table 1.  The black line indicates equality, and fiducial 
gray lines are drawn at $\pm 0.2$ mag to facilitate interpretation.  NGC 4486A 
deviates because a bright foreground star (see Kormendy \etal 2005) is imperfectly 
removed from the Hyperleda photometry.
}

\eject

\figurenum{7}

\centerline{\null} \vskip 3.3truein

\includegraphics{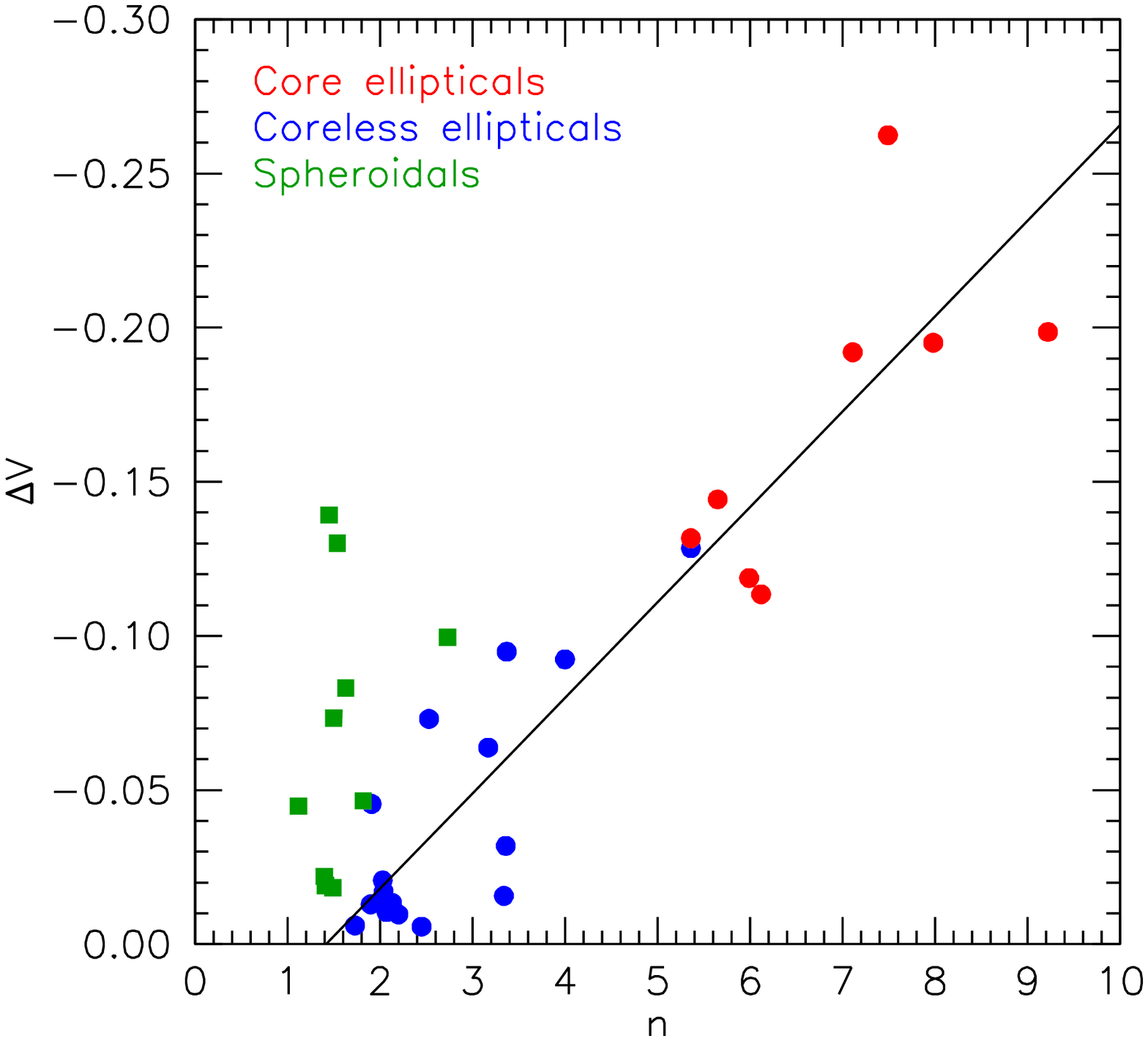}

\figcaption[]
{Corrections to convert our measured $V$-band galaxy magnitudes interior to the outermost
elliptical isophotes in Figures 11{\ts}--{\ts}32 to \hbox{almost-total} magnitudes interior 
to a surface brightness of $\sim$\ts29.7 $V$ mag arcsec$^{-2}$ for core Es and out to an
arbitrarily faint surface brightness for coreless Es and Sphs.  Each correction is 
calculated by integrating the extrapolation of our S\'ersic function fit with the 
ellipticity fixed at the value in the outermost observed isophotes.  The corrections
depend on $n$ as expected: larger $n$ means brighter, more extended outer halos and 
therefore larger $\Delta V$.  The correction is larger for Sph galaxies than for Es of
the same S\'ersic index in part because Sph galaxies have low surface brightnesses at 
small radii (Figures 34\ts--\ts36), so the relative contribution from large radii is relatively 
large.  In addtion to this effect, the scatter results mostly from the fact 
that our observations reach different limiting surface brightnesses in different galaxies; 
$\Delta V$ is small (large) when our photometry is deep (shallow).   However, the
scatter for ellipticals is small.  We use a least-squares fit to the E points 
({\it straight line\/}) only to note that the RMS scatter about the line is 0.027 mag arcsec$^{-2}$. 
}

\centerline{\null}

\figurenum{8}

\vskip 3.55truein

\includegraphics{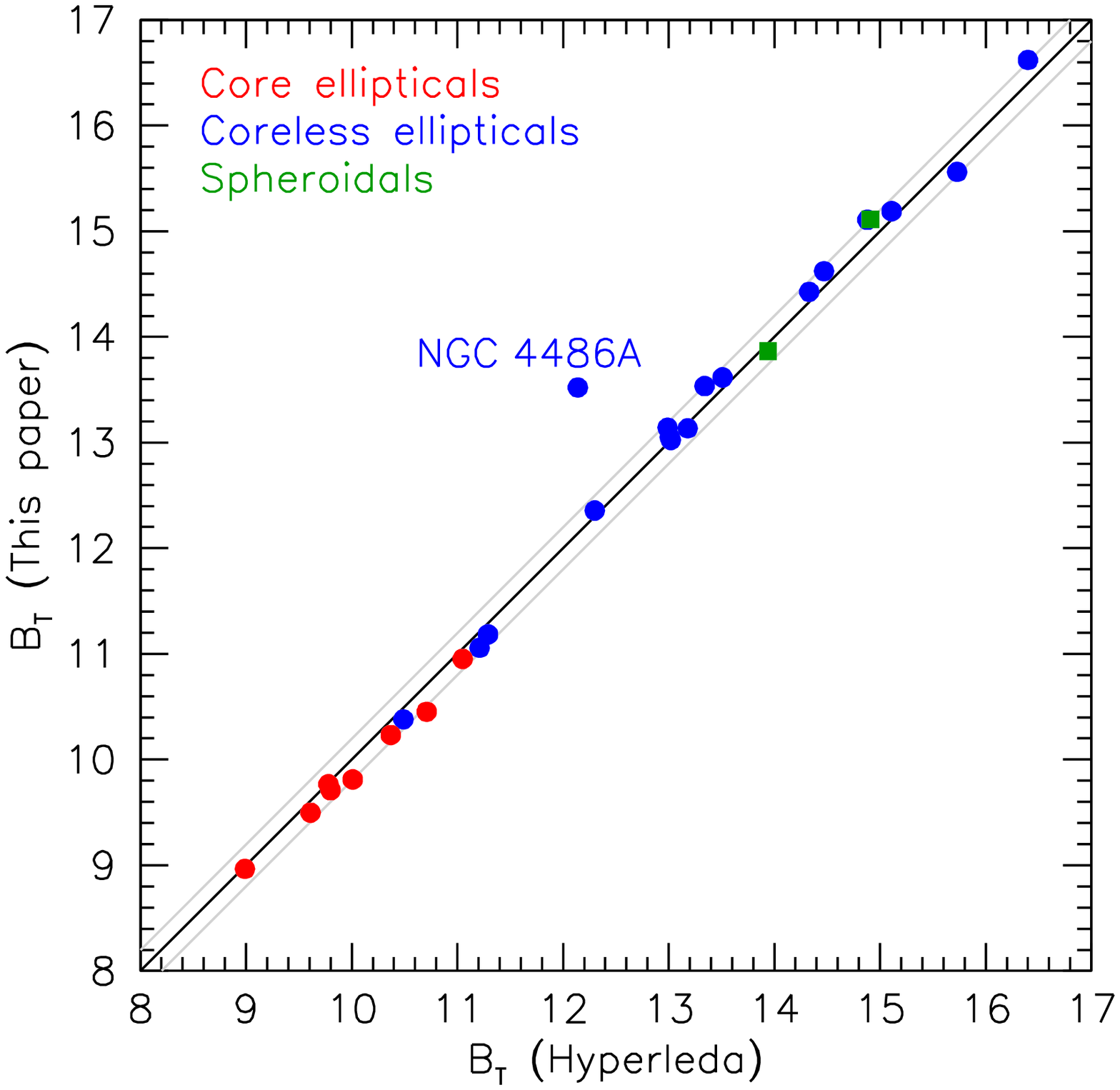}

\figcaption[]
{Comparison of our extrapolated, ``total'' galaxy magnitudes with total  
magnitudes from Hyperleda.  Our $V$-band magnitudes from Table 1 have been 
corrected individually with the $\Delta V$ values plotted in Figure 7 and
converted to $B$ as in Figure 6.   The black line indicates equality, and fiducial 
gray lines are drawn at $\pm 0.2$ mag to facilitate interpretation. 
}
\vskip 10pt

      We concluded in \S\ts7.2 that S\'ersic functions fit the \hbox{major-axis} brightness
profiles of our E and Sph galaxies very well, including the outermost points 
that we trust in our photometry.  Ellipticals are hot stellar systems; 
they cannot easily have sharp features in their brightness profiles.  It is
therefore reasonable to estimate corrections from our isophotal magnitudes
to nearly total magnitudes by integrating extrapolations of our S\'ersic function 
fits, as long as we do not need to extrapolate too~far.  Figure 7 shows such
magnitude corrections $\Delta V$.
They capture most of the missing light.  This is especially true for small-$n$ 
systems: their outer profiles cut off steeply, so their corrections are 
small.  The $\Delta V$ values also are reasonable for giant ellipticals with large
S\'ersic indices.  Their corrections are larger and more uncertain, but we 
already approach the intracluster background light (e.{\ts}g.)~in our profiles
of M{\ts}87 and NGC 4406 (see Mihos \etal 2005 and note that we include several 
isophotes from that paper in our profiles).   At radii not much larger than 
these, total magnitudes become ill defined, because stars there do not ``belong''
exclusively to the galaxy under study but also feel the gravitational potential 
of the cluster and especially of the nearest neighbors.

      Figure 8 plots total magnitudes $B_T = V + \Delta V + (B - V)_T$ from
our photometry versus values from Hyperleda.  The scatter is remarkably small
and the agreement is remarkably good, given that both sources have measurement
errors and that Hyperleda data are very heterogeneous.  The small systematic
differences now have exactly the sense that we would expect.  Hyperleda aperture
magnitudes are extrapolated to total magnitudes using mean growth curves for each 
galaxy type; for ellipticals, the growth curves are based on $n = 4$ de Vaucouleurs 
(1948) laws (Prugniel \& H\'eraudeau 1998).  One of the main conclusions of this 
paper will be that core ellipticals have $n > 4$ whereas almost all coreless 
ellipticals have $n \leq 4$.  Therefore our total magnitudes should be slightly 
brighter than Hyperleda's for core galaxies and slightly fainter than Hyperleda's 
for coreless galaxies.  This is exactly what Figure 8 shows.  For 8 core ellipticals 
plus NGC 4621 (a coreless galaxy which, in exception to the above conclusion, has 
$n = 5.36$) but omitting M{\ts}87 and NGC 4406 (see below), the average difference is
$$ {<}B_T - B_{T,\rm Hyperleda}{>} = -0.116 \pm 0.026~. \eqno{(5)} $$
For 5~coreless ellipticals with $3 < n < 5$ (i.{\ts}e., bracketing $n = 4$),
$$ {<}B_T - B_{T,\rm Hyperleda}{>} = +0.064 \pm 0.080~. \eqno{(6)} $$
For 12 coreless Es and 2 Sphs (``E'' in Hyperleda) with $n < 3$, 
$$ {<}B_T - B_{T,\rm Hyperleda}{>} = +0.056 \pm 0.033~. \eqno{(7)} $$
Equations (5)\ts--\ts(7) imply that our photometric system is consistent with the
heterogeneous but large database in Hyperleda; recall that our zeropoints were
estimated to be good to $\pm$ 0.05 mag.  For correless ellipticals and for Sph 
galaxies, our corrections $\Delta V$ should be accurate roughly to the RMS = 0.028 
mag in Figure 7.  It is unlikely that they are much worse than $\pm$ 0.05
mag even for giant ellipticals, although one cannot be certain about
extrapolations.  We therefore adopt the individual corrections plotted in Figure 
7 for these galaxies to get total magnitudes $V_T$ and hence total absolute 
magnitudes $M_{VT}$ in columns 9 and 11 of Table 1, respectively.

      Three ellipticals in Table 1 require special attention and were omitted 
from the above statistics.  NGC 4486A has a bright star superposed near its
center that is imperfectly removed from the Hyperleda photometry.  The galaxy
is therefore an outlier in Figures 6 and 8.  However, our {\it HST\/} photometry
should be unaffected by the star, so we corrected $V$ to $V_T$ as normal using our 
S\'ersic fit to the profile.  Second, the giant elliptical NGC 4406 in the main chain 
of galaxies near the center of the Virgo cluster is surrounded on all sides by other 
galaxies.  Either because these are imperfectly removed from the photometry or because
the profile is affected by tides from its neighbors, NGC 4406 has an outer profile
that cuts off strongly compared to the outward extrapolation of the inner S\'ersic
fit (Figure 12).  Therefore the normal magnitude correction is not valid.  Based
on a S\'ersic fit to the steep outer profile, we derive $\Delta V = -0.03$.  
Finally, M{\ts}87 almost certainly contains a faint cD halo (\S\ts7.4).  We should not
include intracluster light in $M_{VT}$.   Based on Figure 7 and on the two fits in 
Figure 50, we adopt $\Delta V = 0$.

      The total absolute magnitudes that result from the above procedures are
used throughout this paper.  Including zeropoint errors but not distance errors, we 
conservatively estimate that $M_{VT}$ has errors of $\sim$\ts0.07 mag for galaxies
with $n < 4$, $\sim$\ts0.1 mag for galaxies with $n \geq 4$, and 0.2 mag for M{\ts}87.

\subsection{The cD Halo of M{\ts}87}

\lineskip=0pt \lineskiplimit=0pt
 
      M{\ts}87 = NGC 4486 is the {\it second\/}-brightest galaxy in Virgo.  However, 
it is the central giant elliptical in the cluster, and it is surrounded by an 
enormous X-ray halo which shows that the galaxy is at the bottom of a deep potential 
well (e.{\ts}g., 
Fabricant \& Gorenstein 1983;
B\"ohringer \etal 1994; 2001;
Forman \etal 2007).
In richer clusters, such galaxies are often cDs (Matthews, 
Morgan, \& Schmidt 1964; Morgan \& Lesh 1965), i.{\ts}e., giant ellipticals that 
have extra light at large radii in an enormous halo that belongs more to the cluster 
than to the central galaxy.  ``Extra light'' with respect to what?  The answer 
is best quantified by Schombert (1986, 1987, 1988).  He showed that E 
profile shapes depend on luminosity; he constructed template mean profiles in 
different luminosity bins, and he identified as cDs those giant Es that have 
extra light at large radii with respect to the template that best fits the 
inner parts of the profiles.  Recasting this statement in the language of 
S\'ersic functions, cD galaxies are giant Es that have cluster-sized extra 
light at large radii with respect to the outward extrapolation of a S\'ersic 
function fitted to the inner profile.  cD halos are believed to consist of
stars that were stripped from individual galaxies by collisions (Gallagher
\& Ostriker 1972; Richstone 1976).

      Whether M{\ts}87 is a cD has been uncertain.  This appears to be settled by the 
remarkably deep photometry by Liu \etal (2005) and Mihos \etal (2005).~Both are included 
in Figure~11.  Liu and collaborators, like de Vaucouleurs \& Nieto (1978) and others, 
conclude that M{\ts}87 is a cD.  We agree, but not for the reasons given in their papers.
They conclude that the profile of M{\ts}87 shows extra light at large radii with respect 
to an $r^{1/4}$ law fitted to the inner parts.  This is true, but it is true for all 
galaxies that have S\'ersic $n > 4$.  As reviewed in \S\ts3 and confirmed 
again in this paper, essentially all giant ellipticals have $n > 4$. The evidence
that M{\ts}87 has a cD halo is more indirect.  It is shown in Figure 50.  A 
S\'ersic function fits the whole profile with entirely acceptable residuals outside 
the core (RMS = 0.0448 mag; see the top panels in Figure 50).  However, 
$n = 11.8^{+1.8}_{-1.2}$ is formally much larger than in any other galaxy in our sample.  
When the outer end of the fit range is decreased below $\sim$\ts900$^{\prime\prime}$, 
$n$ drops rapidly.  By construction, such fits have extra light at large radii.  
An example is shown in the bottom panels of Figure 50.  Fitting the profile out to 
419$^{\prime\prime}$ results in S\'ersic index $n = 8.9^{+1.9}_{-1.3}$ that is more 
consistent with the values for the other giant ellipticals in Virgo.  If this fit is 
adopted, then the galaxy has a faint extra halo at large radii.  It is similar to 
but (by construction) fainter than the cD halos advocated by Liu and de Vaucouleurs.  
We emphasize thata this fit is not unique.  It is an interpretation, not a proven result. 
However, based on such fits we do suggest that M{\ts}87 is marginally a cD galaxy. And 
we regard the detection of intracluster light by Mihos \etal (2005) as definitive proof.

     These results are consistent with Oemler's (1976) conclusion 
that cD envelope luminosity $L_{\rm env}$ depends strongly on cluster luminosity, 
\hbox{$L_{\rm env} \propto L_{\rm cluster}^{2.2}$.}  The total luminosity of Virgo 
is near the low end of the range for clusters that contain cDs.  That M{\ts}87 is
a weak cD is interesting in its own right, but it plays no direct role in this 
paper. Either set of fit parameters in Figure 50 is comfortably consistent with
the fundamental plane correlations discussed in \S\ts8.  Our estimate of the amount
of missing light that defines the core is essentially unaffected.  And $n$ is robustly
larger than 4, consistent with our conclusion that S\'ersic index participates in the
E\ts--{\ts}E dichotomy.

\subsection{Comments on Individual Ellipticals}

\lineskip=0pt \lineskiplimit=0pt

      Profile properties that are common to many galaxies are discussed in \S\ts9.  
Here and in \S\ts7.6, we comment on galaxies whose classification (E versus S0) 
has been uncertain.  When we assign a different morphological type to a galaxy than 
the catalog types (columns 3 and 4 in Table 1), we give the reasons.  This 
section involves details; readers who are interested in our main science results 
can jump directly to \S\ts8.

\figurenum{9}

\vskip 3.35truein

\includegraphics{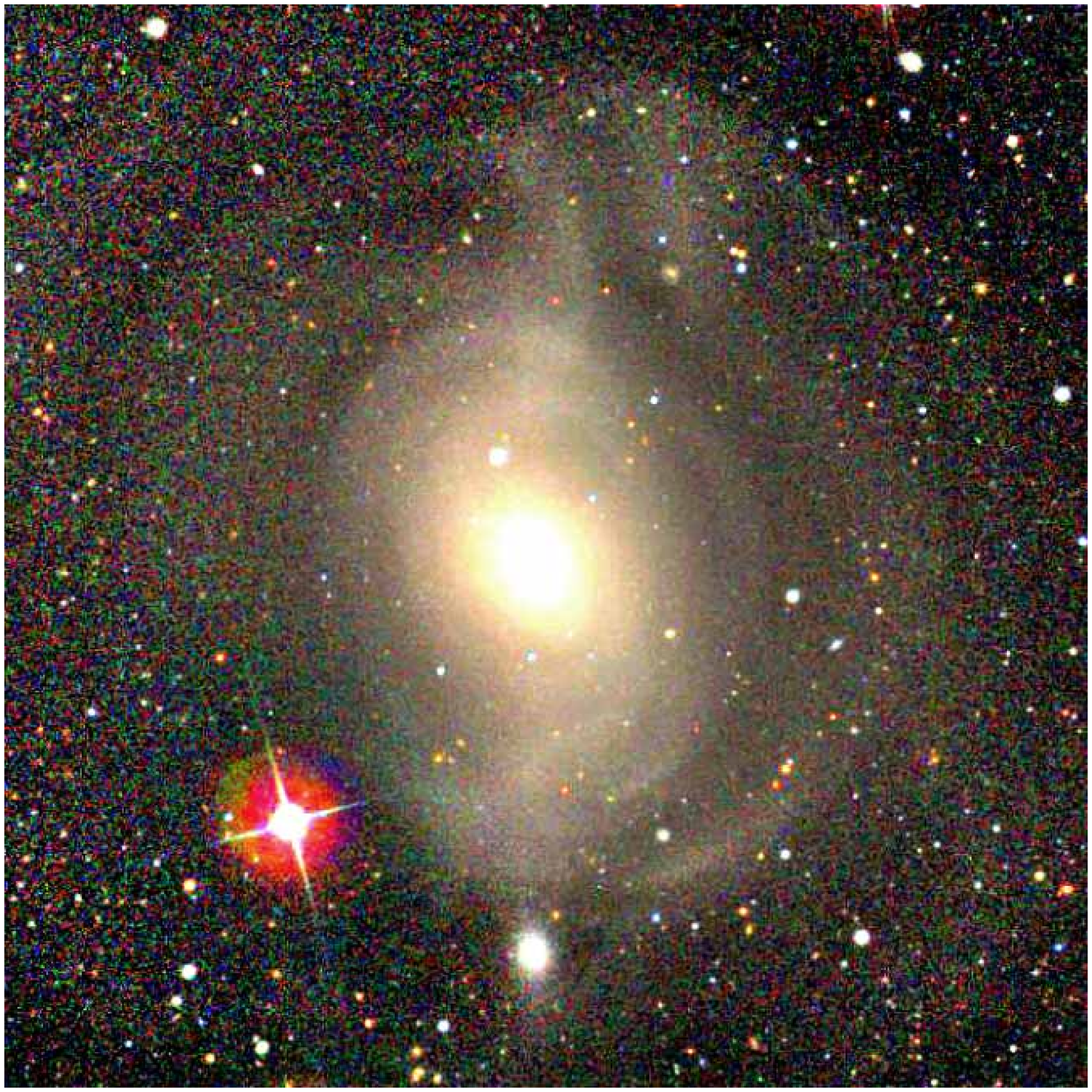}

\figcaption[]
{Contrast-enhanced ${g}{r}{i}$-band color image of NGC 4382 from the SDSS online
site {\tt http://www.wikisky.org}.  Strong fine-structure features are signs
that the galaxy has not finished relaxing after a recent merger.
}
\vskip 10pt

      NGC 4382 is classified as SA0$^+$pec in RC3.  Figure 14 shows
that it has a very unusual brightness profile.  It has extra light 
at intermediate radii, but the $a_4$ profile indicates that a slight disky
distortion at smaller radii disappears here.  This suggests that 
the extra light is not an S0 disk.  Also, when the profile is decomposed 
into a S\'ersic function bulge and an exponential disk, the disk parameters are 
very abnormal (cf.~Freeman 1970).  Finally, the galaxy is asymmetric
and shows fine-structure features indicative of a recent merger (Fig.~9).  
Schweizer \& Seitzer (1992) quantify such features for 69 E and S0 
galaxies; only three galaxies, two of them obvious mergers-in-progress, 
have larger fine-structure indices than does NGC 4382.  The galaxy gets bluer 
and shows enhanced H$\beta$ and depressed Mg{\ts}b spectral lines near the center 
(Fisher \etal 1996; Lauer \etal 2005; Kuntschner \etal 2006), consistent with a
younger stellar population.   We conclude that the galaxy is an 
elliptical -- a recent (damp?)~merger remnant that has not fully settled into equilibrium.  
Aguilar \& White's (1986) $n$-body simulations show that tidal 
stretching and shocking can produce features like the ``extra halo'' in 
Figure 14.  Similarly, Navarro's (1990) $n$-body similations show that 
merger remnants relax violently from the center outward, with waves in the
density (cf.~Fig.~14) that propagate outward during the relaxation process.

      NGC 4406 is classified as S0$_1$(3)/E3 in the VCC and E3 in RC3.
We see no sign of an S0 disk in the surface brightness or $a_4$ profiles 
(Fig.~12).  In particular, $a_4$ shows boxy -- not disky -- isophotes at 
large radii.  The galaxy is zooming through Virgo at $\sim 1400$ km s$^{-1}$,
and it is bracketed closely by NGC 4374, by the pair NGC 4435 $+$ 
NGC 4438, and by many other, not much smaller galaxies.  Its isophotes 
overlap at large radii with those of the adjacent galaxies 
(Kormendy \& Bahcall 1974; Mihos et al.~2005), so the outermost 
profile is uncertain.  This, or else the non-equilibrium tidal distortion 
that can result from a rapid encounter with its neighbors 
(Aguilar \& White 1986) could account for the slightly
non-S\'ersic profile at large radii and for the unusually large
value of $n = 10.27^{+0.49}_{-0.35}$.  Note that the profile is very 
concave-upward in Figure 12.

      NGC 4459 is classified S0 in the VCC and RC3 because of its 
nuclear dust ring. Figure 16 shows no evidence of a stellar disk
in the form of profile departures from a S\'ersic function.
The isophotes are not disky.  We classify the galaxy as an elliptical.

\subsection{Comments on Individual S0s}

      NGC 4318 is classified ``E?''~in the RC3 and E4 in~the~VCC.  However,
its brightness, ellipticity, and position angle profiles show a strongly two-component 
structure (Fig.~32).  The outer component has a disky signature ($a_4 > 0$)
and an exponential profile (Fig.~32). This suggests that the galaxy is an S0.
  
      We can check this by measuring the rotation velocity and velocity dispersion of the
outer component.  Simien \& Prugniel (1997, 1998) took spectra of NGC 4318 using the 
1.93~m telescope of the Observatoire de Haute-Provence.  The latter paper used a 
dispersion was 52 km s$^{-1}$ pixel$^{-1}$ and got a central velocity dispersion of 
$\sigma_0 = 77 \pm 17$ km s$^{-1}$.  The former paper got a maximum rotation velocity 
of $75 \pm 20$ km s$^{-1}$, but the observations did not clearly reach a flat part of the 
rotation curve (Fig.~10).  We therefore remeasured 
NGC 4318 with the LRS spectrograph (Hill \etal 1998) on the 9.2 m Hobby-Eberly Telescope.  
The slit PA was 65$^{\circ}$, the slit width was 1\farcs5, and the exposure time was
900~s.  The standard spectrum was a mean of the spectra of the K0~III stars $\eta$ Cyg 
and HD 172401.  The results are the open squares in Figure 10.  Our dispersion, 
116 km s$^{-1}$ pixel$^{-1}$, is substantially worse than that of Simien \& Prugniel, 
so their velocity dispersion measurements are more reliable than ours.  But our $S/N$ is 
higher, so we reach the $V \simeq$ constant part of the rotation curve.  We 
adopt our measurement of the maximum rotation velocity, $V_{\rm max} = 82.4 \pm 2.3$ 
km s$^{-1}$.  Then, $V_{\rm max}/\sigma_0 = 1.07 \pm 0.24$.  For an ellipticity of 
$\epsilon = 0.35$ in the outer component, the ``oblate line'' in the 
$V_{\rm max}/\sigma_0$\ts--\ts$\epsilon$ diagram (Binney 1976, 1978a, b; Illingworth 1977; 
Kormendy 1982) implies that 
an isotropic, oblate spheroid should have  $V_{\rm max}/\sigma_0 = 0.73$.  The outer 
component of NGC 4318 rotates $(V_{\rm max}/\sigma_0)^* = 1.46 \pm 0.32$ times faster than this.
In practice, we should use a mean velocity dispersion inside approximately the half-light 
radius; from Figure 10, this would be smaller than $\sigma_0$.  Moreover, since the outer
velocity dispersion is small and the $S/N$ of the Simien \& Prugniel measurements is low, 
the true velocity dispersion may be even smaller.  Therefore the outer 
component clearly rotates more rapidly than an isotropic oblate spheroid with the 
observed flattening.  This is a disk signature.

\figurenum{10}

\vskip 2.5truein

\includegraphics{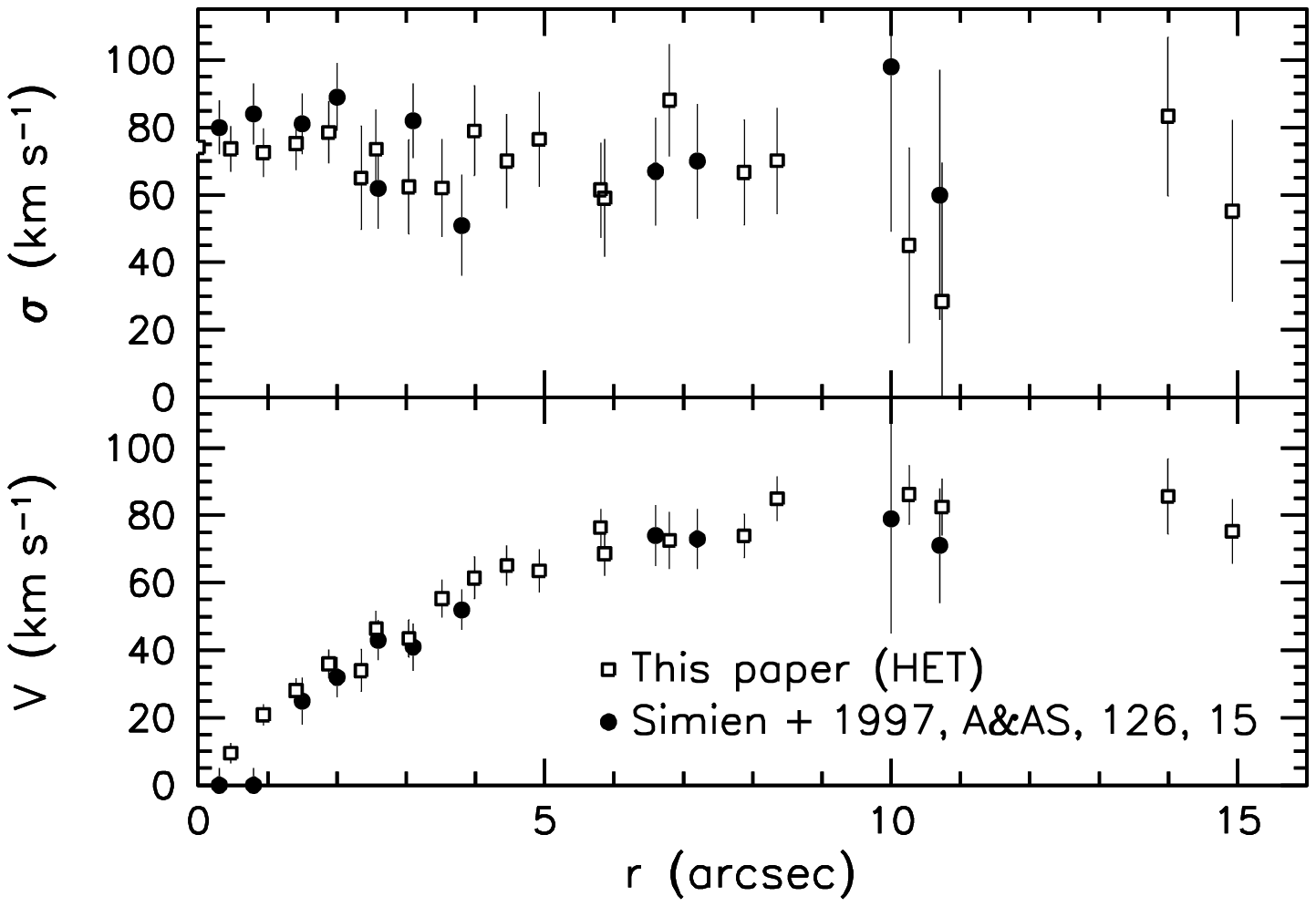}

\figcaption[]
{Absorption-line rotation curve $V(r)$ and velocity dispersion profile $\sigma(r)$
along the major axis of NGC 4318. 
}
\vskip 10pt

      Taking all these signs together, we identify NGC 4318 as
a low-luminosity S0 galaxy.  Figure 32 shows a decomposition into a 
S\'ersic function bulge and an exponential disk.  The bulge has an entirely normal
S\'ersic index of $n = 2.1 \pm 0.4$.

      NGC 4489 is classified E in the RC3 and S0 in the VCC.
It~appears in our photometry to consist of two components (Fig.~31).  The galaxy
is reasonably isolated.  
It is very round, so the $a_4$ profile is not informative.  We classify 
it as an S0, but this is uncertain.  There 
is a sharp isophote twist of $\sim 80^\circ$ between the ``bulge'' and 
the ``disk'' implied by the profile  decomposition in Figure~31.  Given 
suitable structure and viewing geometry, this could be consistent with 
either an E or an S0 classification.

      NGC 4564 is classified E in the RC3 and E6 in the VCC, but the 
brightness profile has the two-component structure of a bulge plus disk, 
and the $a_4$ profile shows a strong disky distortion at the radii
of the extra light (Fig.~31).  This is clearly an almost-edge-on~S0.
Scorza \etal (1998) observed a similar $a_4$ profile; by decomposing 
the two-dimensional brightness distribution into an elliptical galaxy 
component with exactly elliptical isophotes and a disk that accounts 
for the observation that $a_4 > 0$, they estimated that the bulge-to-total 
luminosity ratio is 0.71.  This is probably an underestimate, 
because low-luminosity, coreless ellipticals have isophotes
that are intrinsically disky, and all of the disky distortion was
ascribed to the S0 disk in the decomposition.  The disk of NGC 4564 
is also detected in the doppler asymmetry in the spectral line profiles
(Gauss-Hermite moment $h_3$: Halliday et al.~2001).

      NGC 4660 is classified E:~in the RC3 and E/S0 in the VCC, but it is
a bulge-dominated S0.  Figure 30 shows that extra light above
an almost-$r^{1/4}$ brightness profile coincides with a 
maximum in the $\epsilon$ profile and a very disky value of $a_4$.
These features are well known (Bender \etal 1988; Rix \& White 1990;
Scorza \& Bender 1995); a photometric decomposition implies that the 
disk contains $\sim 1/4$ of the light.  As in NGC 4564, the spectral line 
profiles of NGC 4660 show the kinematic signature of a dynamically cold, 
rapidly-rotating component added to a dynamically hot, slowly 
rotating component (Bender \etal 1994; Scorza \& Bender 1995).

\vfill\eject

%%%% Page 1 -- NGC 4472, NGC 4486 = M87 %%%%%%%%%%%%%%%%%%%%%%%%%%%

\begin{figure*}[b] 

\figurenum{11}

\vskip 9.0truein
 
\includegraphics{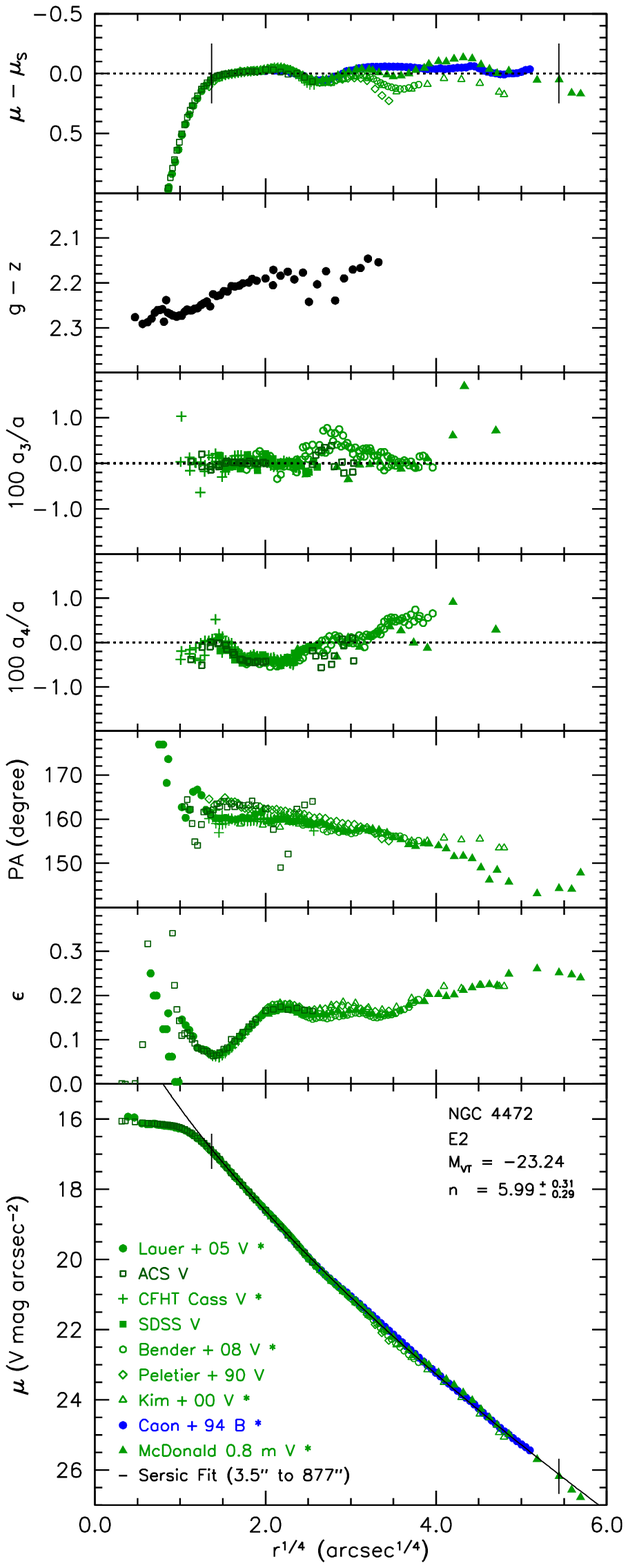}

\includegraphics{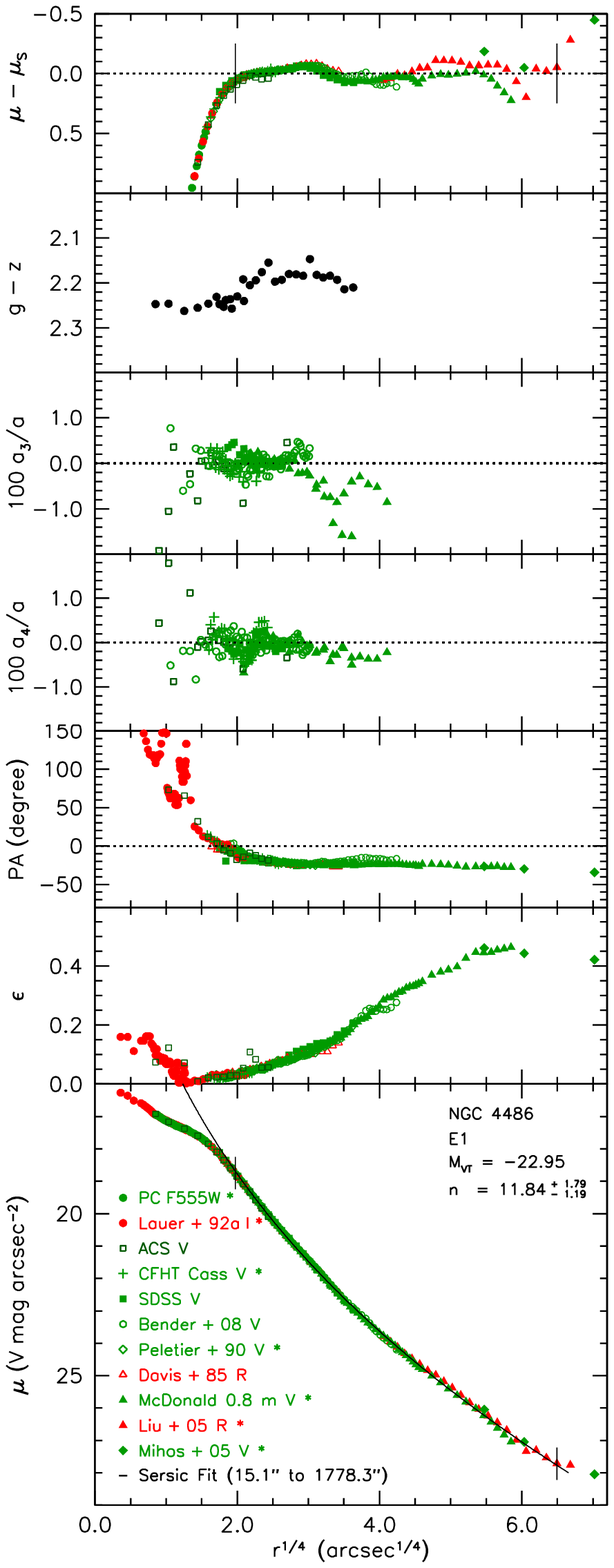}

\figcaption[]
{Composite brightness profiles of Virgo cluster elliptical galaxies ordered by total
absolute magnitude $M_{VT}$ (Column 11 of Table 1).  For each galaxy, the panels show, 
from bottom to top: 
surface brightness $\mu$, 
ellipticity $\epsilon$, 
position angle PA of the major axis east of north, 
the isophote shape parameters $a_4$ and $a_3$ (as percentages of the major-axis radius a $\equiv$ $r$), 
the $g - z$ color profile from {\it HST\/} ACS and from the SDSS, and 
the deviations of the individual profiles in the bottom panel from the best S\'ersic 
   function fit shown by the black curve ($n$ is in the key).  
The S\'ersic function is fitted between the vertical dashes crossing the profiles in the top and
bottom panels.  Note that $a_4 > 0$ implies disky isophotes and $a_4 < 0$ implies boxy isophotes. 
The profile data are color-coded so that blue corresponds to $B$ band, green corresponds 
to $g$ or $V$ band, red corresponds to $R$ or $I$ band, and brown corresponds to $H$ or $K$ band.  
These are the  brightest core galaxies in Virgo.}
\end{figure*}

\eject\clearpage

%%%% Page 2 -- NGC 4649, NGC 4406 %%%%%%%%%%%%%%%%%%%%%%%%%%%

\figurenum{12}

\begin{figure*}[b] 

\vskip 9.0truein

\includegraphics{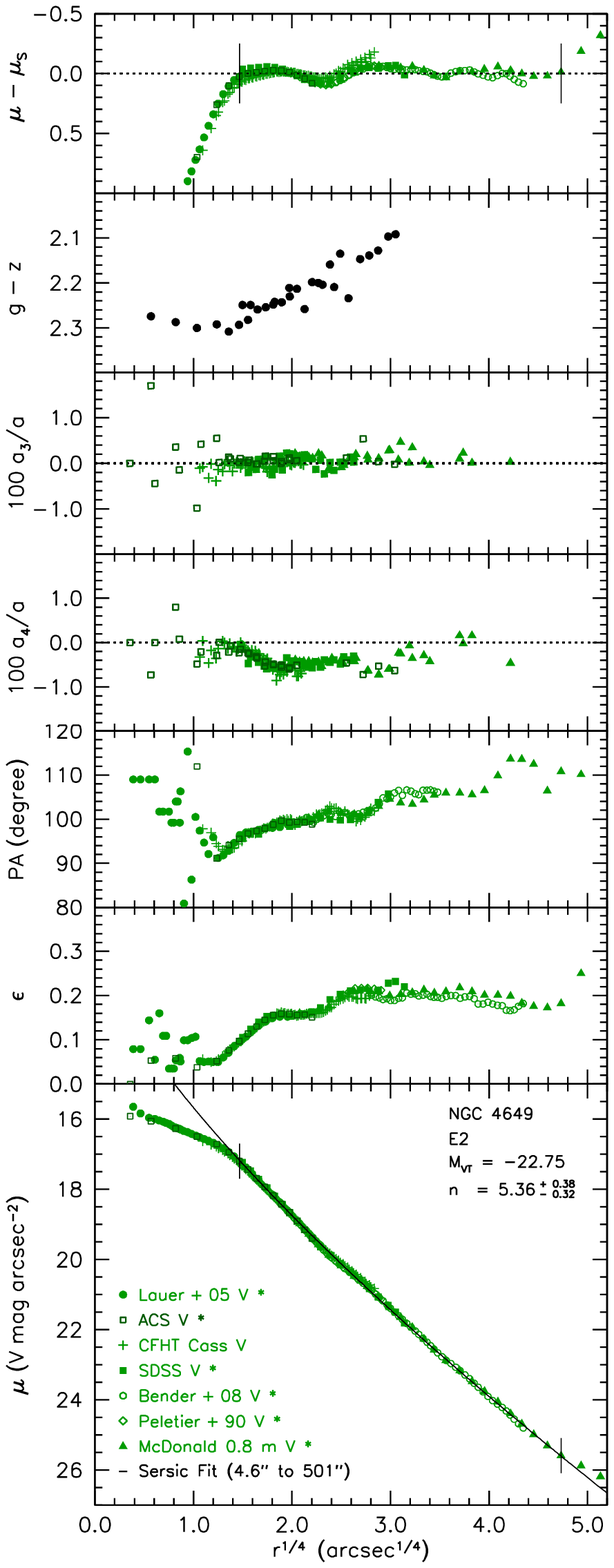}

\includegraphics{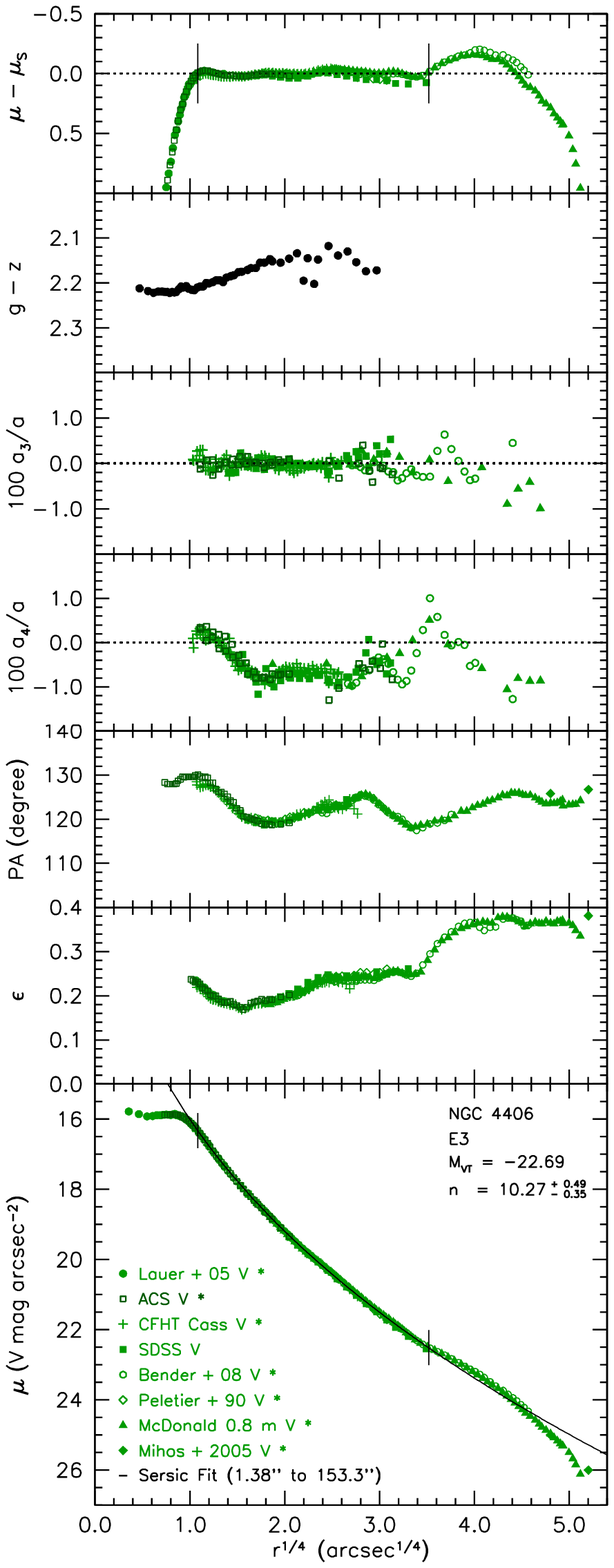}

\figcaption[]
{Photometry of Virgo cluster core ellipticals.  Inside the core of NGC 4406, the surface brightness 
drops slightly toward the center, making this a ``hollow core'' galaxy (Lauer \etal 2002).  The outer
profile of NGC 4406 is affected by many bracketing galaxies (see the text and Fig.~1 in Elmegreen
\etal 2000).}
\end{figure*}

\eject\clearpage

%%%% Page 3 -- NGC 4365, NGC 4374 %%%%%%%%%%%%%%%%%%%%%%%%%%%

\figurenum{13}

\begin{figure*}[b] 

\vskip 9.0truein

\includegraphics{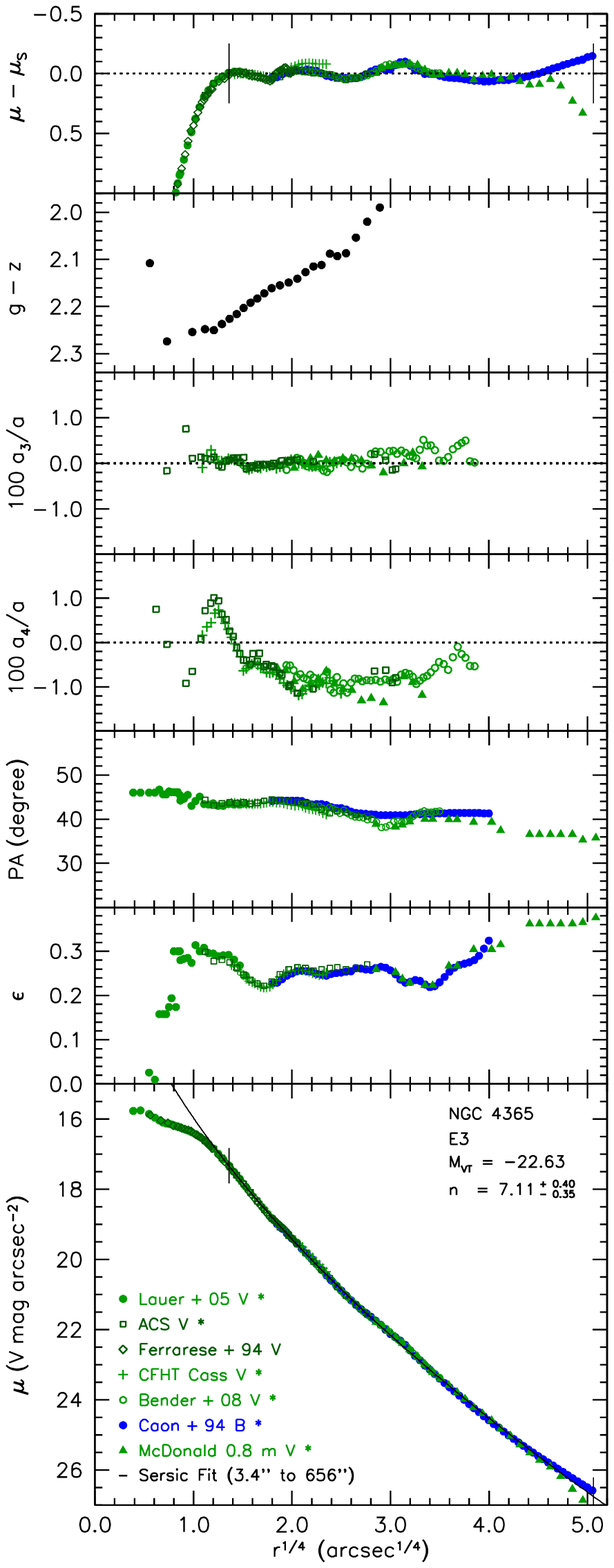}

\includegraphics{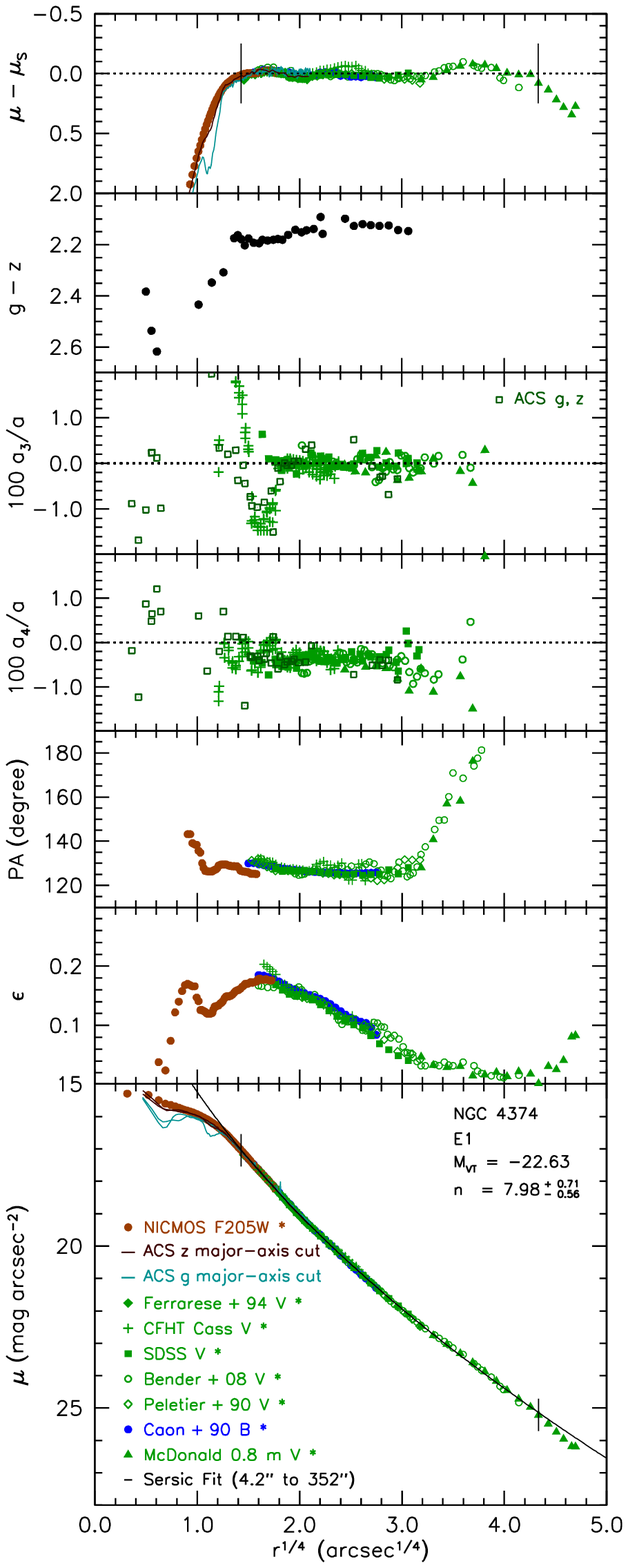}

\figcaption[]
{Photometry of Virgo cluster ellipticals with cuspy cores.  For NGC 4374, the ACS $g$-band (folded) 
cut profile illustrates the well known dust features
(V\'eron-Cetty \& V\'eron 1988; 
Jaffe \etal 1994;
van Dokkum \& Franx 1995; 
Bower \etal 1997; 
Ferrarese \etal 2006a, 
and references therein), but the {\it HST\/} NICMOS F205W profile is almost unaffected by absorption.}
\end{figure*}

\eject\clearpage

%%%% Page 4 -- NGC 4261, NGC 4382 %%%%%%%%%%%%%%%%%%%%%%%%%%%

\figurenum{14}

\begin{figure*}[b] 

\vskip 9.0truein

\includegraphics{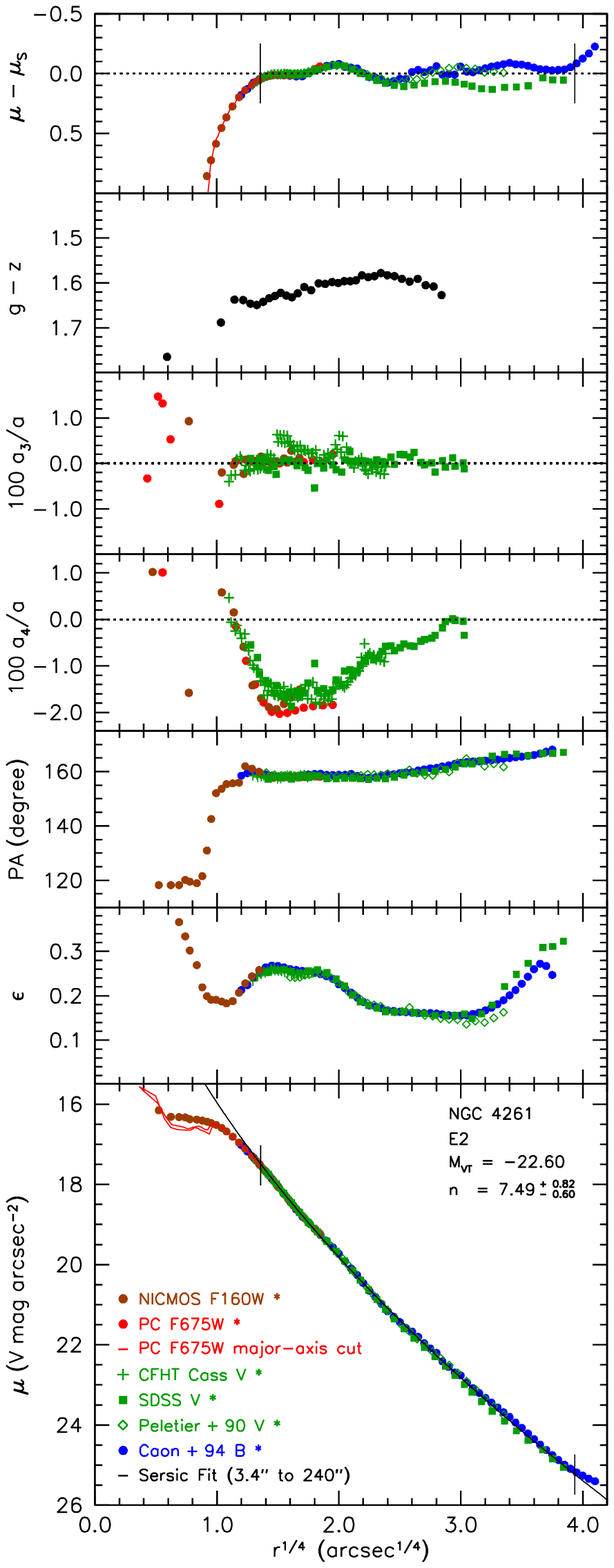}

\includegraphics{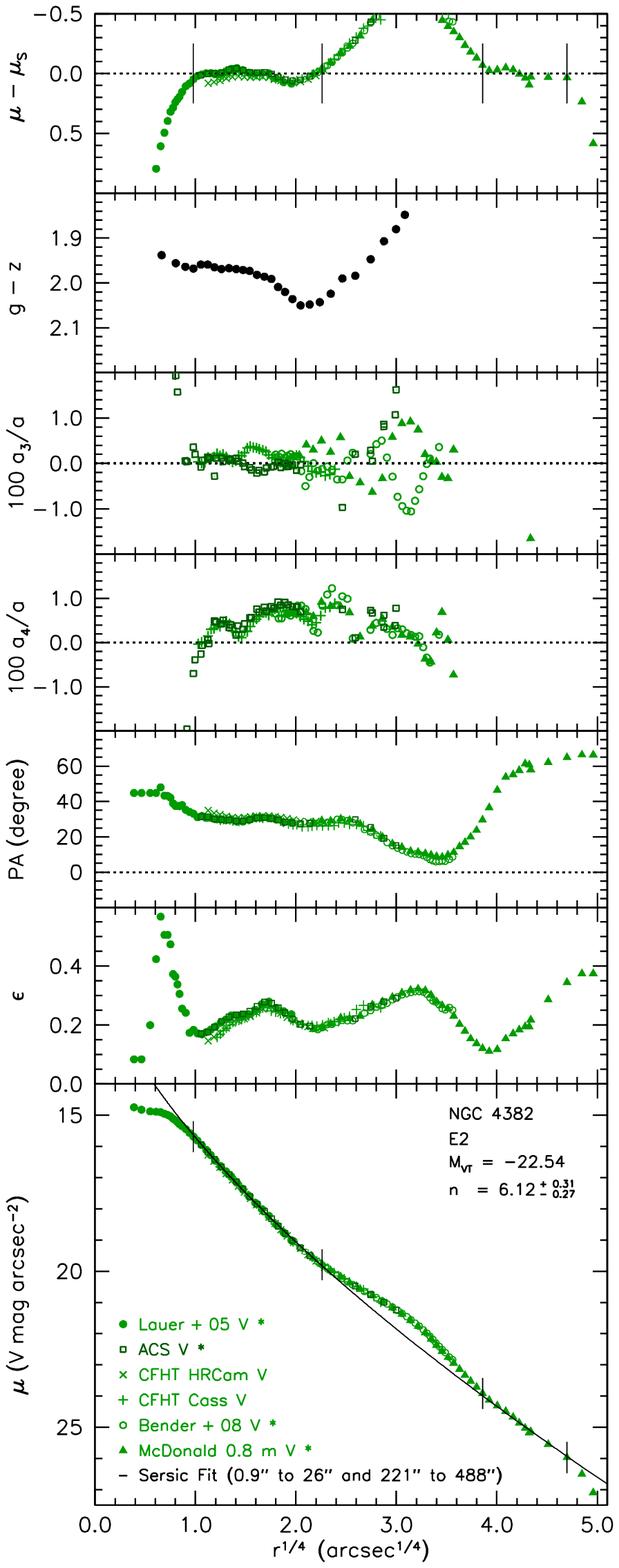}

\figcaption[]
{Photometry of elliptical galaxies with cuspy cores.  
The central dust disk of NGC 4261
(Kormendy \& Stauffer 1987;
M\"ollenhoff \& Bender 1987a,~b;
Jaffe \etal 1993, 1994,  1996;
van Dokkum \& Franx 1995;
Ferrarese, Ford, \& Jaffe 1996;
Martel \etal 2000)
is evident in the (folded) PC F675W cut profile.  However, the NICMOS F160W profile is almost 
unaffected by absorption; the identification of the core is not in doubt.  NGC 4261 is in the 
background of the Virgo cluster ($D = 31.6$ Mpc, Tonry \etal 2001).  NGC 4382 has a complicated 
profile that we interpret as the signature of an unrelaxed recent merger (\S\ts7.5).  Two 
alternative S\'ersic fits to the galaxy are discussed in Appendix A; all are consistent with 
the fundamental plane projections discussed~in~\S\ts8.
}
\end{figure*}

\eject\clearpage

%%%% Page 5 -- NGC 4636, NGC 4552 %%%%%%%%%%%%%%%%%%%%%%%%%%%

\figurenum{15}

\begin{figure*}[b] 

\vskip 9.0truein

\includegraphics{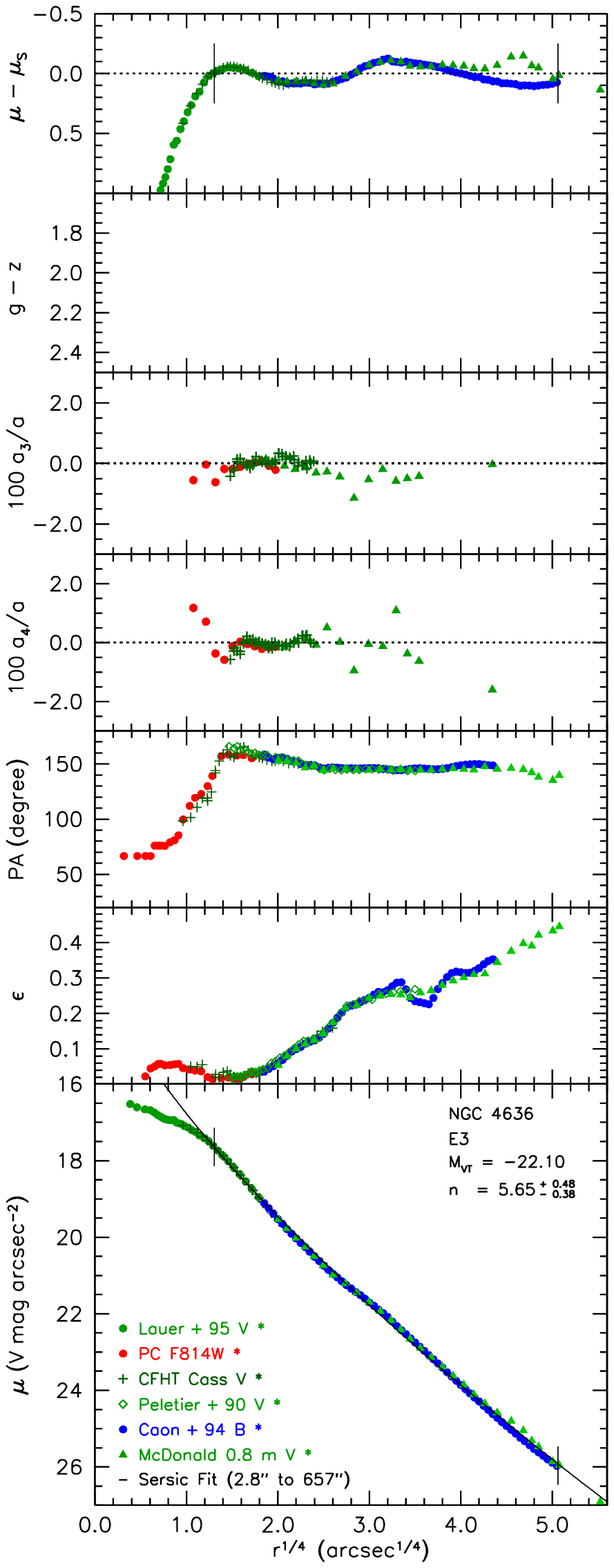}

\includegraphics{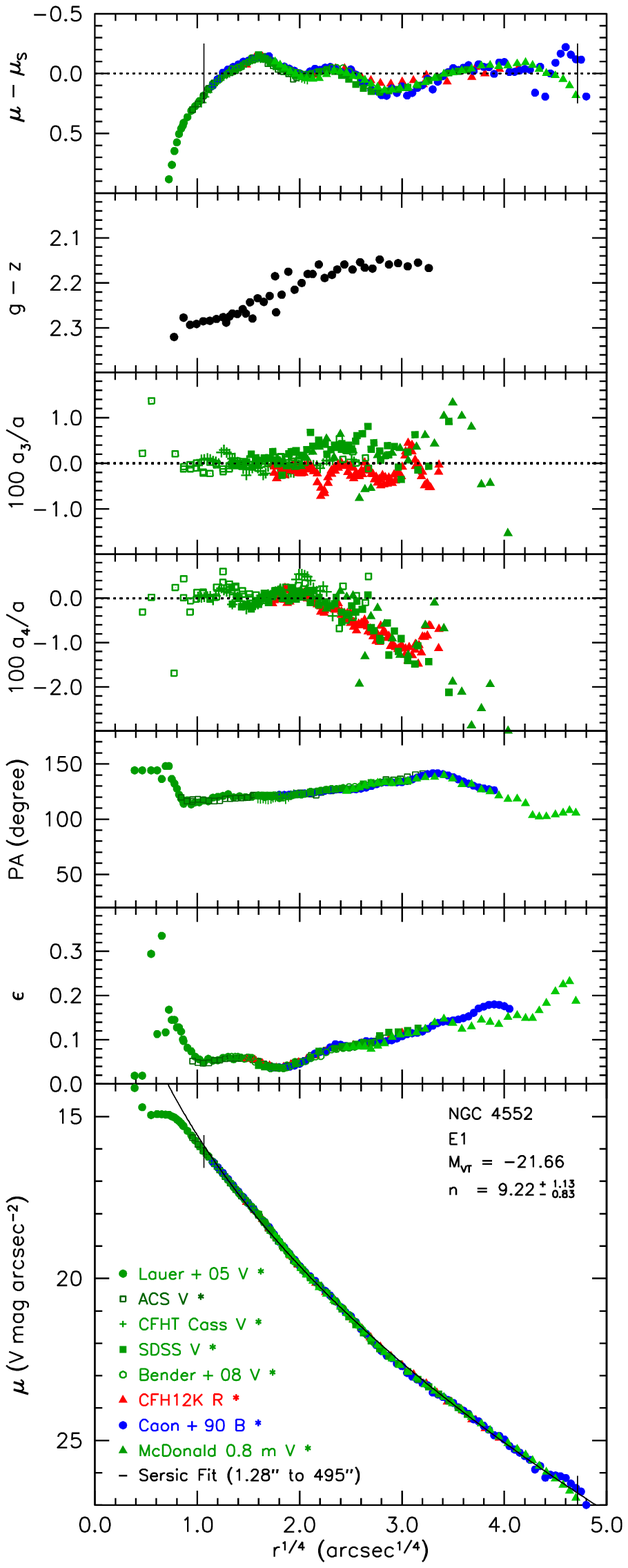}

\figcaption[]
{Photometry of the lowest-luminosity core ellipticals in the Virgo cluster.
}
\end{figure*}

\eject\clearpage

%%%% Page 6 -- NGC 4621, NGC 4459 %%%%%%%%%%%%%%%%%%%%%%%%%%%

\figurenum{16}

\begin{figure*}[b] 

\vskip 9.0truein

\includegraphics{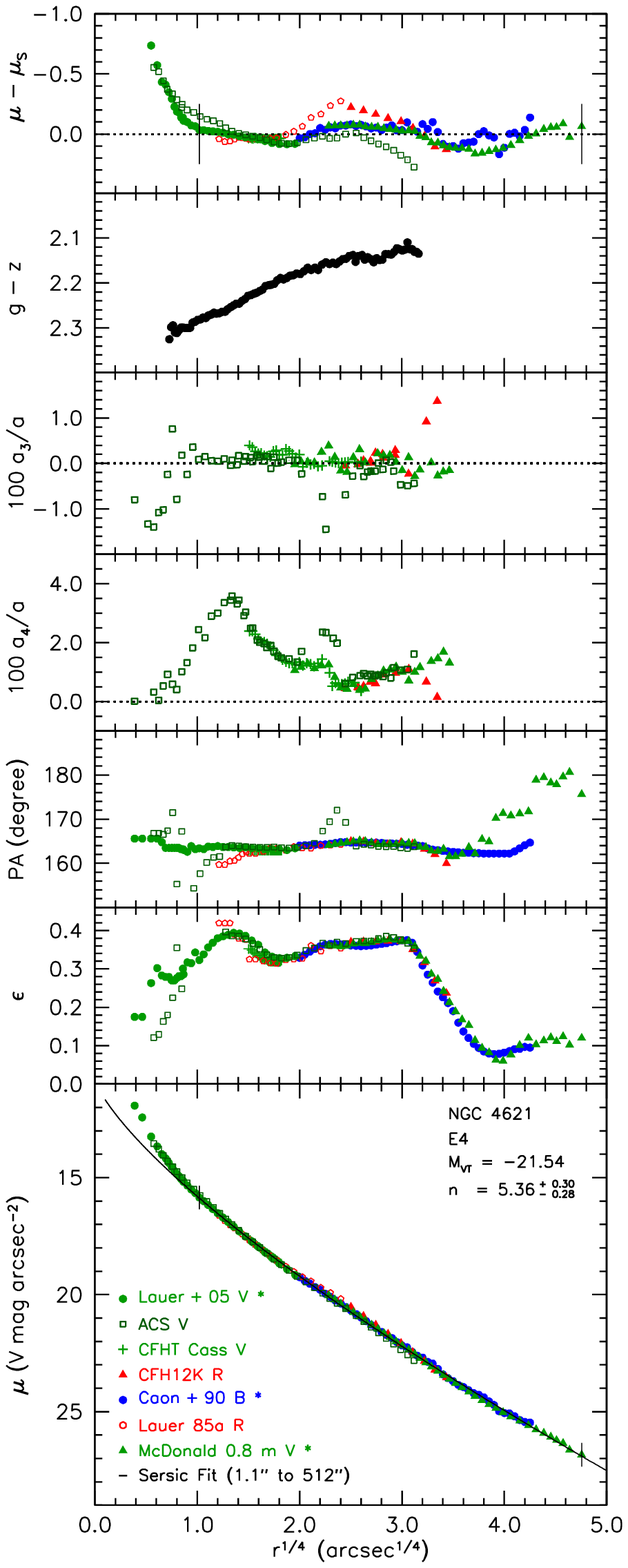}

\includegraphics{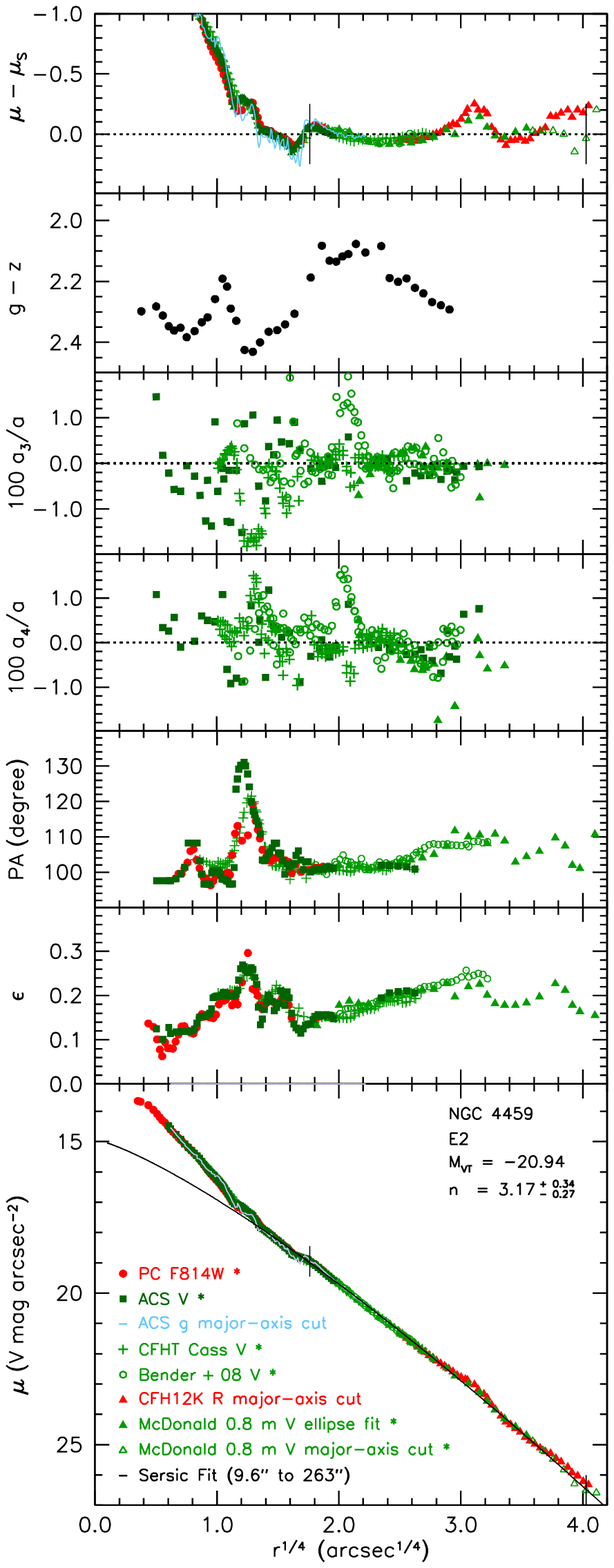}

\figcaption[]
{\lineskip=0pt \lineskiplimit=0pt
Photometry of the highest-luminosity extra light ellipticals in the Virgo cluster
NGC 4621 is the exception to the correlation between $n$ and core properties 
discussed in \S\ts9.  NGC 4459 has a prominent dust disk at $1^{\prime\prime}$ 
\lapprox \ts$r$ \lapprox $10^{\prime\prime}$ (e.{\ts}g., 
de Vaucouleurs 1959;
Sandage 1961; 
Sandage\& Bedke 1994;
Sarzi et al.~2001;
Ferrarese et al.~2006a). 
Therefore, a major-axis $g$-band cut profile is shown as well as the ellipse fit results.  
It shows that the dust absorption is only $\sim$ 0.3 mag deep and is easily 
avoided.  The profile is fitted only exterior to the dust disk; the S\'ersic index is 
robustly less than 4.  There is substantial extra light near the center for any S\'ersic 
fit to the profile outside the dust disk.
}
\end{figure*}

\eject\clearpage

%%%% Page 7 -- NGC 4473, 4478 %%%%%%%%%%%%%%%%%%%%%%%%%%%

\figurenum{17}

\begin{figure*}[b] 

\vskip 9.0truein

\includegraphics{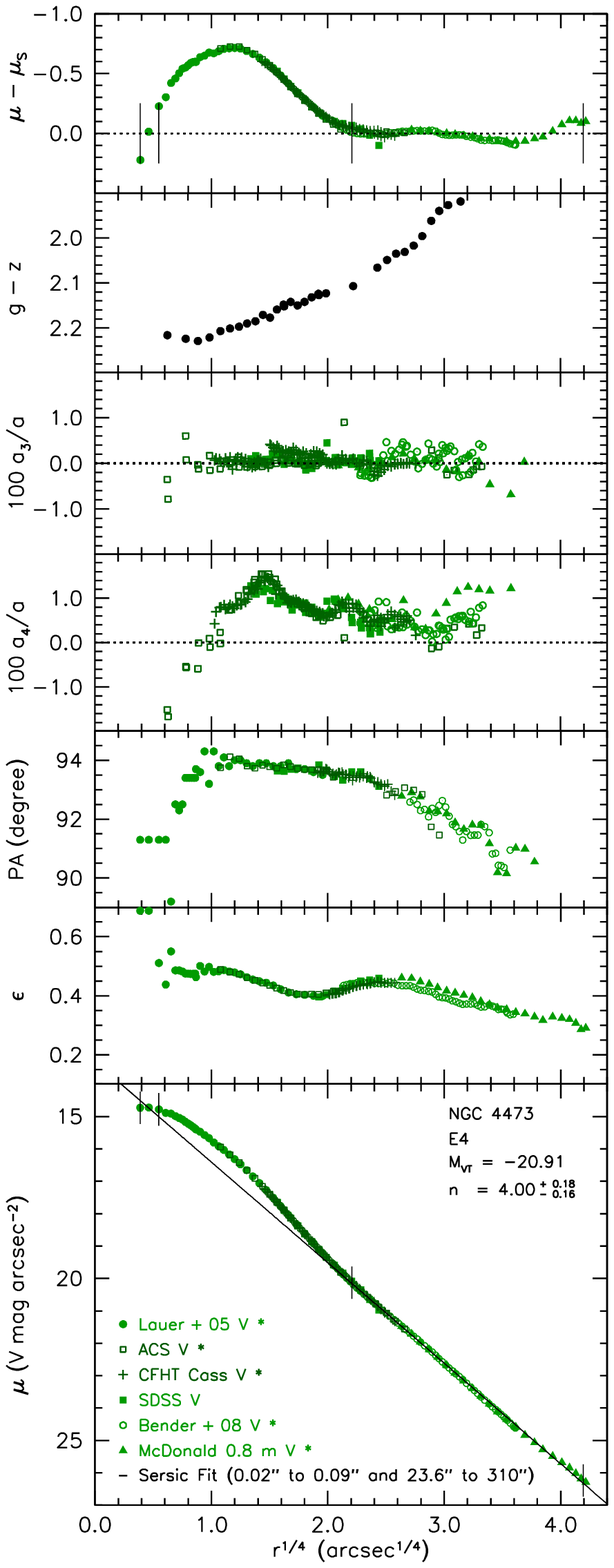}

\includegraphics{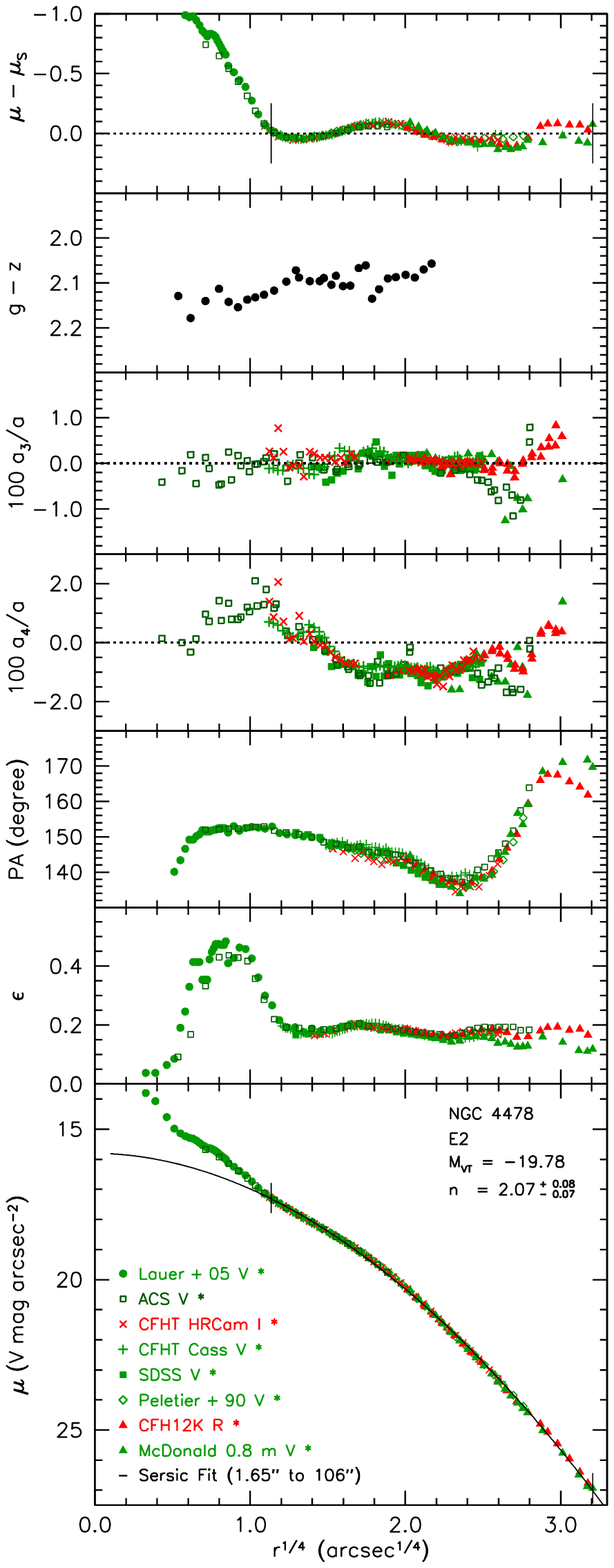}

\figcaption[]
{Photometry of Virgo cluster ellipticals with extra light near the center.  In NGC 4473,
this takes the unusual form of a counter-rotating stellar disk 
(Cappellari \etal 2004; 
Cappellari \& McDermid 2005;
Cappellari et al.~2007; 
see \S\ts9.2 here for discussion).
}
\end{figure*}

\eject\clearpage

%%%% Page 8 -- NGC 4434, NGC 4387 %%%%%%%%%%%%%%%%%%%%%%%%%%%

\figurenum{18}

\begin{figure*}[b] 

\vskip 9.0truein

\includegraphics{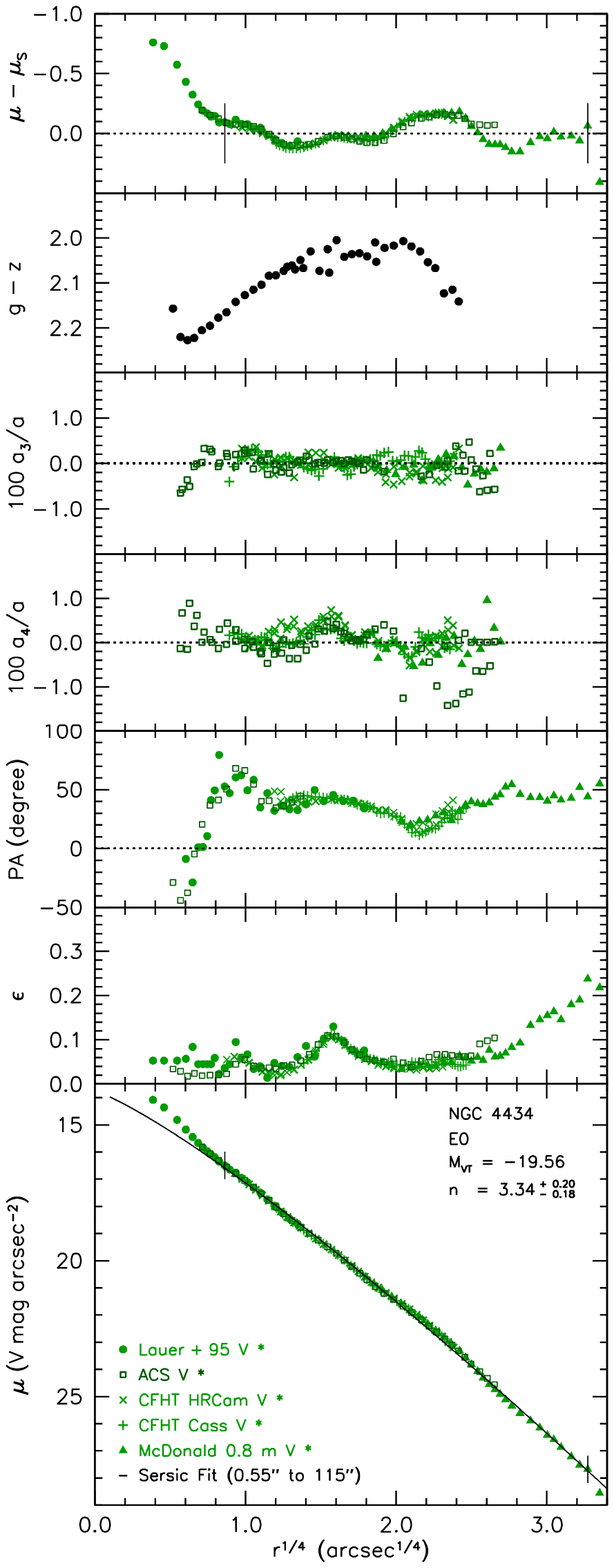}

\includegraphics{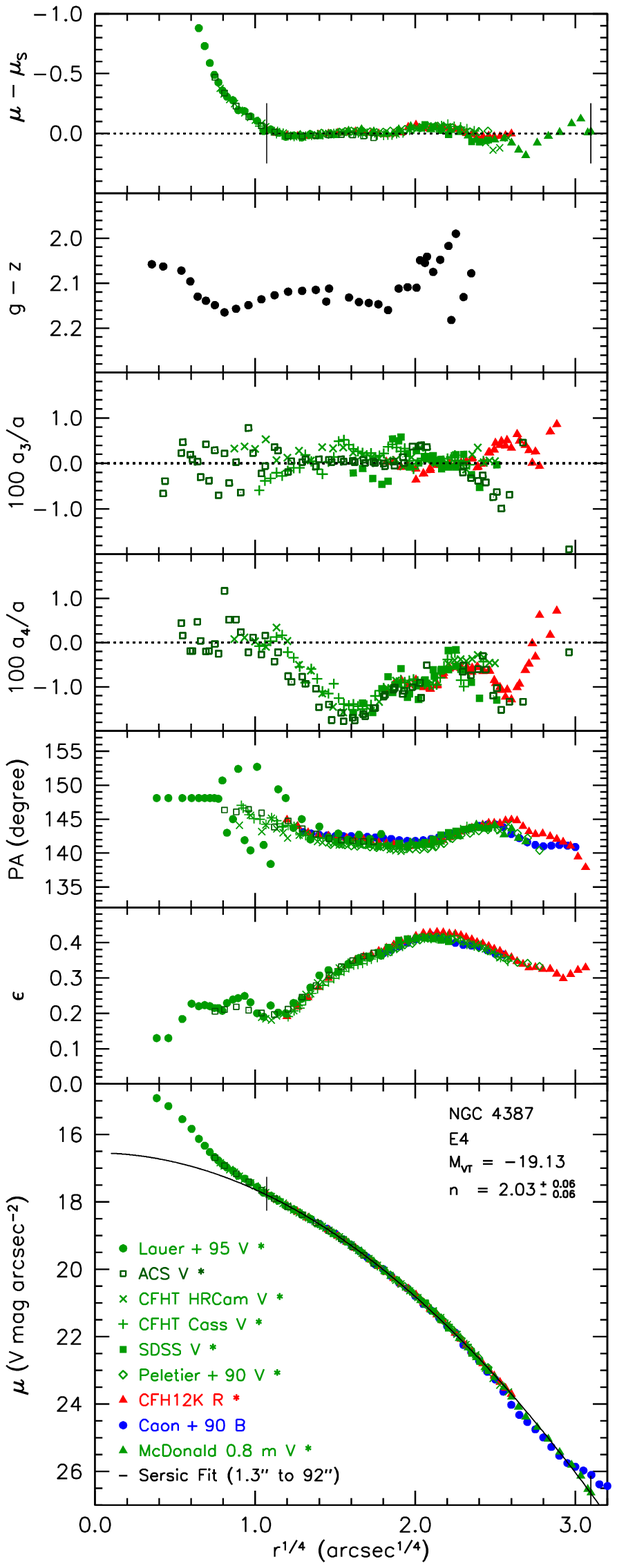}

\figcaption[]
{Photometry of extra light ellipticals.  NGC 4434 is in the background of the Virgo 
cluster ($D = 22.4$ Mpc, Mei \etal 2007), but it behaves like other faint ellipticals. 
}
\end{figure*}

\eject\clearpage

%%%% Page 9 -- NGC 4551, NGC 4458 %%%%%%%%%%%%%%%%%%%%%%%%%%%

\figurenum{19}

\begin{figure*}[b] 

\vskip 9.0truein

\includegraphics{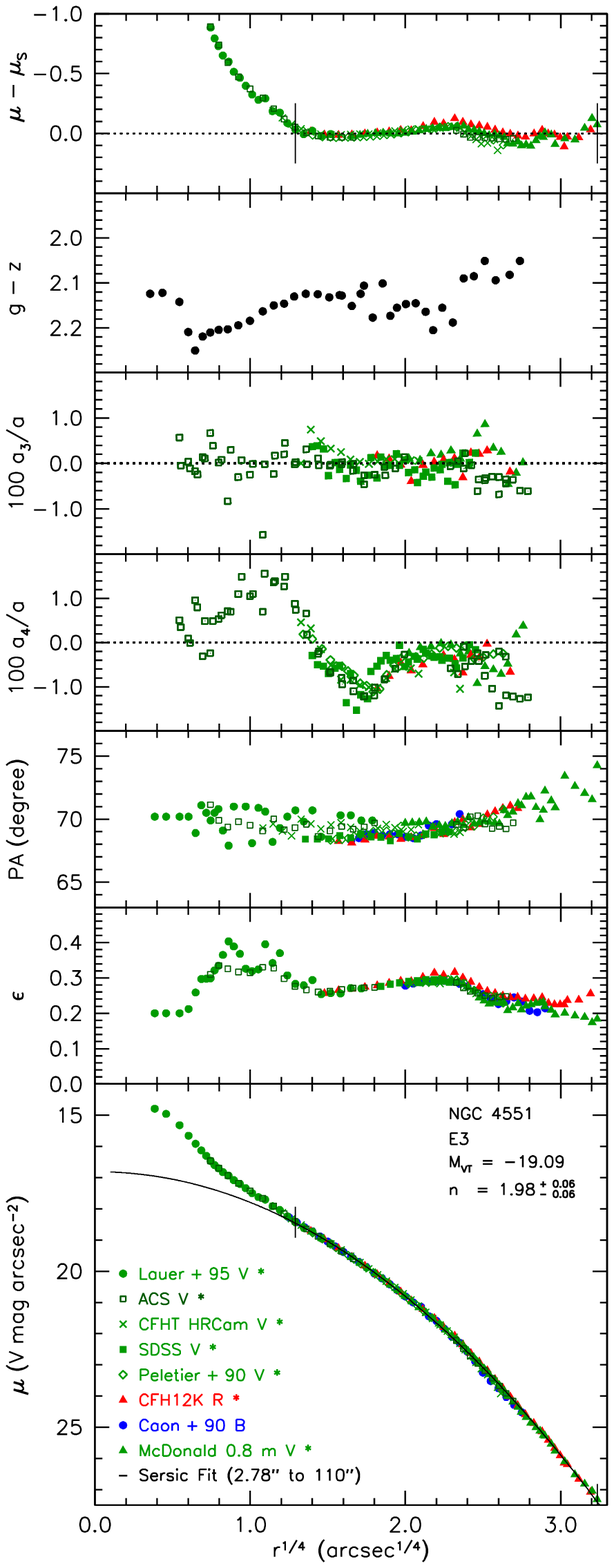}

\includegraphics{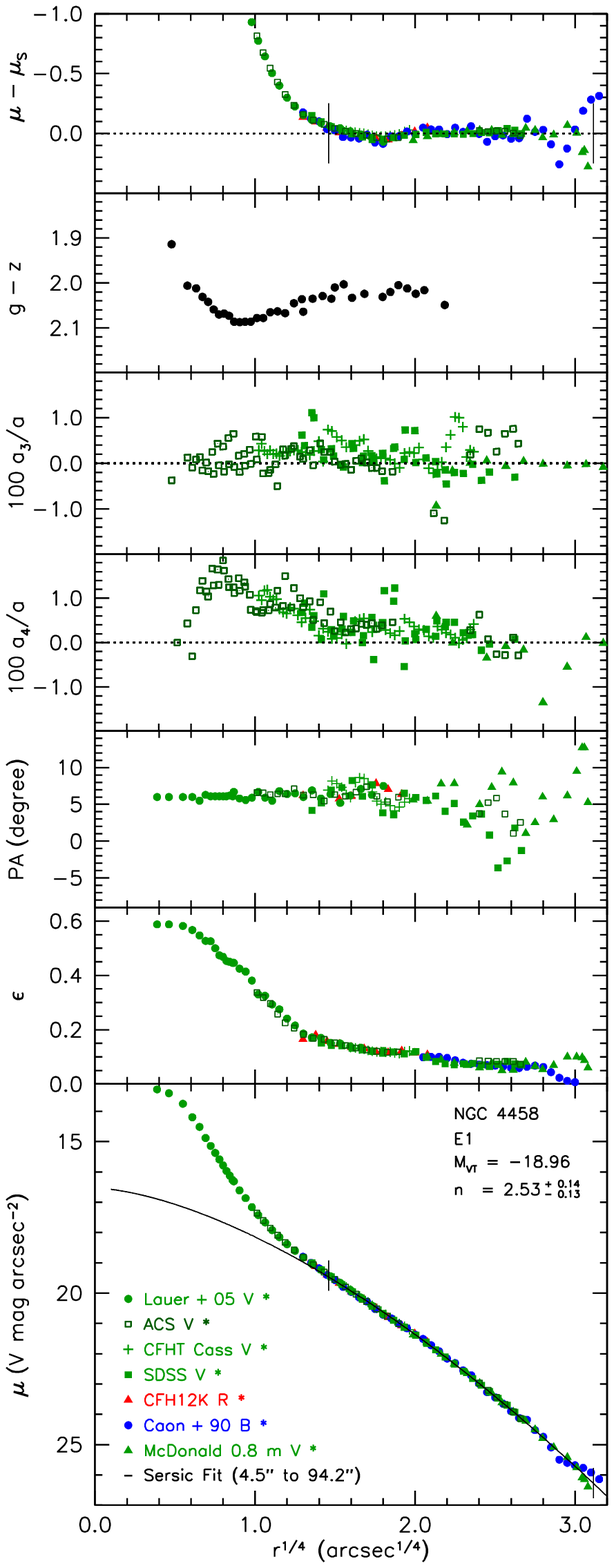}

\figcaption[]
{Photometry of Virgo cluster ellipticals with central extra light.  Note that
the extra light component in NGC 4458 -- like that in M{\ts}32 (Fig.~3) -- is especially well
resolved spatially.  ``Extra light'' is very different from ``nuclei'', that is, tiny nuclear 
star clusters such as that in M{\ts}33 (Kormendy \& McClure 1993; Lauer \etal 1998;
see \S\ts9.7 here and Hopkins \etal 2008b for further discussion).
}
\end{figure*}

\eject\clearpage

%%%% Page 10 -- NGC 4486A %%%%%%%%%%%%%%%%%%%%%%%%%%%

\figurenum{20}

\begin{figure*}[b] 

\vskip 9.0truein

\includegraphics{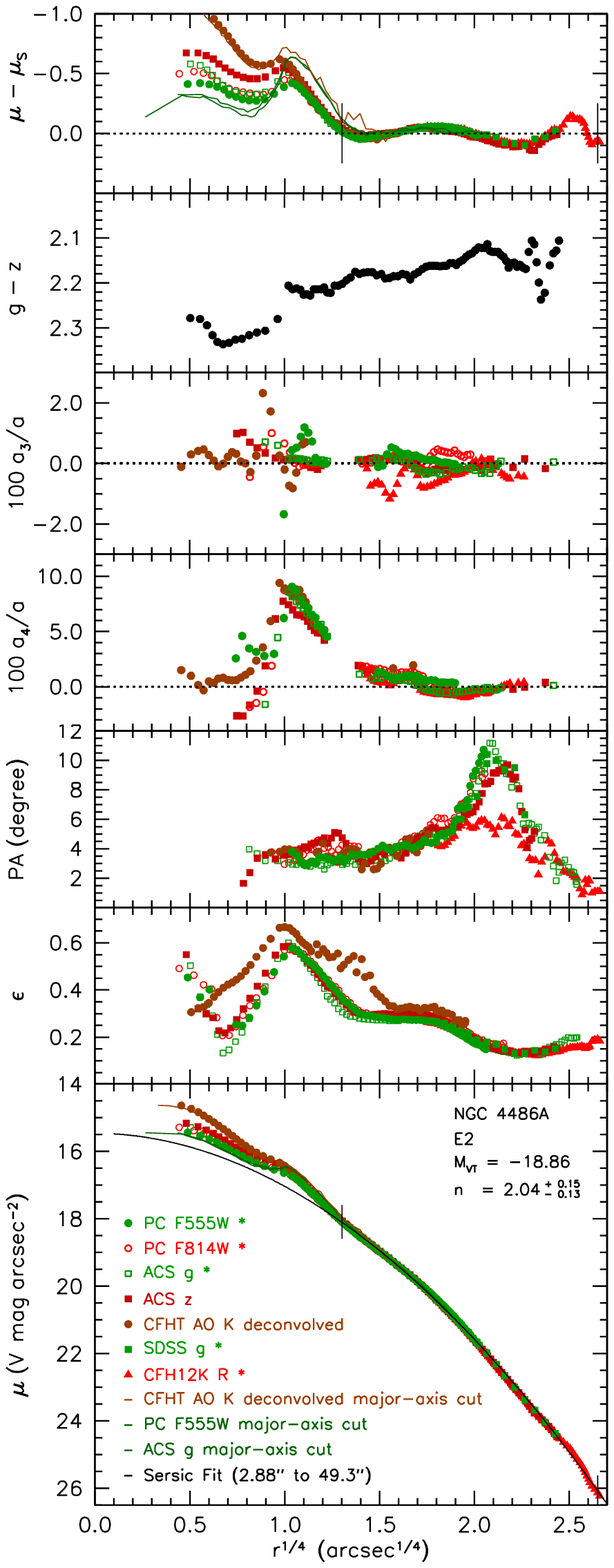}

\includegraphics{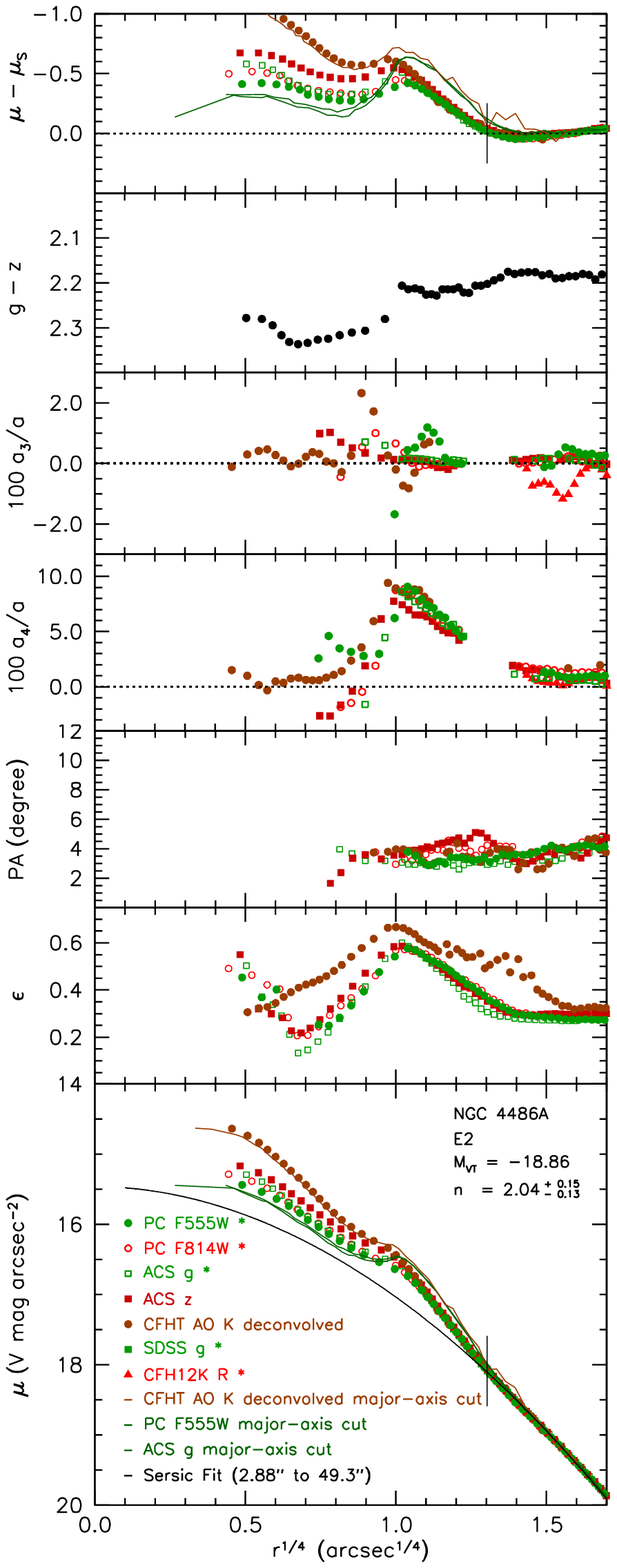}

\figcaption[]
{Photometry of Virgo cluster dwarf elliptical NGC 4486A plotted to show the overall 
profile ({\it left\/}) and an expanded region near the center ({\it right\/}).  
The major-axis cut profiles derived from the HST PC (F555W) and ACS ($g$-band)
provide (and are illustrated with) independent $V$-band zeropoints.  We adopt the mean 
of these two zeropoints.  The amount of extra light at the center is underestimated 
by the $V$ profiles, because the extraordinarily strong nuclear disk (note that $a_4/a$ 
reaches almost 10\ts\%) has an embedded, edge-on dust lane ar radii 
$r$ \lapprox \ts$1^{\prime\prime}$ (Kormendy et al.~2005).  The absorption is more 
obvious in major-axis cut profiles ({\it lines\/}) than in ellipse-fit profiles
({\it points\/}).  Also, as expected, the absorption is strongest in $V$ and $g$,
less strong in ACS $z$, and least strong in the CFHT adaptive optics $K$-band image.  
But the kink in the profile at $1^{\prime\prime}$ suggests that there is some 
absorption even in $K$ band (see Kormendy et al.~2005 for further discussion).
The electronic tables provide both a pure $V$-band profile and one that has 
a $V$-band zeropoint but the $K$-band profile substituted at $r^{1/4} \leq 1.1$.
\lineskip=-6pt \lineskiplimit=-6pt
}
\end{figure*}

\eject\clearpage

%%%% Page 11 -- NGC 4515, NGC 4464 %%%%%%%%%%%%%%%%%%%%%%%%%%%

\figurenum{21}

\begin{figure*}[b] 

\vskip 9.0truein

\includegraphics{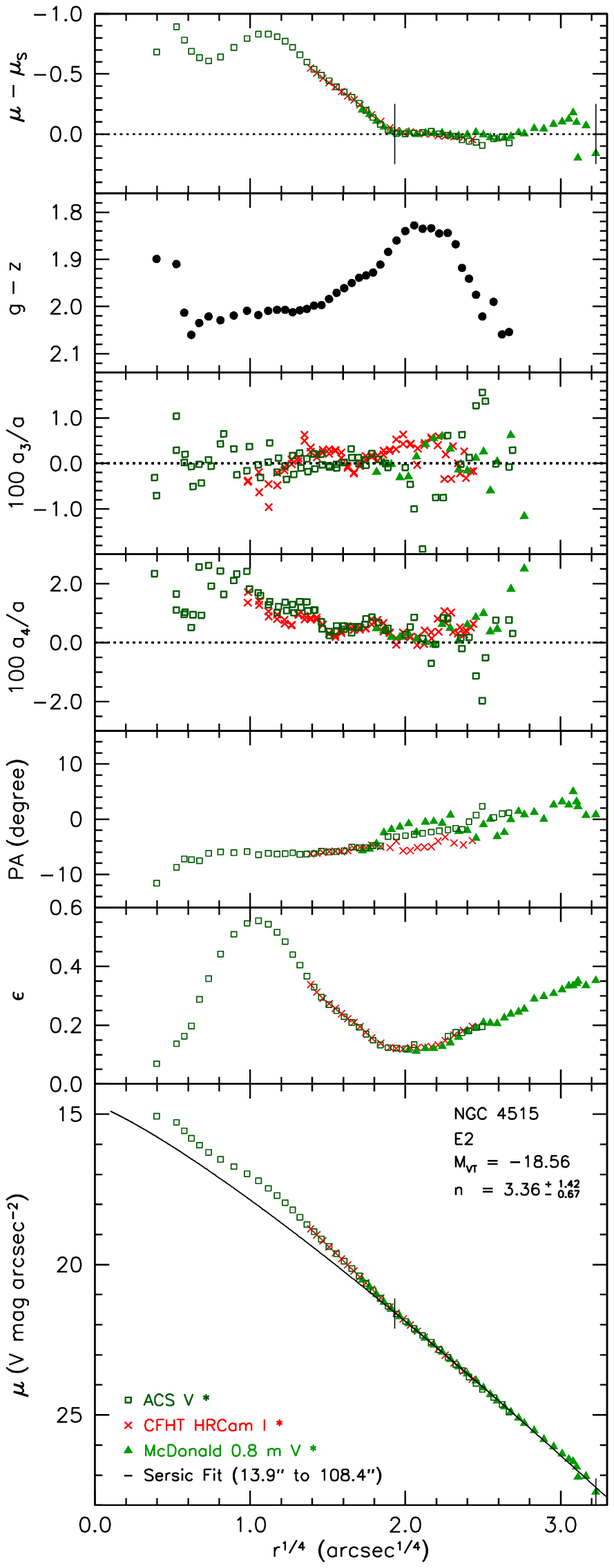}

\includegraphics{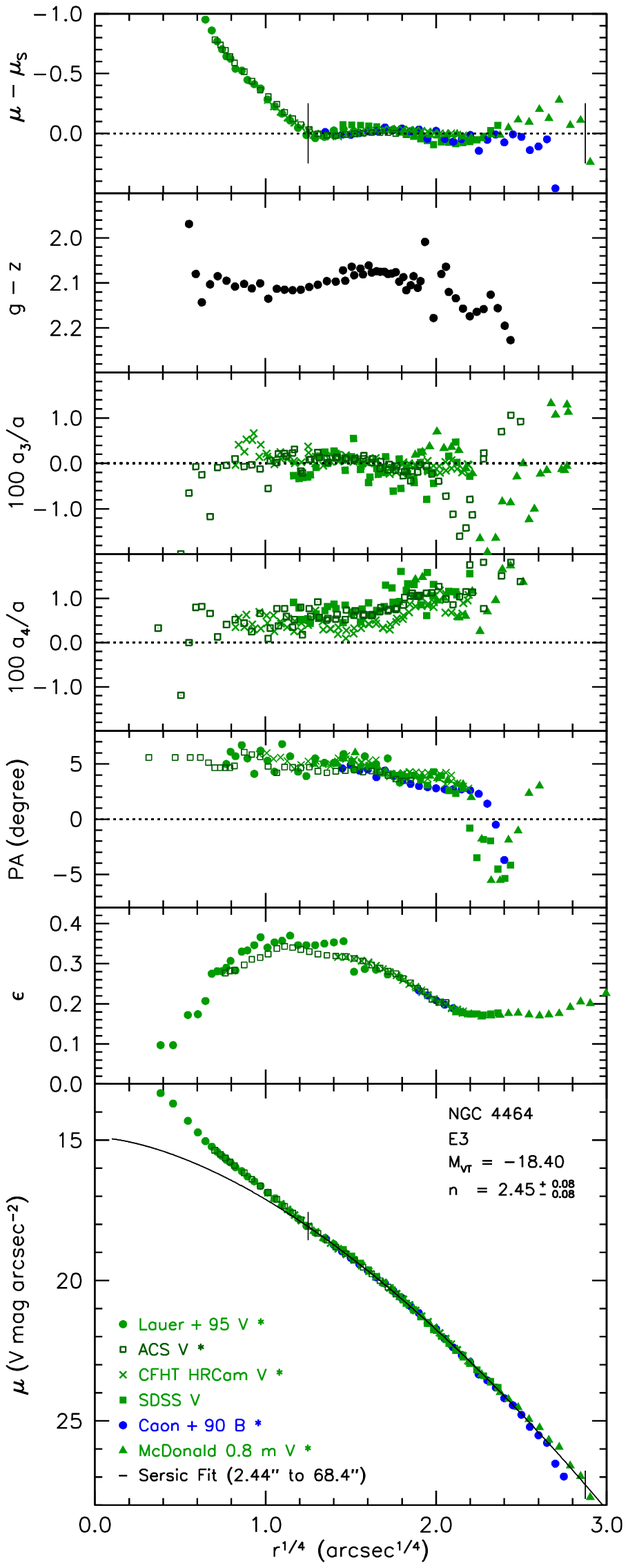}

\figcaption[]
{ Photometry of Virgo cluster ellipticals with central extra light.  For NGC 4415, 
the choice of fit range is discussed in Figures 62 and 63 (Appendix A).  These show
two alternative fits to the major-axis profile and a fit to the minor-axis profile.
For NGC 4464, the PA glitch at $r^{1/4} \simeq 2.4$ is probably not real.
}
\end{figure*}

\eject\clearpage

%%%% Page 12 -- NGC 4486B, VCC 1871 %%%%%%%%%%%%%%%%%%%%%%%%%%%

\figurenum{22}

\begin{figure*}[b] 

\vskip 9.0truein

\includegraphics{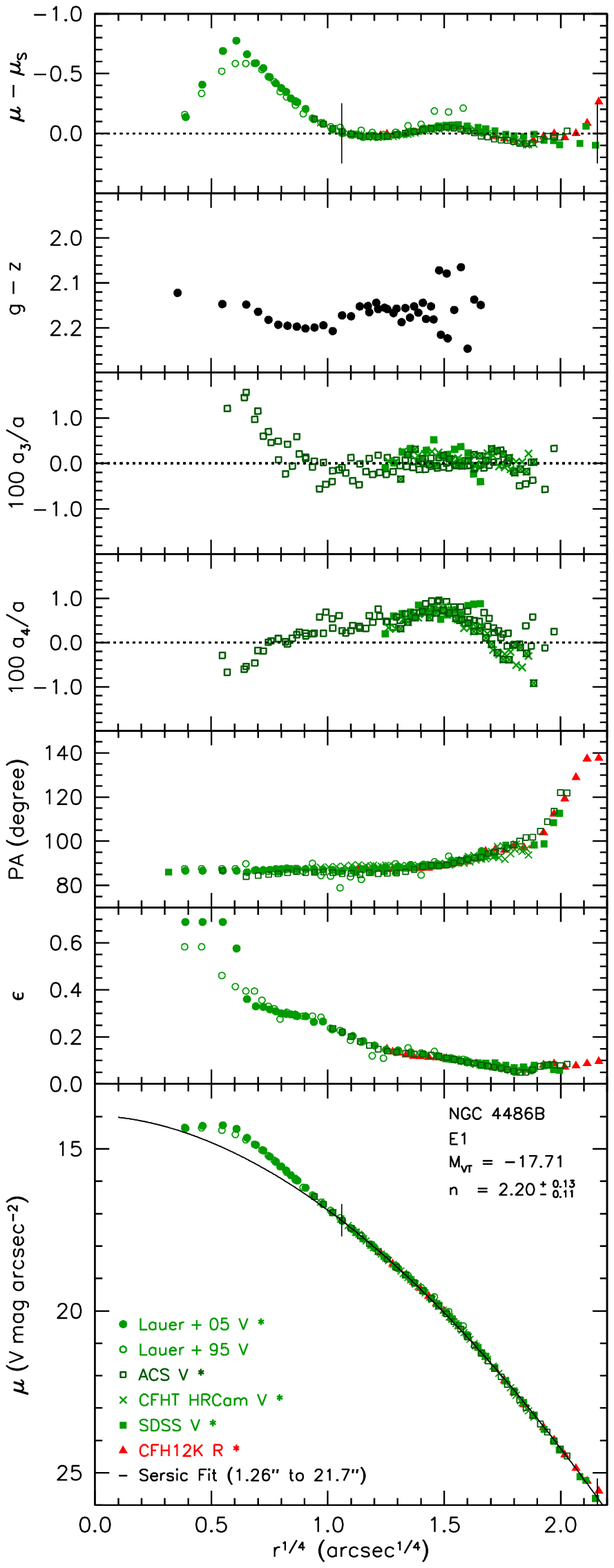}

\includegraphics{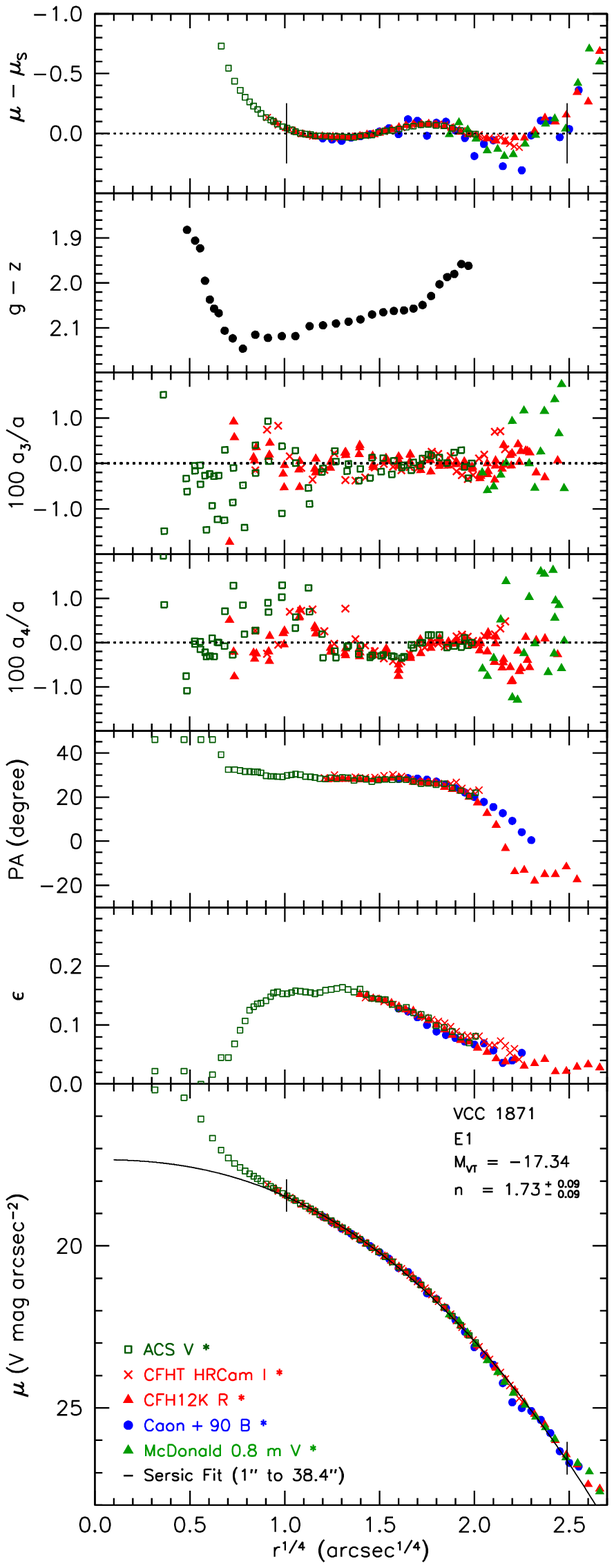}

\figcaption[]
{Photometry of Virgo elliptical galaxies.  These are typical M{\ts}32-like, 
dwarf ellipticals with extra light near the center. NGC 4486B does not -- 
contrary to appearances -- have a core; the central flattening of the profile is 
an effect of the double nucleus (Lauer \etal 1996).  Exterior to the double nucleus, 
the profile shows extra light, as usual for a low-luminosity elliptical.  The isophotes
of NGC 4486B twist toward M{\ts}87 at large radii.  This appears
to be real and not an effect of the overlapping isophotes of the larger galaxy.
We had to model and subtract the overlapping light from M{\ts}87, but it varies on
such a large scale, and NGC 4486B is so small, that it is routine to produce images
that have flat sky surrounding the smaller galaxy.  Four images from three telescopes
give consistent PA measurements.
}
\end{figure*}

\eject\clearpage

%%%% Page 13 -- NGC 4467 = VCC 1192, VCC 1440 %%%%%%%%%%%%%%%%%%%%%%%%%%%

\figurenum{23}

\begin{figure*}[b] 

\vskip 9.0truein

\includegraphics{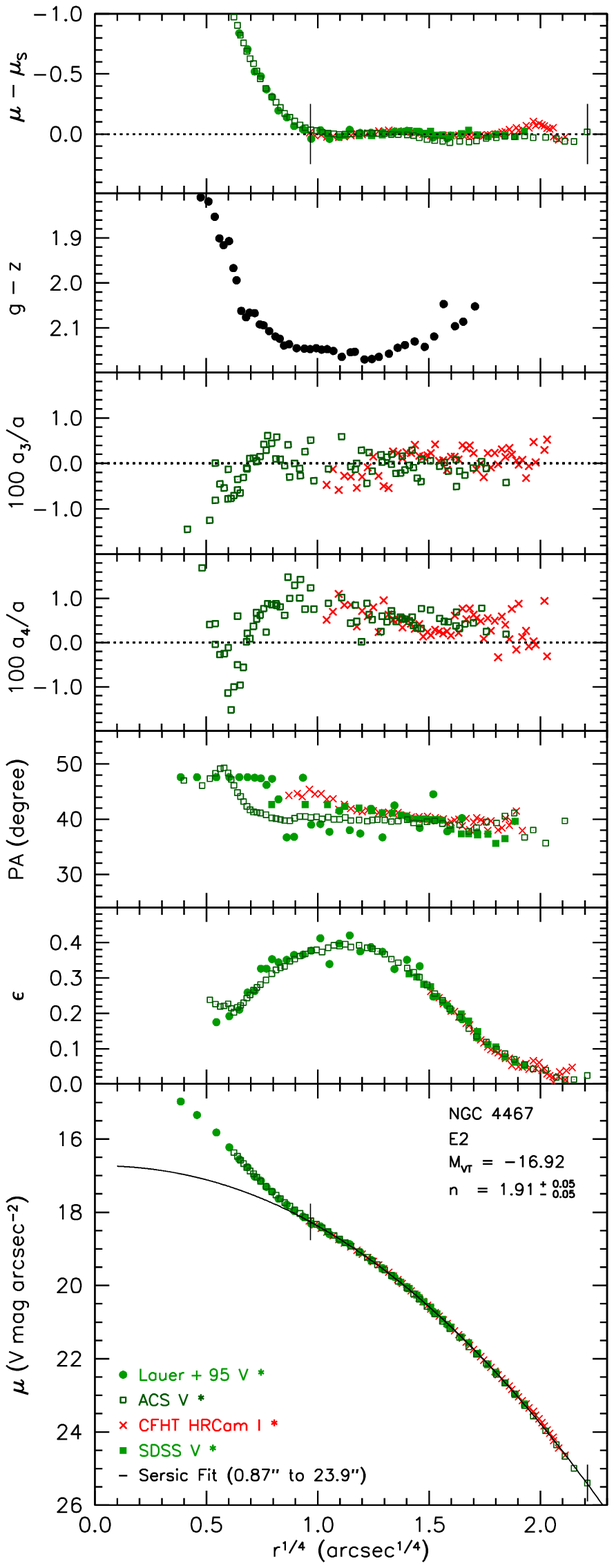}

\includegraphics{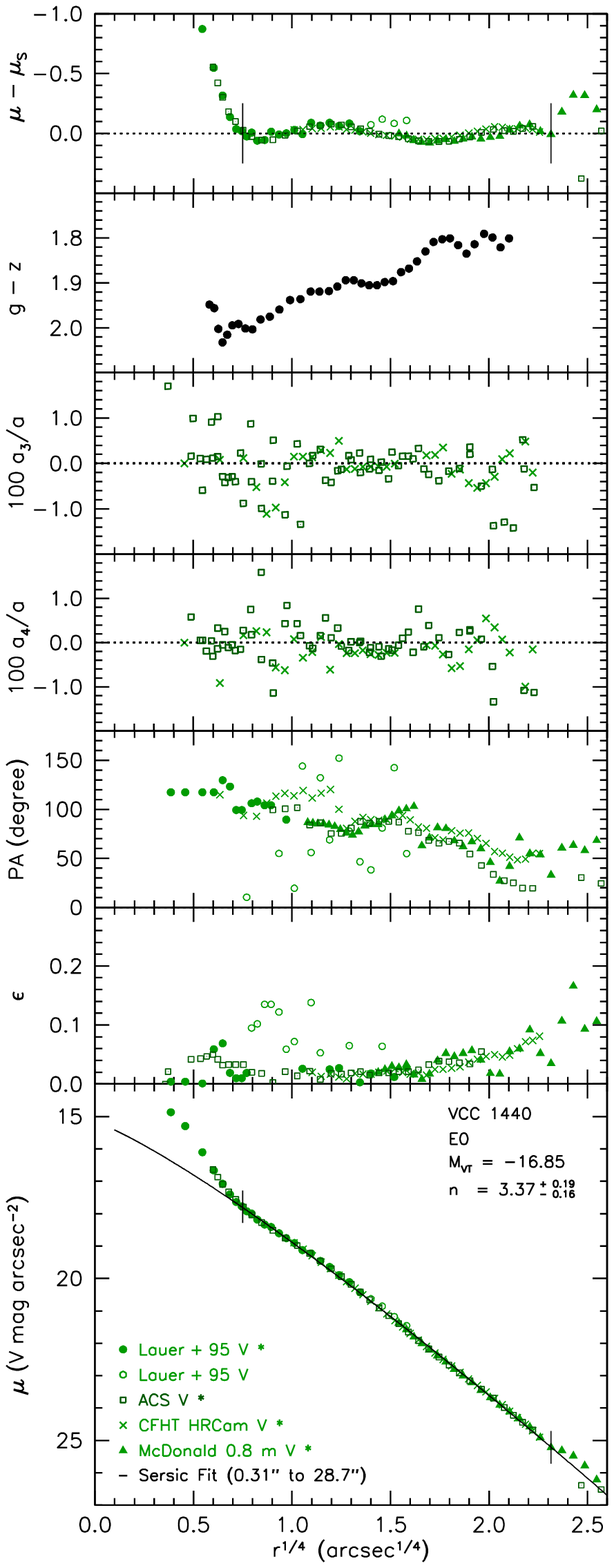}

\figcaption[]
{Photometry of Virgo cluster ellipticals.  These galaxies and the two on the next
page are the faintest ellipticals known in the Virgo cluster.  All four have extra
light near the center with respect to the inward extrapolations of well defined S\'ersic
function fits to the outer profiles.  These galaxies are very similar to M{\ts}32;
recall that M{\ts}32 has $M_{VT} = -16.69$ and $n = 2.82 \pm 0.07$.}
\end{figure*}

%%%% Page 14 -- VCC 1627, VCC 1199 %%%%%%%%%%%%%%%%%%%%%%%%%%%

\figurenum{24}

\begin{figure*}[b] 

\vskip 9.0truein

\includegraphics{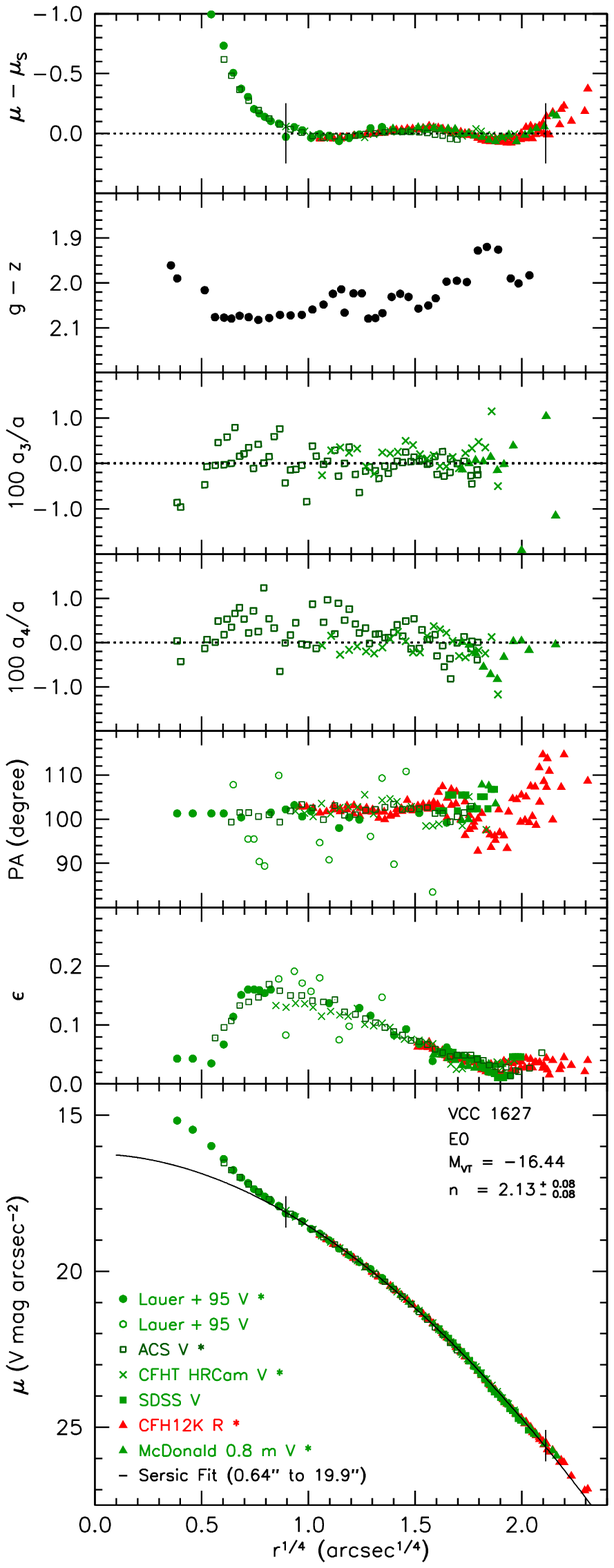}

\includegraphics{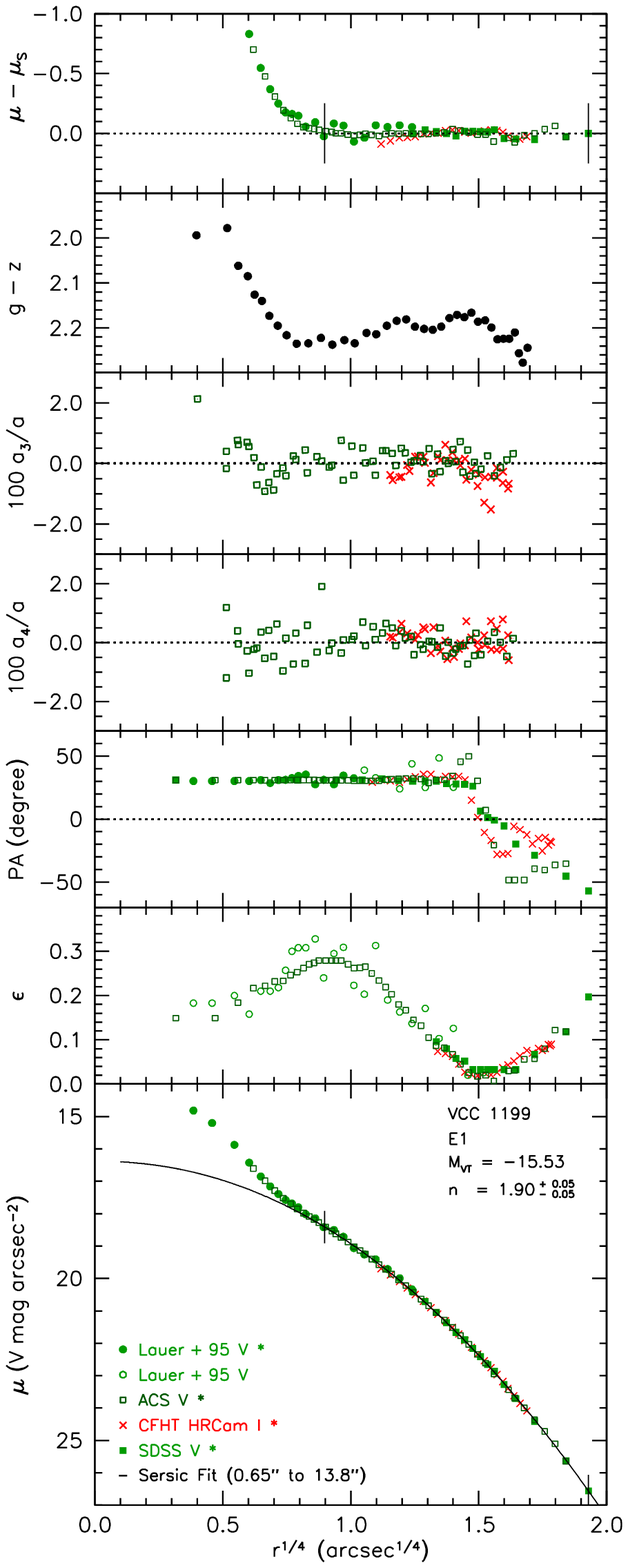}

\figcaption[]
{Photometry of faintest known ellipticals in the Virgo cluster.  They are slightly 
fainter than M{\ts}32, which has $M_{VT} = -16.69$ and $n = 2.82 \pm 0.07$.
}
\end{figure*}

%%%%%%%%%%%%%%%%%%%%%%%%%%%%%%%%%%%%%%%%%%%%%%%%%%%%%
%                                                   %
% Figure 6: Spheroidal galaxies                     %
%                                                   %
%%%%%%%%%%%%%%%%%%%%%%%%%%%%%%%%%%%%%%%%%%%%%%%%%%%%%

%%%% Page 15 -- NGC 4482, VCC 1087 %%%%%%%%%%%%%%%%%%%%%%%%%%%

\figurenum{25}

\begin{figure*}[b] 

\vskip 9.0truein

\includegraphics{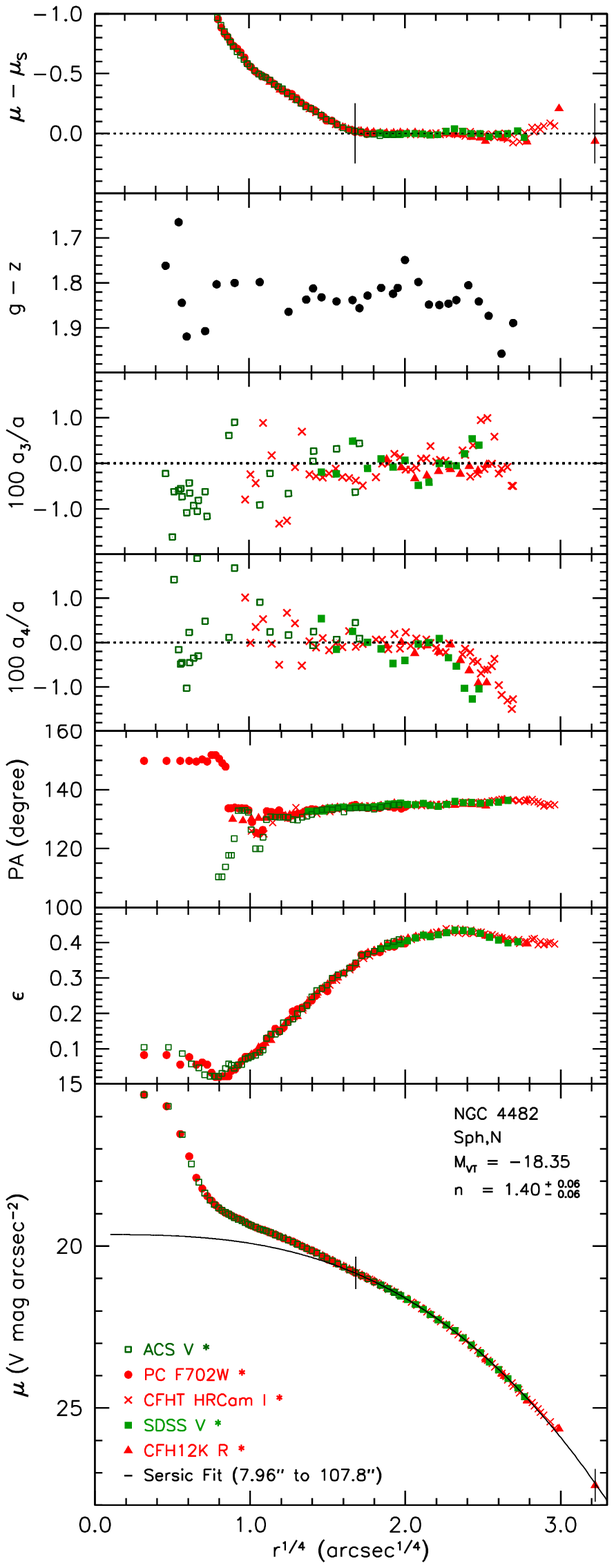}

\includegraphics{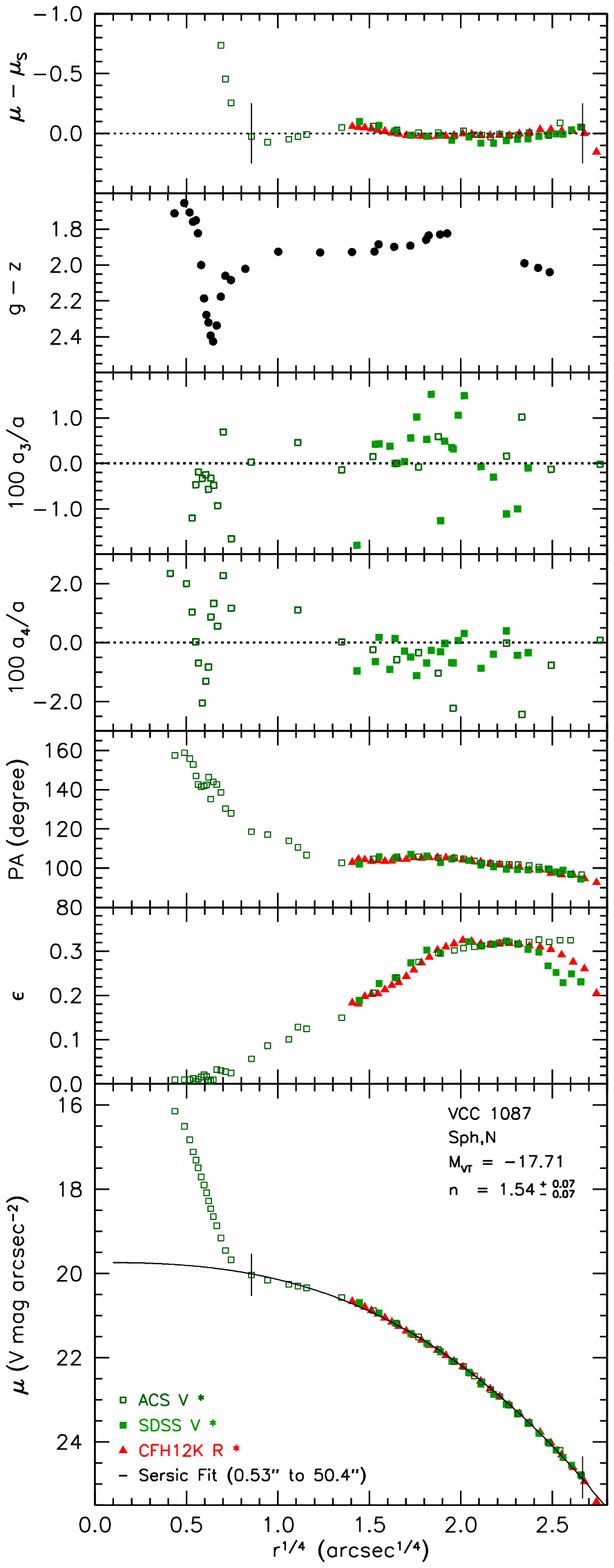}

\figcaption[]
{Composite brightness profiles of Virgo cluster spheroidal (Sph) galaxies
ordered by total absolute magnitude $M_{VT}$.  Symbols, parameters,
and color coding are as in Figures 11 -- 24.
}
\end{figure*}

%%%% Page 16 -- VCC 1355, VCC 1910 %%%%%%%%%%%%%%%%%%%%%%%%%%%

\figurenum{26}

\begin{figure*}[b] 

\vskip 9.0truein

\includegraphics{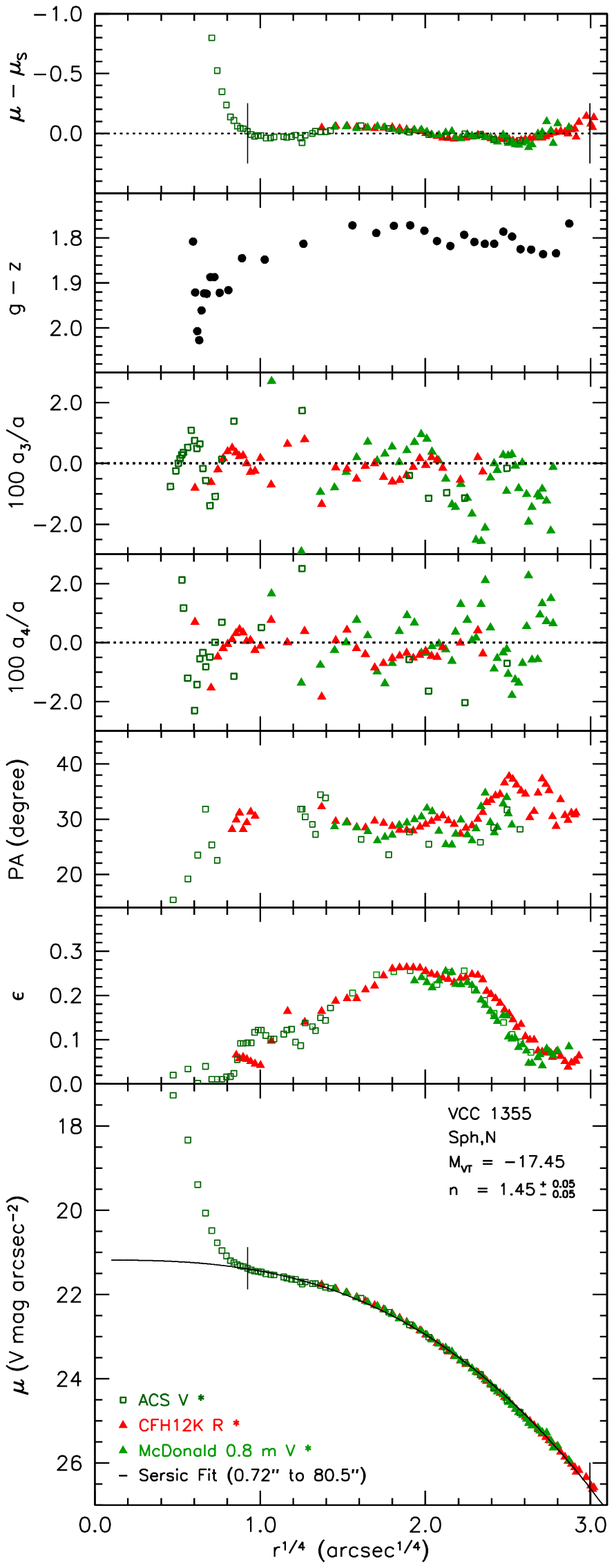}

\includegraphics{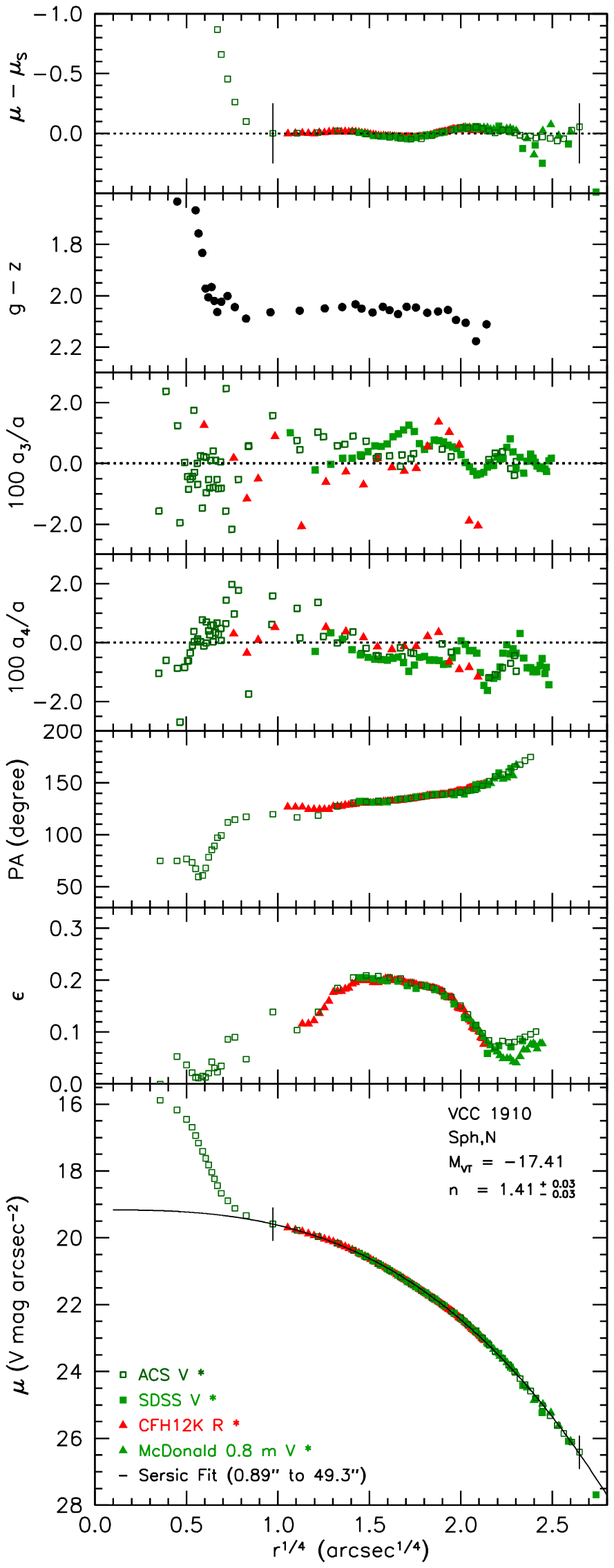}

\figcaption[]
{Photometry of Virgo cluster Sph galaxies.  In VCC 1910, the outermost part of 
the PA twist and the outer rise in $\epsilon$ may be spurious (caused by PSF 
overlap with a nearby star).
}
\end{figure*}

\eject\clearpage

%%%% Page 17 -- VCC 1431, VCC 1545 %%%%%%%%%%%%%%%%%%%%%%%%%%%

\figurenum{27}

\begin{figure*}[b] 

\vskip 9.0truein

\includegraphics{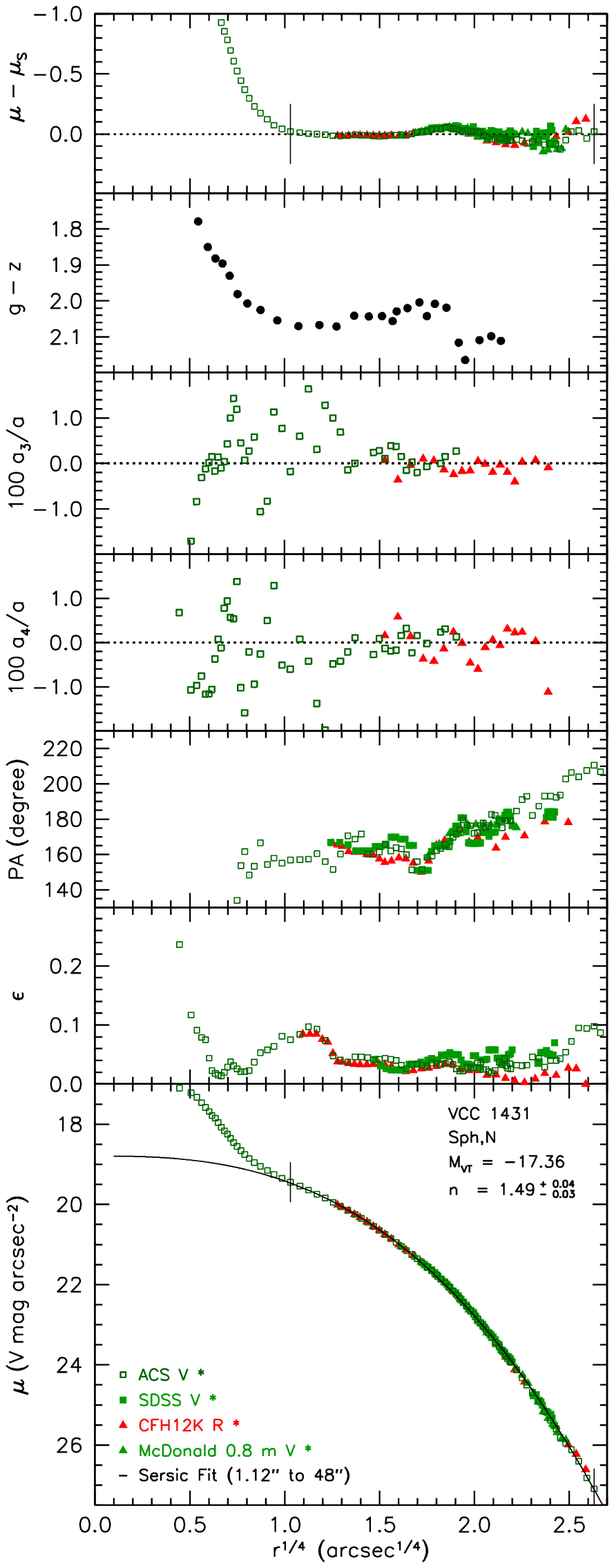}

\includegraphics{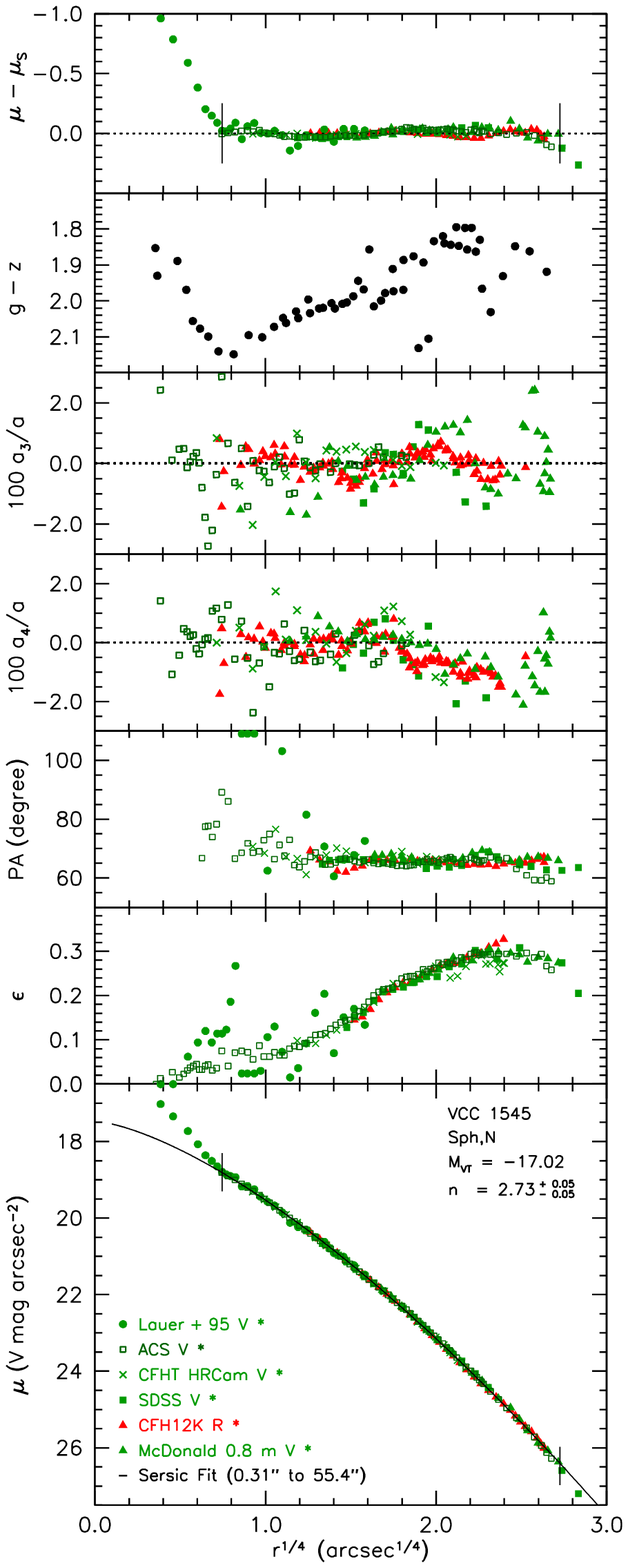}

\figcaption[]
{Photometry of Virgo cluster Sph galaxies.  
}
\end{figure*}

\eject\clearpage

%%%% Page 18 -- VCC 1407, VCC 1828 %%%%%%%%%%%%%%%%%%%%%%%%%%%

\figurenum{28}

\begin{figure*}[b] 

\vskip 9.0truein

\includegraphics{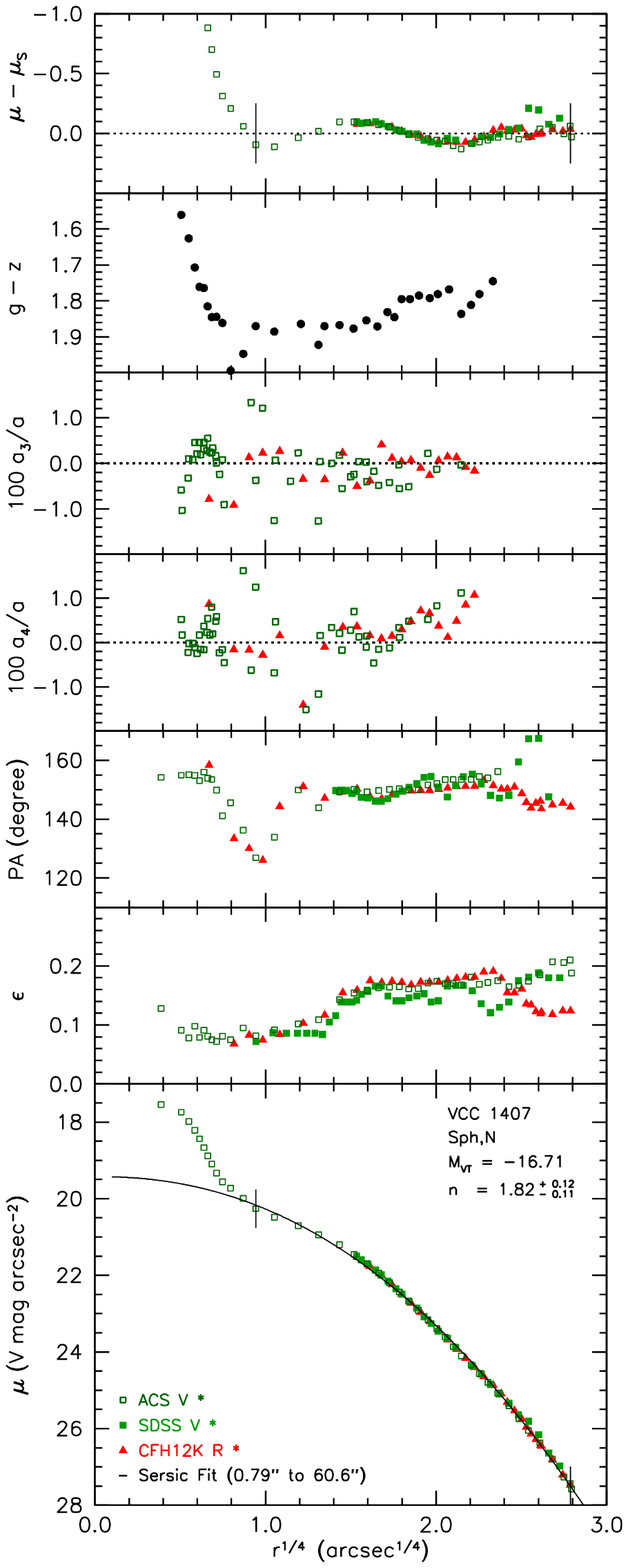}

\includegraphics{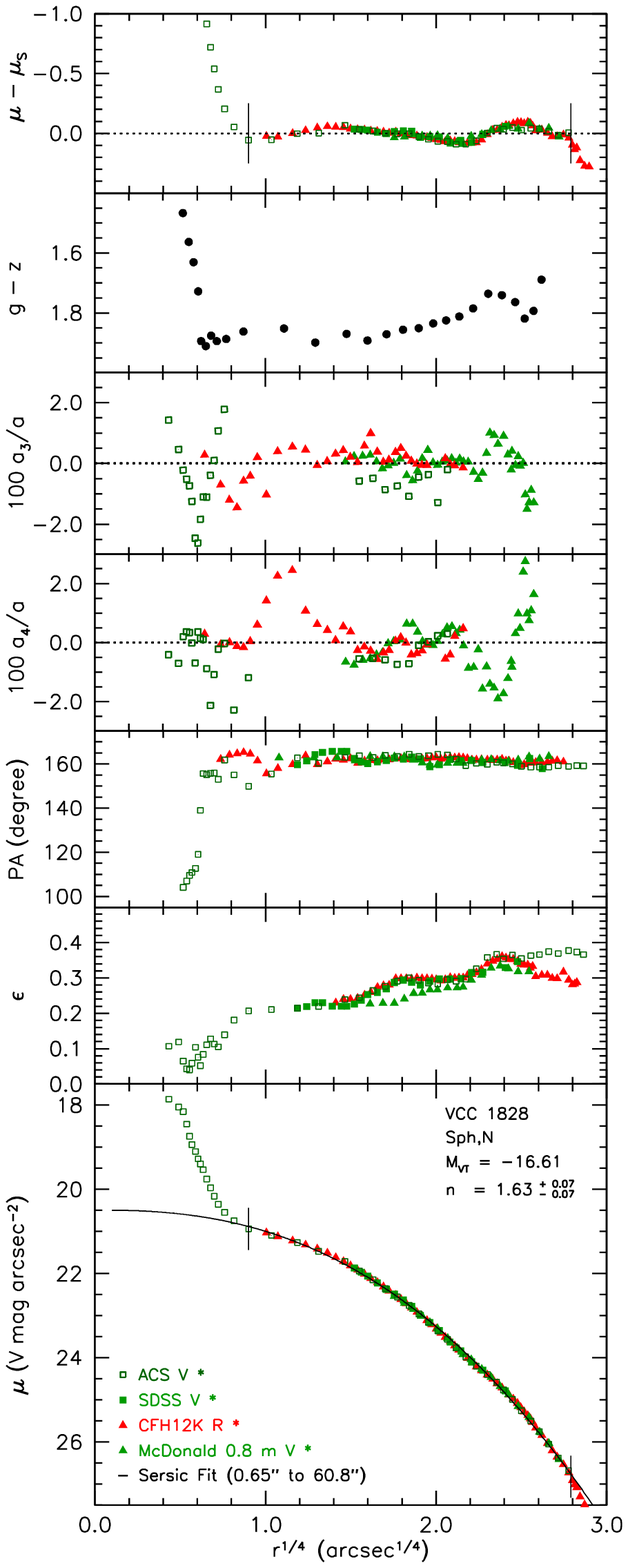}

\figcaption[]
{Photometry of Virgo cluster Sph galaxies.  The outer CFH12K $R$-band
profile of VCC 1828 is not accurate because the galaxy falls on one of the poor CCD chips of
the mosaic: the sky values are mottled and the sky subtraction is not as accurate as normal
}
\end{figure*}

\eject\clearpage

%%%% Page 19 -- VCC 1185, VCC 1489 %%%%%%%%%%%%%%%%%%%%%%%%%%%

\figurenum{29}

\begin{figure*}[b] 

\vskip 9.0truein

\includegraphics{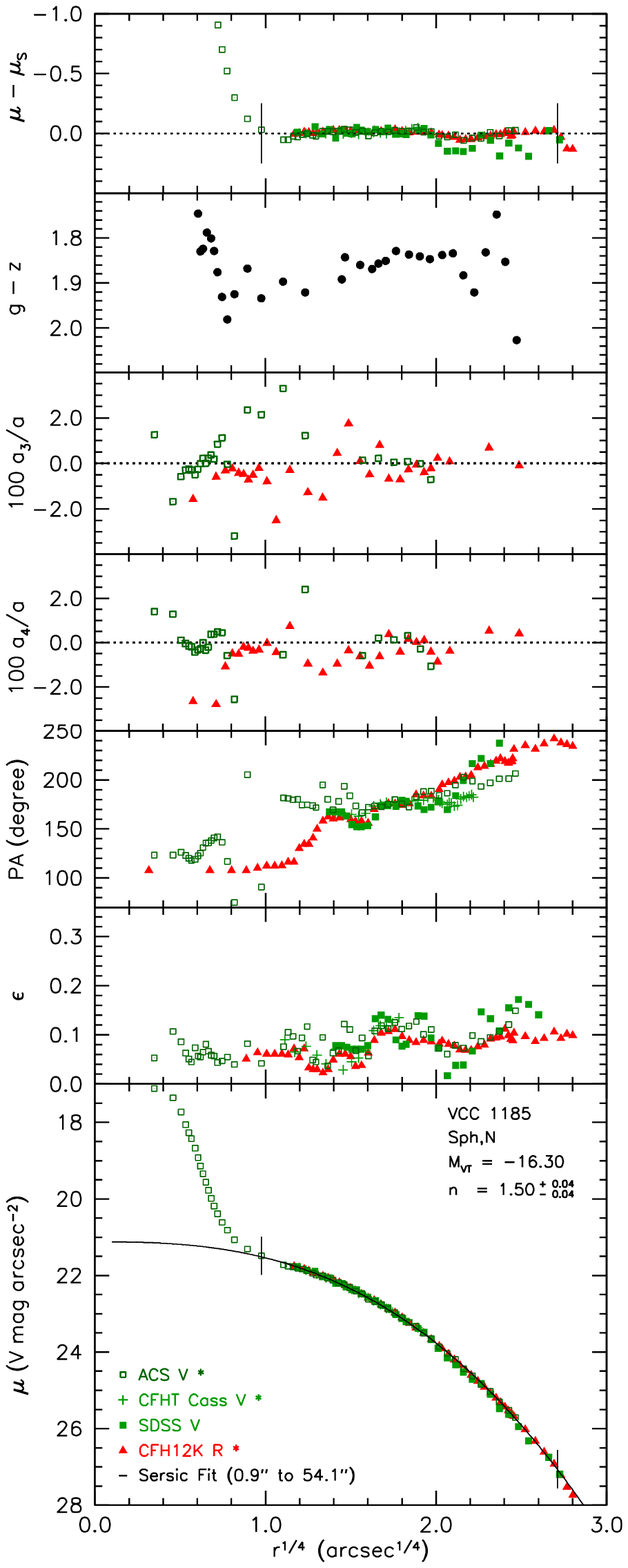}

\includegraphics{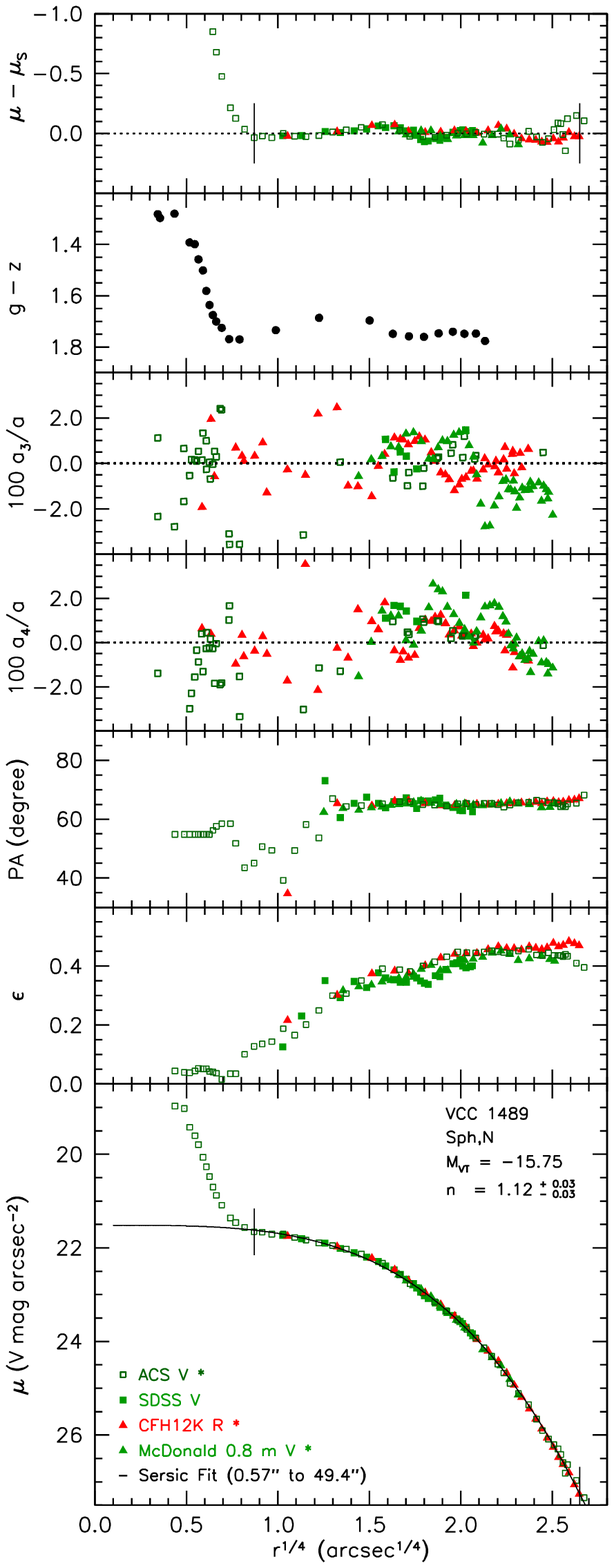}

\figcaption[]
{Photometry of Virgo cluster Sph galaxies.  VCC 1185 and VCC 1489 have almost
the same absolute magnitudes as the faintest dwarf ellipticals in our sample,
VCC 1627 ($M_{VT} = -16.44$), 
VCC 1199 ($M_{VT} = -15.53$), and 
M{\ts}32 ($M_{VT} = -16.69$).  But the spheroidals have very different brightness 
profiles than the ellipticals.  Contrast especially the faint extrapolated central 
surface brightness of the S\'ersic fits to
VCC 1185 ($\mu = 21.12$ $V$ mag arcsec$^{-2}$) and
VCC 1489 ($\mu = 21.52$ $V$ mag arcsec$^{-2}$) with the 100-times brighter values for
VCC 1627 ($\mu = 16.24$ $V$ mag arcsec$^{-2}$) and
VCC 1199 ($\mu = 16.38$ $V$ mag arcsec$^{-2}$) and the still brighter value in
M{\ts}32 ($\mu = 13.42$ $V$ mag arcsec$^{-2}$).
The dichotomy between E and Sph galaxies is particularly clearcut in central parameters
(\S\ts2.1 and Figures 34 -- 36), although it is also seen in global parameters (Figures 37 and 38).
VCC 1489 is the lowest-luminosity Sph galaxy in our sample, which favors spheroidals
that most resemble M{\ts}32-like ellipticals.  Nevertheless, it is brighter than the 
majority of spheroidals in the Virgo cluster (see Figures 34, 37, and 38).
}
\end{figure*}

\eject\clearpage

%%%%%%%%%%%%%%%%%%%%%%%%%%%%%%%%%%%%%%%%%%%%%%%%%%%%%
%                                                   %
% Figure 7: S0 galaxies                             %
%                                                   %
%%%%%%%%%%%%%%%%%%%%%%%%%%%%%%%%%%%%%%%%%%%%%%%%%%%%%

%%%% Page 20 -- NGC 4570, NGC 4660 %%%%%%%%%%%%%%%%%%%%%%%%%%%

\figurenum{30}

\begin{figure*}[b] 

\vskip 9.0truein

\includegraphics{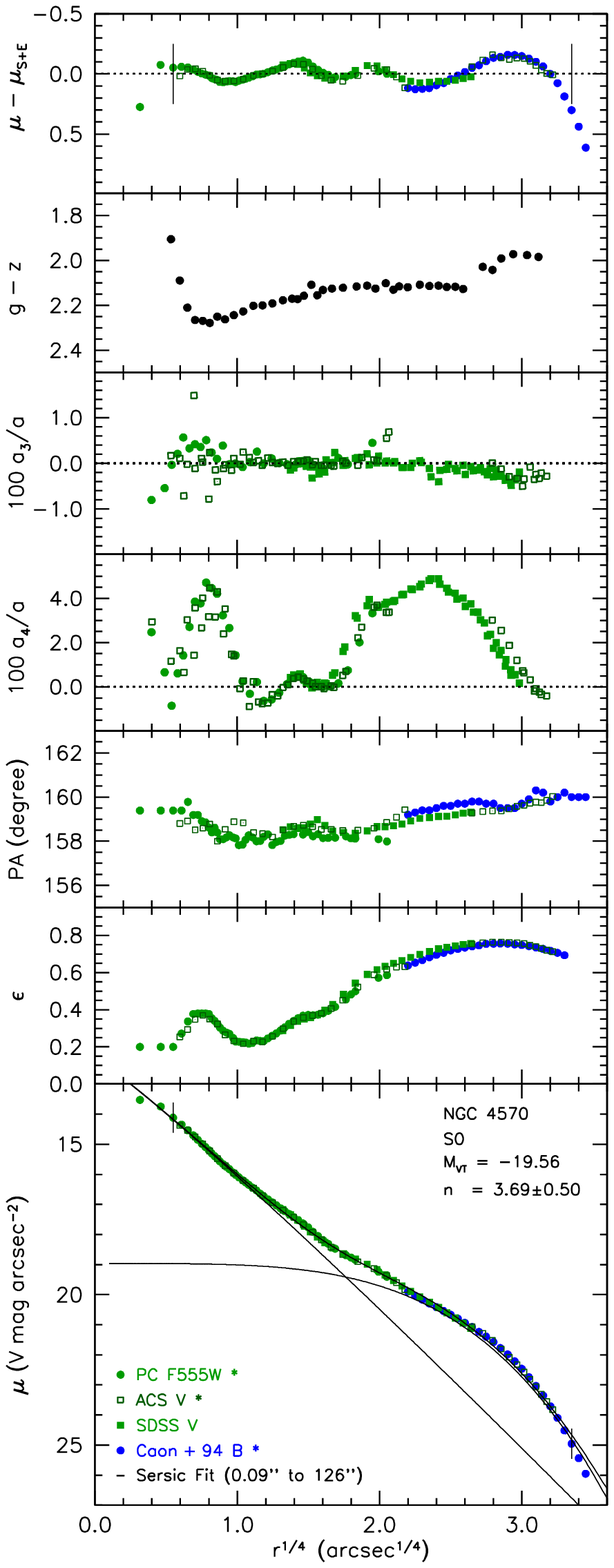}

\includegraphics{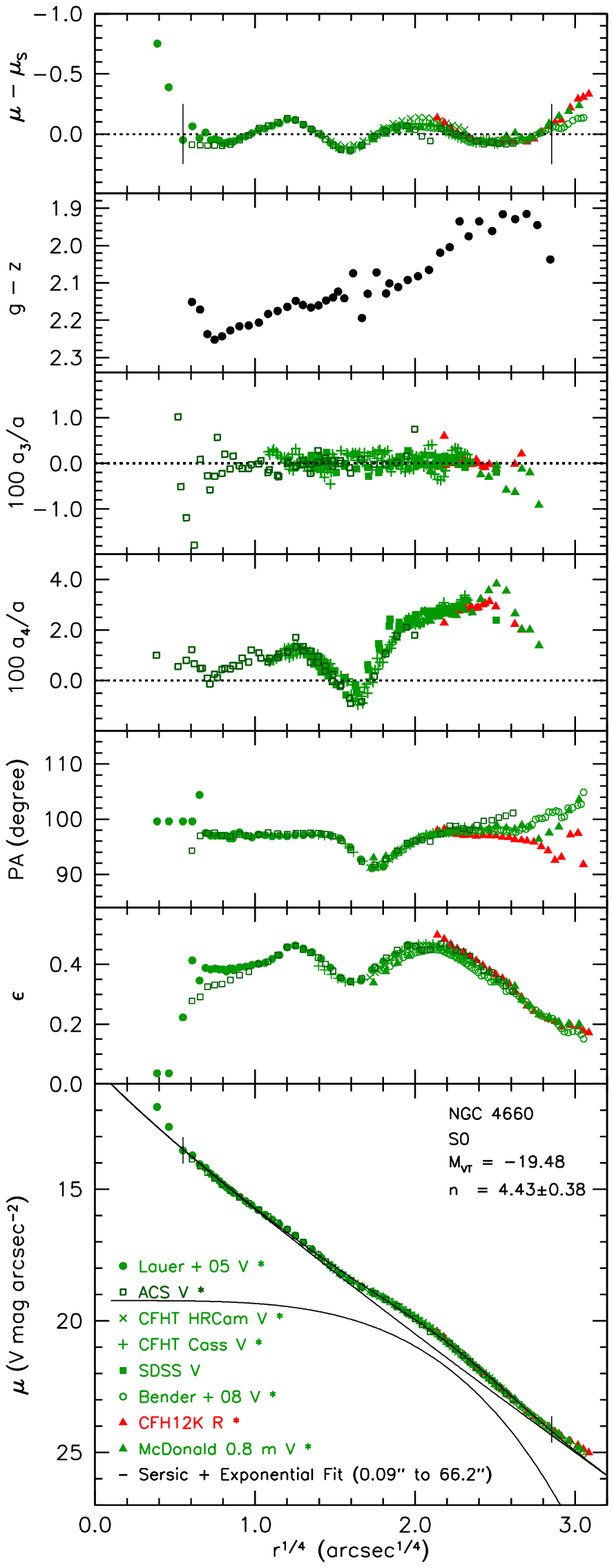}

\figcaption[]
{Photometry of Virgo cluster S0 galaxies. Symbols, parameters, and color coding 
are as in Figures 11 -- 29.   The absolute magnitudes quoted in the keys of
Figures 30 -- 32 refer to the bulge only (see notes to Table 1).  Note the obvious 
disk signatures in the $a_4$ profiles of both galaxies.  Both galaxies are highly 
inclined. In contrast, NGC 4489 (next page) is almost round and shows no $a_4 > 0$ 
disk signature.  NGC 4570 obviously looks like an edge-on S0 in images and is normally 
classified as such.  But NGC 4660 is a good example of an S0 galaxy that is traditionally
misclassified as an elliptical (Table 1).  Its disk contributes relatively little 
light, and the galaxy is seen far enough from edge-on so that the disk is evident 
mostly from the $a_4$ profile.  The S0 nature of NGC 4660 was established by Rix 
\& White (1990) and by Scorza \& Bender (1995).
}
\end{figure*}

\eject\clearpage

%%%% Page 21 -- NGC 4564, NGC 4489 %%%%%%%%%%%%%%%%%%%%%%%%%%%

\figurenum{31}

\begin{figure*}[b] 

\vskip 9.0truein

\includegraphics{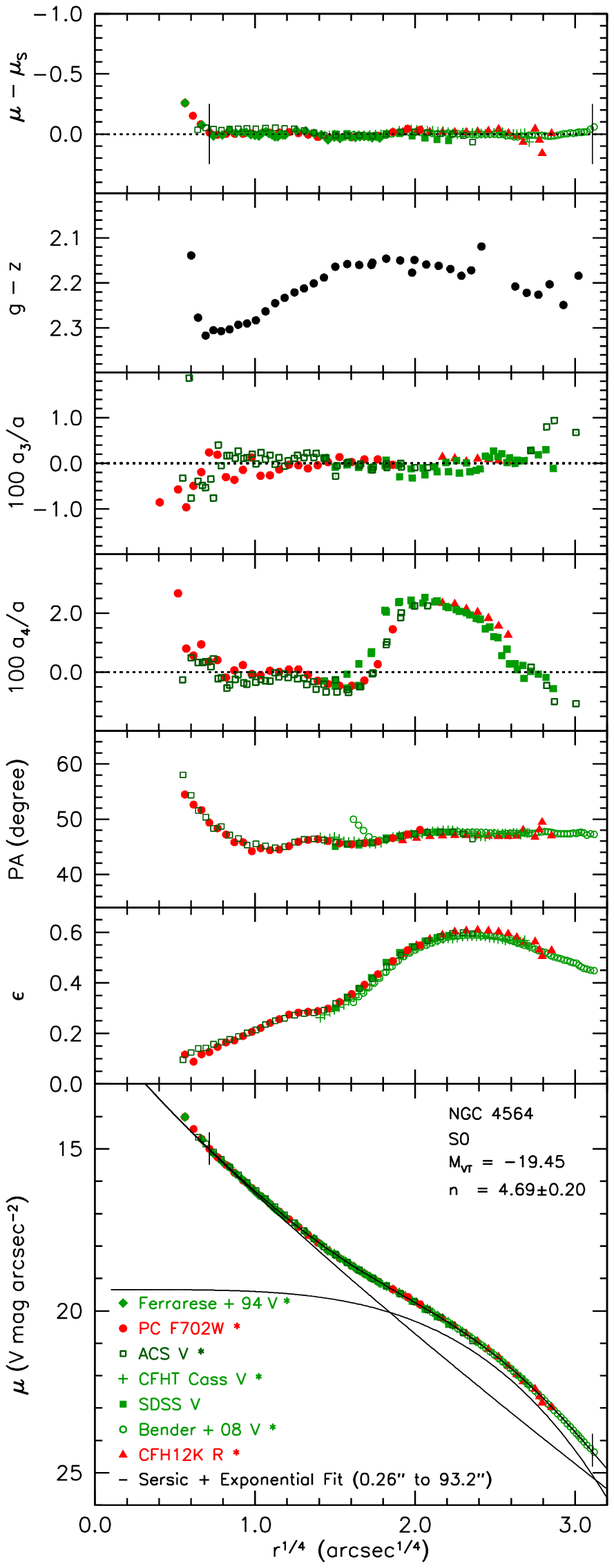}

\includegraphics{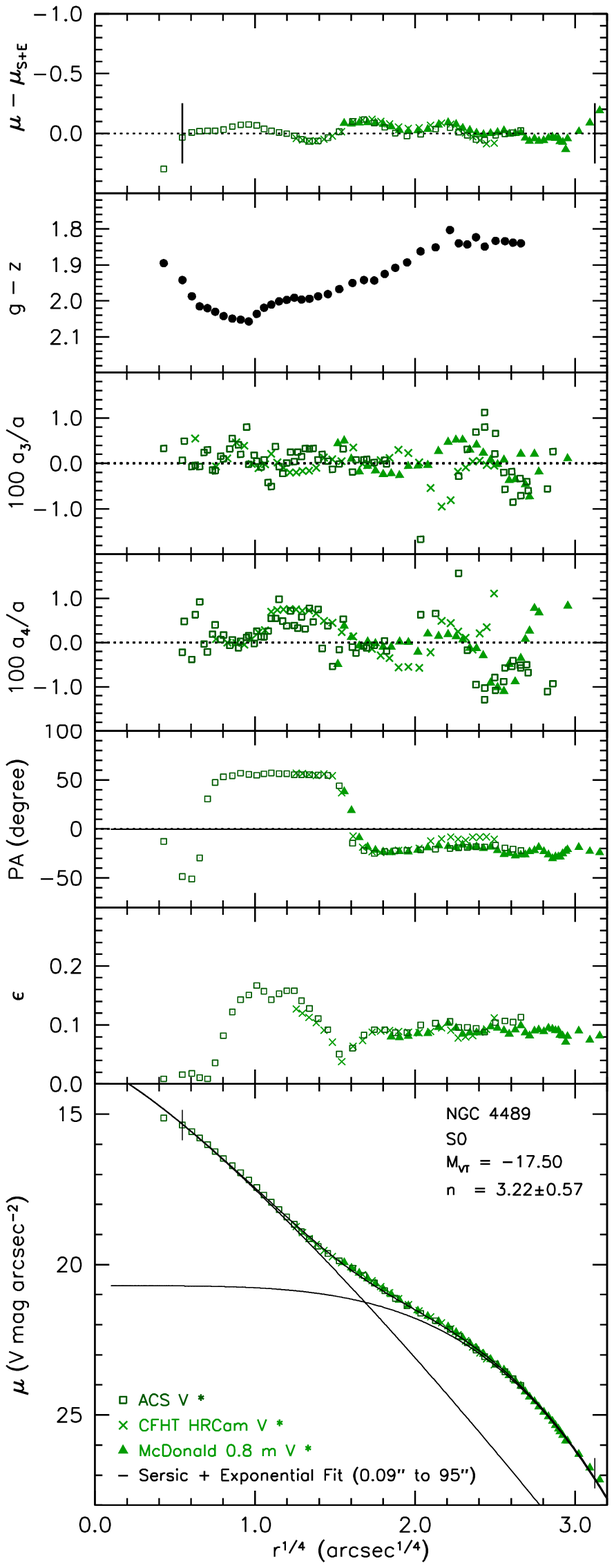}

\figcaption[]
{Photometry of Virgo cluster S0 galaxies.  Note that the highly inclined galaxy 
NGC 4564 shows a strong disky $a_4 > 0$ signature, but the much rounder, presumably 
nearly face-on galaxy NGC 4489 does not (see also Bender \etal 1989; Kormendy \& Bender 1996).
}
\end{figure*}

\eject\clearpage

%%%% Page 22 -- NGC 4318 as S0 %%%%%%%%%%%%%%%%%%%%%%%%%%%

\figurenum{32}

\centerline{\null}\vskip 9.00truein

\includegraphics{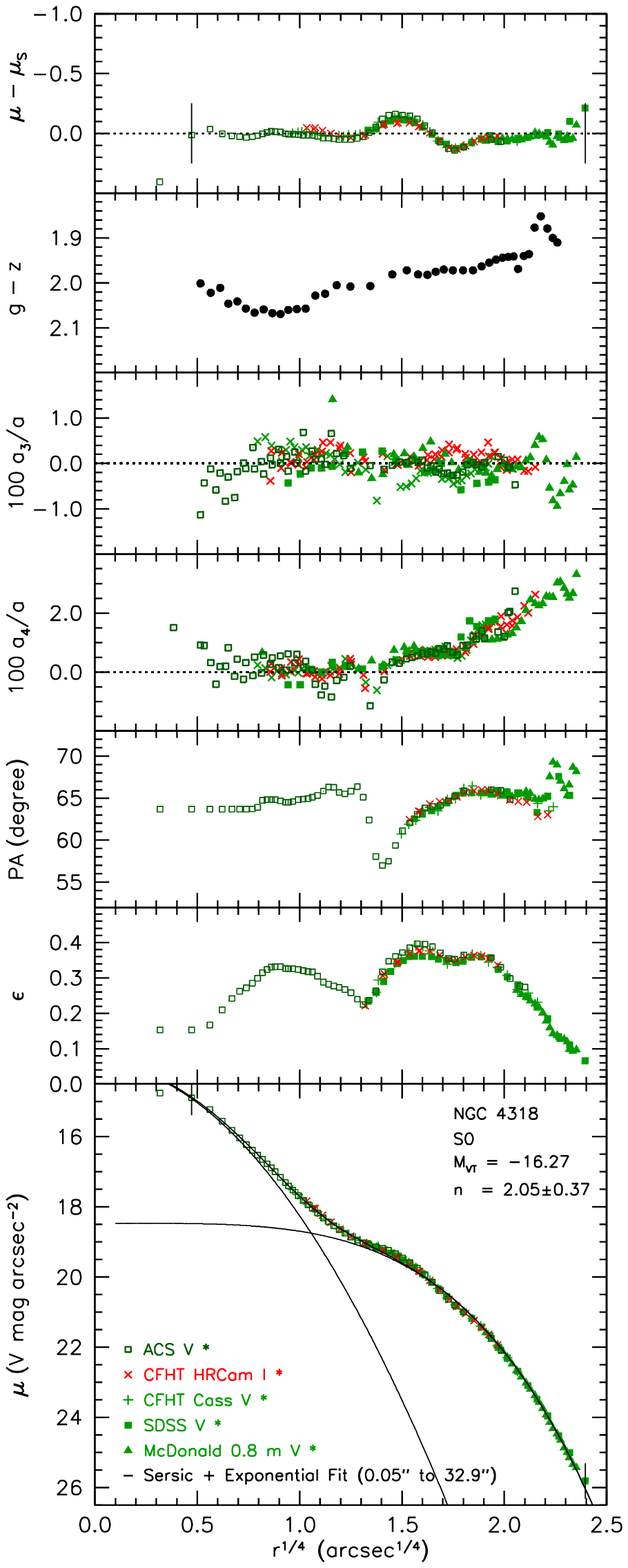}

\figcaption[]
{Photometry of Virgo cluster S0 galaxies.  NGC 4318 is a good example of a tiny S0 
galaxy that is easily misclassified as an elliptical.  High-resolution photometry
is required to distinguish the small bulge, and spectroscopy is required to verify
that the outer component is a disk (see \S\ts7.6).}

\section{Photometry Results.~~I.\\Parameter Correlations and the
         Dichotomy Between Elliptical and Spheroidal Galaxies}

      One principal result of this paper is to verify the dichotomy between 
elliptical and spheroidal galaxies (\S\ts2.1) with modern, accurate photometry. 
This is done in Figures 34 -- 39.  It is a necessary step in refining our sample 
of elliptical galaxies.

      Challenges to the E -- Sph dichotomy are based mostly on two claims, (1)
that the correlation between S\'ersic index~$n$ and galaxy luminosity is 
continuous from spheroidals through ellipticals, and (2) that other parameter
correlations are continuous between spheroidals and low-luminosity ellipticals.
With more accurate parameter measurements, we can better test these claims.
We agree with (1) but not with (2).

      Figure 33 shows the correlation between $n$~and~$M_{VT}$.  Blind to the
E{\ts}--{\ts}Sph distinction (Figures 34{\ts}--{\ts}39), we would conclude
that the \hbox{$n$\ts--\ts$M_{VT}$} correlation is continuous over all 
luminosities.  But this does not prove that E and Sph galaxies are related. 
If they are different, then Figure~33 just tells us that the \hbox{$n$\ts--\ts$M_{VT}$} 
correlation is not sensitive to the physics that makes them different.  
There are other, similar correlations.  Viewed morphologically blindly, E, Sph, 
and even Im galaxies are continuous in the correlations between metallicity 
and galaxy luminosity or velocity dispersion (Bender 1992; Bender 
\etal 1993; Mateo 1998; Tremonti \etal 2004; Veilleux \etal 2005).  Again, 
this does not mean that E, Sph, and Im galaxies are the same.  The conclusion
is that gravitational potential well depth and not the details of galaxy structure
governs the degree to which metals returned to the interstellar medium during 
stellar evolution are retained by a galaxy (Dekel \& Woo 2003).  So all galaxies 
roughly satisfy the same metallicity{\ts}--{\ts}luminosity correlation.   Looking at the 
correlations with morphology in mind, Mateo (1998) and Grebel (2004) find that 
Sph galaxies are slightly more metal-rich than Im galaxies of the same luminosity.
Similarly, ellipticals generally have higher S\'ersic indices than spheroidals of 
the same luminosity.

\vfill

\includegraphics{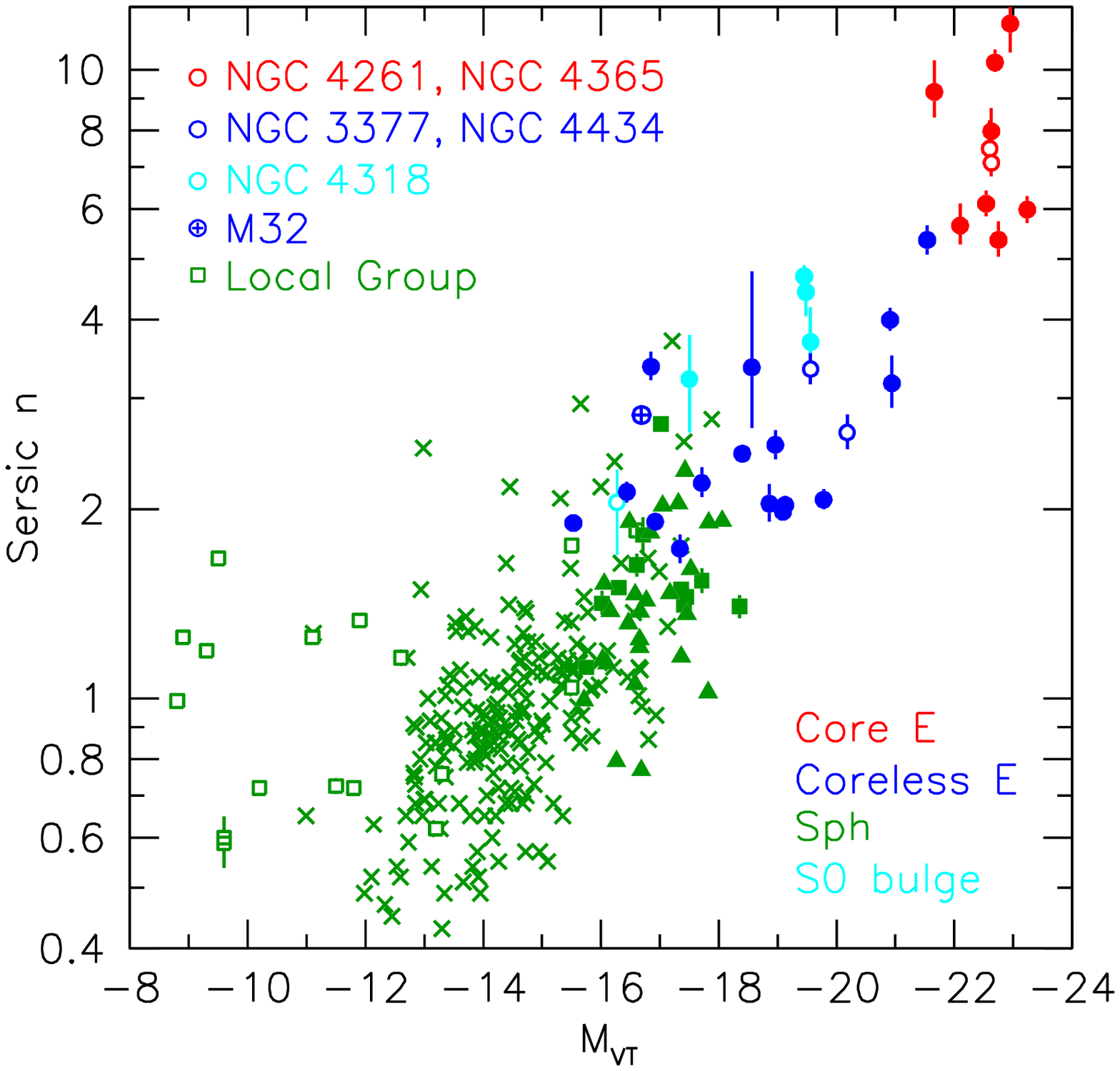}

\figurenum{33}

\figcaption[]
{Correlation between S\'ersic index $n$ and $M_{VT}$: red, blue, green, and 
turquoise points show our core Es, extra light Es, Sph galaxies, and S0 bulges. 
Green triangles show all spheroidals from Ferrarese \etal (2006a) that are not 
in our sample.  Crosses show all spheroidals from Gavazzi \etal (2005) that are
not in our sample or Ferrarese's.  Open squares are for  Local Group spheroidals 
(Caldwell 1999; Jerjen, Binggeli, \& Freeman 2000).  Open symbols 
refer to galaxies that are not Virgo cluster members.
}
\eject

\centerline{\null} 

\figurenum{34}

\vskip 4.94truein

\includegraphics{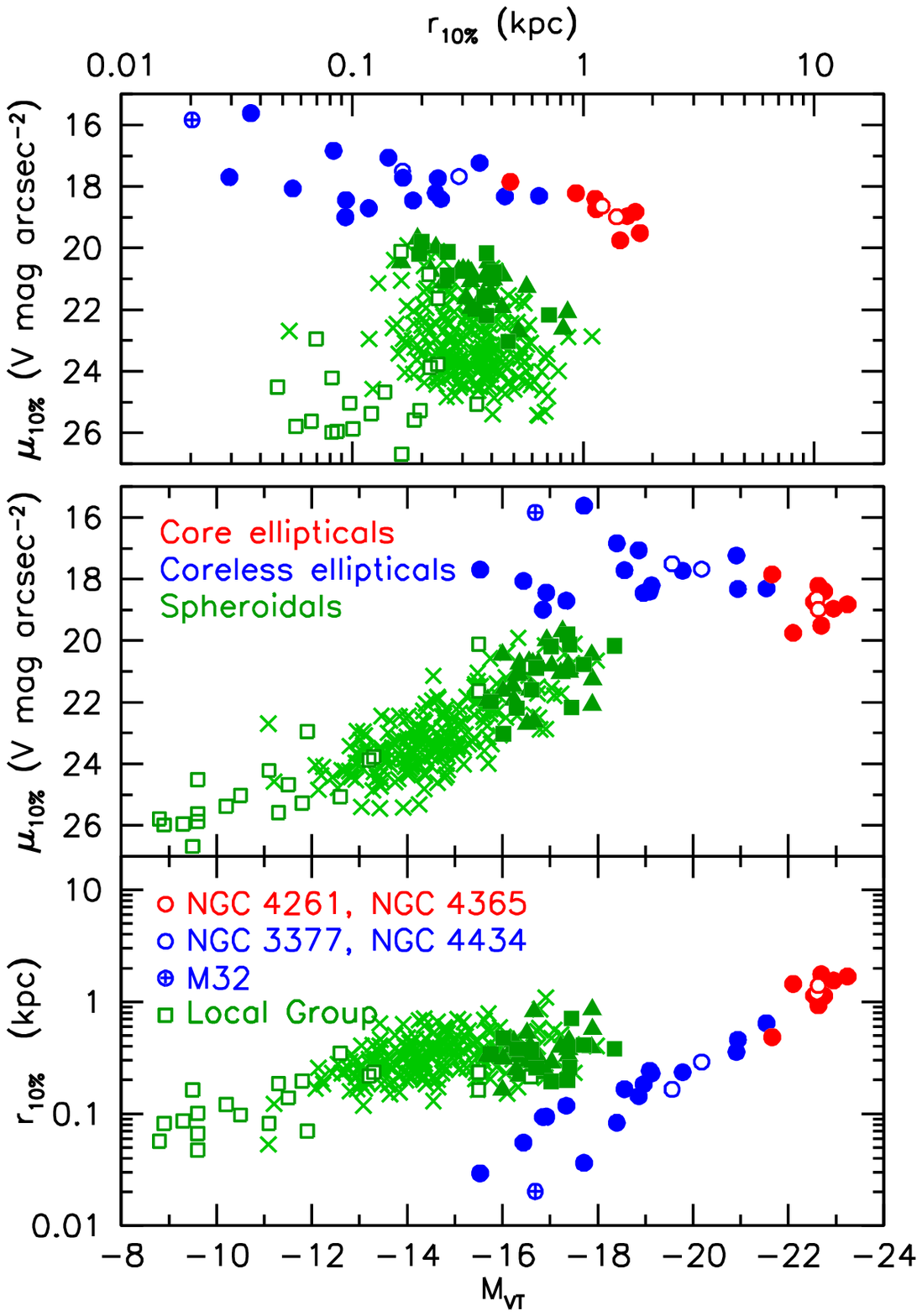}

\figcaption[]
{``Central'' parameter correlations for the main bodies of elliptical and 
spheroidal galaxies. Here $r_{10\%}$ is the major-axis radius of the 
elliptical isophote that contains 10\ts\% of the light of the galaxy
and $\mu_{10\%}$ is the surface brightness at that radius corrected
for Galactic extinction.  The 10-\%-light radius
is approximately the smallest radius that is outside the nucleus
in Sph galaxies and outside cores and extra light in Es. The center panel
shows $\mu_{10\%}$ versus total $V$-band absolute magnitude $M_{VT}$.  
The bottom panel shows $r_{10\%}$ versus $M_{VT}$.  The symbols are as
in Figure 33.  Open squares are Local Group spheroidals from Mateo (1998) 
and from McConnachie \& Irwin (2006).
}

\vskip 6.7pt

     To distinguish galaxy types, we need to use all parameter correlations.
We need to find out which ones are sensitive to formation physics.~Given how the
\hbox{E\ts--{\ts}Sph} dichotomy was discovered, we expect that some of the relevant 
correlations will involve nearly-central surface brightnesses and radii. 
Figure~34 shows such correlations.~We also show in Figures 35 and 36 that
E and Sph galaxies can be distinguished by their qualitatively different surface 
brightness profiles, and in Figures 37 and 38 that we reach similar conclusions 
using global parameters.

\lineskip=0pt \lineskiplimit=0pt

     The top panel of Figure 34 shows the surface brightness $\mu_{10\%}$
at the isophote that contains 10\ts\% of the light of the galaxy versus 
the radius $r_{10\%}$ of that isophote (Table 1).  The central panel shows $\mu_{10\%}$
versus $M_{VT}$.  It is analogous to Figure 1, which shows values  or limits at 
the smallest radii reached by the observations.  Here, we prefer to measure parameters 
at the \hbox{10-\%-light} radius, even though they are less sensitive to the 
E{\ts}--{\ts}Sph distinction than are parameters measured at smaller radii.  There 
are two reasons.  First, these parameters are completely insensitive to PSF smoothing.  
Second, they measure nearly central properties of the main bodies of the galaxies 
outside the radii of extra or missing light near the center.~\ts~Our conclusions 
are not sensitive to the choice of the fraction 10\ts\%; for example

\figurenum{35}

\centerline{\null} \vfill

\includegraphics{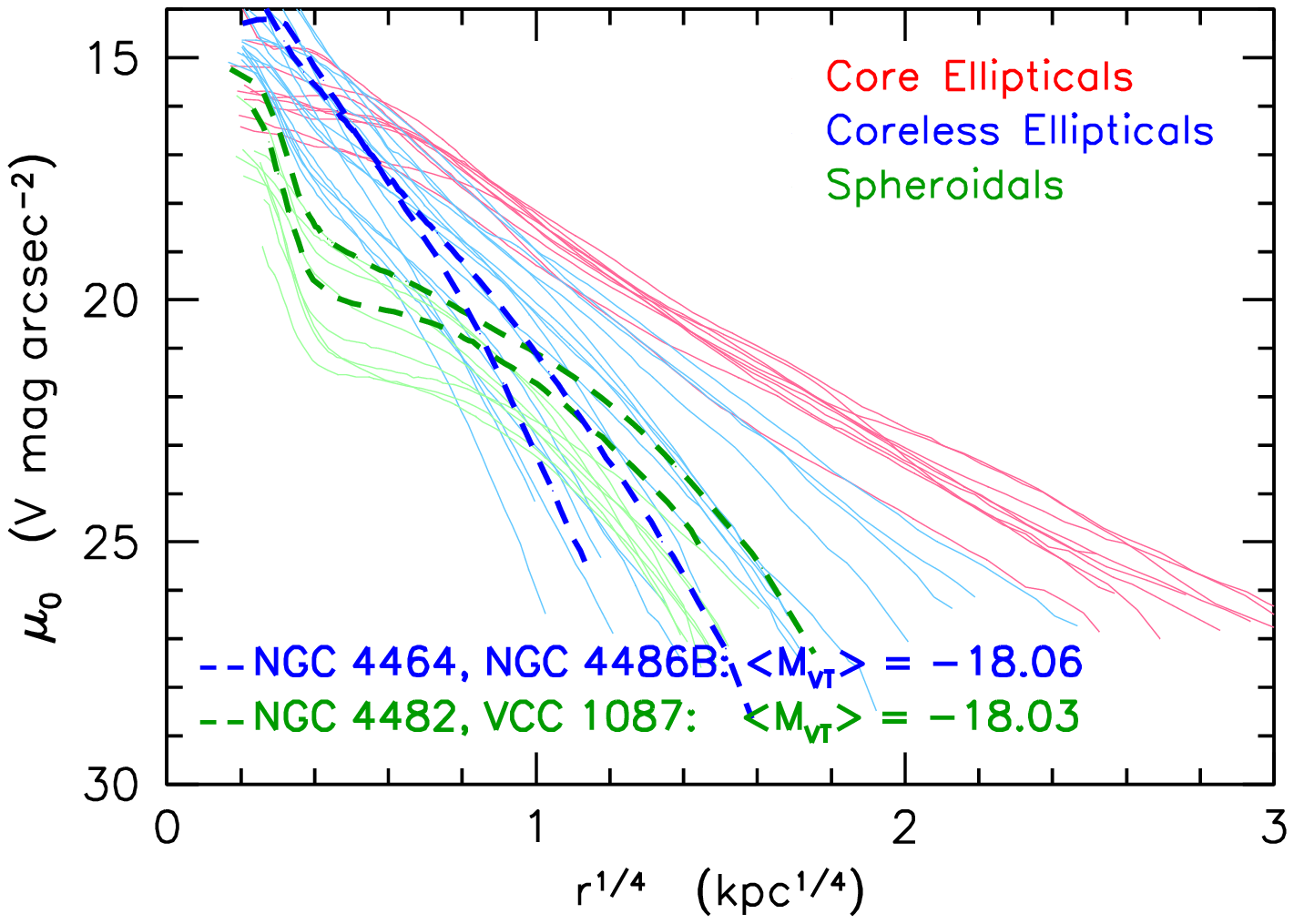}

\figcaption[]
{Major-axis profiles of all E and Sph galaxies in our sample corrected for Galactic 
absorption and scaled so that radius is in kpc.  Plotted with thick dashed lines 
are the profiles of the two brightest Sph galaxies in our sample and the two extra 
light ellipticals that have nearly the same mean $M_{VT}$.
}

\figurenum{36}

\centerline{\null} 

\centerline{\null} \vfill

\includegraphics{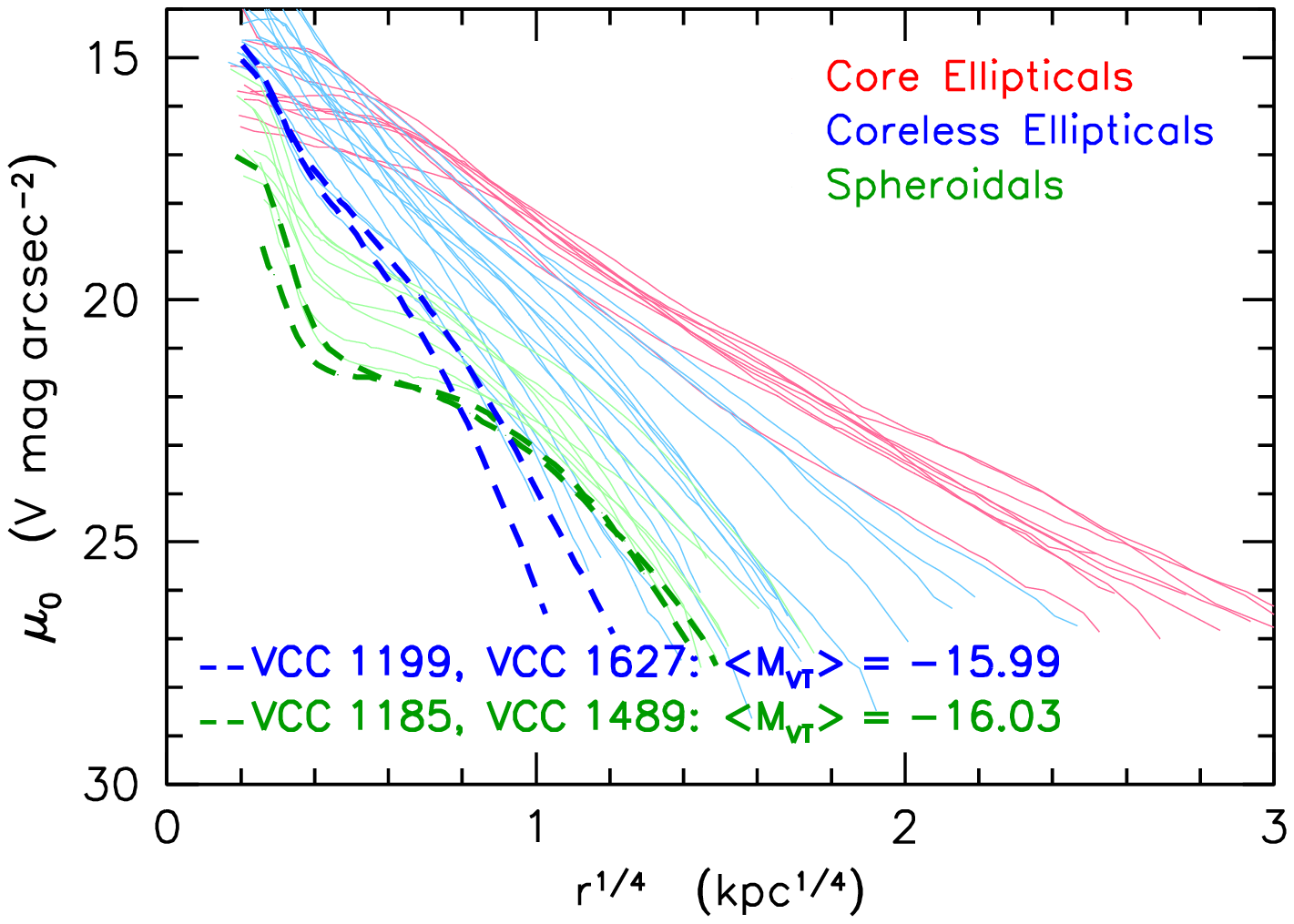}

\figcaption[]
{Major-axis profiles of all E and Sph galaxies in our sample corrected for Galactic 
absorption and scaled so that radius is in kpc.  Plotted with thick dashed lines 
are the two faintest Sph galaxies amd the two faintest extra light ellipticals in our 
sample.  They happen to have the same mean $M_{VT}$.
}

\vskip 10pt

\noindent 
5\ts\% gives similar results.  We calculated $r_{10\%}$ and $\mu_{10\%}$ for our 
galaxies directly from the photometry without using analytic fitting functions and 
without interpreting the profiles.

      {\it All panels of Figure~34 show two distinct, often nearly perpendicular 
sequences of galaxies}, as in Kormendy (1985b, 1987b).  The high-density sequence consists
only of ellipticals.  The other sequence initially consisted mostly of spheroidals
(called dE or dS0 in Binggeli \etal 1985, Gavazzi \etal 2005, and Ferrarese 
\etal 2006a) plus a few galaxies that were classified by Binggeli \etal (1985) 
as low-luminosity, M{\ts}32-like Es.  We included all of these, because we did 
not know which were E and which were Sph -- or, indeed, whether the two types could
be distinguished -- until Fig.~34 -- 38 were constructed.  We included as many 
\hbox{E{\ts}--{\ts}Sph} transition objects identified by other authors as we could. 
Our sample is strongly biased in favor of spheroidals that are most like ellipticals. 
Despite this bias, {\it the E and Sph sequences are clearly distinct.  The differences between 
E and Sph galaxies do not depend on how we measure parameters; E and Sph profiles are 
qualitatively different (Fig.~35 and 36).}  We therefore use Figure 34 to reclassify as Sph
the few galaxies that have parameters in the Sph sequence but that were called E 
by other authors (Table 1).

\eject

\centerline{\null}

\figurenum{37}

\vskip 4.93truein

\includegraphics{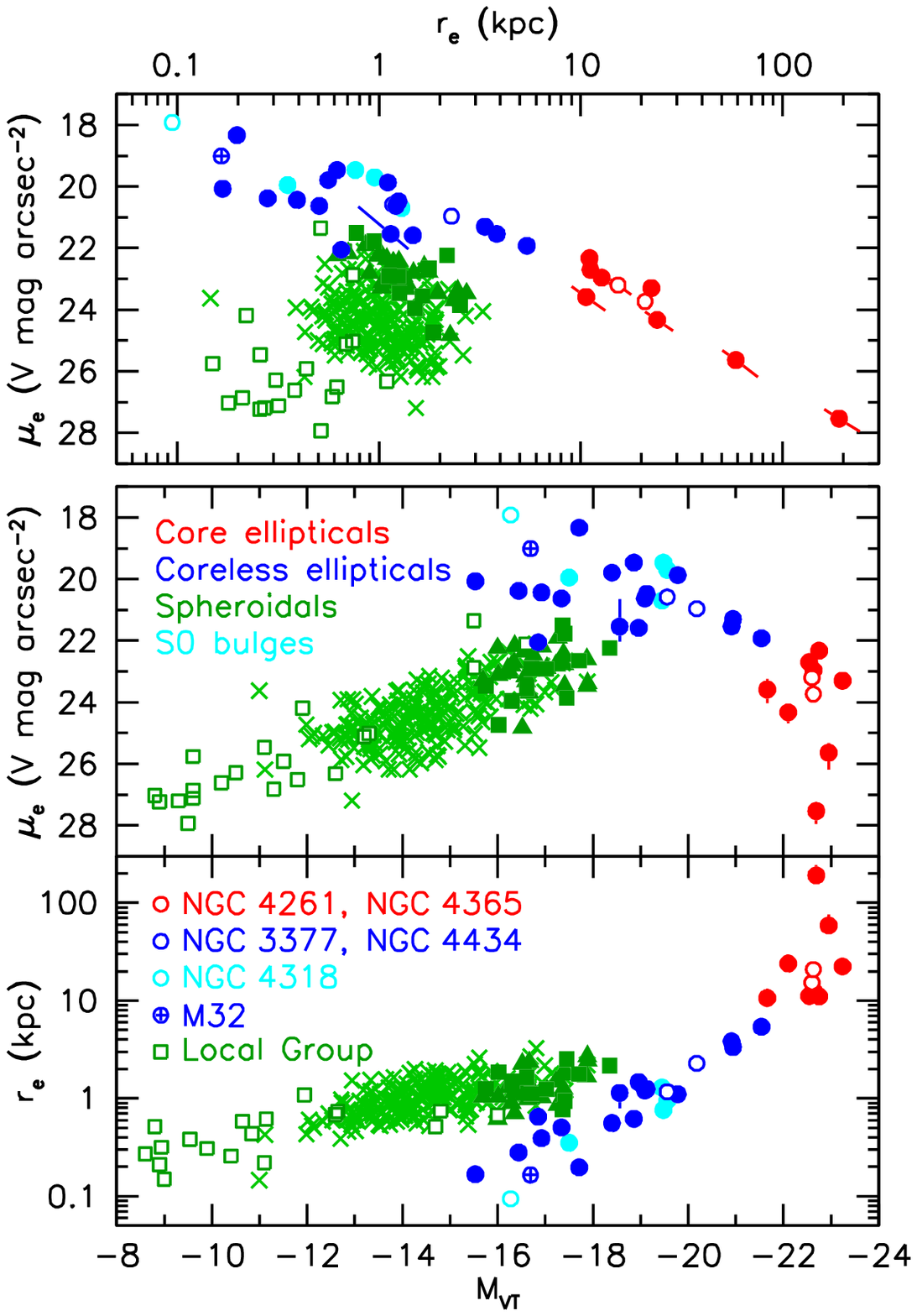}

\figcaption[]
{Global parameter correlations for elliptical and spheroidal galaxies and for S0 bulges. 
The panels are analogous to those in Figure 34, but $r_e$ is the effective radius that 
contains 50\ts\% of the light of the galaxy and $\mu_e$ is the surface brightness at 
$r_e$.  These are the parameters of the S\'ersic fits to the major-axis profiles
(Table 1); as a result, we can include S0 bulges, which require a profile decomposition 
that is based in a S\'ersic fit to the bulge.  Otherwise, the symbols are as in Figures 
33 and 34.  The E and Sph points in our sample have error bars; most are too small to be visible.  
The blue point among the green points in Figures 37 and 38 is for VCC 1440.  It is clearly 
classified E in Figure 34, but its position is symptomatic of the fact that the Sph 
sequence approaches the E sequence near its middle (not its faint end).
}

\vskip 18.4pt

      Figures 37 and 38 are analogous to Figure 34 but show global parameters (Table 1).  
Figure 37 is based on S\'ersic fits to the major-axis profiles.  Figure 38 is based on
integrations of the brightness profiles and  is independent of fitting 
functions. The  top panels show effective brightness versus effective radius --
the Kormendy (1977) % $\mu_e$ -- $r_e$  
relation.  It shows the fundamental plane 
close to edge-on.  The bottom panels show the correlations of $\mu_e$ and $r_e$ with total 
or (for S0s) bulge absolute magnitude.

      Figures 37 and 38 further confirm the distinctions illustrated in Figures~1 and 
34{\ts}--{\ts}36 between elliptical and spheroidal galaxies.  Our results are clearcut 
because we have a large range in $M_{VT}$ and because we have accurate brightness 
profiles over large radius ranges.  We can derive accurate galaxy parameters,
so we can see that the scatter in the \hbox{$\mu_e$\ts--\ts$r_e$} correlation for ellipticals 
is small.~This confirms the  fundamental plane results of Saglia \etal (1993) and J\o rgensen 
\etal (1996).  The scatter increases slightly toward the faintest galaxies.  This is expected,
because they form in fewer mergers than do giant galaxies, so the details of different
merger histories matter more.

\centerline{\null}

\figurenum{38}

\vskip 4.93truein
 
\includegraphics{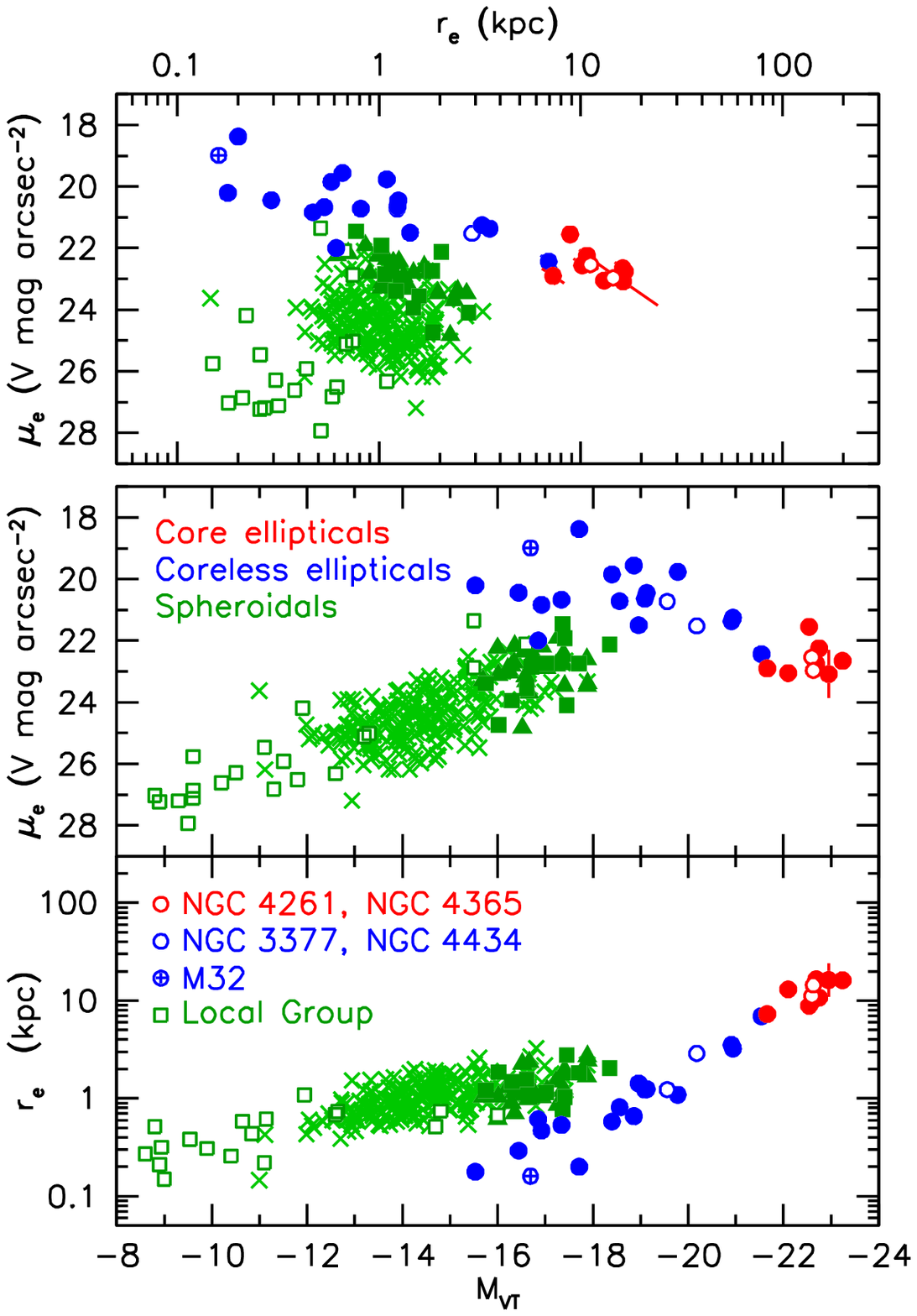}

\figcaption[]
{Global parameter correlations for elliptical and spheroidal galaxies.   Symbols 
are as in Figures 34 and 37.  Effective surface brightnesses $\mu_e$ and major-axis
effective radii $r_e$ are calculated by integrating isophotes with the observed brightness 
and ellipticity profiles out to half of the total luminosity.  S0 bulges are omitted, 
because bulge-disk decomposition requires assumptions that we do not wish to make for
this figure -- either that the bulge and disk profiles have pre-chosen analytic functional
forms or that ellipticity is constant for each of the components.  Thus for our sample, 
$r_e$, $\mu_e$, and $M_{VT}$ are independent of S\'ersic fits.  For the other samples, the 
parameters are based on S\'ersic fits and are corrected to the major axis when necessary.
\vskip 1pt}

\vskip 15pt

      The scatter in Figures 37 and 38 is small enough and the spatial resolution of 
{\it HST\/} photometry is good enough to show that the lowest-luminosity Virgo ellipticals 
extend the elliptical galaxy correlations continuously and with almost no change in slope 
from typical giant ellipticals all the way to M{\ts}32.  That is, M{\ts}32 is a normal, 
tiny -- and hence ``dwarf'' -- elliptical galaxy.

      Most important, the sequence of ellipticals is well enough defined so we can see 
with confidence that the Sph sequence approaches it not at its faint end but 
rather near the middle.  It is not the case, as suggested by Graham \& Guzm\'an (2003), 
Graham \etal (2003), and Gavazzi \etal (2005), that E and Sph galaxies define a single 
set of correlations from which giant ellipticals deviate only because they have cores.
Cores are ``missing'' $\sim 1 \pm 1$\ts\% of the galaxy light (Table 1);
they have negligible effects on global parameters.

      This confirmation of the E{\ts}--{\ts}Sph dichotomy is not new; it is just better 
defined by our photometry.  The middle panels of Figures 37 and 38 can be compared with 
Figure~1a and the bottom panels can be compared with Figure 1b in Binggeli \& Cameron (1991). 
They did not have {\it HST\/} photometry, so the faint part of their E sequence is not 
well defined and the degree to which M{\ts}32 is a normal dwarf elliptical is not obvious. 
Nevertheless, they, too, interpreted\footnote{They have since changed their minds (Jerjen
\& Binggeli 1997).} their results as indicative of a ``dichotomy [that] appears strongest 
in the King core parameter diagrams but [that] is basically {\it model-independent\/}'' 
(their emphasis).   Bender \etal (1992) also emphasized that, in addition to elliptical 
galaxies, ``a second major sequence is comprised of dwarf ellipticals\footnote{Bender 
\etal (1992) use the terminology of the Sandage-Binggeli Virgo cluster survey papers 
(Sandage \& Binggeli 1984; Binggeli \etal 1985, and references therein) in which 
bright spheroidals are called ``dwarf ellipticals'' (dEs).} and dwarf spheroidals.
These systems populate an elongated locus running at right angles to the main 
elliptical locus'' in the $\kappa$ fundamental plane parameters defined in their 
paper.  The different loci of E and Sph galaxies in parameter space can also be seen
in many other papers (e.{\ts}g., Capaccioli, Caon, \& D'Onofrio 1993; Chilingarian \etal 2007,
2008), including the ones that criticize the existence of the dichotomy.  How clearly it 
is seen depends on sample selection (particularly on whether low-luminosity Es are included) 
and on the spatial resolution available for the lowest-luminosity ellipticals (see Appendix B).

      The E -- Sph dichotomy is also evident in their different luminosity functions.
Our figures, Ferrarese \etal(2006a), Kormendy \& Bender (1994), and Binggeli \& Cameron 
(1991) show that a small number of Sph galaxies closely approach the E sequence as defined 
by global parameters.  They are rare -- the luminosity function of Sph galaxies falls 
rapidly toward higher luminosities at $M_{VT} < -18$ (Sandage, Binggeli, \& Tammann 1985a, b). 
But it rises dramatically toward lower $L$, as hinted at by the samples in Figures 34, 37, 
and 38 until they reach tiny dwarfs that are rarely studied outside 
the Local Group.  In contrast, the luminosity function of elliptical
galaxies has a broad maximum near where the Sph sequence approaches it and is bounded at 
both bright and faint magnitudes.  Dwarfs like M{\ts}32 and giants like M{\ts}87 are rare. 
These results are clearly demonstrated in Sandage \etal (1985a, b) and reviewed in Binggeli, 
Sandage, \& Tammann (1988).  Sandage \etal (1985b) conclude that the difference in 
luminosity functions ``suggests that dE's do not form a continuum with the giant E's but
rather [that they] form separate families'' as argued by Wirth \& Gallagher (1984) and by 
Kormendy (1985).  Binggeli \etal (1988) reach the same conclusion.

      We believe that the E{\ts}--{\ts}Sph dichotomy is a secure result.  

      Nevertheless, by using the word ``dichotomy'', we do do not mean to imply there is 
an empty gap between their sequences in global parameter space.  A few galaxies are close 
enough to both sequences so that their classifications are uncertain.  This is most evident
for VCC 1440, which is clearly in the E sequence in Figure 34 but which plots among the 
Sph galaxies in Figures 37 and 38.  What does this mean? 

      These galaxies are not a problem for the developing scenario of E and Sph formation.  
For example, in galaxy harrassment, it is not unreasonable to expect that gas dissipation, 
inflow, and star formation will be most vigorous in the biggest Sph progenitors.  These 
events may not be completely different from the starbursts that accompany dissipative mergers.  
The same may be true for the biggest starbursts in blue compact dwarfs.  So it is reasonable 
that E and Sph galaxies have fundamentally different formation mechanisms but that a few of 
the biggest Sphs end up not too different from some ellipticals. 

\centerline{\null}
\centerline{\null}

\section{Photometry Results.~~II.\\Brightness Profiles of Elliptical Galaxies}

      This section presents our results on the systematic properties of 
the brightness distributions of elliptical galaxies.  Interpretations 
are discussed in \S\S\ts10\ts--\ts12.   

\subsection{S\'ersic Profiles of the Main Bodies of Elliptical Galaxies}

\lineskip=-4pt \lineskiplimit=-4pt

      Figures 11\ts--\ts24 in \S\ts7 and Figures 49\ts--\ts67 in Appendix A show that S\'ersic 
functions fit the major-axis brightness profiles of the main bodies of elliptical galaxies 
remarkably well.~This is a resounding confirmation of the studies reviewed~in~\S\ts3.  
With the improved accuracy and dynamic range provided by composite profiles, we now see 
quantitatively how well this single, three-parameter fitting function works.  Appendix A
provides details.  For 9 giant ellipticals with cores (omitting NGC 4382), 
S\'ersic functions fit the major-axis profiles with a mean RMS dispersion of $0.042 \pm 0.006$ 
mag arcsec$^{-2}$ over a mean surface brightness range of $\Delta\mu_V = 8.7 \pm 0.4$ mag arcsec$^{-2}$. 
For the 16 extra light ellipticals (omitting NGC 4515), S\'ersic functions fit the major-axis
profiles with a mean RMS dispersion of $0.045 \pm 0.005$ mag arcsec$^{-2}$ over a mean 
$\Delta\mu_V$ that is also $8.7 \pm 0.4$ mag arcsec$^{-2}$.  That is, S\'ersic functions fit 
the brightness profiles to 4\ts\% (sometimes 2\ts\%) over a range of 3000 (sometimes $>$\ts10,000) 
in surface brightness.

      This result is remarkable because there is no astrophysical basis for the S\'ersic function.
We know no reason why violent relaxation, dissipation, and star formation should conspire --
surely in different ways in different galaxies -- to produce so simple and general a density 
profile.  We note in \S\ts10.4 that merger simulations make profiles that are more nearly S\'ersic
functions than $r^{1/4}$ laws. The reasons why S\'ersic functions work so well may deserve 
further investigation.

      Even if we do not have an explanation, the empirical result that S\'ersic functions 
fit well has an important consequence.  It allows us confidently to identify and interpret 
departures from these fits.  Otherwise -- if the best analytic representation of the profile 
were only marginally applicable, with profile wiggles above and below that function seen in most 
galaxies and at many radii -- the use of an analytic fitting function would be nothing more than 
fancy numerology. 

      We discuss departures from S\'ersic profiles in \S\S\ts9.2{\ts}--{\ts}9.7.  

\subsection{Cuspy Cores in Giant Ellipticals:\\
            ~~~~~~The Definition of Cores}

      Cores occur in all of the 10 brightest ellipticals in our sample; eight are in Virgo and 
two are in the background.  Our faintest core galaxy is NGC 4552 at $M_{VT} = -21.66$.  We find
no cores in fainter galaxies; our brightest coreless galaxy is NGC 4621 at $M_{VT} = -21.54$. 
The perfect separation at $M_{VT} = -21.6$ between core and coreless galaxies is a fortuitous 
feature of our sample (see below).  Nevertheless, the degree to which one concludes that core 
and coreless galaxies overlap in galaxy luminosity is affected by the definition of what 
constitutes a core:

      We define a core as the central region in a bulge or elliptical galaxy where the brightness
profile breaks away from and drops below a S\'ersic function fitted to the outer profile.  This is 
the definition adopted by Kormendy (1999): ``Elliptical galaxies are divided into two types: 
galaxies with steep profiles that show no breaks in slope or that have extra light at small radii 
compared to a S\'ersic function fit and galaxies that show a break from steep outer profiles to 
shallow inner profiles.''  Figure 3 in that paper demonstrates that the breaks in the projected 
profiles of cores correspond to real breaks in the deprojected profiles.  This confirms analyses 
of the Nuker galaxies by Gebhardt  \etal (1996) and by Lauer \etal (2007b).  Similar definitions 
of cores based on profile breaks have recently been adopted by
Graham \etal (2003), 
Trijillo \etal (2004), and
Ferrarese \etal (2006a).

      The Nuker team definition is different: a galaxy has a core if the inner slope of 
a Nuker function fit (Equation 1) is $\gamma < 0.3$ (Kormendy \etal 1994; Lauer \etal 1995, 2002, 
2005, 2007b; Byun \etal 1996; Faber \etal 1997).  This definition is not different in spirit from 
ours.  It is also based on the detection of an inner, downward break in the profile from an outer 
power law, which fits profiles well just outside the break radius $r_b$.  Most profiles wiggle: 
a fit of Equation~(1) almost always spits out a value of $r_b$.  A quantitative 
criterion was needed to decide when the break was strong enough to justify the identification of a core.  
There is no {\it a priori\/} way to choose a numerical criterion.  The decision to use $\gamma < 0.3$
was based on the observation that $\gamma$ values are bimodal and that there is physics in this.
Is there any collision between the Nuker definition and ours?

      The answer is ``no'', because both definitions are designed to capture the same physics.
They agree on most galaxies.  They disagree on a few objects.  But both definitions occasionally
produce unphysical results, if they are applied blindly, without taking other information into
account.  The objects involved tend to be the ones on which the two definitions disagree. 
We illustrate this with a few examples.

      The most remarkable example is NGC 4473.  Lauer \etal (2005, 2007b) classify it as a 
core galaxy; Ferrarese \etal (1994) reached the same conclusion based on a related definition.  
We can do so, too: Figure 58 ({\it top\/}) in Appendix A shows an excellent fit of a S\'ersic 
function with RMS = 0.043 mag arcsec$^{-2}$ between 2\farcs9 and 311$^{\prime\prime}$ radius.
The fit has $n = 6.1 \pm 0.4$ and implies a core.  It looks consistent with our other core 
fits except that the onset of the core is more gradual than normal as $r \rightarrow 0$.
There is no operational reason to discard this fit.  Indeed, it is substantially nicer
than the fit that we adopt (Figure 58, {\it bottom\/}), which has RMS = 0.070 mag 
arcsec$^{-2}$ over a much smaller radius range.
This fit gives $n = 4.00^{+0.18}_{-0.16}$ and no core.  Instead, there is ``extra light'' interior 
to 23$^{\prime\prime}$.  Why do we prefer the inferior-looking fit?  The reason is that
SAURON observations show that the galaxy contains a counter-rotating embedded disk: added
to the main galaxy, it results in a large apparent velocity dispersion along the major axis 
but not above and below it (Emsellem \etal 2004; Cappellari \& McDermid 2005; Cappellari et al.~2004, 2007). 
Figure 5 in Cappellari et al.~(2007) shows that the counter-rotating disk is important from 
small radii out to $19^{\prime\prime}$  but not at larger $r$. It is associated with a strong
disky signature in Figure 17.  The counter-rotating disk is presumably the result of a late accretion.  
It does not contain much mass, and it has nothing to do with the basic structure of the galaxy.
We therefore fit the profile from $r \simeq 24^{\prime\prime}$ outward, excluding the counter-rotating 
disk (see Figure 58).  As a result, our interpretation changes.  With
the $n = 6.1$ fit, it would have been an unusually faint core galaxy with profile systematics 
that disagree strongly with Figure 40.  There are well known 
virtures to the application of analysis machinery without premature interpretation.  
But in this case, the addition of kinematic information dramatizes how apparent 
virture can lead one astray.  We adopt the $n = 4$ fit in Figure 17 and Table 1.
Then NGC 4473 is a slightly unusual extra light elliptical.

      NGC 4486B (Figure 22) is a simpler example.  The double nucleus (Lauer \etal 1996)
makes the major-axis profile flatten out near the center.~So the Nuker definition says
that the galaxy has a core (Lauer \etal 1996, 2005, 2007b; Faber \etal 1997).  Of course,
the complication of the double nucleus was known.  Interestingly, Figure 22 now shows 
that the central profile flattening and double nucleus are features in an extra light 
component (see below) that is very well defined.
 
      Finally, consider NGC 4458 (Figure 19).  Lauer \etal (1996) call it a power law galaxy
based on {\it HST\/} WFPC1 photometry.  Based on higher-resolution WFPC2 data, Lauer \etal
(2005) see a small core.  Figure 19 shows that the galaxy has a remarkably clearcut extra light
component.  But at the center, the profile clearly flattens.  This may be an example of an
interesting phenomenon that is allowed but not predicted by the formation scenario suggested
in this paper.  Suitable tuning of the relative timescales of merger-induced starbursts (which, 
we suggest, make extra light components) and the orbital decay of binary black holes (which,
we suggest, scour cores) might make it possible to grow a core in an extra light galaxy.
The disadvantage of the Nuker definition of cores is that, without using the whole
profile, it misses the fact that NGC 4458 {\it also\/} contains an extra light component.

     One advantage of our definition is that it eliminates confusion about the existence of 
cores in Sph galaxies.  Trujillo \etal (2004) and Ferrarese \etal (2006a) criticize the Nuker 
definition because it ``identifies'' cores in Sph galaxies: most of them have S\'ersic indices 
$n \sim 1$, so they have shallow profiles with $\gamma < 0.3$ near the center.  As a result, 
the $\gamma$ -- $M_V$ correlation is not monotonic.  Trujillo \etal (2004) note that this 
could be interpreted as part of a dichotomy between E and Sph galaxies, but they do not
believe in this dichotomy, so they interpret it as a shortcoming of the Nuker definition.  
We show in Kormendy (1985b, 1987b) and in \S\ts8 here that the E -- Sph dichotomy is real.
So the issue of almost-flat central profiles in Sph galaxies is moot anyway.  Sph 
structure is related to disk structure -- disks have $n \sim 1$ profiles, too (Freeman 1970)
-- neither are related to E structure.  In addition, Sph profiles generally show no breaks;
they are well fitted by single S\'ersic functions at all radii outside their nuclei (Figures 25 -- 29).  
By our definition, they would not have cores even if they were related to ellipticals.

      Finally, we return to the luminosity overlap between core and coreless galaxies, 
$\Delta M_V \sim 2 \pm 0.5$ mag (Faber \etal 1996; Ravindranath \etal 2001; Laine 
\etal 2003; Lauer \etal 2007b).  With the above tweaks in core classification and distances 
based on surface brightness fluctuations (Tonry \etal 2001; Mei \etal 2007), 
the overlap region in the Faber \etal (1996) sample -- which we can study in detail -- is reduced 
to $\sim 0.7$ mag.  But it is certainly not zero: NGC 3379 is robustly a core galaxy with 
$M_V \simeq -20.9$ and NGC 4621 is robustly a coreless galaxy with $M_V = -21.5$.  The larger 
sample of Lauer \etal (2007b) shows overlap mainly at $-20.5 \gtrsim M_V \gtrsim -23$
(Figure 48 here).  These clasifications have not been repeated with the
present definition, but we find in \S\ts12.3.1 that NGC 6482 is an extra light galaxy with
$M_V \simeq -22.3$.   There are interesting hints that ``poor galaxy groups can harbor more 
luminous power law galaxies than clusters'' (Quillen \etal 2000; see also Faber \etal 1996). 
We agree:~the unusually bright coreless galaxies NGC 6482 and NGC 4125 (Figure 48) are in poor
environments.  On the other hand, some power law galaxies are also brightest cluster members.  
The environmental dependence of the E -- E dichotomy deserves further investigation.  We will
address this in a future paper.

\vskip -10pt \centerline{\null}

\subsection{Extra Light Near The Centers of Coreless Ellipticals}

      NGC 4621 ($M_{VT} = -21.54$) to VCC 1199 ($M_{VT} = -15.53$), that is, all the
faint ellipticals in our sample, do not have cores.  They are called ``power law'' ellipticals 
in Nuker team papers, because their profiles are approximately featureless power laws 
over the relatively small radius range studied in those papers.  

      One of the main results of this paper is that these galaxies do not have simple, 
almost featureless S\'ersic profiles at~all~$r$.  Instead, {\it all Virgo ellipticals
that do not show cores have extra light near the center above the inward extrapolation 
of S\'ersic functions fitted to their main bodies.\/}  These galaxies behave exactly 
like the extra light galaxies that are discussed in Kormendy (1999) and illustrated in Figure 3 
here.  Therefore the results of Kormendy (1999) are not a fluke that applies only to a few, 
unusual galaxies.  Extra light near the center is a general feature of coreless ellipticals. 

      This adds a new feature to the E -- E dichotomy.  Table 1 lists the amount of light 
``missing'' or ``extra'' with respect to the inward-extrapolated S\'ersic fit expressed 
as a percent of total luminosity.  Core Es are missing 0.17\ts--\ts4.2\ts\% of their starlight
near the center.  The mean is 1.15\%; the median is 0.84\ts\%, and the quartiles are 0.22\ts\% 
and 1.52\ts\%.  Coreless ellipticals have 0.27\ts\% to 12.6\ts\% extra light near the center.
The median is 2.3\ts\%; the quartiles are 1.3\ts\% and 5.6\ts\%.  The range is larger than the 
range of missing light in core ellipticals.

      Diagnostic of formation processes, extra light often has disky characteristics.  
It has $a_4$~$>$~0 in 
NGC 4458 and NGC 4478 (see also Morelli et al.~2004), 
NGC 4464,
NGC 4467, 
NGC 4473, 
NGC 4486A (see also Kormendy \etal 2005),  
NGC 4515, 
NGC 4551 (see also Lauer et al.~1995), 
NGC 4621, 
VCC 1627, and 
VCC 1871. 
The isophotes remain disky well into the S\'ersic part of the profile; in fact, they are 
sometimes most disky there and not in the ``extra light'' part of the galaxy.   

      The extra light is neutral ($a_4 \simeq 0$) or boxy ($a_4 < 0$) in 
NGC 4459 (which, however, has an embedded dust disk), 
NGC 4434, 
NGC 4387 (which otherwise is boxy), 
NGC 4486B (see below), 
VCC 1199, and
VCC 1440.~NGC 4434 and VCC 1440 are almost round; the observed correlation of $a_4$ with apparent
flattening implies that ellipticals are either boxy or disky when seen edge-on but have
nearly elliptical isophotes when seen face-on (Bender \etal 1989; Kormendy \& Bender 1996).
So these galaxies have no leverage on the question of whether extra light is disky. In 
NGC 4486B, the extra light includes the double nucleus (Lauer \etal 1996).  Tremaine (1995) 
interprets the analogous double nucleus of M{\ts}31 as an eccentric disk. Statler \etal (1999),
Kormendy \& Bender (1999), 
Statler (1999),
Peiris \& Tremaine (2003), and 
Bender \etal (2005) 
discuss observational evidence in favor of this model.

       We conclude that extra light is usually disky.  Ferrarese et al.~(1994)
reach a more extreme conclusion:~they suggest that 
all power law galaxies are coreless because of central disks.  Lauer et al.~(1995) 
disagree; they show non-disky examples.  We do, also.  Nevertheless, 
the frequent observation that the extra light is disky is a sign that it was produced 
by dissipation.

\subsection{Kinematic Subsystems in Core and Extra Light Galaxies}

      Another clue to galaxy formation is the observation that cores and
extra light are often associated with kinematic subsystems that are decoupled from the 
rest of the galaxy.  We distinguish kinematic subsystems that are misaligned with the 
photometric axes from cold, disky subsystems that corotate with the rest of the galaxy. 
The latter are evidence for dissipative formation, although they do not tell us whether the 
gas that formed the disk was internal or accreted.  In contrast, kinematic misalignments 
do not necessarily imply dissipative formation, but they have traditionally been interpreted 
as accretions.  Work by the SAURON team now shows that this is not always correct:

      Core Es with kinematically decoupled, misagligned centers include 
NGC 4365, NGC 4382,  NGC 4406, NGC 4472, and NGC 4552
(Wagner \etal 1988; 
Bender 1988b; 
Jedrzejewski \& Schechter 1988; 
Franx \etal 1989b; 
Bender \etal 1994;
Surma \& Bender 1995;
Davies \etal 2001;
de Zeeuw \etal 2002;
McDermid \etal 2006;
Krajnovi\'c \etal 2008).
The most thoroughly studied subsystem is in NGC 4365.  Its central sturcture is disky (Figure 13) 
and rapidly rotating ($V/\sigma \sim 1.4$; Surma \& Bender 1994).  
The main body shows minor-axis rotation (Wagner \etal 1988) and so is triaxial (Statler 
\etal 2004).  NGC 4406 shows similar kinematic decoupling (Bender 1988b, Bender \etal 1994) and
minor-axis rotation (Wagner \etal 1988; Jedrzejewski \& Schechter 1989; Franx \etal 1989b).  

      The observation of disky isophotes and $V/\sigma \sim 1.4$ is normally interpreted as an 
argument for dissipative formation.   However, van den Bosch \etal (2006) model two-dimensional 
SAURON kinematic and photometric observations and show that the almost-90$^\circ$ decoupled 
central rotation ``is not dynamically distinct from [the triaxial structure of] the rest of the 
galaxy.''  Its stars are \hbox{metal-rich,} \hbox{$\alpha$-element} overabundant, and old 
(Surma \& Bender 1995).  Davies \etal (2001) remark that ``the decoupled core and the main body of 
the galaxy have the same luminosity-weighted age, $\sim$\ts14{\ts}Gyr, and the same elevated 
magnesium-to-iron ratio. The similarity of the stellar populations in the two components suggests 
that the observed kinematic structure has not changed substantially in 12 Gyr.''  There is no need
to postulate late accretion of a cold component; major mergers can make decoupled 
kinematic subsystems (Jesseit \etal 2007; Naab \etal 2007b).  Kinematic subcomponents in 
core galaxies appear to be no problem for our picture that these galaxies were made in dry mergers.  

      Still, it would be surprising if late accretions did not occasionally build a nuclear disk 
in what used to be a core E despite ``protection'' (\S\ts12.3) from X-ray
gas halos.  NGC 4621 may be an example.  A more obvious example is NGC 5322 (Bender 1988b, 
Rix \& White 1992; Scorza \& Bender 1995).  The presence of an edge-on dust disk (Lauer \etal 
1995, 2005) guarantees that the subcomponent was formed dissipatively.  

      Extra light ellipticals with distinct kinematic subsystems include NGC 4473 (\S\ts9.2) and
the following.  
NGC 4458  has a rapidly rotating center at $r$ \lapprox \ts0\farcs5; at 
$r > 5^{\prime\prime}$, the galaxy rotates slowly or in the opposite direction
(Halliday et al.~2001; 
Emsellem \etal 2004;
Krajnovi\'c \etal 2008). 
NGC 4458 is one of the clearest examples of extra light (Figure 19); it reaches out to $r > 4^{\prime\prime}$.  
Similarly, in NGC 4387, Halliday et al.~(2001) suggest that a central decrease in $\sigma$ implies a 
rotationally supported subsystem that is confirmed by
Emsellem \etal (2004) and
Krajnovi\'c \etal (2008).  
NGC 4621 has a rapidly rotating, disky center, as suggested by Bender (1990) 
and now beautifully shown by two-dimensional SAURON spectroscopy
(Emsellem \etal 2004).~Figure 16 shows its disky $a_4$ signature.  
In all three galaxies, SAURON two-dimensional maps of H$\beta$ line strength reveal no
difference in age between the decoupled center and the rest of the galaxy (Kuntschner \etal 2006).

      These results further imply that extra light components form dissipatively.  Usually 
(but not always) the stellar population indicators suggest that the central  extra light 
structures formed approximately at the same time as the rest of the galaxies' stars.

\centerline{\null}

\figurenum{39}

\vskip 3.3 truein

\includegraphics{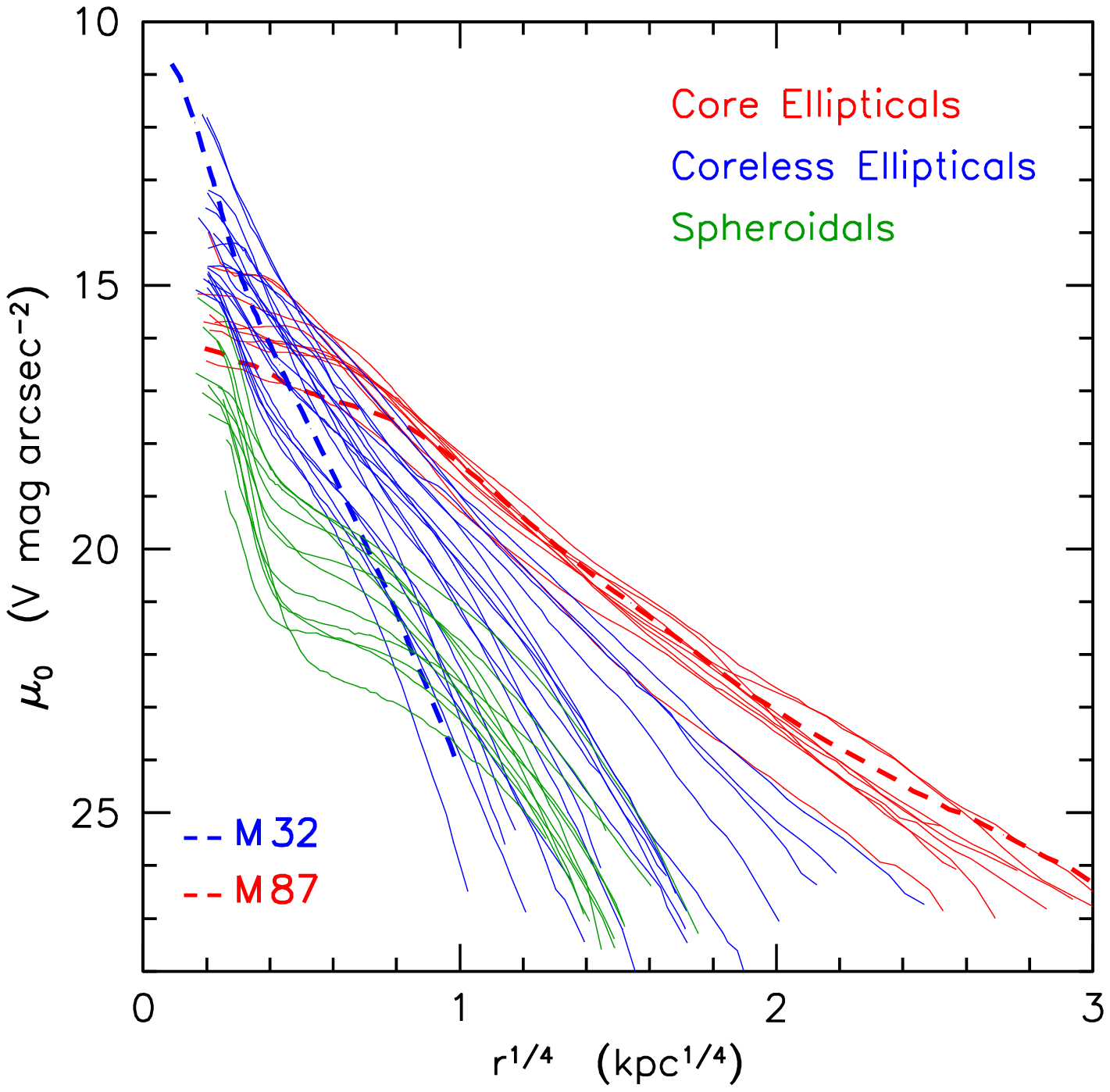}

\figcaption[]
{Major-axis profiles of all elliptical and spheroidal galaxies in our sample scaled so 
that radius is in kpc.  The brightness profiles are corrected for Galactic absorption.
The fiducial galaxies M{\ts}32 and M{\ts}87 are plotted with thick dashed lines.
The same profiles are shown in Figures 35 and 36; the emphasis here is on comparing
the two kinds of ellipticals.  As in Figures 34, 37, 38,and 41, M{\ts}32 is a normal 
example of the lowest-luminosity ellipticals.
}

\vskip 15pt

\subsection{The E{\ts}--{\ts}E Dichotomy\\Illustrated by Scaled Brightness Profiles}

      The dichotomy between core and extra light ellipticals is illustrated further 
in Figures 39 and 40.  Figure 39 shows all profiles in our sample scaled together 
so that radius is~in~kpc.  Because core ellipticals have $n > 4$ and extra light 
ellipticals have $n \lesssim 4$, their profiles curve apart at large radii.  A larger 
fraction of the light lives at large radii in core Es, so $r_e$ is larger and $\mu_e$ 
is {\it fainter\/} than in extra light Es (Fig.~37,~38).  But at almost all metric 
radii outside the core, {\it core ellipticals have higher surface brightnesses than do 
extra light ellipticals at the same metric radius.\/}  This is important, because 
\hbox{$n$-body} models of galaxy mergers predict that the surface brightness in the 
merger remnant is higher than the surface brightness of either progenitor at essentially 
all radii (Hopkins \etal 2008b).  Binary BH core scouring is the exception to this
prediction, and the relatively low absolute surface brightnesses in cores with respect
to extra light is clear in Figures~39~and~40.  The important conclusion from 
Figure 39 is that surface brightnesses in core galaxies are high enough so that they
can be products of dry mergers of extra light ellipticals (but see \S\ts11.1).

      Figure 40 shows all of our elliptical galaxy profiles scaled together at 
approximately the radius where the central core or extra light gives way to the outer 
S\'ersic profile.  Because the profiles of extra light Es break upward while core 
profiles break downward near the center, the core and extra light profiles are well 
separated from each other at small radii.  The present sample shows a fortuitously clean 
separation between core and coreless galaxies; larger samples show a few intermediate cases
(Rest \etal 2001;
Ravindranath \etal 2001;
Lauer \etal 2005, 2007b).  We have not yet checked whether these remain ambiguous
with the present definition of cores.  In any case, the distinction between galaxies with 
and without cores remains robust (Gebhardt \etal 1996; Lauer \etal 2007b). 

\centerline{\null}

\figurenum{40}

\vskip 3.3 truein

\includegraphics{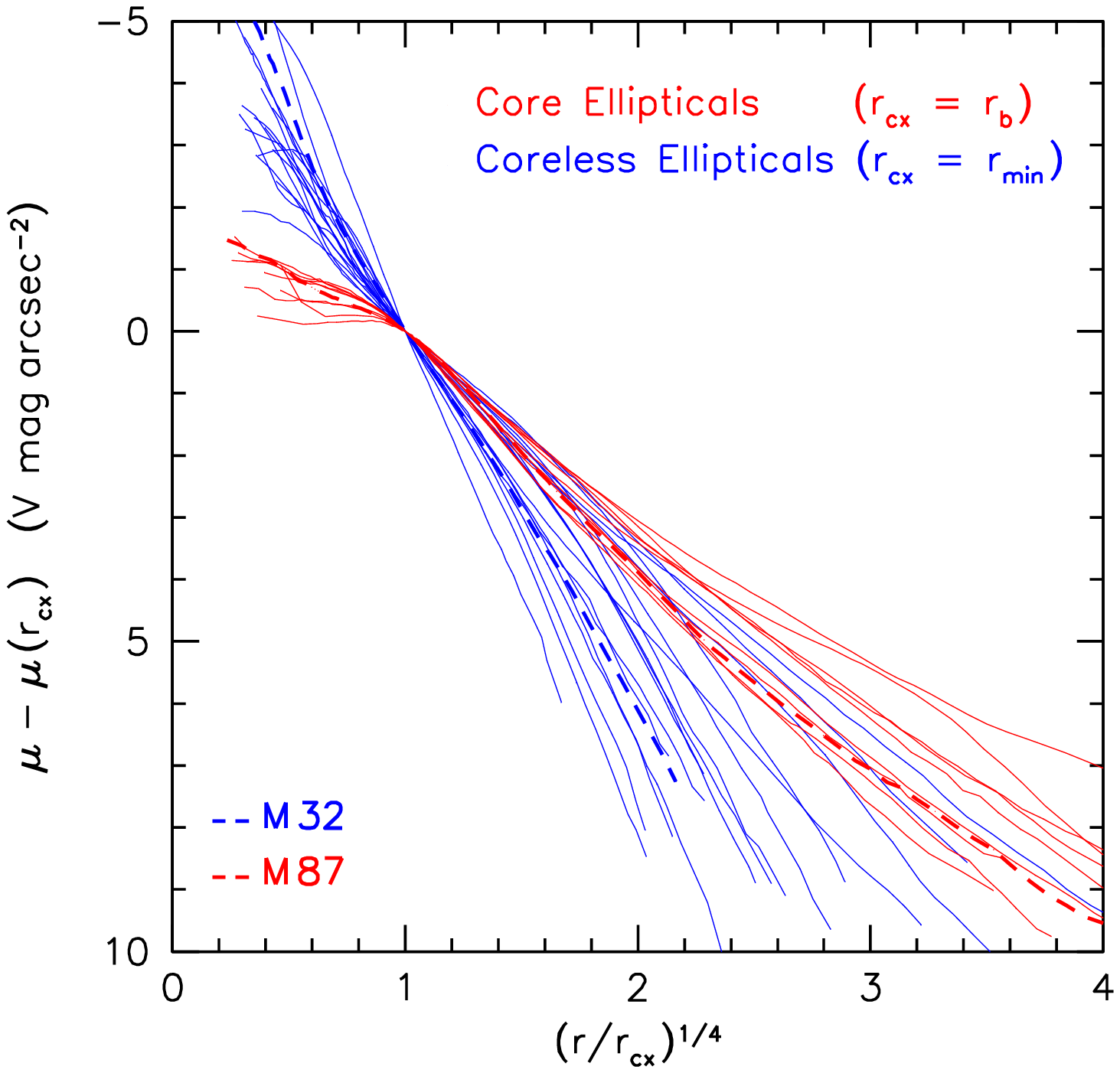}

\figcaption[]
{Major-axis profiles of all of our ellipticals scaled together to
illustrate the dichotomy between core and coreless ellipticals.  Core
ellipticals are scaled together at $r_{cx} = r_b$, the break radius given
by the Nuker function fit in Lauer \etal (2007b).~Coreless ellipticals
are scaled together at the minimum radius $r_{\rm min}$ that was used in our 
S\'ersic fits; interior to this radius, the profile is dominated
by extra light above the inward extrapolation of the outer S\'ersic fit.
}

\vskip 15pt

\subsection{Profile Shape Participates in the E -- E Dichotomy}

\lineskip=-4pt \lineskiplimit=-4pt

      Figure 41 shows again that S\'ersic index $n$ participates in the 
E{\ts}--{\ts}E dichotomy.   Also, E and Sph galaxies are well separated.

\figurenum{41} 

\vfill

\includegraphics{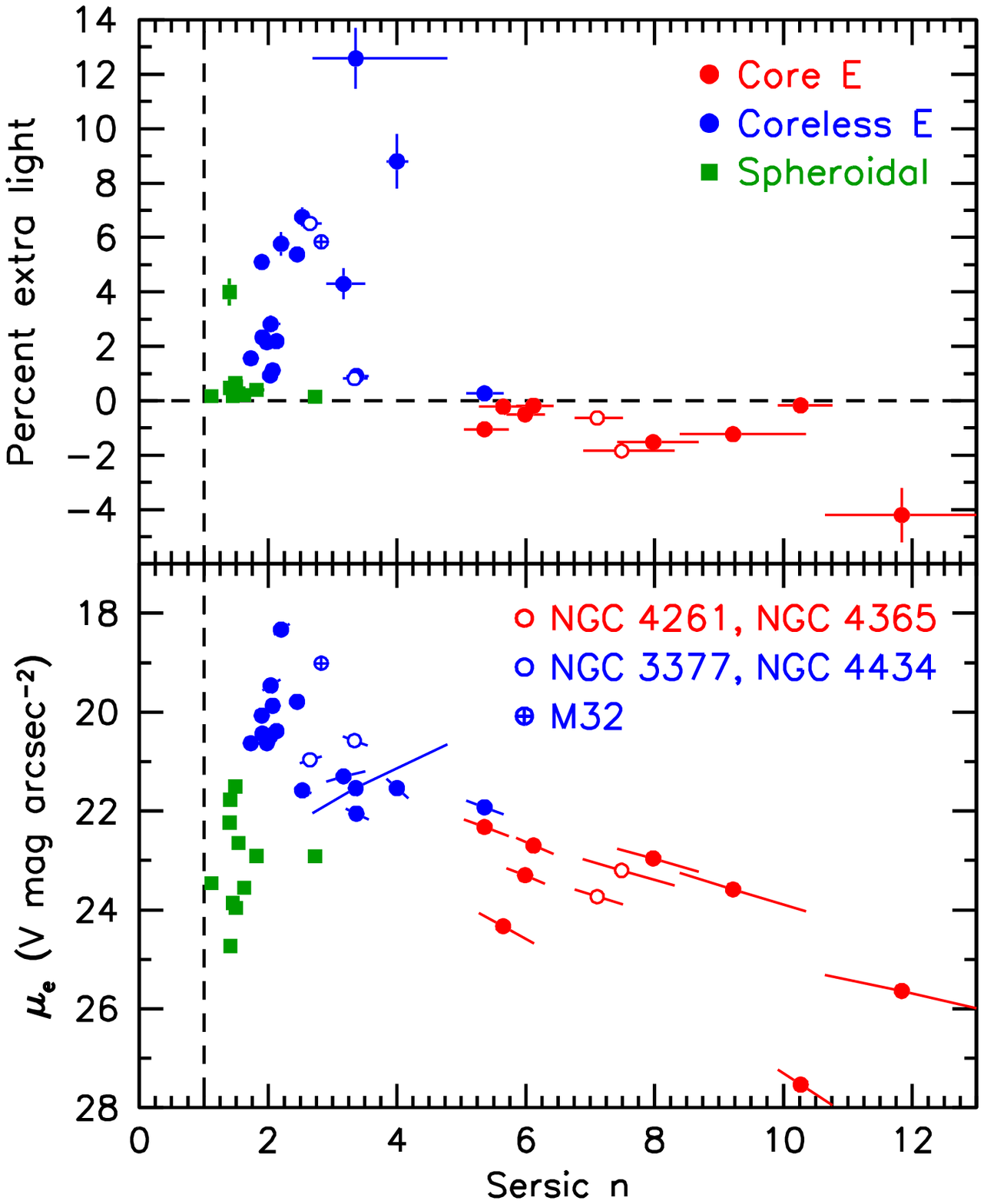}

\figcaption[]
{({\it top\/}) Percent of the total $V$-band luminosity that is ``missing'' 
in core galaxies or ``extra'' in coreless galaxies compared to the inward 
extrapolation of the outer S\'ersic fit.  ({\it bottom\/}) Effective surface 
brightness $\mu_e$ versus S\'ersic index $n$.  The symbols are as in Figures 34, 
37, and 38.
}

\eject

      Figure 41 ({\it top\/}) shows the amount of central extra light above the inward 
extrapolation of the outer S\'ersic fit as a percent of total galaxy luminosity.  
It is negative (light is ``missing'') for core galaxies.   The amount of extra 
light is calculated by integrating the two-dimensional brightness distribution 
of the galaxy non-parametrically from the center to the inner limit $r_{\rm min}$ of the 
S\'ersic function's radial fit range.  From this luminosity, we subtract the integral of 
the fitted S\'ersic function over the same radial range.  In the latter integral, the 
ellipticity $\epsilon$ of the S\'ersic function is kept fixed at $\epsilon(r_{\rm min})$.
Error bars are estimated by substituting plausible (usually small) extrapolations of the outer 
$\epsilon(r)$ profile into the region of the extra light.  These are internal errors only;
e.{\ts}g., the effects of changing the S\'ersic fits within the ranges allowed by their error
bars are not taken into account.  As a result, the error bars in the top panel of Figure 41
are not formally coupled.  The error bars in the bottom panel are coupled; they can be 
correlated or anticorrelated (see Figures 49{\ts}--{\ts}72 in Appendix A).   All points 
in Figure 41 have error bars, but most are too small to be seen.  Table 1 lists the plotted data.

     Note: For M{\ts}87, we used the bottom fit in Figure 50; i.{\ts}e., the one that allows 
for a cD halo.  The top fit in Figure 50 provides the upper error bar on the amount of missing 
light.  That is, for M{\ts}87, the error bars are dominated by the choice of S\'ersic fit.

      Figure 41 demonstrates again that all core galaxies in Virgo (percent extra light $<$ 0)
also have S\'ersic indices $n > 4$.  All of the coreless ellipticals (percent extra light
$>$ 0), have $n \leq 4$ except NGC 4621.  We will use this result in \S\ts10.3.  

      The bottom panel of Figure 41 shows effective brightness against S\'ersic index.  
Ellipticals form a well defined sequence with core and extra light galaxies largely separated.
NGC 4621 is an exception to the E{\ts}--{\ts}E dichotomy: it has $n > 4$ but is disky and has
a little extra light near the center.  Otherwise, this Virgo cluster sample shows the dichotomy 
cleanly, and profile shape in the form of S\'ersic $n$ participates in it.  

      Spheroidal galaxies are well separated from ellipticals in both panels.  As in
Figures 34 -- 39, they have smaller~$n$~and lower central and effective surface 
brightnesses than extra light ellipticals.  This is consistent with their similarity in
parameter correlations to galaxy disks (Kormendy 1985b, 1987b).  

\subsection{Nuclei -- Unrelated to Extra Light and Supermassive BHs}

\lineskip=-4pt \lineskiplimit=-4pt

      Nuclei in spheroidal galaxies are very different from extra light in elliptical galaxies.  
Hopkins \etal (2008b) show that they have almost orthogonal parameter correlations.  Here, 
Figure~41 shows that nuclei contain a much smaller fraction of the total galaxy light.
NGC 4482 (green point at 4\ts\% in the top panel) looks like -- but is not -- an exception; 
the Sersic fit in Fig.~25 fails at relatively large radii, and the extra light interior 
to this is included in the 4\ts\%.  However, the nucleus in NGC 4482 is similar in light 
fraction to the nuclei of other spheroidals.  All of our Sph galaxies are nucleated, and 
the nuclei all contain similar fractions of the galaxies' light.  The mean light fraction of 
our Sph nuclei is $0.33 \pm 0.06$ \%.  The analogous fraction for extra light Es is much larger 
and has a much larger range. 

      Several authors note that nuclei make up roughly the same fraction of spheroidal 
galaxy stellar masses as supermassive BHs do of their host bulges ($\sim$\ts0.13\ts\%: Merritt 
\& Ferrarese 2001; Kormendy \& Gebhardt 2001).  These authors plot BH and nuclear mass 
against galaxy absolute magnitude and find a single, continuous correlation
(C\^ot\'e \etal 2006;
Wehner \& Harris 2006;
Ferrarese \etal 2006b;
Graham \& Driver 2006).
They suggest that nuclei and BHs are related -- a galaxy contains either a nucleus or a BH, and 
perhaps nuclei evolve into BHs.   We confirm the observational conclusion but suggest that it is 
an accident.  Nuclei constitute a canonical fraction of some Sphs, but others contain no nuclei 
(Sandage \etal 1985;
Binggeli \etal 1985, 1987;
C\^ot\'e \etal 2006).
In late-type galaxies, nuclear absolute magnitudes correlate with total magnitudes, but only weakly 
(Carollo, Stiavelli, \& Mack 1998; B\"oker \etal 2004). Furthermore, BHs exist even in bulgeless disks
(Filippenko \& Ho 2003;
Barth \etal 2004;
Greene \& Ho 2004, 2007;
Peterson \etal 2005;
Greene, Barth, \& Ho 2006;
Shields \etal 2008;
Barth, Greene, \& Ho 2008;
Thornton \etal 2008;
see Ho 2008 for a review),
but BH masses correlate very little with their host disks (Kormendy \& Gebhardt 2001). Finally, 
some galaxies -- including ones with classical bulges -- clearly contain both BHs and nuclei.  
Sometimes the BH mass is much larger than that of the nucleus
(NGC 3115:        Kormendy \etal 1996b);
                  sometimes the BH mass is similar to that of the nucleus
(M{\thinspace}31: Light, Danielson, \& Schwarzschild 1974; 
                  Dressler \& Richstone 1988;
                  Kormendy 1988;
                  Lauer \etal 1993; 1998;
                  Kormendy \& Bender 1999;
                  Bender \etal 2005;
NGC 4395:         Filippenko \& Ho 2003;
                  Peterson \etal 2005);
                  and sometimes the BH mass appears to be less than that of the nucleus
(NGC 1042:        Shields \etal 2008).
We believe that there is no observational reason to suspect more of a physical relationship between 
nuclei and BHs than the generic likelihood that both are fed with gas from the disk.

\section{Interpretation: Wet versus Dry Mergers}
\vskip 4pt

\subsection{Black Hole Scouring of Cuspy Cores in Giant Ellipticals}

      Figure 41 shows that a typical core E is missing 1\ts$\pm$\ts1\ts\% of its 
starlight near the center with respect to the inward extrapolation of a S\'ersic 
function fitted to the outer profile.  Implicit in this statement is the hypothesis 
that these ellipticals would have had S\'ersic profiles if not for the process that 
excavates cores.  This is consistent with Figure 39, which shows how representative 
dry-merger progenitor profiles would ``fill'' core profies, and with the canonical 
explanation of how cores form:  

      Understanding cores is nontrivial.  Observed core parameter relations 
show that, in higher-luminosity ellipticals, the break in the profile that 
defines the core occurs at larger radius $r_b$ and fainter surface brightness
$I_b$ (see Faber et al.~1997 for {\it HST\/} core parameter correlations and
Kormendy 1984, 1985b, 1987a, b; Lauer 1985a, b for the analogous ground-based
results).  Mergers generally preserve the  
highest-density parts of their progenitors.  Therefore, when ellipticals
or bulges that satisfy the core parameters correlations merge, this
tends to destroy the correlations (Kormendy 1993; Faber et al.\ 1997).
Fluffy cores in high-luminosity ellipticals are not a natural
consequence of hierarchical clustering and galaxy merging. 

      A possible solution to this problem is the suggestion that cores form via the 
orbital decay of binary supermassive black holes 
(Begelman \etal 1980; 
Ebisuzaki \etal 1991; 
Makino \& Ebisuzaki 1996; 
Quinlan 1996; 
Quinlan \& Hernquist 1997; 
Faber \etal 1997;
Milosavljevi\'c \& Merritt 2001; 
Milosavljevi\'c et al.~2002; 
Makino \& Funato 2004;
Merritt 2006; 
Merritt, Mikkola, \& Szell 2007).  BH binaries form naturally in the galaxy mergers that 
are believed to make ellipticals.  Their orbits decay -- the binaries get harder~-- by 
flinging stars away.   These stars are deposited into a large volume at large radii or are 
ejected from the galaxy; either way, they have little effect on the outer profile.  As stars 
are removed from the small volume near the BHs, the central surface brightness decreases.
In this way, the decaying binary excavates a core.  The effect of a series of mergers is
cumulative; if the central mass deficit after one merger is a multiple $f$ of the BH mass 
$M_\bullet$, then the mass deficit after $N$ dissipationless mergers should be $M_{\rm def}
\simeq N f M_\bullet$.  If this picture is correct and if $f$ can be predicted from theory 
or simulations, then a measure of the observed mass deficit tells us roughly how many 
dissipationless mergers made the galaxy.

      One problem is that $f$ is not well known.  Milosavljevi\'c \& Merritt (2001) estimate 
that $f \simeq 1$ to 2.  Milosavljevi\'c~et~al. (2002) get $Nf \simeq 5$ for formation in a 
hierarchy of mergers.  Until recently, the most accurate $n$-body simulations was that of
Merritt (2006), who concluded that $f \simeq 0.5$.  Past observations of mass deficits 
depended on the functional form used to extrapolate the outer profile inward to the center;
they are larger for Nuker function extrapolations 
(Milosavljevi\'c \& Merritt 2001; 
Milosavljevi\'c et al.~2002; 
Ravindranath \etal 2002)
and smaller for S\'ersic function extrapolations.  As it became clear that S\'ersic 
extrapolations are both well supported by the data and intrinsically conservative (see
Figure 1 in Graham 2004), observations converged on values of 
$Nf \equiv M_{\rm def}/M_\bullet$ between 1 and 2;
most commonly, $M_{\rm def}/M_\bullet \simeq 2$, and values as large as 4.5 are rare
(Graham 2004; 
Ferrarese \etal 2006a;
Merritt 2006).
The conclusion was that these are consistent with galaxy formation by several 
successive dry mergers.

      With more accurate profiles, we can better measure mass deficits.  However, 
only giant ellipticals have deficits; small ellipticals have mass excesses.  
So Figure 42 separately shows central stellar mass deficits ({\it lower panel\/}) 
and mass excesses ({\it upper panel\/}) against $M_\bullet$.  Lines are drawn at 
$M_{\rm def}/M_\bullet = 1$, 5, 10, and 50.   Large symbols denote galaxies with 
dynamical BH detections; for these, the BH mass and stellar mass-to-light ratio are 
taken from the BH discovery paper.  Small symbols denote galaxies without dynamical BH 
detections.  Then $M_\bullet$ is derived from the correlation between $M_\bullet$
and $\sigma$ (Ferrarese \& Merritt 2000; Gebhardt \etal 2000) as fitted
by Tremaine \etal (2002).  The estimated error in $\log {M_\bullet}$ is 0.3.
In constructing Figure 42, we converted light excesses (Table 1) to mass excesses
using mass-to-light ratios $M/L_V \propto L^{0.36}$ fitted to the SAURON sample of 
Cappellari \etal (2006) including M{\thinspace}32.  The zeropoint is $M/L_V = 6.07$ 
at $M_V = -21.6$, i.{\thinspace}e., the divide in Table 1 between core and extra light
ellipticals.  Our error estimate in $\log {M/L_V}$ is 0.153, the RMS scatter of the 
above fit.  This is consistent with the results of Cappellari \etal (2006), who work 
in $I$ band.

\lineskip=-4pt \lineskiplimit=-4pt

      We adopt the Cappellari \etal (2006) $M/L$ ratios because they are based on the 
most accurate, three-integral models applied to the most detailed, two-dimensional 
SAURON~data.  Also, the resulting $M/L$ ratios correlate well with values based on 
stellar population models, although there is an offset that may imply a dark matter
contribution or a problem with the stellar initial mass function used in the
population models.  The choice of $M/L_V$ critically affects the derived mass excesses, 
so independent checks are welcome.  Many are available.  They include additional 
$M/L$ values based on three-integral models (Gebhardt \etal 2003, 2007; Thomas \etal
2007), two-integral models of galaxies observed to very large radii (Kronawitter \etal 
2000; Gerhard \etal 2001), and two-integral models of large galaxy samples (e.{\thinspace}g., 
van der Marel 1991).  All authors generally agree well with the steep $M/L_V$ -- $M_V$ 
correlation that we derive from the Cappellari data.  Significant caveats still need 
exploration.  For example, dynamical $M/L_V$ values may include a dark matter 
contribution that depends on $M_V$.  Also, triaxiality is not included in the
dynamical models and may depend on $M_V$.  But the mass-to-light ratios that we use 
in what follows are the most robust ones that are currently available in the literature.

\figurenum{42} 

\centerline{\null}

\vskip 3.9truein

\includegraphics{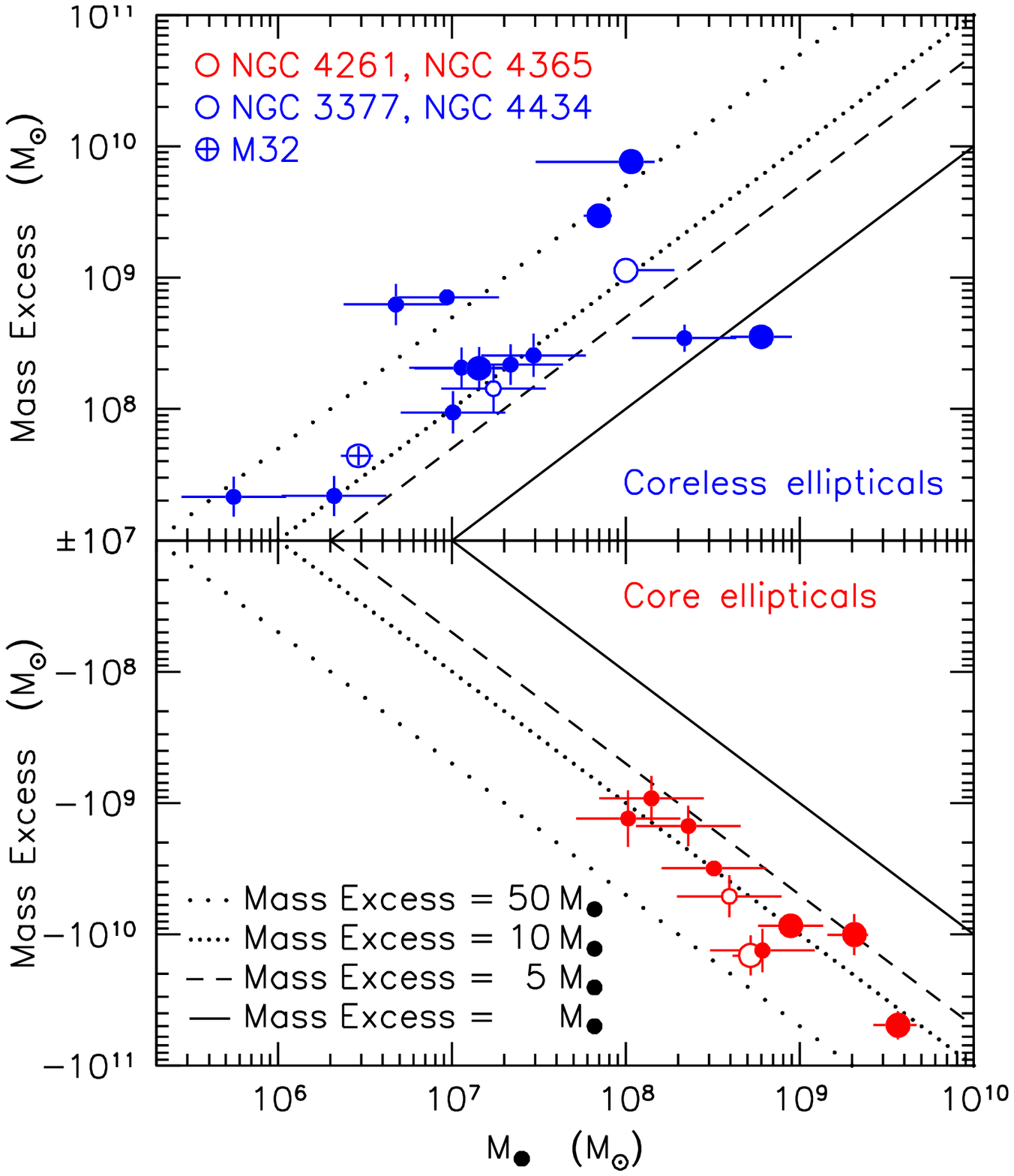}

\figcaption[]
{Total stellar mass that is ``missing'' (in cores, {\it lower panel}) or ``extra'' 
(in coreless galaxies, {\it upper panel\/}) as a function of black hole mass. 
Large and small symbols denote galaxies with and without dynamical BH detections, 
respectively.  NGC 4486B has the smallest excess 
in the upper panel because $M_\bullet$ is unusually large (Kormendy \etal 1997).  
\lineskip=-4pt \lineskiplimit=-4pt
}

\vskip 15pt

      The mass deficits $M_{\rm def}$ that we derive for core galaxies are larger than 
published values, partly because our $M/L_V$ values are larger and partly as a result
of more accurate photometry.  They are also remarkably uniform, and -- although the
sample is small -- they show no offset between galaxies with and without dynamical
BH detections. In Figure 42, the unweighted mean 
$<$\null$\log {M_{\rm def}/M_\bullet}$\null$>$ = $1.04 \pm 0.07$.  The weighted mean is
$<$\null$\log {M_{\rm def}/M_\bullet}$\null$>$ = $1.07 \pm 0.08$.  That is,
$M_{\rm def}/M_\bullet \simeq 11$ with an error in the mean of about 18\ts\%.
The smallest value is $4.9^{+2.4}_{-1.9}$ for NGC 4649, and the largest value
is $28^{+13}_{-10}$ for NGC 4261.  These values are very large in comparison to the 
Merritt (2006) prediction that $M_{\rm def}/M_\bullet \simeq 0.5$ per major merger.
However\ts:

      Two recent results help to explain such large $M_{\rm def}/M_\bullet$ values:

      First, with a more accurate treatment of the late stages of binary BH mergers, 
Merritt, Mikkola, \& Szell (2007) find that $M_{\rm def}/M_\bullet$ can be as large as 
$\sim 4$ per merger.  Then our results are reasonably consistent with estimates
(Faber 2005;
van Dokkum 2005;
Bell \etal 2006)
that several dissipationless mergers produced the bright end of the ``red sequence''
part of the color bimodality of galaxies observed by the Sloan Digital Sky Survey
(Strateva \etal 2001; 
Hogg \etal 2002, 2004; 
Kauffmann \etal 2003a, b;
Blanton \etal 2003, 2005;
Baldry \etal 2004)
and by the COMBO-17 survey (Bell \etal 2004). 
If present-day galaxies provide any guide to the properties of merger progenitors
(and they may not -- see \S\ts11.1), then it is essentially required that
galaxies as big as M{\ts}87 formed in several successive dry mergers.  Giant
ellipticals are so big that plausible immediate progenitors are cold-gas-poor galaxies.

      Second, an additional process has been proposed to make large-$M_{\rm def}/M_\bullet$ 
cores (Merritt \etal 2004; Boylan-Kolchin, Ma, \& Quataert 2004; Gualandris \& Merritt 2008).  
Coalescing binary BHs emit gravitational radiation anisotropically; they recoil at velocities
comparable to galaxy escape velocities.  If they do not escape, they decay back to the 
center by dynamical friction.  In the process, they throw away additional stars.  Gualandris 
\& Merritt (2008) estimate that they can excavate as much as $M_{\rm def}/M_\bullet \sim 5$ 
{\it in addition to the mass that was already scoured by the pre-coalescence binary.}  
Any conclusions to be reached from Figure 42 necessarily depend 
on our choice of S\'ersic functions as our models for unscoured merger remnants.  But it 
appears that our observations present no problem for the idea that cores in giant 
ellipticals are made by a combination of the above two BH scouring mechanisms acting over 
the course of one or more successive dry mergers.

\subsection{Extra Light in Low-Luminosity Ellipticals: Implications 
               for Black Hole Scouring and AGN Energy Feedback}
 
      Figure 42 ({\it upper panel\/}) shows, for coreless galaxies, the 
central stellar mass excess above the inward extrapolation of the outer
S\'ersic profile.  Five galaxies ({\it large symbols\/}) have dynamical BH detections,
M{\ts}32,
NGC 3377,
NGC 4459, 
NGC 4486A (Nowak \etal 2007), and
NGC 4486B (see Kormendy 2004 for additional references).  BHs and extra light 
are not mutually exclusive.  In fact, if essentially all bulges and ellipticals 
contain BHs (Magorrian \etal 1998), then the other extra light ellipticals are 
likely to contain BHs, too.~They are included in Fig.~42 with BH masses 
from the $M_\bullet$ -- $\sigma$ relation.  The median of 
$\log {M_{\rm def}/M_\bullet}$ is 1.120 (quartiles 0.955, 1.608);
i.{\thinspace}e., median $M_{\rm def}/M_\bullet$ = 13 (quartiles 9, 41).
The mean is $<$\null$\log {M_{\rm def}/M_\bullet}$\null$>$ = 1.159 $\pm$ 0.150
or $<$\null$M_{\rm def}/M_\bullet$\null$>$ = $14^{+6}_{-4}$.

      What are the implications of the extra light for our picture of core
formation by binary BH scouring?  We emphasize: {\it Extra light ellipticals 
satisfy the $M_\bullet$ -- $\sigma$ correlation as well as do core ellipticals.  
We believe that they formed in mergers.  These mergers cannot all have involved 
at least one pure-disk, black-hole-less galaxy.  Why, then, do coreless ellipticals 
have extra light, not missing light, in their centers?  Why did core scouring by 
binary black holes fail?}  We suggest an answer based in part on the observations
in \S\S\ts9.3 and 9.4 that point to dissipational formation of coreless ellipticals.
{\it We suggest that core scouring is swamped by the starburst that results
from the rapid infall of gas that occurs in a wet merger} (e.{\ts}g., Mihos \& 
Hernquist 1994).  The mass excesses in coreless Es tend to be somewhat larger than
the mass deficits in core Es, when both are expressed as multiples of the BH mass.
Our measurements of mass excesses may be slight underestimates (\S\ts10.3). 
This suggests that it is relatively easy for new stars to swamp any core scouring 
that may have occurred.  We pursue the possible starburst formation
of the extra light in the next subsection.

      First we note an implication for energy feedback from active galactic nuclei 
(AGNs).  A popular hypothesis to explain why giant ellipticals stopped making stars
after $<$\ts1 Gyr (Bender 1996, 1997; Thomas et al.~1998, 1999, 2005)
is that AGN feedback quenched star formation 
(Springel, Di Matteo, \& Hernquist 2005;
Scannapieco, Silk, \& Bouwens 2005;
Bower et al.~2006;
De Lucia et al. 2006).
We suggest in \S\ts12.3 that AGN feedback is fundamental to the creation of
the E -- E dichotomy.  Here we note that such feedback can easily quench the 
star formation that -- we suggest (\S\ts10.3) -- makes the extra light in coreless 
galaxies.  This implies that the importance of AGN energy feedback is a 
strong function of galaxy and BH mass.  It may have regulated the formation of giant 
ellipticals, but it cannot have quenched all star formation in coreless ellipticals 
if our interpretation of the extra light is correct.

\vskip -15pt \centerline{\null}

\subsection{Dissipative Merger Formation of Extra Light in Low-Luminosity Ellipticals}

      This brings us back to the explanation of the extra light in coreless 
ellipticals.  As reviewed in \S\ts4.2, Kormendy (1999) found the extra
light component in three ellipticals that span the luminosity range
over which this paper shows it to occur.  In M{\thinspace}32 and NGC 3377, the extra
light was well resolved by {\it HST\/} photometry.  The brightness profiles
of all three galaxies closely resemble the density profiles of ellipticals produced 
in simulations of gas-rich mergers (Mihos \& Hernquist 1994:~Fig.~4 here).
The gas sinks rapidly to the center during the merger; the resulting starburst 
produces an ``extra'' component of young stars that are clearly distinct 
from the S\'ersic profile ($n < 4$) of the mostly dissipationless part of the 
merger remnant.  Mihos \& Hernquist (1994) were concerned that such two-component
density profiles were not consistent with the observations.
After further simulations confirmed these results, Mihos \& Hernquist 
(1996, see p.~660) remarked, ``Perhaps more worrisome are the stellar residuals 
of the nuclear starbursts. \dots~The light profile of the starburst population 
does not join `seamlessly' onto that of the old stars in the remnant but is instead 
manifest as a luminosity `spike', in apparent disagreement with the core properties 
of massive ellipticals (see, e.{\ts}g., Lauer et al.~1995).  What is the significance
of this result for the merger hypothesis?''  Kormendy (1999) pointed out that the 
results of the gas-rich-merger simulations look just like the two-component 
profiles observed in the above galaxies and suggested that the inner component
was produced, as in the Mihos \& Hernquist paper, in the merger starburst.

      Note that this explanation does not require the extra light to be young.
If the merger happened long ago, the age difference between the main
body and the extra light would be hard to detect.  Worthey (2004) 
observed a stellar population gradient in M{\ts}32 (age 4 to 6 Gyr 
at $r$ \lapprox 5$^{\prime\prime}$ and 8 to 10 Gyr at larger radii), 
although he saw no discontinuity at the radius of the break between the extra light
and main body of the galaxy.  This is consistent with the present formation 
picture.  However, it would be reasonable to expect that, in a large sample, 
at least some central components should have younger stellar populations than 
the rest of the galaxy.  This is observed 
(Lauer \etal 2005;
Kuntschner \etal 2006;
McDermid \etal 2006).

      In our sample, we find extra light in all coreless galaxies.  Like Kormendy (1999), 
{\it we suggest that the extra light in low-luminosity 
elliptical galaxies generally formed as in the Mihos \& Hernquist (1994) models; that is, 
in the starburst that accompanies the merger that made the elliptical.}  Alternatives
exist and almost certainly happened in some galaxies.  A few
extra light components in large ellipticals could be the remnants of the compact and dense
centers of dissipationlessly accreted small ellipticals (Kormendy 1984; Balcells \& Quinn 
1990), provided that they were too massive to be lifted by BH binaries.  However, 
the frequent observation that the extra light is disky and rapidly rotating argues that it
usually forms dissipatively (Scorza \& Bender 1995; \S\S\ts9.3 and 9.4 here).  So the more
likely alternative is that a few extra light components formed via 
accretions of gas-rich dwarfs (\S\ts11.2).

      More recent simulations of gas-rich mergers also produce an extra component
near the center as a result of merger-induced starbursts (e.{\ts}g., Springel 
2000).  We illustrate two of these.

      Springel \& Hernquist (2005) ran a merger simulation in which the progenitors were dark
matter halos containing gas disks but no stars.  They included star formation according to a
Schmidt (1959) -- Kennicutt (1998a, b) law, energy feedback from supernovae, and thermal 
evaporation of cold gas clouds.  The density distribution of the merger remnant is shown 
in Figure 43.  Stars that form in an early close passage later relax violently in the merger
and produce an almost-$r^{1/4}$-law, elliptical-galaxy-like component; they call this the 
``spheroid'' and we label it \hbox{``bulge $\approx$ E''} in Figure 43.  Inspection of 
their Figure 3 shows that this component is, except near the center, a S\'ersic function 
with $n < 4$.  During the merger, much of the remaining gas falls to the center, and a
starburst produces a more compact ellipsoid that Springel \& Hernquist call the ``bulge'' 
and that we label ``extra light'' in Figure 43.  Gas that survives the merger settles into
a new disk that forms stars slowly; this disk has an exponential stellar density 
distribution and is labeled ``disk'' in Figure 43.  Because the progenitor galaxies 
contained no stellar disks that could be heated and destroyed in the merger, the final extra 
light{\ts}:{\ts}bulge{\ts}:{\ts}stellar disk mass ratios, 0.55{\ts}:{\ts}0.22{\ts}:{\ts}0.23, 
are much different than they are in real galaxies.  Nevertheless, the merger remnant has
the qualitative character that we see in our data.  The non-disk part of the remnant 
consists of an elliptical-galaxy-like part that satisfies an $n < 4$ S\'ersic function plus 
extra light at the center that gives the sum a two-component look.  Enough gas survives the 
merger to make a new disk.  We observe S0 galaxies that have such disks, a bulge that satisfies
a S\'ersic function, and sometimes extra light (Fig.\ts30\ts--\ts32).

\vskip 0.15truein
\figurenum{43}

\centerline{\psfig{file=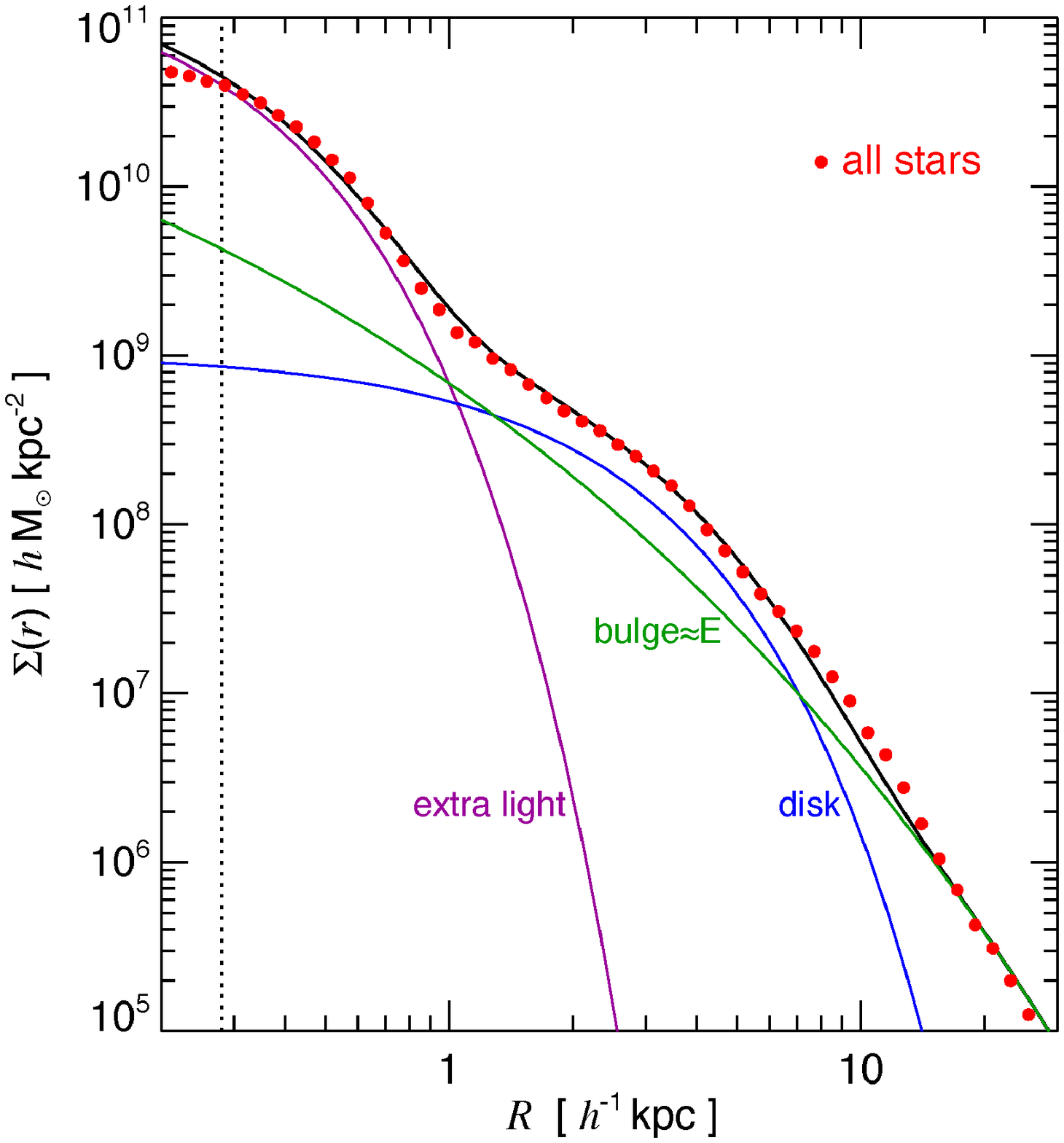,width=8.83cm,angle=0}}

\figcaption[]
{Surface density distribution of the remnant of a gas-rich merger adapted from
Fig.~3 of Springel \& Hernquist 2005, astro-ph/0411379 version; the vertical dotted
line is the resolution of the simulation.  The progenitors contained no stars,
only gas disks embedded in dark halos.  Stars that formed in the first, pre-merger 
encounter later relaxed violently into the density distribution labeled ``bulge $\approx$ E''; 
it is a S\'ersic function with $n < 4$.  During the merger, two-thirds of the remaining
gas falls to the center, undergoes a starburst, and makes the density distribution labeled
``extra light''.  The remaining gas settles into a new star-forming disk whose stellar 
density profile is labeled ``disk''.  Note that the ellipsoidal part of the galaxy; that is,
the sum of the bulge and extra light, has a two-component density profile like those in
Figures 16 -- 24 but with more extra light than is seen in the observations.
}

\vskip 0.6cm

\centerline{\null}

\figurenum{44}

\begin{figure*}[t] 

\vskip 4.2truein

\includegraphics{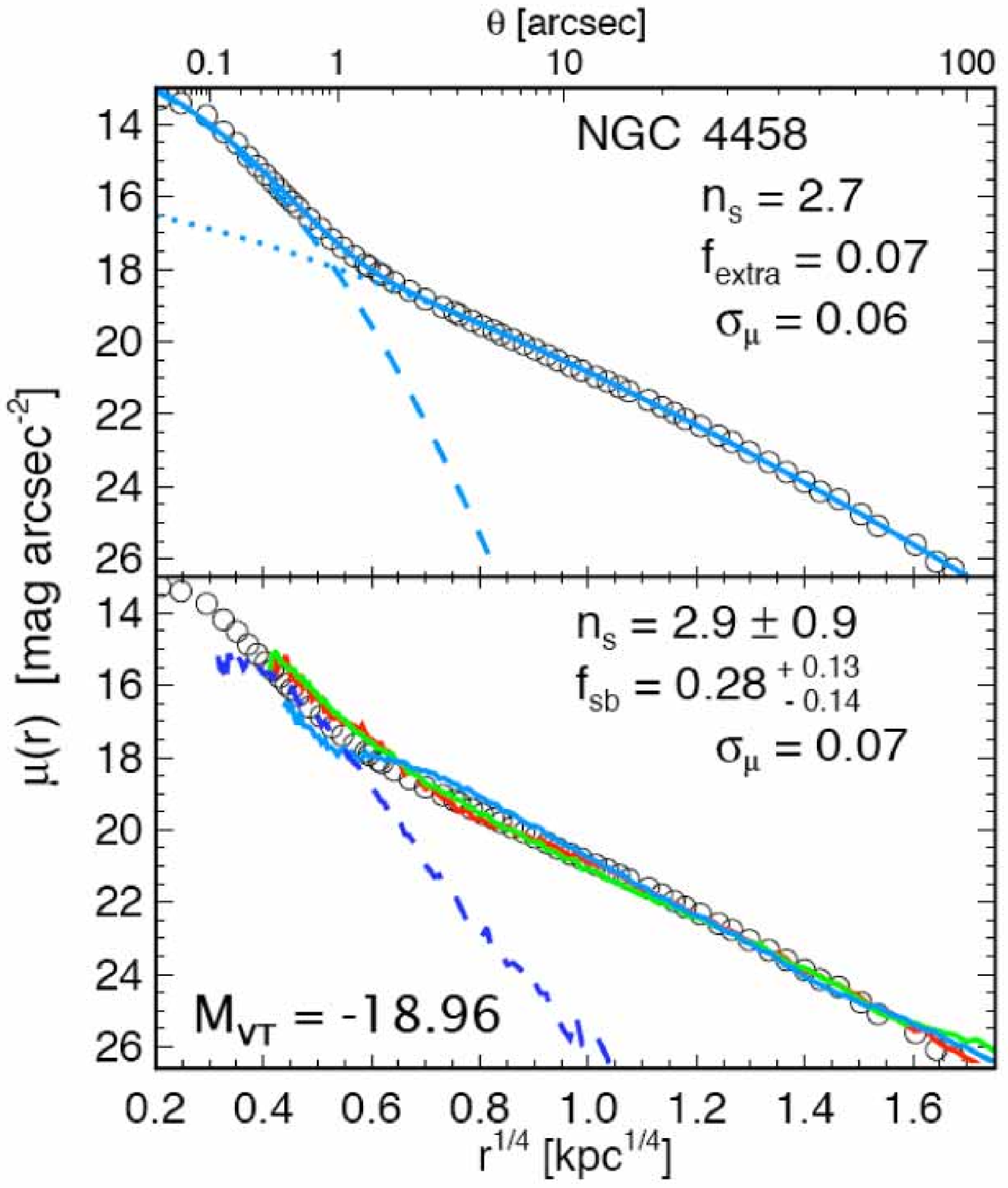}

\includegraphics{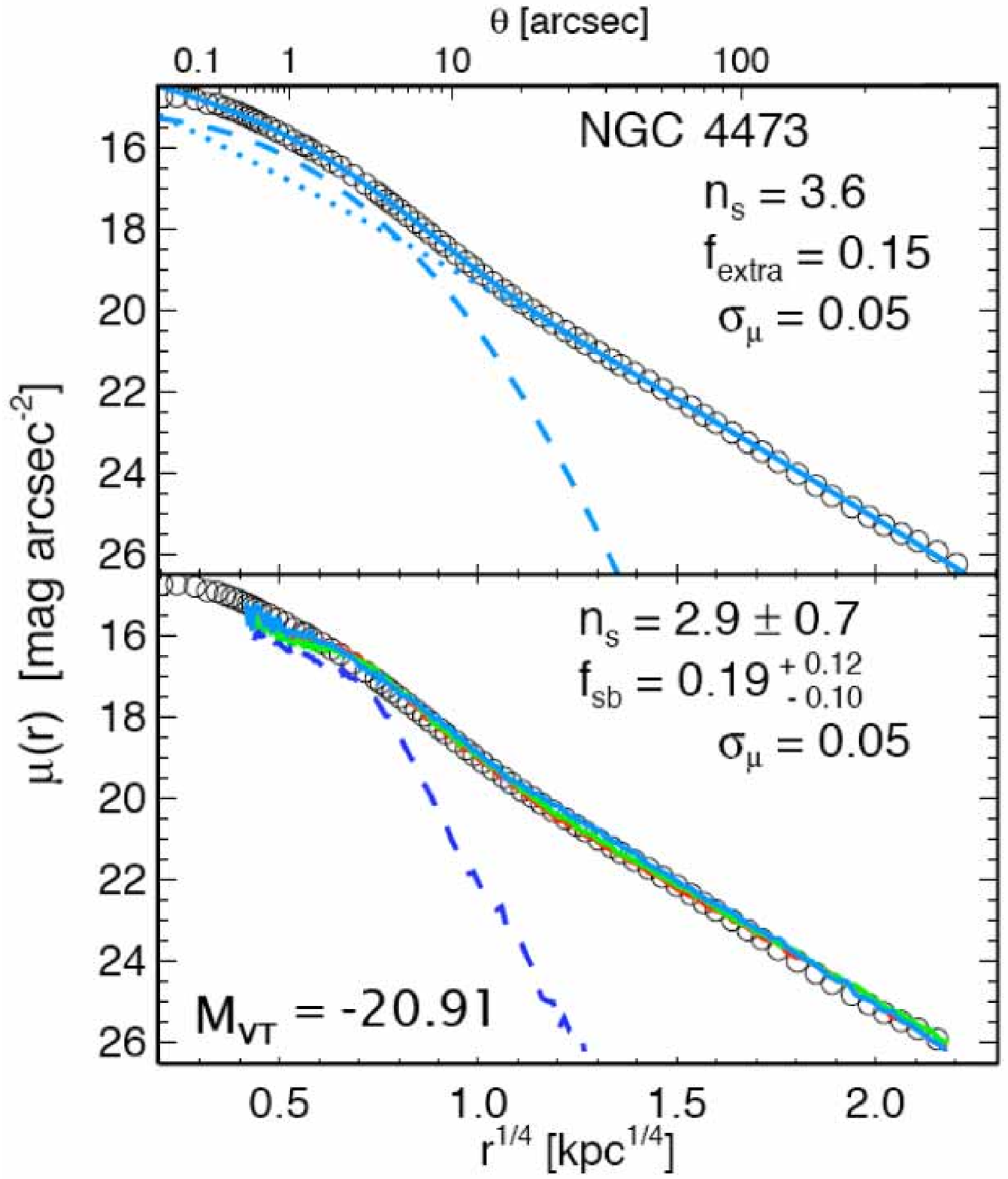}

\figcaption[]
{Stellar density profiles of the remnants of dissipative starburst mergers from Hopkins \etal (2008b). 
The {\it top panels\/} show brightness profiles from the present paper ({\it open circles\/}) 
decomposed into two S\'ersic functions; the S\'ersic index $n_S$ of the main body, the fractional 
contribution $f_{\rm extra}$ of the central extra light, and the RMS deviations of the fit (mag
arcsec$^{-2}$) are given in the key.  The corresponding values from our analysis are
$n = 2.53^{+0.14}_{-0.13}$, $f_{\rm extra} = 0.068 \pm 0.004$, and RMS = 0.0295 mag
arcsec$^{-2}$ for our fit to the main body of NGC 4458 and $n = 3.9$, $f_{\rm extra} = 0.09$,
and RMS = 0.048 mag arcsec$^{-2}$ for our decomposition of NGC 4473 (Figure 59).
The {\it bottom panels} show, in different colors, density profiles from the three library
simulations that best fit the galaxy profiles ({\it open circles\/}).  Also, the {\it blue dashed 
line\/} shows the starburst extra light component formed in the best-matching simulation.  The 
range of main-body S\'ersic indices for various viewing geometries is given in the key, 
together with the percent mass contribution $f_{sb}$ of the starburst in these three 
simulations and the RMS deviations of the fits.
}
\end{figure*}

\vskip 10pt

      Cox et al.~(2005) simulated dissipative mergers with a more detailed treatment
of radiative cooling, star formation consistent 
with a Schmidt-Kennicutt law, and energy feedback from massive stars and 
supernovae.  The progenitor galaxies were realistic approximations to Sbc galaxies, 
both structurally and in terms of gas content.  Moreover, the progenitor disks 
were constructed to have reasonable Toomre (1964) stability parameters $Q$ and 
realistic star formation rates; this required careful tuning of the prescriptions 
for star formation and energy feedback.  A range of parameters that bracket 
realistic Sbcs was explored to investigate the robustness of the conclusions.  
Star formation rates were very sensitive to the details of energy feedback.  
However, the density profiles of the remnant ellipticals proved to be relatively
insensitive to the energy feedback and gas physics (e.{\ts}g., equation of state).  
They confirm that star formation in gas that is dumped close to the center by the
merger builds a distinct central component in density that is brighter than 
the inward extrapolation of the density profile of the main body of the remnant.  
How much of the extra component was built by star formation and how much was the 
remnant of the progenitor bulges depends on the energy feedback; less efficient
feedback results in more star formation near the center.  If there is too much
feedback, the extra component cannot form.

      Since the submission of this paper, the most comprehensive simulations of dissipative 
mergers are a series of papers by Hopkins \etal (2008a, b, c, d, e) that are motivated 
directly by the present results and by similar observations of mergers in progress by 
Rothberg \& Joseph (2004, 2006).  They construct libraries of gas-rich merger 
simulations in which merger-induced starbursts make extra light components.  They match 
these up with galaxy observations -- including ours -- and they explore both wet and dry 
mergers in great detail.  They make substantial progress beyond this paper.  A review of 
this progress is beyond the scope of the present paper.  But it is important to connect up 
their results and ours, especially because they are based in part on the same observational data.

      Figure 44 shows two examples of model results from Hopkins \etal (2008b).  The 
{\it top panels\/} show decompositions of our profiles into two S\'ersic functions; the 
purpose is to estimate the fraction $f_{\rm extra}$ of the luminosity that is in the extra
light.  The {\it bottom panels\/} match the observed profiles with the best fitting results 
from their simulation library.  Unlike the interpretations of the extra light in the top panels
and in the present paper, the simulations have known fractions $f_{\rm sb}$ in their starbursts.
The extrapolation of the starburst component into the region dominated by the main body of the 
galaxy is not necessarily matched by the machinery in the top panel, but on the whole, the 
decompositions and the models give similar results for the starbursts.  That is, the behavior 
of the models fitted to the data in Figure 44 are entirely consistent with the formation
picture discussed in the present paper.  Since all details of the models are known, Hopkins 
\etal (2008a\ts--{\ts}e) can explore how models look from different viewing geometries and 
demonstrate that the results are consistent with observations of boxy and disky isophote distortions.

      It is instructive to compare the extra light fractions derived by Hopkins \etal
(2008b) with our estimates.  For 10 of the 18 extra light galaxies in common, the agreement
is very good; individual ratios of $f_{\rm extra}$ divided by our values range from 0.65 to 1.27
and average $1.00 \pm 0.08$.  For six more, the ratio ranges from 1.7 to 4.2 and averages 
$2.85 \pm 0.40$.  For the other two, the ratio is 18 for NGC 4434 and 12 for NGC 4486.

      These results are expected.  On the whole, the decomposition procedure in Hopkins \etal 
(2008b) is reasonable,\footnote{We cannot similarly confirm the decompositions of core galaxies 
in Hopkins \etal (2008c).  As stated in that paper, it is true in principle that ``all core 
galaxies are extra light galaxies, too'' in the sense that their merger progenitors may have
included extra light ellipticals.  If the extra light is not scoured away by binary BHs, 
it survives and contributes to the steep central brightness profiles of giant Es that, 
together with their shallow halos, gives them their large S\'ersic indices.  But looked 
at quantitatively, it is not clear how much extra light survives.  Most S\'ersic fits to core 
Es have small residuals whose profiles in Figures 11\ts--\ts15 and 49\ts--\ts56 show 
no significant upward wiggles just outside the core that are suggestive of extra light.  
NGC 4636 residuls allow a two-component structure, but the culprit is more likely to be 
an outer disk (illustrative decomposition in Fig.~55).  The giant elliptical whose residual 
profile most allows both a core and extra light is NGC 4472 (Fig.~49, {\it bottom\/}).  
However, this figure also shows brightness profiles of candidate extra light progenitors, 
NGC 4459 (which is one of the brightest) and NGC 4458 (which is typical in luminosity but which 
has an unusually large amount of extra light).  In both galaxies, the extra light lives at radii 
that are inside the core of NGC 4472.  Also, the amount of light that is missing in the
core of NGC 4472 has an absolute magnitude of $M_{V,\ts{\rm def}} = -17.5$.  The amount of 
extra light in NGC 4459 and NGC 4458 is $M_{V,\ts{\rm extra}} = -17.5$ and $-16.0$, respectively.
If present-day, extra light ellipticals in Virgo are the dry merger progenitors of giant core Es, 
then the stars in the extra light components are preferentially scoured away during core formation.  
Also, these galaxies may not be typical merger progenitors (\S\ts11.1).  Finally, the Hopkins \etal 
(2008c) decompositions of core galaxies into extra light components and main bodies have larger 
residuals than the present single-S\'ersic fits.  These decompositions are an interpretation that
is worth investigation within a well articulated formation picture.  But they are not required 
by the present profile data.}
as they demonstrate by comparing to model results.  The decomposition
is particularly robust for galaxies like those in Figure 44 that have bright and well resolved 
extra light and hence good ``leverage'' on both components.  A decomposition tends to give 
larger fractions of extra light than our estimation procedure.  This is expected, because
we made no decomposition; instead, we fitted the main body of each galaxy and added up the 
central light above this fit to estimate the extra component.  This almost certainly 
underestimates the starburst component slightly.  On the other hand, we did not make 
decompositions (except for NGC 4473 in Figure 59), because nothing in the residual profiles 
in Figures 16\ts--\ts24 and 57\ts--\ts67 demands them.  Indeed, our one-component S\'ersic fits 
often have smaller residuals than the two-component decompositions in Hopkins \etal (2008b).
Nevertheless, both the above comparisons and the tests done in the above paper show that
the decompositions are reasonable interpretations of the data.  The most questionable cases
(e.{\thinspace}g., NGC 4434 and NGC 4486A)  are ones where the wiggles in the extra
light profile formally cause the decomposition procedure to fit very shallow extra light
components.  These few objects have little influence on the conclusions. 

      So the conclusions from dissipative merger simulations are robust.  Some details
of remnant structure depend on gas physics and energy feedback.  
But the simulations very generally predict an extra component near the center that 
is produced by the merger starburst.  Authors of the early papers that showed this worried 
about whether these extra components are realistic or a problem, because they had not been
observed in the published brightness profiles of most ellipticals.

      Our results appear to settle this issue, at least for ellipticals in 
the Virgo cluster.  Extra light is almost ubiquitous in coreless ellipticals.  
Cores are believed to be scoured~by~binary~BHs.  The suggestion is that the last major merger 
that made core ellipticals was dry, whereas the last major merger that made coreless ellipticals
was wet and included a substantial central starburst.  

\subsection{S\'ersic Index as a Galaxy Formation Diagnostic}

      One of the clearest conclusions of this paper is that galaxy profile shape
as parametrized by the S\'ersic index participates in the E{\ts}--{\ts}E dichotomy.
This changes our view of the well known correlation that $n$ increases with galaxy 
luminosity.  Figure~33 shows that  $n$ does correlate with $M_{VT}$ in Sph galaxies.
But elliptical galaxies do not show a continuous correlation.  Instead, our observations 
show two clumps of points: core Es have $n > 4$ but no correlation of $n$ with $M_{VT}$, 
and extra light galaxies have $n \simeq 3 \pm 1$ but little correlation between $n$ 
and $M_{VT}$.  NGC 4621 is the exception; it behaves like a core galaxy that 
(e.{\thinspace}g.)~has had its core filled by a late accretion.  

      Signs of this behavior have been evident from the beginning.  The S\'ersic 
indices in Caon \etal (1993) are, on the whole, very accurate (see Figure 74 in Appendix
A3), and they already show two clumps of points in $n$ -- $r_e$ plots.  Also, Caon 
\etal (1993) note that ``boxy galaxies have larger $n$ than disky~galaxies''.  D'Onofrio 
\etal (1994) presciently comment that ``it is hard to understand whether there is a global 
trend of [$n$] with [$\log {r_e}$] or whether instead there are two distant clusters of 
points \dots~corresponding to the two galaxy families, and not presenting any correlation
between [$n$] and [$\log {r_e}$] within itself, but the relative positions of which mimic 
the global trend.''  Their galaxy families are closely related to our \hbox{E{\ts}--{\ts}E}
dichotomy.  In the same vein, Graham \etal (1996) see no correlation of $n$ with luminosity
for brightest cluster galaxies (their Fig.~8), although they see an $n$ -- $r_e$ correlation 
(their Fig.~11) that may be the product of parameter coupling (their Fig.~3).  In truth, the 
main reason why people have come to believe in an \hbox{$n$\ts--\ts$M_{V}$} correlation 
appears to be that they included Sphs -- which have nothing to do with ellipticals -- and
that the $n$ -- $M_V$ dichotomy was sometimes blurred by measurement errors.

      What do we learn from our S\'ersic index results?

      A hint can be seen in the earliest simulations constructed~to investigate the 
kinds of mergers that make realistic ellipticals.  van~Albada (1982) is remembered 
(Binney \& Tremaine 1987) for having shown that larger amounts of dynamical violence -- that is, 
larger collapse factors and lumpier initial conditions -- produce ellipticals with more nearly 
$r^{1/4}$-law profiles.  A closer look at his figures shows that van Albada's merger remnants
are more consistent with S\'ersic functions than with $r^{1/4}$ laws.  They depart from 
$r^{1/4}$ laws such that $n < 4$ for gentle collapses or mergers, whereas $n > 4$ for violent
collapses or mergers.  This is not surprising, because large collapse factors give some stars
total energies that are nearly zero.  That is, they fling stars into extended halos with $n > 4$. 
The hint is that giant, core ellipticals, which have $n > 4$, formed with more dynamical 
violence than small, coreless ellipticals, which have $n < 4$.  Tiny ellipticals have S\'ersic 
indices $n \sim 2$ that are not much higher than $n \simeq 1$ in exponential disks.  Little 
splashing of stars to large radii is required to make these profiles, although large amounts
of dissipation are needed to turn low-density disks into high-density ellipticals 
(Carlberg 1986; Kormendy 1989; Nipoti, Londrillo, \& Ciotti 2003; Hopkins \etal 2008a, b, c, e).

\centerline{\null} % One blank line

\centerline{\null} % One blank line

\centerline{\null} % One blank line

\centerline{\null} % One blank line

      A comparison of our results with simulations of galaxy mergers and their remnants
shows good agreement with the above picture.  The simulated merger remnants in Figure 4 have
S\'ersic function profiles with $n < 4$.  Examination of Figure~3 in Springel \& Hernquist 
(2005) shows that the old stars in the remnant (''bulge $\approx$ E'' in Figure 43  here)
have a S\'ersic profile with $n < 4$.  This is not obvious in Figure 43 because the radius
scale is logarithmic.  Extensive simulations of binary mergers by Naab \& Trujillo (2006) also
tend to produce $n \sim 3$ to 4.   The remnants in Figure 44 have $n \simeq 3$.  Hopkins \etal 
(2008b) emphasize that ``the outer shape of the light profile in simulated and observed
systems (when fit to properly account for the central light) does not depend on mass, with a 
mean outer S\'ersic index $\sim$ 2.5.''  We emphasize the same point; excluding NGC 4621, our
extra light Es have an unweighted mean S\'ersic index of $2.51 \pm 0.17$ and little 
dependence on $M_{VT}$.   So there is excellent consistency between observations of extra 
light galaxies and simulations in which these galaxies were made in a single merger of 
plausible, gas-rich progenitor galaxies.  {\it We conclude that the structure of extra light 
galaxies was created by only a few major mergers.}  

      In contrast, core Es have much larger $n$ values that likely are produced by many 
successive mergers, lots of merger violence, and -- plausibly -- later heating and minor
galaxy accretion.  Simulations of binary dry mergers show only a little redistributioon of
energy outward, i.{\ts}e., a small increase in $n$ (Hopkins \etal 2008c).  However,
repeated minor mergers cause $n$ to evolve toward larger values (Bournaud, Jog, \& Combes 
2007).  The dynamical violence inherent in hierarchical clustering naturally heats the
outer halos of giant galaxies; an extreme version of this process is the blending of the
outer parts of certain giant ellipticals into their cD halos of cluster debris (Gallagher
\& Ostriker 1972; Richstone 1976; Dressler 1979; Kelson \etal 2002).  Nevertheless, 
further study of exactly what combination of physical processes gives core ellipticals
their large S\'ersic indices would be worth while.

\vfill

\section{Complications}

      This section highlights complications in our results.  They do not 
threaten our conclusions, but they deserve further work.

\centerline{\null}  %  One blank line

\centerline{\null}  %  One blank line

\centerline{\null}  %  One blank line

\centerline{\null}  %  One blank line

\eject

\subsection{Today's Extra Light Ellipticals Are Not The\\~~~~~~~~Merger Progenitors Of 
            Most Core Ellipticals}

      Some small core ellipticals may be dry merger remnants of today's
extra light ellipticals.  But these cannot be the merger progenitors of
most core Es.  Figures 45 and 46 show why.

\vskip 0pt

\figurenum{45}

\centerline{\psfig{file=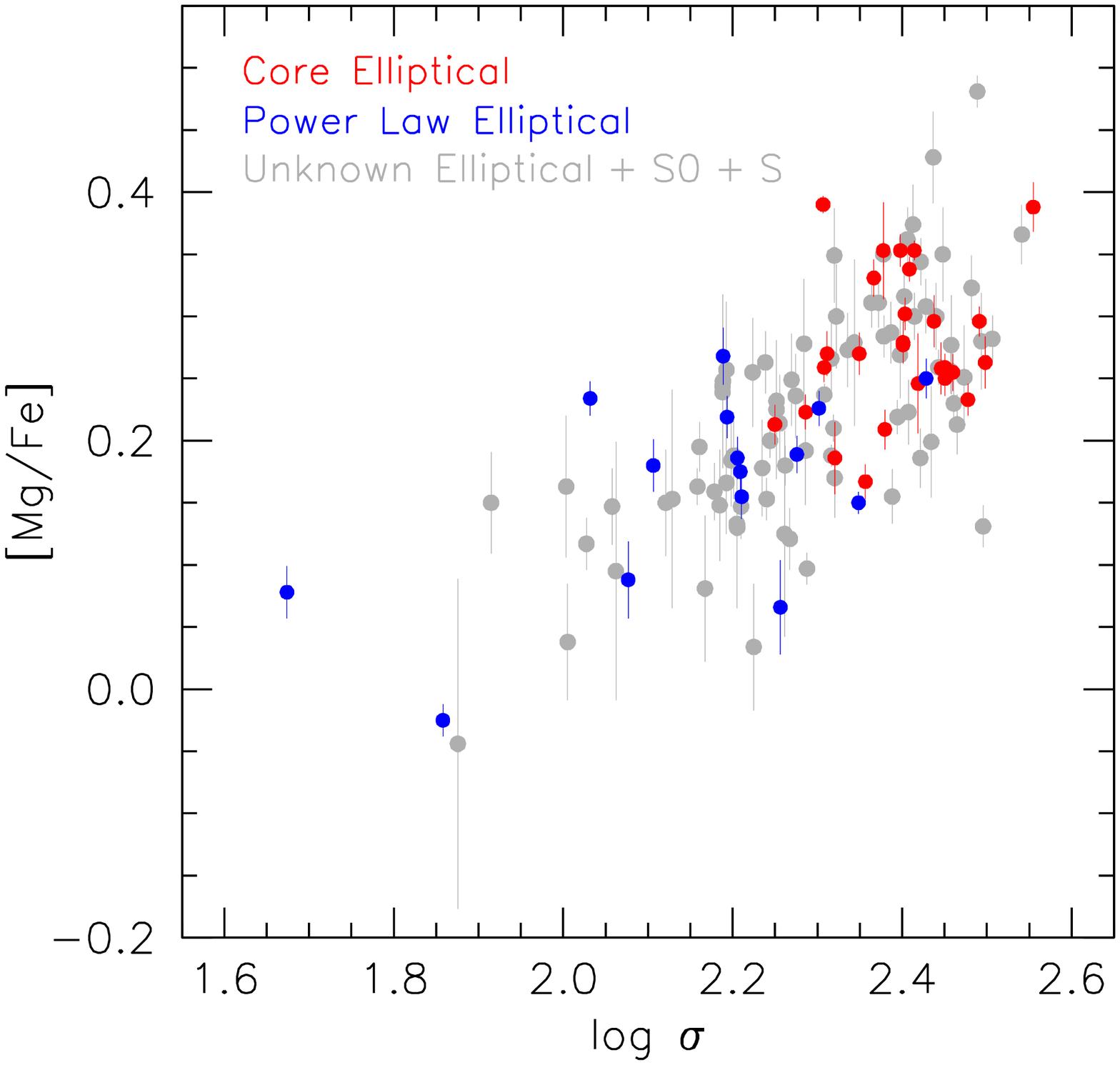,width=8.6cm,angle=0}}

\vskip -12pt

\figcaption[]{Alpha element overabundance in $\log$ solar units versus 
velocity dispersion in km s$^{-1}$ (data from Thomas et al.~2005).  Red and blue points 
denote core and power law ellipticals classified here or by Lauer et al.~(2007b). 
}

\vskip 1pt

\figurenum{46}

\centerline{\psfig{file=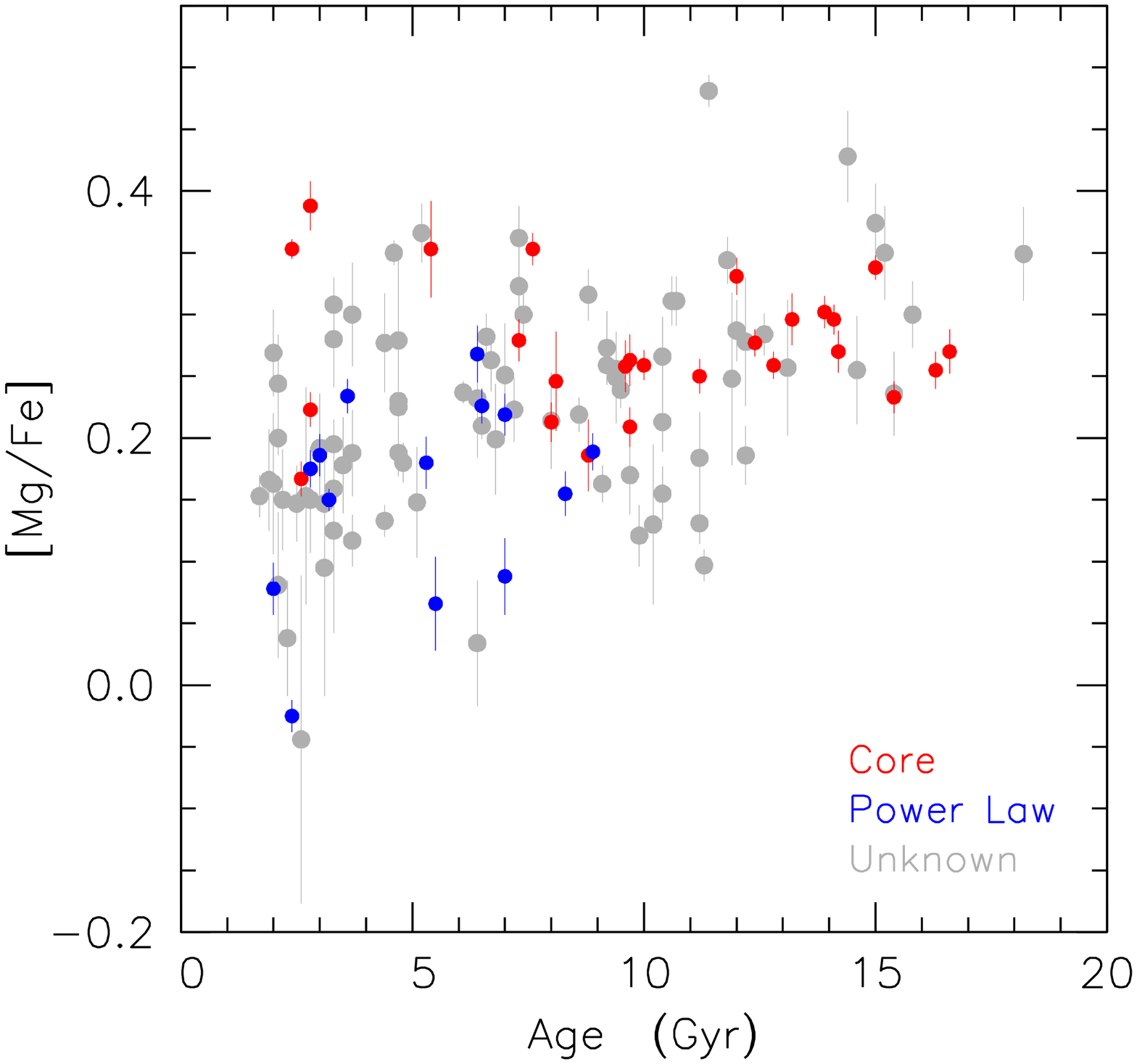,width=8.6cm,angle=0}}

\vskip -10pt

\figcaption[]{Alpha element overabundance versus the relative age of the stellar
population for the sample in Figure 45 (data from Thomas et al.~2005).}

\vskip 10pt

      Figure 45 shows the well known correlation between alpha element overabundance
and galaxy velocity dispersion.  Galaxies that have cores are shown in red, while galaxies 
that have coreless central profiles (``power law'' in Lauer et al.~2007b or ``extra light'' 
here) are shown in blue.  We know that cores predominate in giant Es whereas extra 
light is the rule in low-luminosity Es.  We also know that luminosity correlates with
velocity dispersion (Faber \& Jackson 1976).  So it is not surprising that core and power law
galaxies occupy different, slightly overlapping parts of the [Mg/Fe] -- $\sigma$
correlation.  However, this correlation and the Faber-Jackson relation have substantial
scatter, so the above result is not guaranteed.  In fact, Figure 45 demonstrates that 
[Mg/Fe] enhancement participates in the E -- E dichotomy.  This is an important new 
result.

      It has implications for the merger formation of ellipticals.  Alpha 
element overabundances tell us the timescales on which the stars formed.  Alpha
elements like Mg are produced soon after starbursts when massive stars die 
as supernovae of Type II.  They get diluted by Fe produced by Type I supernovae
starting \lapprox \ts1 Gyr later.  After that, [$\alpha$/Fe] can never be very enhanced again.  
So, large [$\alpha$/Fe] favors short star formation timescales (Worthey, Faber, \& Gonzalez 
1992; Terndrup 1993; Matteucci 1994; Bender \& Paquet 1995; Thomas et al.~1999, 2002, 2005).

      Therefore Figure 45 implies that the stars in core Es formed over shorter times
than did the stars in power law Es.  Neither the observed [Mg/Fe] values nor the inferred 
star formation timescales can be altered by dry mergers.  If the formation of core Es 
included any star formation, this is likely to decrease [Mg/Fe] further.  So Figure 45 is
consistent with the hypothesis that some small core Es are dry merger products 
of the biggest power law ellipticals.  But today's power law Es cannot be the progenitors of 
most -- and especially not the biggest -- core Es.

      Similarly, $n$-body simulations of dry binary mergers robustly predict that $\sigma$
in the remnant is similar to $\sigma$ in the progenitors (see Hopkins et al.~2008c, who 
also review previous results).  Core galaxies generally have larger $\sigma$ than power
law galaxies.  Either their progenitors were not like present-day power law galaxies or the 
mergers were not like those that were modeled.

      Finally, Fig.~46 shows [Mg/Fe] versus relative age.  Core and power law ellipticals
overlap only slightly.  Stellar population ages are part of the E -- E dichotomy (Nipoti 
\& Binney 2007).  

      Again, the progenitors of most core ellipticals must have been different from today's 
power law  ellipticals.~The latter are mostly younger than the former.  Dry mergers cannot age stars.

      These results threaten neither the merger picture nor our conclusion that core and
extra light Es were made, respectively, in dry and wet mergers.  However, they do provide
clues about the details of the formation processes.  Physics that is missing from our
present picture but that almost certainly affected the formation of core ellipticals 
includes:

      (1) The merger progenitors that made core ellipticals may have been different 
from all galaxies seen today (e.{\ts}g., Naab \& Ostriker 2008; Buitrago \etal 2008; 
van der Wel \etal 2008).  They could have included an earlier generation of power law ellipticals, 
provided that essentially all of them were used up.

      (2) Quasar-mode AGN feedback (e.{\ts}g., Cattaneo et al.~2008b) is ignored but is
believed to have  whittled the high-mass end of the galaxy mass function down from the shallow 
slope predicted from the cold dark matter fluctuation spectrum to the much steeper form observed
(Binney 2004).  If it could do this, it is easy to believe that it could affect the internal 
structure of galaxies.

      (3) We consider only mergers of two galaxies with each other.  In the early Universe,
 many galaxies may have merged simultaneously.  This affects the structure of the remnant and 
can change the prediction that $\sigma$ is unchanged by a dry merger.

      These comments should not be interpreted as criticisms of published formation models.
Galaxy formation is complicated and not fully constrained by observations.  Modeling it is 
a step-by-step process.  Impressive progress has been made by including gas dissipation,
star formation, and energy feedback, most recently by Hopkins et al.~(2008a, b, c).  We hope
that the observational constraints discussed here will provide input for the next generation 
of formation models.

\subsection{Do Minor Mergers Build Extra Light Components?}

      We suggest that extra light was made in starbursts triggered by major mergers.  The connection 
between extra light in S\'ersic-function ellipticals and simulations of dissipative 
mergers is one of the main results of this paper.  However, an alternative possibility is that extra 
light was built out of gas that trickled in during minor mergers. These must happen (e.{\ts}g., NGC 4473; 
\S\ts9.2).  In some ellipticals, dust has settled into well defined, major-axis disks at small radii, 
where dynamical clocks run quickly, but remains irregular at large radii, where clocks run slowly 
and galaxies remember accretion geometries for a long time.~An example is NGC 315 (Kormendy \& 
Stauffer 1987; Verdoes Kleijn et al.~1999; Capetti et al.~2000).    

      However, there are signs that minor accretions did not build the extra light in most
ellipticals.  Often it is as old as the rest of the galaxy (Kuntschner \etal 2006; 
\S\ts9.4 here).  Also: 
{\it The extra light participates in a dichotomy of physical properties that mostly involves 
global structure.}  Global rotation, isophote shape, and flattening (E3 for coreless Es 
but E1.5 for~core~Es; Tremblay \& Merritt 1996) are not likely to be affected by minor accretions. 
We expect that minor accretions occasionally affect central structure.  But the above arguments 
suggest that they are not the main source of the extra light.

\subsection{Uncertainties in Profile Results}

      S\'ersic indices are affected by a number of factors that are not taken into account 
in the fitting errors listed in Table 1.

      First, Figures 11 -- 32 illustrate major-axis profiles, and the S\'ersic indices 
in Table 1 also apply to major-axis profiles.  We made this choice because we wanted 
as much radial leverage as possible in distinguishing central and global 
properties and in recognizing and decomposing bulges and disks.  Since ellipticity 
profiles are not flat, mean- and minor-axis profiles have slightly different S\'ersic 
indices than those along the major axis. However, they agree on the essential question
of whether $n < 4$ or $n > 4$ (Fig.~63).  Since S\'ersic index measures how much the 
outer profile is extended compared to the inner profile, and since an extended outer halo 
is a natural consequence of dynamical heating (splashing) during violent relaxation, 
it is reasonable to expect that the major-axis profile is the one that is most 
sensitive to the physics that we wish to explore.  

      Second, we measure $V$-band surface brightness profiles and use them as proxies for 
projected stellar densities.  That is, we assume that mass-to-light ratios are 
constant with radius.  The color gradients illustrated in Figures 11 -- 32 show that 
this is not quite true.  Converting $g - z$ colors shows that $V - K$ typically varies 
by a few tenths of a mag arcsec$^{-2}$ over the S\'ersic part of the profile.  Near-infrared,
$K$-band profiles are insensitive population differences.  Applying $V - K$ colors to the
observed profiles would change $n$ by small amounts but would not change the dichotomy 
that we find between coreless ellipticals with $n < 4$ and core ellipticals with $n > 4$.

      A more serious issue is dark matter.  Its importance must depend on radius.
It is remarkable that there is so much regularity in the light profiles when we do not 
take dark matter into account.  The correlations that we observe are clearcut.
But it will be important to investigate how the stellar structure of galaxies is affected
by halo structure and dynamics.

      Finally, we need to keep in mind that our results are derived almost entirely
from galaxies in the Virgo cluster.  Work on a larger sample is in progress to check 
whether ellipticals in other environments are similar to those in Virgo.

\section{Elliptical Galaxy Formation}

\subsection{Summary: New Features of the E -- E Dichotomy}

      We have measured and assembled composite surface photometry from as many
sources as possible for all 24 known elliptical galaxies in the Virgo cluster
plus three background ellipticals.  Because their classifications were unclear 
at the start of the program, we also included 5 galaxies that proved to be S0s 
and 10 galaxies that proved to be spheroidals.  Composite photometry over large 
dynamic ranges provides improved control of systematic problems such as sky 
subtraction errors.  We can derive more accurate profile parameters and use them 
to investigate galaxy formation.  Our conclusions are as follow:

      S\'ersic functions fit the brightness profiles of the main bodies of 25 of 
our 27 ellipticals to within $\simeq 0.04$ mag arcsec$^{-2}$ over a mean surface 
brightness range of $8.7 \pm 0.4$ mag arcsec$^{-2}$.  In 5 of the largest-dynamic-range 
galaxies, the fit range is 10.3 -- 11.5 mag arcsec$^{-2}$, i.{\ts}e., factors of
13,000 to 40,000 in surface brightness.  As a result, we can reliably identify 
departures from S\'ersic functions that are diagnostic of formation processes.  

      The distinction between cuspy core ellipticals and galaxies without 
cores is well known and clearly evident in our data.  We base the distinction 
on inner departures from outer S\'ersic profiles rather than on the slope of the 
projected brightness profile at small radii as in Nuker papers (Lauer \etal 1995,
2005, 2007b), but both kinds of analysis machinery usually identify the same 
galaxies as having cores.

      Our results reveal new aspects of the dichotomy (\S\ts2.2) into two kinds 
of elliptical galaxies: 
(1) Giant-boxy-core ellipticals have stellar populations that mostly are old and
enhanced in $\alpha$ elements.   Their main bodies have S\'ersic indices $n > 4$, 
uncorrelated with $M_{VT}$.  The light that is ``missing'' in cores with respect 
to the inward extrapolation of the outer S\'ersic profile corresponds to a stellar
mass -- in our sample -- about 11 times as big as the masses of the central BHs.
(2) Lower-luminosity, disky-coreless ellipticals generally are made of younger stars
than are core ellipticals.  Their stellar populations also are less enhanced or even
Solar in $\alpha$ element abundances.  Their main bodies have S\'ersic indices
$n \leq 4$ almost uncorrelated with $M_{VT}$.  And they do not have featureless, 
nearly power-law central profiles; rather, they show distinct profile breaks and,
interior to them, extra light with respect to the inward extrapolation of their 
outer S\'ersic profiles.  Previously called ``power law'' ellipticals, we  refer 
to them as ``extra light ellipticals''.  The amount of extra light is a larger and 
more varied fraction of the total light of the galaxy than is the missing light 
that defines cores.  A small number of exceptions to all aspects of the dichotomy 
are observed.  The dividing line between the above types is at absolute magnitude 
$M_{VT} \simeq -21.6$ and is not sharp.

\subsection{How The E -- E Dichotomy Arose}

      We suggest that core and extra light ellipticals formed in dissipationless (``dry'') 
and dissipational (``wet'') mergers, respectively.  

      This idea is not new.  The need for dissipation to make the high phase-space
and mass densities of low-luminosity Es has been recognized for a 
long time (Ostriker 1980; Carlberg 1986; Gunn 1987; Kormendy 1989; Kormendy \& Sanders 1992);
it has been connected with the merger picture from the beginning (Toomre \& Toomre
1972).  So, for example, Faber \etal (1997) concluded that ``Disky [power law] galaxies, 
including their high central densities, suggest final mergers that were gas rich.''  

      Our observations further strengthen this picture.  Numerical simulations of 
dissipative mergers that include star formation and energy feedback predict extra, dense 
central components just like the ones that we observe.~We interpret the extra light as a 
``smoking gun'' that points to dissipational formation.  It frequently has disky structure 
and kinematic decoupling that are natural consequences of dissipative mergers.  Extra light 
profiles like those that we see in old ellipticals have also been observed in 
mergers-in-progress (Rothberg \& Joseph 2004, 2006).   Some simulations
suggest further that larger S\'ersic indices $n$ are produced by more violent mergers.
{\it Thus numerical simulations and our observations both lead to a picture in 
which the last merger that made coreless galaxies was relatively gentle and wet,
while the last merger that made core galaxies was relatively violent and dry.}     

      Because:~in the absence of supermassive BH, mergers of coreless
galaxies tend to make coreless galaxies.  Therefore, Faber \etal (1997) pointed
out that ``arguments concerning [the formation of] boxy [core] galaxies 
are less clear: the global kinematics of these galaxies suggest final mergers that 
were gas poor, but forming and preserving cores in such models may be difficult.''  
To solve this problem, the key realization has been that cores may be excavated by 
binary BHs.  This idea, once radical and ad hoc, has become 
mainstream as we have found a BH in every well-observed elliptical.  If we 
believe that ellipticals form by major mergers, then these must generally make 
BH binaries.  Black hole scouring, far from being ad hoc, becomes inevitable.  
While the BHs are well separated; they sink individually by dynamical 
friction against the background stars.  The light distribution of the galaxy is not 
affected, because the BHs have a small fraction of the mass of the galaxy.  
But as soon as the BH separation $2R$ is small enough so that the total stellar mass
at $r$ \lapprox \ts$R$ is comparable to the mass of the BHs, 
they must affect the stellar density profile.  After several dry mergers, the stars 
that they have flung to larger radii add up to several times the combinedBH  masses.
The excavated cores can even be hollow, and a few hollow 
cores have been observed (Lauer \etal 2002).  Faber \etal (1997) showed that observed 
core properties are reasonably consistent with core scouring.

      But an important problem remains unsolved. The puzzle is no longer, ``How can 
cores form?''~but rather, ``How can core excavation by binary black holes be prevented?''  
Faber \etal (1997) ask the same question and propose the same answer that we do: 
``\dots~if cores are formed by merging binary BHs, why do power law galaxies \dots~not
have cores?  BHs appear to be just as common in power law galaxies (Kormendy \& 
Richstone 1995).  Perhaps power laws can be regenerated by star formation from 
fresh gas supplied by the latest merger.  However, to avoid being ejected by the 
BH binary, the new stars must form {\it after} the BH binary shrinks, which poses 
a timing problem if BHs sink to the center more slowly than gas.''

      Our observations suggest the same solution.  The extra stellar masses in coreless
ellipticals tend to be larger than BH masses.  BH binaries cannot fling most of it away. 
We suggest that central starbursts associated with dissipative mergers have swamped BH 
scouring and filled in any cores.  This reduces the timing problem discussed by Faber 
\etal (1997).  It may not prevent the occasional late formation of a new core if the 
BH binary survives the starburst.  In fact, several extra light ellipticals show signs 
of tiny cores {\it in the extra light\/}.~NGC\ts4458 is the best example (Figure 19).  
The interplay between star formation and BH scouring is likely to be complicated.  
Any over-simplistic interpretation is likely to suffer exceptions.

       Meanwhile, $n$-body simulations that seek to reproduce orbit structure, rotation,
and isophote shapes are most successful when disky Es are made in wet mergers and boxy Es 
are made in dry mergers (Naab \etal 1999; Naab \& Burkert 2003; Naab, Khochfar, \& Burkert 2006; 
Naab, Jesseit, \& Burkert 2006; Burkert, Naab, \& Johansson 2007).  Making extreme, non-rotating
Es is still a challenge (Naab \etal 2007a); the solution may be a succession of mergers 
of several galaxies at once.  So:

      {\it How\/} the differences between the two kinds of ellipticals
arose appears well established by observations and simulations.  {\it Why\/} they arose
is the subject of the next section.

\subsection{Why The E -- E Dichotomy Arose}

\subsubsection{X-Ray-Emitting Gas and AGN Energy 
               Feedback\\~~~~~~~~~~~~Create the E -- E Dichotomy}

      The key observations prove to be two aspects of the \hbox{E{\thinspace}--{\thinspace}E} 
dichotomy that are shown in Fig.~47.~Bender et al.~(1987, 1989) discovered (1) that boxy 
ellipticals tend to be radio-loud while disky ellipticals do not, and (2) that boxy 
ellipticals mostly contain \hbox{X-ray-emitting} gas while disky ellipticals do not.  
These correlations were not understood; most subsequent discussions did not mention 
them but rather concentrated on the structural and dynamical differences between the two 
kinds of ellipticals.  Now the X-ray and radio correlations take center stage.

      We suggest that X-ray-emitting gas that is kept hot by AGN feedback
is the reason why giant-boxy-core ellipticals formed dissipationlessly.~In contrast, disky-extra 
light ellipticals and their merger progenitors are too low in mass to hold onto hot gas.  Also, we
suggest that AGN feedback is weaker in these galaxies; they experienced either weak 
feedback (\S\ts10.2) or positive feedback (Silk 2005).  As a result, dissipative starbursts were 
possible. Figure 47 provides the connection between \hbox{X-ray} gas, AGN physics, and the E -- E dichotomy.

\figurenum{47}

\vskip 4.2truein

\includegraphics{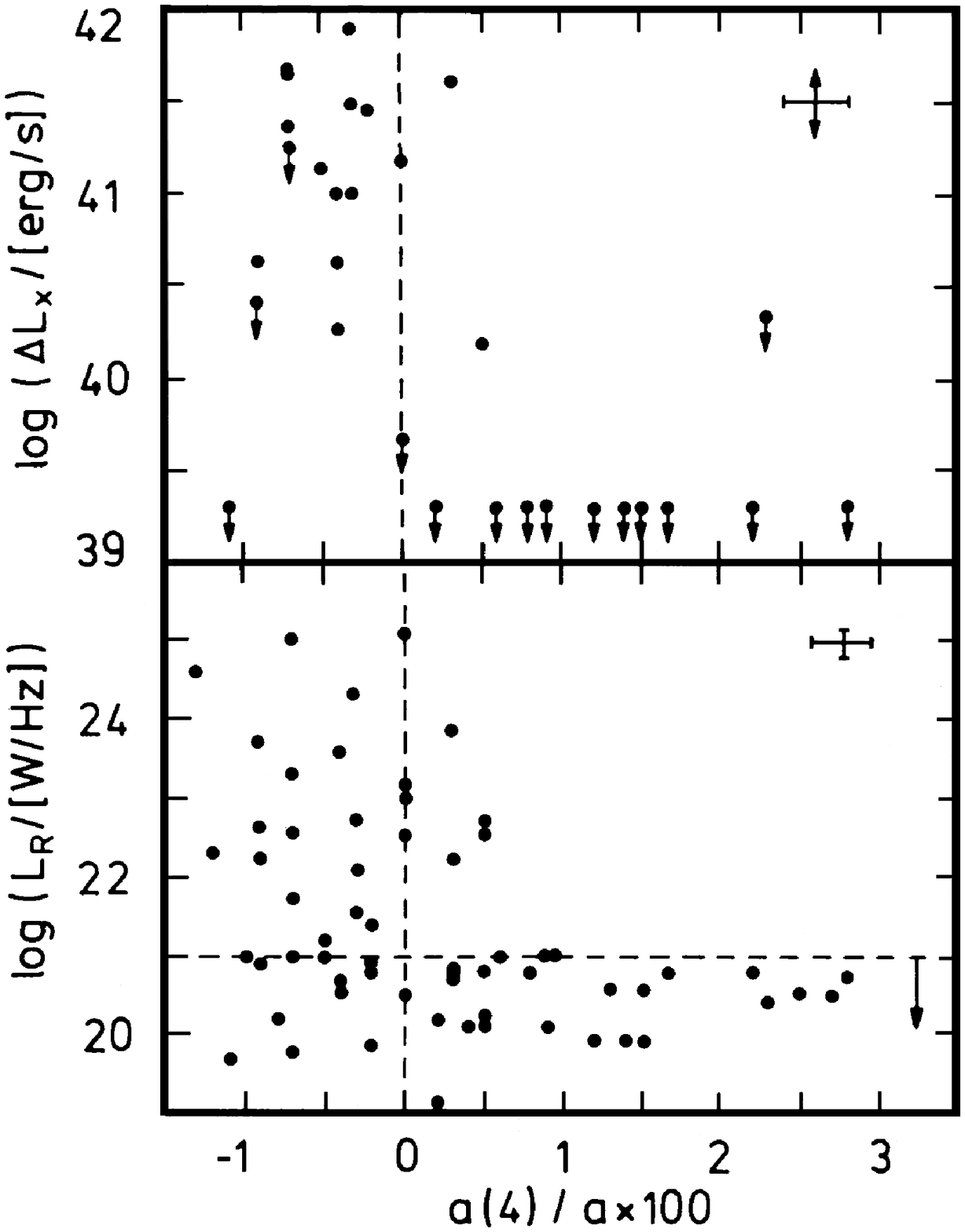} 

\figcaption[]
{Correlation of X-ray emission from hot gas ({\it top\/}) and radio emission
({\it bottom\/}) with isophote shape parameter $a_4$ of elliptical galaxies
(from Bender et al.~1989).  Boxy isophotes have $a_4 < 0$; disky isophotes
have $a_4 > 0$.  }

\vskip 15pt

      BH binding energies are enormous; if only a small fraction of the energy
released in making them is fed back into gaseous protogalaxies, the effect
on galaxy formation is profound (Ostriker \& Ciotti 2005).  Silk \& Rees (1998)
make a compelling case that AGN feedback has a major effect on the formation of 
giant galaxies.  Their arguments, the results of galaxy formation models
(reviewed by Cattaneo \etal 2008b), and \S\ts10.2 here suggest that AGN 
feedback is a strong function of galaxy luminosity.  But the introduction of 
feedback into formation models is {\it ad hoc\/} -- it is tuned to solve 
specific problems, but we do not understand the underlying physics.  And AGNs 
are episodic, with long ``down times'' between short periods of activity.  How can
we be sure that an AGN is switched on every time we need one (e.{\thinspace}g.)~to 
quench star formation when gas-rich galaxies are accreted by old, 
$\alpha$-element-enhanced ellipticals?  Therefore:

      A welcome watershed in the credibility of AGN feedback was a workshop on 
``The Role of Black Holes in Galaxy Formation and Evolution'' (Potsdam, Germany; 
Sept.~2006).  McNamara \& Nulsen (2007) and Cattaneo~et~al.~(2008b) 
provide reviews.  The above problems are plausibly solved if AGN energy is fed 
into X-ray-emitting gas in giant galaxies and galaxy clusters.  As emphasized by Best 
(2006; cf.~Kauffmann, Heckman, \& Best 2008), feedback requires a working surface.
Hot gas provides that surface.  We suggest that it stores AGN energy and 
smooths out the episodic nature of the energy input.  It quenches star formation in 
accreted, gas-rich galaxies before that star formation threatens the observation that 
stars in giant Es are old (Binney 2004; Dekel \& Birnboim 2006; Nipoti \& Binney 2007).
Can radio AGNs keep hot gas hot?  We~are~not~sure.  But 
{\it Chandra\/} and {\it XMM-Newton\/} observations make a strong case that central 
radio sources heat the X-ray gas in clusters of galaxies.  Examples include the
Perseus cluster (B\"ohringer \etal 1993; Fabian \etal 2000, 2003, 2006, 2008; Sanders 
                 \& Fabian 2007); Hydra A (McNamara \etal 2000);
Abell 2052 (Blanton, Sarazin, \& McNamara 2003);
M{\ts}87 (Forman \etal 2005); and
MS0735.6$+$7421 (McNamara \etal 2005).
Evidence for shock fronts, bubbles, and compression waves are
signs that energy outflow in jets is redistributed more isotropically into the hot
gas.  The evidence that jets heat gas within galaxies as well within
clusters is less direct.~Best et~al. (2006) conclude that ``the radio sources
which give rise to the bulk of radio source heating are low-luminosity sources
which tend to be compact and more confined to the host galaxy.''  Diehl \& Statler (2008)
also find evidence for AGN feedback within normal Es.  These observations 
make AGN heating of hot gas more believable.  {\it We assume that, for AGN feedback to work,
a galaxy needs both an X-ray gas halo and sporadic AGN activity.}  

\lineskip=-2pt \lineskiplimit=-2pt

      Figure 47 shows that both features are common in boxy and rare in disky galaxies.
This is confirmed by Balmaverde \& Capetti (2006), Capetti \& Balmaverde (2006), and
Ellis \& O'Sullivan (2006).  Almost equivalently, both features are 
common in big and rare in small galaxies (O'Sullivan \etal 2001; Ellis \& 
O'Sullivan 2006; Best \etal 2005; Pasquali \etal 2008).  First consider radio AGN heating.  
Best \etal (2005) show that the fraction $f_{\rm radio-loud}$ of galaxies 
that are radio-loud increases dramatically with increasing stellar mass $M_*$,
$f_{\rm radio-loud} \propto M_*^{2.5}$.  In particular, $f_{\rm radio-loud}$ $>$
1\thinspace\% at $M_* > 10^{11}$ $M_\odot$; this is roughly the transition mass
between the two kinds of ellipticals.  At the highest $M_*$, which are generally 
the oldest (Fig.~46), most $\alpha$-element-enhanced (Fig.~45) and most boxy (Fig.~47)
galaxies, $>$\ts30\ts\% of ellipticals are radio-loud.   Not surprisingly, Best and 
collaborators conclude that radio-mode heating is a strong function of galaxy stellar
mass.  Taking jet properties and AGN duty cycles into account, they estimate that 
radio-mode heating scales with central black hole mass as $M_\bullet^{2.2}$.  Therefore 
it is similarly a strong function of $M_*$ and $M_{VT}$ (Faber \& Jackson 1076; 
Tremaine \etal 2002).

      We can also update the connection between hot gas X-ray luminosity and the
E{\ts}--{\ts}E dichotomy.  Pellegrini (1999, 2005) confirms that X-ray luminosity 
participates in the dichotomy.  Like Bender \etal (1989), she sees a correlation 
with $a_4$.  She also finds the corresponding correlations with central profile slope and 
the degree of rotational support.  In addition:

      Figure 48 shows how the total X-ray emission of elliptical galaxies depends on 
stellar luminosity.  It updates Figure 9 in Ellis \& O'Sullivan (2006), which shows
the {\it ROSAT\/} sample of O'Sullivan, Forbes, \& Ponman (2001) coded according to 
whether the galaxies have core or power law profiles.  More profile classifications are
now available.  Also, we can use boxy versus disky structure to distinguish the two
types of ellipticals.  (Occasionally this conflicts with profile classification; then
we use the latter.)  The {\it black line\/} shows the O'Sullivan \etal (2001) estimate of 
the contribution from discrete sources such as X-ray binaries.  The discrete source 
contribution to $L_X$ is proportional to $L_B$ (Fabbiano 2006).
Consistent with Bender \etal (1989, Fig.~47 here), Figure 48 shows that few coreless-disky 
galaxies are detected in X-rays and those that are detected mostly are consistent with 
the discrete source estimate.  In contrast, almost all core-boxy galaxies are detected in
X-rays and show a steep dependence of $L_X$ on $L_B$.  So Figure 48 further confirms
that X-ray luminosity participates in the E{\ts}--{\ts}E dichotomy.  

\vbox{\centerline{\null}
\vskip 3.46truein

\includegraphics{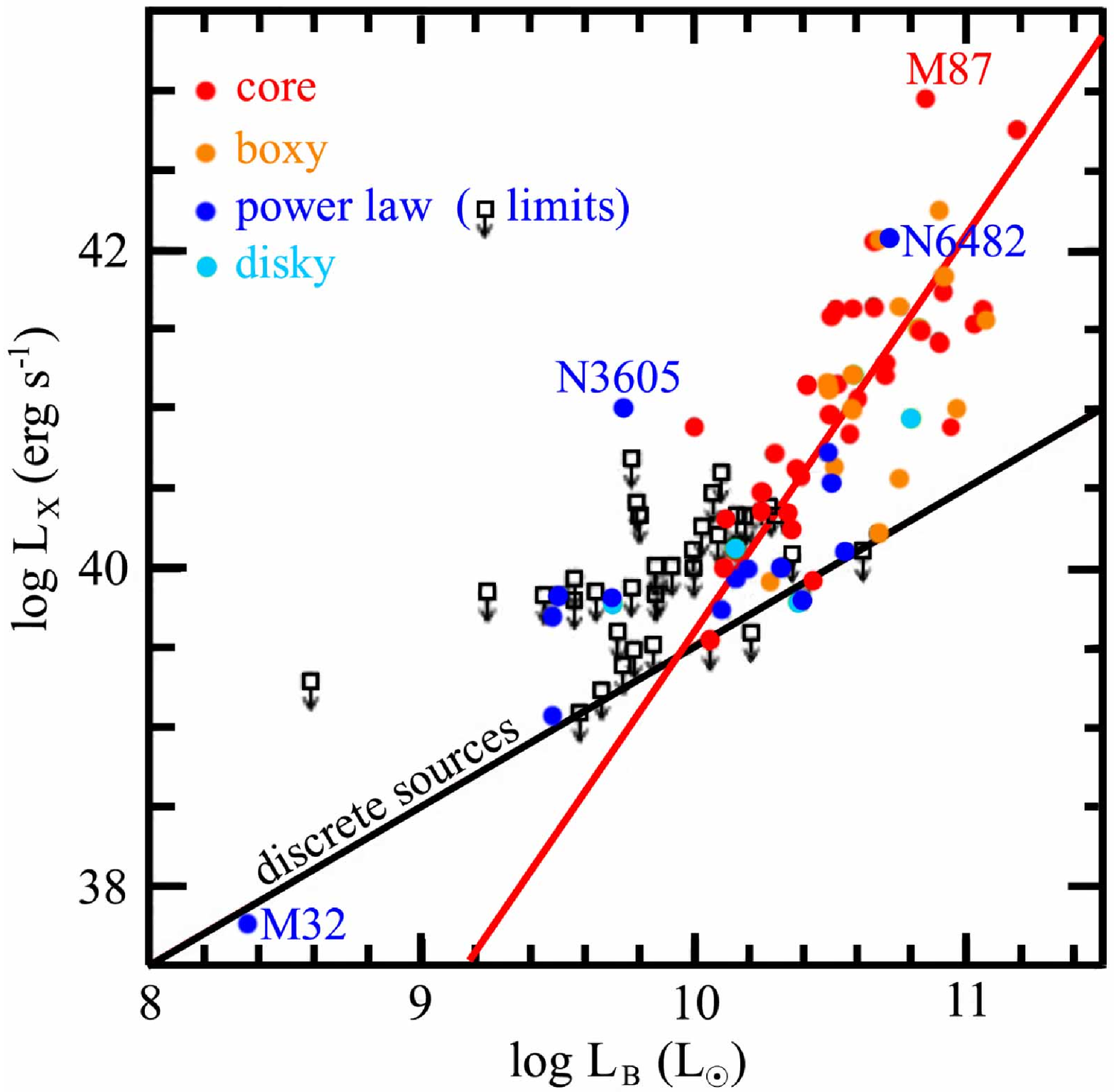}

\figurenum{48}
\figcaption[]
{Total observed X-ray emission versus galaxy $B$-band luminosity (adapted from Fig.~9 
of Ellis \& O'Sullivan 2006).  Detections are color-coded according to the E -- E dichotomy 
(see the key).  New classifications of core and power law profiles are from Lauer \etal 
(2007b) and from our photometry.  Classifications of boxy and disky structure are from Bender 
\etal (1989).  The contribution from discrete sources is estimated by the {\it black line\/} 
(O'Sullivan \etal 2001).  The {\it red line\/} is a bisector fit to the core-boxy points, 
i.{\ts}e., the bisector of regressions of log $L_X$ on log $L_B$ and of log $L_B$ on~log~$L_X$.
Core-boxy ellipticals statistically reach $L_X = 0$ from hot gas at $L_B \simeq 9.94$.  
This corresponds to $M_V \simeq -20.4$, which is about 1 magnitude fainter than the stellar 
luminosity that divides the two kinds of ellipticals.  Core and power law Es clearly overlap
in luminosity, but most core galaxies do and most power law galaxies do not contain 
significant X-ray-emitting gas.  The strongest exceptions, NGC 3605 and NGC 6482, are
discussed in Footnote 13.}
}
\vskip 15pt

      The {\it red line\/} crosses the {\it black line\/} at log $L_B \simeq 9.94$.
That is, the X-ray luminosity from hot gas goes to zero at $M_V \simeq -20.4$.  This is 
about 1 magnitude fainter than the stellar luminosity that divides the two kinds of ellipticals.
Core and power law Es are known to overlap in luminosity (Lauer \etal 1995, 2005, 2007b;
Faber \etal 1997), and this is evident in Figure 48.  But Figure 48 suggests that most core 
galaxies do and most power law galaxies do not contain significant X-ray-emitting gas.  
 
      A few power law galaxies may contain small amounts of X-ray gas, including NGC 4387, 
NGC 4473, NGC 4458, and NGC 4621 from our 
sample\footnote{Nine of 17 extra light galaxies in Table 1 are represented in Figure 48; 
the four detections are discussed in the text; the rest (mostly fainter galaxies) are limits.
Three of our five S0s are represented in Figure 48; all are limits.  All of our core galaxies 
except NGC 4382 are represented in Figure 48; all are detections.  So our conclusions about 
the relevance of hot gas to the E{\ts}--{\ts}E dichotomy are based very significantly 
on X-ray observations of the present Virgo cluster sample.}. 
However, O'Sullivan~et~al. (2001) estimate that the contribution from discrete sources varies 
by a factor of $\sim 4$ from galaxy to galaxy. It is not certain that these galaxies contain
hot gas.  More importantly, NGC 4387 is located between the gas-rich, giant Es NGC 4374 
and NGC 4406.  NGC 4473 is in the chain of Virgo galaxies that has NGC 4374 and NGC 4406 at one 
end.  NGC 4458 forms a close pair with the brighter S0 NGC 4461.  All three galaxies 
benefit from the nearby presence of additional gravitational potential 
wells.\footnote{
\lineskip=0pt \lineskiplimit=0pt
This is also true of NGC 3605, which stands out in Figure 48 as having high $L_X$ at low $L_B$. 
But NGC 3605 lives inside the X-ray halo of the much brighter elliptical NGC 3607.  It is not 
clear that NGC 3605 perturbs the X-ray contours of NGC 3607 (Fabbiano \etal 1992).  At best, 
measuring a separate X-ray luminosity for NGC 3605 is tricky.  But also, NGC 3605 benefits from 
the deep potential well of the bigger galaxy.  So rather than being an exception to our conclusions,
it is a good example of the importance of high mass in retaining hot gas.  A possible real 
exception is NGC 4125, the highest-$L_X$ disky galaxy in Figure 48.  A not-yet-relaxed merger 
in progress (Schweizer \& Seitzer 1992), the observation of nuclear dust (Rest \etal 2001; 
Lauer \etal 2005; Draine \etal 2007) -- which prevents us from classifying the central profile -- 
suggests that the merger involved some cold gas.  The disky structure may be temporary, and 
the X-ray luminosity may be temporarily enhanced.  However, the galaxy may settle down to be a 
weak exception to our conclusions; that is, the remnant of a merger that was at least damp in 
a galaxy that ends up luminous enough to contain some X-ray gas.  Finally, NGC 6482 is not a 
problem in terms of $L_X$ ($L_B$ is certainly high enough), but it is a {\it bona fide\/} exception
to the usual luminosity at which the E-E dichotomy happens.  It is very disky (Bender \etal 1989).
From archival {\it HST\/} images, we find that it has a extra light and a normal small S\'ersic 
index of $2.4 \pm 0.2$.  It is an example of a ``fossil group'' (Khosroshahi, Jones, \& Ponman 2004).
We interpret it as the fossil of the merger(s) of several progenitors that were too low in mass 
to have hot gas and that therefore could merge dissipatively.   After the merger, the remnant 
is much more massive than normal remnants of wet mergers.  Given that ellipticals have a
great variety of merger histories, we expect a few exceptions to all aspects of the E -- E dichotomy, 
including the luminosity at which it happens.  That is, it seems inevitable that a few outliers like NGC 6482
will have formed in rare variations on the merger theme.
}
And NGC 4621 has $M_{VT} = -21.54$.  It is not surprising if these four galaxies 
contain a little hot gas.  It is also consistent with our formation picture: Any 
merger progenitors of these galaxies were less luminous and less able to hold hot gas; 
it is plausible that hot gas could be retained only after a merger made a deep enough 
potential well.  Also, from stellar population data, the wet mergers that made these 
galaxies took place long ago, when the Virgo cluster was less well formed than it is now.  
This highlights an unavoidable uncertainty in our picture: We interpret the formation physics 
in terms of X-ray gas that is observed now, but that formation took place long ago.  Since then,
hot gas content, heating mechanisms, and cooling rates may have evolved.  Connecting
present-day observations with a formation picture depends on our assumption that mass controls
X-ray gas content.  It is supported by the conclusion that AGN heating rates currently balance
cooling rates, so steady state is possible (Best \etal 2006, 2007a, b).

\lineskip=0pt \lineskiplimit=0pt

      What we find compelling is this: {\it The transition luminosity between galaxies that 
should contain X-ray gas and those that should not can be estimated from theory and tested for
consistency with observations using semi-analytic models.~The results agree with the observed X-ray
transition luminosity found above and with the observed E{\ts}--{\ts}E transition luminosity.}  
Birnboim \& Dekel (2003) and Dekel \& Birnboim (2006, 2008) present theoretical arguments and
Kere\v{s} \etal (2005) find in SPH simulations of gas accretion in hierarchical clustering
that, when gas falls into shallow potential wells, the dynamics are gentle, the inflow stays cold, 
and it makes star-forming disks.  In contrast, when gas accretes onto giant galaxies, a shock 
develops, the gas is heated to the virial temperature, and star formation is quenched.
Dekel and Birnboim (2006, 2008) propose that the gas is maintained at this hot temperature 
by the heating caused by additional accretion; AGN feedback is an alternative
heat source (Best \etal 2006; Best 2007a, b).  The transition between galaxies with and 
without X-ray gas is expected to occur at the dark matter halo mass at which the 
hot gas cooling time is comparable to the infall time.  Dekel~\& Birnboim estimate that 
this happens at $M_{\rm crit} \simeq 10^{12}$~$M_\odot$.  Kere\v{s} \etal 
(2005) get $10^{11.4}$ $M_\odot$.  Implementing $M_{\rm crit}$ 
quenching proves to allow semi-analytic models of galaxy formation to reproduce the color 
bimodality of galaxies (``red sequence'' versus ``blue cloud'') as a function of redshift 
(Cattaneo \etal 2006, 2008a, b).  Using a baryon-to-total mass ratio of 1/6 (Komatsu et al.~2008),
$M_{\rm crit} = 10^{12}$\ts$M_\odot$ implies a stellar mass of $M_* = 1.7 \times 10^{11}$\ts$M_\odot$.  
With a $M/L_B \simeq 8$ 
(\S\ts10.1), this corresponds to $M_B = -20.3$ or $M_V = -21.3$.  This is almost exactly
the absolute magnitude that divides our faintest core galaxy (NGC 4552, $M_{VT} = -21.66$) 
from our brightest extra light galaxy (NGC 4621, $M_{VT} = -21.54$).  The dividing luminosity 
in Figure 48 is formally a factor of 3 fainter, but $L_X$ is significantly higher than the 
discrete source estimate only at $M_B \lesssim -20.6$ (log $L_B \gtrsim 10.4$).  This is 
remarkable agreement.

\subsubsection{ULIRGs as Ellipticals in Formation:\\
               Do Supernovae Control Dwarf Galaxy Evolution Whereas 
                       AGNs Control Giant Galaxy Evolution?}

\lineskip=-2pt \lineskiplimit=-2pt

      Are low-luminosity ellipticals gas-free?  If so, why?  Gas shed by dying stars is 
just as large a fraction of small galaxies as it is of large ones, and galaxies fill quickly
with recycled gas (Ciotti et al.\ts1991; Ostriker \& Ciotti~2005).  We suggest that the 
answer to the first question above is a resounding ``yes and no''.  

      First the ``yes'' part: Published work and present results suggest that {\it the energy
feedback that controls galaxy evolution changes fundamentally from supernovae in small 
galaxies to AGNs in large ones}.  We have argued that AGN feedback gets more important at 
higher galaxy masses.   At the highest masses, the case for AGN feedback is compelling 
(Cattaneo \etal 2008b).  In dwarfs, it is difficult to doubt the importance of supernova-driven 
baryonic blowout as  one process that gives extreme dwarfs their low baryon densities and that 
converts irregulars into spheroidals (\S\ts2.1, \S\ts8; Dekel \& Silk 1986).  Very general arguments 
imply that supernova feedback gets less important at higher galaxy masses (e.{\thinspace}g., 
Dekel \& Silk 1986; 
Somerville \& Primack 1999, who review earlier work;
Benson \etal 2000, 2003;
Garnett 2002;
Dekel \& Woo 2003;
Ostriker \& Ciotti 2005;
Veilleux, Cecil, \& Bland-Hawthorn 2005).  

      Provided that star formation is rapid, Dekel \& Woo (2003) find that supernovae can 
unbind the remaining gas if the stellar mass is $M_* \lesssim 3 \times 10^{10}$ $M_\odot$.~This 
agrees remarkably well with the mass $M_* \sim 5 \times 10^{10}$ $M_\odot$ at which the
\hbox{$L_X$\ts--\ts$L_B$} {\it red~line\/} in Figure 48 crosses the estimate of $L_X$ from 
discrete~sources.  I.{\thinspace}e., Dekel \& Woo suggest that supernovae can drive gas out of galaxies 
over just the mass range where Figure 48 shows that no hot gas is seen.~However, a starburst
is necessary so that many supernovae go off together.~Absent a starburst, Dekel \& Woo 
assume that supernovae merely regulate 
star formation.  Like Dekel \& Silk (1986) and consistent with Garnett (2002), they use 
supernova-driven baryon ejection and supernova-regulated star formation to explain the low-luminosity, 
low-surface-brightness sequence of spirals, irregulars, and spheroidals whose Sphs form  
one side of our E{\ts}--{\ts}Sph dichotomy.

      Fundamental to the physical picture that we suggest in this paper is a merger-induced 
starburst that makes the extra light component in coreless galaxies.  This may be the rapid star 
formation event that Dekel \& Woo need in order that supernovae can clean low-mass merger remnants 
of their gas.  

      Doing so is not a trivial issue:

      Ultraluminous infrared galaxies (ULIRGs) are mergers-in-progress 
(Joseph \& Wright 1985; 
Sanders \etal 1988a, b;
Sanders \& Mirabel 1996;
Rigopoulou \etal 1999;
Dasyra \etal 2006a) 
that are prototypes of the formation of ellipticals in the local universe (Kormendy \& Sanders 1992).  
They are rich in gas and dust.  Their structural parameters are consistent with the fundamental plane 
(Kormendy \& Sanders 1992; 
Doyon \etal 1994;
Genzel \etal 2001;
Tacconi \etal 2002;
Veilleux \etal 2006;
Dasyra \etal 2006a, b).  
Stellar velocity dispersions $\sigma \simeq 100$ to 230 km s$^{-1}$ show that local 
ULIRGs are progenitors of moderate-luminosity ellipticals; i.{\thinspace}e., the disky-coreless 
side of the E{\thinspace}--{\thinspace}E dichotomy and {\it not\/} boxy-core ellipticals
(Genzel \etal 2001;
Tacconi \etal 2002;
Dasyra \etal 2006b, c).
So {\it ULIRGs are consistent with 
our formation picture: they are merger-induced starbursts that are making $\sigma \sim 160 \pm 60$ 
km s$^{-1}$ (hence coreless-disky) ellipticals}.  After much debate about what
energy source dominates ULIRGs (Joseph 1999; Sanders 1999), it has become
clear that starbursts dominate energetically in almost all cases
(Lutz \etal 1996;
Genzel \etal 1998;
Downes \& Solomon 1998;
Joseph 1999;
Rigopoulou \etal 1999;
Genzel \etal 2000;
Tran \etal 2001;
Spoon \etal 2007;
Netzer \etal 2007;
Vega \etal 2008;
Nardini \etal 2008).
ULIRGs are rare locally, but they get more common rapidly with increasing redshift 
(Sanders \& Mirabel 1996; 
Le Floc'h \etal 2005).
This is consistent with the protracted overall star formation histories of disky-coreless but not 
boxy-core Es (see Renzini 2006 for a review).  On the other hand, the timescales of individual 
starbursts in ULIRGs are a few tens of millions of years
(Lutz \etal 1996;
Genzel \etal 1998),
not much longer than the lifetimes of the stars that die as supernovae and short enough for Dekel \& Woo's
argument.  ULIRGs are exactly the ellipticals-in-formation that we propose.  That's the good news.  
Here is the bad news:

      It is a big step to understand how these intermediate-mass mergers-in-progress lose their gas,
as they must do if they are to form extra light ellipticals.  A plausible 
picture is this: (1) star formation in the infalling gas in a merger efficiently converts 
much of the gas into stars, (2) the observed, strong winds from ULIRGs -- driven mainly by hot  
stars and supernovae -- are beginning the process of gas ejection (see Veilleux \etal 2005 for a review), 
and (3) Dekel \& Woo's argument tells us the mass range over which this process will 
ultimately be successful.  Their estimate is consistent with our conclusion that a change in dominance 
from supernova to AGN feedback happens over a range of several magnitudes between 
$M_V \simeq -20.4$ and $-21.6$.

      This helps:~We have come to think that all dissipative~mergers are like ULIRGs.  Because of 
their extraordinary infrared luminosities, they deservedly attract attention.  
But there exist many less spectacular dissipative mergers with easily enough central star formation 
for our picture but less of a gas cleaning problem (e.{\thinspace}g.,
Schweizer 1980, 1982, 1987, 1989, 1996, 1998;
Hibbard \etal 1994;
Hibbard \& van Gorkom 1996).
It is not necessary always to be {\it soaking\/} wet.

      A caveat is the possible ``no'' answer above.  Gas may not be completely absent in low-$L$ 
galaxies; it may just get too cool to radiate \hbox{X-rays}.  After all, there are strong 
reasons~to~believe that a warm-hot intergalactic medium surrounds even small galaxies (e.{\ts}g.,
Danforth \& Shull 2008; see Bregman 2007 for a review).  But the good correlation of $L_X$
with the E -- E dichotomy suggests that a small amount of hot gas in low-luminosity Es (Ho 2008)
is no problem for our formation picture.  Still-smaller galaxies that were their merger 
progenitors can easily have contained the cold gas necessary to make wet mergers wet. 

\subsubsection{Perspective}

      In summary, we suggest that X-ray gas prevented star formation in the last mergers that made
giant-boxy-core Es.  And we suggest that AGN feedback is the
main process that keeps hot gas hot.  Thus, $M_{\rm crit}$ quenching is the fundamental reason 
{\it why\/} the E -- E dichotomy arose.  It is not necessary that both merger progenitors lacked 
cold gas, since hot gas can prevent star formation even when some cold gas is present.  
Metaphorically, there are three ways to be dry: water can be absent, frozen, or steam.  This 
section was about steam.

      Our picture of the formation of elliptical galaxies is closely similar to that
advocated by Dekel and Cattaneo and collaborators on theoretical and modeling grounds and by
Faber (2005) and Faber \etal (2007) based on observations of SDSS and distant galaxies.  Their
picture of ``$M_{\rm crit}$ quenching'' of star formation was developed to explain specific 
observational puzzles, mainly the color bimodality of galaxies and the surprising observation
that the biggest ellipticals formed their stars quickly and long ago.  Much effort has gone 
into showing that it explains the properties of galaxies as a function of redshift. 
These are important accomplishments.  They account for the well deserved popularity of 
this formation picture. 

      Our results lead to the same bottom line via a different route.  Independently of the 
above work, this paper has developed an observational picture of what it means to be an 
elliptical galaxy.  We confirm that ellipticals form a well defined structural sequence -- 
distinct from that of spheroidal galaxies -- with a luminosity function that is bounded at 
low $L$ approximately by M{\ts}32 and at high $L$ by M{\ts}87 and by still brighter cD galaxies. 
Ellipticals formed via major mergers; this was known.  We have added to the evidence that 
ellipticals come in two varieties that have interpretably different properties.  Among these 
are the distinction into core galaxies, which (if scoured by binary black holes)
require dry mergers, and ``extra light'' ellipticals, where the extra light is a ``smoking gun''
that implies dissipative formation.  This strengthens the conclusion -- otherwise not 
new -- that the reason for the E{\ts}--{\ts}E dichotomy is dry versus wet mergers.  Why there is 
such a dramatic wet-versus-dry distinction and why it depends on galaxy mass was not known.
Also, while it was known that the E{\ts}--{\ts}E dichotomy includes the presence or not of X-ray gas and
the importance or not of radio AGNs, the relevance of these observations was not understood.
We connect them into a coherent picture in which the X-ray dichotomy is central to our 
understanding of {\it why\/} the E{\ts}--{\ts}E dichotomy developed.  Fundamental to the explanation
is a transition from supernova-driven energy feedback in small galaxies to AGN feedback in 
large ones.  We suggest that X-ray gas is the essential agent that makes dry mergers dry and 
that AGN feedback is important only in giant galaxies and keeps hot gas hot.  The essential 
property that allows a galaxy to retain an X-ray halo is mass.  The mass necessary for
the observations that we have discussed is exactly the critical mass in the $M_{\rm crit}$ 
quenching picture.  The two pictures have converged ``for free.''  

\subsection{Context: Summary of Elliptical Galaxy Formation}

\lineskip=-4pt \lineskiplimit=-4pt

      Our results contribute to a picture of elliptical galaxy formation that now
encompasses a broad range of phenomena.  Hierarchical clustering (White \& Rees 1978) 
leads to galaxy mergers that scramble disks and make ellipticals (Toomre~1977; 
Schweizer 1989).~Merger progenitors usually contain gas; gravitational torques drive it 
to the center (Barnes \& Hernquist 1991, 1996) and feed starbursts (Mihos \& Hernquist 
1994, 1996).  ULIRGs are local examples of dissipative mergers.  With intermediate masses, 
their descendants correspond to the extra light--disky part of the E{\ts}--{\ts}E dichotomy.  
Observations (reviewed in \S\ts12.3.2) and theoretical models (Kauffmann \& Haehnelt 2000; 
Hopkins \etal 2005a,{\ts}b; 2006a,{\ts}b) imply that ULIRGs are related to quasars.
The consequences for galaxy evolution are not clear.  AGNs are
seen to be more important in more luminous ULIRGs 
(Lutz \etal 1998;
Genzel \etal 2000;
Tran \etal 2001;
Farrah \etal 2002;
Veilleux \etal 2006;
Schweitzer \etal 2006;
Netzer \etal 200).
But most ULIRGs are energetically dominated by starbursts.  It is clear that merger-induced
starbursts like those discussed in this paper as the origin of ``extra light'' in coreless
ellipticals have not been prevented by AGN feedback; nor do the papers reviewed in the previous
section find any correlation of AGN importance with the dynamical stage (early or late) of the
host merger.  Altogether, it appears likely that
quasar energy feedback has a major effect on the formation of bright ellipticals
(Silk \& Rees 1998; 
Ciotti \& Ostriker 2001;
Ostriker \& Ciotti 2005) 
but not faint ellipticals (this paper).  This helps to explain why supermassive black holes 
correlate with bulges (Kormendy \& Richstone 1995; Ferrarese \& Merritt 2000; 
Gebhardt \etal 2000; Tremaine \etal 2002) but not disks (Kormendy \& Gebhardt 2001) -- bulges 
and ellipticals are made in mergers, but disks are  not.  So, while many details remain to be
worked out, our picture of the formation of extra light--disky ellipticals is 
becoming well articulated.  Now our understanding of core-boxy ellipticals is
catching up.  Critically important is the observation that essentially all 
of their star formation happened quickly and long ago (Bower \etal 1992; Bender 1996, 
1997; Thomas \etal 1999, 2005; Bernardi \etal 2003; Renzini 2006).  We know little about their
merger progenitors.  Nevertheless, parallel investigations of the theory of gas accretion 
during hierarchically clustering (Birnboim \& Dekel 2003; Dekel \& Birnboim 2006, 2008), 
simulations of the accretion (Kere\v{s} \etal 2005), semi-analytic models of galaxy 
formation including energy feedback (Cattaneo \etal 2006, 2008a), observations 
of galaxy evolution with redshift (Faber 2005; Faber \etal 2007), and archaeology of 
galaxy structure (this paper) have converged on an ``$M_{\rm crit}$ picture'' in which
total mass $M$ is the main parameter that controls galaxy evolution.  Only at
$M > M_{\rm crit}$ can galaxies create, continually reheat, and hold onto hot gas halos 
at X-ray temperatues; they keep them hot via a combination of AGN feedback and cosmological
infall, and they use them to quench star formation and make subsequent mergers dry. 
We show that this picture accounts naturally for the observed dichotomy of 
elliptical galaxies into dry merger remnants that contain cores and wet merger 
remnants that contain extra central components that are the signatures of merger starbursts.
Merger simulations that are motivated by these results and that incorporate 
the above physics do remarkably well in reproducing the different properties of core
and extra light ellipticals (Hopkins \etal 2008a, b, c, d, e).

\section{The E -- Sph Dichotomy}

       Fundamental to the above discussion is the conclusion that elliptical and 
spheroidal galaxies are physically different.  This result, presciently guessed by Wirth 
\& Gallagher (1984), demonstrated by Kormendy (1985b, 1987b), and confirmed
by Binggeli \& Cameron (1991) and Bender \etal (1992), has been much criticized in
recent years.  With high-dynamic-range brightness profiles, we show in Figures 34\ts--\ts38 
that the \hbox{E\ts--{\ts}Sph} dichotomy is real.  In correlations such as effective 
brightness versus effective radius and effective brightness versus absolute magnitude, 
ellipticals and spheroidals form almost perpendicular sequences.  These sequences approach
each other at $M_{VT} \simeq -18$, near the maximum of the luminosity function for 
ellipticals but at a luminosity where spheroidals are rare.  The dichotomy is not a result
of a biased sample; in fact, our sample is biased in favor of finding the spheroidals 
that are most like ellipticals.  

      This result is critically important to our understanding of galaxy formation.
Consider the contrary: If spheroidal galaxies and all ellipticals except those with 
cores formed a continuous Sph{\ts}--{\ts}E sequence in parameter space, then that sequence would
be completely different from the fundamental plane discovered by Djorgovski \& Davis 
(1987) and Faber \etal (1987) and studied by many others (e.{\ts}g., Bender \etal 1992, 1993).
That Sph{\ts}--{\ts}E sequence would be almost perpendicular to our fundamental plane,
$ r_e \propto \sigma^{1.4 \pm 0.15} ~ I_e^{-0.9 \pm 0.1}$.
Its interpretation that structure is controlled by the \hbox{Virial theorem,
$r_e \propto \sigma^2 ~ I_e^{-1}$}, modified by small nonhomologies would be
wrong.  A Sph{\ts}--{\ts}E sequence would be inconsistent with the well established result that 
the scatter in the E -- E fundamental plane is small (Saglia \etal 1993; J\o rgensen \etal 1996).
Merger~simulations (Boylan-Kolchin \etal 2006; Robertson \etal 2006; Hopkins \etal 2008d,{\ts}e) reproduce
the E{\ts}--{\ts}E fundamental plane, not a set of \hbox{Sph{\ts}--{\ts}E} correlations.  Equating spheroidals 
with low-luminosity ellipticals would imply that they formed similarly, but we are confident 
that ellipticals formed by mergers, and we believe that dwarf spheroidals cannot have formed 
by mergers (Tremaine 1981).  Continuous Sph -- E correlations are inconsistent with almost 
everything that we know about galaxy formation.

      However, our results confirm that elliptical galaxies of both types together define 
the classical fundamental plane in which lower-luminosity galaxies have smaller $r_e$ and 
brighter $I_e$ (Kormendy 1977) all the way from giants like M{\ts}87 to dwarfs like M{\ts}32.
Spheroidals overlap this sequence in luminosity, but much below the brightness of M{\ts}32 
($M_{VT} = -16.7$), where we find no  ellipticals, their luminosity functions rise steeply 
all the way to the faintest galaxies known ($M_{VT} > -9$).  Along this sequence, visible 
matter densities decrease rapidly with decreasing galaxy mass, consistent with the progressive 
loss of more and more baryons as gravitational potential wells get shallower and as supernovae 
get more effective in ejecting gas (e.{\ts}g., Dekel \& Silk 1986; Dekel \& Woo 2003).  For 
our overall understanding of galaxy formation, confirmation of the E{\ts}--{\ts}Sph dichotomy 
is the most important result of this paper.

\vfill\eject

\acknowledgments

      This paper has dominated JK's and DBF's research for more than three years,
and it has been a recurring theme for JK and RB for periods as long as 25 years.
During this long gestation, many people have helped us and deserve our 
sincere thanks.  Most importantly, we thank the anonymous referee for an 
exraordinarily helpful~report.  This report, our own inclinations, and input from 
several readers (notably Karl Gebhardt and Tod Lauer) motivated the addition of 
Appendix A on S\'ersic function fitting machinery, as well as numerous other 
improvements.  We are grateful to  
Sandra Faber,
Luis Ho, 
Tod Lauer, 
Joe Silk, and
Scott Tremaine
for helpful comments on the MS.  Incisive questions by Sandy Faber particularly 
helped us to refine the conclusions of \S\ts12.  JK thanks
Josh Barnes,
Philip Best,
Andi Burkert,
Michele Cappellari,
Andrea Cattaneo,
Avishai Dekel, 
Bill Forman,
Karl Gebhardt, 
Lars Hernquist, 
Philip Hopkins,
Milos Milosavljevi\'c,
Richard Mushotzky, and
Jerry Ostriker
for helpful conversations.  DBF thanks Roberto Saglia for his detailed introduction
to the Bender/Saglia surface photometry code as implemented in {\tt MIDAS\/}.  Finally,
it is appropriate here to thank George Djorgovski, whose emphasis on thoroughness on 
many occasions (e.{\ts}g., Djorgovski \& King 1986) reinforced our own and had a 
significant effect on this paper.

      The CFH12K observations of the Virgo cluster were made in collaboration with 
Richard Wainscoat; we are grateful to him for agreeing to let us use the images here.  
We thank Ying Liu for sending the $R$-band, major-axis profile of M{\ts}87 as derived 
from his multiband photometry.  This work makes extensive use of the data products
from the {\it Hubble Space Telescope\/} image archive and from the Two Micron All Sky 
Survey (2MASS).  2MASS is a joint project of the University of Massachusetts and the 
Infrared Processing and Analysis Center/California Institute of Technology, funded by 
NASA and the NSF.  We also made extensive use of the digital image database of the 
Sloan Digital Sky Survey (SDSS).  Funding for the SDSS and SDSS-II has been provided 
by the Alfred P.~Sloan Foundation, the Participating Institutions, the National Science 
Foundation, the U.~S.~Department of Energy, the National Aeronautics and Space 
Administration, the Japanese Monbukagakusho, the Max Planck Society, and the Higher
Education Funding Council for England.  The SDSS is managed by the Astrophysical Research 
Consortium for the Participating Institutions. The Participating Institutions are the 
American Museum of Natural History, Astrophysical Institute Potsdam, University of Basel, 
University of Cambridge, Case Western Reserve University, University of Chicago, Drexel 
University, Fermilab, the Institute for Advanced Study, the Japan Participation Group, 
Johns Hopkins University, the Joint Institute for Nuclear Astrophysics, the Kavli 
Institute for Particle Astrophysics and Cosmology, the Korean Scientist Group, the 
Chinese Academy of Sciences (LAMOST), Los Alamos National Laboratory, the 
Max-Planck-Institute for Astronomy (MPIA), the Max-Planck-Institute for Astrophysics 
(MPA), New Mexico State University, Ohio State University, University of Pittsburgh, 
University of Portsmouth, Princeton University, the United States Naval Observatory, 
and the University of Washington.

      The NGC{\ts}4318 spectrum was obtained with the Marcario Low Resolution Spectrograph 
(LRS) and the Hobby-Eberly Telescope~(HET). LRS is named for Mike Marcario of High 
Lonesome Optics; he made optics for the instrument but died before its 
completion.  LRS is a project of the HET partnership and the 
Instituto de Astronom\'\i a de la Universidad Nacional Aut\'onoma de M\'exico.  The HET
is a project of the University of Texas at Austin, Pennsylvania State University, 
Stanford University, Ludwig-Maximilians-Universit\"at M\"unchen, and Georg-August-Universit\"at 
G\"ottingen.  The HET is named in honor of its principal benefactors, William P.~Hobby and 
Robert E.~Eberly.

      This work would not have been practical without extensive use of the NASA/IPAC 
Extragalactic Database (NED), which is operated by the Jet Propulsion Laboratory and
the California Institute of Technology under contract with NASA.  We also used the 
HyperLeda electronic database (Paturel \etal 2003) at {\tt http://leda.univ-lyon1.fr} and 
the image display tool SAOImage {\tt DS9} developed by Smithsonian Astrophysical Observatory.
Figure 9 was adapted from the WIKISKY image database ({\tt www.wikisky.org}).
Finally, we made extensive use of NASA's Astrophysics Data System bibliographic services.

      JK's ground-based CFHT photometry was supported by NSF grant AST--9219221.  Also,
NSF support under grant AST-0607490 during the late stages of this work is gratefully 
acknowledged.   JK's HST photometry work as a member of Nuker team program GO-06099 was 
supported by NASA through grants from the Space Telescope Science Institute (STScI). 
STScI is operated by AURA under NASA contract NAS5-26555.  MEC was supported in part by 
a generous and much appreciated donation to McDonald Observatory by Mr.~Willis A.~Adcock.
Finally, we acknowledge that this multi-year research program would not have been possible 
without the long-term support provided to JK by the Curtis T.~Vaughan, Jr.~Centennial 
Chair in Astronomy.  We are most sincerely grateful to Mr.~and Mrs.~Curtis T.~Vaughan, 
Jr.~for their continuing support of Texas astronomy.  It is a pleasure to dedicate this paper 
to them.

\vskip 10pt

{\it Facilities:} HST: WFPC1;
                  HST: WFPC2;
                  HST: ACS;
                  HST: Nicmos;
                  CFHT: Cassegrain camera;
                  CFHT: HRCam;
                  CFHT: 12K camera;
                  SDSS: digital image archive;
                  McDonald Observatory: 0.8 m telescope

\vfill\eject

\onecolumn

\section*{Appendix A}

\section*{S\'ERSIC FUNCTION FITS TO THE ELLIPTICAL AND SPHEROIDAL GALAXIES}

      Appendix A documents our S\'ersic (1968) function fits to the major-axis brightness profiles 
$I(r)$ of elliptical and spheroidal galaxies (Figures 49\ts--\ts72).  We test the robustness of
our fits to changes in the adopted fit range.  We provide a summary (Figures 73) with which users
of S\'ersic functions can judge whether or not the dynamic range of their profile data are adequate
for reliable fits.  

The S\'ersic function is
$$
I(r) = I_e~{\rm dex}~\biggl\lbrace-b_n\biggl[\biggl({r \over r_e}\biggr)^{1/n} - 1\biggr]\biggr\rbrace~,
\eqno{ (A1) } 
$$ 
where $b_n$ is chosen so that $r_e \equiv$ ``effective radius'' contains half of the total
light of the model profile and $I_e \equiv$ ``effective brightness'' 
is the surface brightness at $r_e$.  Over the range of S\'ersic indices $0.5 \leq n \leq 16.5$, 
numerical integration gives the approximation formula, $b_n \simeq 0.868{\thinspace}n - 0.142$ 
(Caon et al.~1993).  That paper, Ciotti 1991), Graham et al.~(1996), Ciotti \& Bertin (1999),
Trujillo \etal (2001), and Graham \& Driver (2005) discuss S\'ersic functions in detail.  They have 
become popular machinery to describe the profiles of E and Sph galaxies and to derive parameters $n$,
$r_e$, and $\mu_e \equiv -2.5 \log{I_e}$, for structural analyses such as fundamental plane studies.  

      This Appendix concentrates on two aims that are not discussed in previous
literature.  We illustrate each fit, including $\chi^2$ ellipses in the fit parameters.  
These provide realistic error estimates (\S{\thinspace}A1) that take the (often very strong) 
parameter coupling into account.  Second (\S{\thinspace}A2), we explore the robustness of
the fits to changes in the range of radii that are fitted.  This is important
because neither profile measurement errors nor errors associated with any failure
of the function to describe the profiles are random.  Fits can change substantially
depending on whether particular wiggles in the profile are included or not.
{\it How much dynamic range in a galaxy brightness profile is required to get a robust 
S\'ersic fit?}   With accurate profiles over large dynamic ranges, we can answer this
for the Virgo cluster sample.  The results should be useful as a general guide to 
interpreting the reliability of published and future S\'ersic fits.
 
      Figures 49\ts--\ts72 illustrate the fits.  Consistent with \S\ts4.1, we fit S\'ersic
functions over the largest radius ranges over which they agree with the composite 
major-axis profiles.  Fit tolerances are determined from the profile measurement errors 
implied by the scatter at each radius of the individual measurements illustrated
in Figures 11 -- 29 (top and bottom panels) and from the function fitting errors to
the mean profile in Figures 49\ts--\ts72 (top-left panels here).  In general, the latter errors
dominate.  The median RMS of the 27 fits is 0.040 $V$ mag arcsec$^{-2}$.  The mean RMS is
0.046 $V$ mag arcsec$^{-2}$, and the dispersion in RMS values is 0.019 $V$ mag arcsec$^{-2}$.
S\'ersic functions fit the main parts of the profiles of both E and Sph galaxies
astonishingly well over large ranges in surface brightness.

      Of course, the above RMS values depend on our decisions on where to cut the
fit ranges at small and large radii.  At large radii, we prefer to keep deviations to
 $<$\ts0.1 mag arcsec$^{-2}$; as judged from the agreement between different sources,
this is approximately the estimated profile error at large radii.  However, in some cases, 
slightly larger deviations are accepted if doing so greatly increases the radius range of 
the fit.  Our aim is to have the S\'ersic fit describe as large a fraction of the total
light of the galaxy as possible, consistent with measurement errors.   Note that in almost 
all galaxies, the fit does not fail at large radii; rather, the profile ends where the 
signal-to-noise becomes too low, where sky subtraction becomes insecure, or where we 
reach the edge of the detector field of view.  {\it For most galaxies, the S\'ersic fits 
accurately describe the major-axis profiles over radius ranges that include $\sim 93$\ts\% 
to 99\ts\% of the light of the galaxies\/} (see Figure 41).  

      At small radii  $r$, the deviations of the profiles from the best fits become large
and systematic, and they do so quite suddenly as $r$ decreases.  This indicates the 
presence of cores or extra light.  Again, we cut the fit range where these deviations 
become comparable to the measurement errors.  We tend to be slightly conservative: we 
often include radii where the fit departures associated with cores or extra light are 
starting to become apparent, again in order to include as much of the galaxy in the
fit as possible.

      For a few galaxies, small radius ranges near the center are excluded because of dust 
absorption.  These do not significantly affect the fit results.  Also, for a few galaxies,
parts of the profile are excluded where large fit errors are associated with non-equilibrium 
structures that can be identified on physical grounds.  These are discussed in \S\thinspace{A.3}.

\section*{A1. PARAMETER ERRORS ESTIMATED VIA $\chi^2$ ELLIPSES}

      Figures 49\ts--\ts72 show two fits each, i.{\thinspace}e., the top and bottom halves
of each page.  For each fit, the left two panels show the mean profile points, the
fit range, the fit (as a solid curve), and the residuals $\Delta \mu$ from the fit,
together with the RMS within the fit range.  Figures 11\ts--\ts29 are corresponding plots
that show all of the original data sources.  The middle column of each figure here shows
the $\chi^2$ ellipses and lists the fit parameters.  The quoted parameter errors 
are the half-widths of each $\chi^2$ ellipse in that parameter.  The right-hand
columns of figures explore robustness to changes in the fit range; they are discussed
in \S{\thinspace}A2.

     The best-fit  Sersic models were derived by minimizing
$$ \chi^2\; = \; \frac{1}{N_{\rm ind}-3}\sum_{i=1}^{N_{\rm data}}{\frac{[\mu_i(r_i)-\mu_{S}(r_i)]^2}{\sigma_{\mu,i}^2}} \eqno{  (A2)  }$$
where $\mu_i(r_i)$ are the observed surface brightnesses at radii $r_i$ with measurement errors 
$\sigma_{\mu,i}$, and $\mu_S(r_i)$ is the surface brightness of the Sersic model, equation (A1), at
$r_i$.   Also, $N_{\rm data}$ is the number of data points, and $N_{\rm ind}$ is the number of 
independent data points.  Estimating $N_{\rm ind}$ has always been one of the central uncertainties
in profile fit error analysis.  We are helped by the fact that we average many independent data 
sets from different telescopes and profile measurement techniques.  On the other hand, closely
spaced data points near galaxy centers are coupled by PSF smoothing; adjacent data points at 
large radii usually suffer from similar problems with large-scale flat fielding and sky 
subtraction, and it is common for profile measurement software to smooth images at large
radii in order to improve S/N and to compensate for problems with masked or
removed foreground stars and background galaxies.  Therefore it is unrealistic to believe that
all data points in our tabulated mean profiles are independent.  After experimentation with
the data sets for individual galaxies, we adopt the somewhat conservative assumption that
$N_{\rm ind} \approx N_{\rm data}/2$.

\vfill\eject

     The other uncertainty in applying equation (A2) is the estimation of $\sigma_{\mu,i}$.
Inherent in $\chi^2$ minimization is the assumption that the errors in the fitted data points
are random and uncorrelated.  The residual plots show that both assumptions are almost always 
violated.  A few profile wiggles are produced artificially when (for example) one profile
data set starts to deviate from the others and, at some radius, suddenly gets omitted from
the average.  But examination of Figures 11 -- 29 shows that most profile wiggles are real~--
they look the same in many data sets.  They represent failures of the S\'ersic function to
describe the profiles at the few-percent level.  Such failurs are in no sense unexpected.  On
the contrary, it is surprising that S\'ersic functions work as well as they do.  Nevertheless,
the wiggles in the residual profiles -- and, to a lesser extent, scatter in the residual profiles
that is indicative of more-or-less random measurement errors -- represent the ultimate limit on
the accuracy of the S\'ersic fits.  We use the RMS scatter of the fits (see the keys of Figures 
49\ts--\ts72) as our estimate of $\sigma_{\mu,i}$.  As long as this RMS scatter -- although partly
systematic -- is a few hundredths of a mag arcsec$^{-2}$ and therefore comparable to profile
measurement errors, this choice is reasonable and unlikely to lead us far astray.  Nevertheless,
the need for this choice of $\sigma_{\mu,i}$ means that our error analysis is necessarily approximate.

     The rest of the job is engineering.   The $\chi^2$ minimum was determined with a simple grid search
technique using three steps of successive refinement.  Providing error estimates for the Sersic parameters
that reflect the fit quality in a meaningful way is tricky, because the errors of the three Sersic parameters 
can be strongly coupled.  Then the usual marginalized 1-$\sigma$ errors corresponding to $\Delta\chi^2=1$ 
around the minimum are misleadingly optimistic. We therefore decided to provide more realistic estimates
for the fit uncertainties, namely the sizes of the three-dimensional 1-$\sigma$ error ellipsoids 
{\it as projected onto the parameter axes}. These ellipsoids are defined by $\Delta\chi^2=3.53$ 
(Press \etal 1986, Chapter 14.5, ``Confidence Limits on Estimated Model Parameters'').  The two-dimensional 
projections of the error ellipsoids are shown in the middle columns of plots in Figures 49\ts--\ts72.
The corresponding parameter errors are listed in the keys above the plots.  Note that these are calculated
directly by interpolation in the $\chi^2$ arrays, whereas the $\chi^2$ ellipses are calculated ``on the fly''
by the {\tt sm} contouring code.  As a result, the illustrated $\chi^2$ ellipses do not agree perfectly 
with the (more reliable) tabulated errors.  Note also that extremely thin and elongated $\chi^2$ ``bananas'' 
sometimes break up into isolated islands when the {\tt sm} contouring program has trouble with the interpolation.

      The error estimates listed in the keys above the $\chi^2$ ellipses in Figures 49\ts--\ts72 
are included in Table 1 and used in our analysis.

      These error estimates are consistent with the results of our fit range tests as discussed in \S\thinspace{A.2}.

\section*{A2. ROBUSTNESS OF S\'ERSIC FITS TO CHANGES IN THE RADIAL FIT RANGE}

      Two kinds of fits are shown in Figures 49\ts--\ts72.  Most illustrations show the adopted fit for 
each galaxy (e.{\thinspace}g., top fit for NGC 4472 in Figure 49).  A few alternative interpretations 
with different radial fit ranges are included to illustrate specific scientific points (e.{\thinspace}g.,
bottom fit for NGC 4472 in Figure 49).  These are discussed in the text, but their parameters are not 
included in Table 1.

      For the adopted fits but not for the illustrative fits, the right-hand panels in Figures 49\ts--\ts72
test the effect of changing the outer radius $r_{\rm max}$ of the fit range from the adopted value 
$r_{\rm max,adopted}$ listed in the key of the large panel.  As a function of $r_{\rm max}/r_{\rm max,adopted}$, 
they show how the RMS residuals and the fit parameters (e.{\ts}g., $r_{\rm e}$) change from the adopted value 
(e.{\ts}g., $r_{\rm e,adopted}$) listed above the middle panels of $\chi^2$ ellipses.  The outer end of 
the fit range is changed by one tabulated profile point at a time, moving inward from the outermost tabulated 
point past the adopted point $r_{\rm max,adopted}$ (frequently the same as the outermost point) and on toward
smaller $r$ until the fit deviates drastically from the adopted one.  For every choice of $r_{\rm max}$, a
S\'ersic fit is made and its results are illustrated .   The plotted error bars are the half-widths of 
the $\chi^2$ ellipses corresponding for that particular fit to the ones illustrated in the middle columns 
for the adopted fit.  That is, the error bars take parameter coupling into account.

      Examination of the fit range tests shows that our adopted S\'rsic fits are very robust for
almost every galaxy:

      Sometimes the outermost data points (beyond $r_{\rm max,adopted}$) deviate suddenly
above or below the adopted fit and would change that fit noticeably if included.  But these points are
very vulnerable to sky subtraction or flat-fielding errors.  We include these points in the tabulated
profile in part because they result in more realistic total magnitudes but also so that readers can see
our profile calculations begin to fail where they get difficult.  We have no problem in discarding these
points from the S\'ersic fits.

      More fundamental issues are these: As $r_{\rm max}$ is decreased, which wiggles in the 
composite profiles should we include in the fits?  Are the fits sensitive to these choices?  How much
can we shrink the fit range and still derive reliable S\'ersic parameters?  That is, how much dynamic
range in galaxy profiles is necessary for the confident use of S\'ersic function fitting machinery?

      Also, do the fit range tests support our error bars?

      The figures provide clear answers to these questions.  Fits derived with $r_{\rm max} \equiv f r_{\rm max,adopted} 
< r_{\rm max,adopted}$ differ by $\leq 1\ \sigma$ from the adopted fits in to about $f = 0.50$.  More precisely, 
the limiting $f$ has a mean of $0.48 \pm 0.03$ (dispersion = 0.15) and a median of 0.50 (quartiles = 0.37, 0.59).
For somewhat smaller $r_{\rm max}$, the derived parameters still change only slightly as different profile 
wiggles are successively omitted from the fit.  Of course, as $r_{\rm max}$ is decreased, the RMS gets smaller,
because the program struggles to fit fewer profile wiggles.  Also, the parameter error bars grow, because their 
derivation is based on fewer data points.  But {\it the changes in the parameters are consistent 
with the error bars given by the adopted fit.  This confirms that our error analysis is realistic even though 
the profile fit errors are more systematic than random.}  No conclusions of this paper are vulnerable to 
modest changes in fit ranges.

      Eventually, as $r_{\rm max}$ is decreased well below $0.5\ r_{\rm max,adopted}$, the fits begin to 
deviate more significantly from the adopted ones.  This is a sign that the dynamic range has become dangerously
too small; i.{\thinspace}e., that a very few profile wiggles are ``torquing'' the fit unrealistically.
The degree to which this is a problem depends on S\'ersic $n$.  That is, the dynamic range in profile
data that are needed for robust S\'ersic fits depends on $n$.  We summarize both the dynamic range that we have 
in the present data and the reduced dynamic ranges that gives fiducial errors in the S\'ersic parameters in 
Figure 73.

\vfill\eject

\figurenum{49}

\centerline{\null} \vfill
 
\includegraphics{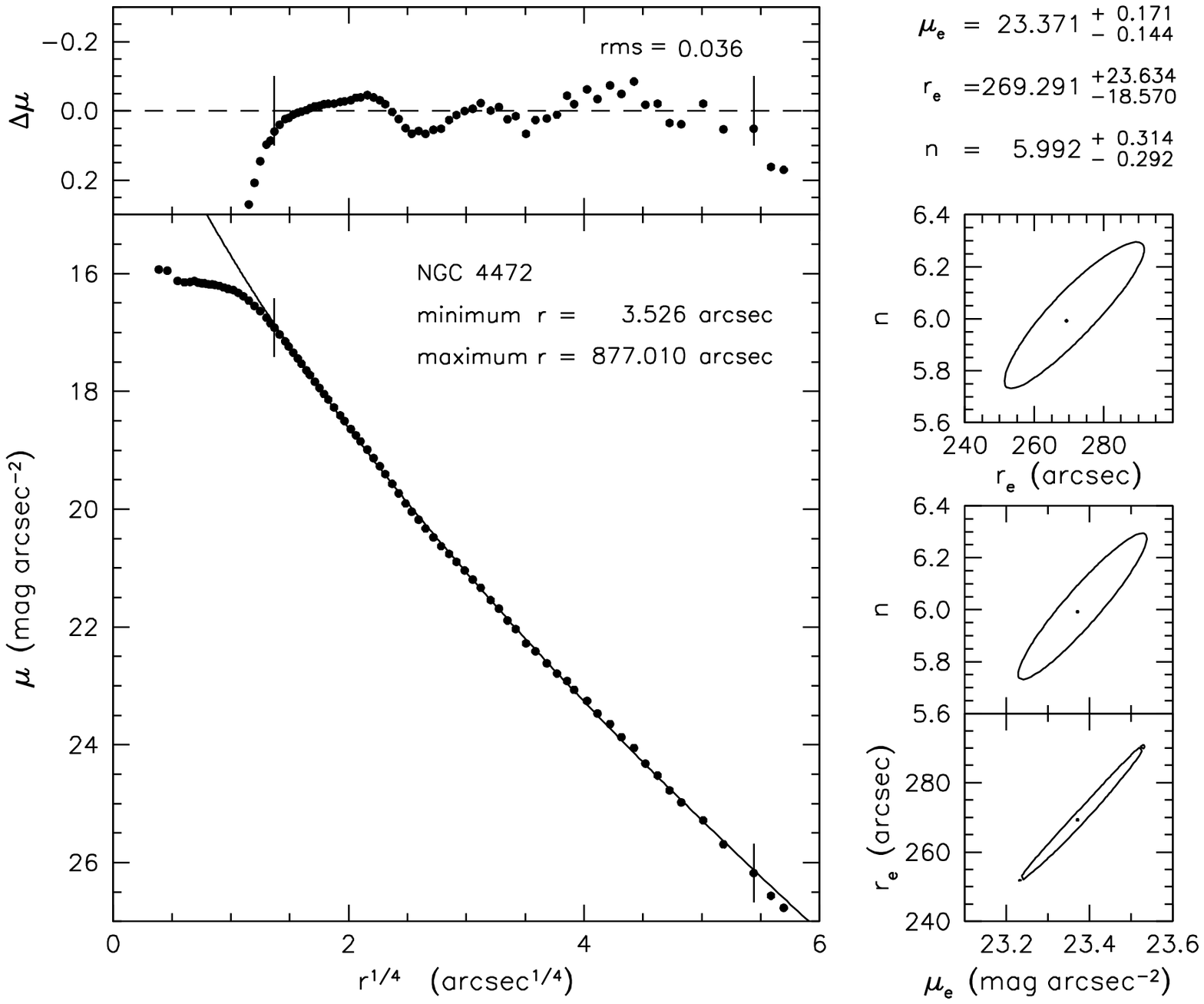}

\includegraphics{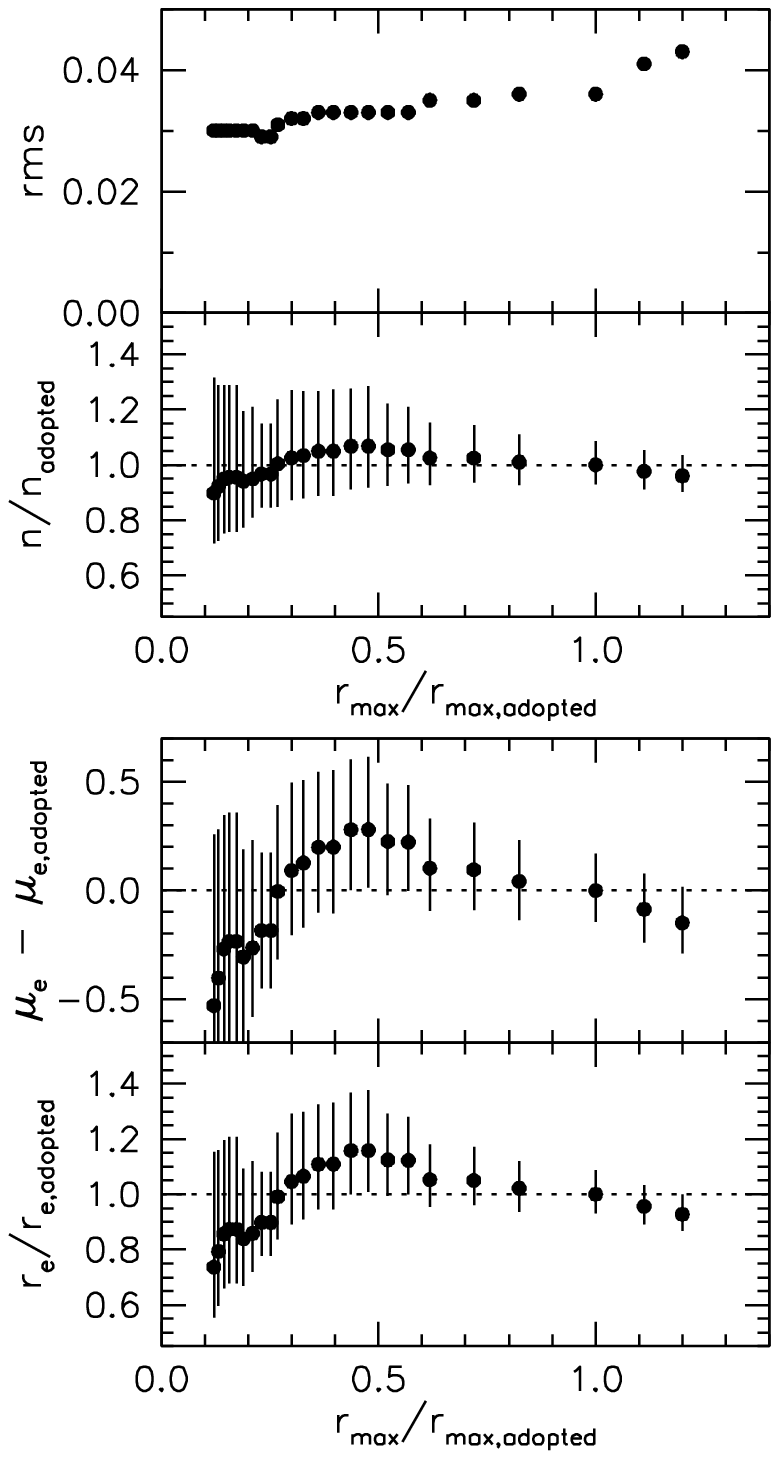}

\includegraphics{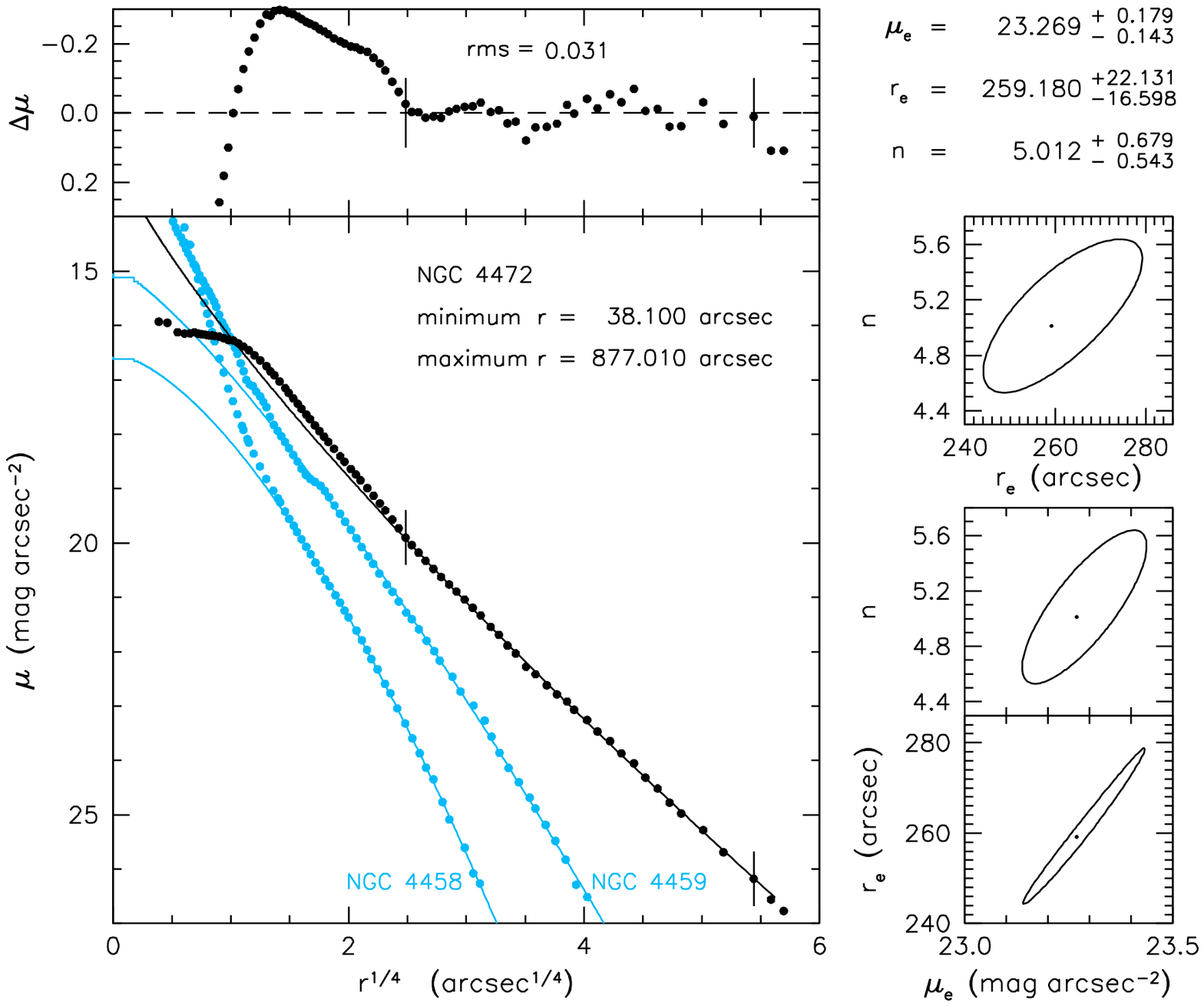}

\figcaption[]
{S\'ersic (1968) function fits to the major-axis profile of NGC 4472 (also fits to
NGC 4458 and NGC 4459, for comparison, at the bottom).  This figure and the ones that
follow show all known elliptical-galaxy members of the Virgo cluster in order of
decreasing luminosity, followed by our spheroidal galaxies, 
also in order of decreasing luminosity.  In this and the following figures,
the large panel shows the fit (solid curve) to the profile used in all calculations; 
it is the average of the individual profiles illustrated in Figures 11 -- 29, 
as discussed in the text.  The top-left panel shows the deviations of the profile 
from the fit and lists the RMS deviation in magnitudes.  In both panels, the fit 
range is shown by vertical dashes.  The fit parameters are listed in the middle at
the top.  The small panels in the middle show the three-dimensional, 1-$\sigma$ $\chi^2$
contours projected into two dimensions.  They illustrate the parameter coupling. 
Appendix~A shows two kinds of fits, the adopted fits for all galaxies 
(e.{\ts}g., at top) and, for some galaxies, one or more additional fits that are designed 
to illustrate specific astrophysical issues discussed in the text (e.{\ts}g., bottom fit here).  
For the final fits but not for the illustrative fits, the right-hand panels test the effect of 
changing the outer radius $r_{\rm max}$ of 
the fit range from the adopted value $r_{\rm max,adopted}$ listed in the key of the large
panel.  As a function of $r_{\rm max}/r_{\rm max,adopted}$, they show how the fit RMS
and the fit parameters (e.{\ts}g., $r_{\rm e}$) change from the adopted value (e.{\ts}g., 
$r_{\rm e,adopted}$) listed above the middle panels.  The NGC 4458 and 4459 profiles
are discussed in \S\ts10.3, Footnote 11.
}

\eject\clearpage

\figurenum{50}

\centerline{\null}

\vfill

\includegraphics{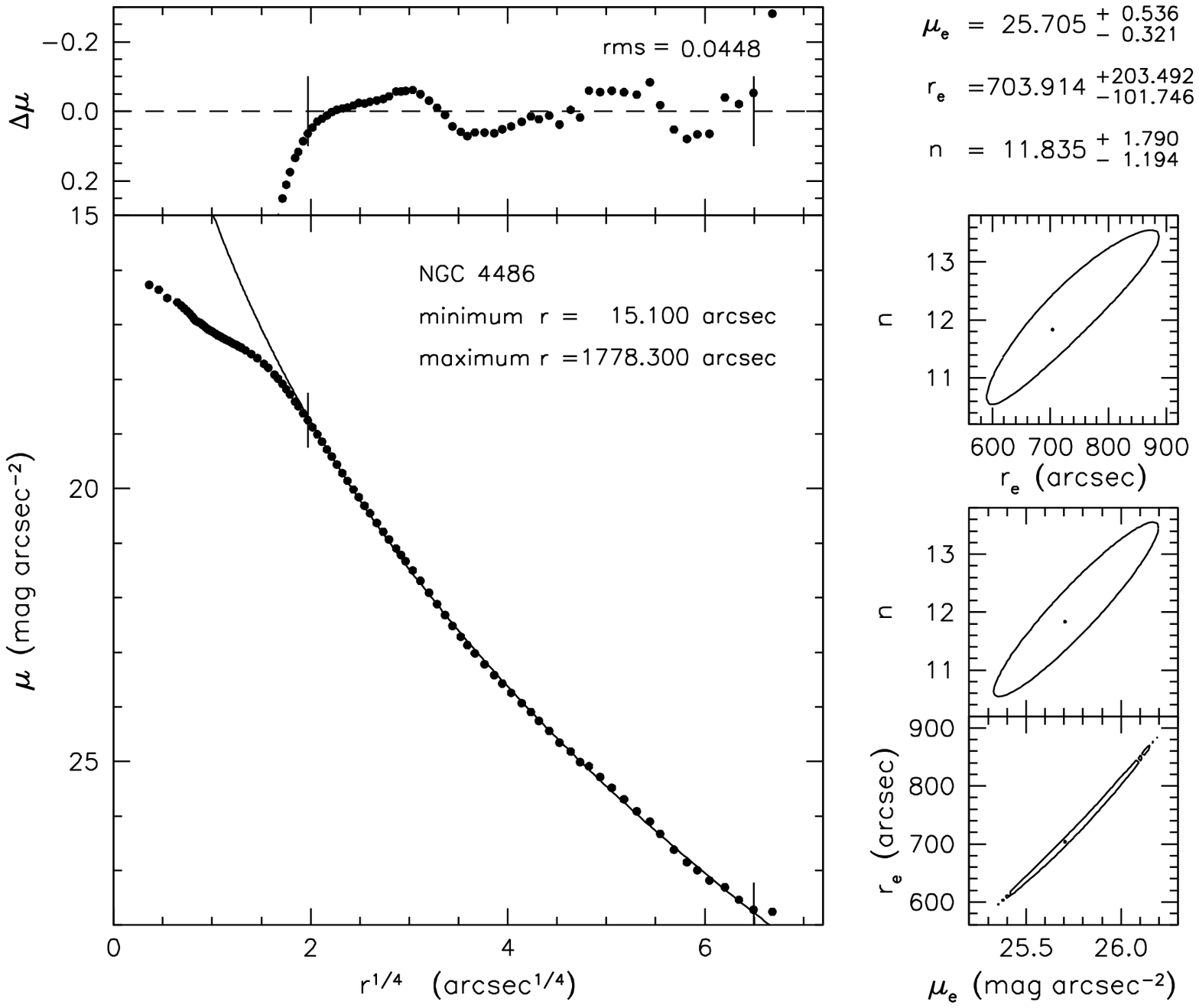}

\includegraphics{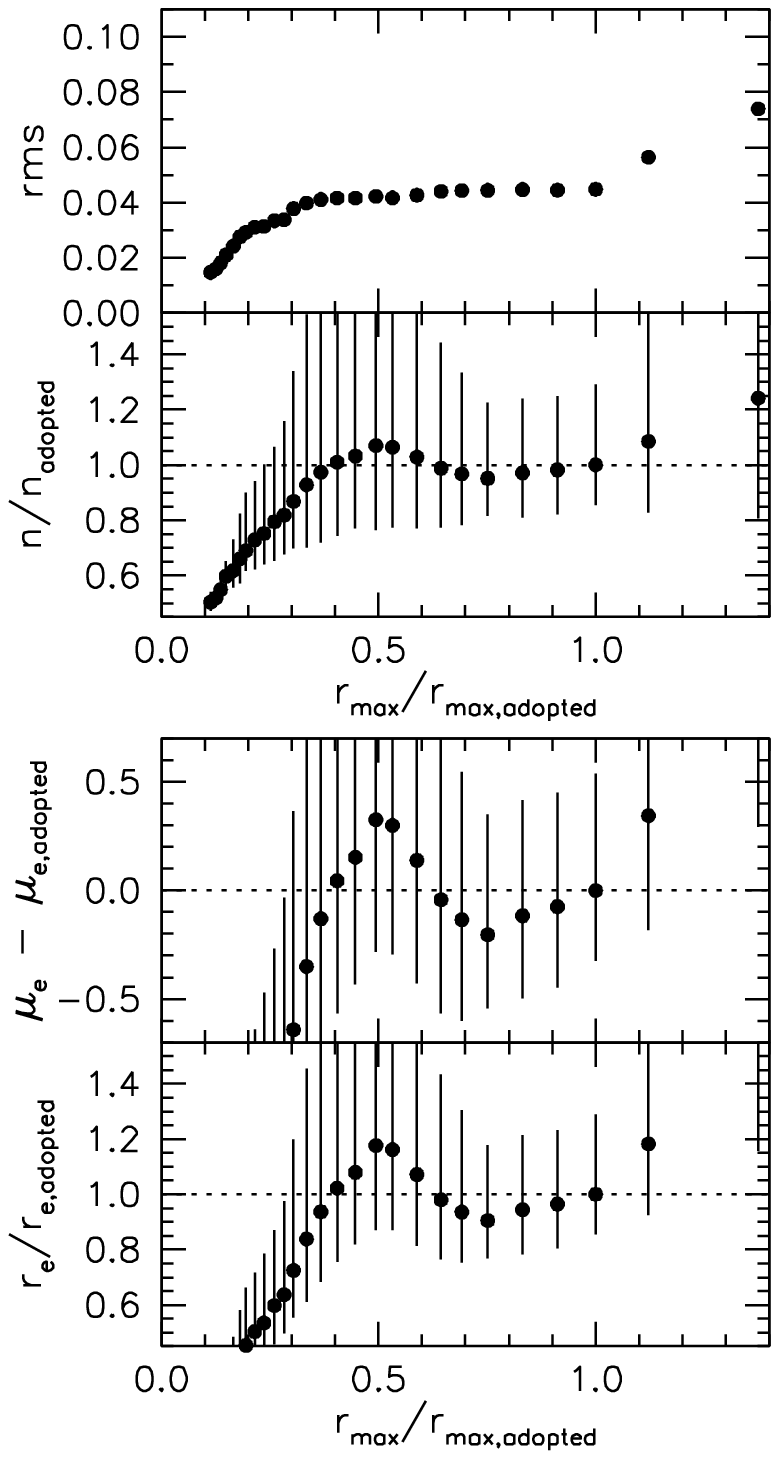}

\includegraphics{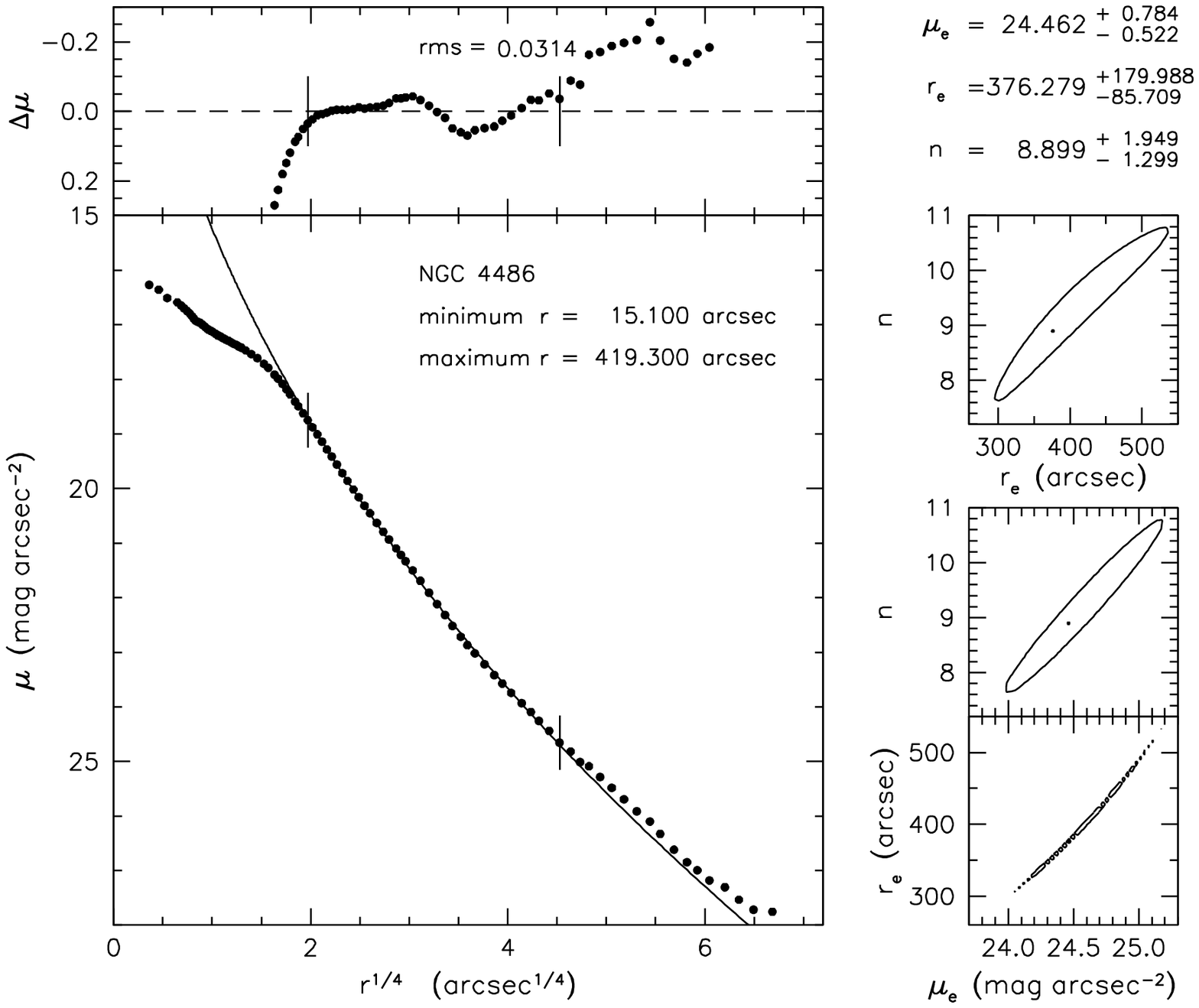}

\figcaption[]
{S\'ersic function fits to the major-axis profile of NGC 4486 (M{\ts}87).
The layout is as in Figure 49. In some $\chi^2$-contour figures here and on
the following pages, the {\tt sm} contouring routine has difficulty with the thinnest 
$\chi^2$ contours.  They are plotted as distinct ``pearls'' but of course are continuous.
The contours also are approximate when they have sharp, pointed ends.  The extraordinarily
large $n$ value in the upper fit may be due to the inclusion of a low-surface-brightness
cD halo.  At the bottom, we illustrate a plausible fit over a smaller radius range that
excludes such a halo.}

\eject\clearpage

\figurenum{51}

\centerline{\null}

\vfill
 
\includegraphics{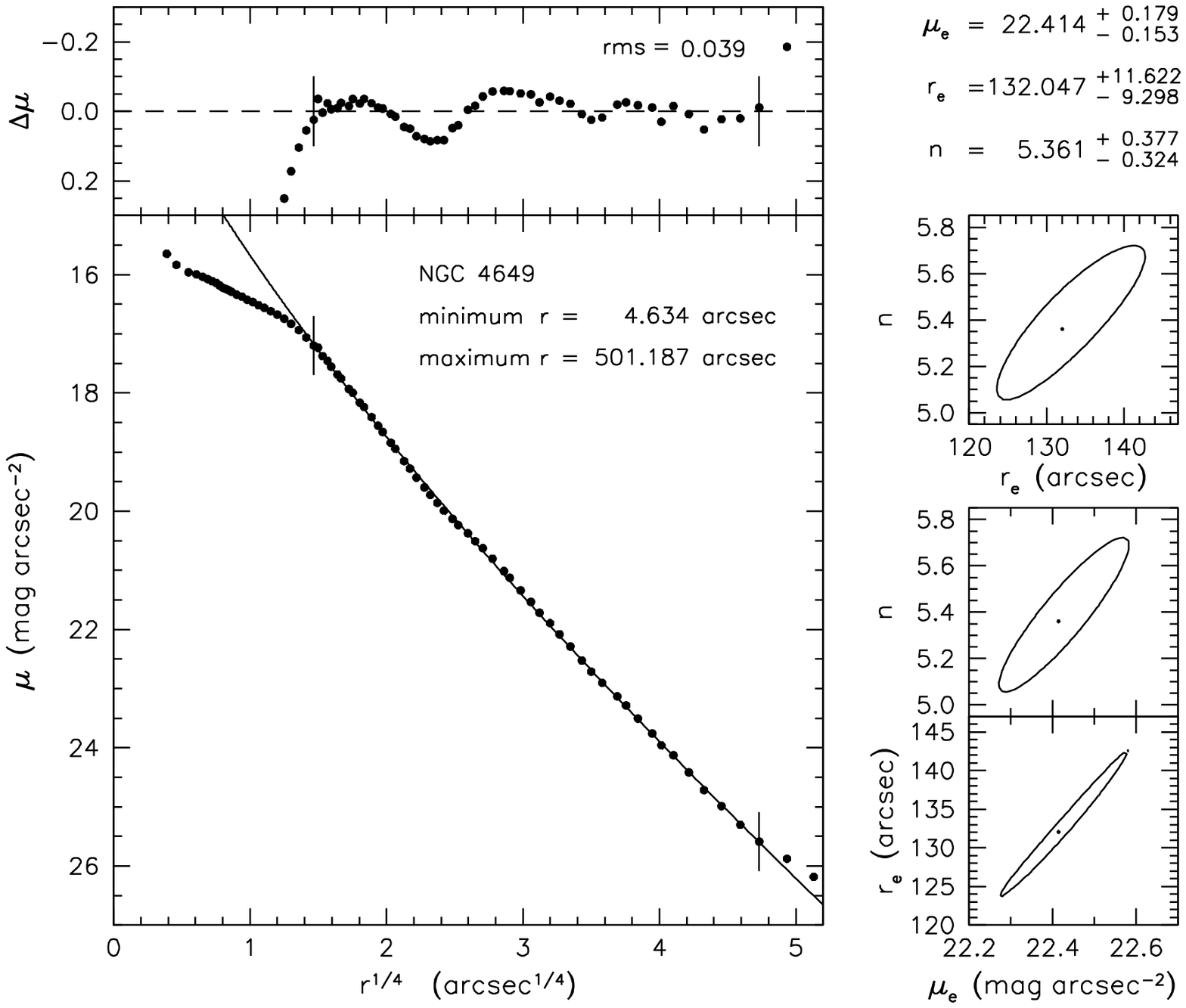}

\includegraphics{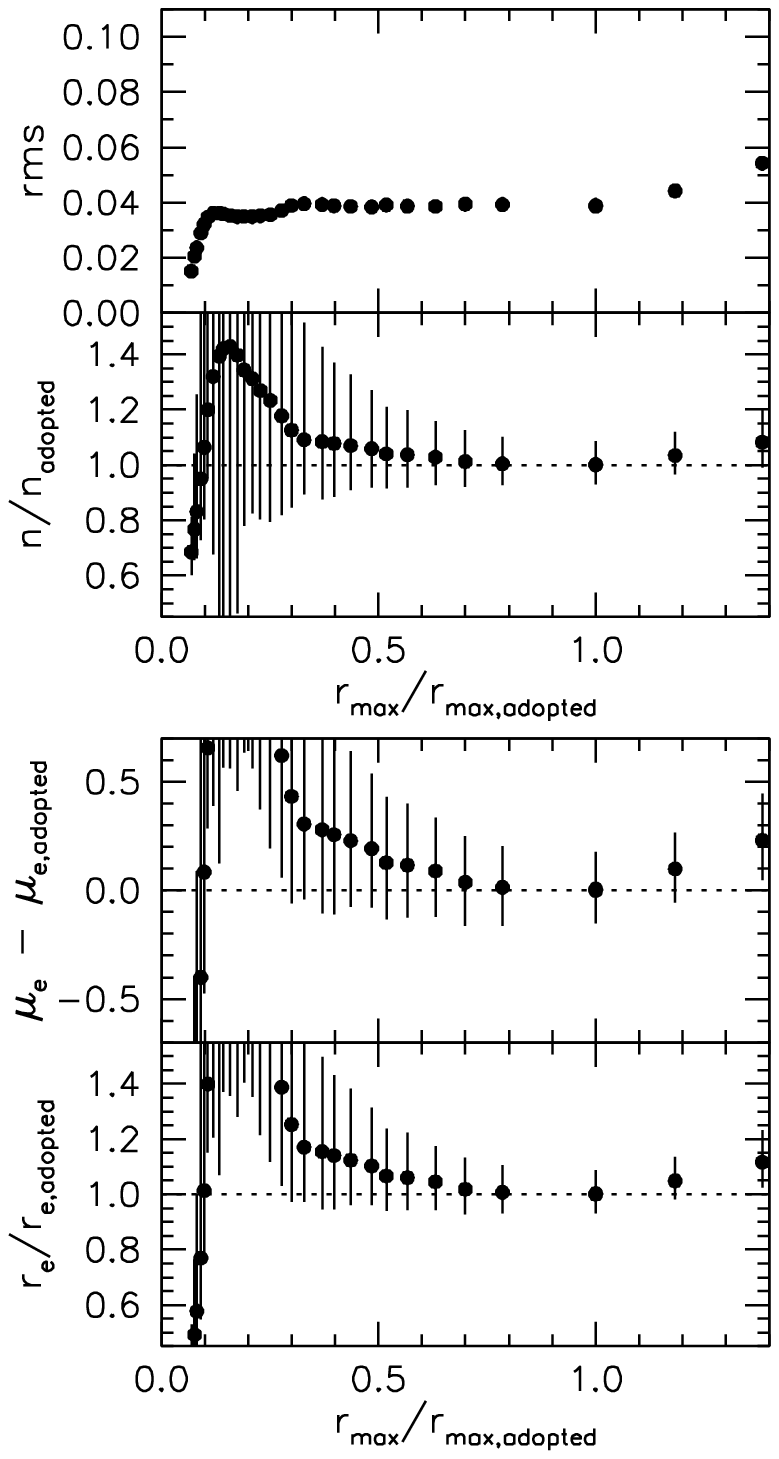}

\includegraphics{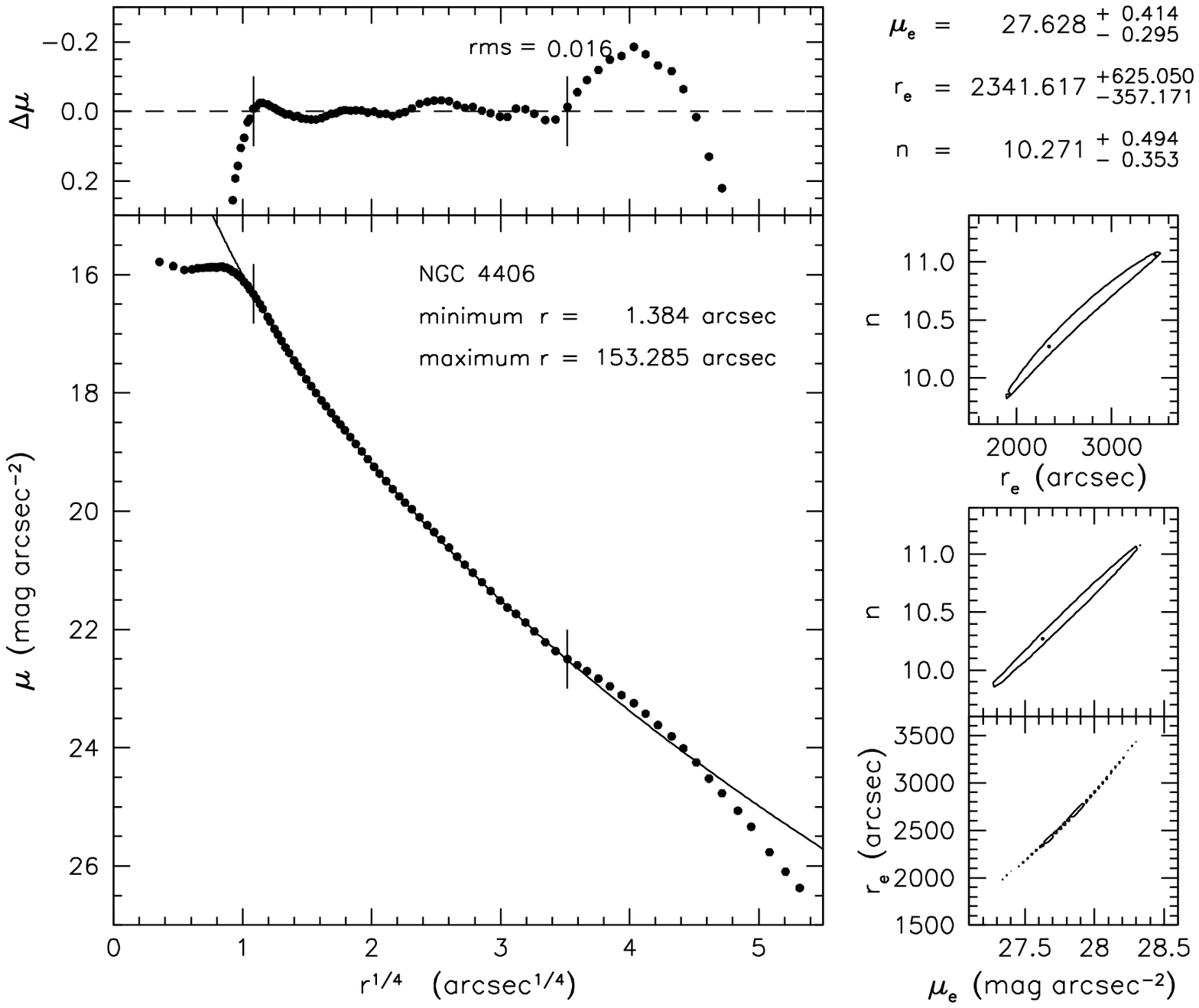}

\includegraphics{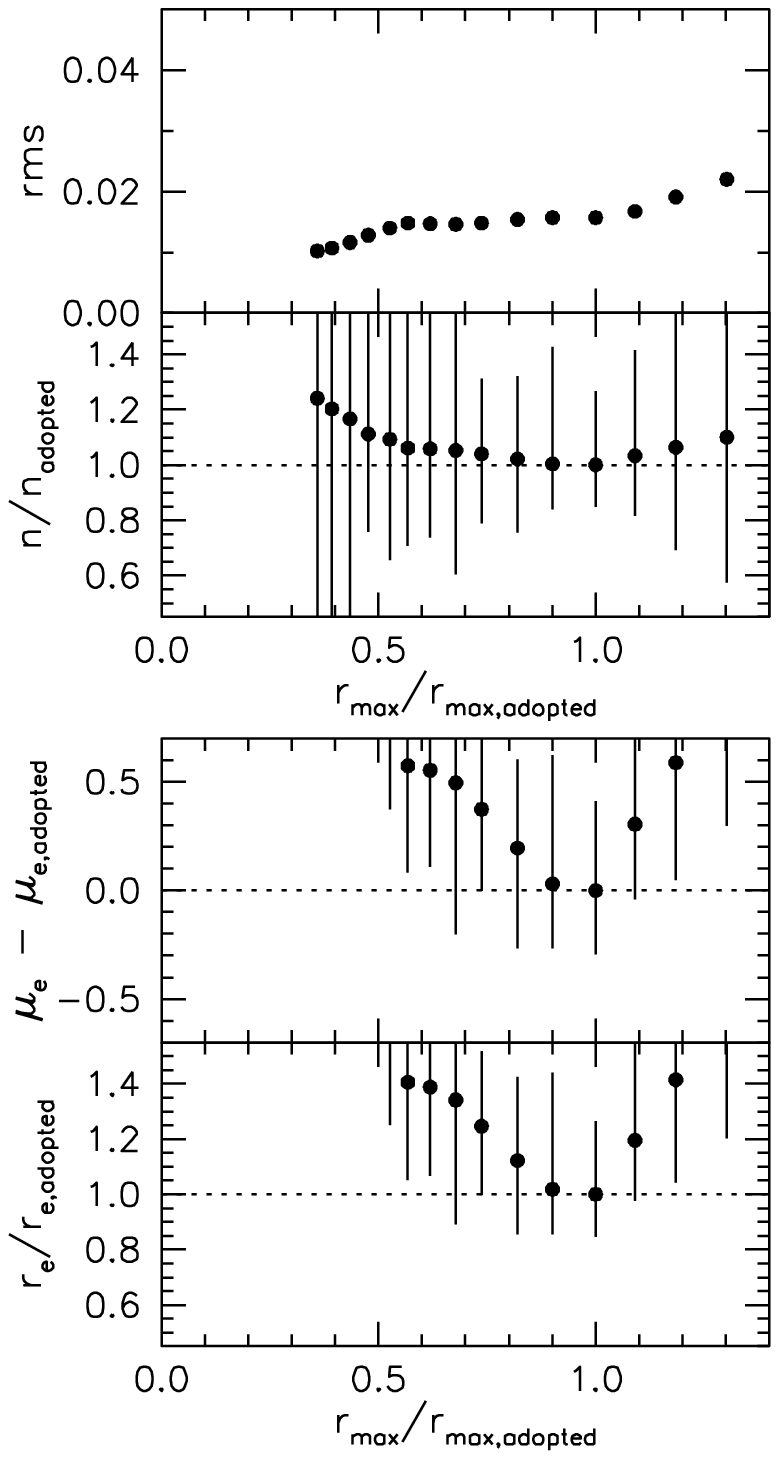}

\figcaption[]
{S\'ersic function fits to the major-axis profiles of NGC 4649 and
NGC 4406.  The layout is the same as in Figure 49.  Note the extraordinarily strong 
parameter coupling in the NGC 4406 fit.  This is characteristic of fits with large 
S\'ersic indices.}
%\end{figure*}

\eject\clearpage

\figurenum{52}

\centerline{\null} 
\vfill

\includegraphics{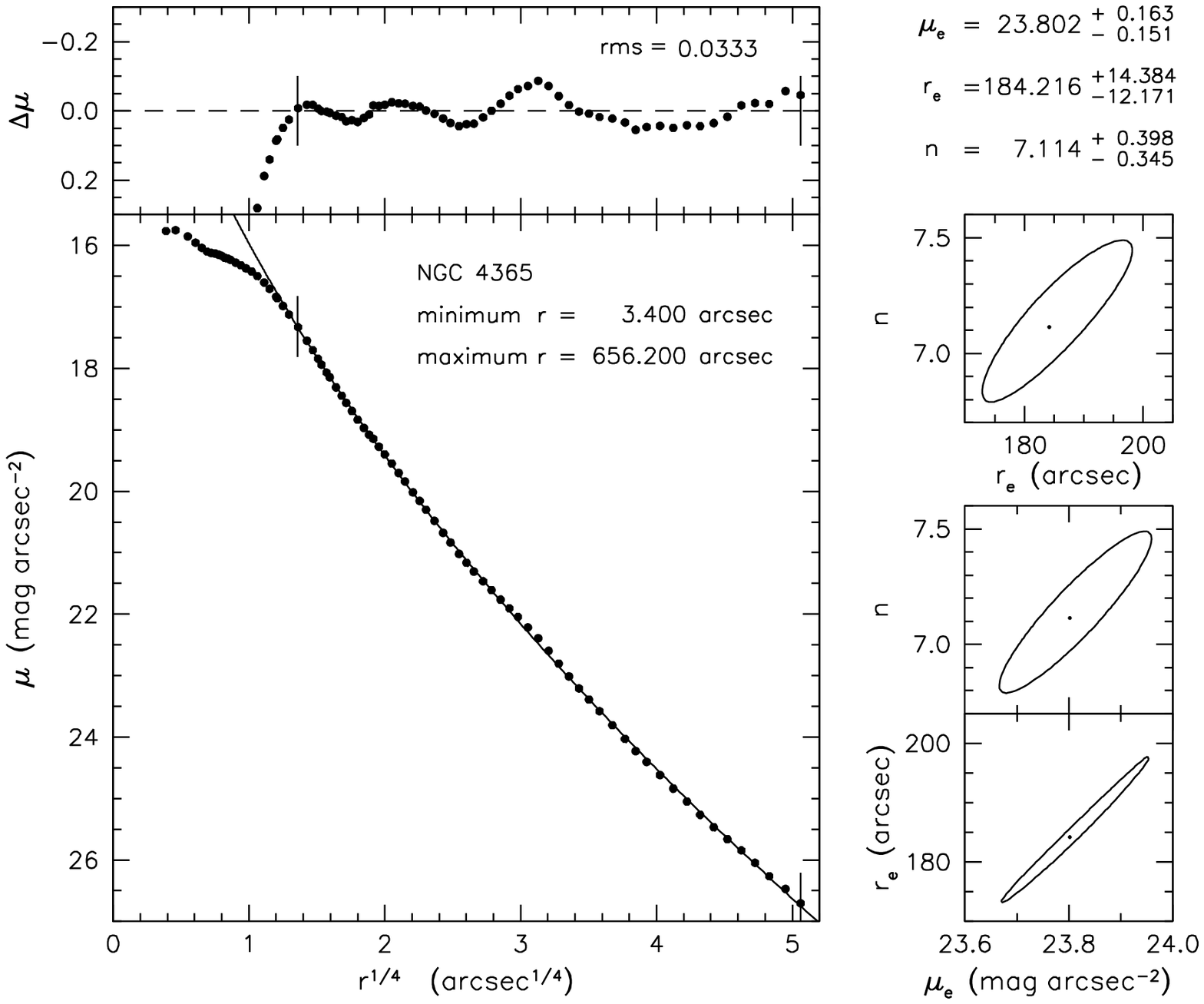}

\includegraphics{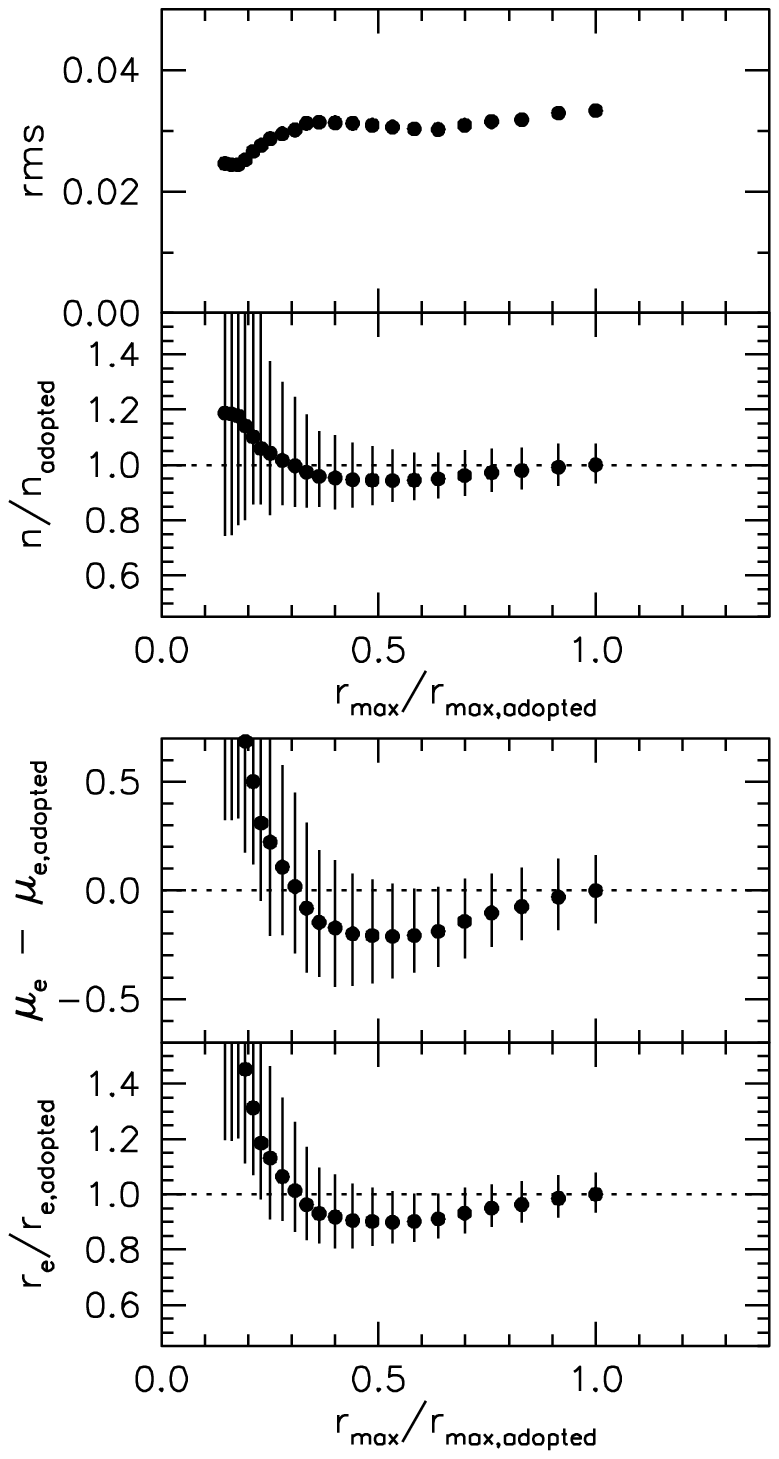}

\includegraphics{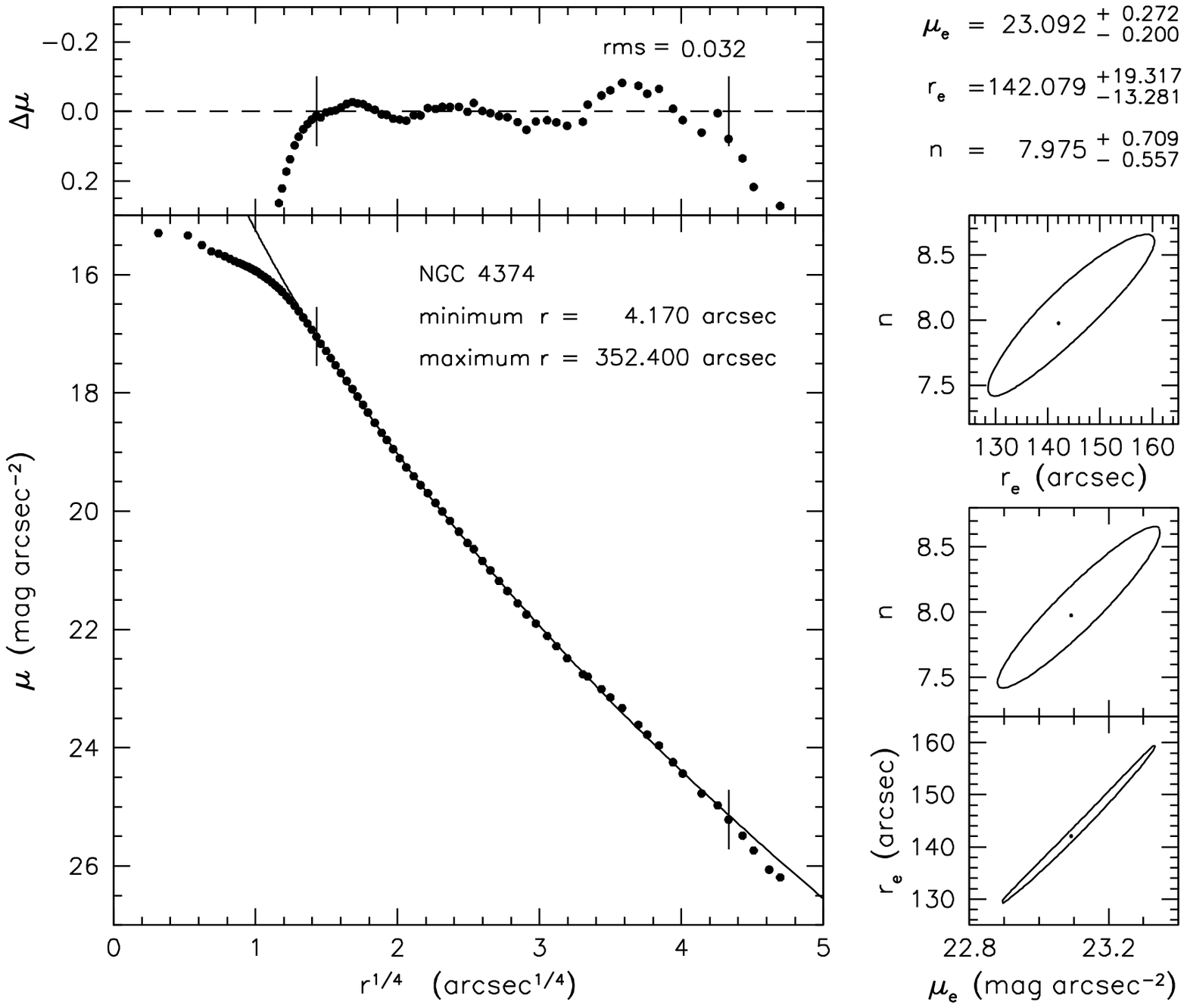}

\includegraphics{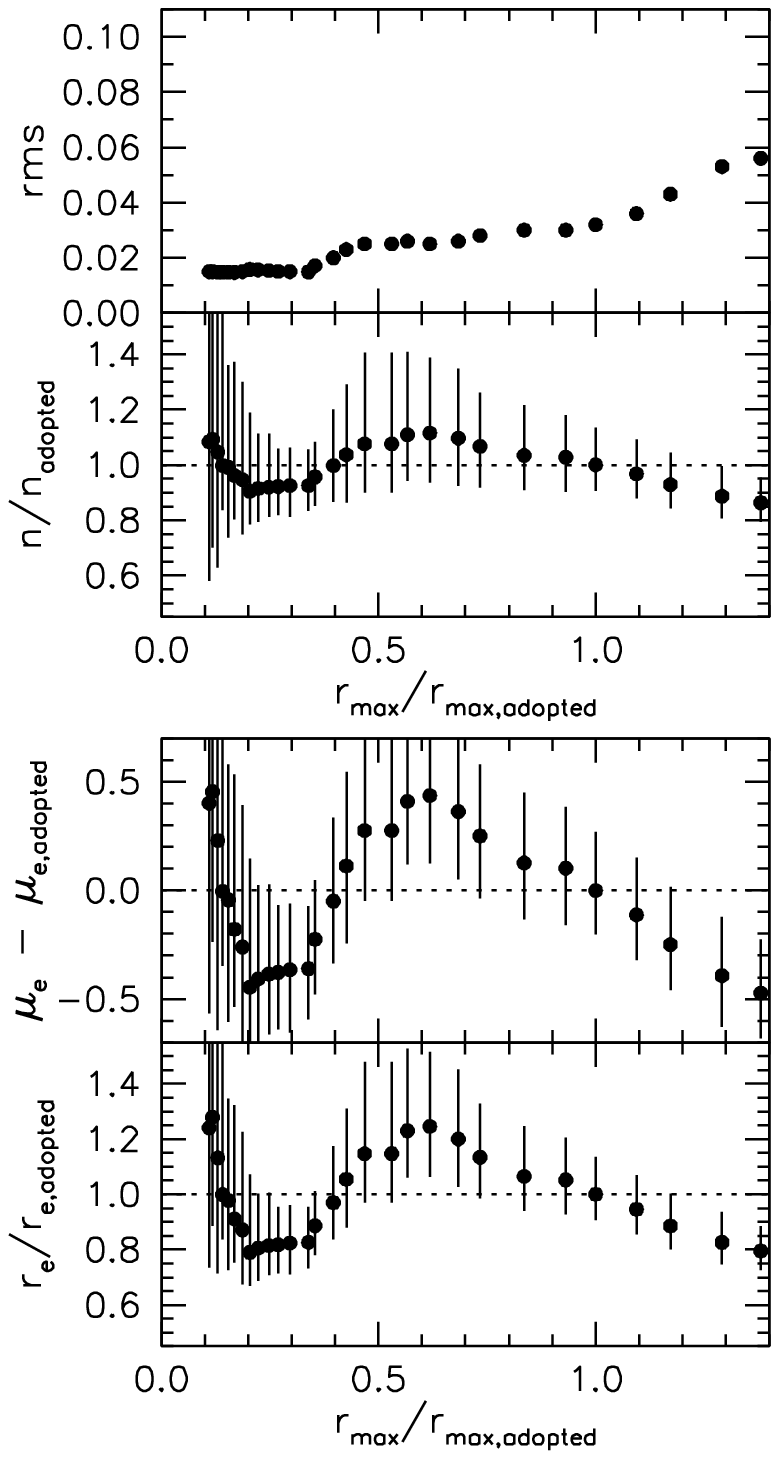}
%\begin{figure*}[b]
\figcaption[]
{S\'ersic function fit to the major-axis profiles of NGC 4365 and NGC 4374.
The figure layout is the same as in Figure 49.}

\eject\clearpage

\figurenum{53}

\centerline{\null} 
\vfill

\includegraphics{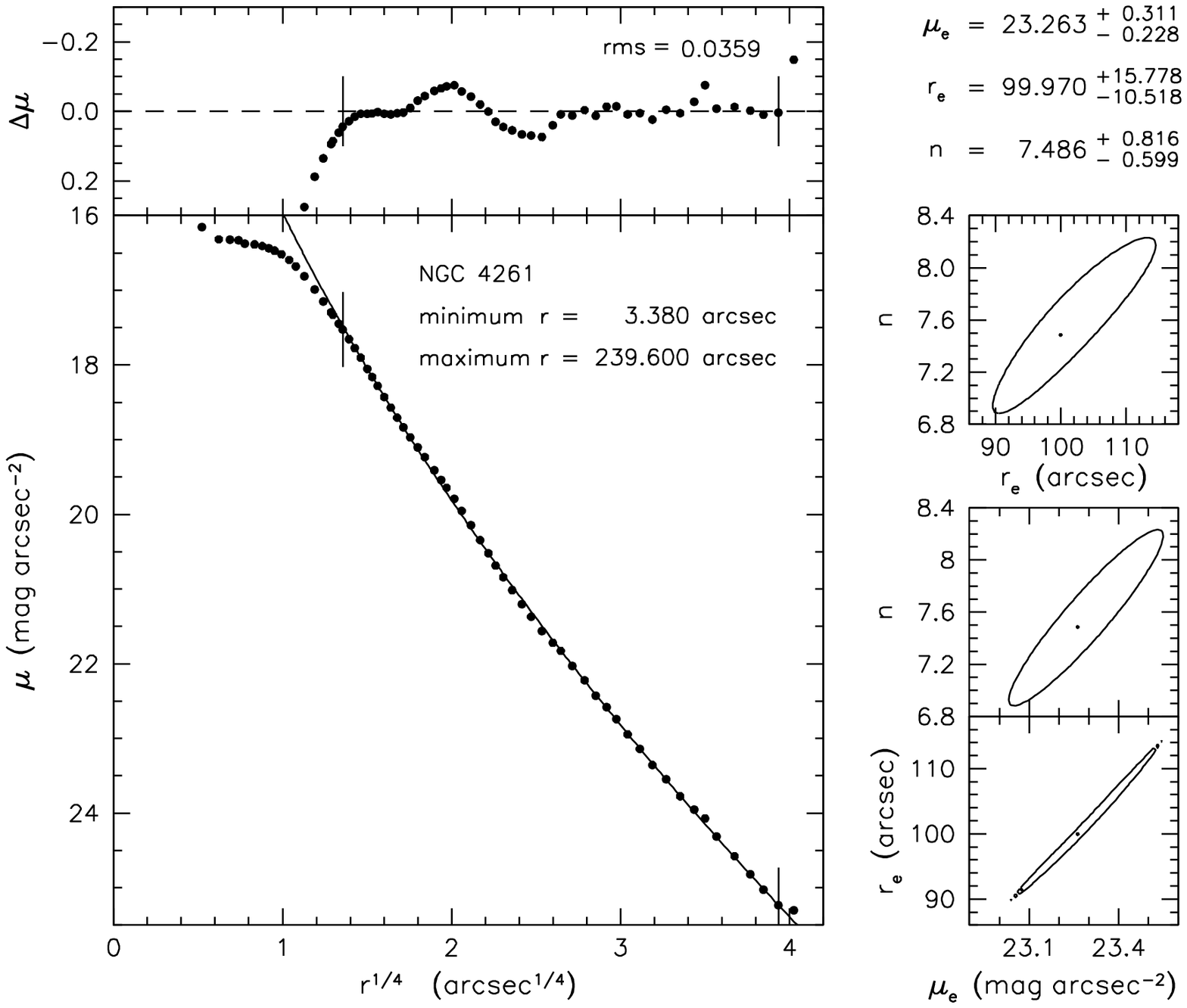}

\includegraphics{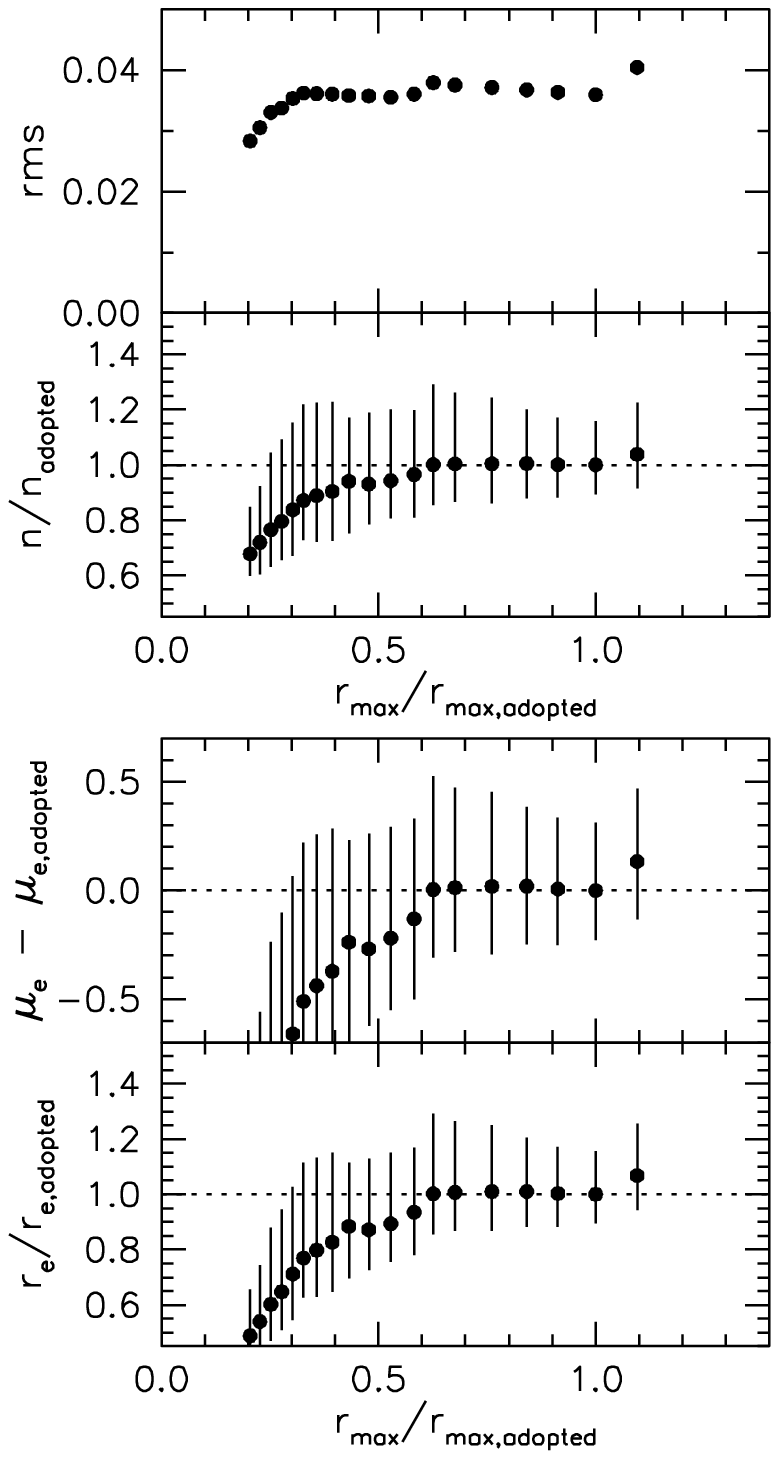}

\includegraphics{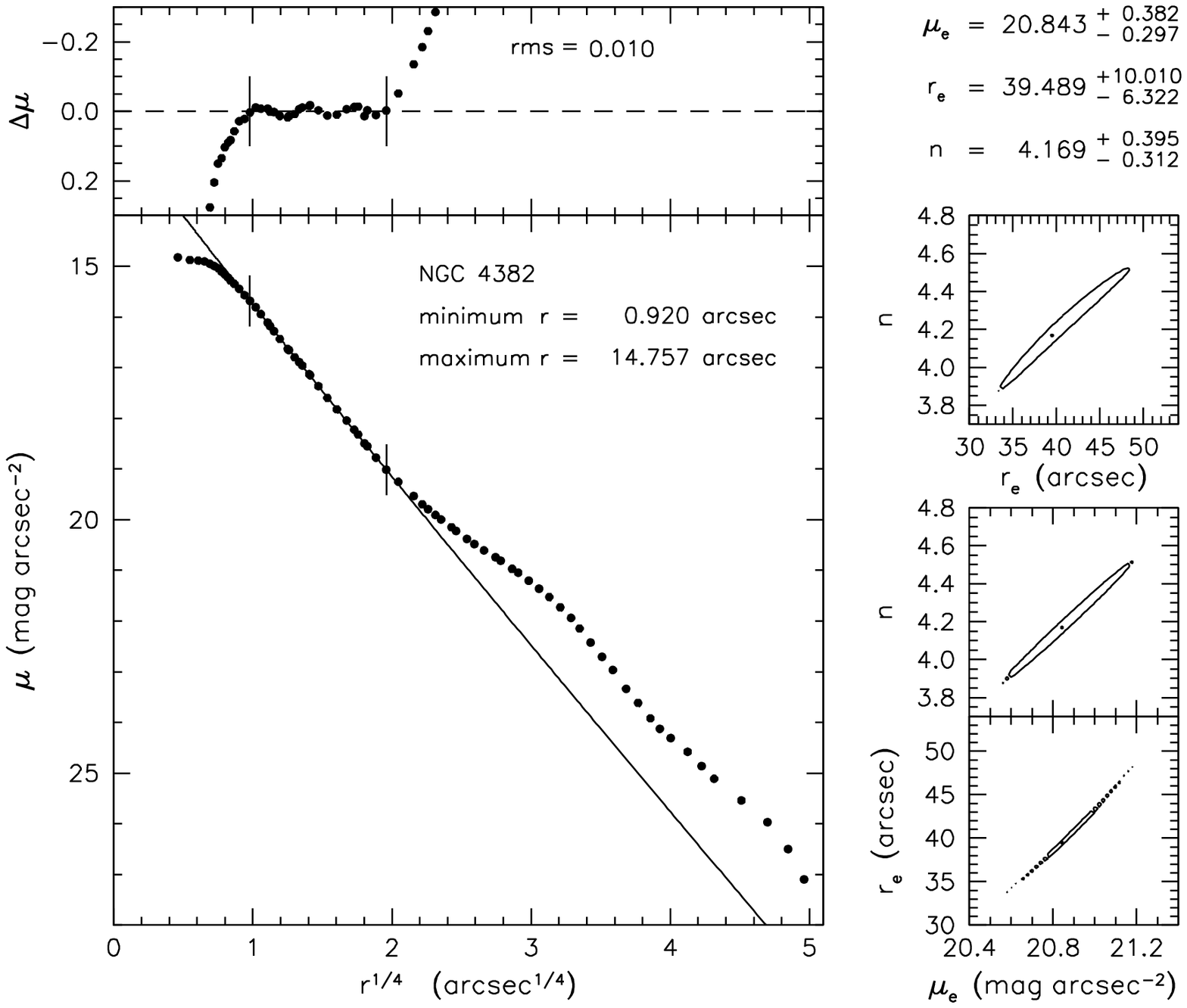}

\includegraphics{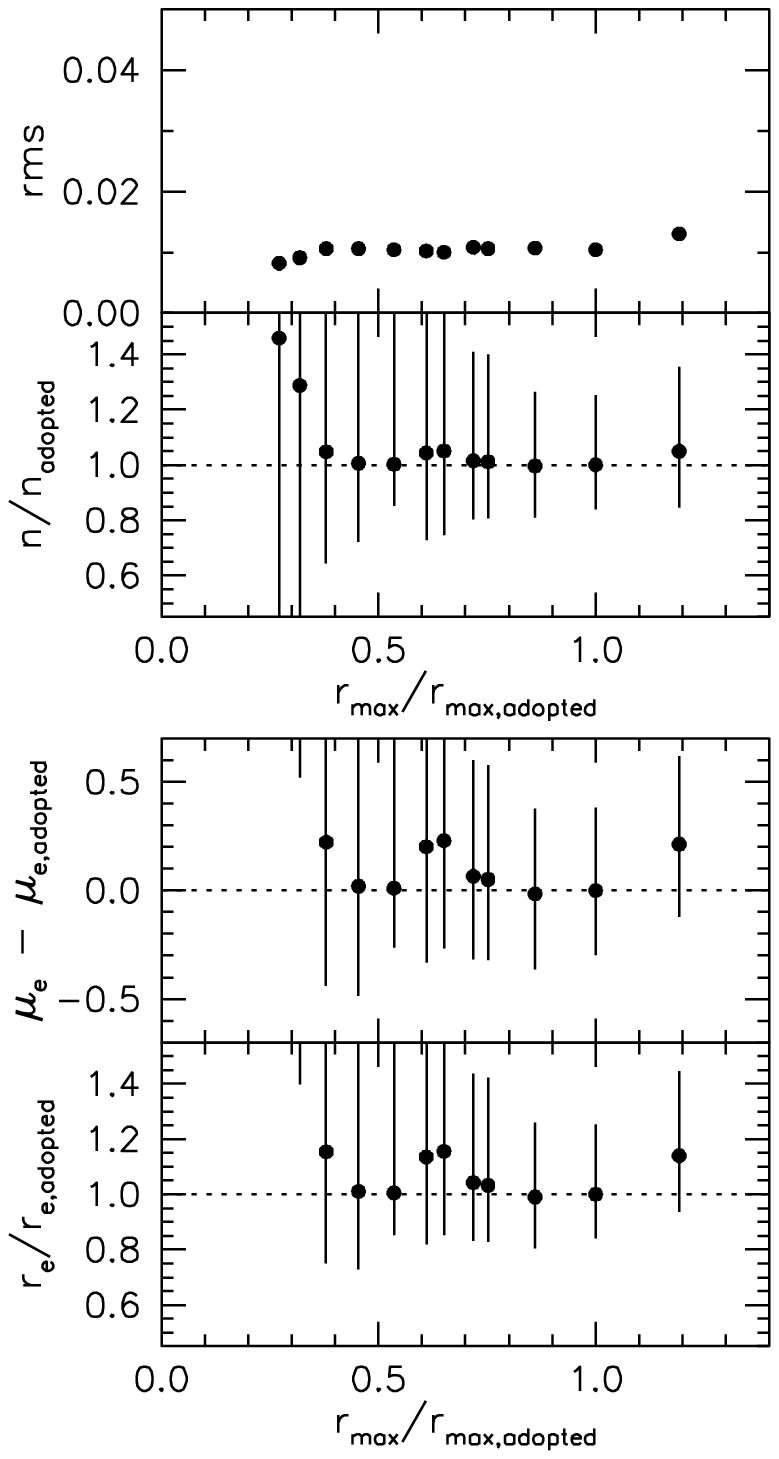}

\figcaption[]
{S\'ersic function fits to the major-axis profiles of NGC 4261 ({\it top\/})
and an illustrative fit to the inner part of the profile of NGC 4382 ({\it bottom\/}).
The adopted fit to the profile of NGC 4382 is shown on the next page.
The layout is as in Figure 49.  Note that NGC 4261 is in the background of the Virgo
Cluster (see distances in Table 1).}

\eject\clearpage

\figurenum{54}

\centerline{\null} 
\vfill

\includegraphics{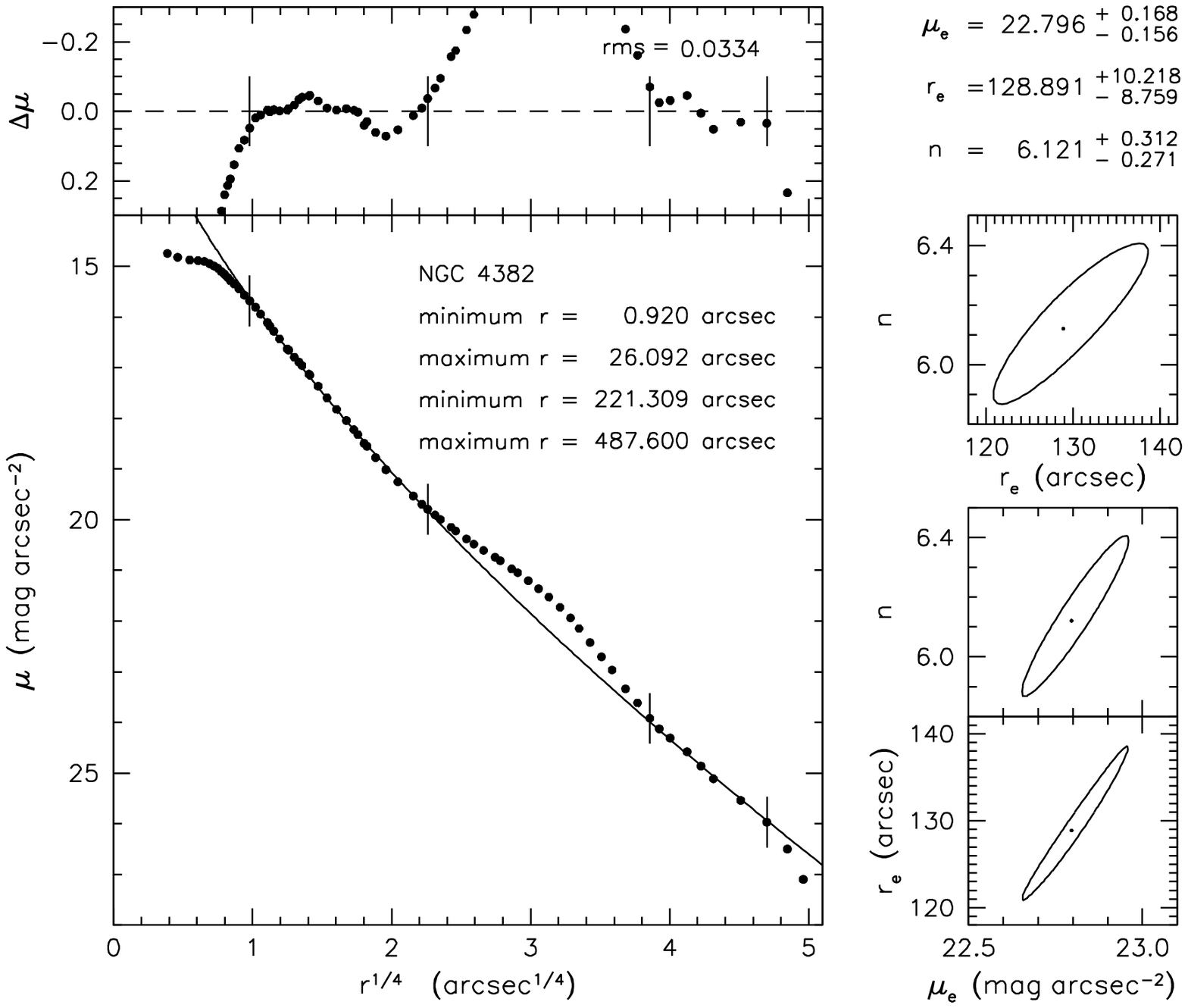}

\includegraphics{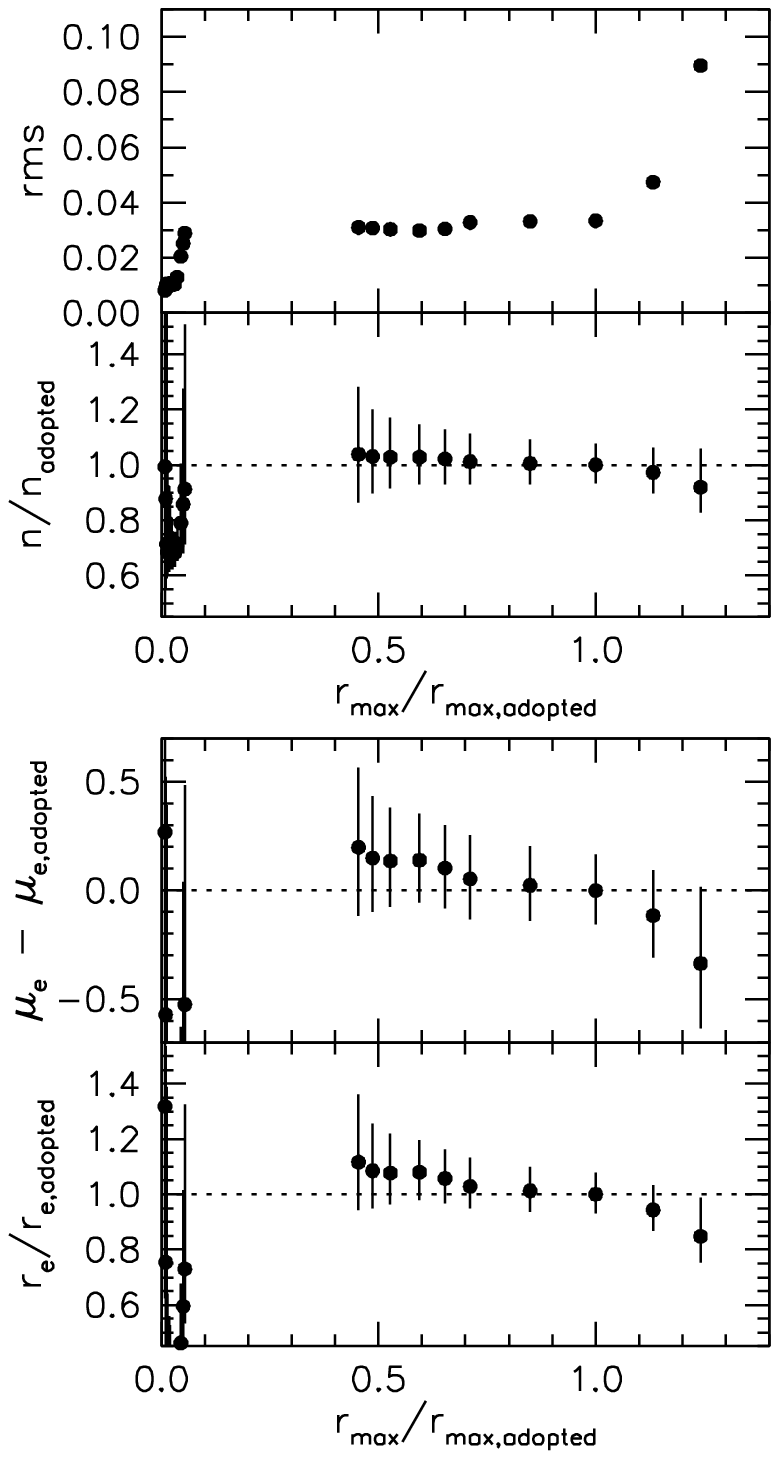}

\includegraphics{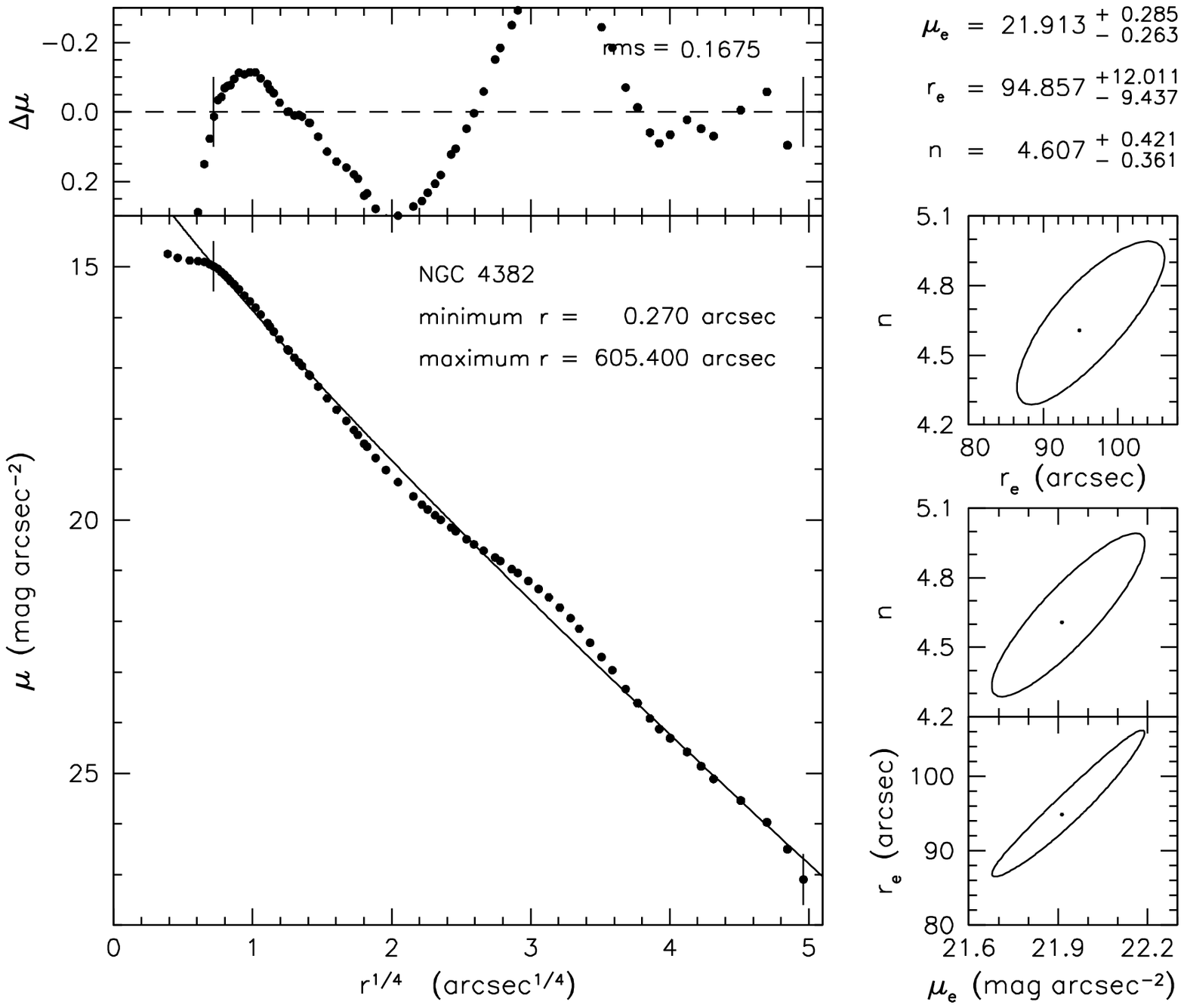}

\includegraphics{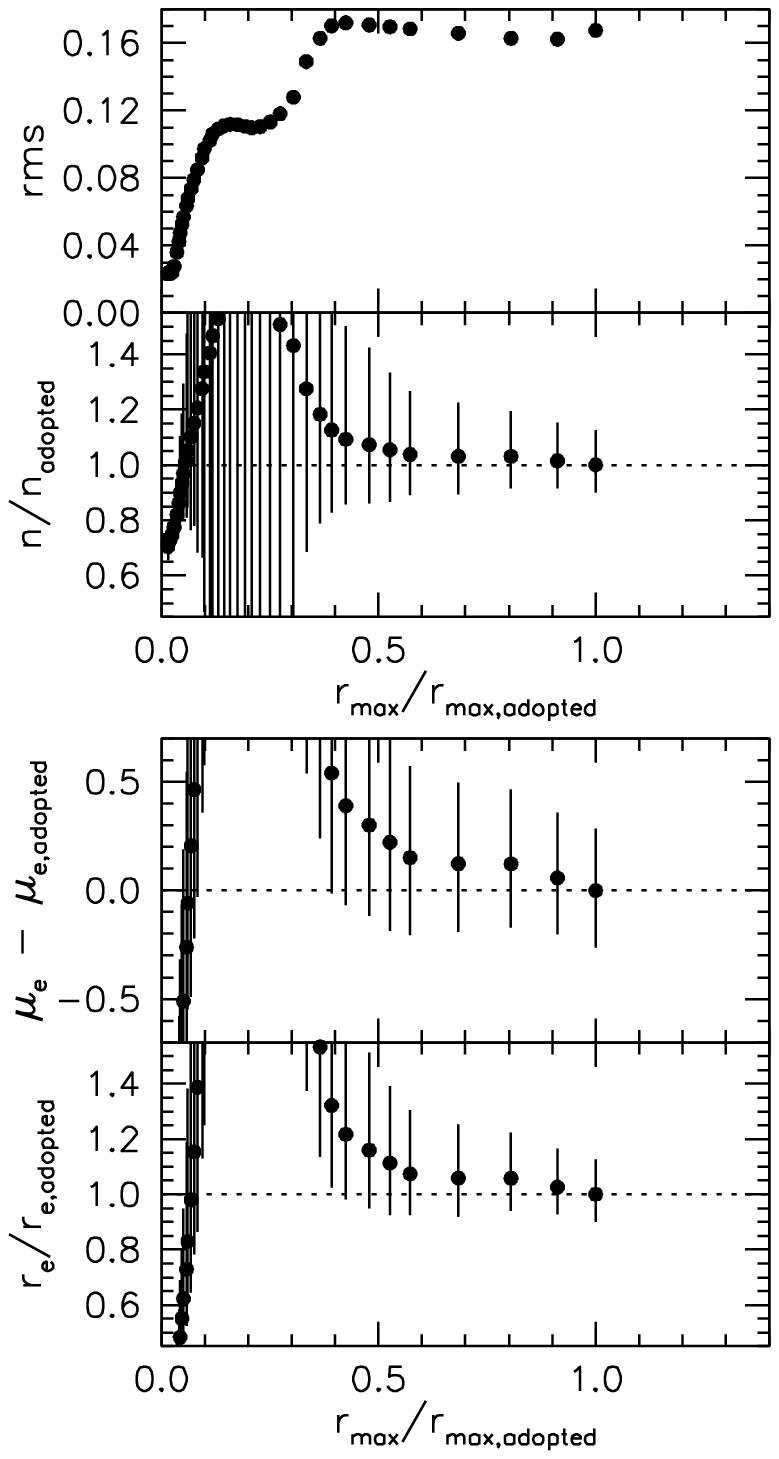}

\figcaption[]
{Alternative S\'ersic function fits to the major-axis profile of NGC 4382.  
The layout is as in Figure 49.  The top panels show a fit to the inner and outer profile 
omitting intermediate points between 28$^{\prime\prime}$ and 202$^{\prime\prime}$ 
inclusive.  This is the adopted fit whose parameters are listed in Table 1.  The 
bottom panels show an overall fit, giving triple weight to the points  at 
$202^{\prime\prime} \leq r \leq 552^{\prime\prime}$ to ensure a good fit at large
radii.
}

\eject\clearpage

\figurenum{55}

\centerline{\null} 
\vfill
 
\includegraphics{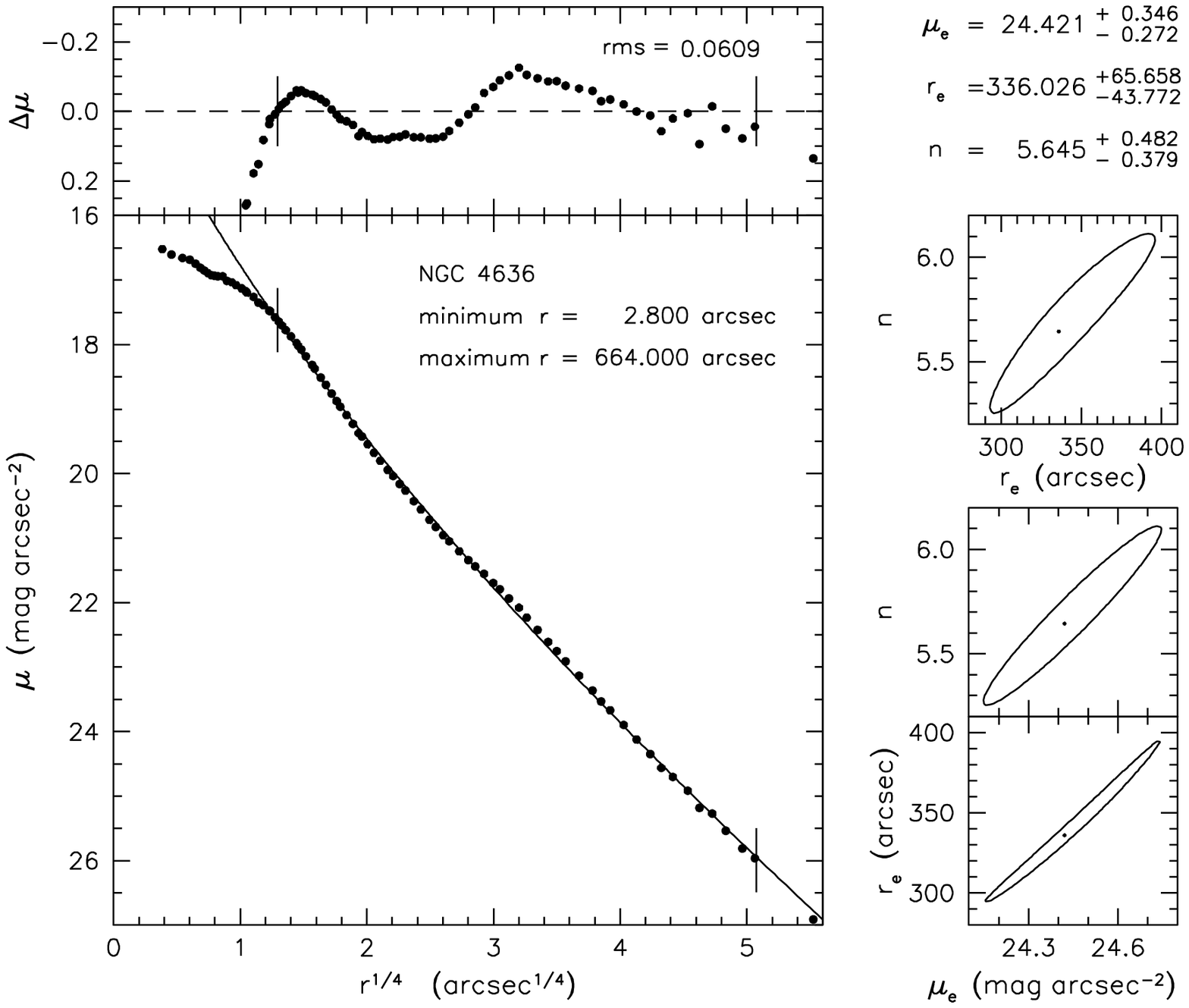}

\includegraphics{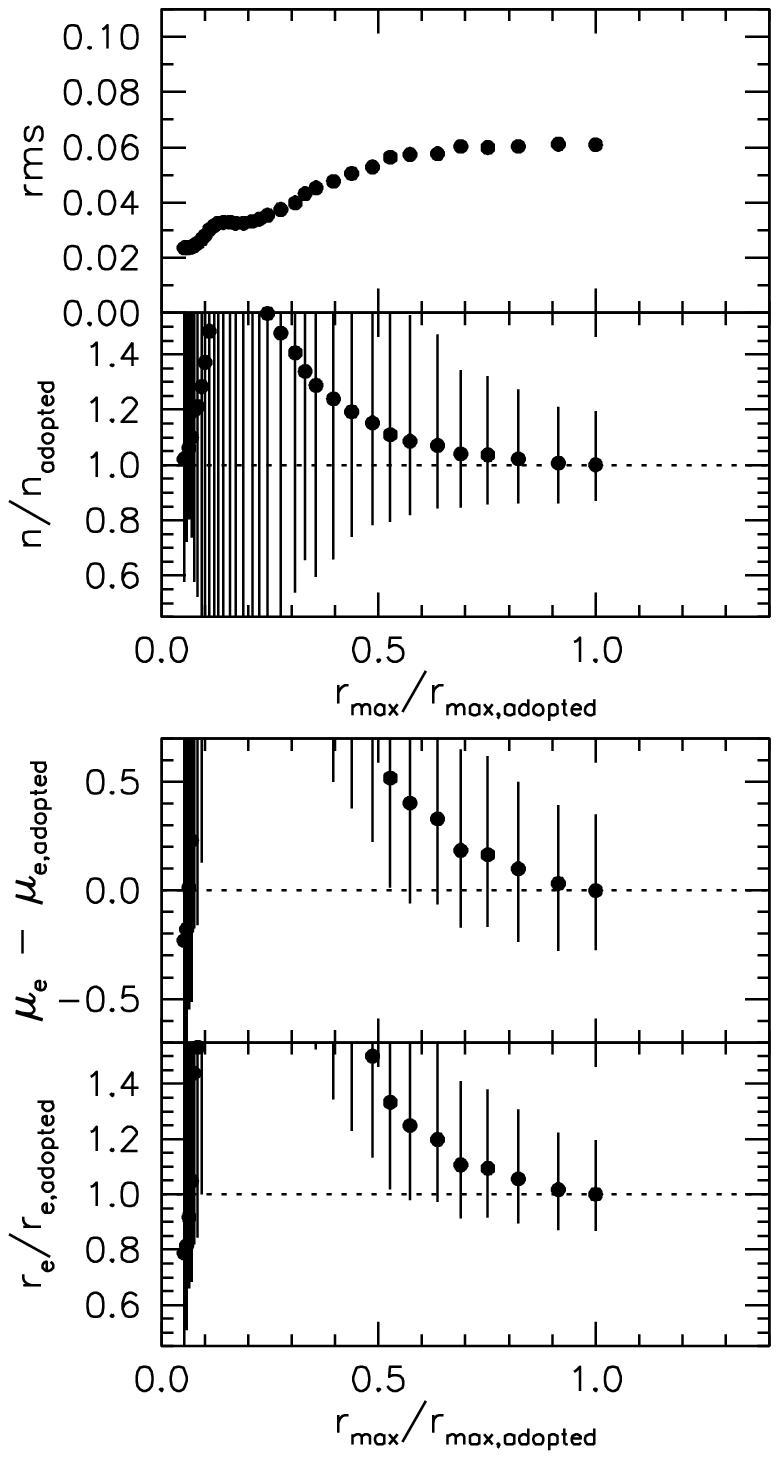}

\includegraphics{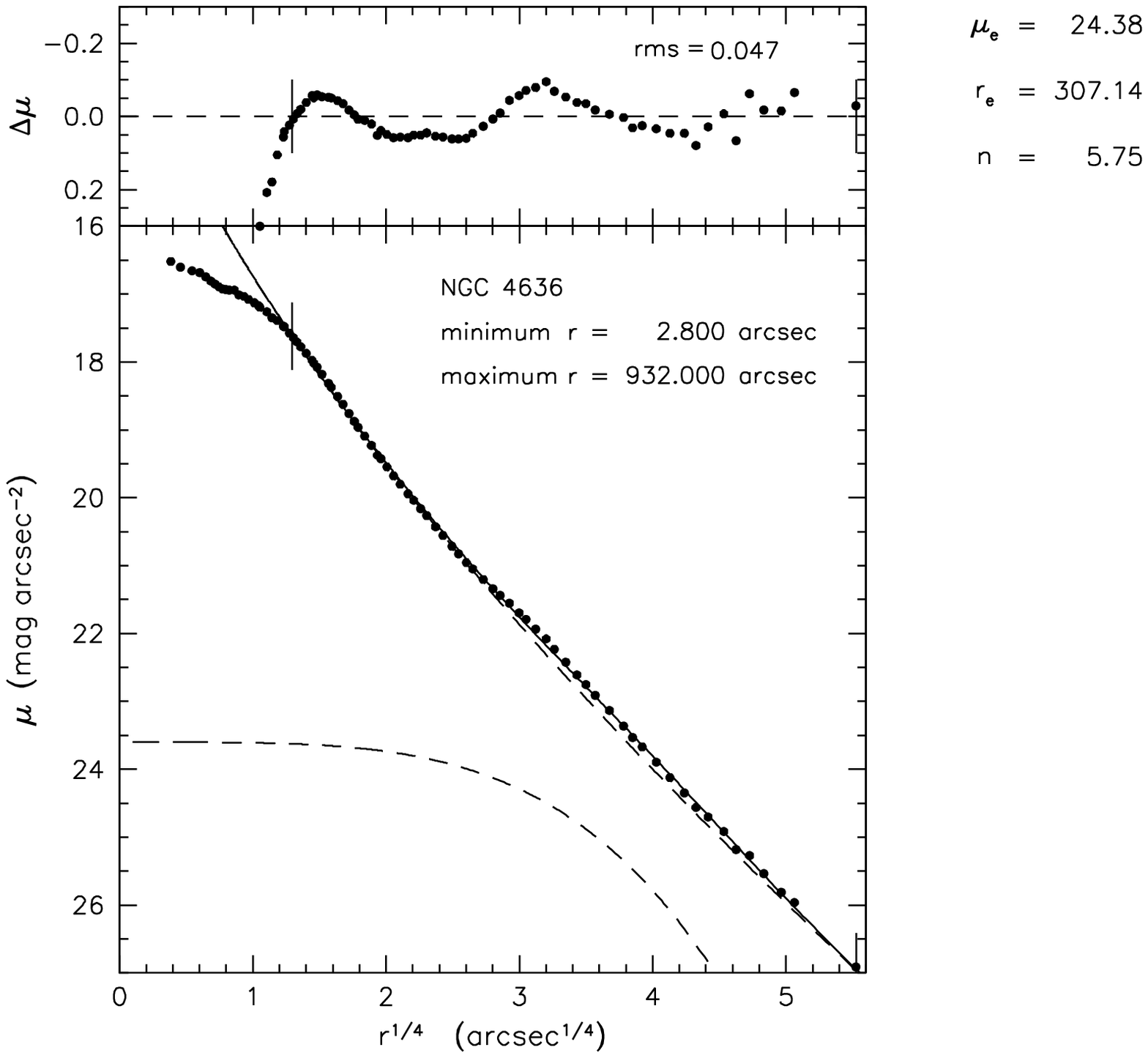}

\figcaption[]
{S\'ersic function fits to the major-axis profile of NGC 4636.  The layout is as 
in Figure 49.  The adopted fit is at the top.  The RMS residual is slighly larger than
normal, mostly because of a profile wiggle that is centered at $r^{1/4} \simeq 2.8$.  The
form of the wiggle (the model is too bright just inside the above radius and too faint 
just outside this radius) suggests the possibility that NGC 4636 may be a 
bulge-dominated S0, i.{\thinspace}e., a face-on version of NGC 3115 (Hamabe 1982, Fig.~5a).
Therefore, the bottom panels show a decomposition into a S\'ersic function bulge plus an 
exponential ``disk'' represented by the upper and lower dashed curves, respectively. 
Their sum is the solid curve.  It fits the observed profile marginally better than does 
the adopted pure S\'ersic fit, but the difference is not significant.  In particular, 
the wiggle in the residual profile is not much reduced by the decomposition, because it 
happens over a smaller radius range than the exponential can accommodate.  Thus there is no 
compelling evidence that NGC 4636 is an S0.  In any case, the ``disk'' in the lower fit 
contributes only 8\thinspace\% of the total light, so the bulge parameters given by the 
decomposition are almost the same as those given by the adopted fit (see the keys). 
}

\eject\clearpage

\figurenum{56}

\centerline{\null} 
\vfill
 
\includegraphics{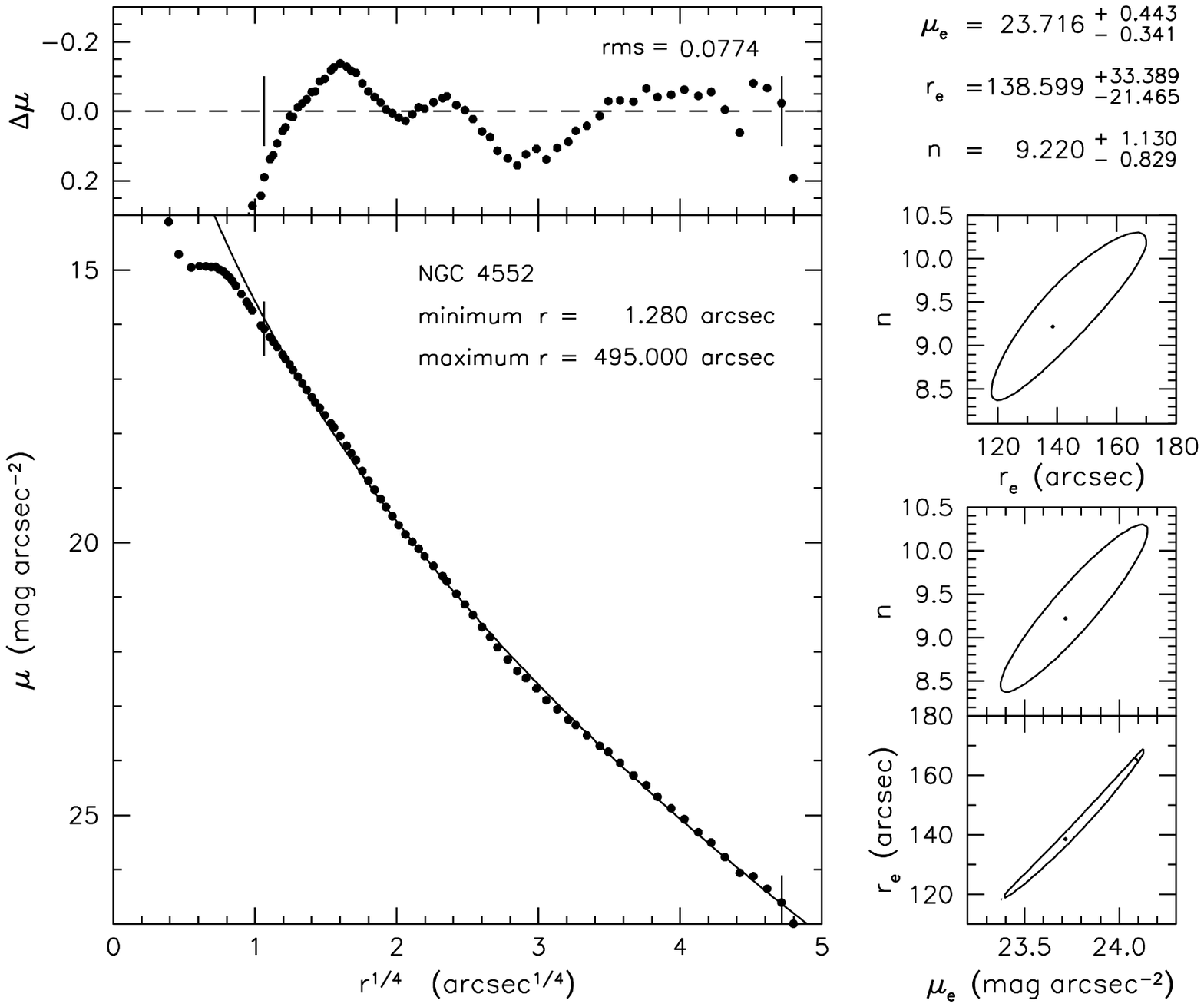}

\includegraphics{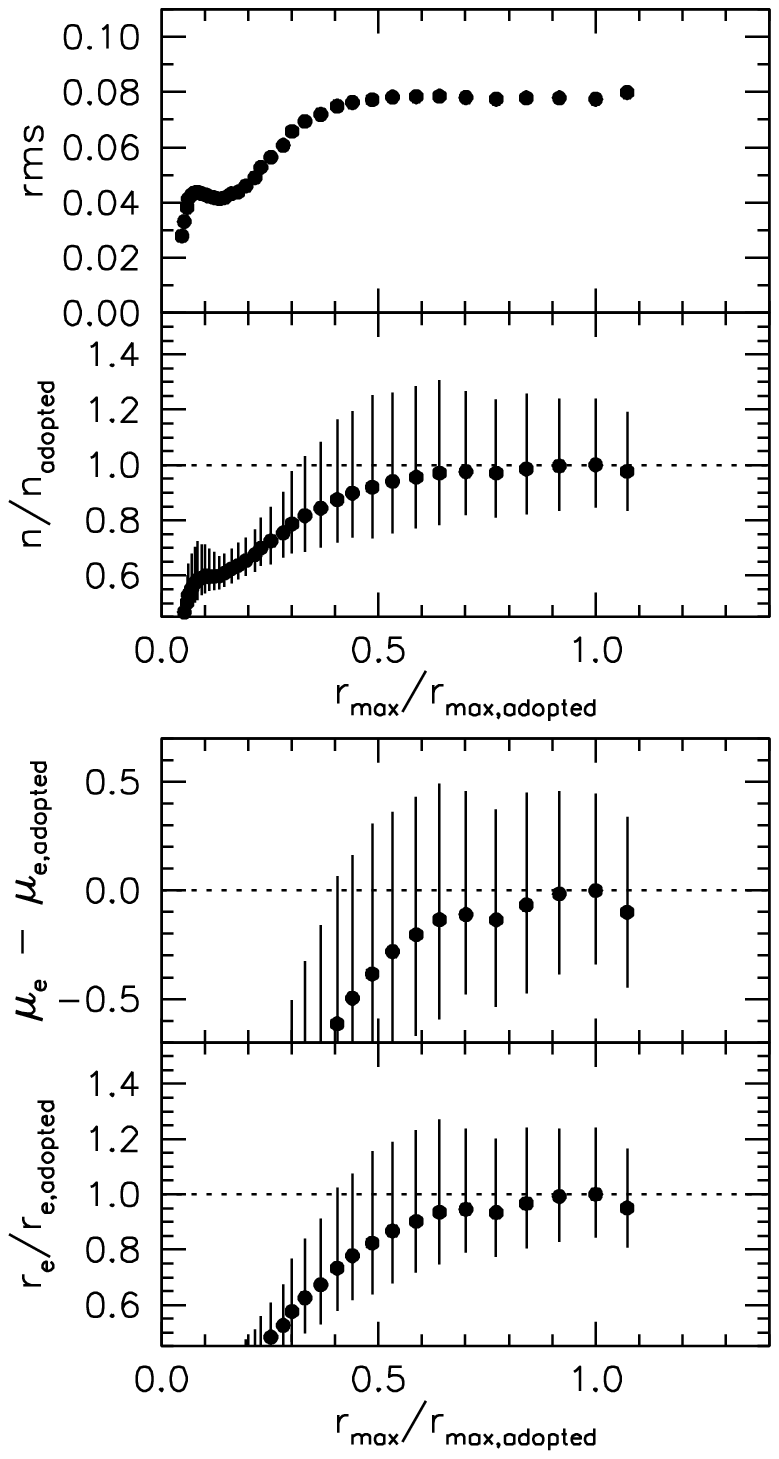}

\includegraphics{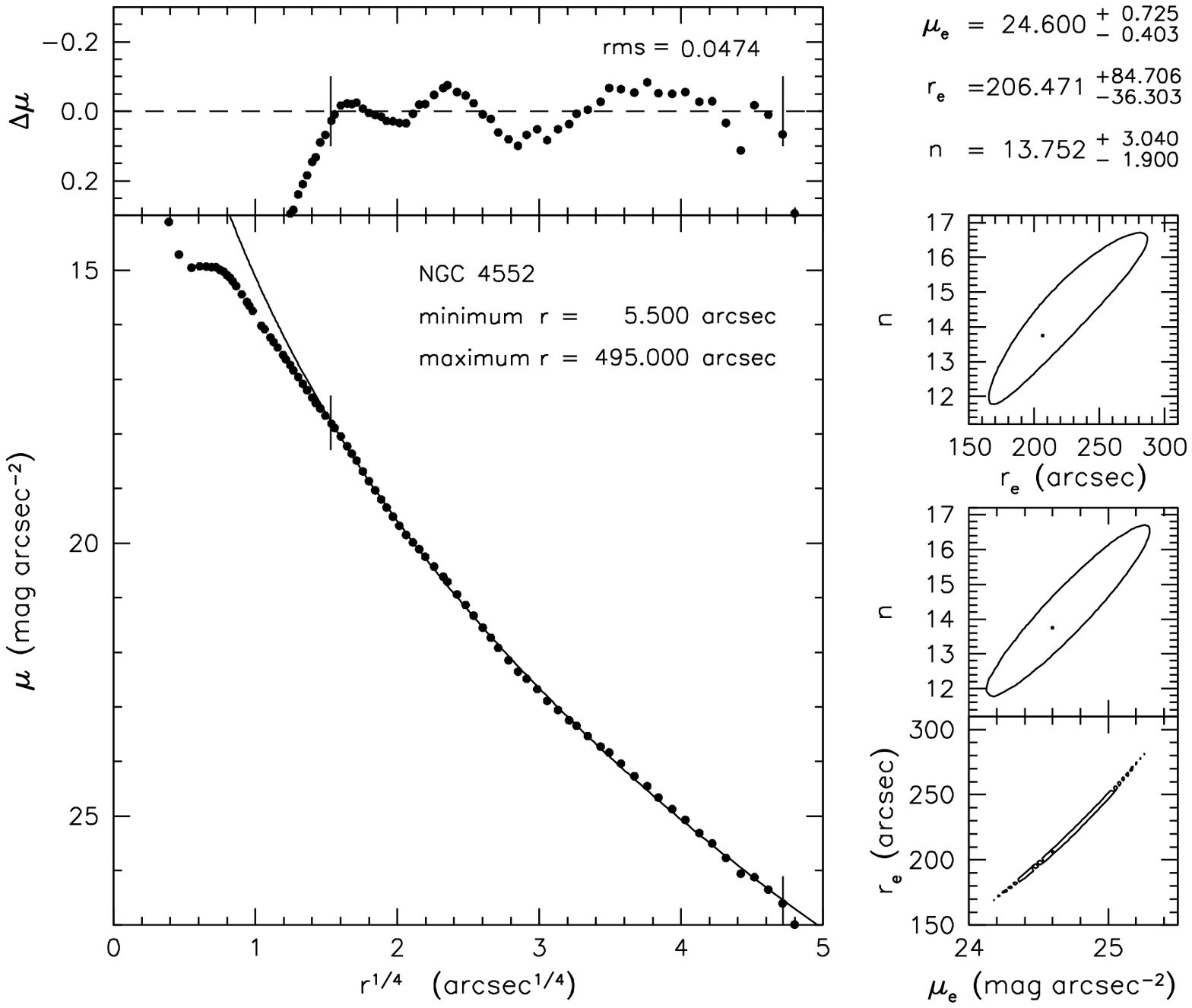}

\figcaption[]
{S\'ersic function fits to the major-axis profile of NGC 4552.  The layout is as 
in Figure 49.  The adopted fit (top) has a higher-than-normal RMS residual and a slightly
concave-upward residual profile.  It is possible that too much of the core region was 
included in the fit.  Therefore, the bottom fit uses a restricted radius range; it results
in smaller and non-systematic residuals.  The resulting core-within-a-core structure is
intriguing but highly unusual.  This fit may be an overinterpretation of the profile wiggles.  
We therefore adopt the top fit.  The bottom fit is discussed in \S{\thinspace}A3 and used 
in Figure 74.  Note that, at absolute magnitude $M_{VT}= -21.66$, NGC 4552 is the 
lowest-luminosity core elliptical in Virgo.}

\eject\clearpage

\figurenum{57}

\centerline{\null} 
\vfill

\includegraphics{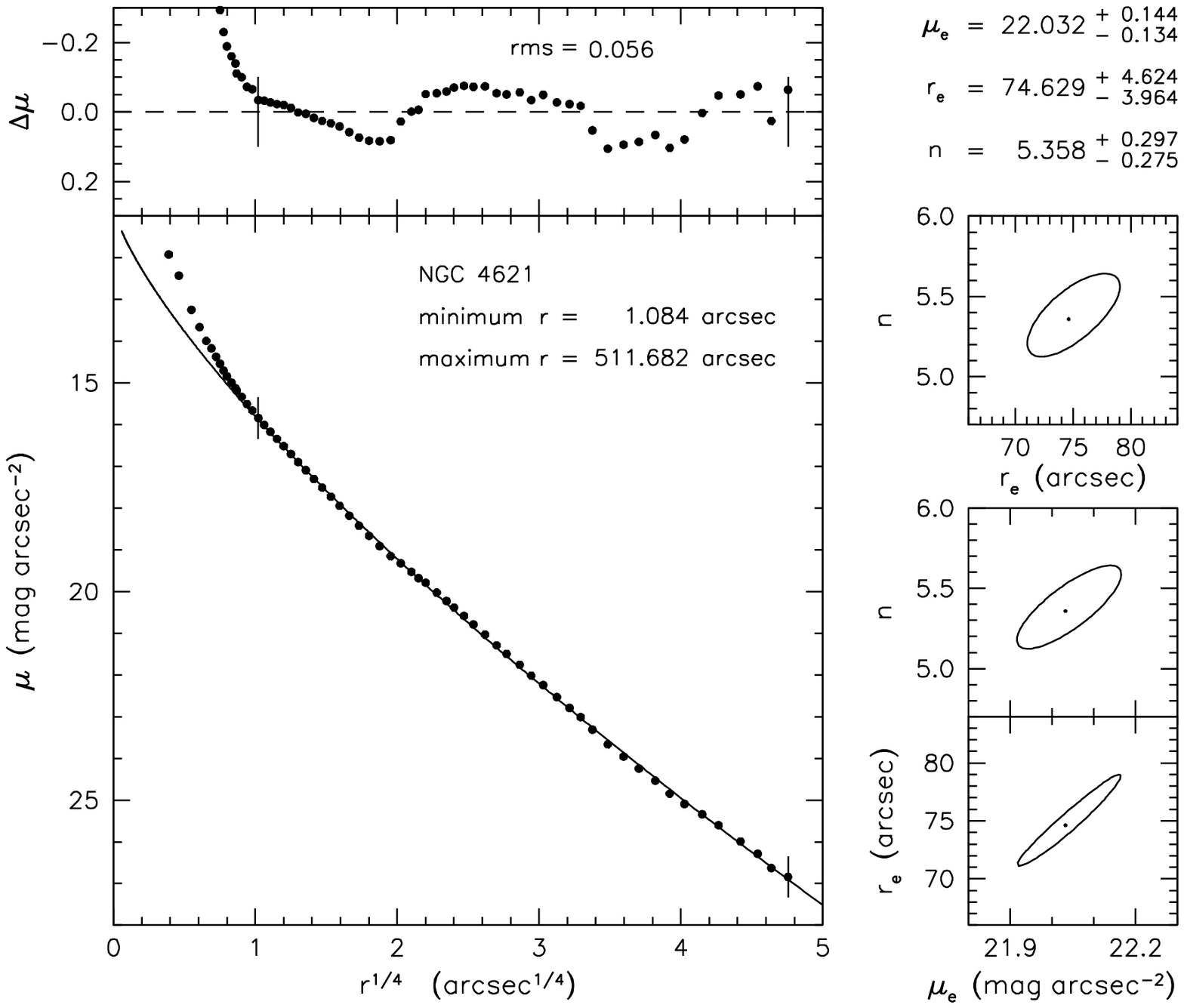}

\includegraphics{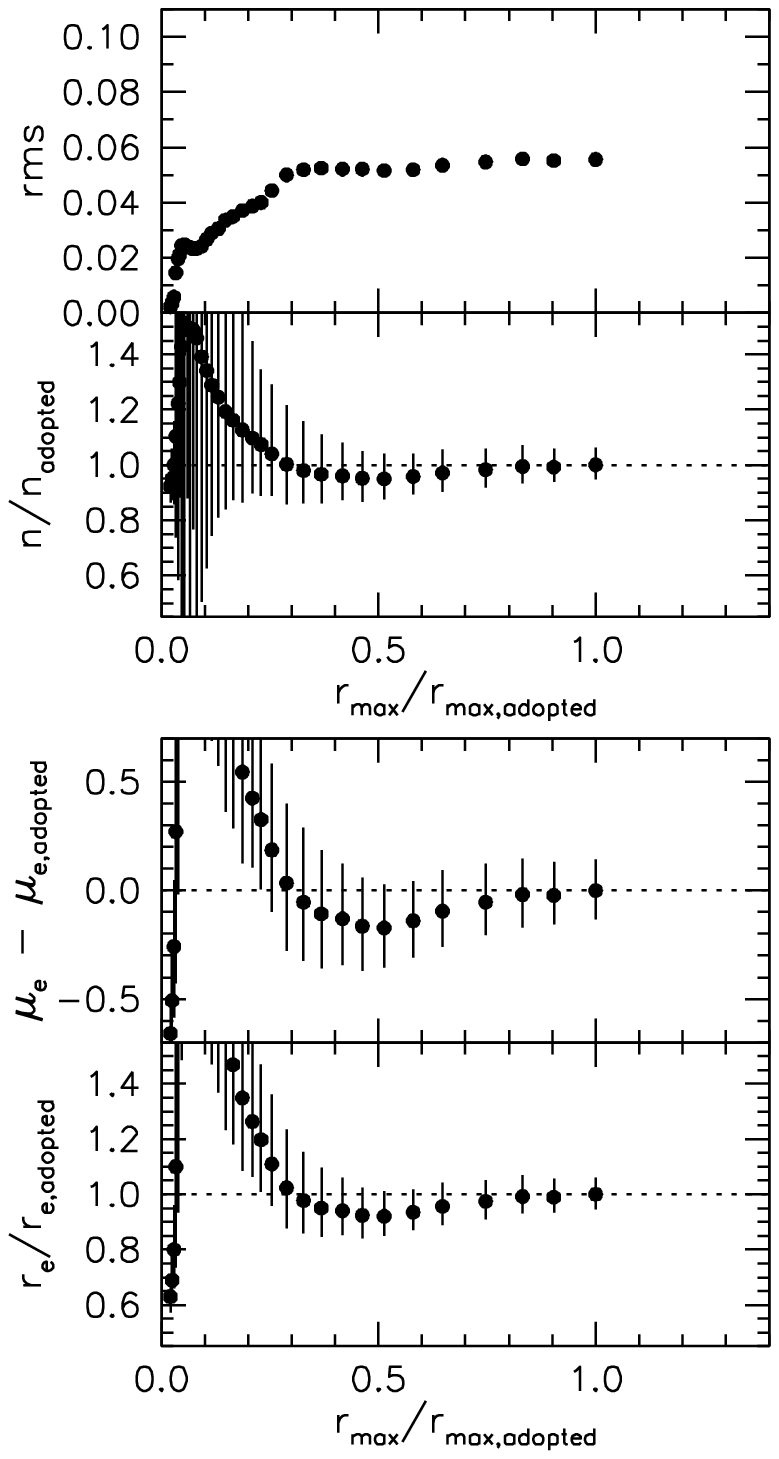}

\includegraphics{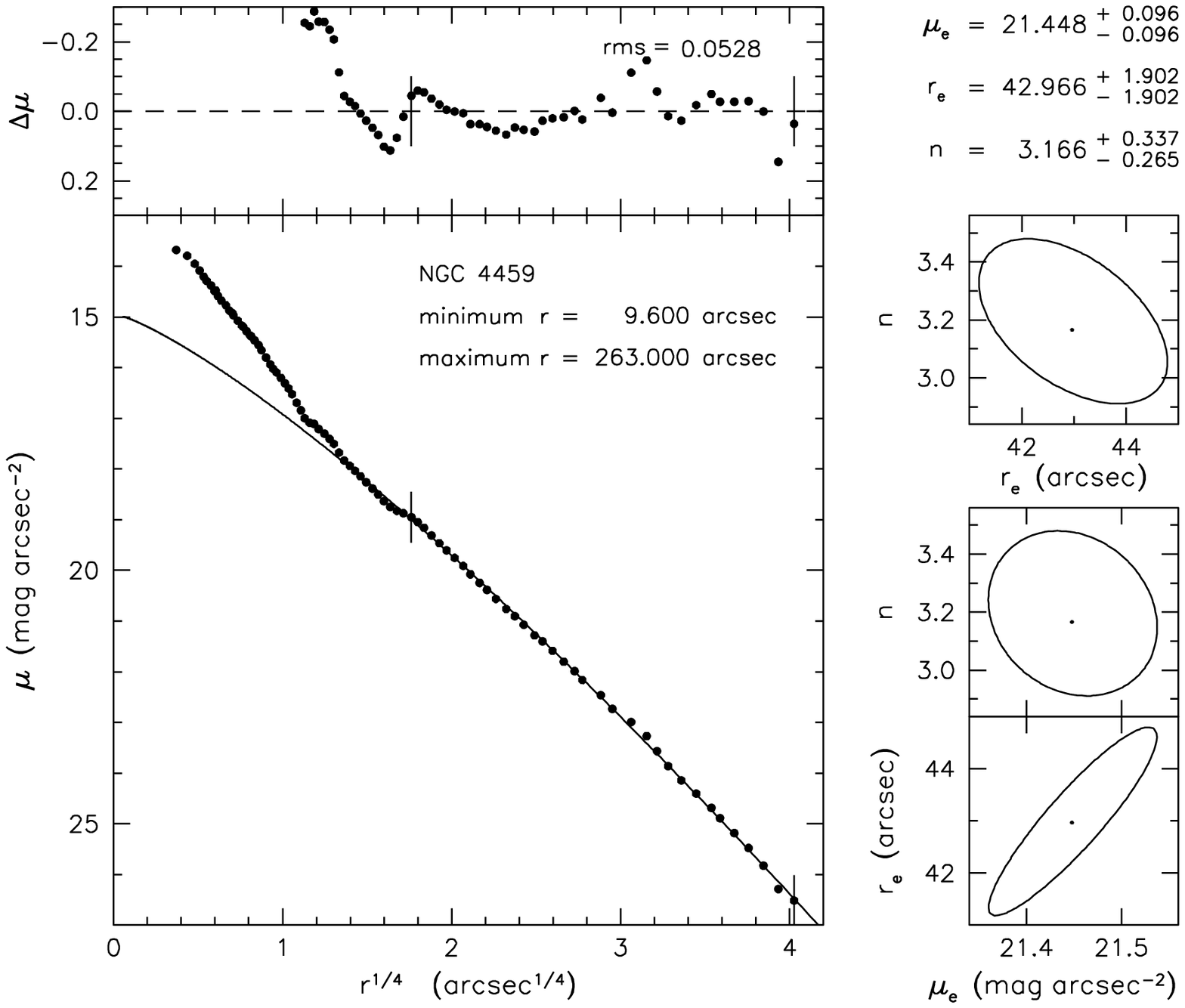}

\includegraphics{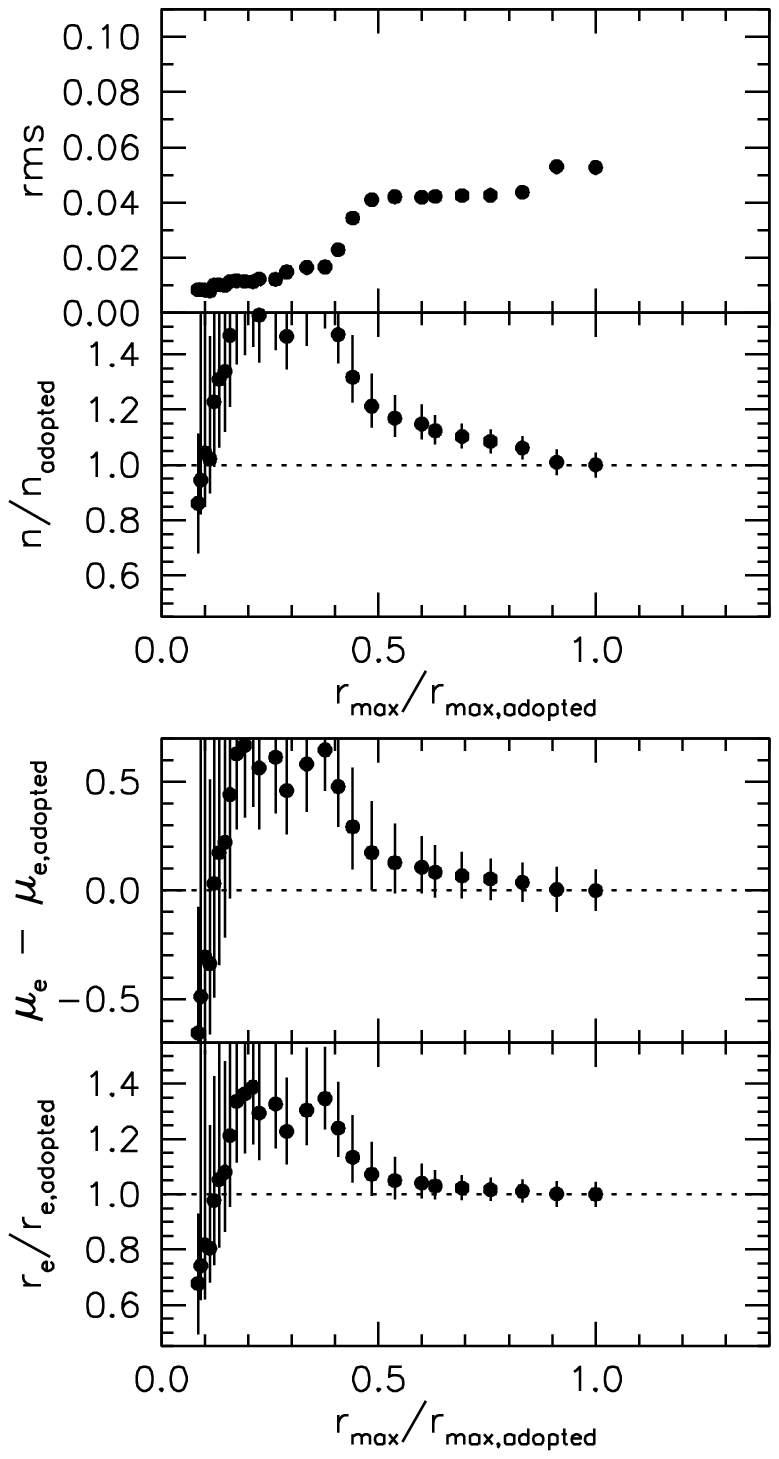}

\figcaption[]
{S\'ersic function fits to the major-axis profiles of NGC 4621 and NGC 4459.
The layout is as in Figure 49.  In larger samples, core and power law galaxies overlap
in luminosity and NGC 4621 is in the overlap region (Faber \etal 1997).  More accurate
individual distances based on surface brightness fluctuations imply a luminosity such
that {\it in the Virgo cluster\/} the separation between core and extra light ellipticals
is fortuitously clean.  At $M_{VT} = -21.54$, NGC 4621 is the brightest extra light
elliptical in the cluster.  NGC 4459 has a prominent dust disk between $r \sim 1^{\prime\prime}$
and 9\farcs6 (e.{\thinspace}g., Sandage 1961; Sandage \& Bedke 1994;  Ferrarese \etal 2006a); 
it is easily identified in the profile and has been omitted from the fit.  The outer part of 
the galaxy is a very clean S\'ersic function with $n < 4$ and no sign of an S0 disk.  
With respect to this fit -- and in spite of any dust absorption -- NGC 4459 clearly has 
extra light near the center.
}

\eject\clearpage

\figurenum{58}

\centerline{\null} 
\vfill

\includegraphics{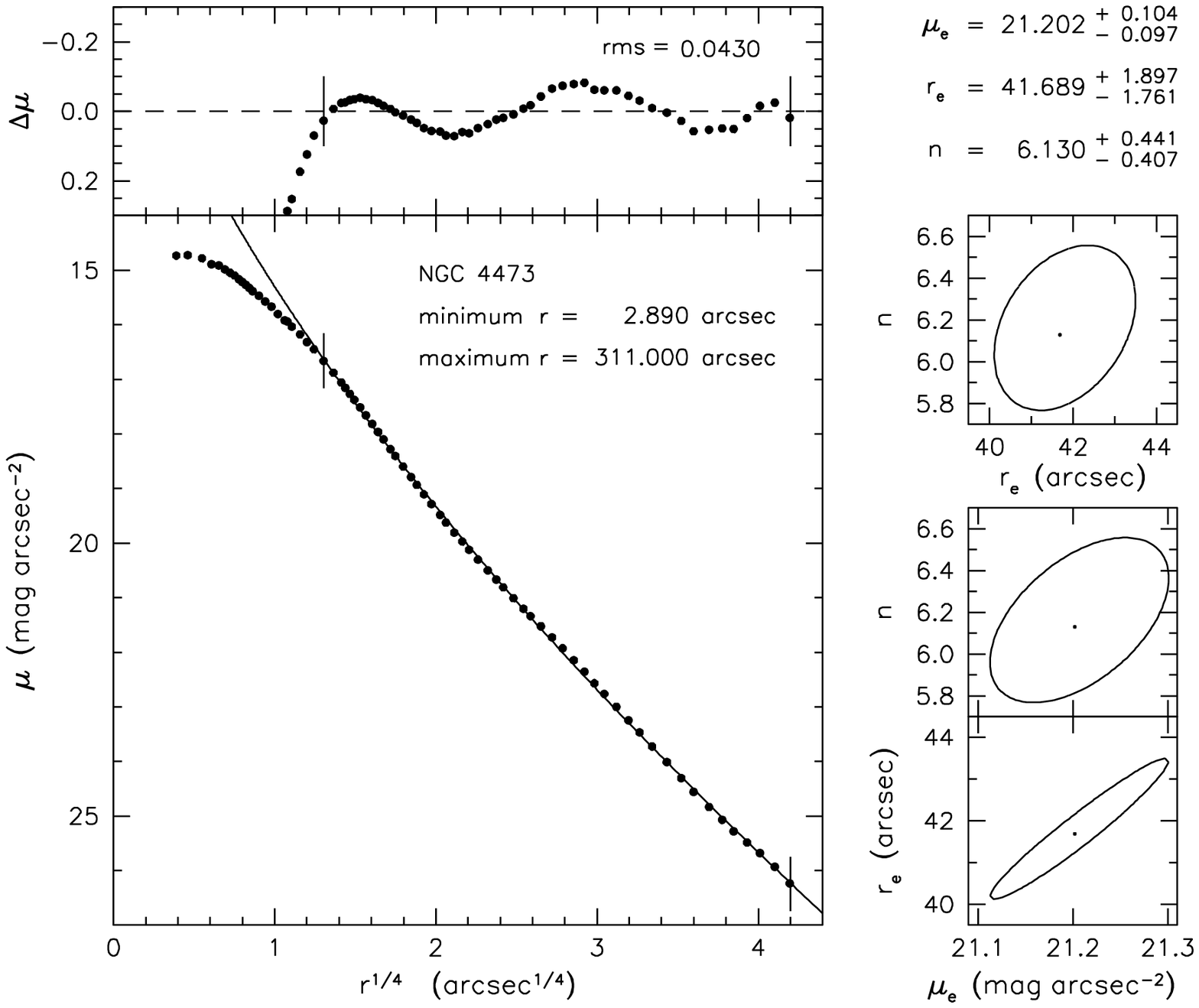}

\includegraphics{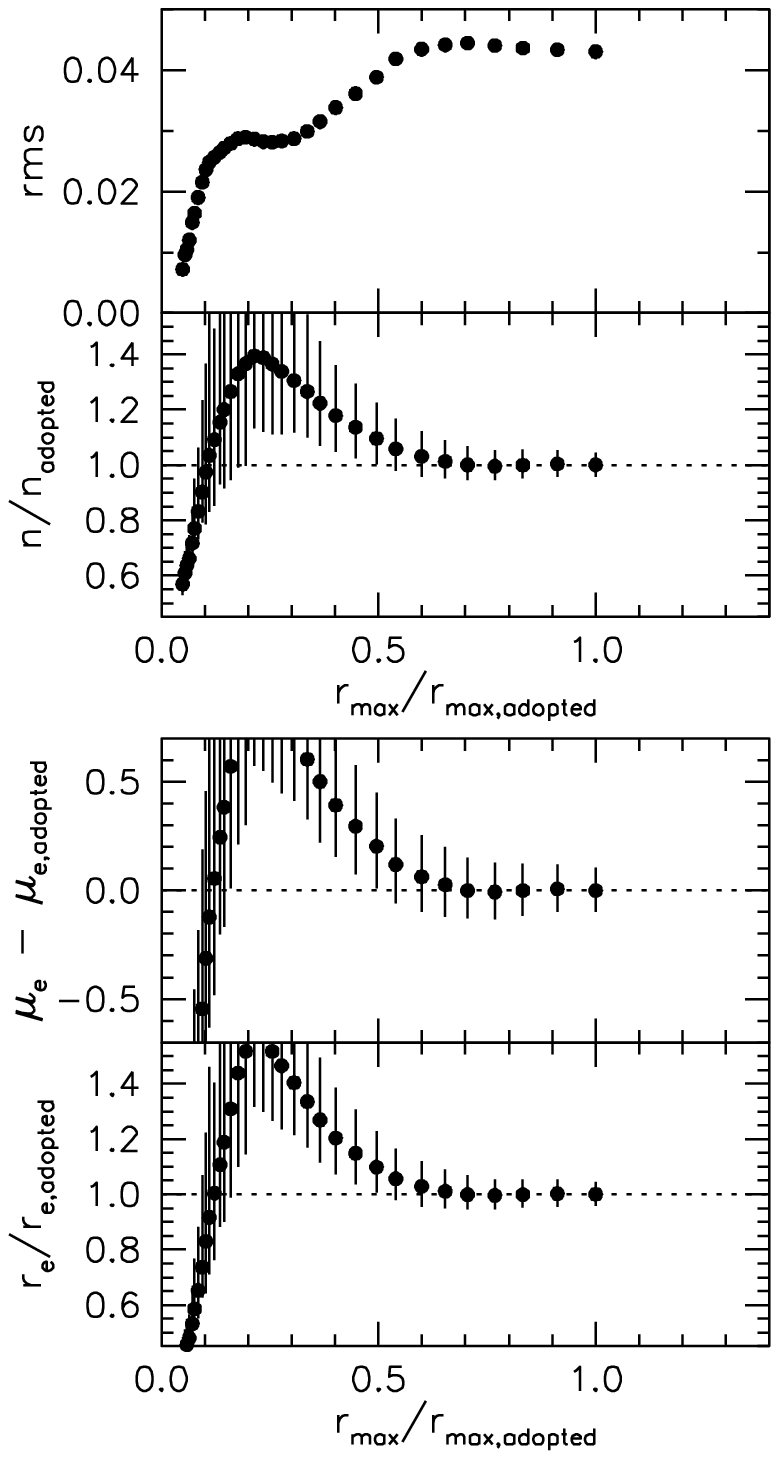}

\includegraphics{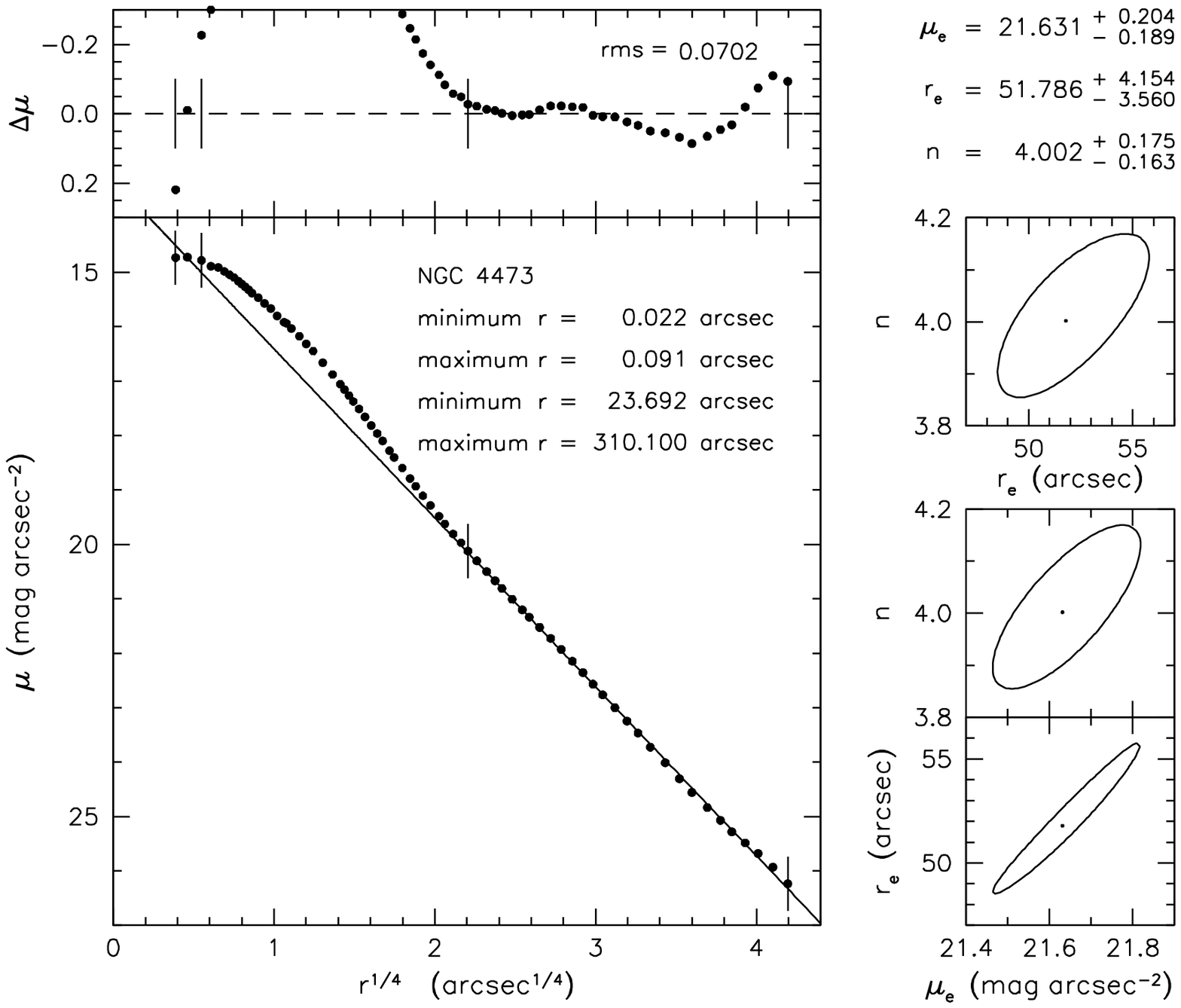}

\includegraphics{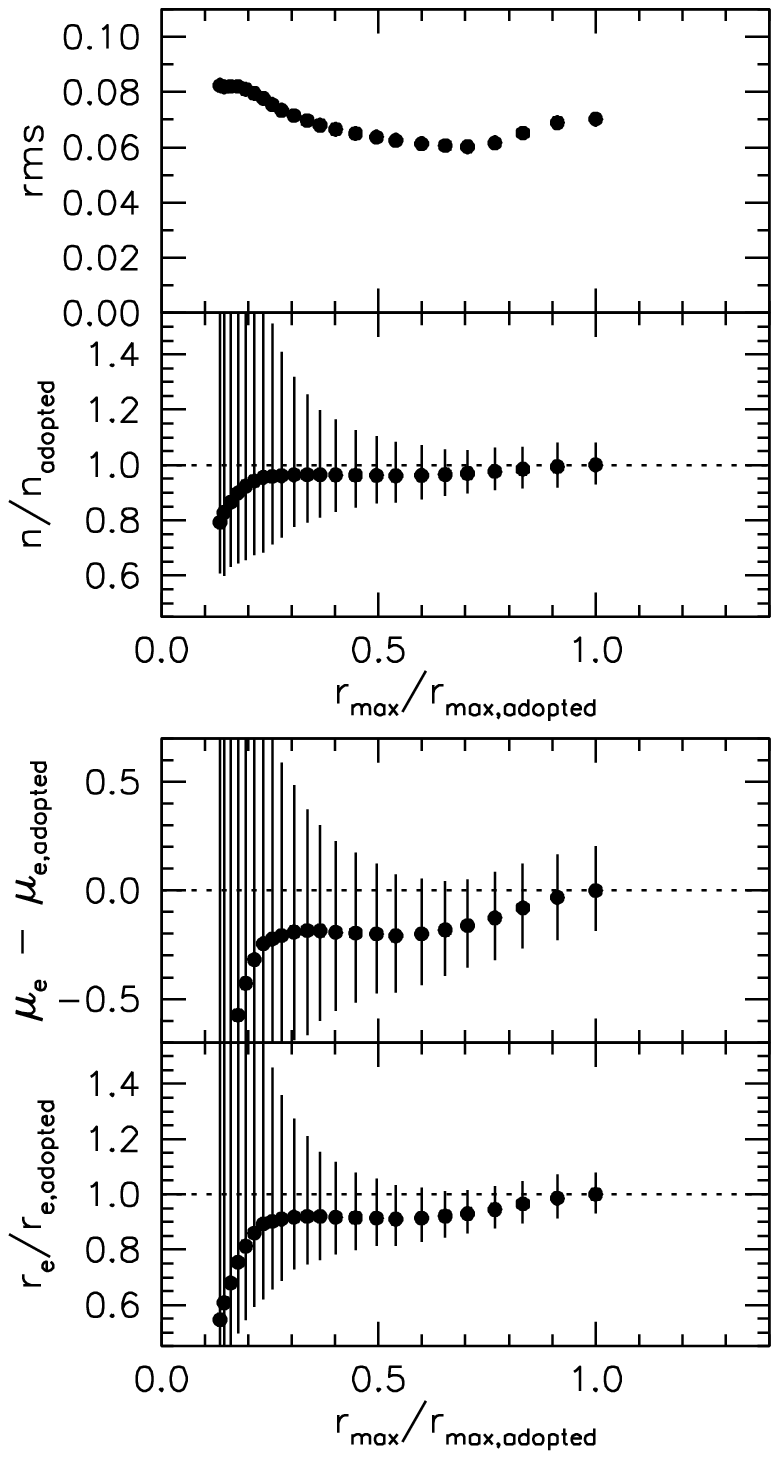}

\figcaption[]
{S\'ersic function fits to the major-axis profile of NGC 4473.  The layout is
as in Figure 49.  NGC 4473 is a tricky case.  It dramatically illustrates the danger
of purely ``operational'' analysis -- in this case, least-squares fit of a S\'ersic
function that minimizes profile residuals -- without taking other observations and their
physical implications into account.  The top fit looks beguilingly good, better 
than the bottom fit.  If it were adopted, we would
conclude that the galaxy has a core and a S\'ersic index $n > 4$.  However, we adopt the
bottom fit.  The reason is that SAURON observations show that the galaxy has
a counter-rotating embedded disk (Cappellari \& McDermid 2005; Cappellari et al.~2004, 2007, 
see \S{\thinspace}9.5 here).  Figure 5 in Cappellari et al.~(2007) shows that the 
counter-rotating disk is important from small radii out to $r \simeq 19^{\prime\prime}$ 
(that is, to $r^{1/4} \simeq 2.1$) but not at larger radii.  We therefore fit the profile 
from $r \simeq 23\farcs7$ outward, excluding the counter-rotating disk.  The inner edge 
of the fit range is determined by where the residuals from the outer S\'ersic fit start to 
grow large, but they are consistent with the Cappellari results.  We also include three points 
near the center to provide stability to the fit.  Since stars in the embedded disk pass in 
front of the center, the surface brightness there is higher than that of the main body of the 
galaxy.  Therefore the true S\'ersic index is smaller than the value, $n \simeq 4.0 \pm 0.17$, 
that we derive.  To illustrate this, we show in Figure 59 a decomposition of the profile into
a S\'ersic function main body and an exponential fit to the extra light.
\lineskip=-2pt \lineskiplimit=-2pt
}

\eject\clearpage

\figurenum{59}

\centerline{\null} 
\vfill

\includegraphics{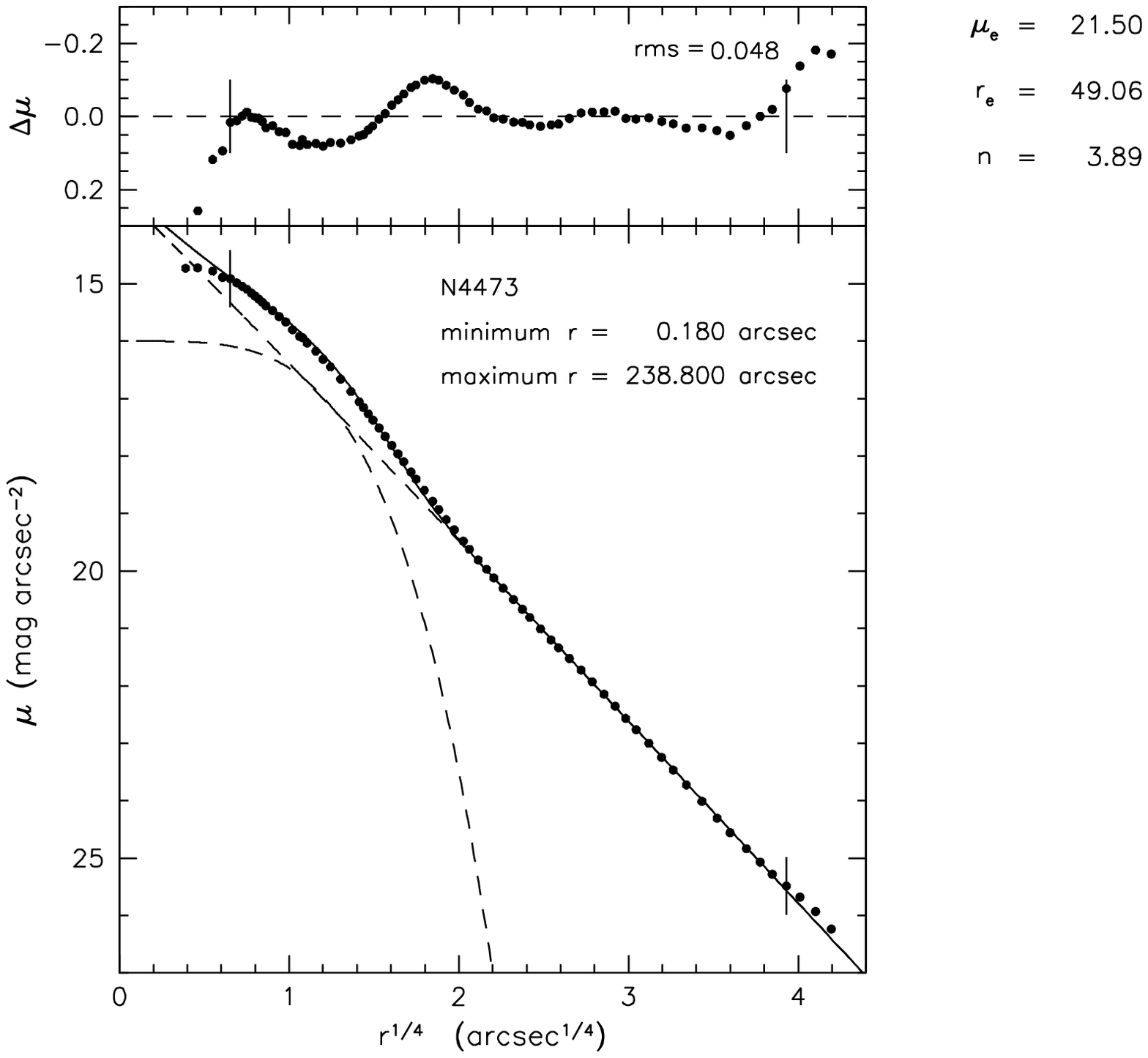}

\includegraphics{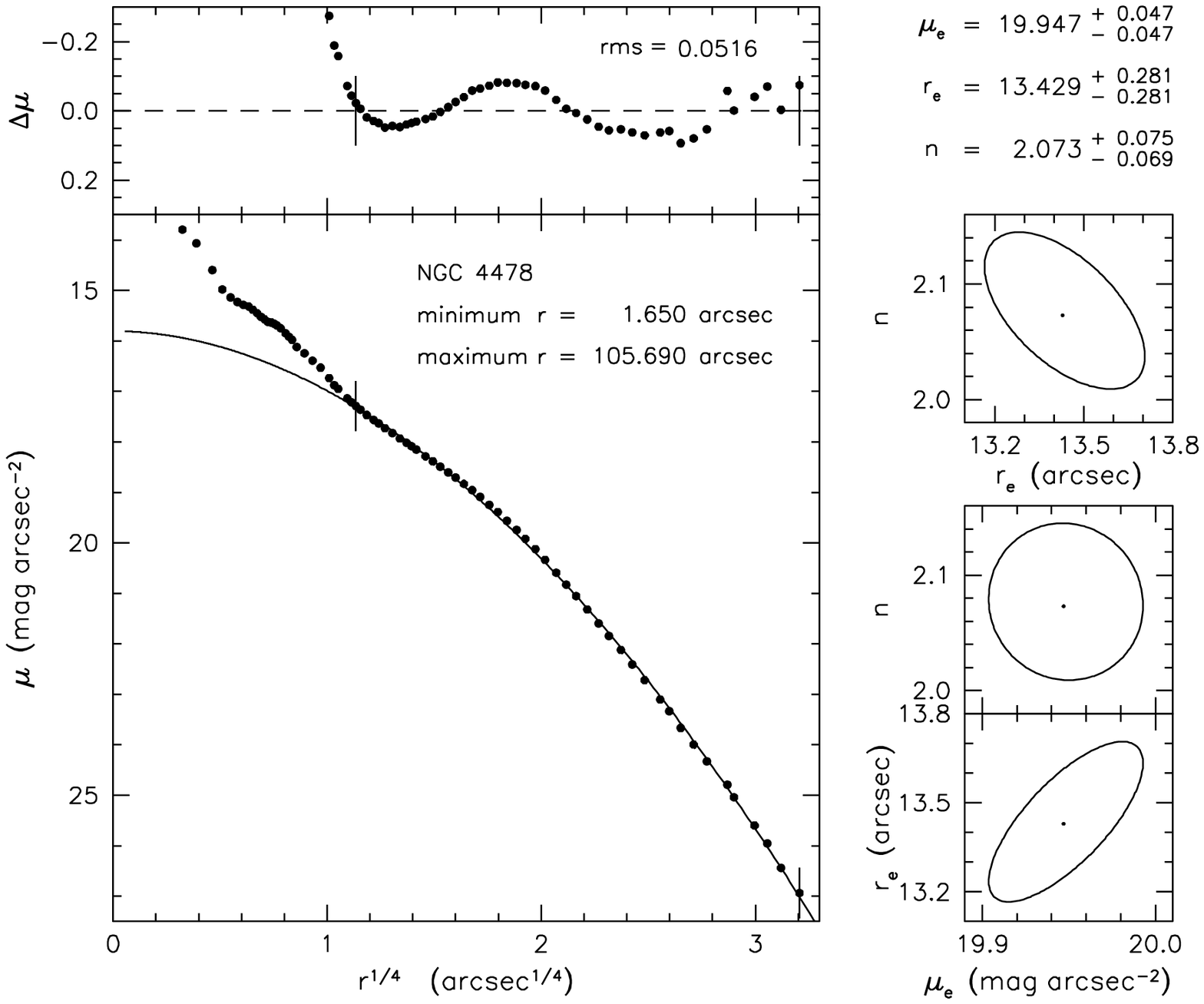}

\includegraphics{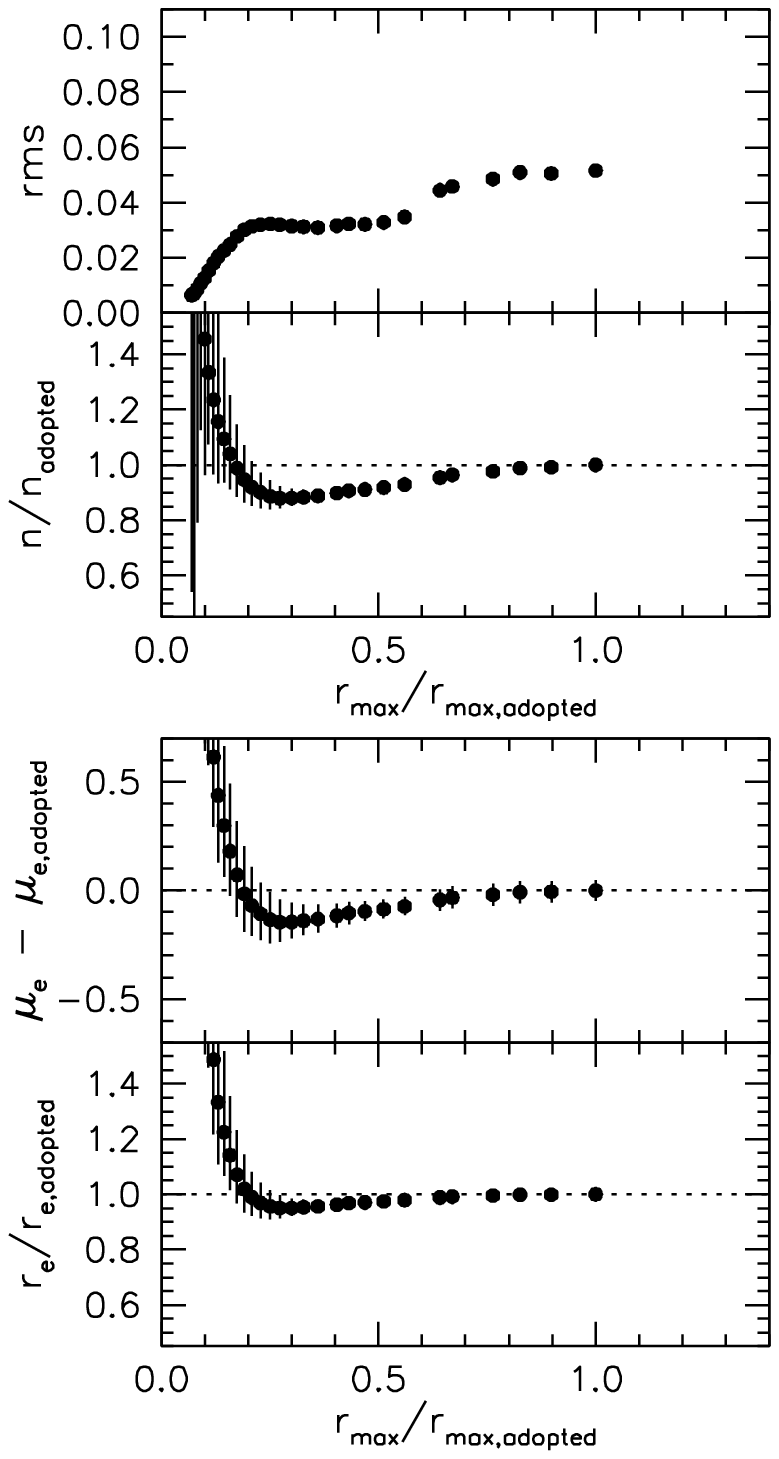}

\figcaption[]
{The {\it top panels\/} show a decomposition of the major-axis profile of NGC 4473 into
an inner exponential fitted to the extra light (in essence, the counter-rotating disk) and
an outer S\'ersic function.  The parameters of the main body of the galaxy are almost
unchanged from the fit in Figure 58, but $n$ drops slightly below 4, as expected.  This
decomposition is directly comparable to the Hopkins \etal (2008b) decomposition reproduced
here in Figure 44.  It gives a fractional contribution of the extra light of 9.1\ts\%,
compared with 15\ts\% for the brighter and shallower disk fit by Hopkins.
The {\it bottom panels\/} show our S\'ersic function fit to the major-axis profile
of NGC 4478.  The layout is as in Figure 49.}

\eject\clearpage

\figurenum{60}

\centerline{\null} 
\vfill

\includegraphics{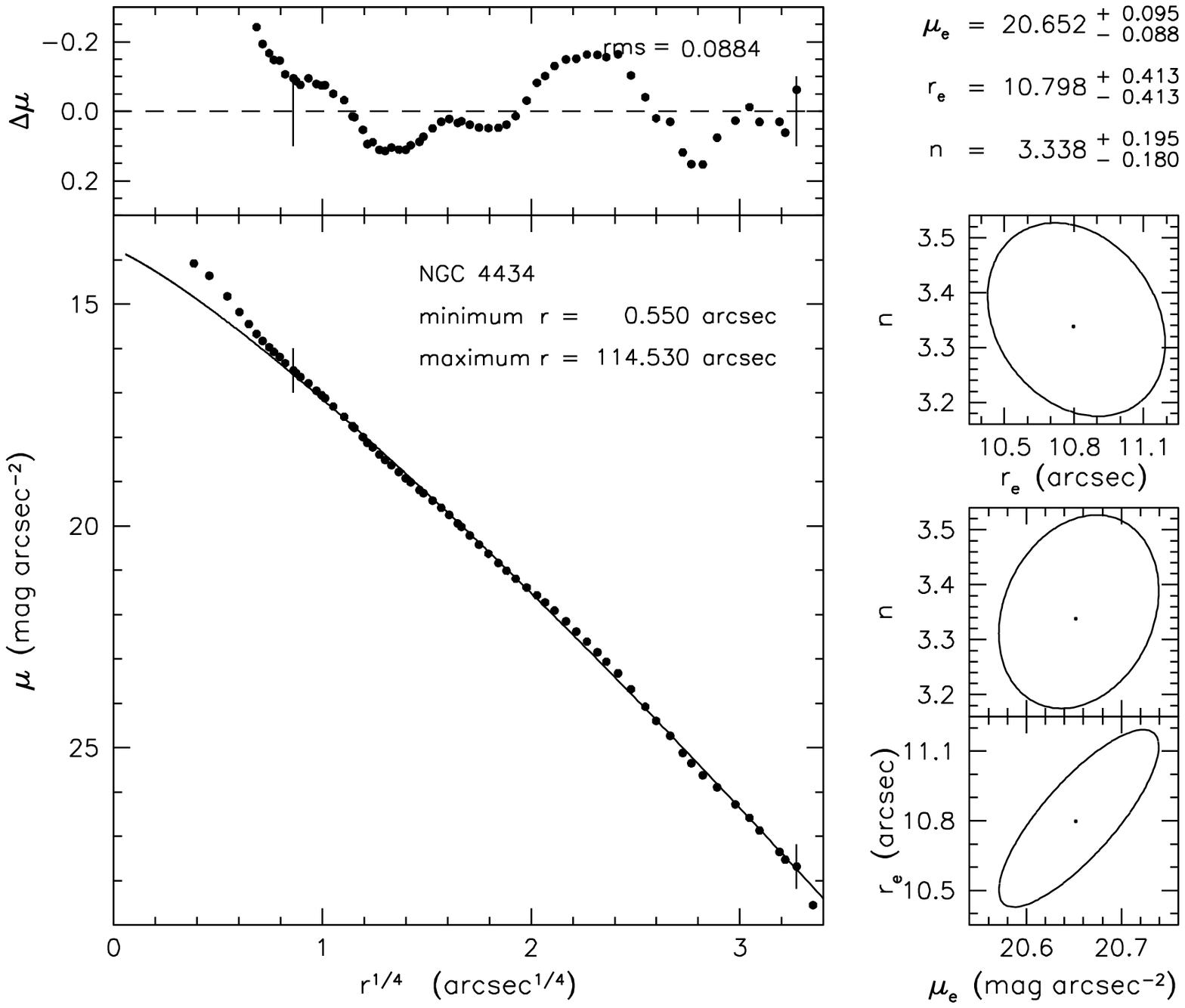}

\includegraphics{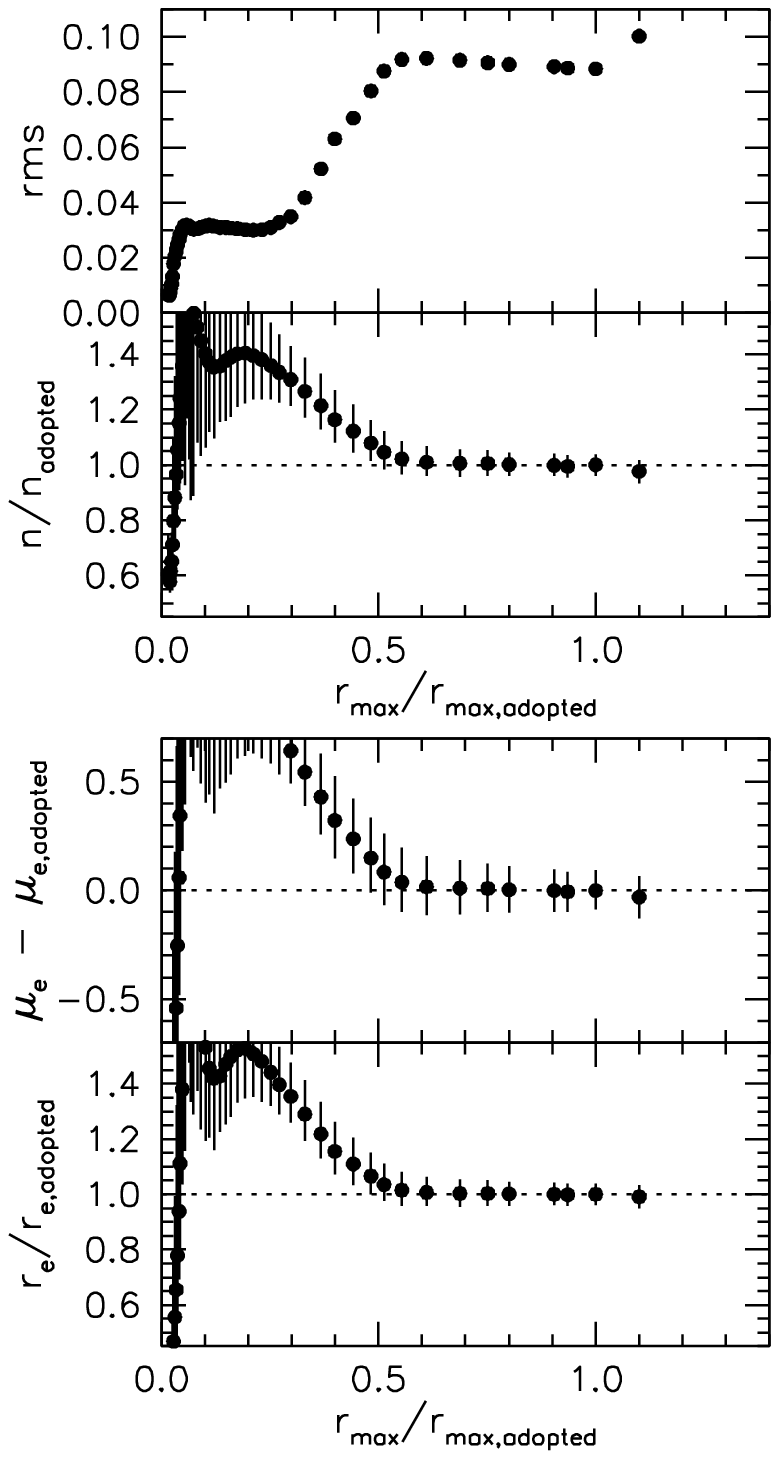}

\includegraphics{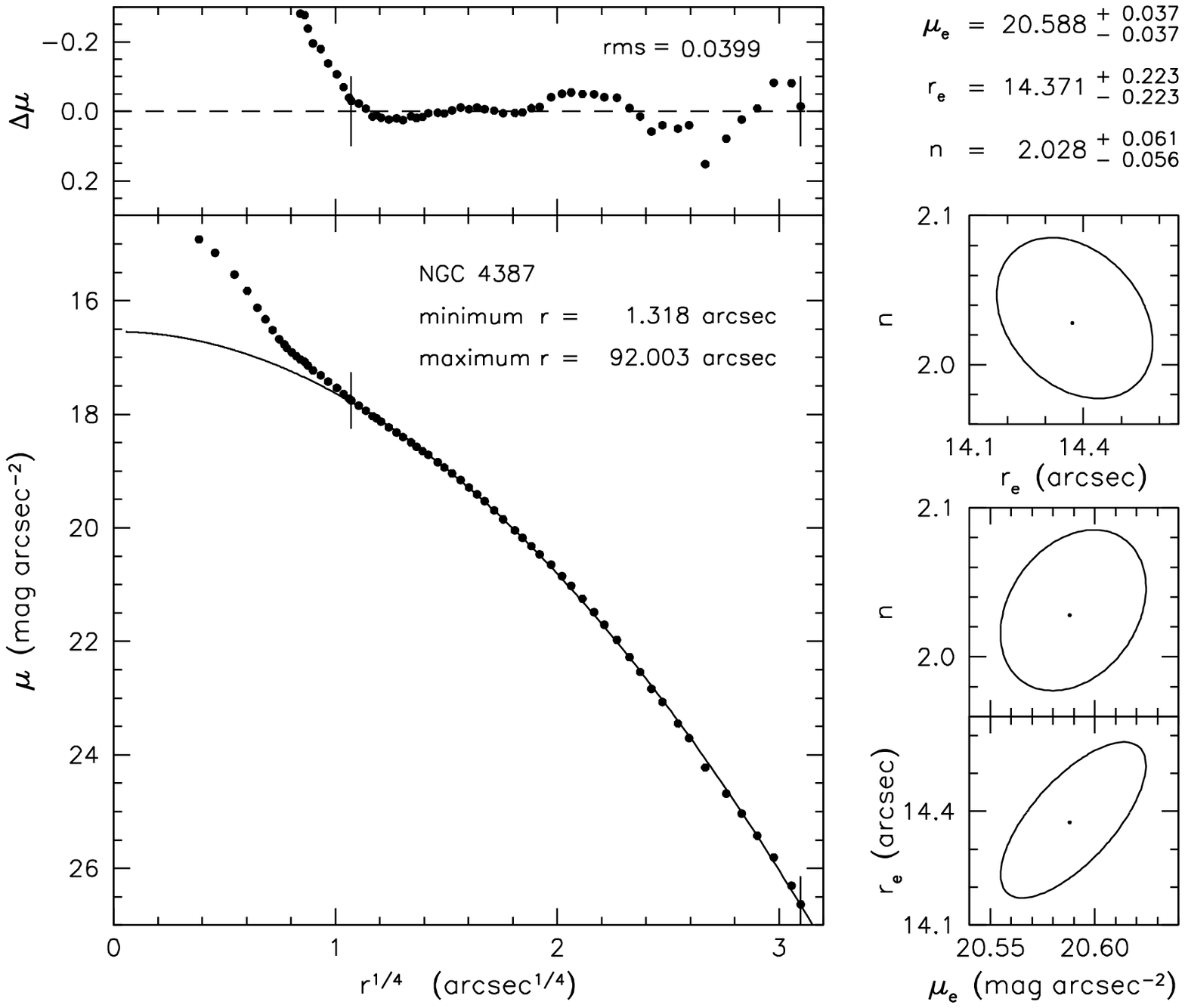}

\includegraphics{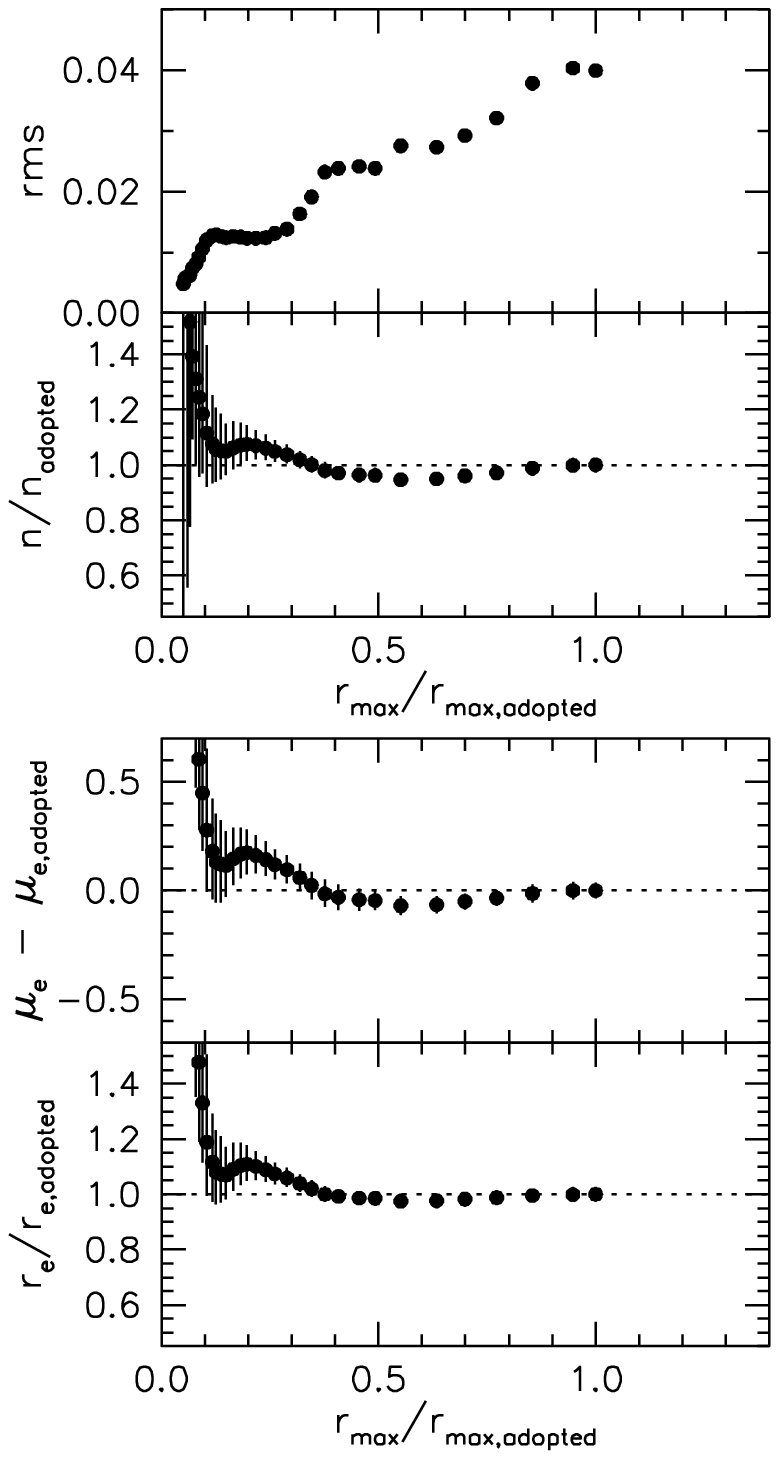}

\figcaption[]
{S\'ersic function fits to the major-axis profiles of NGC 4434 and NGC 4387.
The layout is as in Figure 49.}

\eject\clearpage

\figurenum{61}

\centerline{\null} \vfill

\includegraphics{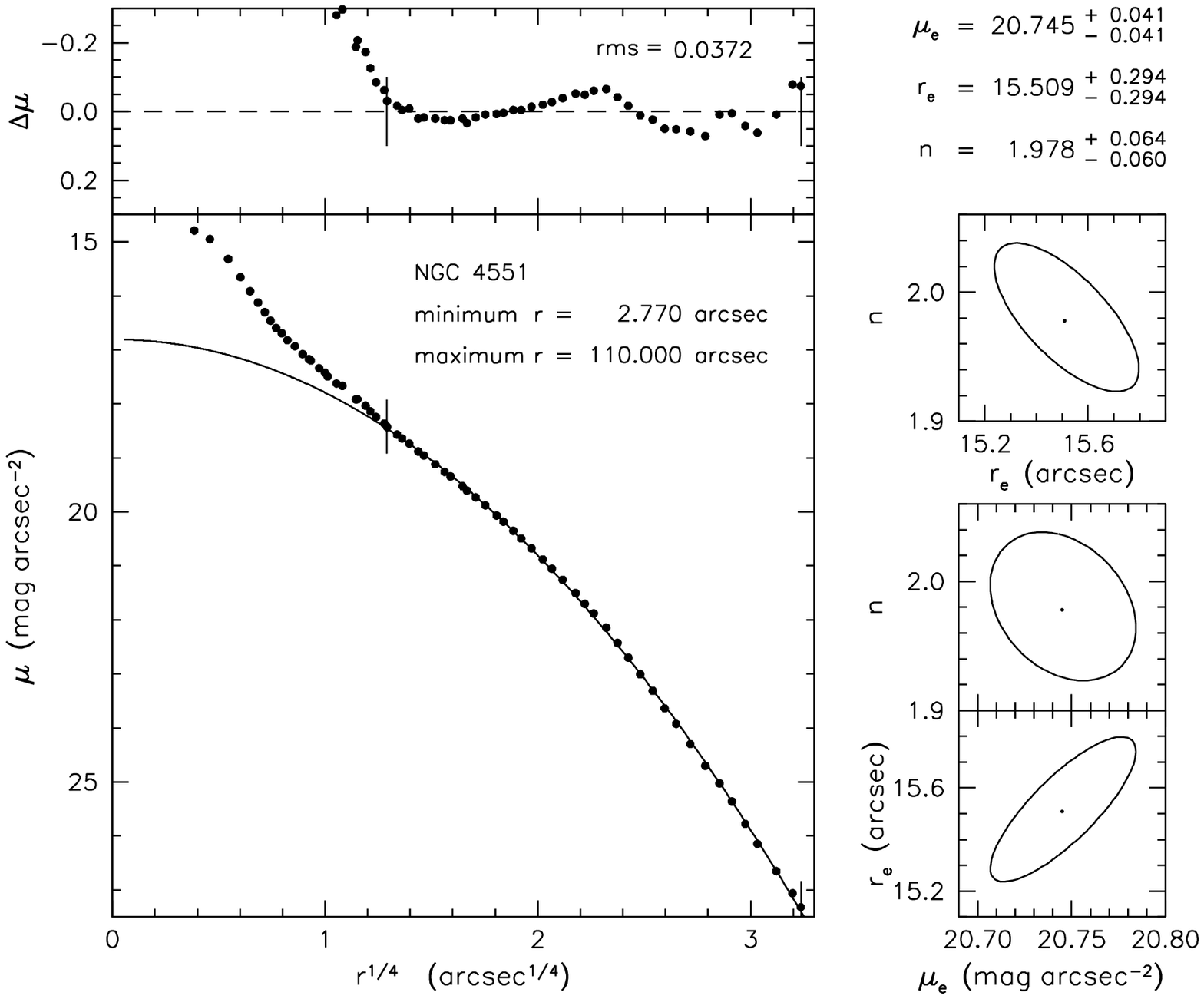}

\includegraphics{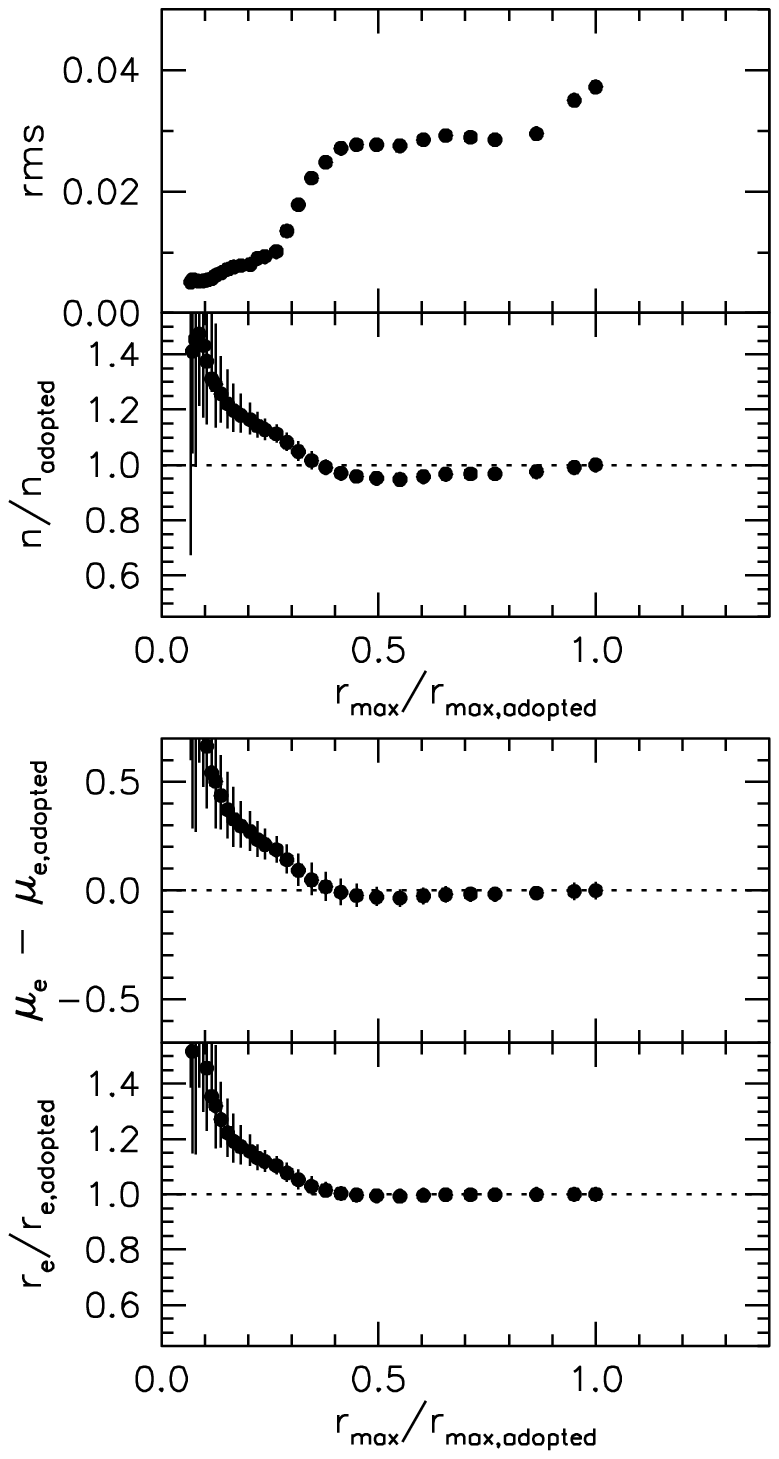}

\includegraphics{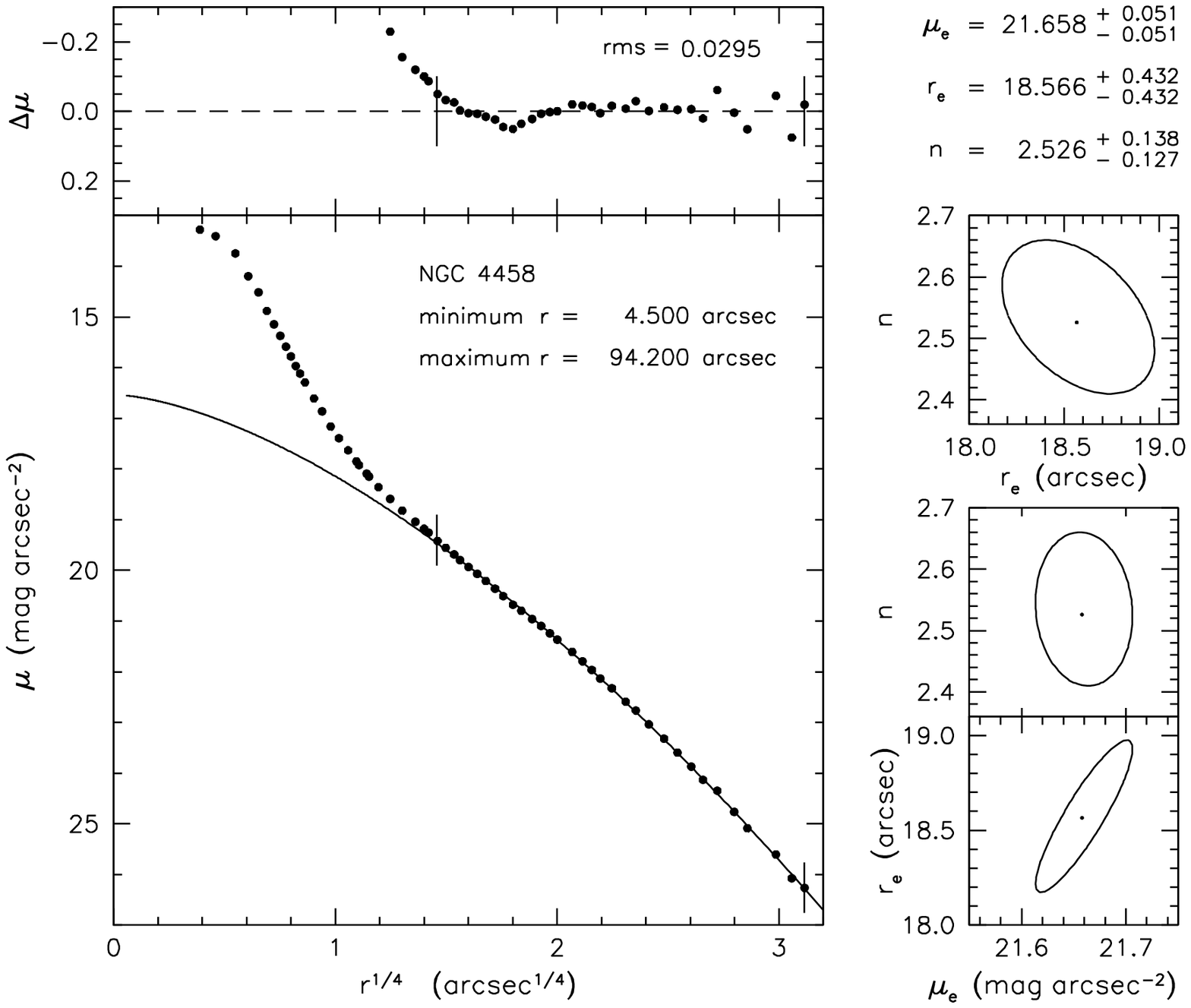}

\includegraphics{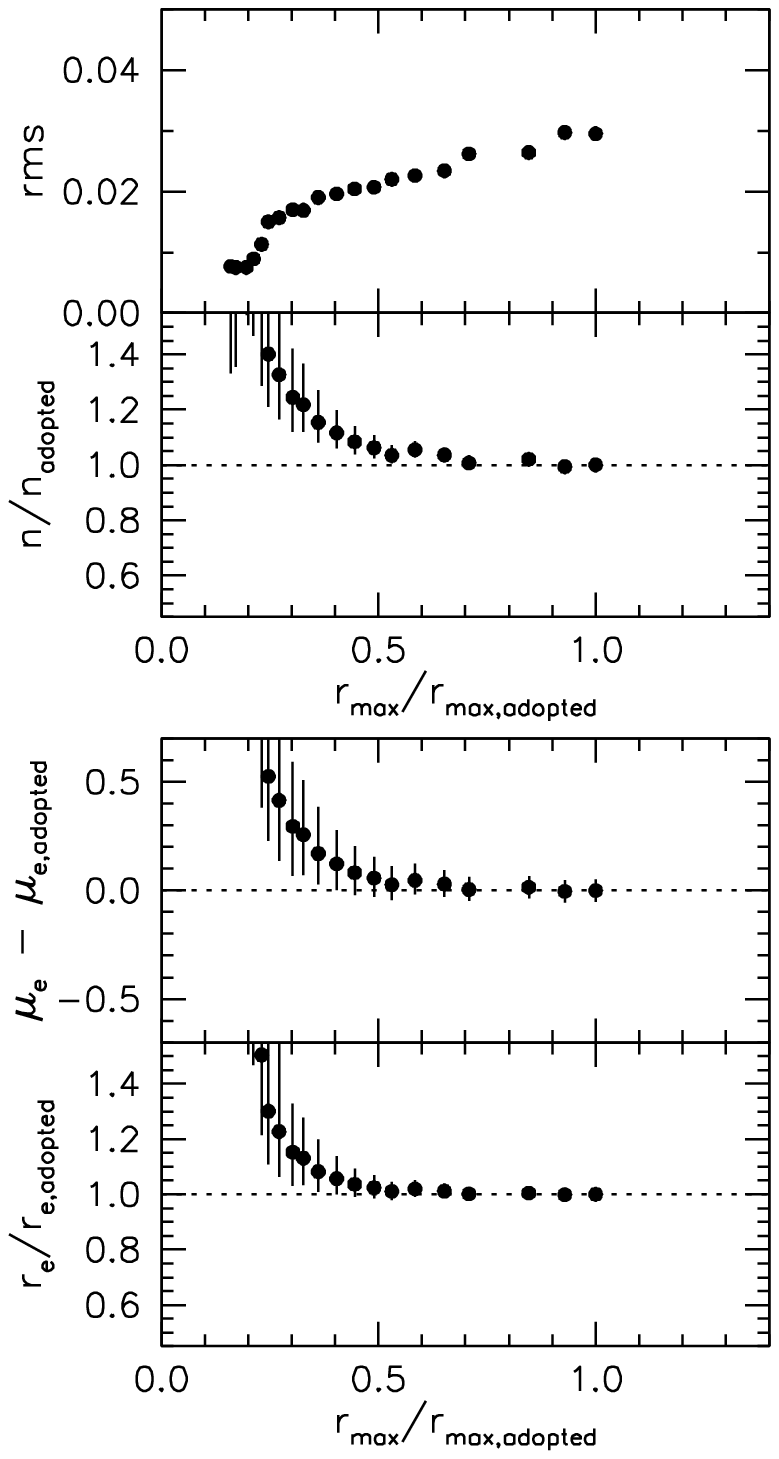}

\figcaption[]
{S\'ersic function fits to the major-axis profiles of NGC 4551 and NGC 4458.  
The layout is as in Figure 49. 
}

\eject\clearpage

\figurenum{62}

\centerline{\null} \vfill

\includegraphics{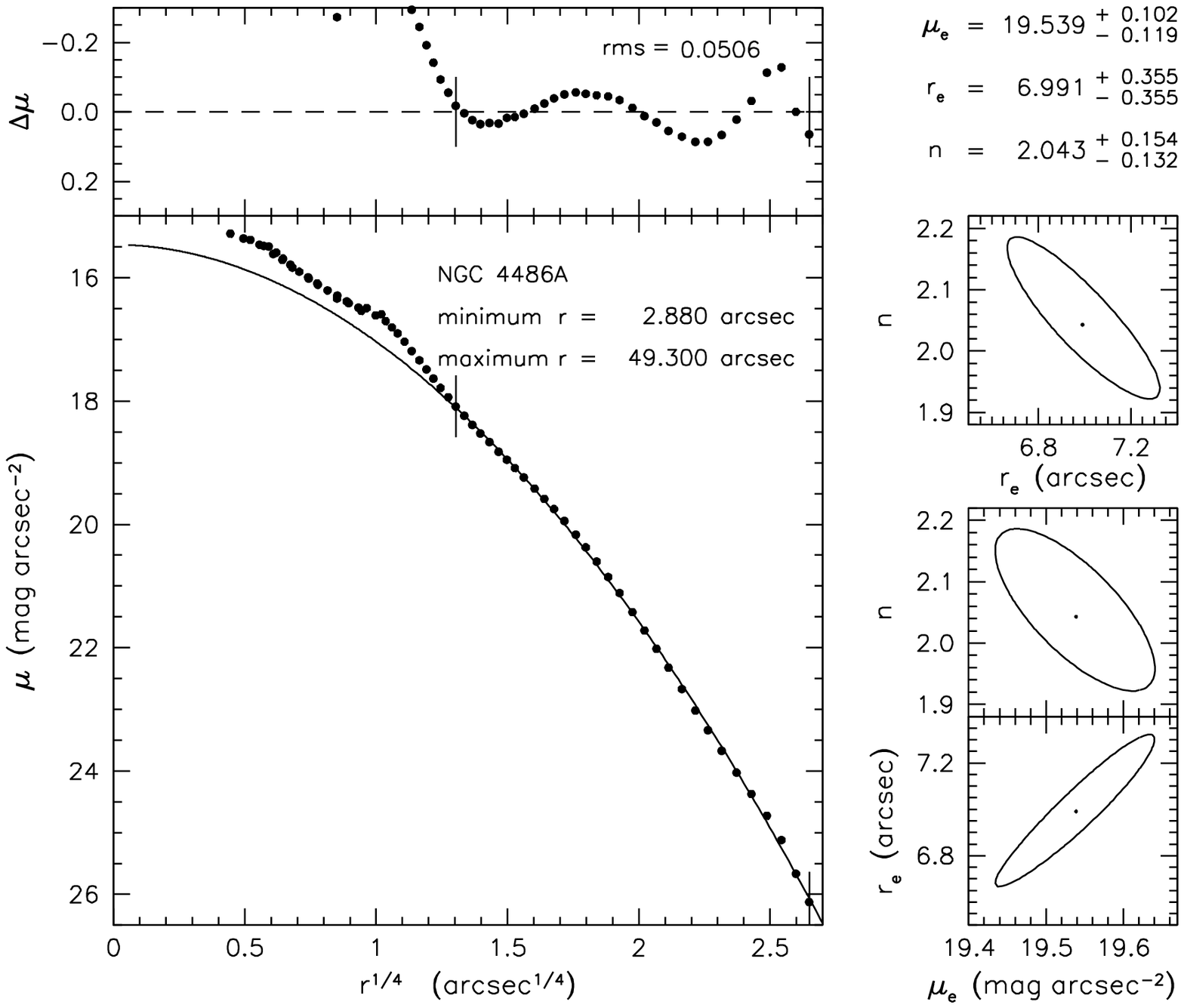}

\includegraphics{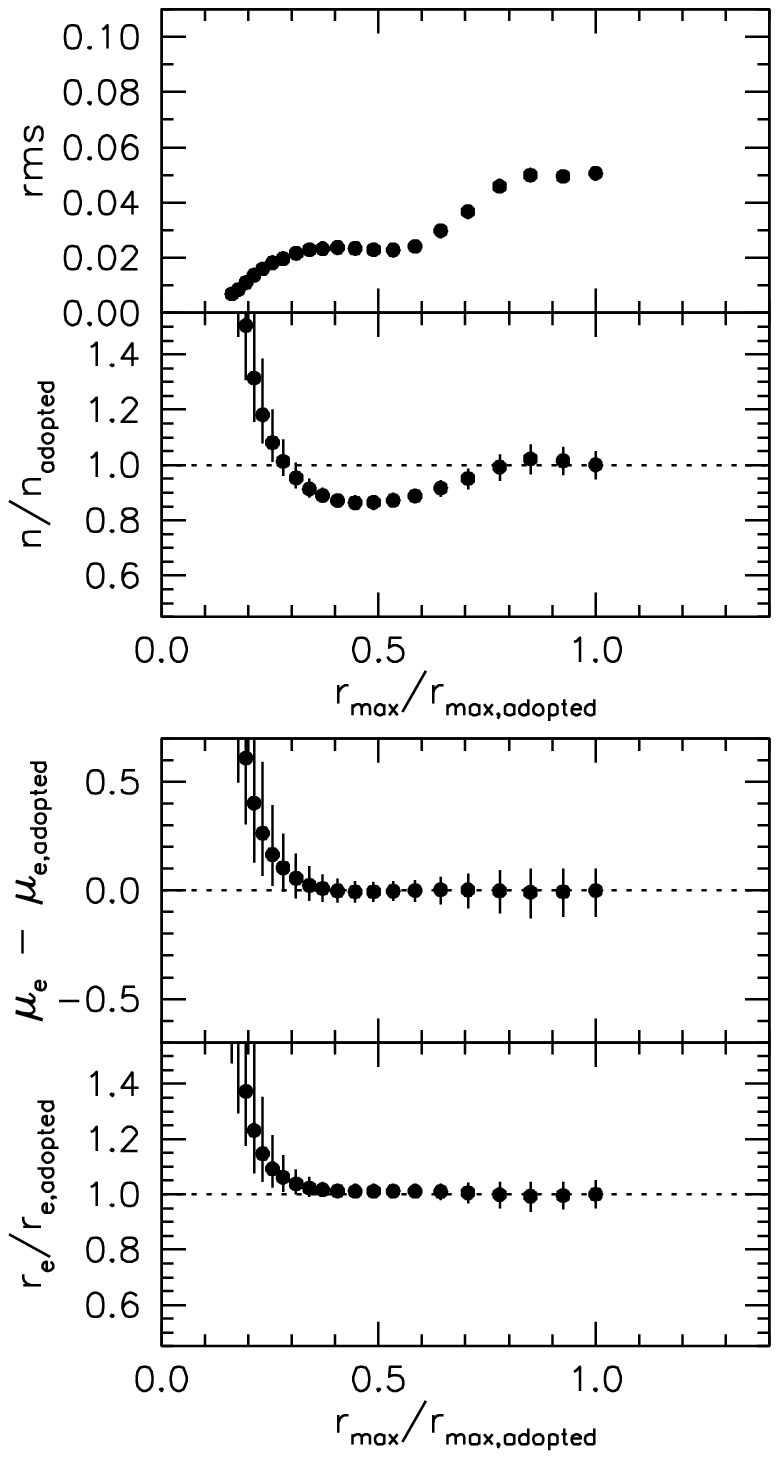}

\includegraphics{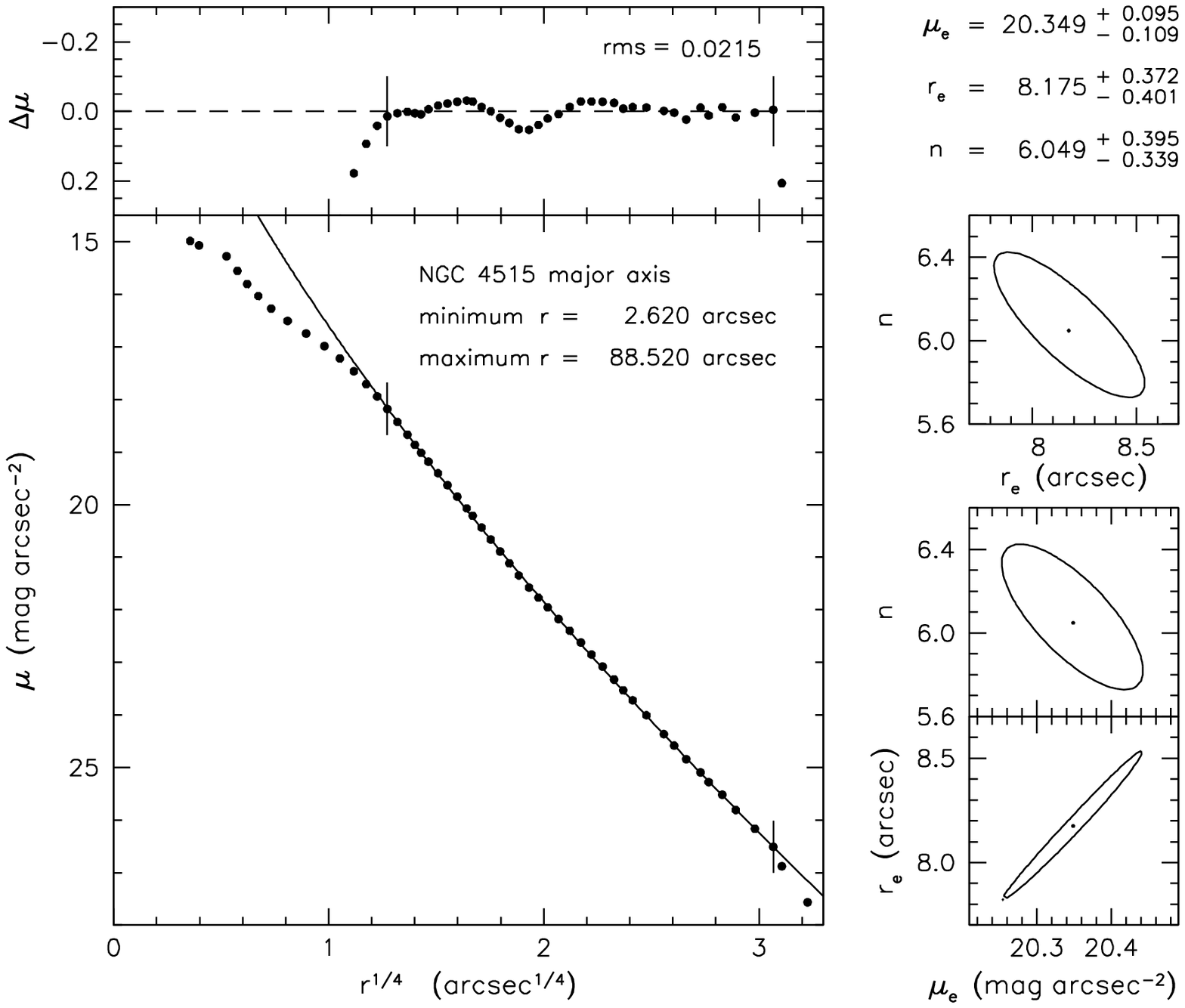}

\figcaption[]
{The {\it top panels\/} show our S\'ersic function fit to the major-axis profile of NGC 4486A.
The extra light is a particularly obvious nuclear disk bisected by a strong dust lane (see Figure 
20 here and Kormendy et al.~2005) that produces the kink in the profile at $\sim$\ts$1^{\prime\prime}$.
The {\it bottom panels\/} show a S\'ersic fit to the major-axis profile of NGC 4515.  This is 
superficially an excellent fit, with small RMS deviations over a large radius range and a 
canonical combination of an apparent core (albeit with an unusually steep profile) and a 
S\'ersic index $n > 4$.  However, we do not adopt this fit.  The reasons -- and our adopted fit --
are given in Figure 63.
\lineskip=-2pt \lineskiplimit=-2pt
}

\eject\clearpage

\figurenum{63}

\centerline{\null} \vfill

\includegraphics{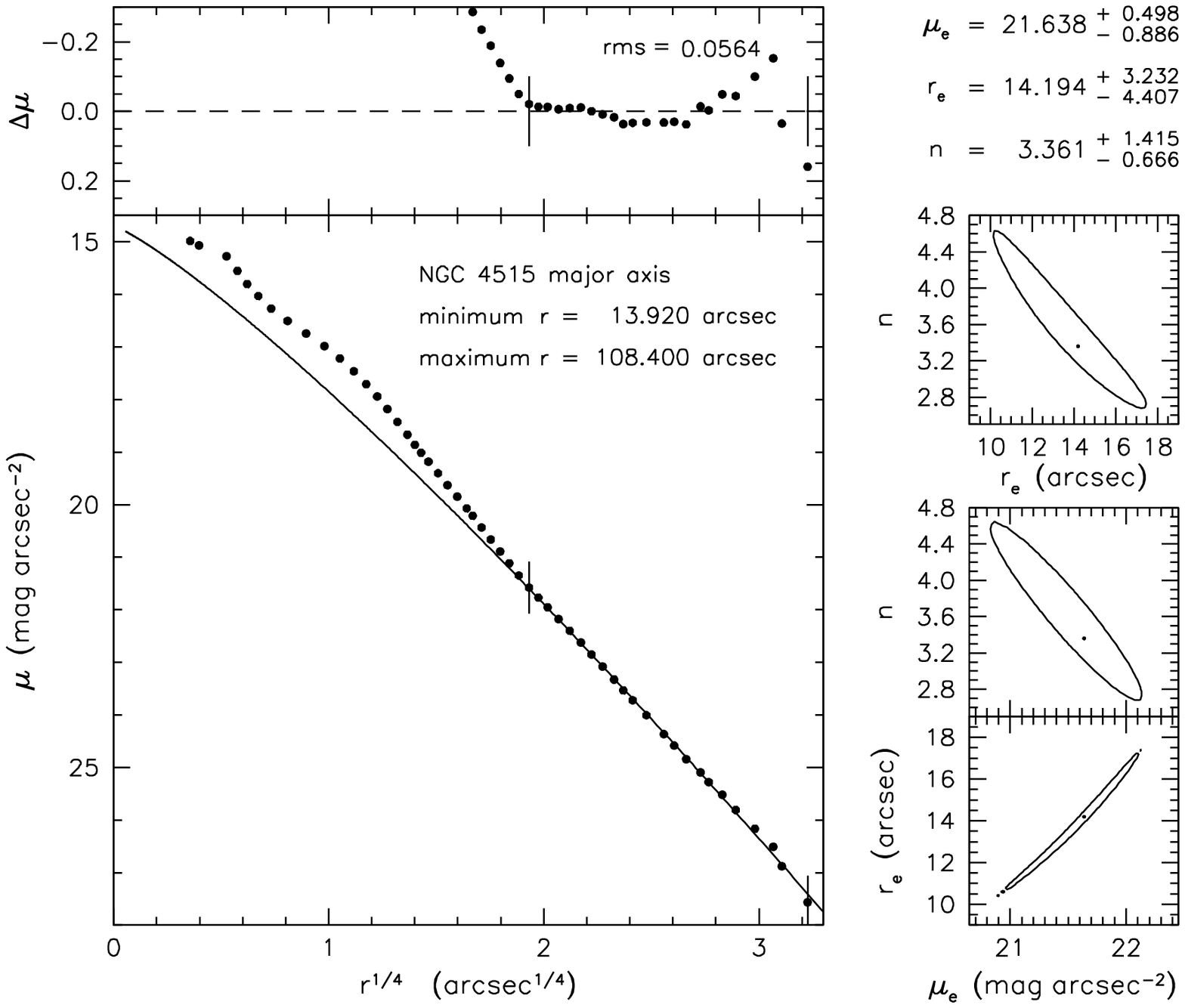}

\includegraphics{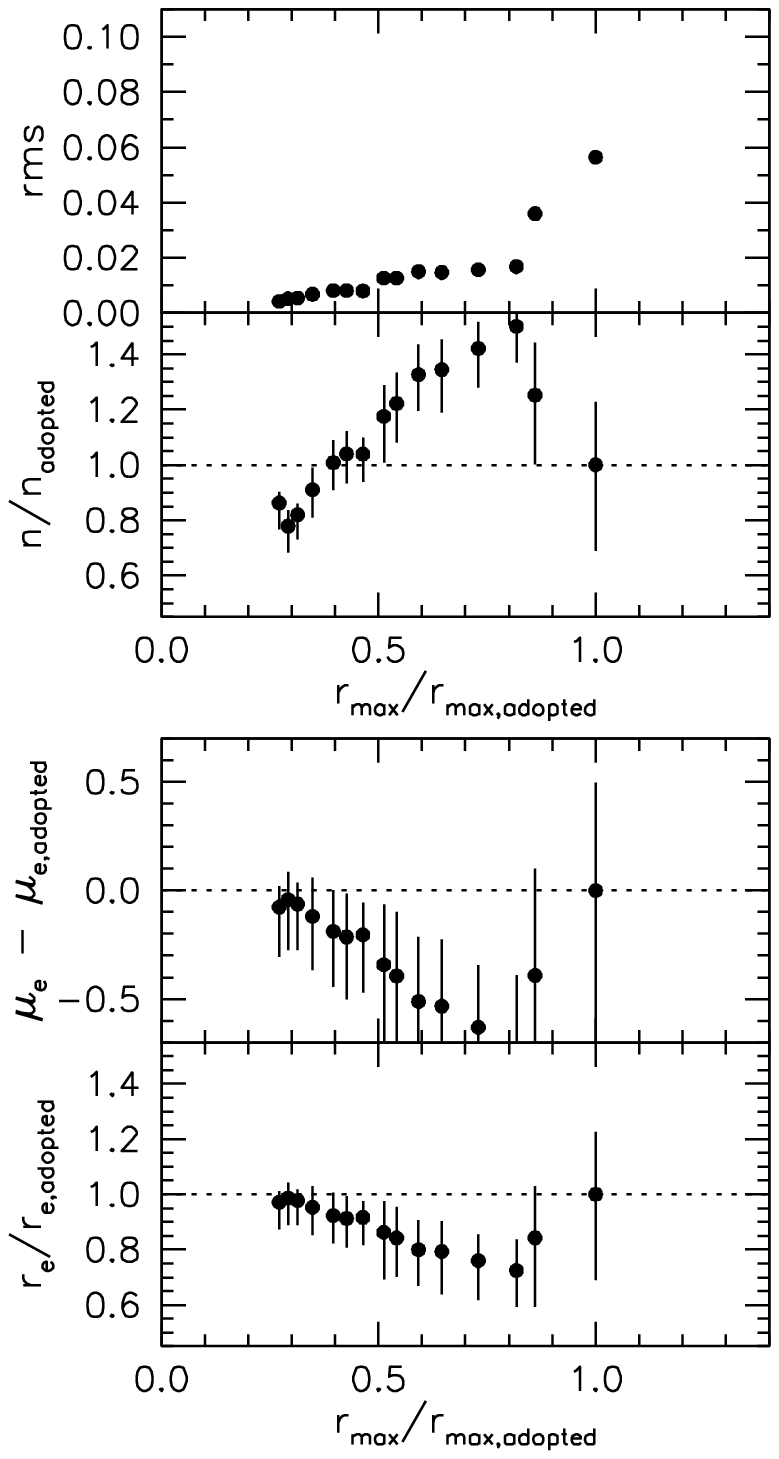}

\includegraphics{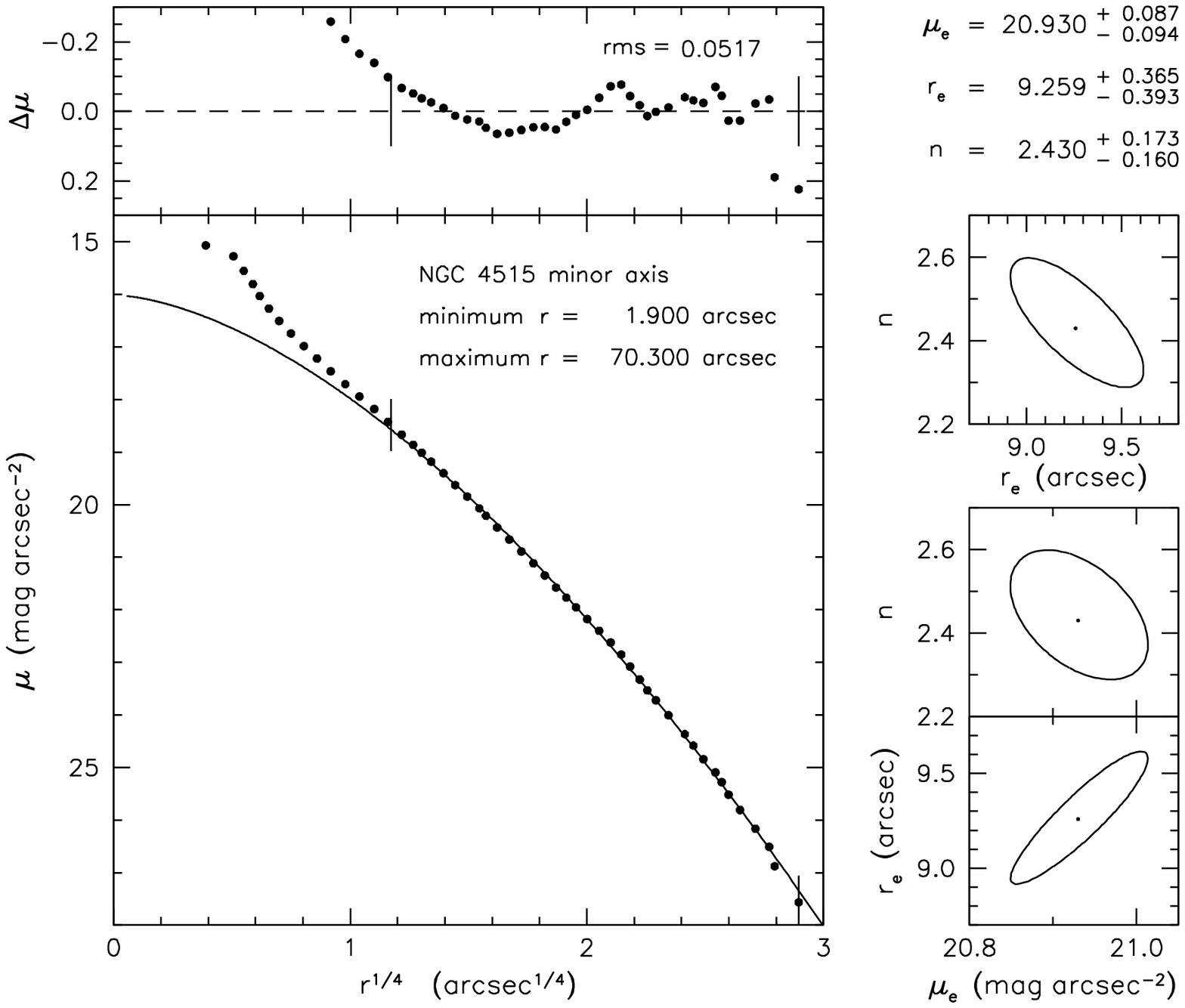}

\includegraphics{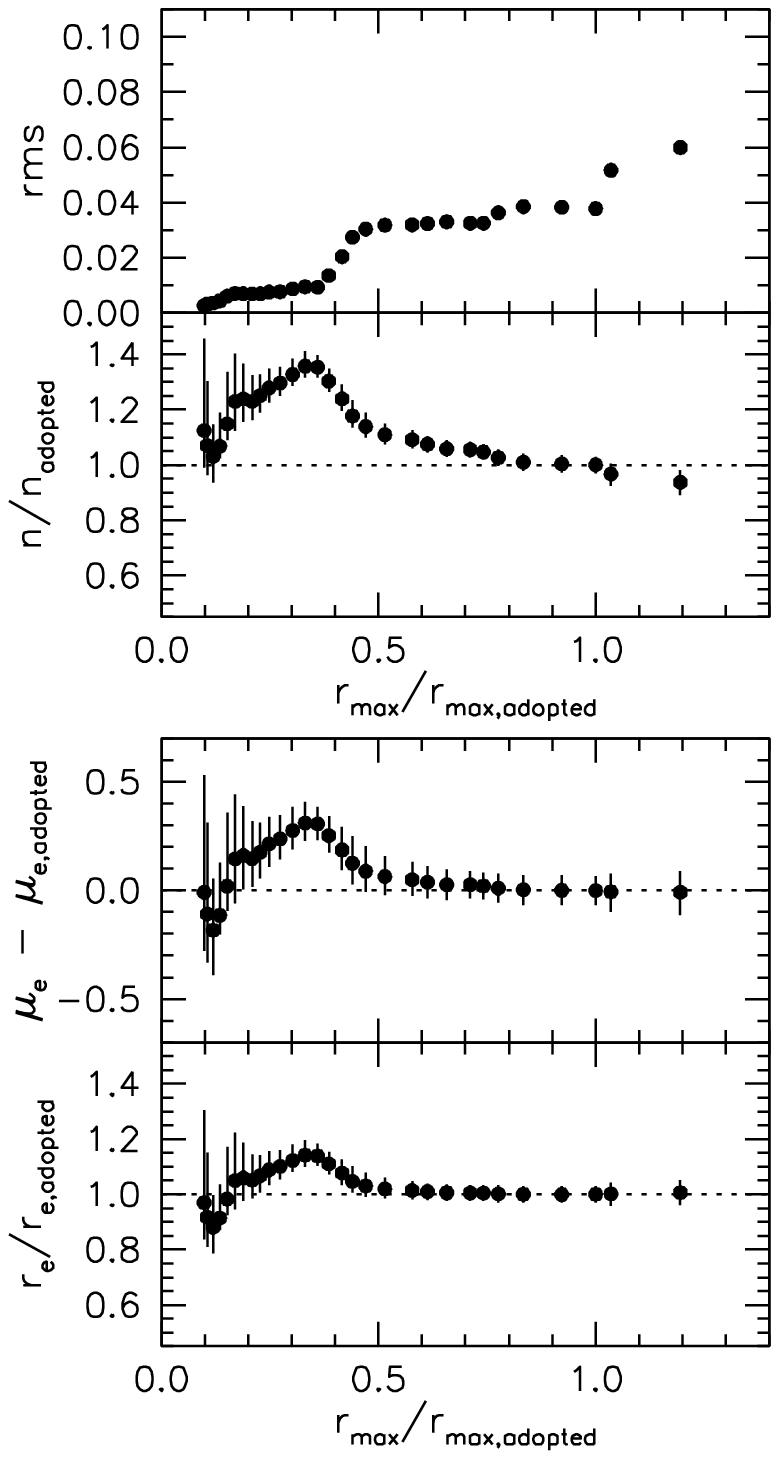}

\figcaption[]
{The {\it top panels\/} show our adopted S\'ersic fit to the major-axis profile of NGC 4515
In our sample, this galaxy is the trickiest one to interpret.  It is similar to NGC 4473.  
The ellipticity and $a_4$ profiles show the signature of an extended nuclear disk (Fig.~21). 
But this disky central region shows almost no rotation ($V_{\rm rot} \lesssim 20$ km s$^{-1}$),
a moderately high velocity dispersion ($\sigma \sim 90$ km s$^{-1}$) and hence an unusually 
low ratio of $V_{\rm rot}/\sigma$ for a low-luminosity elliptical (Bender \& Nieto 1990). 
It would be interesting to look for counter-rotation.  Given this situation, we are not 
persuaded by the superficially excellent fit in Figure 62.  Instead, we adopt the top fit
here, which omits the central disky structure.  Is this reasonable?  For an answer, we resort 
to the minor-axis profile  ({\it bottom panels\/}).  In all of our other galaxies, the major- 
and minor-axis profiles consistently both give $n < 4$ or both give $n > 4$.  The minor-axis 
profile of NGC 4515 confirms that $n < 4$ and that extra light is detected.  For this reason,
we adopt the interpretation in the upper panels.
\lineskip=-2pt \lineskiplimit=-2pt
}

\eject\clearpage

\figurenum{64}

\centerline{\null} \vfill

\includegraphics{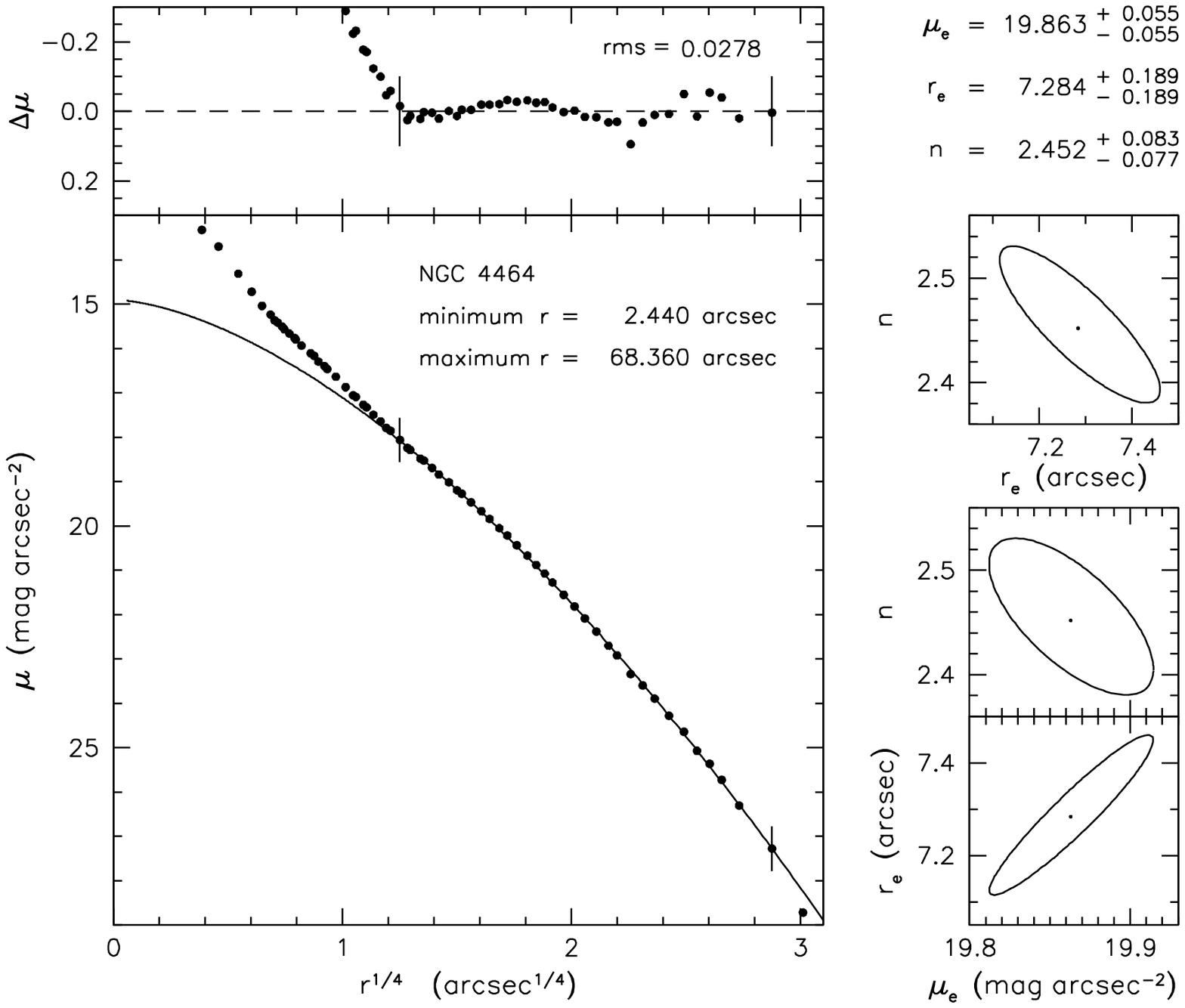}

\includegraphics{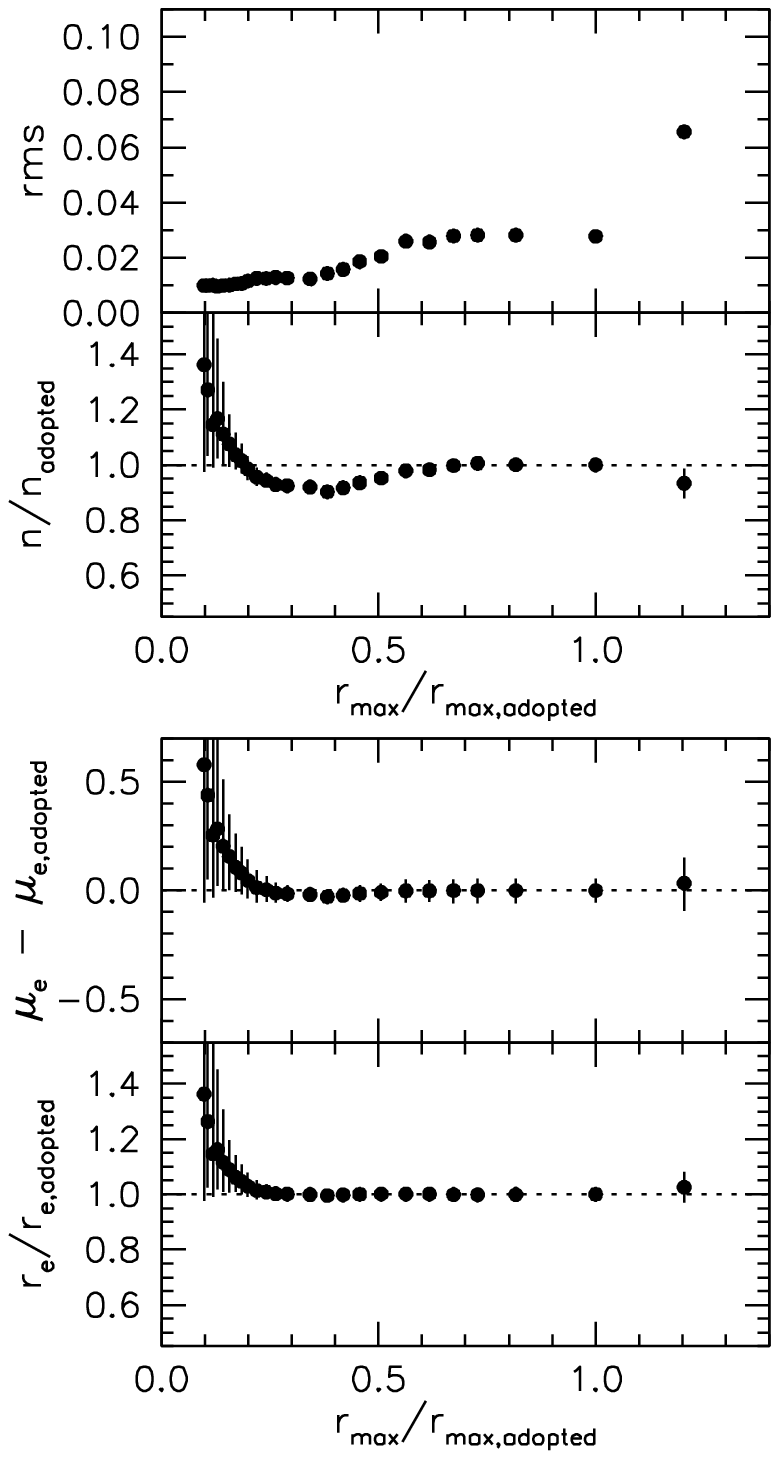}

\includegraphics{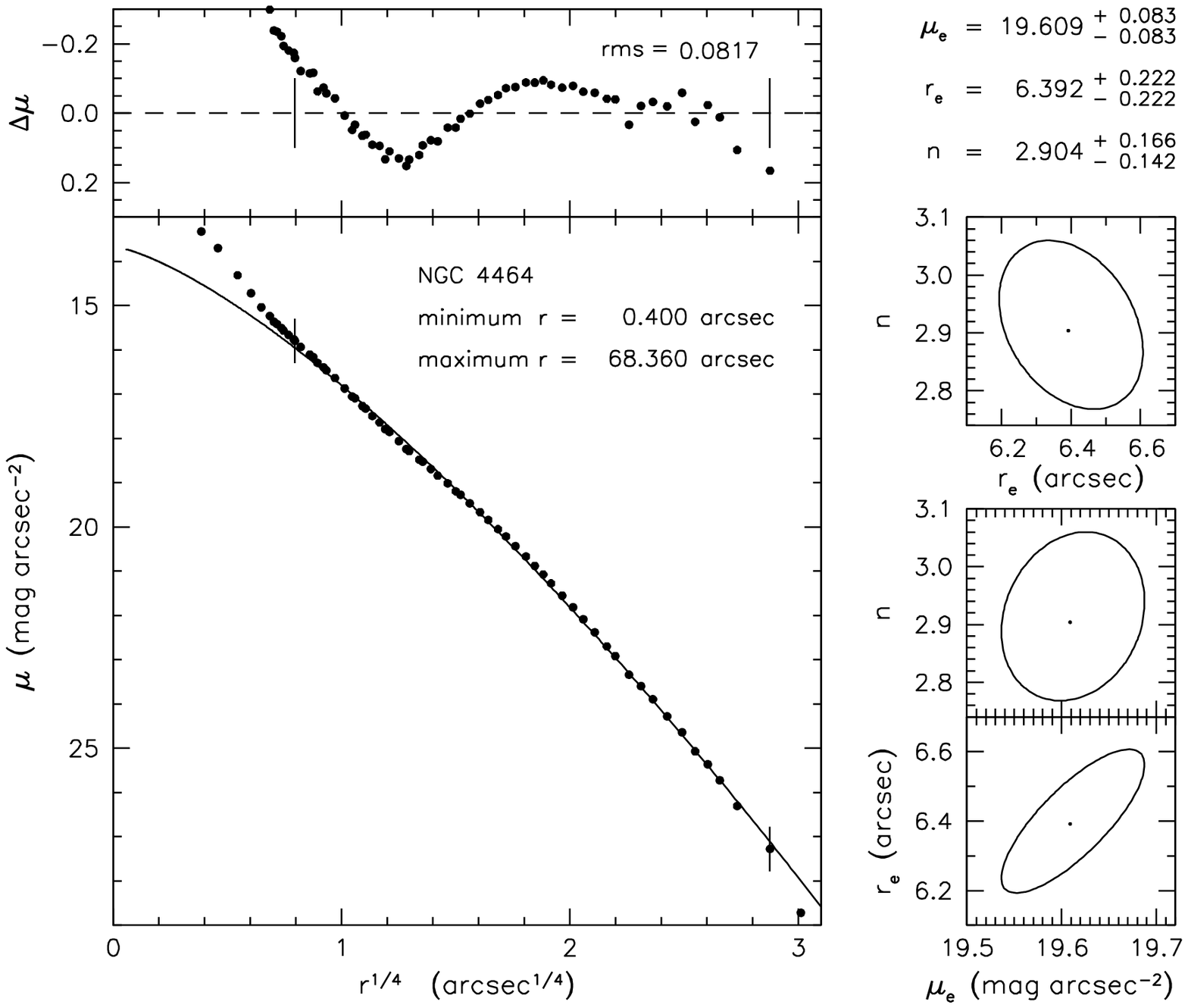}

\figcaption[]
{We use NGC 4464 to illustrate the robustness of our choice of the inner end of our fit range.
That is, we use it as an example of how including extra light in the S\'ersic fit 
produces systematic residuals that are unacceptable.  The {\it top panels\/} show our
adopted fit.  In it, the upward residual produced by the extra light appears to start quite 
suddenly interior to the minimum radius 2\farcs44 used in the fit.  But the change in curvature
of the actual profile is subtle.  Could we extend the fit to smaller radii?  The {\it bottom
panels\/} show that the answer is ``no''.  If we add additional profile points inward to 0\farcs40,
the resulting fit -- while not extremely bad -- has residuals that are
substantially larger than our measurement errors.  More tellingly, the residual profile still 
shows a strong kink at 2\farcs4, and it is systematically curved in a way that implies that
we have included extra light in the fit.  Therefore this fit is not acceptable.  We emphasize 
the importance of the high accuracy and dynamic range of our profile data.  Without it, we would
be much less sure that the upper fit is valid while the lower fit is not.  On the other hand, 
note that our scientific conclusions that $n < 4$ and that there is extra light are robust 
enough to be evident in both fits.  Also, the parameters derived from the bottom fit would
not significantly change our fundamental plane parameter correlations (Figures 34, 37, and 38).
}

\eject\clearpage

\figurenum{65}

\centerline{\null} \vfill

\includegraphics{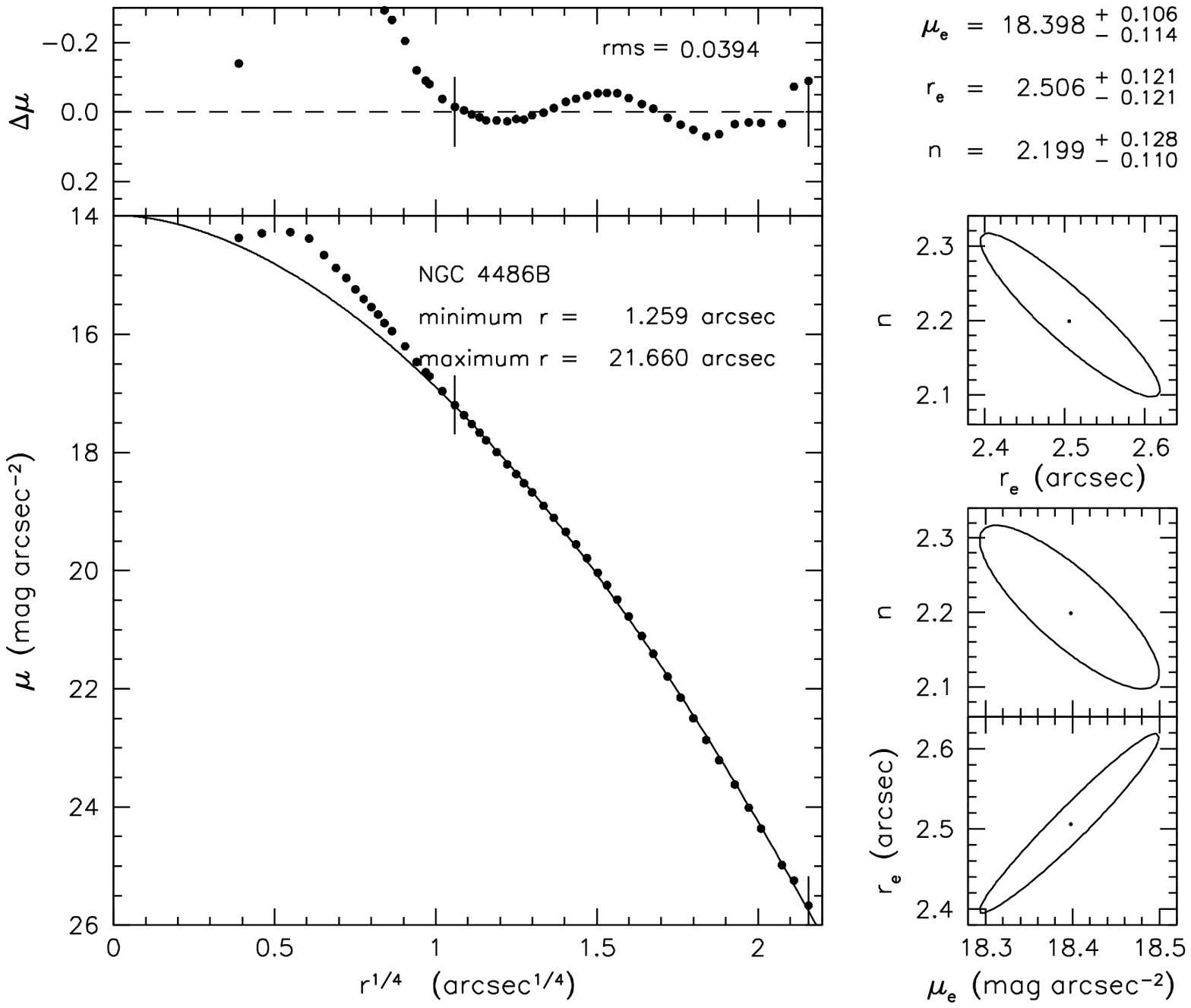}

\includegraphics{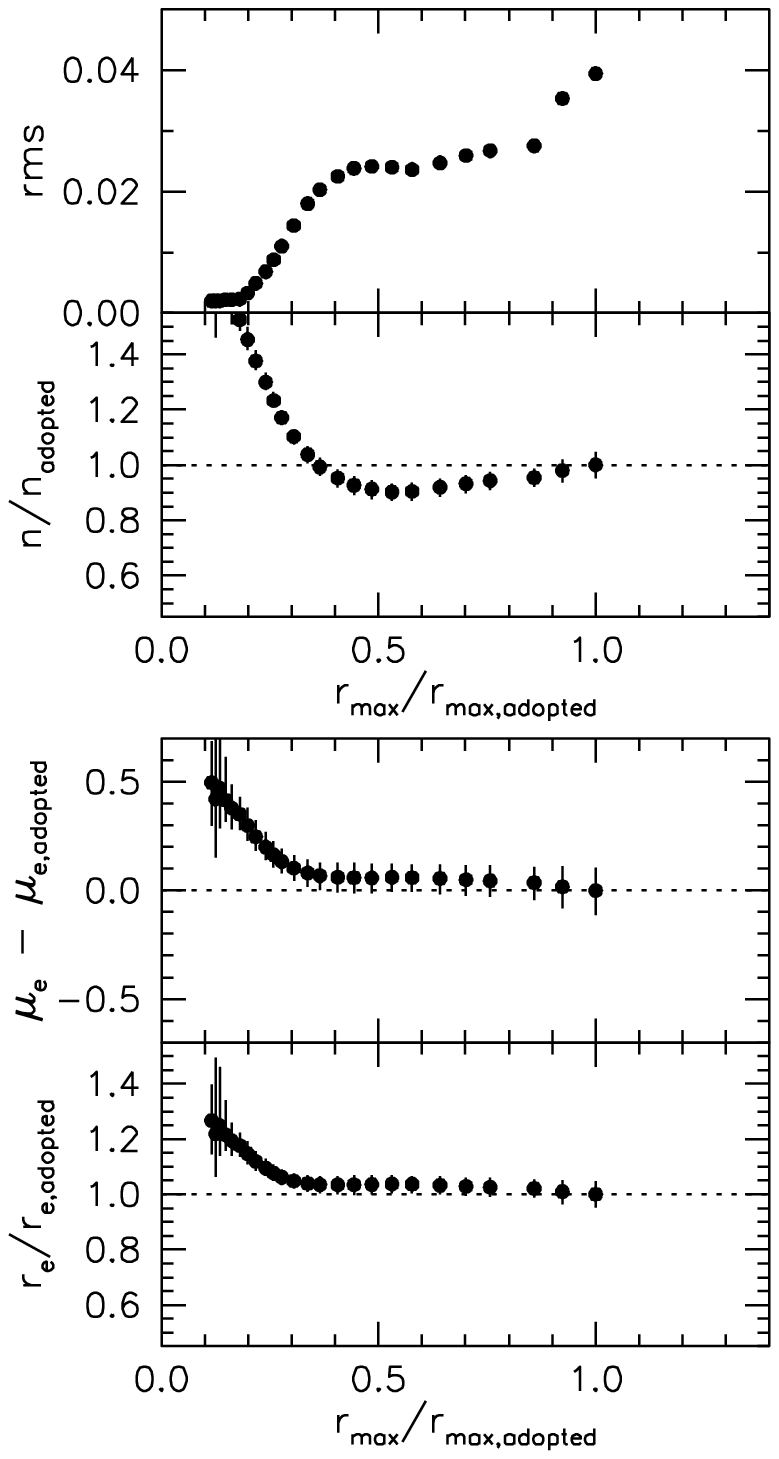}

\includegraphics{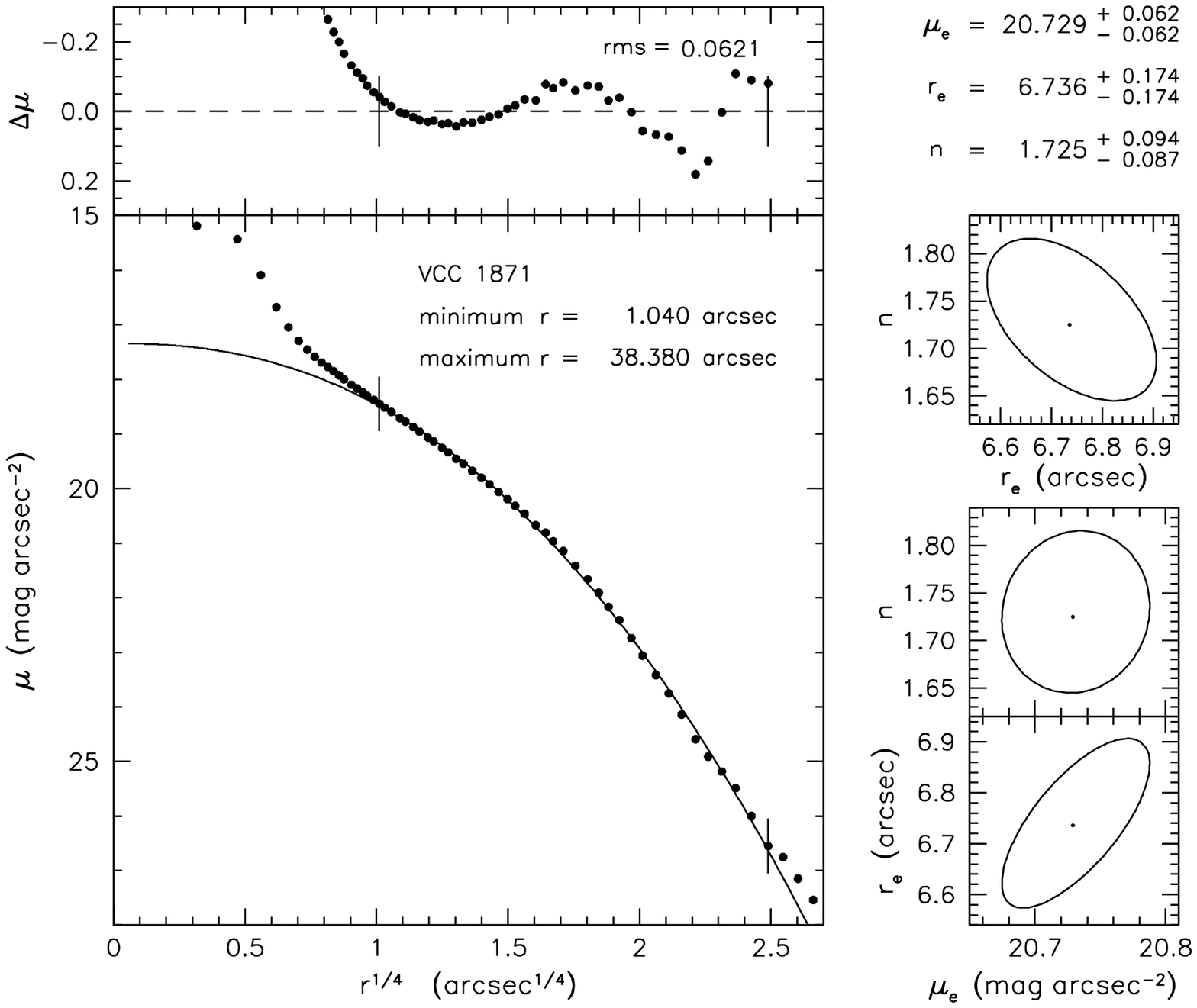}

\includegraphics{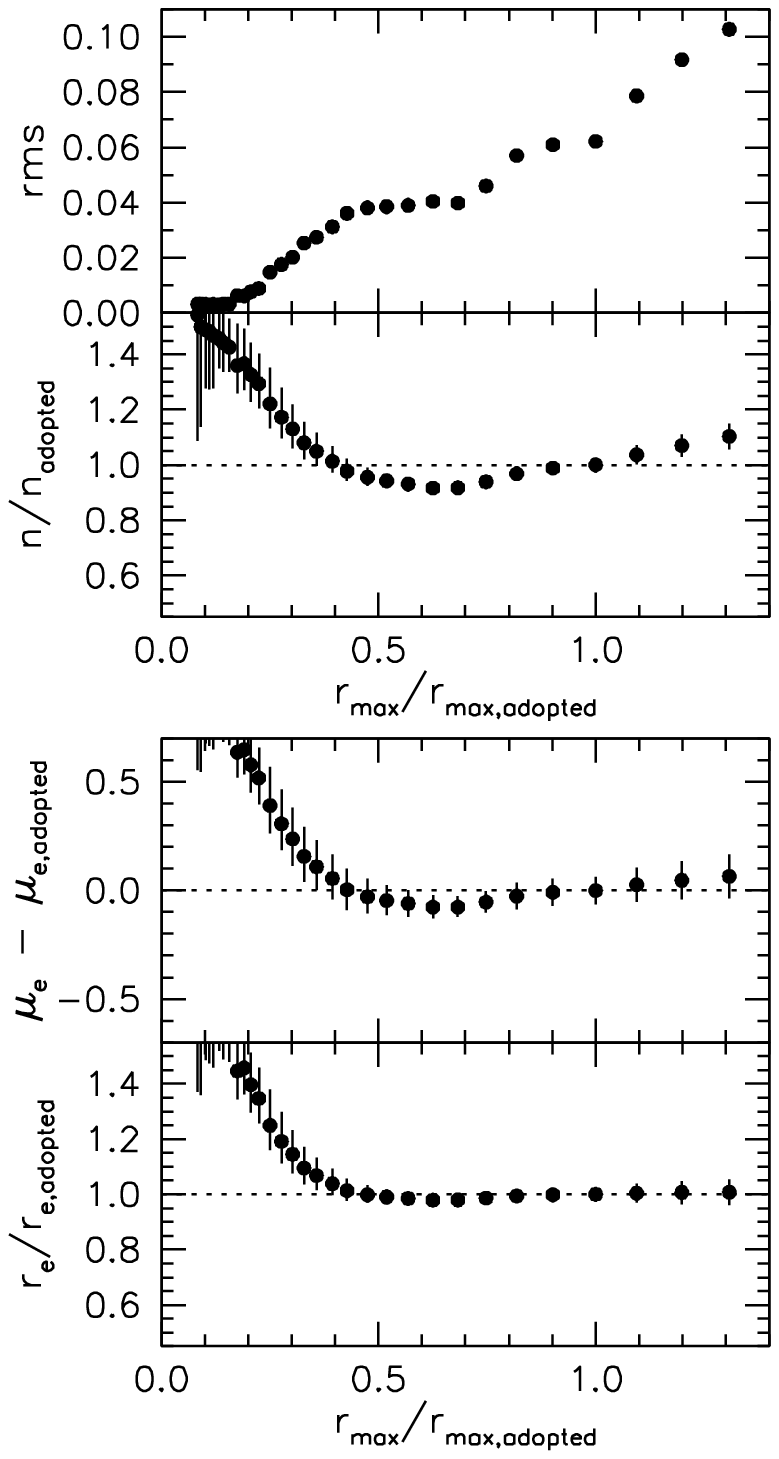}

\figcaption[]
{S\'ersic function fits to the major-axis profiles of NGC 4486B and VCC 1871.  When the 
well known double nucleus of NGC 4486B (Lauer \etal 1996) is measured with a program that
fits elliptical isophotes, the result looks like a core profile.  However, the double 
nucleus actually is a feature inside the extra light component of a normal, tiny elliptical 
with a normal S\'ersic function profile and a robust value of $n = 2.20^{+0.13}_{-0.11}$.
}

\eject\clearpage

\figurenum{66}

\centerline{\null} \vfill

\includegraphics{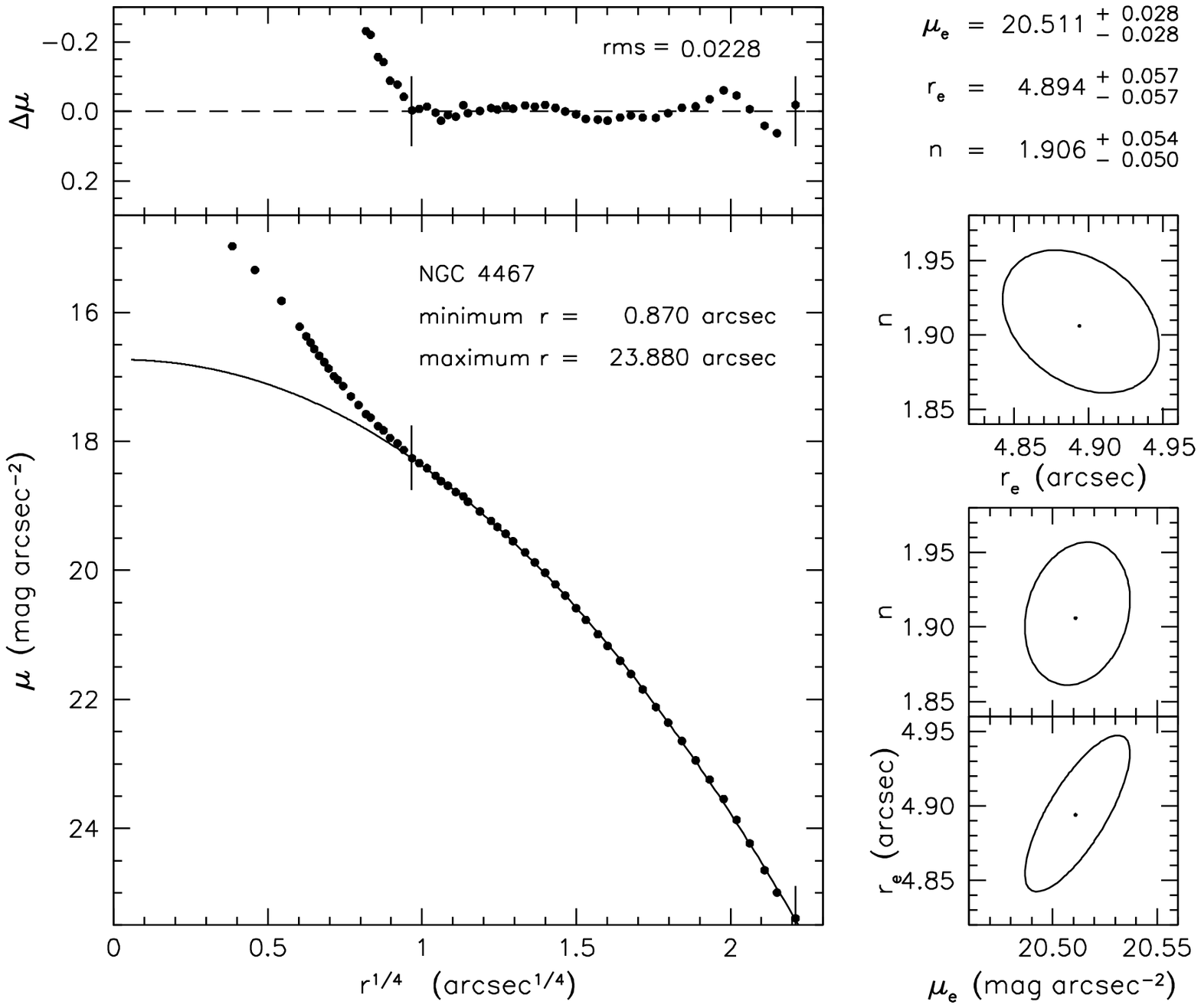}

\includegraphics{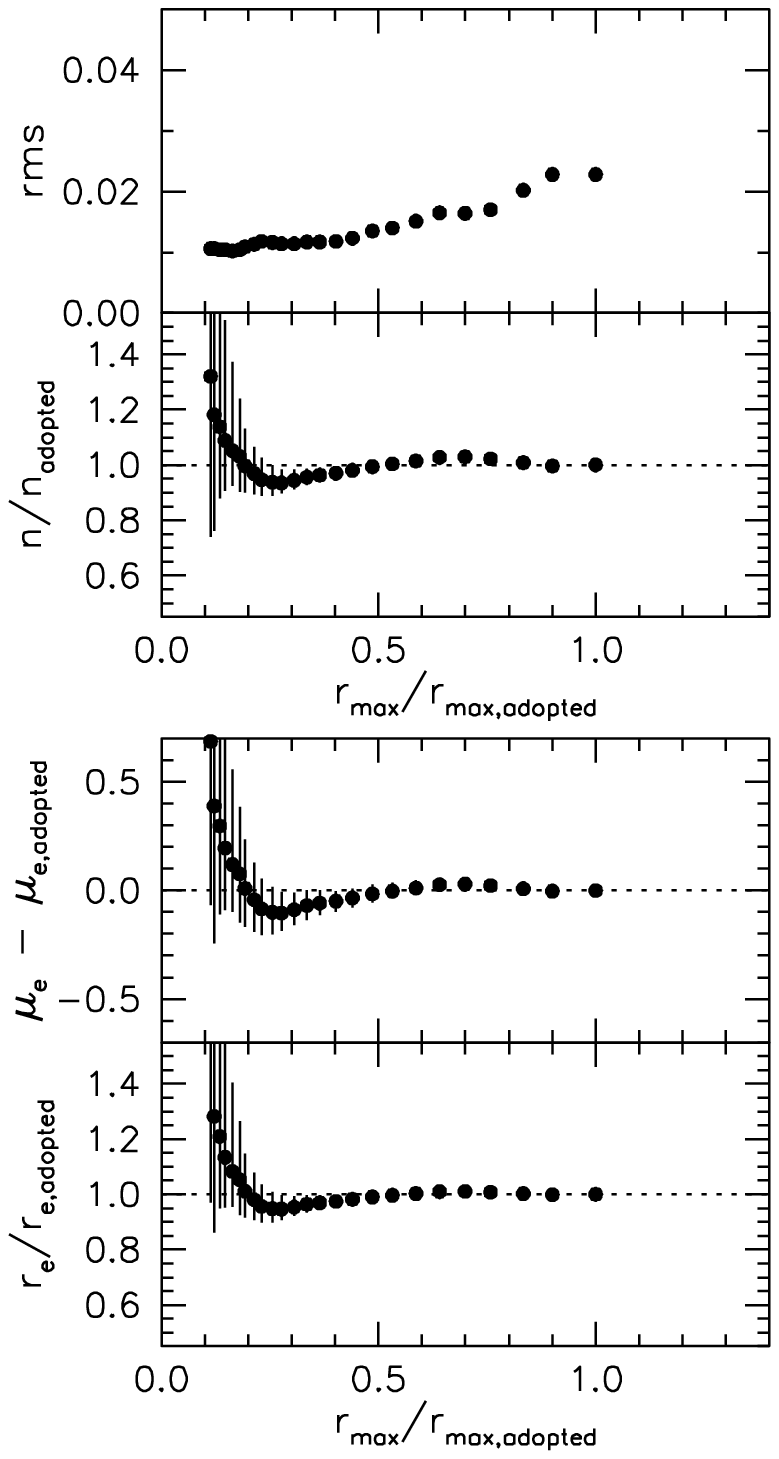}

\includegraphics{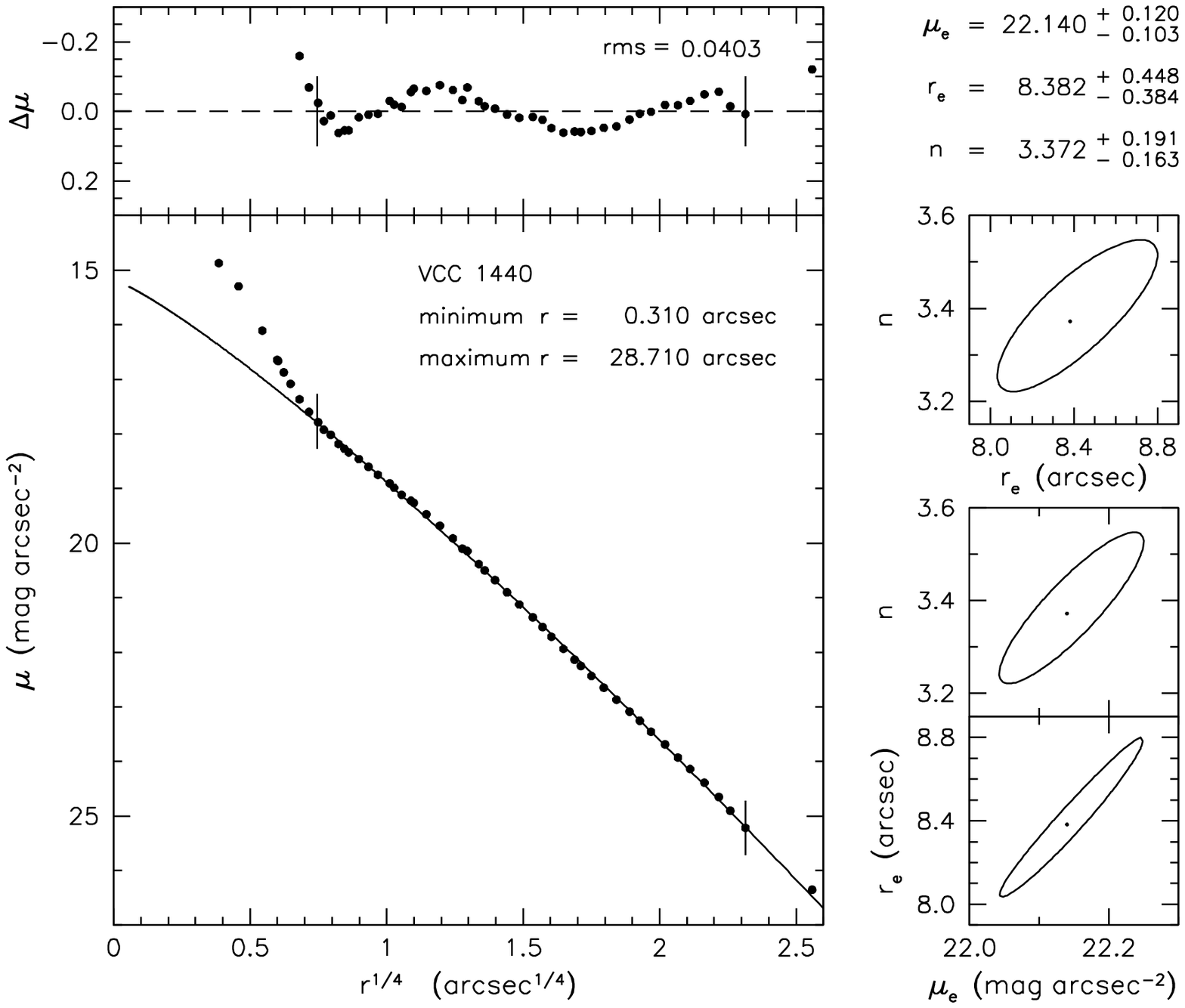}

\includegraphics{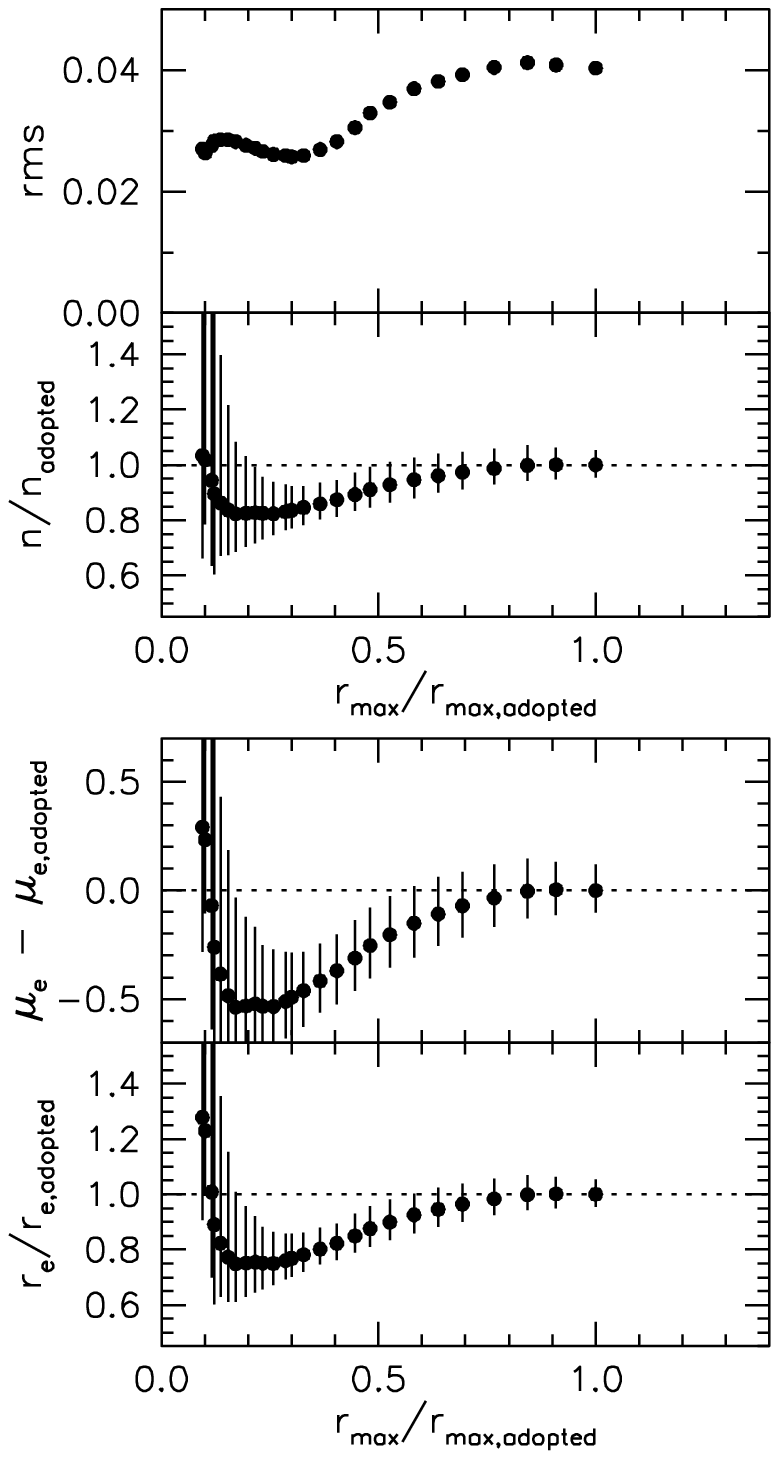}

\figcaption[]
{S\'ersic function fits to the major-axis profiles of NGC 4467 and VCC 1440.
They have $M_{VT} = -16.92$ and $-16.85$, respectively.  That is, they are Virgo cluster
analogs of M{\ts}32 ($M_{VT} = -16.69$).  
}

\eject\clearpage

\figurenum{67}

\centerline{\null} \vfill

\includegraphics{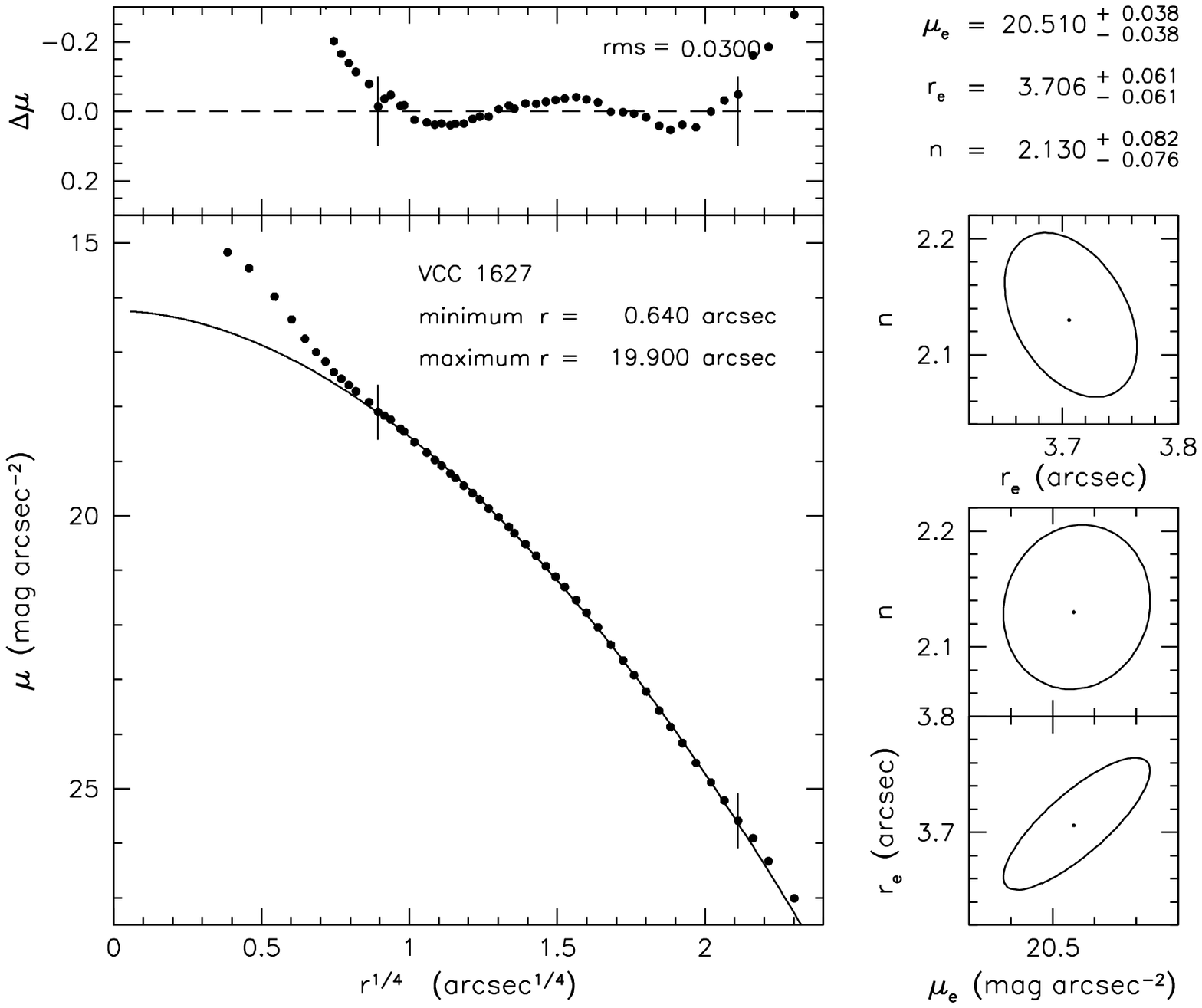}

\includegraphics{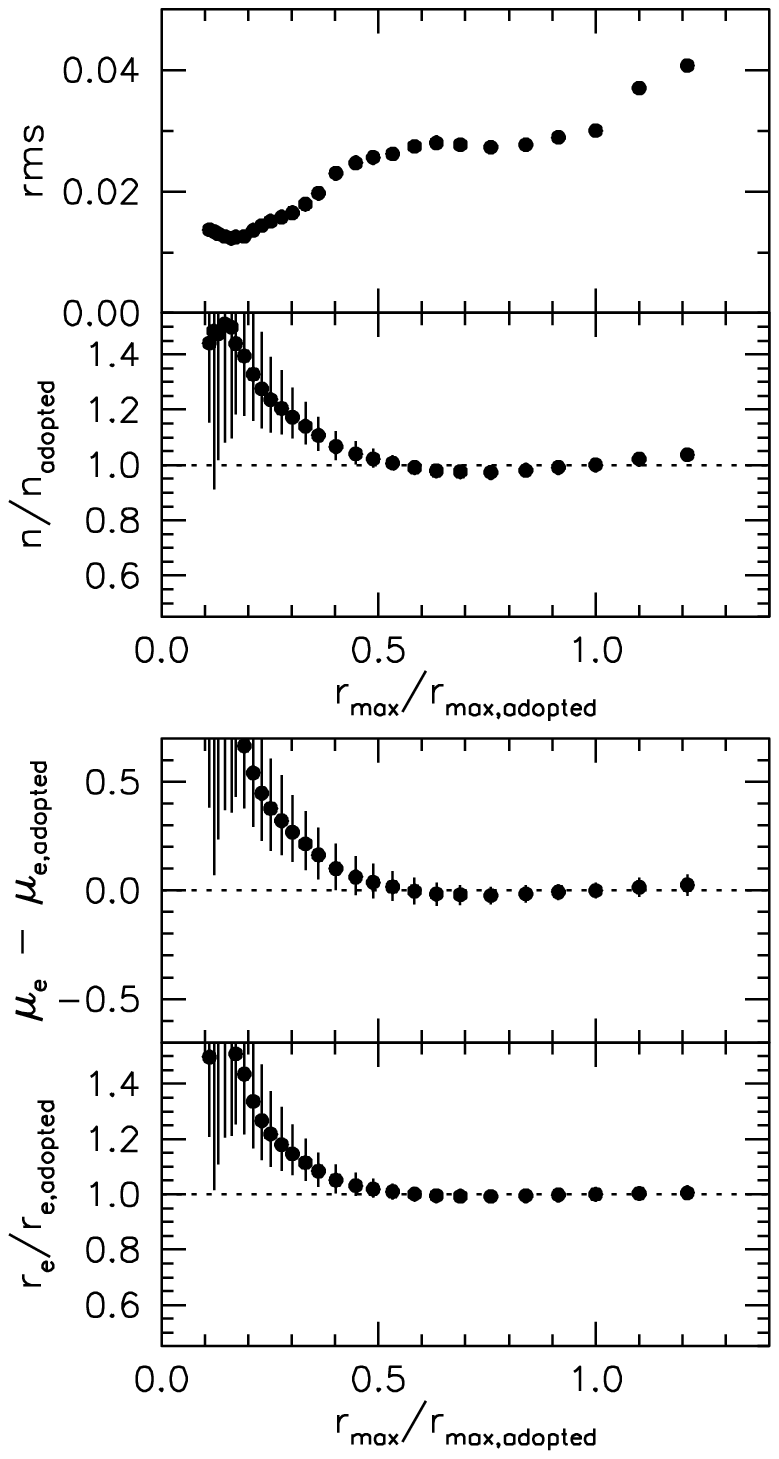}

\includegraphics{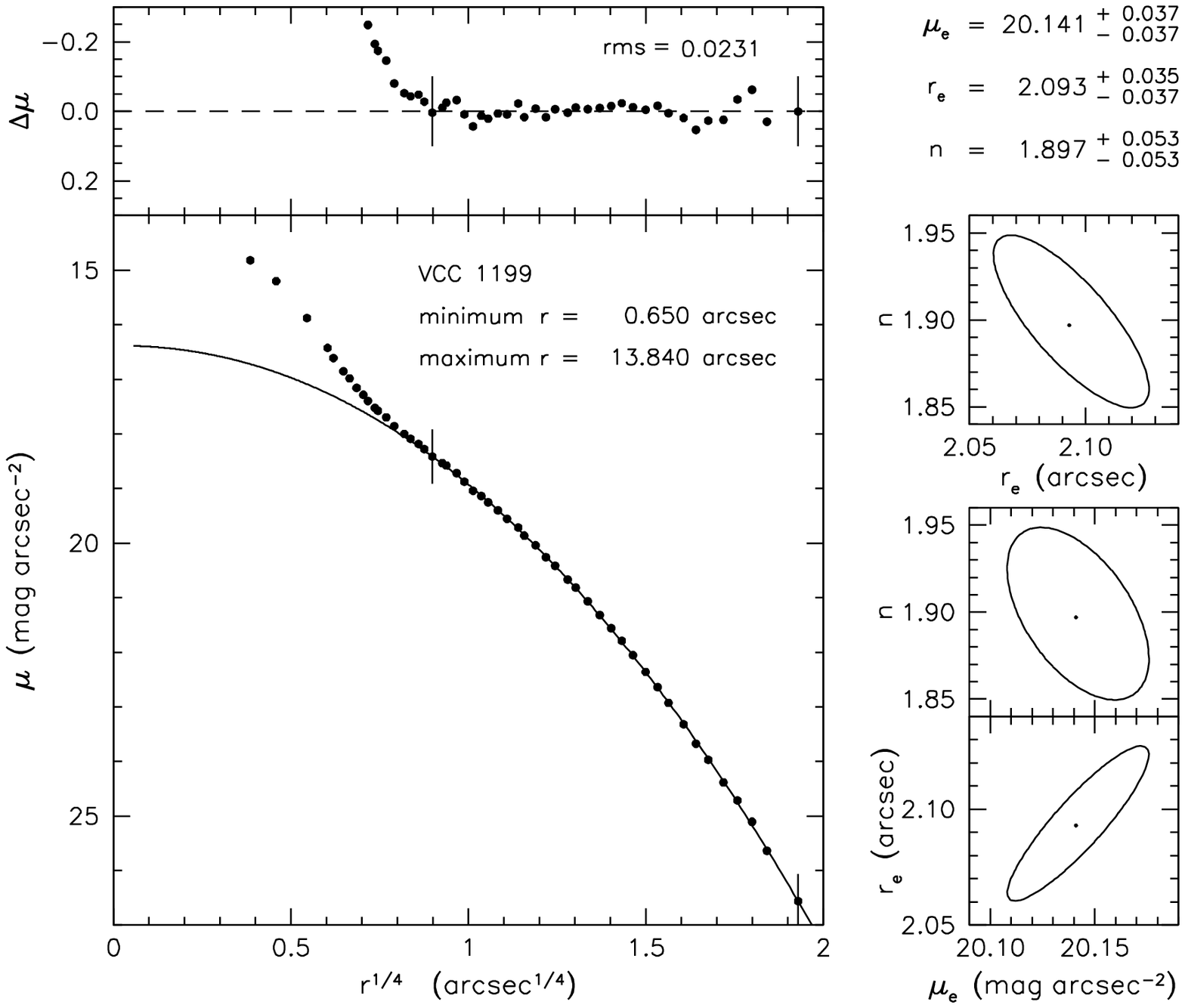}

\includegraphics{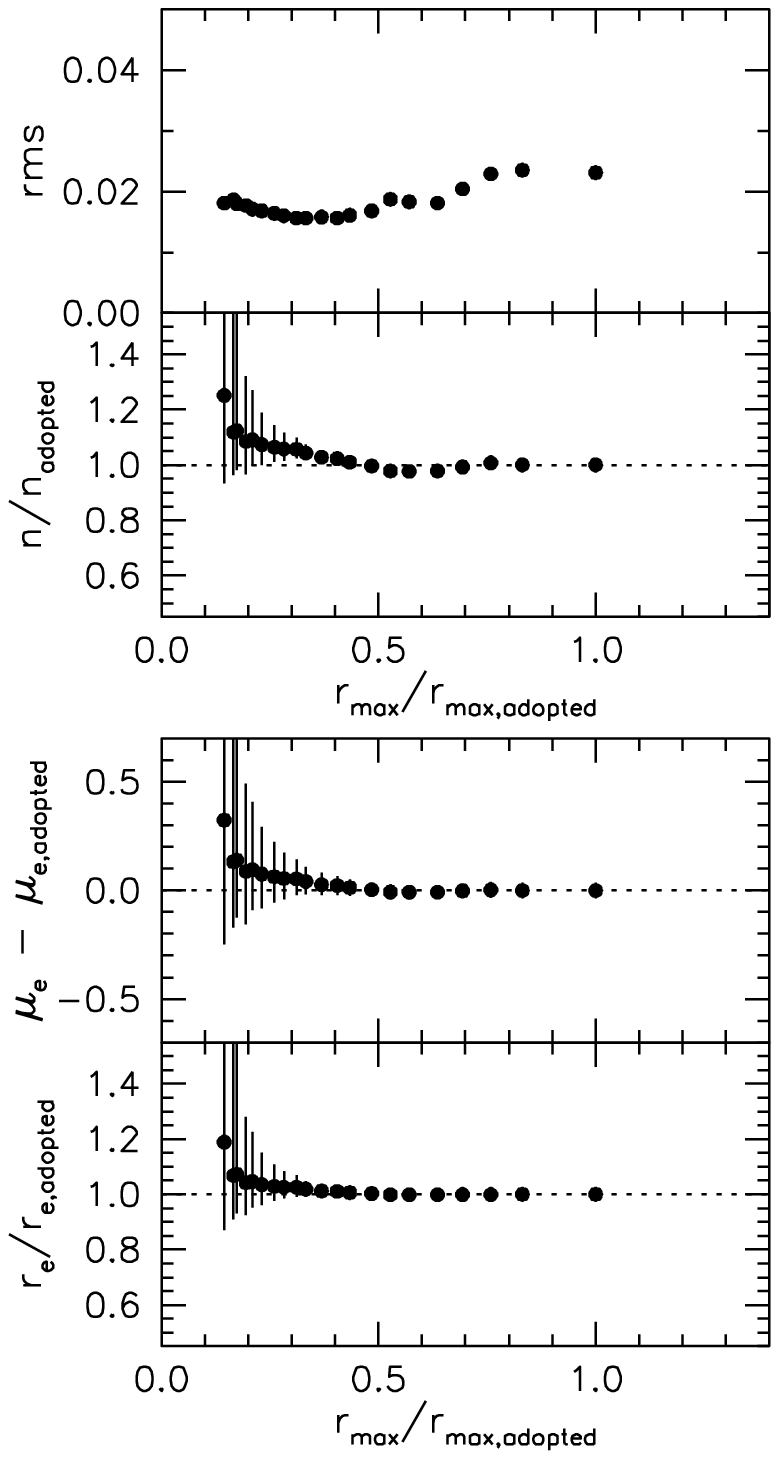}

\figcaption[]
{S\'ersic function fits to the major-axis profiles of VCC 1627 and VCC 1199.  The layout
is as in Figure 49.  At $M_{VT} = -16.44$ and $-15.53$, respectively, these are the 
lowest-luminosity (known) true ellipticals in Virgo.  VCC 1199 is about 1 mag fainter 
than M{\ts}32 ($M_{VT} = -16.69$).
}

\eject\clearpage

\figurenum{68}

\centerline{\null} \vfill

\includegraphics{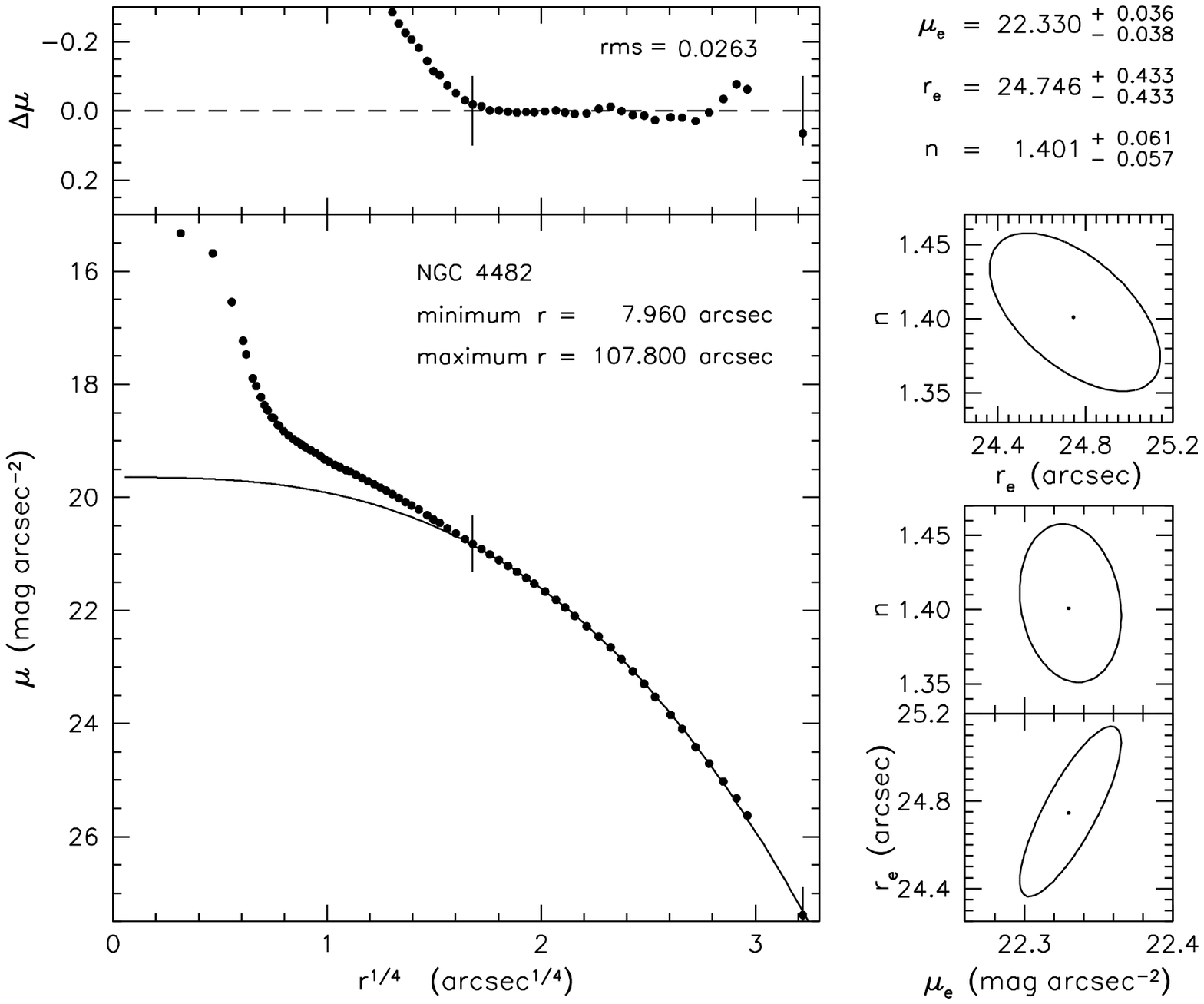}

\includegraphics{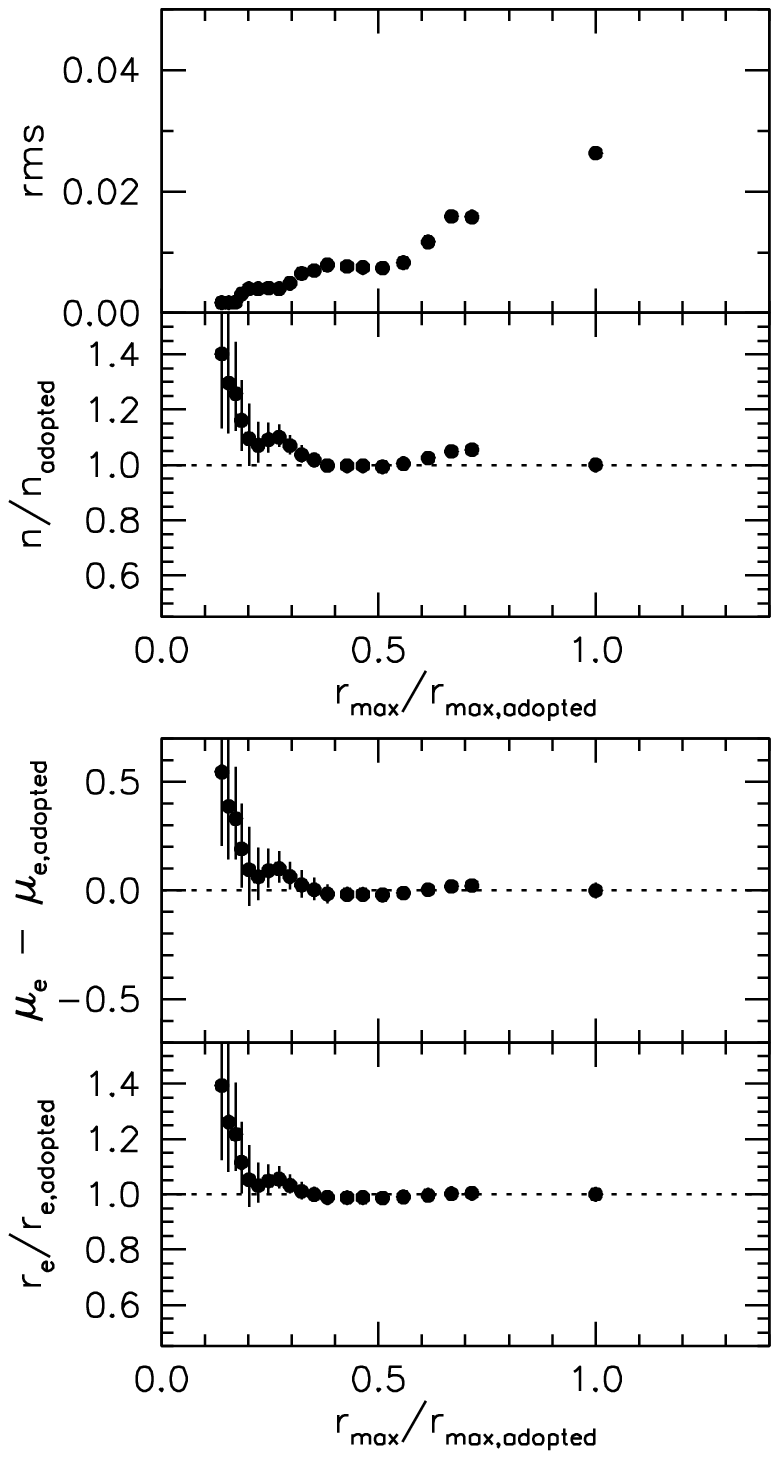}

\includegraphics{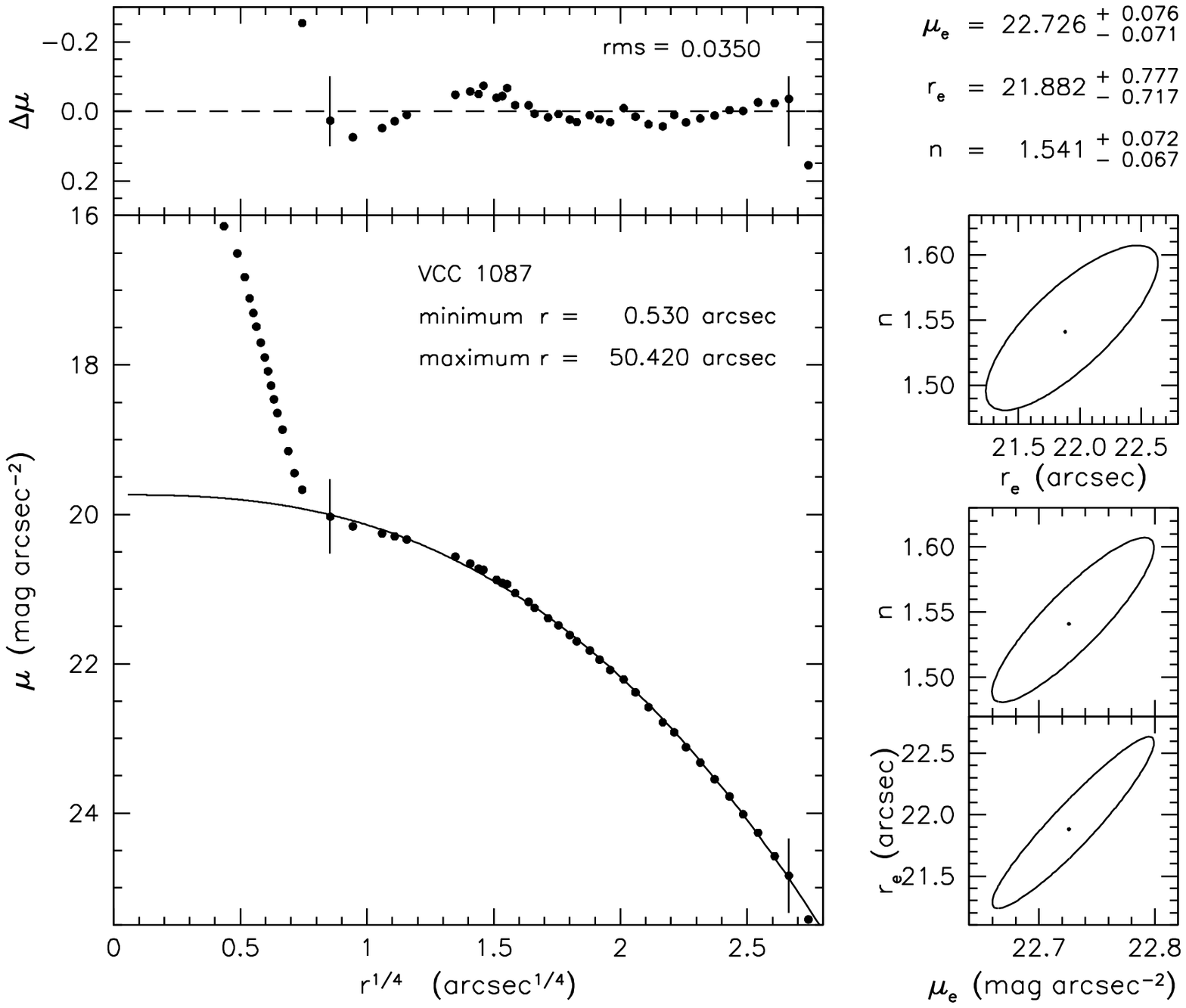}

\includegraphics{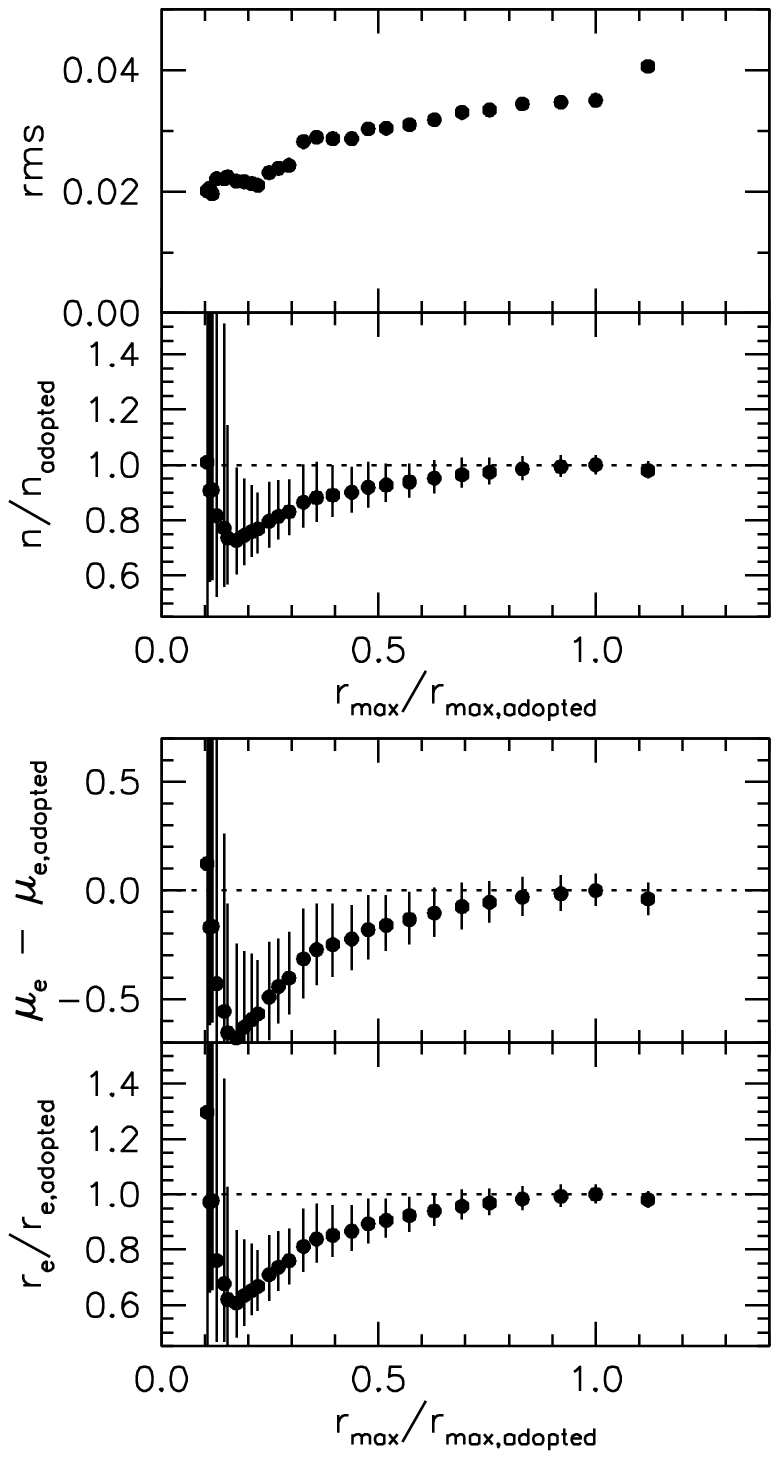}

\figcaption[]
{S\'ersic function fits to the major-axis profiles of NGC 4482 and VCC 1087, 
the brightest spheroidal galaxies in our sample.  The layout is as in Figure 49.
Spheroidals show signs of more complication in their profiles than do ellipticals.
The inner part of NGC 4482 outside the prominent nuclear star cluster is not fit
by a S\'ersic function.  The fits for VCC 1087 (this page), VCC 1355 (Fig.~69), and
VCC 1407 (Fig.~71) show features similar to those of the ``Type~II'' exponential 
profiles discussed by Freeman (1970).  Our S\'ersic fits have excellent to good, 
small RMS residuals.  But the profile data are accurate enough to show subtle
systematic curvature in the residuals.  The form of the curvature is such that a
S\'ersic function with a {\it slightly\/} higher $n$ would fit better at large $r$.
But then the inner profile outside the nucleus would drop below the inward
extrapolation of the outer S\'ersic fit, exactly as in a ``Type II exponential''.  
This is a subtle similarity to disk galaxies that we note in addition to the more
obvious similarities revealed by the fundamental plane correlations (Figures 1, 34, 
37, and 38; \S\S\ts2.1 and 8).}

\eject\clearpage

\figurenum{69}

\centerline{\null} \vfill

\includegraphics{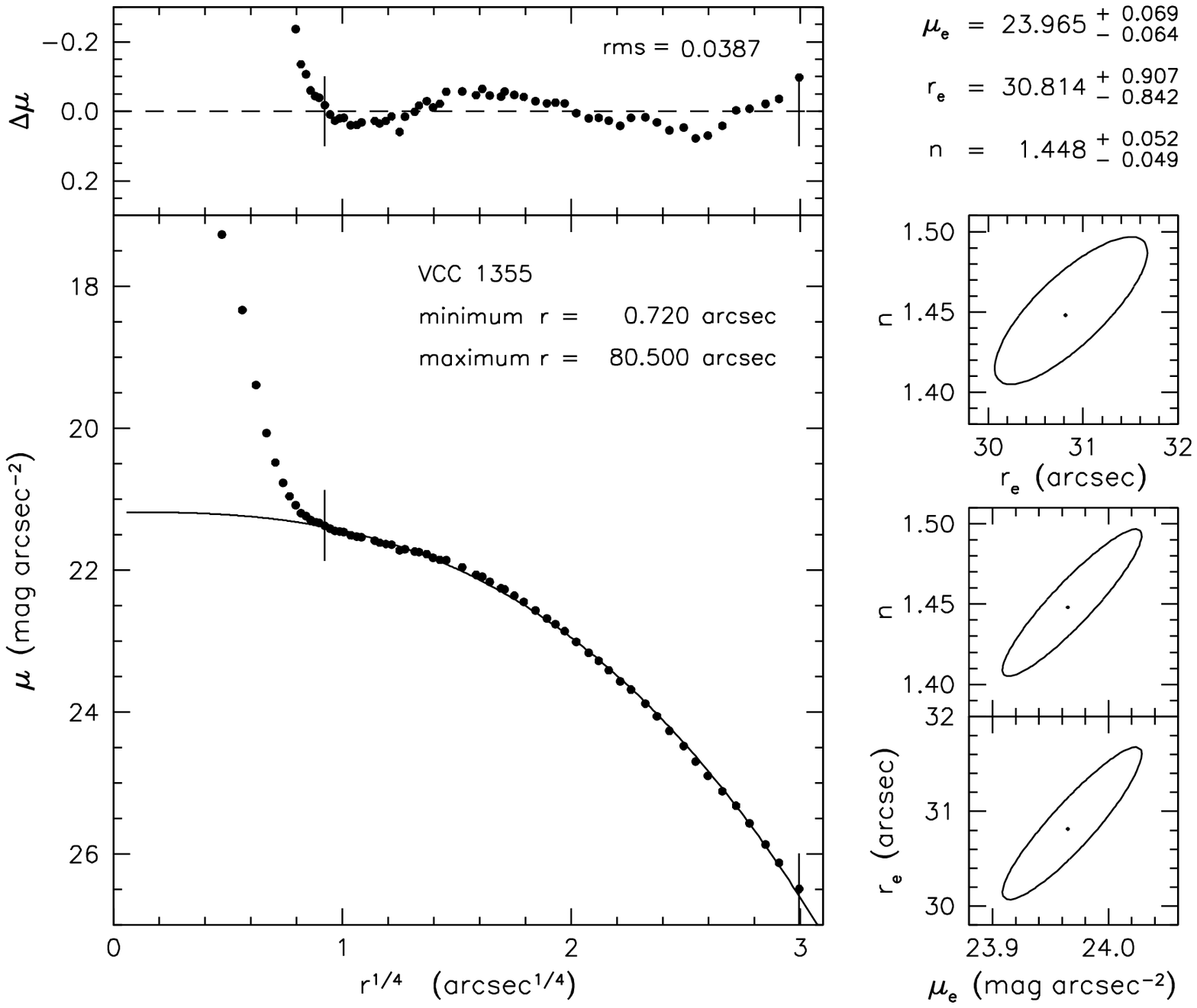}

\includegraphics{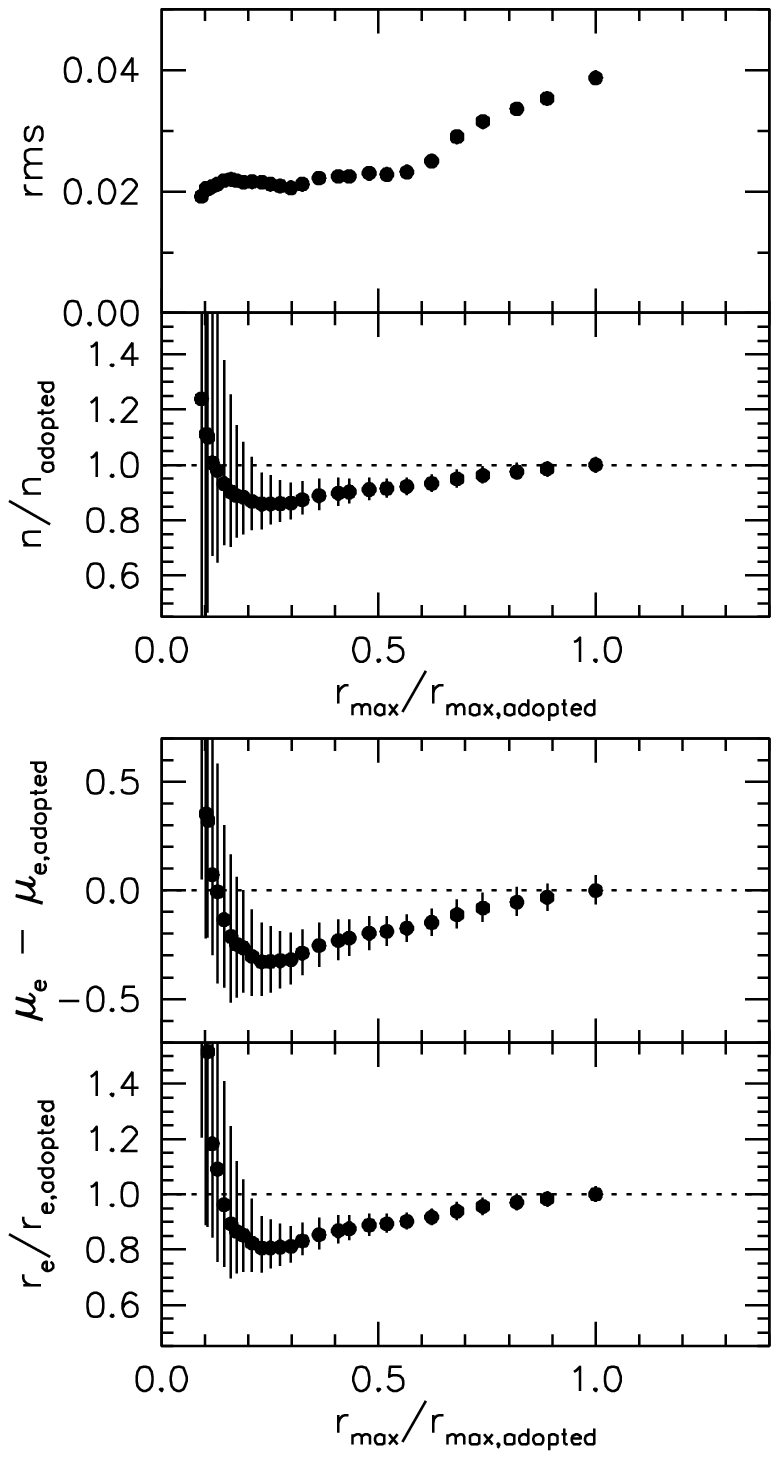}

\includegraphics{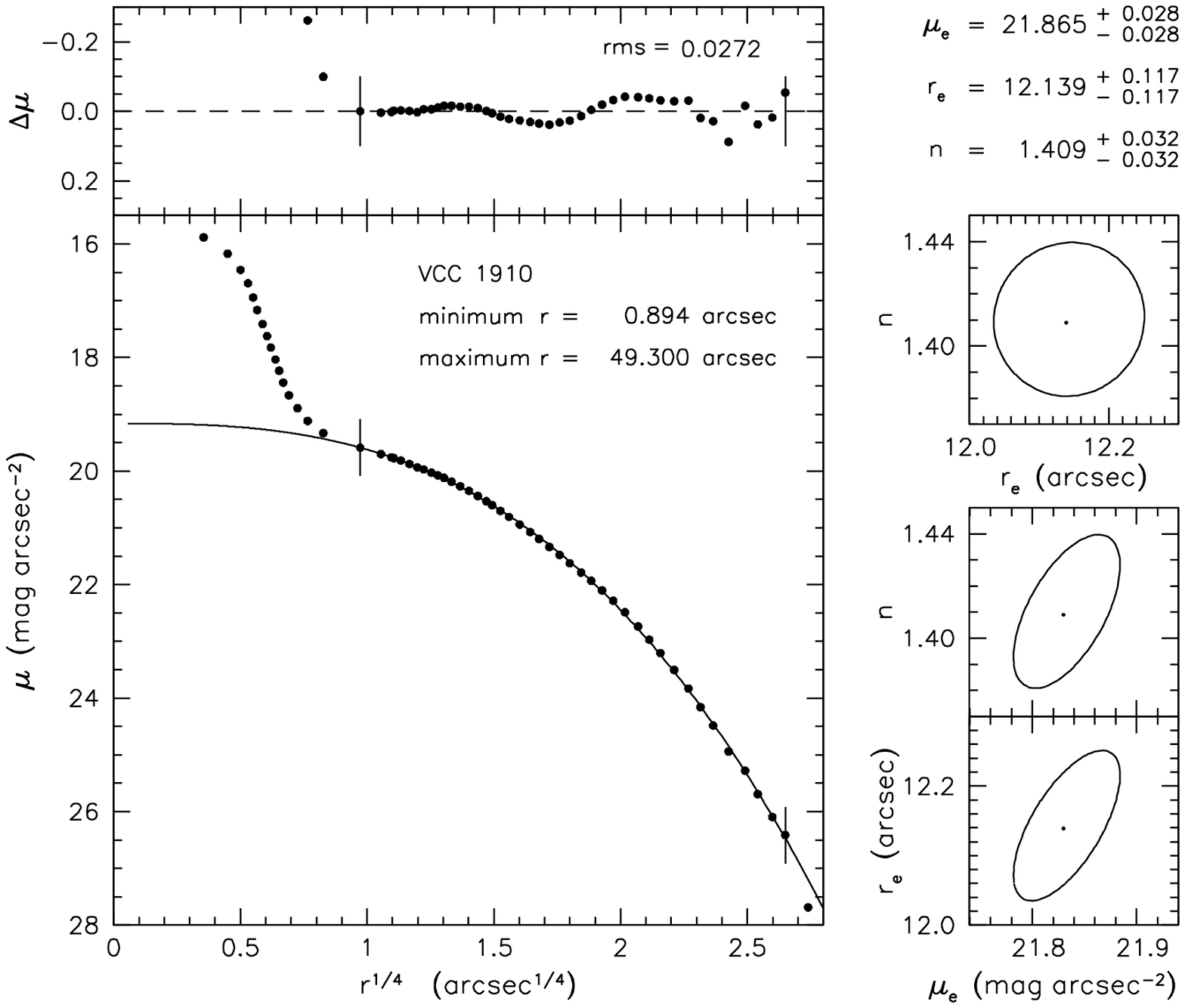}

\includegraphics{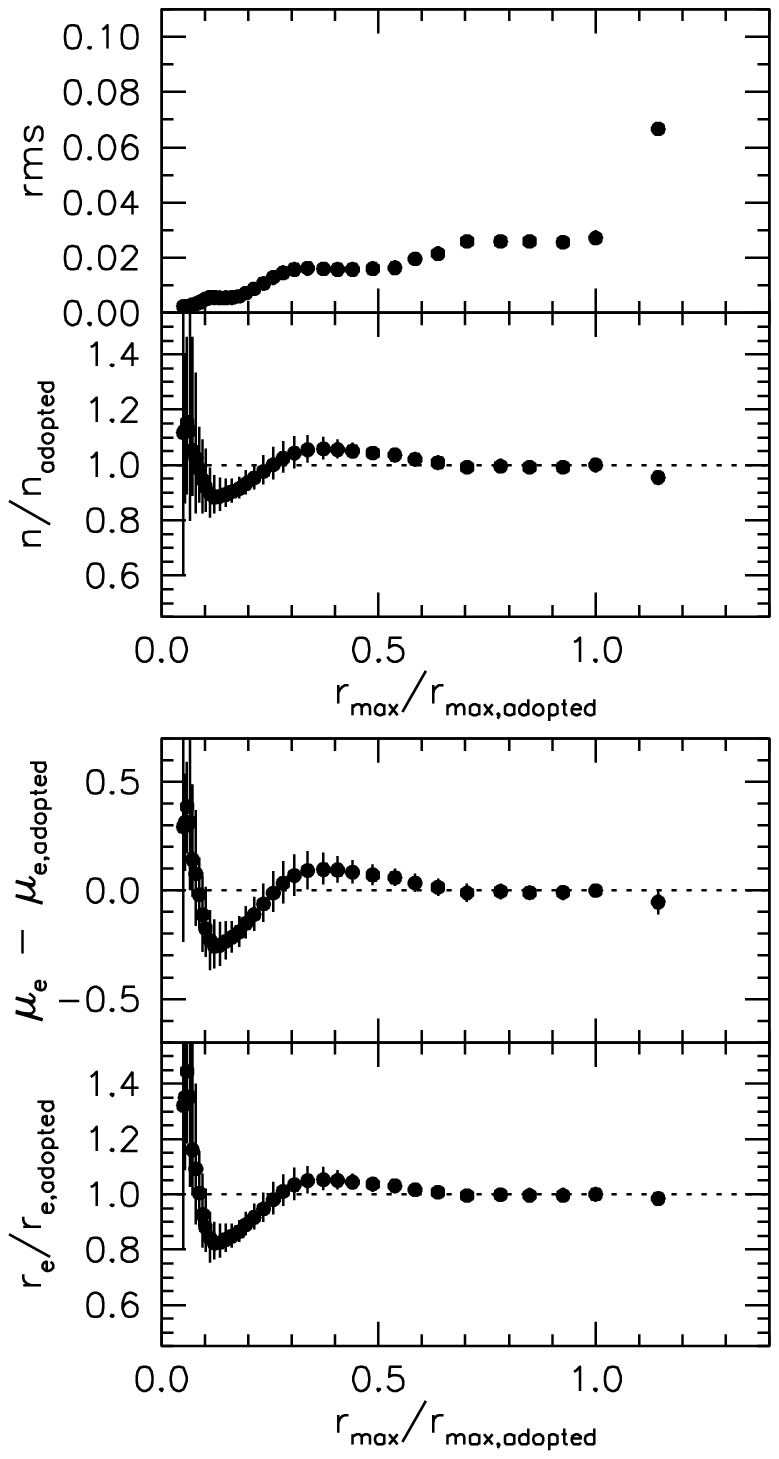}

\figcaption[]
{S\'ersic function fits to the major-axis profiles of the spheroidal galaxies
VCC 1355 and VCC 1910.  The layout is as in Figure 49.  VCC 1355 shows a hint of
``Type II S\'ersic function'' behavior (see the caption to Figure 68).
}

\eject\clearpage

\figurenum{70}

\centerline{\null} \vfill

\includegraphics{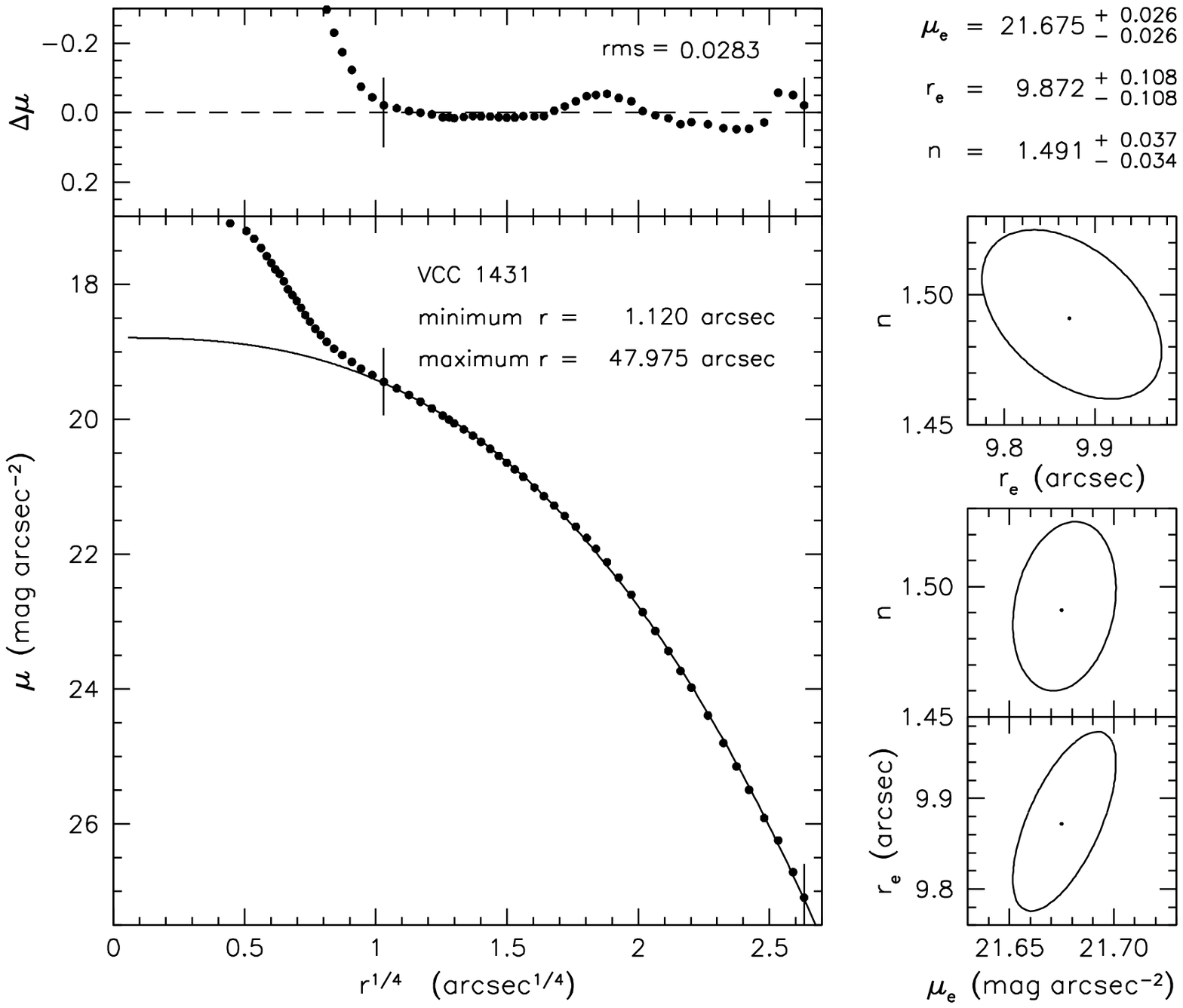}

\includegraphics{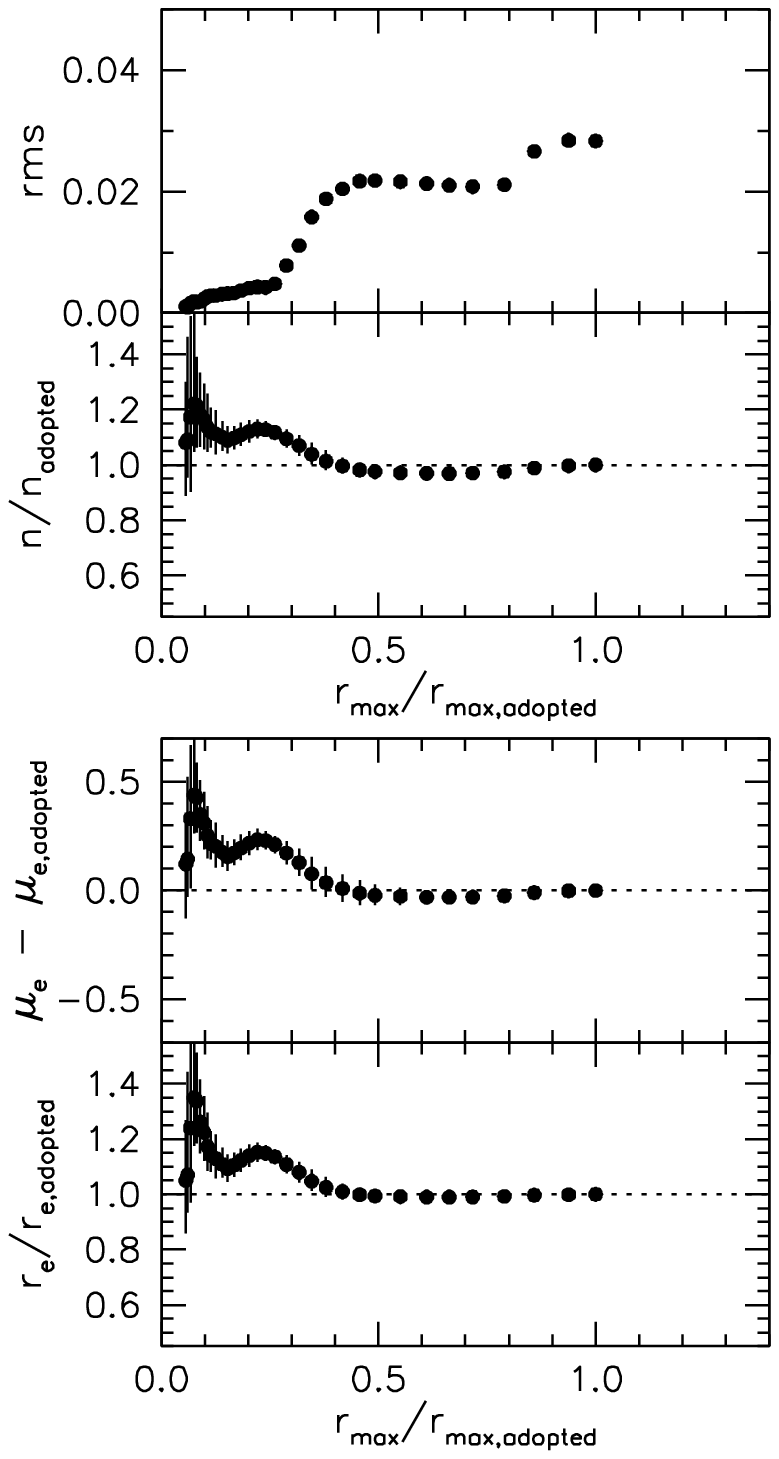}

\includegraphics{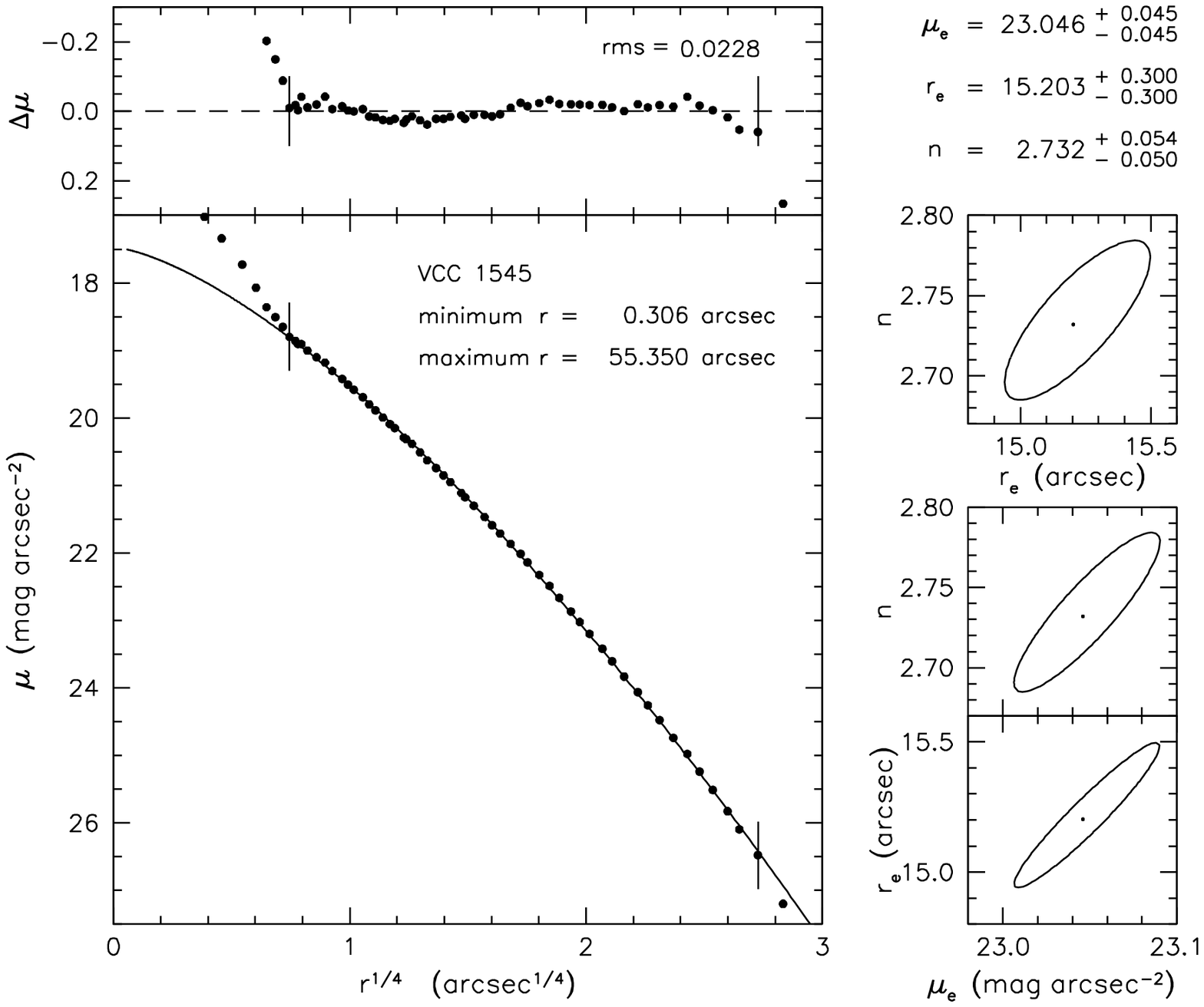}

\includegraphics{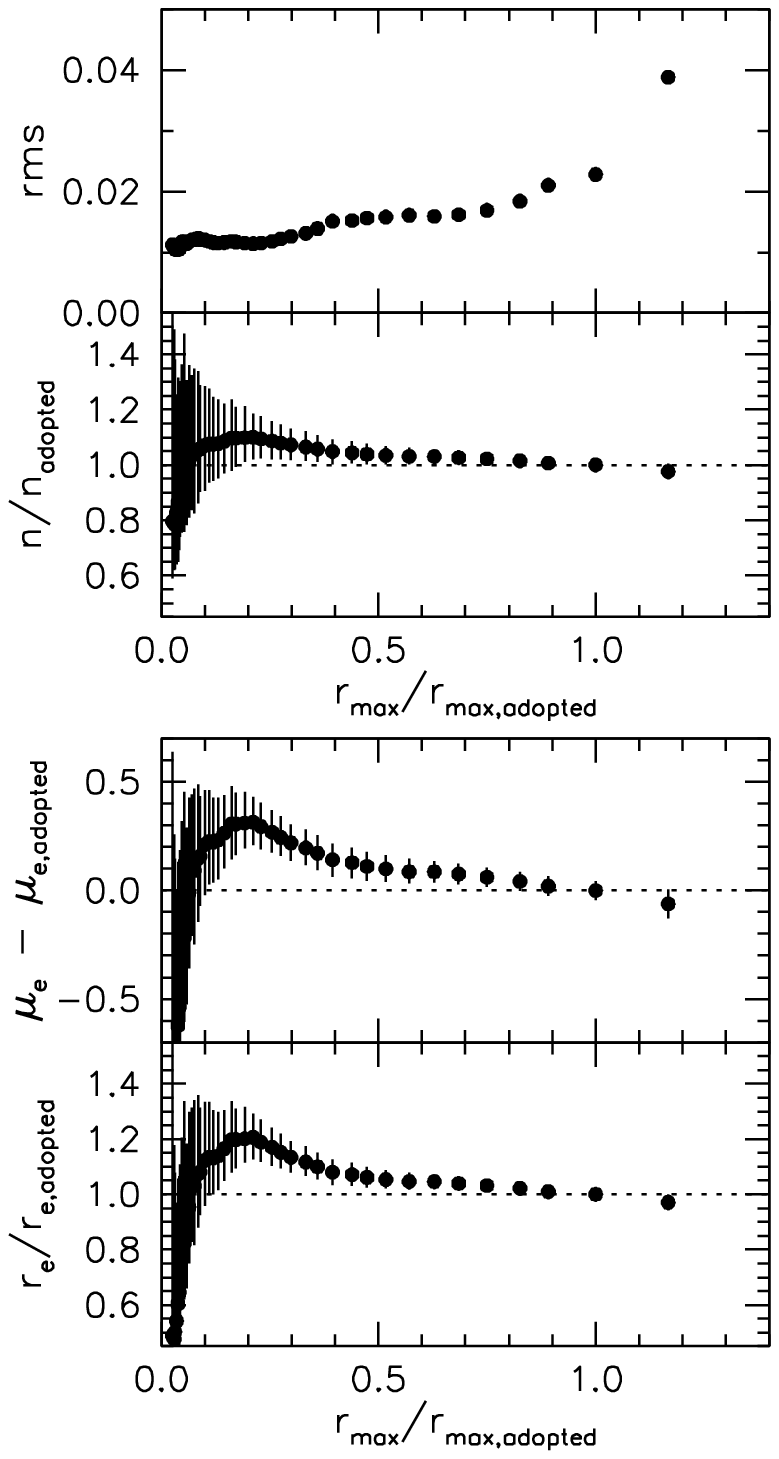}

\figcaption[]
{S\'ersic function fits to the major-axis profiles of the spheroidal galaxies 
VCC 1431 and VCC 1545.  The layout is as in Figure 49.  
}

\eject\clearpage

\figurenum{71}

\centerline{\null} \vfill

\includegraphics{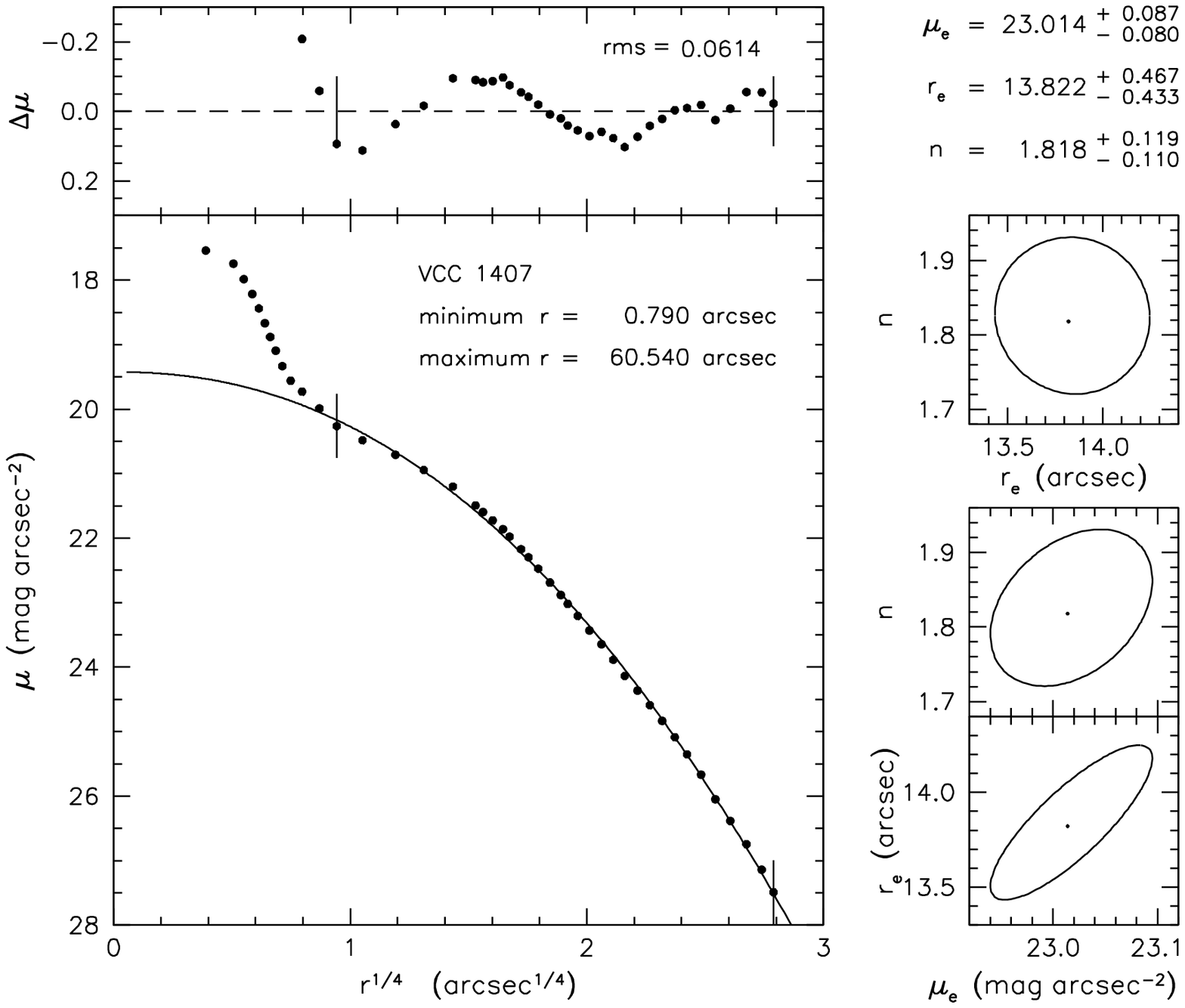}

\includegraphics{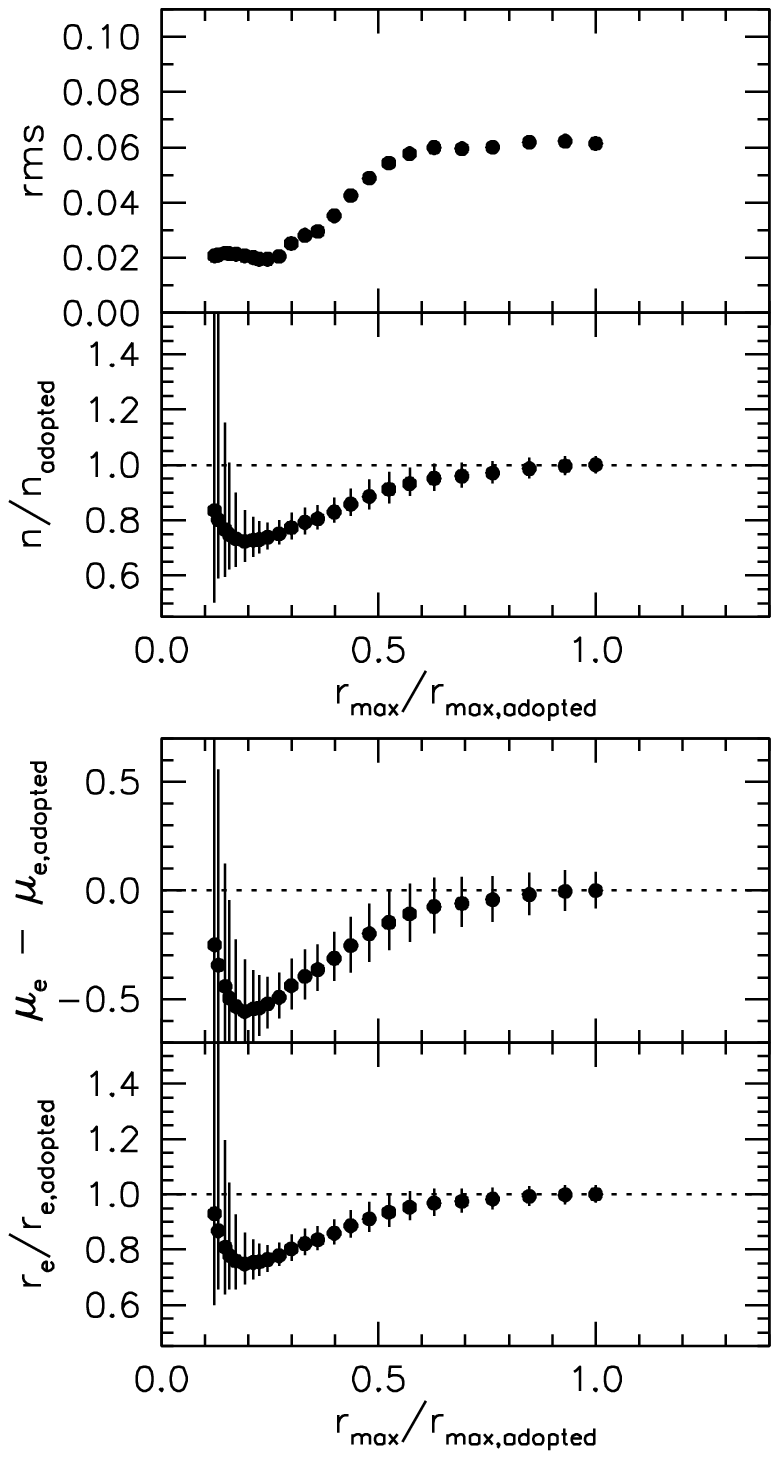}

\includegraphics{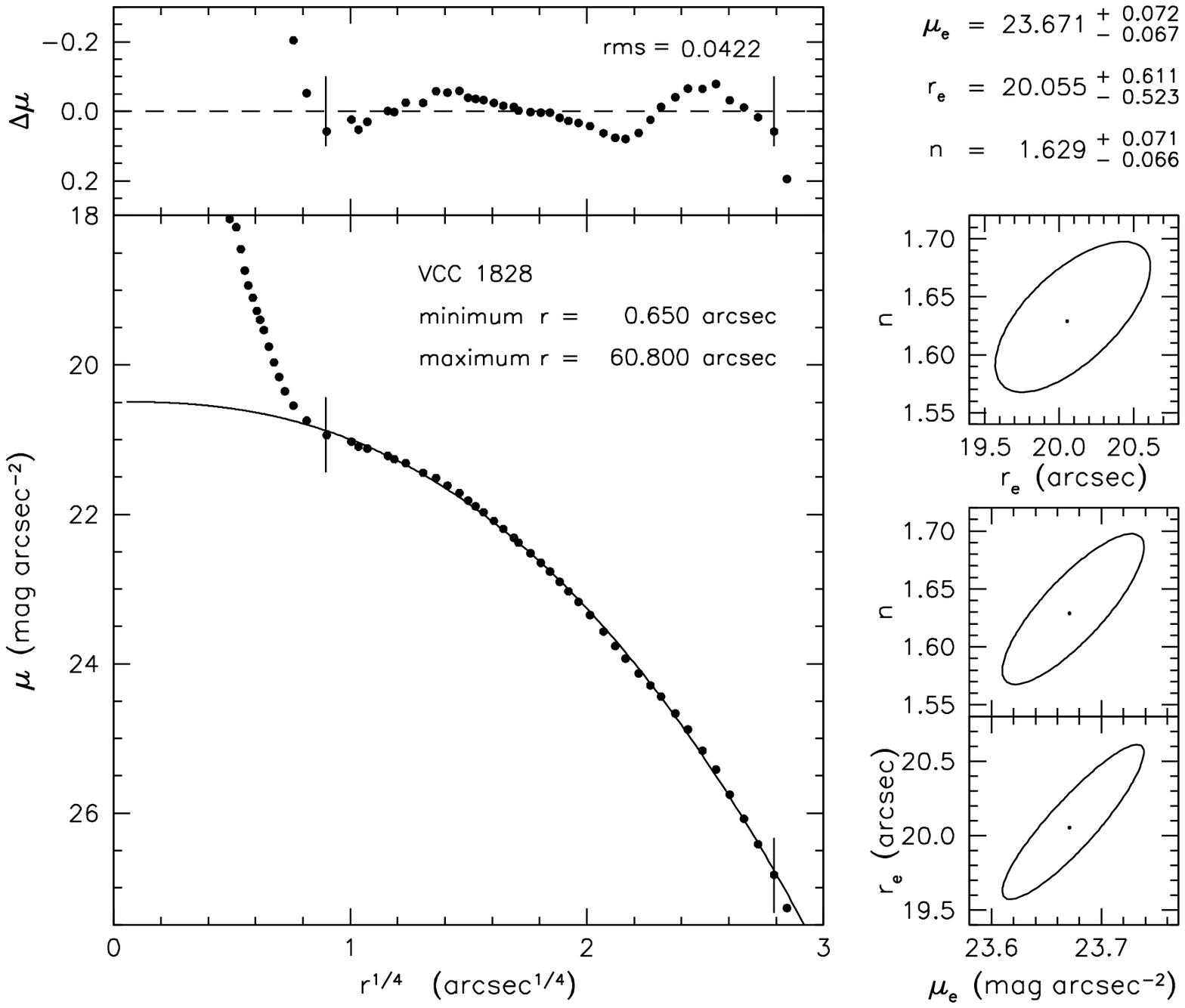}

\includegraphics{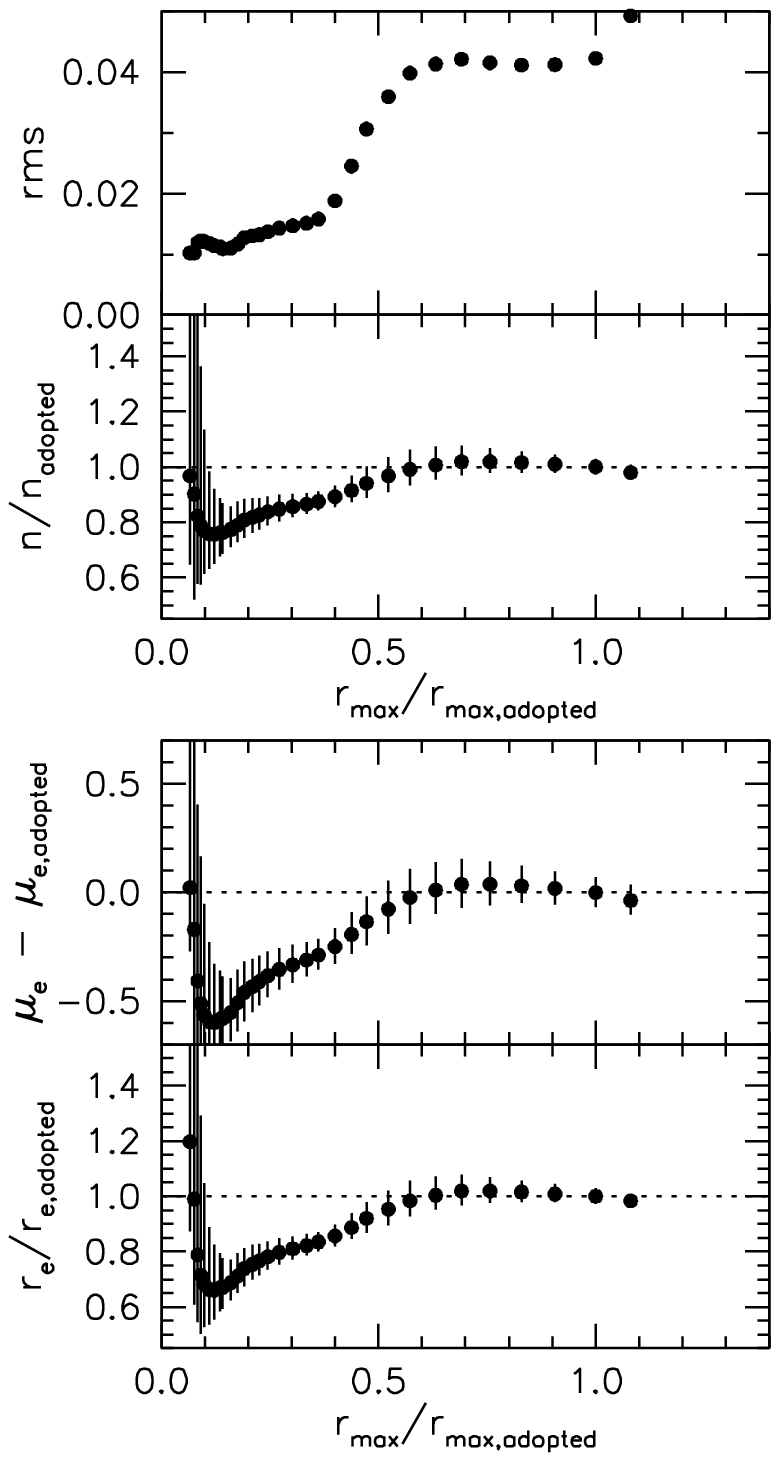}

\figcaption[]
{S\'ersic function fits to the major-axis profiles of the spheroidal galaxies
VCC 1407 and VCC 1828.  The layout is as in Figure 49.  VCC 1407 shows a hint of
``Type II S\'ersic function'' behavior (see the caption to Figure 68).  With
$M_{VT} = -16.71$ and $-16.61$, respectively, these galaxies have almost the
same luminosity as M{\ts}32 ($M_{VT} = -16.69$), but they have much lower S\'ersic
indices and central surface brightnesses. 
}

\eject\clearpage

\figurenum{72}

\centerline{\null} \vfill

\includegraphics{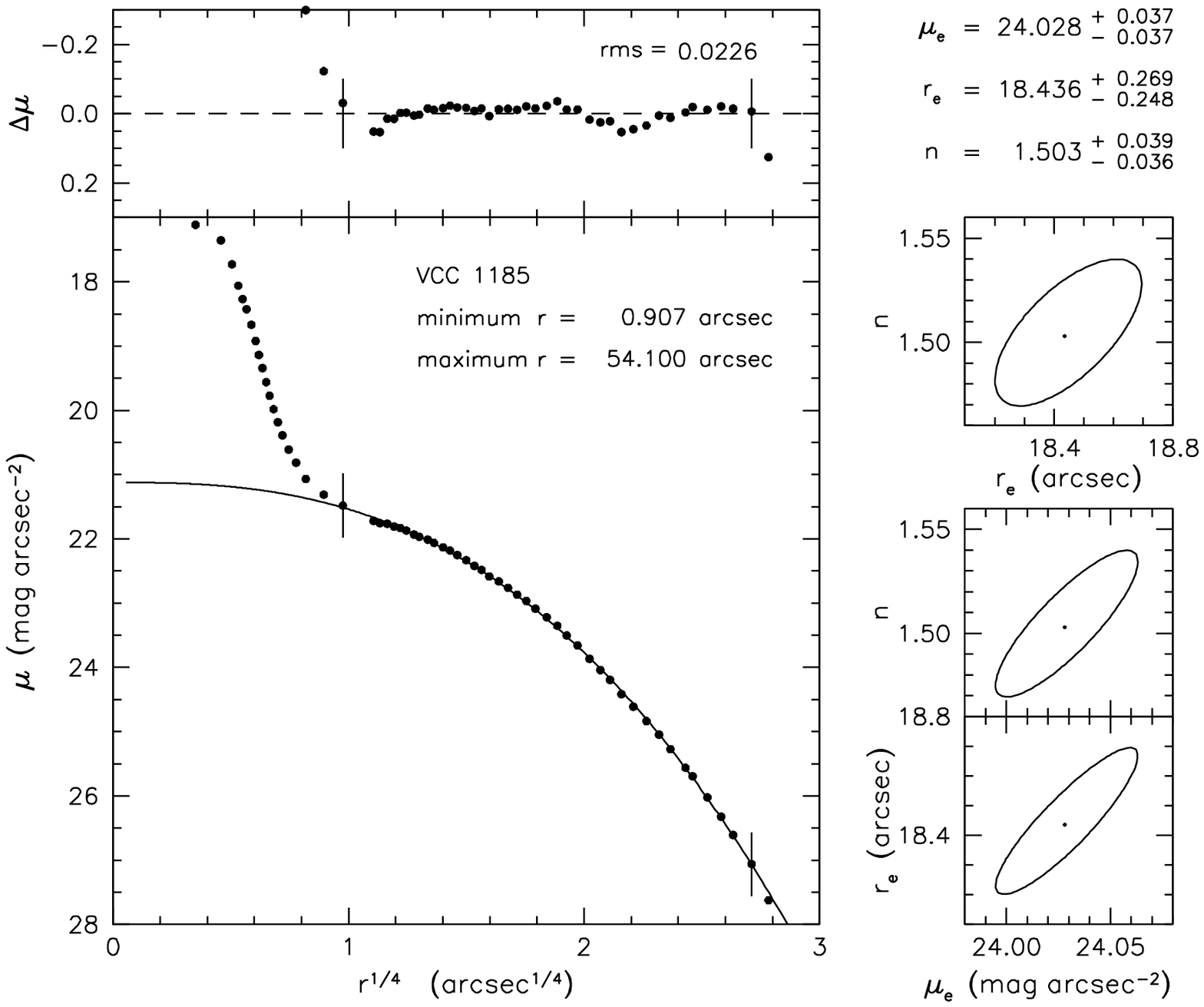}

\includegraphics{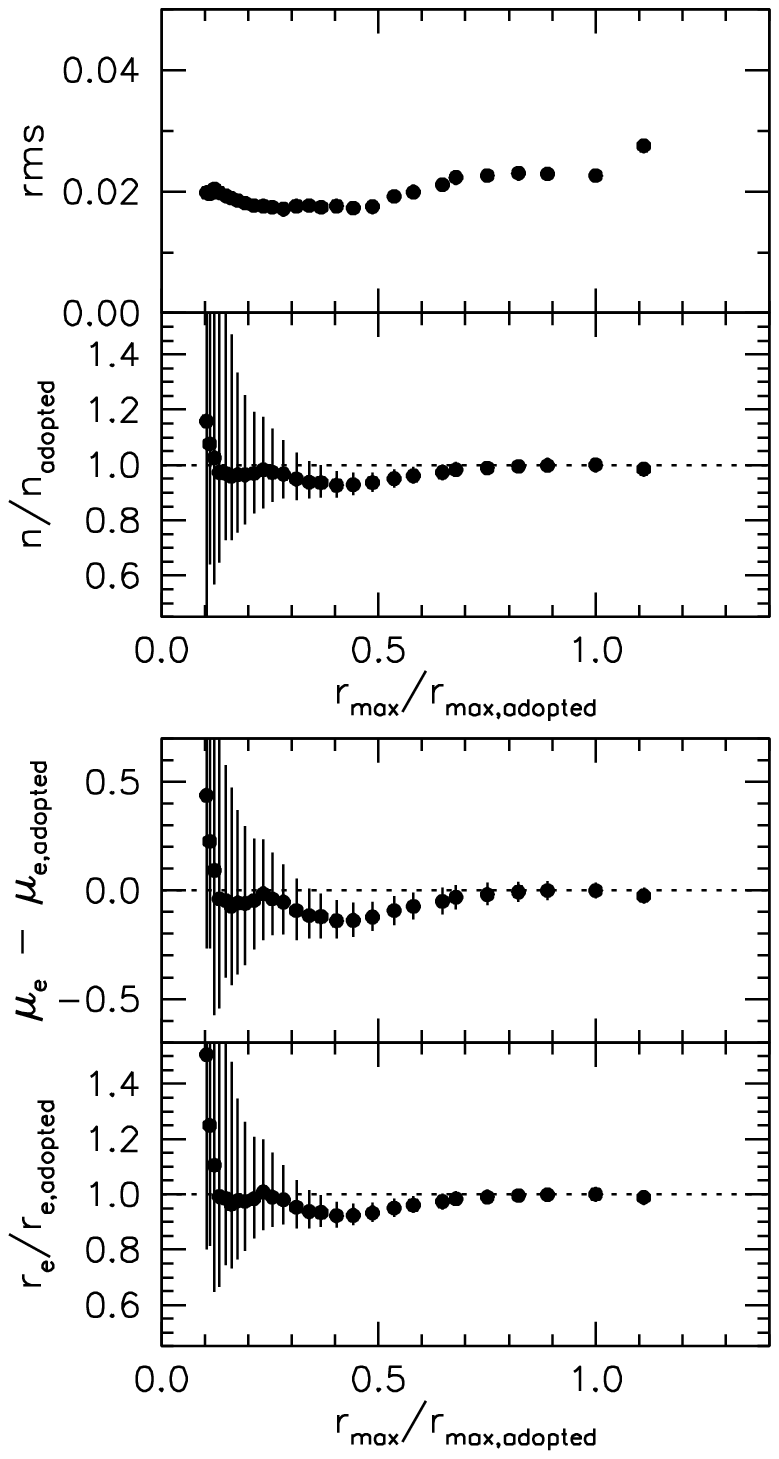}

\includegraphics{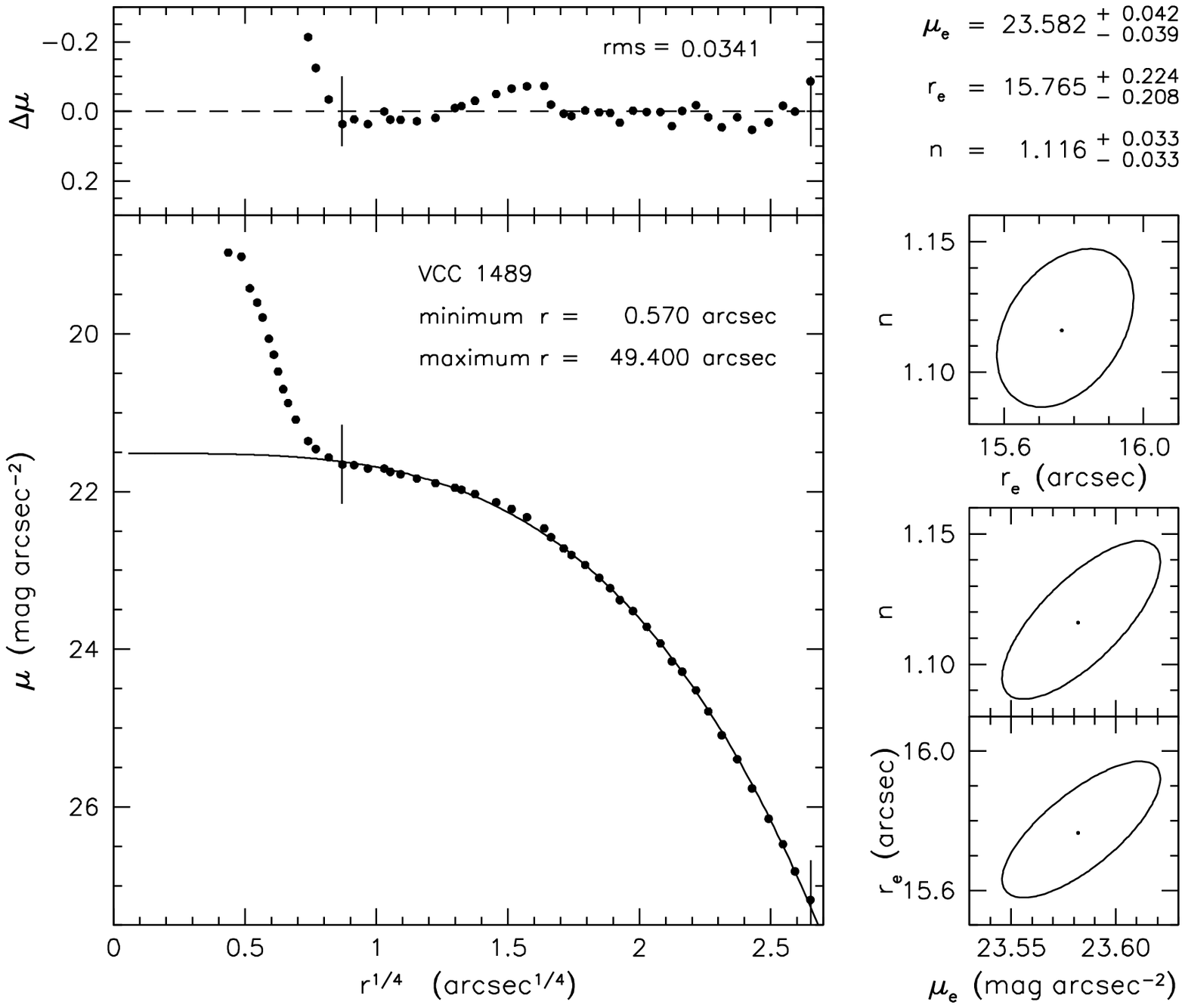}

\includegraphics{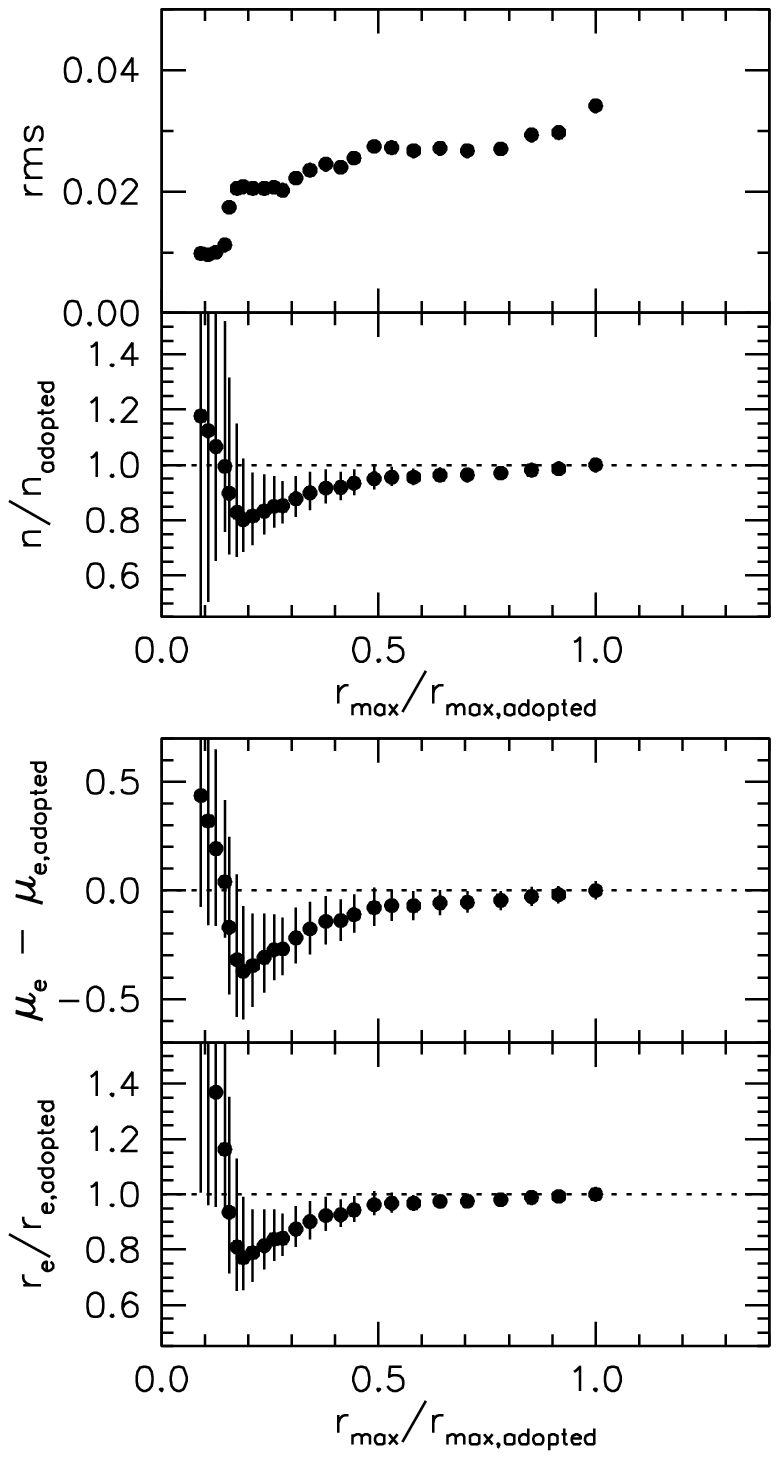}

\figcaption[]
{S\'ersic function fits to the major-axis profiles of the spheroidal galaxies
VCC 1185 and VCC 1489.  The layout is as in Figure 49.
}

\eject\clearpage

\figurenum{73}

\centerline{\null} \vskip 3.55truein

\includegraphics{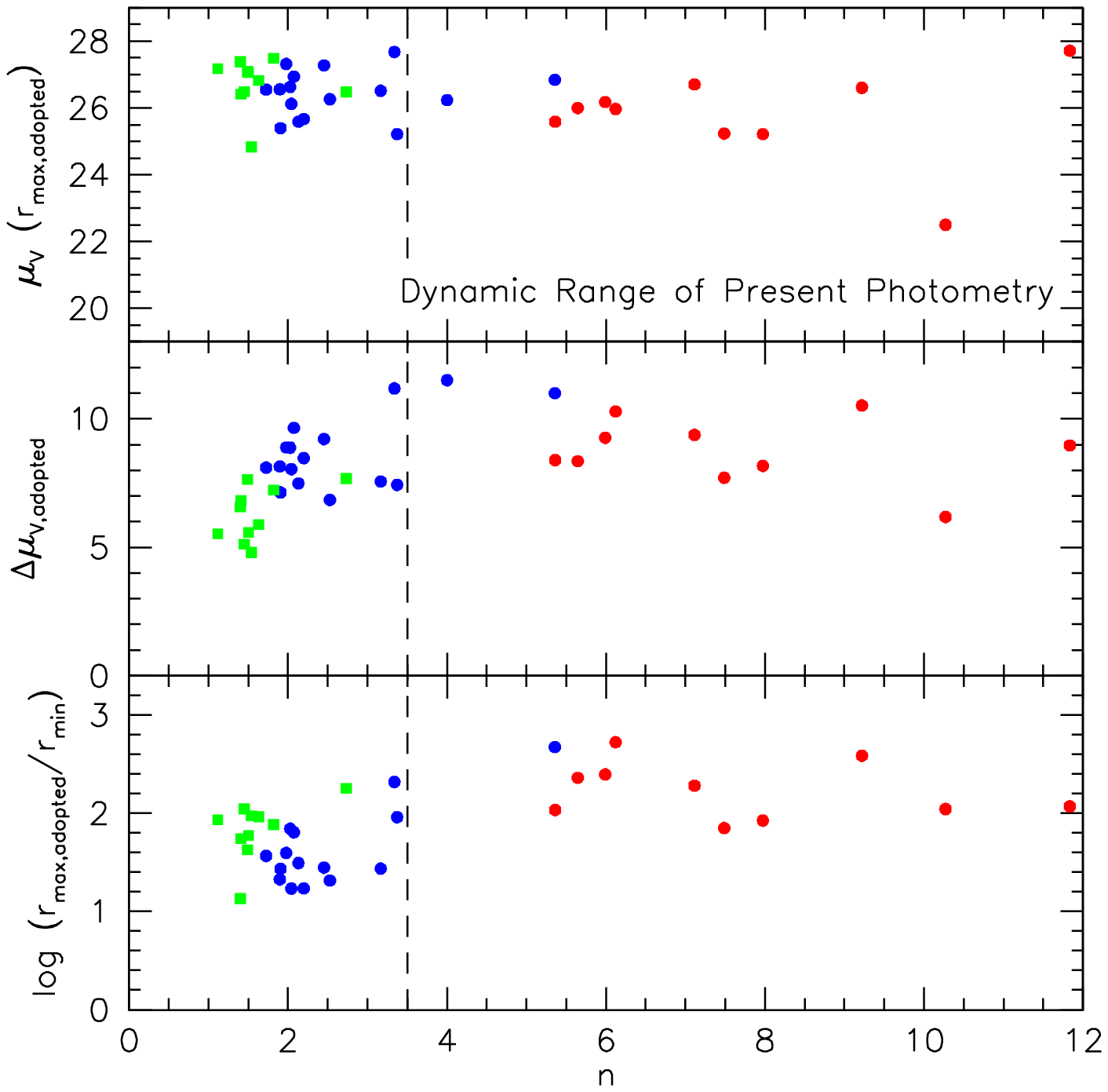}

\includegraphics{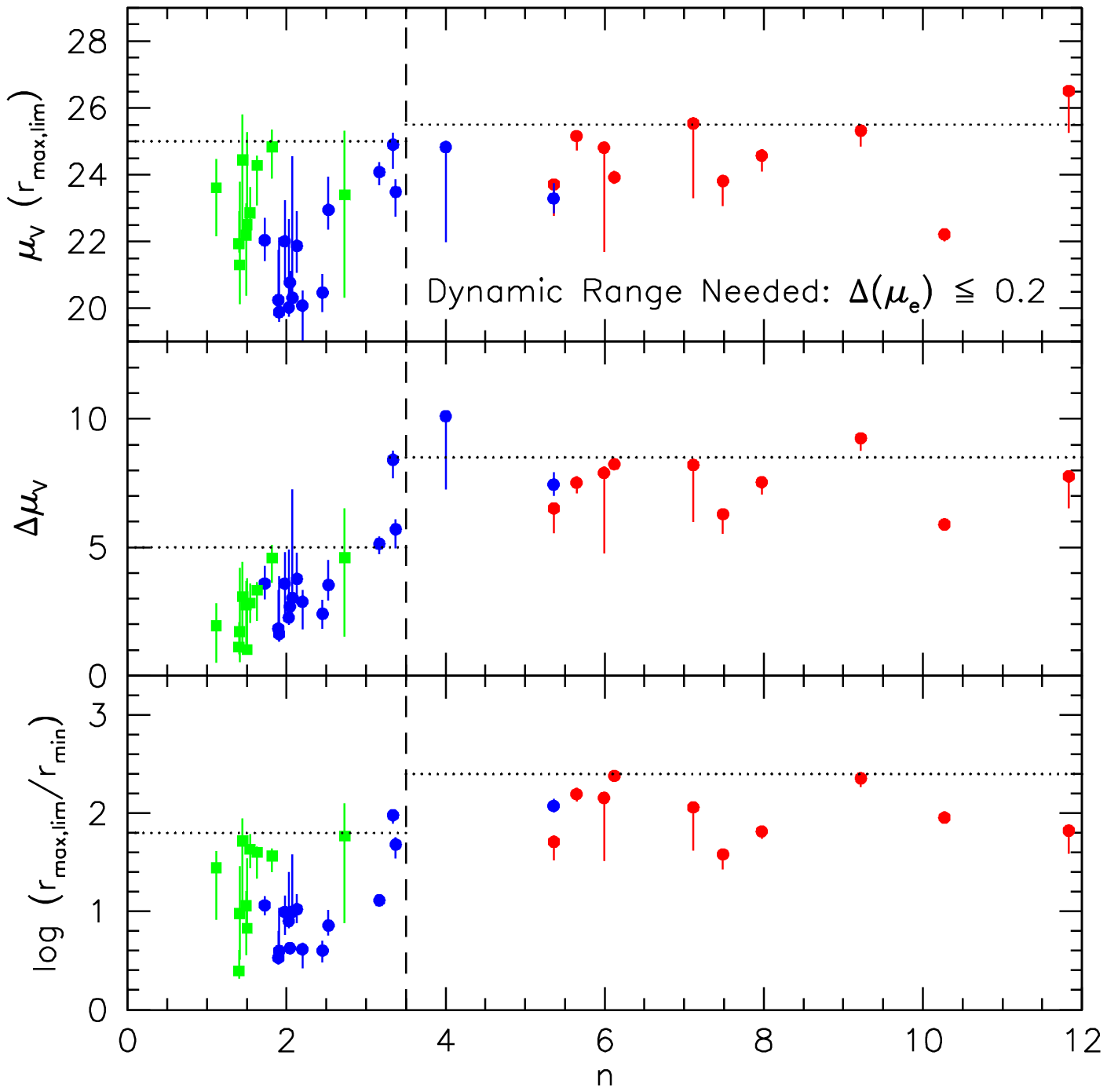}

\figcaption[]
{The left panels illustrate the dynamic range of the profile points used in our adopted
S\'ersic function fits; red points are for core ellipticals,  blue points are for extra light
ellipticals, and green points are for spheroidals.  The right panels illustrate the reduced 
dynamic range that would, with the present, high-quality profile data, give S\'ersic parameters 
that differ from our adopted ones by $\Delta(\mu_e) = 0.2$ mag arcsec$^{-2}$, a factor of 1.12 in 
$r_e$, and a factor of 1.10 in $n$ (see text).  Upward-pointing ``error bars'' end at the minimum 
dynamic range required to give S\'ersic fits that agree with our adopted ones to $1\ \sigma$.  
Downward-pointing ``error bars'' end at the dynamic range required to give parameters that agree 
with our adopted ones to $\Delta(\mu_e) = 0.40$, a factor of 1.24 in $r_e$, and a factor of 1.19 
in $n$.  The top panels show the faint limit of the surface brightness range included in 
our fits ({\it left\/}) or required for $\Delta(\mu_e) \leq 0.2$ mag arcsec$^{-2}$ ({\it right\/}).  
The middle panels show the surface brightness range of the
profile data used in our fits ({\it left\/}) or required for $\Delta(\mu_e) \leq 0.2$ mag 
arcsec$^{-2}$ ({\it right\/}).  The bottom panels show the corresponding ratio of the radius 
of the outermost profile point included in the fit to the radius of innermost profile point 
included in the fit.  The right-hand plots provide conservative criteria by which users of S\'ersic 
functions can judge whether the dynamic range of their data is sufficient for robust fits (see 
text for caveats).  Approximate target dynamic ranges are indicated by horizontal dotted lines 
and depend somewhat on S\'ersic index.  For example ({\it middle-right panel\/}), for giant, core
galaxies, which generally have $n \gtrsim 4$, it is almost always safe to have a surface brightness 
range of 8.5 mag arcsec$^{-2}$ from just outside the core, where the fit becomes acceptable, to large 
radii, where the fit stops being good and/or where sky subtraction becomes a problem.  In contrast, 
S\'ersic fits are much more benign when $n < 3.5$, and progressively smaller surface brightness or
radius ranges are sufficient, always assuming that the profile data are high enough in quality. 
One could choose a target dynamic range $\Delta\mu_V$ that decreases with $n$.  We adopt the 
simpler approach of noting ({\it dotted line\/}) that $\Delta\mu_V \gtrsim 5$ mag arcsec$^{-2}$ 
is essentially always safe.
\lineskip=-2pt \lineskiplimit=-2pt
}

\vskip 10pt

      Figure 73 (left) summarizes the large dynamic range of our observations.  Our S\'ersic fits
generally reach 25{\thinspace}--{\thinspace}27.5 $V$ mag arcsec$^{-2}$.  In many cases, the fit 
range extends to the faint limit of our photometry; in some cases, it ends where sky subtraction 
errors or overlapping objects affect the profiles.  The S\'ersic function almost never fails
dramatically to fit low surface brightnesses.  NGC 4406 is the main exception, but the outer 
profile may be affected by tidal shocking, or our measurements may be contaminated by the bracketing
galaxies.  The ranges of surface brightnesses that we fitted are shown in the middle-left panel, 
and the corresponding radial fit ranges are shown in the bottom panel.  The inner end of each fit
range is chosen to be where ``missing light'' in cores or extra light above the outer S\'ersic fit 
becomes significant.  The core galaxy with the unusually small $\Delta\mu_{V,\rm adopted}$ is 
NGC 4406, as discussed above.  Nevertheless, 
the inner part of the galaxy is an excellent S\'ersic function, and fit uncertainties do not affect 
our interpretation of fundamental plane correlations.  The same is true of NGC 4382: non-equilibrium 
structure diagnostic of a not-yet-relaxed merger remnant create wiggles in the profile that can be 
fitted in various ways (three S\'ersic fits are shown in Figure 53 and 54), but the plausible ones -- 
the ones that fit large radius ranges -- both lie in the derived parameter correlations.  Our efforts
to compile accurate profiles over large radius ranges have paid off in robust parameters that allow
confident interpretation of the parameter correlations.

      As a tool for users of S\'ersic functions, we provide in Figure 73 (right) three summaries of 
the dynamic ranges needed for fits to the present data to give various fiducual parameter errors.  
They depend somewhat on S\'ersic index, which is not known {\it a priori\/}.  However, the 
dependence on $n$ is weak enough so that a sufficiently good value can be derived with a preliminary fit.
Therefore, we plot results as functions of $n$.  There are two regimes.  Fits that have $n \lesssim 3.5$ 
are very robust; a modest dynamic range is sufficient, and limitations on the fit come mostly  from 
data quality and from decisions about the fit range and not from insufficient dynamic range.  On the 
other hand, when $n \gg 4$, the fit is unstable and a generous dynamic range is necessary in order to get 
reliable results.

      Quantitatively, the right panels of Figure 73 were constructed as follows.  From each 
fit range test (Figures 49\ts--\ts72), we determined the maximum fit radius $r_{\rm max,lim}$ at which
the fitted $\mu_e$ differs from the adopted value by (say) 0.2 $V$ mag arcsec$^{-2}$.  Since the fits
tend to preserve the total magnitude $V = \mu_e - 5\ \log{r_e}$ $+$ constant, an error in $\mu_e$ of
$\Delta(\mu_e) = 0.2$ mag arcsec$^{-2}$ should correspond approximately to $\Delta(\log{r_e}) = 0.04$, 
i.{\thinspace}e., a derived $r_{e,\rm lim} = 1.10\ r_{e,\rm adopted}$ or $r_{e,\rm lim} = (1/1.10)\ r_{e,\rm adopted}$
depending on the sign of $\Delta(\mu_e)$.  The fit range tests confirm that the parameters are coupled
in this way: removing the sign of $\Delta(\mu_e)$, the actual mean $<r_{e,\rm lim}/r_{e,\rm adopted}> = 1.119 \pm 0.004$ 
($\sigma/\sqrt{36}$).  The corresponding error in $n$ is $<n_{\rm lim}/n_{\rm adopted}>\ = 1.096 \pm
0.010$ ($\sigma/\sqrt{36}$).  These are the plotted points in the right panels of Figure 73.  They
show the fit ranges required with our data for 20\thinspace\% errors in effective brightness, 
12\thinspace\% errors in effective radius, and 10\thinspace\% errors in S\'ersic index.  The $\chi^2$
ellipses tell us that the errors are coupled so that fainter $\mu_e$ corresponds to larger $r_e$.

\vfill\eject

      The points in the right-hand panels are plotted with ``error bars'' to show the fit ranges required
for two different choices of $\Delta(\mu_e)$.  The ``error bars'' that point toward larger dynamic 
range show the requirements for $\mu_e$ to agree with our adopted values to within our error bars. 
These fits were discussed earlier in this section.  Corresponding error bars do not appear for many core ellipticals,
because our errors in $\mu_e$ are already larger than the fiducial $\Delta(\mu_e) = 0.2$ mag arcsec$^{-2}$
used for the plotted points.  However, for extra light Es and for spheroidal galaxies, the S\'ersic fits are 
very robust, our $\mu_e$ errors are small, and disagreeing with our adopted fits by only one error bar requires 
a larger dynamic range than disagreeing with our adopted fits by $\Delta(\mu_e) = 0.2$ mag arcsec$^{-2}$.  
In Figure 73, the ``error bars'' that point toward smaller dynamic range show the (easier) requirements for
$\Delta(\mu_e) = 0.4$ mag arcsec$^{-2}$.  The corresponding mean 
$<r_{e,\rm lim}/r_{e,\rm adopted}>\ = 1.239 \pm 0.006$ ($\sigma/\sqrt{35}$) and 
$<n_{\rm lim}/n_{\rm adopted}>\ = 1.189 \pm 0.025$ ($\sigma/\sqrt{35}$).  Only 35 galaxies are included 
in the means because the formal errors on the NGC 4382 fits do not reach 0.4 mag arcsec$^{-2}$ 
before we run out of points inside the annulus that was omitted from the fits.  Again, the
parameter coupling approximately preserves the total luminosity of the S\'ersic function fit.

      In the right panels of Figure 73, the horizontal dashed lines provide conservative estimates
of safe dynamic ranges required to achieve the above parameter accuracies.  The requirements depend 
somewhat on S\'ersic index.  For $n \leq 3.5$, dynamic range requirements are not severe, because
small-$n$ S\'ersic fits are relatively stable.  A range of 5 mag arcsec$^{-2}$ in $\mu_V$, corresponding
to a range of a factor of about 60 in the ratio of the largest radius to the smallest radius fitted is
almost always safe.  Given typical amounts of extra light in the present galaxies, the above values
correspond to a limiting surface brightness of 25 $V$ mag arcsec$^{-2}$.  Note that this is the limiting
surface brightness to which the S\'ersic function still fits adequately; the data may reach (and, in some
of our galaxies, does reach) fainter surface brightnesses at which we no longer trust our sky or 
overlapping galaxy subtraction.  In general, the dynamic range requirements for small-$n$ galaxies 
are not difficult to meet.  Large-$n$ galaxies are more of a challenge.  Sometimes a dynamic range of a
factor of 250 in surface brightness is enough, but other fits are less stable, and a surface brightness
range of 8.5 mag arcsec$^{-2}$ is needed to make essentially all galaxies in the present sample have safe 
fits.  This corresponds to a range of a factor of $\sim$ 250 in radius. 

      We emphasize: {\it Dynamic range is only one requirement to get a good S\'ersic fit.  Equally 
important are the accuracy of the profile data and the decisions that are made about which profile points 
to include in the fit and which to omit because the are interpreted as showing missing light or extra light 
at small radii, S0 disks at intermediate radii, or sky subtraction errors at large radii.\/}  The guidelines 
in Figure 73 are relevant only if the data are comparable in quality to those presented here.   Also, they 
are only guidelines; for some of our galaxies, it is clearly sufficient to have less dynamic range than 
the dashed lines suggest.

      It is important to note a final caveat: One of the main conclusions of this paper is that S\'ersic
functions fit the major-axis brightness profiles of Virgo cluster elliptical galaxies remarkably well.  If this 
proves to be less true of ellipticals in a wider variety of environments -- that is, if their profiles turn
out to be more heterogeneous -- then both the validity of S\'ersic fits as analysis machinery and the 
right-hand panels of Figure 73 as guidelines to required dynamic ranges are compromised.

\section*{A3. ROBUSTNESS OF S\'ERSIC FITS: COMPARISON WITH CAON ET AL.~(1993)}

      We illustrate two examples of the robustness (or not) of S\'ersic fits.  Figure 74 compares our 
results with those of Caon \etal (1993).  Appendix B compares our results with those of Ferrarese \etal (2006a).

      As noted in \S\thinspace3, Caon \etal (1993) were the first to establish the importance of 
S\'ersic functions.  They fitted $B$-band profiles of 52 early-type galaxies.  The profiles were composites
derived from deep Schmidt plates and CCD images of the central regions.  They had large dynamic ranges;
only three Caon fits for galaxies that we have in common do not satisfy the dynamic range requirements 
suggested in Appendix A2 (circled points in Figure 74).  The comparison of their major-axis $n$ values with ours
shows excellent agreement for almost all galaxies.  The differences in $n$ values are very large for three
galaxies and moderately large for three more.  For two of these, Caon \etal (1993) had less dynamic range 
than we found to be adequate.  The rest can readily be understood:

\vfill

\includegraphics{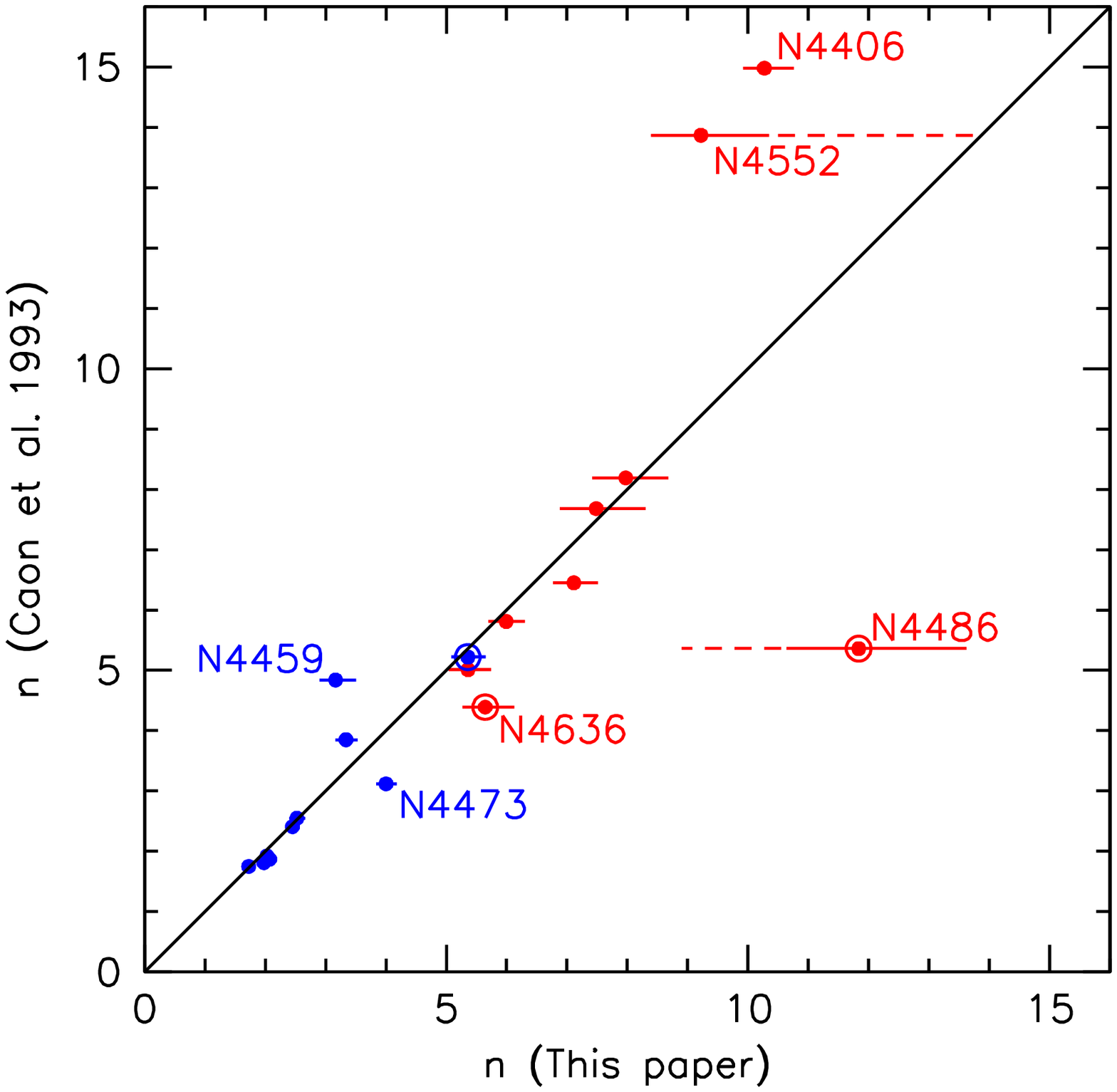}

\figurenum{74}
\figcaption[]
{Comparison of our S\'ersic $n$ indices (Table 1) with those derived by Caon \etal (1993)
for all 9 core ellipticals ({\it red circles\/}) and all 10 extra light ellipticals 
({\it blue circles\/}) that we have in common.  We have no spheroidals in common.  
Dashed lines point to our alternative fits as discussed in the text.  Circled points
indicate that the Caon \etal (1993) fits had less dynamic range than we found to be adequate 
for these galaxies from our fit range tests.}

\eject

      We have noted that NGC4486 (M{\thinspace}87) is a weak cD.  As the fit range is increased, more
cD halo gets included, and $n$ looks larger.  Our adopted fit uses a brightness range of 9.0 $V$ mag 
arcsec$^{-2}$ and gives $n = 11.8^{+1.8}_{-1.2}$.  An alternative fit in Figure 50 includes less cD halo:
the fit range is 6.0 $V$ mag arcsec$^{-2}$ and $n = 8.9^{+1.9}_{-1.3}$.  Caon \etal (1993) had a fit
range of 5.5 $B$ mag arcsec$^{-2}$ shifted away from the cD halo to higher surface brightnesses than 
those that we fit.  Not surprisingly, they got a smaller S\'ersic index, $n = 5.36$.
We also would get a smaller S\'ersic index if we reduced our fit range further.

      NGC 4406 has a profile that is very accurately S\'ersic out to $r = 153^{\prime\prime}$, the outer
end of our fit range.  Beyond this, the profile that we measure turns up suddenly.  If we included
the upturn in our fit, we would get a larger $n$.  Caon \etal (1993) did this: 
they fitted the profile out to 1 mag arcsec$^{-2}$ fainter than we did.  Our composite profile is based 
on two different data sets that agree on the above deviations.  Including the profile upturn in the
S\'ersic fit results in residuals that are not consistent with the accuracy of our profile.

      For NGC 4552, the difference between Caon's fit and ours is a matter of interpretation. 
We cannot prove that one fit is better than the other.  But we can understand the difference. 
The residual plots in Figures 15 and 56 show that, for our chosen {\it inner\/} end of the fit 
range at $r = 1\farcs28$, the residuals look systematically concave-up from $r^{1/4} = 1.6$ ($r =
6\farcs5$) outward.  The residuals are systematic (all data sets in Figure 15 agree) and they are
larger than average.  But they are not outside the range of what is reasonable.   We chose 
$r = 1\farcs28$ as the inner end of our fit range because we wanted to fit as much of the galaxy
light as possible.  However, it could reasonably be argued that we should have chosen a larger
minimum radius.  If we choose $r = 5\farcs5$ (bottom fit in Fig.~56), then the residuals no longer 
look systematic, the
total RMS is reduced from 0.0774 to 0.0474 mag arcsec$^{-2}$, and $n = 13.75^{+3.04}_{-1.90}$.  
This value is at the end of the dashed line from the NGC 4552 point in Figure 74.  It agrees exactly 
with Caon's value.  This is, in fact, exactly how they got their value: their $B$-band fit range 
corresponds to about 17 -- 25.5 $V$ mag arcsec$^{-2}$ in Figures 15 and 56, i.{\thinspace}e., essentially
our modified fit range.  No conclusions in this paper would significantly be changed if we adopted
the modified fit range.  The fundamental plane correlations would have slighly larger scatter, but
the distinction between E and Sph galaxies would look stronger.  The derived amount of missing 
light in the core would be substantially larger, suggestive of rather more than $\sim 3$ dry 
mergers.  

      NGC 4459 is deviant in Figure 74 because Caon \etal (1993) fitted parts of the inner profile
that we, with our more accurate photometry, can confidently recognize as extra light.  That is, the
outer profile that we derive robustly has $n < 4$.  Including extra light as Caon did would 
increase $n$ to be greater than 4 as Caon found.  

      NGC 4473 is tricky because of the embedded counterrotating disk.  Our $n$ is essentially 
fixed by our choice to include a few central points in the fit.  We did this for reasons of 
stability: otherwise small wiggles in the outer profile render the fit unstable 
because then the fit range is too small.  Given the precise fit range chosen by Caon \etal (1993),
the slightly smaller $n$ that they derive is understandable.  Their value is plausible; we
noted earlier that our value of $n$ is an upper limit.

      These few differences have taken a disproportionately large number of words to explain.
In fact, the agreement between Caon's results and ours is excellent.  Note that differences
are not usually the result of dynamic range problems.  Most differences result from different
choices of which profile points to fit, consistent with the discussion in the previous section.

      We used the Caon profiles for some of our galaxies, usually when we had problems with other 
data that we wanted to check.  We did not systematically check all Caon data against our own.  We
were initially reluctant to use their data, partly because the $B$ bandpass is bluer than most
others used in this paper and partly because the outer profiles in Caon \etal (1993) are based
on photographic plates.  In retrospect, Figure 74 shows that we were too conservative:
color gradients are less important than sky subtraction uncertainties at large radii, and the
quality of the Caon \etal (1993) photometry is generally very good.

\section*{Appendix B}

\section*{Comparison With Ferrarese et al.~(2006a)}

      Ferrarese \etal (2006a) present photometry of 100 early-type galaxies in the Virgo cluster 
obtained with the HST as part of the ACS Virgo Cluster Survey (C\^ot\'e \etal 2004).  Their 
data reduction and ours generally agree to the extent that we can check them; e.{\ts}g., their 
$g - z$ colors and ours agree well (\S\ts6.3, Equation 4).  Their paper and ours also agree on some 
results.  E.{\ts}g., in some galaxies, they find central light excesses, although they call them 
``nuclei''.  Most significantly, Ferrarese \etal (2006a) disagree with both dichotomies that are 
the focus of this paper.  Since these dichotomies are our most important results, we concentrate on them.

\section*{B1. The~~E{\ts}--{\ts}Sph~~Dichotomy}

      Ferrarese \etal (2006a, astro-ph/0602297 version) argue against the E{\ts}--{\ts}Sph 
dichotomy: ``Once core galaxies are removed, dwarf and bright ellipticals 
display a continuum in their morphological parameters, contradicting some 
previous beliefs that the two belong to structurally distinct classes.''
Thus they echo papers reviewed in \S\ts2.1.  They consider this to be a solved problem: ``the structural
dichotomy between dwarf and regular ellipticals as advocated by Kormendy
(1985b) was likely the result of observational biases.''    

      We disagree.   Figures 34\ts--\ts38 provide strong confirmation of the E{\ts}--{\ts}Sph 
dichotomy, and Figure 41 illustrates it also.  Kormendy (1985b, 1987b) had few galaxies in the
magnitude range $M_V \sim -16$ to $-17$ (with the present distance scale) where the E and Sph sequences 
overlap, but the sequences were far apart and diverging from each other where they approached 
this magnitude range.  The problem was not sample bias but rather (i) the luminosity functions 
(faint Es and bright Sphs are rare; Sandage \etal 1985) and (ii) spatial resolution (except for 
M{\ts}32, tiny ellipticals were so poorly resolved with ground-based photometry that they could 
not be plotted in the parameter correlation diagrams).  With HST, we can observe M{\ts}32 analogs 
in the Virgo cluster well enough to solve both problems.  Figures 34{\ts}--{\ts}38 have many 
galaxies in the E{\ts}--{\ts}Sph overlap region.

     Moreover, far from being biased in favor of finding the dichotomy, our present sample is 
biased in favor of spheroidals that are similar to small ellipticals.  This was deliberate: 
we targeted galaxies near the E{\ts}--{\ts}Sph transition because we wanted to know whether 
there are intermediate galaxies.  Figures 34{\ts}--{\ts}38 show that we succeeded in mapping
out the transition region: our Sph galaxies ({\it green squares\/}) approach closer to the E 
sequence than do the larger samples of Ferrarese \etal (2006a: {\it green triangles\/}) and 
Gavazzi \etal (2005, {\it green crosses\/}).  Yet the E and Sph sequences remain distinct.

\vfill\eject

     Why did Ferrarese \etal (2006a) not find this result? 
There are three main reasons: (1) Our parameters measurements are more accurate, 
because composite profiles give us larger radius ranges over which to 
fit S\'ersic functions while minimizing systematic errors at low surface 
brightnesses. (2) Ferrarese includes S0 galaxies without doing bulge-disk 
decomposition.  We show 5 large-bulge S0s in Fig.~37, but in general, we omit S0s, 
because we have too little leverage on the bulge parameters.  
Including S0s without doing bulge-disk decomposition is certain to increase
the scatter in the correlations.  This makes it hard to distinguish the
E and Sph sequences where they approach each other. (3) Ferrarese \etal (2006a) 
observed spheroidal galaxies over only a 2 mag range in absolute magnitude, so they had too 
little luminosity leverage to see the {\it sequence\/} of spheroidals in parameter space.
 
     Figure 75 compares S\'ersic parameters derived by Ferrarese \etal (2006a) 
with our measurements.  In many cases, the parameters agree well.  This is
particularly true of Sphs; they are small and have small $n$, so they are 
are well observed with the small ACS field of view.   However, for some 
galaxies, Ferrarese's parameters disagree with ours by much more than our estimated errors.  

\centerline{\null}

\figurenum{75}

\vfill

\includegraphics{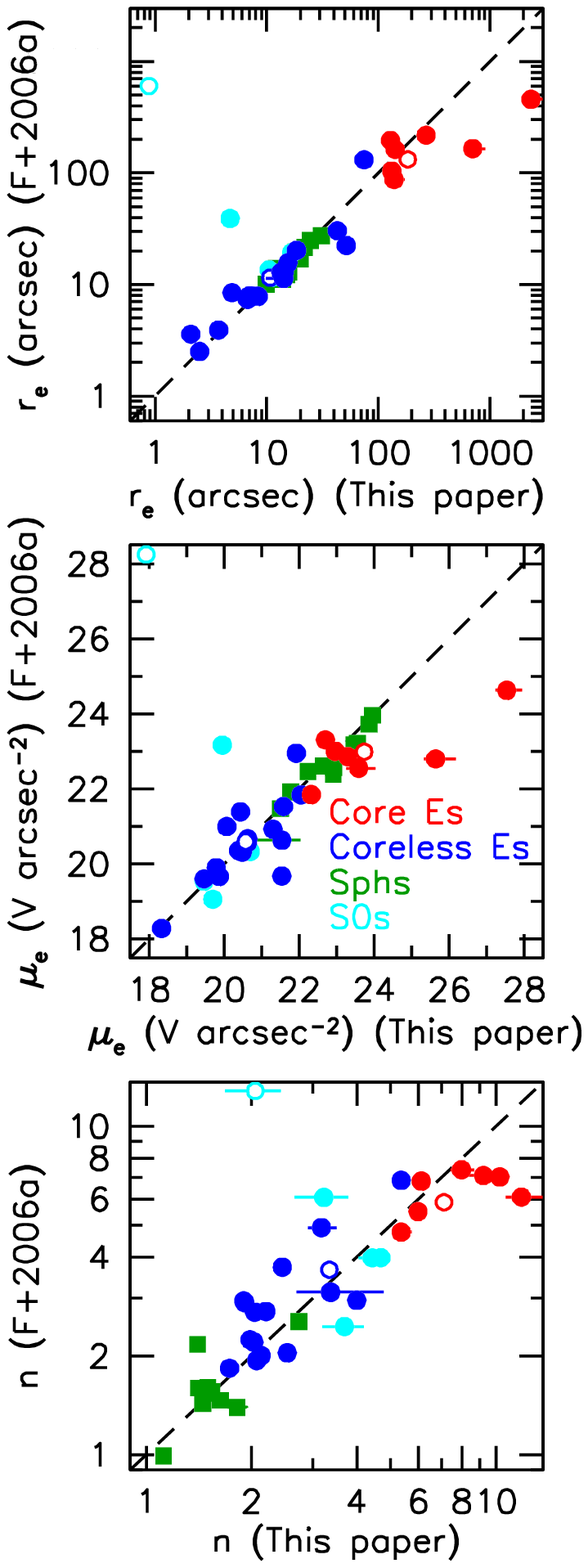}

\figcaption[]
{Comparison of S\'ersic parameters fitted by Ferrarese \etal (2006a) with our
Table 1 values.  The Ferrarese values of $r_e$ are converted from mean axis to major axis 
for consistency with our parameters.  Also, $g$-band $\mu_e$ values are converted to
$V$ band using Equation (3) and $g - z$ values from Ferrarese's Table 4.
The symbols are as in Figures 34 and 37\ts--\ts38.  All of our parameters include error
bars except $r_e$ and $\mu_e$ for bulges.  Most error bars are too small to be visible.}

\eject

\centerline{\null} 

\figurenum{76}

\vskip 4.92truein

\includegraphics{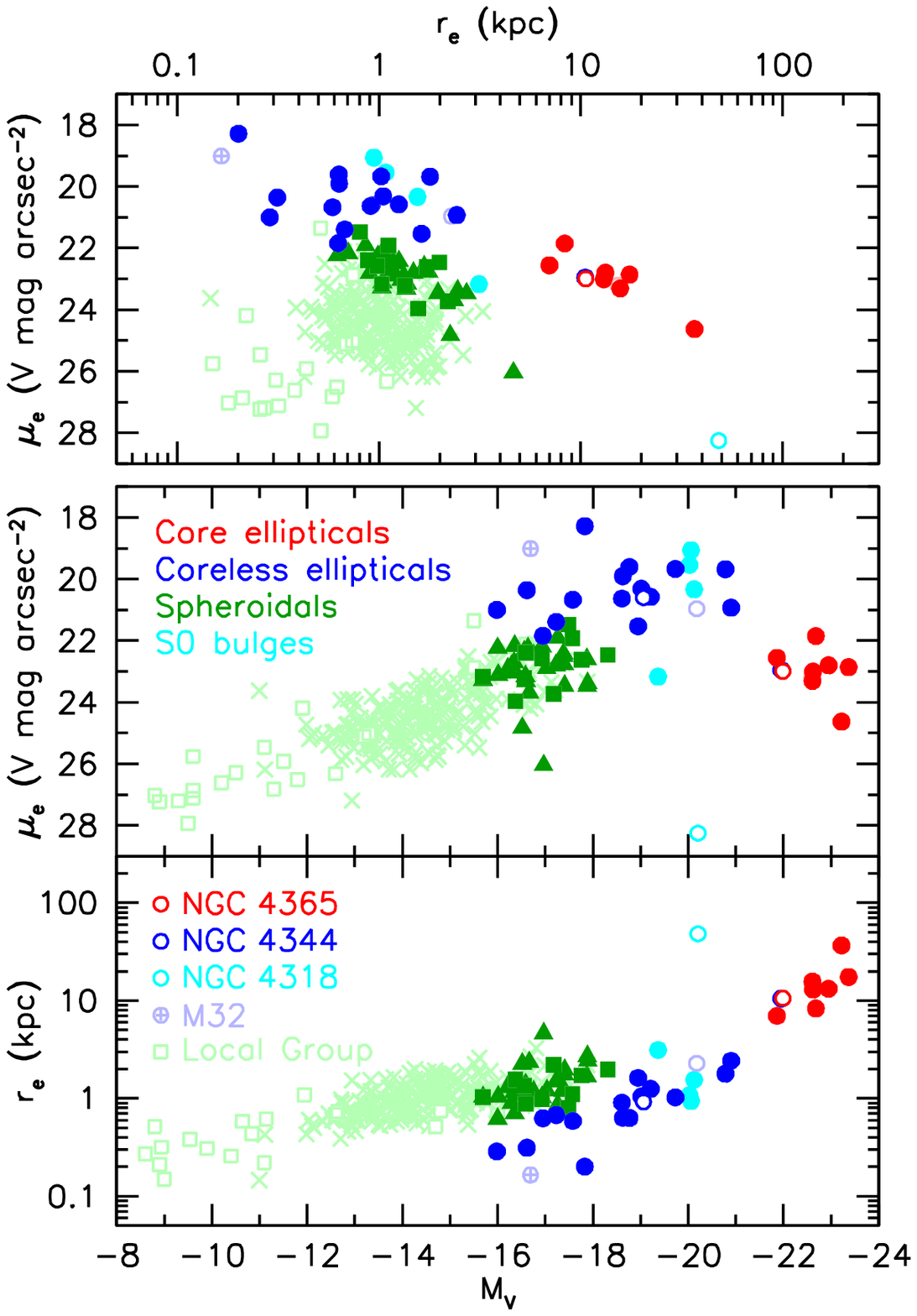}

\includegraphics{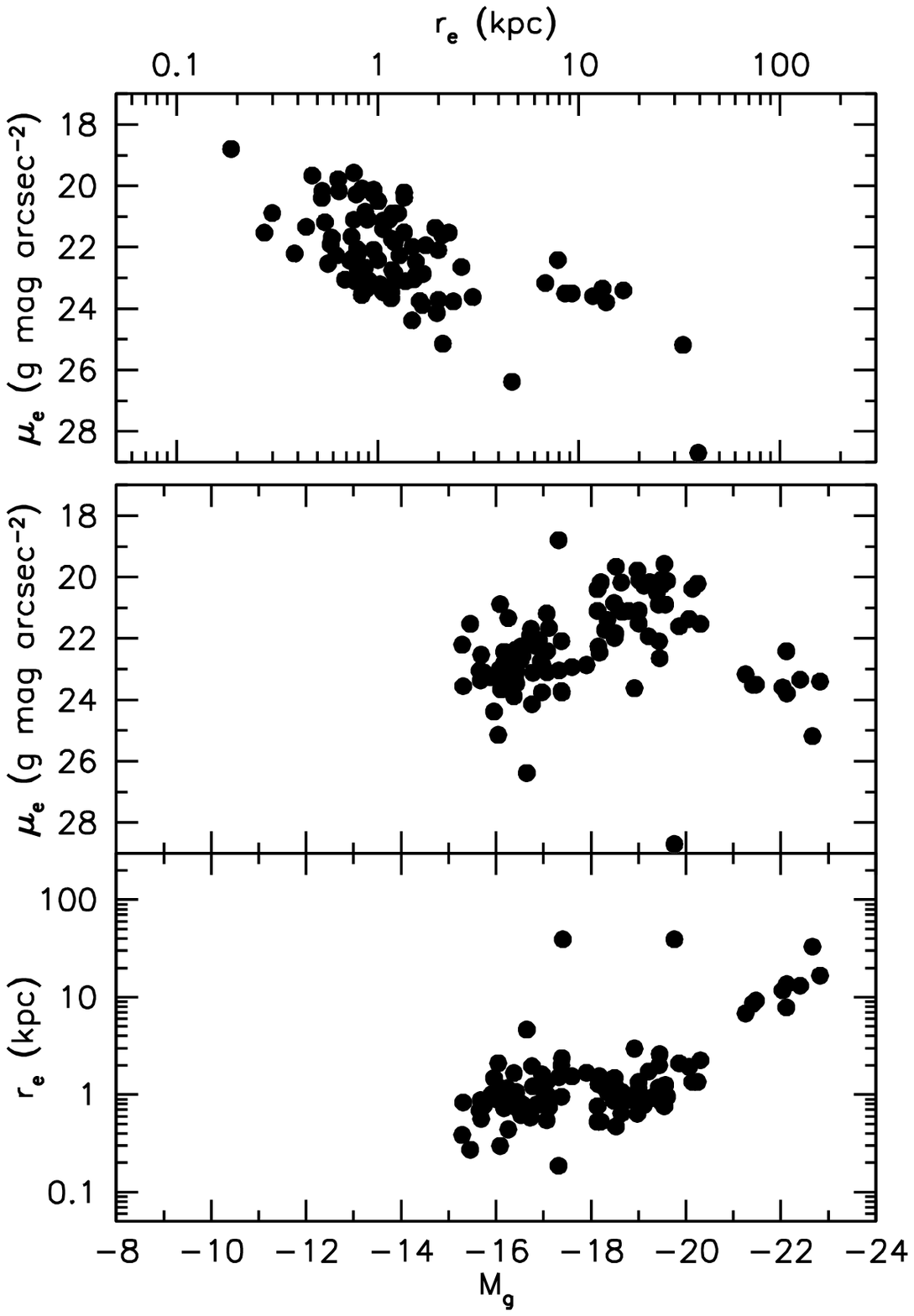}

\figcaption[]
{Global parameter correlations for elliptical and spheroidal galaxies using 
the galaxy sample, classifications, and symbols of our Figure~34 but with all 
parameters as measured by Ferrarese~et~al.~(2006a).  This figure can directly 
be compared with Figure 37.  The two main differences between our analysis and 
that of Ferrarese \etal (2006a) are the treatment of the galaxy sample and the 
accuracy of the parameter measurements.  This figure mainly tests the parameter
measurements, while Figure 77 also tests the effects of sample differences.  
Here, faint symbols show the parameters of galaxies that are not in 
Ferrarese's sample but that are in our sample or in that of Gavazzi \etal (2005) 
or that are in the Local Group.  For consistency with these 
galaxies, Ferrarese's $r_e$ values (their Table 3) have been converted from mean 
axis to major axis by dividing by (1 - $<$\null$\epsilon$\null$>$)$^{1/2}$, where 
$<$\null$\epsilon$\null$>$ values are mean ellipticities from their Table 4.  
The correction is approximate, because S\'ersic $n$ is not the same along the mean
and major axes.  This is insignificant except when $n \gg 4$ and
has no effect on our conclusions.
\lineskip=-2pt \lineskiplimit=-2pt
}

\figurenum{77}
\figcaption[]
{Global parameter correlations using the galaxy sample and parameter 
measurements of Ferrarese \etal (2006a).  Different galaxy types are not 
distinguished.  These are (from top to bottom) panels df, af, and ac of
Figure 116 in Ferrarese \etal (2006a) with our figure orientations and
parameter limits to allow a direct comparison with Figures 37 and 76.
\lineskip=-2pt \lineskiplimit=-2pt
}

\vskip 15pt

      In Figure 75, the very discrepant turquoise point is for the S0 galaxy NGC 4318. 
Ferrarese's $n = 12.8$ fit includes the bulge and the inner part of the disk shown in 
Figure 32.  However, outside the bulge, the disk is a well defined exponential 
($n = 1.11 \pm 0.11$).  The other large discrepancy for an S0 galaxy is NGC 4489.  
But the small number of large discrepancies is not the main problem.

      Figures 76 and 77 test how well Ferrarese \etal (2006a) could see the E{\ts}--{\ts}Sph 
dichotomy with their parameter measurements.  Figure 76 shows Ferrarese's parameters
but our galaxy classifications.  A comparison with Figure 37 tests the effect of differences 
between their parameters and ours for the same sample of galaxies.  One problem is
immediately apparent.  Ferrarese \etal (2006a) get $\mu_e$ values that are 1 mag arcsec$^{-2}$ 
fainter than we do for three extra light ellipticals (Figure 75).  Of these, NGC 4467 and
VCC 1199, are M{\ts}32-like, faint Es that are especially important in Figures 37 and 38.
Their small $r_e$ and (in our data) bright~$\mu_e$ help to define the extension 
of the E sequence toward more compact galaxies, left of where the Sph sequence approaches the 
ellipticals in Figure 37.  Our profiles are based on four data sets each from three different 
telescopes; they agree well (Figures 23 and 24), and they suppoprt robust S\'ersic fits with 
RMS dispersions = 0.02 mag arcsec$^{-2}$ (Figures 66 and 67).  With Ferrarese's 
parameter values, these points lie close to the Sph galaxies in Figure 76, and the 
extension of the E sequence to the left of the Sph sequence is less obvious.

  Also, Ferrarese \etal (2006a) observed Sphs over only the brightest 2 mag of their 
luminosity function.  Without the luminosity leverage provided by the fainter Sphs used 
in Figures 34, 37, and 38 and shown in Figure 76 by the ghostly points, one is not driven 
to conclude that there are separate, nearly perpendicular {\it linear sequences\/} of 
E and Sph galaxies in parameter space.  So the luminosity bias in the Ferrarese sample 
contributes to their inability to distinguish the two types of galaxies.  Nevertheless, 
guided by the ghostly points, it is possible to see the main features of Figure 37 in 
Figure 76.  The ellipticals (blue and red points) define a fundamental plane, and the 
Sph galaxies approach the fundamental plane projection in the top panel near its middle,
not near its end.  A few faint ellipticals would be misclassified using Figure 76, but
Ferrarese \etal (2006a) could have found the distinction between nearly perpendicular 
E and Sph sequences using their parameters.

\centerline{\null}

      (The same is true for Gavazzi \etal 2005.  They argue against the E{\ts}--{\ts}Sph 
dichotomy, but it is apparent in their Figure 10.  Core ellipticals [their {\it dotted 
parallelograms\/}] and faint ellipticals including M{\ts}32 define continuous linear sequences 
in parameter space that are clearly distinct from the sequence of spheroidals 
[mostly {\it open circles\/}].)

      Figure 77 tests the importance of omitting S0 galaxies in Figure 76.
It includes all galaxies in Ferrarese \etal (2006a), using $r_e$ and $\mu_e$ from
their Table 3 and total $g$-band magnitude from Table 4.  Unlike Figure 76, it does 
not use mean ellipticity to estimate major-axis parameters; Figure 77 shows parameters
for the ``mean axis'' at 45$^\circ$ to the major axis.  That is, Figure 77 shows 
(from top to bottom) panels df, af, and ad from Figure 116 of Ferrarese \etal (2006a). 
Comparison of Figure 77 with Figure 76 shows that the inclusion of S0 galaxies further 
increases the scatter in the E fundamental plane.  Given this, and without guidance
from the fainter spheroidals shown as ghostly points in Figure 76, it is easy to 
understand why Ferrarese \etal (2006a) concluded that E and Sph galaxies are 
continuous in parameter space.  Still, it is interesting to note that there are two 
partly distinct clouds of points -- in addition to the core ellipticals -- in the 
middle and bottom panels of Figure 77.

      In summary, we believe that there are three main reasons why Ferrarese \etal 
(2006a) missed the distinction between elliptical and spheroidal galaxies.  
(1) Their parameter measurements are somewhat less accurate than ours, increasing 
the scatter in the E fundamental plane, especially at low luminosities.  Use of a 
single distance to all galaxies contributes marginally to this effect.  (2) Inclusion 
of S0 galaxies increases the scatter in fundamental plane parameter correlations 
for two reasons, first because bulge-disk decomposition was not carried out to measure 
bulge parameters, and second because -- even with decomposition -- there is much less
leverage on bulge parameters than on those of elliptical galaxies.
(3) Since Ferrarese \etal (2006a) observed Sph galaxies over only a limited luminosity 
range and did not include published parameters of tinier galaxies, they had too 
little luminosity leverage to find the {\it sequence\/} of spheroidals in parameter space.
In addition, they did not plot parameters at the 10\ts\%-of-total-light radius, so they 
did not see the much larger separation of the sequences in our Figure 34.

\section*{\it B2. The~~E{\ts}--{\ts}E~~Dichotomy}

      Ferrarese \etal (2006a) also argue against the dichotomy of elliptical 
galaxies into ``core'' and ``power law'' types.  Their most compact
statement is in the astro-ph/0602297 version: ``The widely adopted 
separation of early-type galaxies between `core' and `power law' types 
\dots~prompted by the claim of a clearly bimodal distribution of [inner profile 
slope] values is untenable based on the present study''.  They then rediscover
the dichotomy based on breaks in the surface brightness profiles from steep
S\'ersic functions at larger radii to shallow power laws at small radii:
``In agreement with previous claims, the inner profiles \dots~of eight of
the 10 brightest galaxies, to which we will refer as `core' galaxies, are
lower than expected based on an extrapolation of the outer S\'ersic model,
and are better described by a single power law model.  Core galaxies are
clearly distinct in having fainter central surface brightness \dots~and 
shallower logarithmic slope of the inner surface brightness profile \dots~than
expected based on the extrapolation of the trend followed by the rest of the
sample.  Large-scale, global properties also set core galaxies apart
\dots~.''

      However, cores have long been defined by many authors based on a central break in profile
shape.  As quoted in \S\ts9.2, the Abstract of Kormendy (1999) begins, ``Elliptical galaxies are
divided into two types: galaxies with steep profiles that show no breaks in slope or that have 
extra light at small radii compared to a S\'ersic function fit and galaxies that show a break 
from steep outer profiles to shallow inner profiles.''  We use the same definition.  The
fact that ``large-scale, global properties also set core galaxies apart'' has always been central
to descriptions of the E{\ts}--{\ts}E dichotomy (see the papers listed in \S\ts2.2).

      The Nuker team also defined cores using the profile break:
``At the `break radius' $r_b$ (formerly called the core radius $r_c$),
the steep outer surface brightness profile turns down into a shallow inner
power law'' $I(r) \propto r^{-\gamma}$ (Kormendy \etal 1994) whose slope
is observed to be $\gamma \simeq 0.1 \pm 0.1$.  Lauer \etal (1995)
included the profile slope in the definition, ``We now define a {\it core\/}
to be the region interior to a sharp turndown or break in the steep outer 
brightness profile, provided that the profile interior to the break has 
$\gamma < 0.3$.''  Including or not including a range of $\gamma$ 
values in the definition has, it turns out, only minor effects on 
ones's conclusions.  Our definition based only on the profile break and 
the Lauer's definition that includes $\gamma$ agree on most galaxies (\S\ts9.2).

      And the distribution of central properties robustly shows a dichotomy, even though a few 
intermediate cases are found (Gebhardt \etal 1996; Lauer \etal 2007b; this paper).

      Ferrarese \etal (2006a) are confused by the Lauer \etal (1995) definition 
in part because they treat Sph galaxies as ellipticals.  They state, ``Although the
 brightest [ellipticals] have shallow inner profiles, the shallowest profiles are found in 
faint dwarf systems.''  We discuss this point in \S\ts9.2.  We agree that 
low-luminosity Sph galaxies have S\'ersic $n \simeq 1$, which means that their central 
brightness profiles -- outside any nuclei -- satisfy the $\gamma$ part of Lauer's 
definition.  But most do not show a downward break from the outer S\'ersic profile, 
so {\it Sph galaxies do not satisfy our definition of a core.\/}  
Instead, these galaxies have almost-exponential profiles at all radii, highlighting again 
(see \S\ts2.1) their structural similarity to late-type galaxies.  Section 8 confirms 
that Sph galaxies are not ellipticals.  They should not cause difficulty in the definition 
of cores in ellipticals.

     We emphasize another aspect of the E{\ts}--{\ts}E dichotomy which shows that it 
has physical meaning.  The existence or otherwise of the dichotomy is not just about 
profile analysis.  The distinction between core galaxies and extra light galaxies is 
also a distinction between many global physical properties, including isophote shape, 
the importance of rotation, hence also velocity distributions, and overall flattening.
The discoveries of many of these correlations were based on a successful application of the
Nuker definition of cores (Faber \etal 1997).  Ferrarese \etal (2006a) ignore these successes.
We find additional physical properties that are part of the E{\ts}--{\ts}E dichotomy,
including stellar population ages and $\alpha$ element enhancements (\S\ts11.1). 

      Finally, we note that, when Ferrarese \etal (2006a) detect extra light, they consider it
to be equivalent to nuclei.  They do not mention that Kormendy (1999) already detected extra light 
and interpreted it as the central, distinct stellar component predicted by the Mihos \& Hernquist 
(1994) merger simulations.  Since the submission of this paper, C\^ot\'e \etal (2007) have begun 
to refer to ``extra light'' in low-luminosity Es and to interpret it in the context of the Mihos 
\& Hernquist models.

\centerline{\null}

\end{document}